\documentclass[a4paper,12pt,oneside]{book}
\usepackage[english]{babel}
\usepackage[titles]{tocloft}
\usepackage{graphicx,color}
\usepackage{ae}
\usepackage{amsmath,amssymb,amsfonts,bm,euscript,amscd}
\usepackage{amsthm}
\usepackage{tikz}
\usetikzlibrary{matrix,arrows}
\usepackage{mathtext}
\usepackage{exscale}
\usepackage{latexsym}
\usepackage{slashed}
\usepackage{array,calc,tabularx}
\usepackage[all]{xy}
\usepackage{epsfig}
\usepackage{makeidx}
\usepackage[T1]{fontenc}
\usepackage{mathptmx}
\usepackage{newcent}
\usepackage[cp1251]{inputenc}
\usepackage[active]{srcltx}
\usepackage[pdftex,bookmarks,colorlinks]{hyperref}
\usepackage{ifthen}
\usepackage{pdfpages}

\newtheorem*{hypothesis}{Hypothesis}
\newtheorem*{conclusion}{Conclusion}
\newtheorem*{proposition}{Proposition}
\newcommand{\tr}{\mathrm{Tr}}
\newcommand{\sgn}{\mathrm{sgn}}

\newcommand{\artanh}{\mathrm{artanh}}
\newcommand{\cor}{\mathrm{Cor}}
\newcommand{\rv}{\left|\textrm{0}\right\rangle}
\newcommand{\lv}{\left\langle\textrm{0}\right|}
\newcommand{\vv}{\left\langle{\textrm{0}}|{\textrm{0}}\right\rangle}

\usepackage{fancyhdr}
\fancypagestyle{plain}{%
\fancyhead{}

}

\hypersetup{colorlinks,%
linkcolor=blue,%
urlcolor=blue,%
citecolor=red,%
filecolor=magenta}

\title{
\Huge{{\AE}THEREAL MULTIVERSE}\vspace*{10pt}\\
\textsl{\Large{\mbox{Selected Problems of Lorentz Symmetry Violation}, Quantum Cosmology, and Quantum Gravity}}\vspace*{100pt}\\
}

\author{
\textbf{\L UKASZ ANDRZEJ GLINKA}\vspace*{15pt}\\
\textit{International Institute for}\vspace*{5pt}\\
\textit{Applicable Mathematics \& Information Sciences}\vspace*{5pt}\\
\textit{Hyderabad (India) \& Udine (Italy)}\vspace*{40pt}\\
}

\date{Draft of Book}

\makeindex

\begin{document}
\newpage
\thispagestyle{empty}
\maketitle
\newpage
\thispagestyle{empty}
\frontmatter
\tableofcontents

\newpage
\thispagestyle{empty}

\chapter*{Preface}
\addcontentsline{toc}{part}{Preface}

\begin{flushright}
\textit{A thought is an idea in transit.}\vspace*{10pt}\\
\textsl{Pythagoras}
\end{flushright}

\emph{{\AE}thereal Multiverse} is my personal research attempt to demonstrate the productive realization of the fusion of two fundamental concepts of Antiquity, i.e. {\AE}ther and Multiverse which lay the foundations of Aristotelian and Epicurean--Islamic systems, respectively. It is my deep conviction that this combination enables the fruitful description of physical Reality, which allows to understand constructively the anthropic everything, i.e. all what can be observed and detected by a man and all devices produced by humankind. From my standpoint the constructive proposal for the fusion involves the essential theoretical symbols of the physical Reality: Quantum Mechanics, Quantum Field Theory, General Relativity, and Thermodynamics.

On the one hand this monograph collects advanced developments of certain elementary knowledge of theoretical and mathematical physics. On the other hand, however, the presented deductions are performed step-by-step and often include detailed calculations. In this manner this book is available for the readers interested in development and applications of the fundamental knowledge. Intentionally the content is divided onto two independent lines which have arose in my research work of the years 2006-2010. Part I, \emph{Lorentz Symmetry Violation}, contains 3 chapters and is devoted to discussion of several applications of the noncommutative geometry of the Snyder quantized space-time. Strictly speaking I shall focus in some detail on the basic consequences of what I call the Snyder--Sidharth Hamiltonian constraint, i.e. the modification of Special Relativity arising from the noncommutative geometry. Particularly the Compton effect and the massive neutrinos model are discussed. Part II, \emph{Quantum General Relativity}, contains 8 chapters and is much more inhomogeneous. This part presents my approach to quantum cosmology and quantum gravity, and my point of view on inflationary cosmology. I present the version of quantum geometrodynamics strictly based on the Wheeler--DeWitt equation, which leads to constructive and consistent deductions. Particularly two approaches to entropy calculation, which lead to distinguished formulations of thermodynamics of space quanta {\AE}ther, are discussed.

The greatest motivation to this book is the 2010' book of an experienced mathematical physicist Robert W. Carroll \cite{rcarroll} who expressed his opinion about certain part of my research results\\

\noindent\textit{(\ldots) we sketch some work of L. Glinka et al on thermodynamics and quantum gravity. This involves bosonic strings and quantum f\/ield theory (QFT) and is speculative (but very interesting). (\ldots) we deal with some fundamental articles by L. Glinka for which some general theory is also motivated by theoretical material involving second quantization and Bogoliubov transformations (\ldots) we suggest printing out the latest versions and working from them - we can only give a sketch here and remark that some of the work has a visionary nature which is valuable in itself.}\\

\noindent This book collects all my non-coauthored research results rigorously revisited and enriched by necessary updates. Numerous typos and technical mistakes are improved. The necessity of the improvement follows from my deep conviction that the only mathematical truth results in the constructive theoretical and mathematical physics. This book develops also my philosophical standpoint on theoretical physics involving the philosophical interpretation needed for the new physics.

My grateful acknowledgements belong to numerous senior scientists and scholars who helped me kindly in the various aspects related to this book. Professor Robert W. Carroll granted me by a number of valuable discussions and comments, and mailed a hard copy of his 2010' book. Professor Sir Harold W. Kroto gave matter-of-fact discussions, and included my views into his GEOSET programme. Essential comments and discussions due to Professor Burra G. Sidharth significantly helped in edition of the primary manuscript of this book. Discussions with Dr. Andrej B. Arbuzov and Professor Alessandro De Angelis benefitted during my research work were helpful. Comments from Professor Wojciech H. Zurek were also valuable and constructive.

I dedicate this book to all Friends of mine. Comments and discussion are welcome and invited to \href{mailto:laglinka@gmail.com}{\textbf{laglinka@gmail.com}}.
\begin{flushright}
\L ukasz Andrzej Glinka
\end{flushright}

\newpage
\thispagestyle{empty}

\chapter*{Prologue}
\addcontentsline{toc}{part}{Prologue}

\begin{flushright}
\textit{It is a dogma of the Roman Church that the existence of God can be proved by natural reason. Now this dogma would make it impossible for me to be a Roman Catholic. If I thought of God as another being like myself, outside myself, only inf\/initely more powerful, then I would regard it as my duty to defy him.}\vspace*{10pt}\\
\textsl{Ludwig Wittgenstein}
\end{flushright}

\section*{The U-turn to Antiquity}
In the second decade of the 21th century technological progress def\/ines evolution of humankind. Present civilizations are based on technocratic ideas which are often overly formal. Such a state of things rejects freedom of thinking and focuses on creation of new barriers, called standards and norms. The obstacles follow from promotion of feudal submissiveness toward various traditions. This fight with rationalism results in inhibition of mental growth of individuals, and leads to evident intellectual poverty. The irrationalism ruins efforts of rationalists, establishes and stabilizes illusions and delusions, and effectively results in the permanent damage of individual mentality. Such a regime has been produced the emergence of social multiple disabilities, and above all  a regular de-evolution of humankind.

On the other hand there is detectable the vast U-turn to the ideas of the most fundamental epoch in the history of human reasoning, \emph{Antiquity}. The necessity of such a nontrivial turnabout is fully justif\/ied by the situation. In fact, only the ancient ideas possess the natural ability to reconstruct the most valuable heritage, because of Antiquity is not infected by certain destructive effects of the modern civilization. A productive description of the physical Reality requires fusion of two notions: {\AE}ther, the idea of the Aristotelian system, and Multiverse which lays the foundations for the Epicurean--Islamic system.

\section*{{\AE}ther}

In the mid-1950s, when Albert Einstein abandoned the living nature, Cosmology altered manifestly esoteric and religious countenance, and became a respected scientif\/ic branch. The satisfactory explanation of observational and experimental data consolidated both the physical and philosophical heritage of General Relativity. Riemannian geometry apparently is able to describe a number of astronomical phenomena like Mercury's perihelion precession and light rays deviation in Sun's vicinity, and offers constructive generalization of the Newton universal gravitation. This Einstein's theory has been described visible objects and predicted existence of new physical beings.

Albeit, Einstein's legacy is wide spread \cite{pais} and deserves being called \textit{phenomenology}. Despite Einstein was established as the specialist in molecular physics and thermodynamics, he gave trailblazing description of fast particles, solids, opalescence, emission and absorption of electromagnetic radiation, and corrected the Zero-Point Energy hypothesis of M. Planck. His pioneering approach to the photoelectric effect used the wave-particle duality which became the main stream of the 20th century theoretical physics. He was the follower of unif\/ication of gravity and electromagnetism. Despite the heritage incontrovertibly impacted on the mentality and the character of theoretical physics, Einstein's intellectual growth was diverse. It is an appearance, because of Einstein always tried to describe {\AE}ther.

\subsubsection{The Mythical Nature}

The concept of {\AE}ther is strictly rooted within the mythology of Ancient Greece. Ancient Greeks professed Protogenoi, which are the immortal <<primordial deities>> born in the beginning of Universe. These primary gods were {\AE}ther (Mists of Light, Upper Air), Ananke (Inevitability, Necessity, Compulsion), Chaos (Void, Lower Air), Chronos (Time), Erebus (Mists of Darkness), Eros (Generation), Gaea (Earth), Hemera (Day), Hydros (Water), Nesoi (Islands), Nyx (Night), Oceanus (Ocean), Ourea (Mountains), Phanes (Procreation), Phusis (Nature), Pontus (The Sea), Tartarus (Hell), Tethys (Fresh Water), Thalassa (Sea Surface), Thesis (Creation), Uranus (Heaven), which generated Giants, Titan, Olympian, Oceanic, and Chthonic gods. In Homeric Greek language {\AE}ther $\alpha\iota\theta\eta\rho$ means ''pure, fresh air'' or ''clear sky'', which in Greek mythology is the pure essence where the gods lived and which they breathed. {\AE}ther was a personified idea of the cosmogony professed by ancient Greeks, and considered as one of the elementary substances forming the Universe. According to the Orphic hymns {\AE}ther is the soul of the world emanating all life.

In alchemy and natural philosophy {\AE}ther was originated by Aristotle as <<quinta essentia>>, the cognate chemical f\/ifth element of the heavens are made and unifying the sublunary elements, i.e. Fire, Earth, Air, and Water. Aristotle claimed that the four elements move rectilinearly, and because of orbits of the heavenly bodies are circular and lie on the rotating spheres surfaces, so the physical spheres must be a body. In other words, the celestial motions required an existence of the superior element. Aristotle gave several plausible arguments for existence of {\AE}ther. One of them, called argument from incorruptibility, states that the sublunar elements are easy transiting into each other, but because of the eternal heavens they must be made of a different element. Therefore, existence of {\AE}ther follows from the necessity for a body endowed with natural circular motion. By Aristotle the f\/ifth element is unborn, immortal, and invariable, and its name arises from ancient <<aei thein>> what means eternal motion. Factually, he identif\/ied {\AE}ther with the mind or soul, which is divine and does not corrupt with the earthly elements. Aristotle wrote \cite{aristotle}\\

\noindent\textit{(\ldots) That is why the upper part is moved in a circle, while the All is not anywhere. For what is somewhere is itself something, and there must be alongside it some other thing wherein it is and which contains it. But alongside the All or the Whole there is nothing outside the All, and for this reason all things are in the heaven; for the heaven, we may say, is the All. Yet their place is not the same as the heaven. It is part of it, the innermost part of it, which is in contact with the movable body; and for this reason the earth is in water, and this in the air, and the air in the aether, and the aether in heaven, but we cannot go on and say that the heaven is in anything else.}

\subsubsection{From Boyle \& Newton to Lorentz}
In the 17th century R. Boyle \cite{boyle} treated {\AE}ther as the material substance due to what he called subtle particles. However, in the context of modern physics I. Newton, in the corpuscular theory of light presented in the book \emph{Opticks} \cite{opticks}, proposed the pioneering idea of an {\ae}thereal medium carrying vibrations traveling faster than light, which straightforward intervention results in refraction and diffraction of light. Newton negated the idea proposed by Ch. Huygens, claiming that light travels by {\AE}ther medium, and presented the explanation of the phenomenon of gravitation via the pressure of an atomic {\AE}ther impacting upon matter in abnormal state. Subtle {\ae}thereal molecules entered into matter via the pores, and when approached a physical body became less resilient and rarely distributed. A material body f\/ind itself under the pressure due to {\AE}ther upon all sides. Therefore, between two bodies this pressure is less and rare {\AE}ther distribution causes in compulsory gravitation. Newton, however, did not justify the rarefaction mechanism emerging near matter.

We should reference also so called Le Sage's kinetic theory of gravitation, proposed by N. Fatio de Duillier and G.L. Le Sage, which was {\AE}ther-based constructive explanation of the Newton universal gravitation. Le Sage \cite{lesage} suggested purely mechanical nature of the universal gravitation which in an effect due to streams of ultra-mundane tenuous {\AE}ther corpuscules moving at the speed of light and acting on all matter from all directions. In ''Lucr\`{e}ce Newtonien'' Le Sage expressed the following standpoint\\

\noindent\textit{I am well convinced that since the law governing the intensity of universal gravitation is similar to that for light, the thought will have occurred to many physicists that an ethereal substance moving in rectilinear paths may be the cause of gravitation, and that they may have applied to it whatever of skill in the mathematics they have possessed.}\\

W. Thomson the 1st Baron Kelvin \cite{kelvin} introduced another impacting concept of {\AE}ther. He treated {\AE}ther as an elastic solid medium transmitting the electromagnetic waves, and proposed its mechanical model. Such a proposition led him to explanation of the nature of radiation.

J.C. Maxwell \cite{jcmaxwell} was a follower of so called Luminiferous {\AE}ther, i.e. the cosmic medium physically transmitting light. He applied this notion to deduce equations of electrodynamics, called the Maxwell equations. Maxwell showed that light is an electromagnetic wave, and thought about the physical lines of forces of electric and magnetic f\/ields as the lines within the {\AE}ther. His idea based on the Poisson equation is referred as the Maxwell Vacuum.

Another diverse concepts of {\AE}ther were widely propagated and developed by numerous eminent scientists and scholars at the turn of the 19th and 20th centuries (See e.g. the contributions in the Ref. \cite{otherae}). Exceptionally detailed analysis of the historical development of {\AE}ther till discovery of quantum mechanics was elegantly performed by E.T. Whittaker \cite{whitt}. The best example is the {\AE}ther model investigated by J.J. Thomson, who considered the context of a hypothetical radiation that could be more penetrating then R\"ontgen's X-rays. Another significant investigation applying certain {\AE}ther model was performed by G.H. Darwin in computation of a geometric deviation from Newton's law of universal gravitation.

The most intriguing idea was the concept of {\AE}ther wind called also {\AE}ther drag and {\AE}ther drift. This hypothetic phenomenon has the place when Luminiferous {\AE}ther is dragged by motion of matter or entrained by matter. A.J. Fresnel proposed {\AE}ther wind with partial entraining, which was empirically confirmed by the 1851 experiment of H. Fizeau. Different version of the {\AE}ther drag hypothesis, founded by G. Stokes in 1845, was experimentally conf\/irmed by A.A. Michelson and E.W. Morley \cite{michmor} in 1881 and 1887. Fifteen years after interpretation of the results of the Michelson--Morley experiment as the conf\/irmation of Einstein's Special Relativity, Michelson published the book \cite{michel} in which one can f\/ind several interesting looking reflections\\

\noindent\textit{The standard light waves are not alterable; they depend on the properties of the atoms and upon the universal ether; and these are unalterable. It may be suggested that the whole solar system is moving through space, and that the properties of ether may differ in different portions of space. I would say that such a change, if it occurs, would not produce any material effect in a period of less than twenty millions of years, and by that time we shall probably have less interest in the problem. (\ldots) the vibrations of these particles, or of their electric charges, produce the disturbance in the ether which is propagated in the form of light waves; and that the period of any light wave corresponds to the period of vibration of the electric charge which produces it. (\ldots) the ether itself is electricity; a much more probable one is that electricity is an ether strain - that a displacement of the ether is equivalent to an electric current. If this is true, we are returning to our elastic-solid theory. I may quote a statement which Lord Kelvin made in reply to a rather skeptical question as to the existence of a medium about which so very little is supposed to be known. The reply was: "Yes, ether is the only form of matter about which we know anything at all." In fact, the moment we begin to inquire into the nature of the ultimate particles of ordinary matter, we are at once enveloped in a sea of conjecture and hypotheses - all of great diff\/iculty and complexity. One of the most promising of these hypotheses is the "ether vortex theory," which, if true, has the merit of introducing nothing new into the hypotheses already made, but only of specifying the particular form of motion required. The most natural form of such vortex motions with which to deal is that illustrated by ordinary smoke rings, such as are frequently blown from the stack of a locomotive. Such vortex rings may easily be produced by f\/illing with smoke a box which has a circular aperture at one end and a rubber diaphragm at the other, and then tapping the rubber. The friction against the side of the opening, as the puff of smoke passes out, produces a rotary motion, and the result will be smoke rings or vortices. (\ldots) Investigation shows that these smoke rings possess, to a certain degree, the properties which we are accustomed to associate with atoms, notwithstanding the fact that the medium in which these smoke rings exists is far from ideal. If the medium were ideal, it would be devoid of friction, and then the motion, when once started, would continue indef\/initely, and that part of the ether which is differentiated by this motion would ever remain so. (\ldots) Another peculiarity of the ring is that it cannot be cut - it simply winds around the knife. Of course, in a very short time the motion in a smoke ring ceases in consequence of the viscosity of the air, but it would continue indef\/initely in such a frictionless medium as we suppose the ether to be. (\ldots) Suppose that an ether strain corresponds to an electric charge, an ether displacement to the electric current, these ether vortices to the atoms - if we continue these suppositions, we arrive at what may be one of the grandest generalizations of modern science - of which we are tempted to say that it ought to be true even if it is not - namely, that all the phenomena of the physical universe are only different manifestations of the various modes of motions of one all-pervading substance - the ether. (\ldots) Then the nature of the atoms, and the forces called into play in their chemical union; the interactions between these atoms and the non-differentiated ether as manifested in the phenomena of light and electricity ; the structures of the molecules and molecular systems of which the atoms are the units; the explanation of cohesion, elasticity, and gravitation — all these will be marshaled into a single compact and consistent body of scientif\/ic knowledge. (\ldots) In all probability, it not only exists where ordinary matter does not, but it also permeates all forms of matter. The motion of a medium such as water is found not to add its full value to the velocity of light moving through it, but only such a fraction of it as is perhaps accounted for on the hypothesis that the ether itself does not partake of this motion. (\ldots) The phenomenon of the aberration of the f\/ixed stars can be accounted for on the hypothesis that the ether does not partake of the earth's motion in its revolution about the sun. All experiments for testing this hypothesis have, however, given negative results, so that the theory may still be said to be in an unsatisfactory condition.}

H.A. Lorentz, one of the most eminent theoretical physicists of the turn of the 19th and 20th centuries, also manifestly professed {\AE}ther \cite{halorentz}. In his lectures he straightforwardly supports {\AE}ther. In the lectures delivered at Caltech one finds\\

\noindent\textit{Nowadays we are concerned only with the electromagnetic theory of light, in which there is no longer any discussion of a density or elasticity of the ether. In the electromagnetic theory of light attention is f\/ixed on the electric and magnetic f\/ields that can exist in the "ether". (...) the state of the ether is the same at all points of a plane perpendicular to the direction of propagation, and so the waves may be called plane waves.}\\

\noindent Similarly, Leiden lectures of Lorentz contain the ambiguous opinion\\

\noindent\textit{(\ldots)[W]hether there is an aether or not, electromagnetic f\/ields certainly exist, and so also does the energy of electrical oscillations. If we do not like the name of ‘aether’, we must use another word as a peg to hang all these things upon. It is not certain whether ‘space’ can be so extended as to take care not only of the geometrical properties but also of the electric ones. One cannot deny to the bearer of these properties a certain substantiality, and if so, then one may, in all modesty, call true time the time measured by clocks which are f\/ixed in this medium, and consider simultaneity as a primary concept.}\\

\noindent The Lorentz transformations were deduced on base of {\AE}ther, and Lorentz defended his own standpoint called Lorentz {\AE}ther theory. Lorentz pointed out inconsistency between results of the Michelson--Morley and the Fizeau experiments. Basing on the stationary {\AE}ther arising from the theory of electrons Lorentz removed Luminiferous {\AE}ther drag.

G.F. FitzGerald, who was under the inf\/luence of calculations performed by O. Heaviside \cite{heavi} including deformations of magnetic and electric f\/ields surrounding a moving charge and the effects of it entering a denser medium like e.g. what is today called Cherenkov's radiation, in Science article \cite{fitzg} wrote {\AE}ther-based conclusion\\

\noindent\textit{[T]he length of material bodies changes, according as they are moving through the ether or across it, by an amount depending on the square of the ratio of their velocities to that of light.}\\

Another example is the 1915' paper of W.J. Spillman, in which reasoning is based on the \emph{Fundamental Assumptions} involving {\AE}ther

\textit{\textbf{The ether.} - The ether is assumed to exist in the interstices of matter and in open space. It is assumed to be capable of distortion by f\/inite force, and to oppose such distorting force with an equal opposite force. In other words, every point of the ether has a position which it normally occupies, and when removed from that position tends forcibly to return to it, the force being proportional to the distortion. It is further assumed that distortion at a given point in the ether tends to become
distributed in the surrounding ether according to the law of inverse squares, and that such distribution occurs at a f\/inite rate (the velocity of light).}\\
\indent\textit{\textbf{The electron.} - It is assumed that in the immediate vicinity of the electron there is a region of maximum permanent ether distortion, the distortion at other points in the surrounding ether varying inversely as the square of the distance from the center of the electron, and that pressures are transmitted at a higher velocity through distorted than through non-distorted ether.}\\
\indent\textit{The permanent distortion of the ether in the vicinity of the electron may be conceived of as a pushing back of the ether radially from the center of the electron, as if an impenetrable and inelastic body were injected into the midst of an elastic body; or it may be conceived of as being circular, in two hemispheres facing each other, and opposite in direction in the two hemispheres.}\\
\indent\textit{In explaining the phenomena of inertia, electric currents, magnetism, chemical aff\/inity, and radiant energy, the above alternative assumptions concerning the character of the distortion lead to essentially similar lines of reasoning, but the treatment on the assumption of radial distortion is very much simpler. For this reason it is used here. In the case of static electricity only the assumption of circular distortion, opposite in direction in the two kinds of electric elements, will explain the facts. The extension of the theory to static electricity is left for future treatment. It must be remembered, however, that the development of the theory on the basis of radial distortion differs only in detail, not in principle, from that on the basis of circular distortion.}\\
\indent\textit{\textbf{The Atom.} - The atom is assumed to consist, at least in part, of a Saturnian system of electrons in rapid orbital motion. It will be shown that such orbital motion, with the assumptions here made, would give rise to a pressure in the ether such as Newton showed would account for gravitation.}\\
\indent\textit{Physicists are agreed that the phenomena of inertia, the electric current, magnetism, and possibly also chemical aff\/inity are probably related to each other in such manner that when we f\/ind the explanation of one of them this explanation will also throw light on the others.}

\subsubsection{{\AE}ther Drag}
The most intriguing phenomena related to the concept of {\AE}ther is {\AE}ther drag/drift/wind. Both Lorentz {\AE}ther theory and Einstein's Special Relativity, the theories of stationary {\AE}ther, are interpreted are the theories extraordinary strongly antagonistic to the {\AE}ther wind. The essence of the history of 20th century physics and the milestone of empirical negating of {\AE}ther is the Michelson--Morley experiment which purpose was to detect Earth's motion with respect to Luminiferous {\AE}ther via comparison of speed of light in diverse directions with respect to Earth. On the one hand, the theory of stationary {\AE}ther propagated by H.A. Lorentz this attempt manifestly and straightforwardly negated existence of {\AE}ther drag because of contradiction with the result of the Fizeau experiment. On the other hand, the Michelson--Morley experiment is consistently explained and interpreted within Einstein's Special Relativity with no reference to {\AE}ther drag.

There were performed also another experiments \cite{othexp} having the purpose to verify empirically the {\AE}ther drift: 1903' F.T. Trouton and H.R. Noble, 1908' F.T. Trouton and A.O. Rankine, 1913' G. Sagnac, 1925' A.A. Michelson and H.G. Gale, 1932' R.J. Kennedy and E.M. Thorndike, and 1935' G.W. Hammar. Their results suffered the fate analogous to the Michelson--Morley experiment, i.e. were interpreted as negation of {\AE}ther wind and {\AE}ther in general. Transparent discussion of these results can be found in the book of W. Pauli \cite{pauli}.

The Trouton--Noble experiment was strict realization of the idea due to FitzGerald. He concluded that motion of a charged f\/lat condenser through the {\AE}ther should result in its perpendicular orientation to the motion. The experimenters detected the lack of the relative motion to the {\AE}ther, what was interpreted as the negative result.

The Trouton--Rankine experiment was performed for detection of ''preferred frame'' which would be the syndrome of existence of the Luminiferous {\AE}ther. The measurement assumed that the length contraction produce a measurable effect in the rest frame of the object observed in other frame. Both Special Relativity and Lorentz {\AE}ther theory predicted that length contraction is non measurable. Trouton and Rankine, however, applying the Ohm law and the Maxwell equations theoretically predicted that the effect, change of resistance, is measurable in the laboratory frame. The experimenters used the Wheatstone bridge, and the change of resistance was not detected.

Sagnac applied a rotating interferometer, and observed a dependence of interference fringe position on the angular velocity. The result of this experiment was theoretically predicted in 1911 by M. von Laue \cite{mvlaue}, who proved manifestly its consistence with Special Relativity and numerous models of stationary {\AE}ther, including the Lorentz {\AE}ther theory. Factually, results of this experiment conf\/irmed existence of the stationary {\AE}ther, even in the sense of the Lorentz {\AE}ther theory, but Laue's calculations were interpreted as the ref\/lection of Special Relativity correctness. The Sagnac effect is probably the most positive result for the {\AE}ther drag, and was also treated as the straightforward demonstration of existence of the {\AE}ther. Moreover, the Sagnac effect was constructively explained also within General Relativity which was claimed by Einstein to be the theory of non-physical {\AE}ther.

The Michelson--Gale experiment modif\/ied the Michelson--Morley experiment by application of enlarged Sagnac's ring interferometer. The purpose was to f\/ind out the relation between the Earth rotation motion and the light propagation in the vicinity of the Earth. Similarly as in the case of the Michelson--Morley experiment, the Michelson--Gale version compared the light from a single source after two directional travel. The difference was replacement of the two the Michelson--Morley arms with two different rectangles. The obtained results were compatible with both Special Relativity and models of stationary {\AE}ther. However, Lorentz {\AE}ther theory contradicted the Michelson--Morley experiment, and therefore the Michelson--Gale version was interpreted also as the confirmation of correctness of Special Relativity.

The investigators of the Kennedy--Thorndike experiment also interpreted their results in terms of relativistic effects. They modif\/ied the Michelson--Morley experiment via application of the interferometer's in which one arm is very short in comparison with the second arm. the Michelson--Morley experiment constructively verif\/ied the length contraction hypothesis, whereas in the Kennedy--Thorndike version time dilation was straightforwardly examined. The much shorter arm and maximal stabilization of the apparatus enabled the verify existence of a specif\/ical fringe shift, which theoretically should change of the light frequency. According to theoretical predictions the shift resulted from a change of speed of the Earth with respect to {\AE}ther, would result in changes of time of light travel. Albeit, the shift was not detected, what was interpreted as the conf\/irmation of Special Relativity correctness.

The purpose of the Hammar experiment was emergence of the {\AE}ther drag asymmetry by application of massive lead blocks on both side of only one the Michelson--Morley interferometer's arm. The idea of this investigation was similar to the tests performed by O.J. Lodge. The blocks absence should result in equal affect of both arms by {\AE}ther, while in the blocks presence the one arm should be affected. Hammer reported independence of fringe displacements on the blocks absence/presence, what was the argument against the {\AE}ther wind.

The conclusion is unambiguous. Stigmatization of {\AE}ther and {\AE}ther drag based on application of the Ockham razor. The physical interpretation used \emph{ad hoc} absence of the {\AE}ther drag, and non-existence of {\AE}ther was concluded. The purpose was intentional interpretation of experimental data supporting elimination of {\AE}ther and related phenomena. In other words exclusion of one thing was performed via imposing of absence of another one. However, the matter is much more sophisticated because of in general the things must not be correlated.

It is worth stressing that computed effects of the {\AE}ther drag, obtained from diverse {\AE}ther models, were checked in numerous experiments. Since 1904 D.C. Miller and collaborators \cite{miller} performed over 200,000 perspective empirical investigations devoted to the verif\/ication. Basing on the results, Miller propagated existence of the {\AE}ther wind, which effects were much smaller than predictions due to stationary {\AE}ther. Miller's work has been inspired a lot of scholars and researchers \cite{drift}. In 1955 R.S. Shankland et al \cite{shankland} devaluated the Miller research, and claimed that the only statistical f\/luctuation due to the local temperature conditions and systematic error generated the Miller effect. However, in 1983 W. Broad et al suggested to review with attention the results received by D.C. Miller, and straightforwardly negated the refusal due to Shankland. Interestingly, R.A. M\"uller \cite{muller} constructively explained anisotropies of the Cosmic Microwave Background Radiation via application of the concept of the {\AE}ther drag.

\subsubsection{Einstein's Visions}
We refer for detailed discussion of Einstein's opinions which recently has been performed by L. Kostro in his book \cite{kostro}. It looks like that Einstein as the creator of Special Relativity, against his will, was stigmatized as the killer of {\AE}ther. This opinion was based on the intentional interpretation of the very small fragment of his 1905 Special Relativity paper \cite{korper1}, which can be found in introduction to this paper\\

\noindent\textit{The introduction of a 'luminiferous ether' will prove to be superf\/luous inasmuch as the view here to be developed will not require an <<absolutely stationary space>> provided with special properties, nor assign a velocity-vector to a point of the empty space in which electromagnetic processes take place.}

\noindent Nevertheless, he was also stigmatized as a resurrector of the {\AE}ther on the base of \emph{Sidelights of Relativity} \cite{sidel} in which one f\/inds\\

\noindent\textit{We may say that according to the general theory of relativity space is endowed with physical qualities; in this sense, therefore, there exists an aether. According to the general theory of relativity space without aether is unthinkable; for in such space there not only would be no propagation of light, but also no possibility of existence for standards of space and time (measuring-rods and clocks), nor therefore any space-time intervals in the physical sense. But this aether may not be thought of as endowed with the quality characteristic of ponderable media, as consisting of parts which may be tracked through time. The idea of motion may not be applied to it.}\\

\noindent Also in the article \emph{Concerning the Aether} Einstein supported {\AE}ther\\

\noindent\textit{When we speak here of aether, we are, of course, not referring to the corporeal aether of mechanical wave-theory that underlines Newtonian mechanics, whose individual points each have a velocity assigned to them. (\ldots) Instead of 'aether', one could equally well speak of 'the physical quantities of space'. (\ldots) So we are effectively forced by the current state of things to distinguish between matter and aether, even though we may hope that future generations will transcend this dualistic conception and replace it with a unif\/ied theory, as the f\/ield theoreticians of our day have tried in vain to accomplish. (\ldots) It is usually believed that aether is foreign to Newtonian physics and that it was only the wave theory of light which introduced the notion of an omnipresent medium inf\/luencing, and affected by, physical phenomena. (\ldots) Newtonian mechanics had its ‘aether’ in the sense indicated, albeit under the name ‘absolute space’. To get a clear understanding of this and, at the same time, to explore more fully the concept of aether, we must take a step back. (\ldots) The kinematics, or phoronomy, of classical physics had as little need of an aether as (physically interpreted) Euclidean geometry has. (\ldots) We will call this physical reality which enters the Newtonian law of motion alongside the observable, ponderable real bodies, the aether of mechanics. The occurrence of centrifugal effects with a (rotating) body, whose material points do not change their distances from one another, shows that this aether is not to be understood as a mere hallucination of the Newtonian theory, but rather that it corresponds to something real that exists in nature. (\ldots) The mechanical aether - which Newton called ‘absolute space’ - must remain for us a physical reality. Of course, one must not be tempted by the expression aether into thinking that, like the physicists of the 19th century, we have in mind something analogous to ponderable matter. (\ldots) When Newton referred to the space of physics as ‘absolute’, he was thinking of yet another property of what we call here aether. Every physical thing inf\/luences others and is, it its turn, generally inf\/luenced by other things. This does not however apply to the aether of Newtonian mechanics. For the inertia-giving property of this aether is, according to classical mechanics, not susceptible to any inf\/luence, neither from the conf\/iguration of matter nor anything else. Hence the term ‘absolute’. (\ldots) Viewed historically, the aether hypothesis has emerged in its present form by a process of sublimation from the mechanical aether hypothesis of optics. After long and fruitless efforts, physicists became convinced that light was not to be understood as the motion of an inertial, elastic medium, that the electromagnetic f\/ields of Maxwell’s theory could not be construed as mechanical. So under the pressure of this failure, the electromagnetic f\/ields had gradually come to be regarded as the f\/inal, irreducible physical reality, as states of the aether, impervious to further explanation. (\ldots) While at least in Newtonian mechanics all inertial systems were equivalent, it seemed that, in the Maxwell-Lorentz theory, the state of motion of the preferred coordinate system (at rest with respect to the aether) was completely determined. It was accepted implicitly that this preferred coordinate system was also an inertial system, i.e. that the principle of inertia [Newton’s f\/irst law] applied relative to the electromagnetic aether. (\ldots) No longer was a special state of motion to be ascribed to the electromagnetic aether. Now, like the aether of classical mechanics, it resulted not in the favoring of a particular state of motion, only the favoring of a particular state of acceleration. Because it was no longer possible to speak, in any absolute sense, of simultaneous states at different locations in the aether, the aether became, as it were, four dimensional, since there was no objective way of ordering its states by time alone. According to special relativity too, the aether was absolute, since its inf\/luence on inertia and the propagation of light was thought of as being itself independent of physical inf\/luence. (\ldots) Thus geometry, like dynamics, came to depend on the aether. (\ldots) Thus the aether of general relativity differs from those of classical mechanics and special relativity in that it is not ‘absolute’ but determined, in its locally variable characteristics, by ponderable matter. (\ldots) On the one hand, the metric tensor, which codetermines the phenomena of gravitation and inertia and, on the other, the tensor of the electromagnetic f\/ield appear still as different expressions of the state of the aether, whose logical independence one is inclined to attribute rather to the incompleteness of our theoretical ediface than to a complex structure of reality. (\ldots) But even if these possibilities do mature into an actual theory, we will not be able to do without the aether in theoretical physics, that is, a continuum endowed with physical properties; for general relativity, to whose fundamental viewpoints physicists will always hold fast, rules out direct action at a distance. But every theory of local action assumes continuous f\/ields, and thus also the existence of an ‘aether’.}\\

\noindent Therefore, Einstein's point of view was that {\AE}ther is the core fundament of physics. In fact, he never neglected and negated {\AE}ther existence, and moreover he developed this concept. In the famous book coauthored with L. Infeld \cite{eininf}, he presented development of physics with respect to the concept of {\AE}ther. Several fragments are cited below\\

\noindent\textit{Our picture of ether might very probably be something like the mechanical picture of a gas that explains the propagation of sound waves. It would be much more diff\/icult to form a picture of ether carrying transverse waves. To imagine a jelly as a medium made up of particles in such a way that transverse waves are propagated by means of it is no easy task. (\ldots) Yet we know from mechanics that interstellar space does not resist the motion of material bodies. The planets, for example, travel through the ether-jelly without encountering any resistance such as a material medium would offer to their motion. If ether does not disturb matter in its motion, there can be no interaction between particles of ether and particles of matter. Light passes through ether and also through glass and water, but its velocity is changed in the latter substances. How can this fact be explained mechanically? Apparently only by assuming some interaction between ether particles and matter particles. We have just seen that in the case of freely moving bodies such interactions must be assumed not to exist. In other words, there is interaction between ether and matter in optical phenomena, but none in mechanical phenomena! This is certainly a very paradoxical conclusion! (\ldots) We may still use the word ether, but only to express some physical property of space. This word ether has changed its meaning many times in the development of science. At the moment it no longer stands for a medium built up of particles. Its story, by no means f\/inished, is continued by the relativity theory. (\ldots) For the time being, we shall continue to believe that the ether is a medium through which electromagnetic waves, and thus also light waves, are propagated, even though we are fully aware of the many diff\/iculties connected with its mechanical structure.}

\subsubsection{Theoretical Objections}
It is evident that Einstein's attempts were focused on explanation of a whole Universe via using of the concept of <<non physical>> {\AE}ther. In Special Relativity he introduced the non-Euclidean Minkowski Space-time that agreed with the Lorentz transformations. Recall that Lorentz performed {\AE}ther-based deduction of these transformations. It looks like that Einstein manifestly swept out {\AE}ther under space-time carpet, and unexpectedly reinvented {\AE}ther when his position was consolidated. Possibly, the impacting personality and authority of H.A. Lorentz caused such a situation. Saying <<non physical>> in the context of {\AE}ther looks rather like diplomacy then physics. Unfortunately, physics is often based on diplomatic truth, and diplomacy is limitlessly applied within physics.

The negating, which is irrelevant to General Relativity, manifestly supports Special Relativity. The empirical results, however, can be reinterpreted as the support of {\AE}ther existence and the phenomenon of the {\AE}ther drag. Perhaps the concept of electromagnetic Luminiferous {\AE}ther is incorrect, but in general it does not exclude another form or forms of {\AE}ther. For instance {\AE}ther treated as the primordial cause can exist only initially and must not exist in later stages of Universe evolution. The best example of theoretical investigation of {\AE}ther was done more than 40 years ago. In mid-1920's E. Cartan \cite{ecartan} by application of connections formulated General Relativity in terms of Newtonian dynamics. In 1966 A.M. Trautman \cite{traut} showed by straightforward calculation that the Einstein f\/ield equations are the special case of the Newtonian gravitation equations coupled to a thing which Trautman called Luminiferous {\AE}ther. Soon after these results C.W. Misner in \emph{Gravitation} coauthored with K.S. Thorne and J.A. Wheeler \cite{mtw} axiomatized the Trautman approach to show that Newtonian dynamics consistently joints General Relativity with the Cartan--Trautman {\AE}ther.

P.C.W. Davies \cite{davies} interviewed J.S. Bell, one of founders of quantum physics and originator of Bell's theorem/inequality. Bell straightforwardly expressed the opinion that the concept of {\AE}ther can be very useful tool in resolving the Einstein--Podolsky--Rosen paradox, regarding measurements of microscopic systems, by involving a reference frame in which signals go faster than light. In his view the Lorentz length contraction is correct but inconsistent with Special Relativity, but can result in the theory of {\AE}ther which is consistent with the results of Michelson--Morley experiment. Bell manifestly stated wrongness of rejection of the concept of {\AE}ther from physics, and proposed resurrection of the {\AE}ther because of a number of unsolvable issues is very easy to solve by imaging of existence of {\AE}ther. In \cite{bell} Bell discussed {\AE}ther.

Another attempt, which is good candidate for the model of {\AE}ther, was made by R.P. Feynman \cite{feynman}. Feynman proposed that the partial-differential equations are able to describe classical macroscopic motion of X-ons, i.e. certain very small entities. The medium created by these entities can be treated as the model of {\AE}ther. Similarly the action on distance approach to electrodynamics, proposed by Wheeler and Feynman \cite{whefeyn}, gives great hopes for {\AE}ther.

In this manner the situation of {\AE}ther is non-established. Moreover, intentional interpretation of experimental data, so widely applied by antagonists of {\AE}ther, enables to consider {\AE}ther as a physical being. On the other hand, as suggested J. Bell, {\AE}ther may be a helpful tool in constructive and consistent explanation of numerous phenomena and effects. A number of heightened attempts manifestly rejecting {\AE}ther from description of Nature is based on the methodology which in itself is the selection and propagating of preferred interpretation. In fact, all presented justif\/ications of non-existence of {\AE}ther are easy to straightforward invalidation by suing of the theoretical as well as the empirical arguments. The question is whether the physical truth should be technocratic or diplomatic. Of course, Nature is neither technocratic nor diplomatic, and with no ideological constraints tells what is the truth.

\subsubsection{Dirac {\AE}ther}
In 1951 P.A.M. Dirac, regarded by Einstein the founder of relativistic quantum mechanics and quantum f\/ield theory, concluded existence of {\AE}ther ref\/lecting the nature of four-velocity in the context of theory of electrons following from his new electrodynamics \cite{dirae}\\

\noindent\textit{It was soon found that the existence of an {\ae}ther could not be f\/itted in with relativity, and since relativity was well established, the {\ae}ther was abandoned. (\ldots) If one reexamines the question in the light of present-day knowledge, one f\/inds that the {\ae}ther is no longer ruled out by relativity, and good reasons can now be advanced for postulating an {\ae}ther. (\ldots) at the present time it needs modif\/ication, because we have to apply quantum mechanics to the {\ae}ther. The velocity of the {\ae}ther, like other physical variables, is subject to uncertainty relations. For a particular physical state the velocity of the {\ae}ther at a certain point of space-time will not usually be a well-def\/ined quantity, but will be distributed over various possible values according to a probability law obtained by taking the square of the modulus of a wave function. We may set up a wave function which makes all values for the velocity of the {\ae}ther equally probable. Such a wave function may well represent the perfect vacuum state in accordance with the principle of relativity. (\ldots) A thing which cannot be symmetrical in the classical model may very well be symmetrical after quantization. This provides a means of reconciling the disturbance of Lorentz symmetry in space-time produced by the existence of an {\ae}ther with the principle of relativity. (\ldots) we may very well have an {\ae}ther, subject to quantum mechanics and conforming to relativity, provided we are willing to consider the perfect vacuum as an idealized state, not attainable in practice. (\ldots) We have now the velocity at all points of space-time, playing a fundamental part in electrodynamics. It is natural to regard it as the velocity of some real physical thing. Thus with the new theory of electrodynamics we are rather forced to have an {\ae}ther.}\\

\noindent One year later Nature magazine published interesting looking polemics between L. Infeld and P.A.M. Dirac \cite{indirin}. Infeld jointed the formulas of Dirac in a certain intentional way and used of rather laconic then logical arguments to point out that the new electrodynamics does not need the concept of {\AE}ther if <<all>> its conclusions will be accepted. Infeld, as the typical representative of the scholastics of Soviet block, did not precise what means the word <<all>> in such a context. In other words, factually even Infeld did not accept <<all>> conclusions of the Dirac electrodynamics and selected the only these ones which were adequate for elimination of Dirac {\AE}ther. In this manner Infeld used intentional interpretation to argue his beliefs, and did not focus attention on the physical aspects of Dirac {\AE}ther. Moreover, such a negative opinion was also straightforwardly opposite to the efforts of Albert Einstein which nota bene were supported by Infeld several years earlier. Deduction of the existence of {\AE}ther performed by Dirac used purely formal aspects of the Hamiltonian approach to the Maxwell electrodynamics, i.e. the constraints and the action. Recall that over 25 years earlier Dirac discovered the linkage between classical and quantum mechanics via the Poisson brackets correspondence. Maxwell electrodynamics was unsatisfactory formulated because of the correspondence works for the only Hamiltonian version of a classical theory. Dirac performed the Hamiltonian formulation of Maxwell electrodynamics, and discerned {\AE}ther in this theory. Applying the typical arguments of the Ockham razor based on personal beliefs, L. Infeld manifestly discredited the purposes and efforts due to P.A.M. Dirac.

Such a situation strengthened the Dirac strategy which was a follower of the revisionist approach with respect to even well-established and accepted physical knowledge. The self\/ish and unnatural selection performed by Infeld quickly obtained an adequate and constructive reply due to Dirac. The reply is brief and can be cited entirely\\

\noindent\textit{Infeld has shown how the f\/ield equations of my new electrodynamics can be written so as not to require an {\ae}ther. This is not suff\/icient to make a complete dynamical theory. It is necessary to set up an action principle and to get a Hamiltonian formulation of the equations suitable for quantization purposes, and for this the {\ae}ther velocity is required.}\\
\indent\textit{The existence of an {\ae}ther has not been proved, of course, because of my new electrodynamics has not yet justif\/ied itself. It will probably have to be modif\/ied by the introduction of spin variables before a satisfactory quantum theory of electrons can be obtained from it, and only after this has been accomplished will one to be able to give a def\/inite answer to the {\ae}ther question.}\\

The method of L. Infeld was the tip of the iceberg and ref\/lected the true countenance of the regional standpoint based on the fossilized traditional beliefs. At this time Soviet school of physics dominated European science and the Marxist--Leninist scholastics was one of the most fashionable streams. Such people like P.A.M. Dirac and A. Einstein were the pioneers who wanted to change this manifestly irrational status quo. In fact, the opinion due to L. Infeld about Dirac {\AE}ther straightforwardly crossed the efforts of both the Nobel laureates. Albeit, above all Infeld negated his own opinions published several years earlier together with Albert Einstein. This controversy was too serious for the science of the region and, in fact, resulted in some kind of hackwork within the Polish physics. Evidently seen lack of Nobel Prizes in Natural Sciences in Poland is the most gross syndrome of the fossilized reasoning and approach to science, and labels the civilization stagnation. Recall that Infeld was the only one of 11 signatories to 1955's Russell--Einstein Manifesto who never received a Nobel Prize. His creative efforts with respect to Poland started when in 1950 he left Canada, where by 12 years worked at the University of Toronto. He came back to communist Poland and decided to help in reconstruction of the Polish science which during World War II lost few generations of scholars. Admittedly his dictatorial approach to science resulted in tremendous contribution to the Warsaw school of physics, but transformed this school into the sanctuary of the Soviet scholastics.

\subsubsection{Zero Point Energy \& Planck Scale}

In 1900 Max Planck \cite{planck} published the revolutionary formula for energy of a single vibrating atom. Several years after, in 1913, Einstein together with his another collaborator O. Stern \cite{stern} modif\/ied the Planck formula by involving of the concept of cosmic heat bath. This concept was directly related to the universal frequency f\/ield associated with the Zero Point Energy. From the modern point of view one can say that Einstein and Stern renormalized the Planck energy at at absolute zero temperature via using of the residual oscillating energy. In fact, the cosmic bath heat is the model of {\AE}ther which explains numerous experimental data, like e.g. the Casimir effect, the Lamb shift.

Factually, the Maxwell model of {\AE}ther has never been experimentally refuted. In the context of quantum geometrodynamics due to J.A. Wheeler \cite{geoms} Maxwell vacuum can be regarded as quantum foams, i.e. a subquantum sea of Zero Point Energy f\/luctuations. According to Wheeler space-time warps, called wormholes, follow from tremendous densities of local energy due to the high energetic modes of the universal frequency f\/ield. Wormholes are the tunnels transmit electricity between two separate spatial places or, in more general context, between different universes creating the superspace -- conf\/igurational space of General Relativity. Wheeler proposed to think in terms of mini holes, i.e. primitive charged particles, as the wormholes related to the local space. In his view the electricity goes orthogonally via our universe from a fourth dimension. The mechanism of electron-positron pairs production follows by black and white mini holes.

Interestingly, H. Aspden \cite{aspden} proposed the hadronic model based on near-balanced continuum and quons, i.e. massless {\AE}ther particles giving a charge and condensing electron-positron pairs. Such a line of thinking was prolonged also by H. Puthoff \cite{puthof} who applied quantum theory to redef\/inition of the Zero Point Energy hypothesis.

Quantum mechanics is also referred as a theory of {\AE}ther based on quantum foams, leading to f\/luctuations of small scales which generate quick creations and annihilations of particle pairs. In modern cosmology, the f\/ifth element unifying other ones, called oftentimes Dark Energy or quintessence, has been considered as Zero-Point Field or quantum vacuum and already identif\/ied with {\AE}ther by B.G. Sidharth \cite{sidbooks}. Recently, also Einstein {\AE}ther theory, i.e. generally covariant generalization of General Relativity describing space-time and endowed both a metric and a unit time-like vector f\/ield manifestly violating Lorentz invariance, has became popular (See e.g. papers in the Ref. \cite{aemod}).

\section*{Multiverse}

\subsubsection*{Epicurus and Eastern Cosmologies}

Epicurus \cite{epicurus} was probably the f\/irst ancient philosopher who openly propagated the concept of Multiverse. In his Letter to Herodotus one f\/inds manifestly expressed his standpoint\\

\indent \textit{Moreover, there is an inf\/inite number of worlds, some like this world, others unlike it. For the atoms being inf\/inite in number, as has just been proved, are borne ever further in their course. For the atoms out of which a world might arise, or by which a world might be formed, have not all been expended on one world or a f\/inite number of worlds, whether like or unlike this one. Hence there will be nothing to hinder an inf\/inity of worlds.\\
\indent (\ldots) After the foregoing we have next to consider that the worlds and every f\/inite aggregate which bears a strong resemblance to things we commonly see have arisen out of the inf\/inite. For all these, whether small or great, have been separated off from special conglomerations of atoms; and all things are again dissolved, some faster, some slower, some through the action of one set of causes, others through the action of another.\\
\indent And further, we must not suppose that the worlds have necessarily one and the same shape. For nobody can prove that in one sort of world there might not be contained, whereas in another sort of world there could not possibly be, the seeds out of which animals and plants arise and all the rest of the things we see.}\\

\noindent Al-Qur'an, the holy book of Islam, also directly refers to multiple worlds. S\={u}ratu Al-F\={a}tihah, called <<The Seven Verses of Repetition>>, translated into English language by Haf\/iz Abdullah Yusuf Ali \cite{quran} sounds
\begin{eqnarray}
  &&^{1}\textit{~In~the~name~of~Allah,~Most~Gracious,~Most~Merciful.}\nonumber\\
  &&^{2}\textit{~Praise~be~to~Allah,~the~Cherisher~and~Sustainer~of~the~worlds!}\nonumber\\
  &&^{3}\textit{~Most~Gracious,~Most~Merciful.}\nonumber\\
  &&^{4}\textit{~Master~of~the~Day~of~Judgement.}\nonumber\\
  &&^{5}\textit{~Thee~we~do~worship,~and~Thine~aid~we~seek.}\nonumber\\
  &&^{6}\textit{~Show~us~the~straight~way,}\nonumber\\
  &&^{7}\textit{~The~way~of~those~on~whom~Thou~hast~bestowed~Thy~Grace,}\nonumber\\
  &&^{~}\textit{~those~whose~(portion)~is~not~wrath,~nor~of~those~who~go~astray.}\nonumber
\end{eqnarray}
\noindent One of the thinkers and philosophers straightforwardly inspired by Al-Qur'an was the muslim polymath Fakhr Al-Din Al-Razi. His point of view rejected the Aristotelian-Avicennian single universe revolving around a single world. A. Setia \cite{setia} referred fragments of the unpublished manuscript \emph{al-Matalib al-'Aliyah} of Razi\\

\noindent \textit{It is established by evidence that there exists beyond a void without a terminal limit, and it is established as well by evidence that God Most High has power over all contingent beings. Therefore he the Most High has the power to create a thousand thousand worlds beyond this world such that each one of those worlds be bigger and more massive than this world as well as having the like of what this world has of the throne, the chair, the heavens,and the earth, and the sun and the moon. The arguments of the philosophers for establishing that the world is one are weak, f\/limsy arguments founded upon feeble premises.}\\

\noindent Multiverse is also present in Puranas, the generic texts of Hinduism, Jainism or Buddhism. For example in Bhagavata Purana 9.4.56 one f\/inds the direct reference to multiple universes\\

\noindent\textit{Lord \'{S}iva said: My dear son, I, Lord Brahm\={a} and the other devas, who rotate within this universe under the misconception of our greatness, cannot exhibit any power to compete with the Supreme Personality of Godhead, for innumerable universes and their inhabitants come into existence and are annihilated by the simple direction of the Lord.}

\subsubsection{Modal Realism}
In the most general formulation the Multiverse hypothesis takes into account the scenario in which there exists, numerable or innumerable, collection of multiple possible universes. These worlds may include a whole Nature, the concepts of space, time, matter, light, and even its psychological aspects related to the concept of mind. Multiverse as the scientif\/ic concept was introduced by American psychologist and philosopher W. James \cite{james}, who included into human psychology the influence of divine and mystic experiences. The structure of Multiverse, which in fact defines the nature of a possible universe as well as the various relationships between distinguishable universes, are not rigidly established and manifestly depend on a model of Multiverse.

From the philosophical point of view the concept of Multiverse naturally belongs to the logical system investigated by L. Wittgenstein in his famous \emph{Tractatus Logico-Philosophicus} \cite{wittgen}. In this logic the logical truth is defined as a statement true in all possible worlds or under all possible interpretations, and a fact is only true in this world as it has historically unfolded. This ontological system continues the program investigated by G. Frege \cite{frege}, but manifestly neglects the Frege axiom which semantic form is \emph{all true (and, similarly all false) sentences describe the same state of affairs, that is, they have a common referent}. The pioneering formalization of Wittgenstein's \emph{Tractatus} was performed by Polish logician R. Suszko \cite{suszko}, and resulted in so called Non-Fregean Logic in which there are no theorems asserting how many semantic correlates of sentences there can be. Recently this logic has been expanded by M. Omy\l a \cite{omyla} onto the logic connecting situations and objects. Wittgensteinian metaphysics, however, leads to emergence of identical objects existence in diverse worlds.

Counterpart theory of D.K. Lewis \cite{dlewis} showed that such objects should be regarded as similar rather than identical. Lewis elucidated the role of probability and hypothetical statements. His version of modal realism led to all possible worlds possessing equally realistic character like the actual world. In \emph{Parts of Classes}, Lewis applied the pluralistic approach to the foundations of mathematics. He considered such issues like set theory, the Peano arithmetic, and the G\"odel incompleteness theorems to mereology and plural quantif\/ication. In Lewis's approach such a word like ''actual'' is merely indexing procedure, labeling of position within a world. He proposed also the def\/inition of truth, strictly based on Multiverse nature of modal realism, which states that things are necessarily true when they are true in all possible worlds. Lewis was not the f\/irst philosopher studying possible worlds, but contributed the essential idea about equally concreteness of all possible worlds, and created the concept of the world in which an existence of the object is no more real than an existence of this object in different possible world. Similarly as in the case {\AE}ther the concept of Multiverse manifestly violates the maxim due to English theologian and a member of the the mendicant Order of Friars Minor (Franciscan) William Ockham, called the Ockham Razor. The Ockham Razor says <<no>> to multiply entities, because of the Multiverse hypothesis is beyond being a necessary explanation of the facts which theories want to describe.

Different possible worlds are propagated also by modern American philosopher and logician S.A. Kripke \cite{kripke}. He has described modality via using of a metaphysical route, and employed them to semantics, what resulted in so called Kripke's theory of truth. In this theory a natural language contains its own truth predicate without rising contradiction. Involving the property of partial def\/inition of truth over the set of grammatically well-formed sentences in the language, Kripke recursively showed that a language can consistently contain its own truth predicate. In other words Kripke negated the impossibility of such a situation deduced by A. Tarski \cite{tarski}. In Kripke's view truth predicate adds new sentences to the language and truth is the union of all the elements, i.e. is in turn def\/ined for all of them. Inf\/inite number of steps establishes ''f\/ixed point'' in the language, which can be treated as the fundamental natural language containing its own truth predicate.

Another pluralistic formulations of the problem of truth involve correspondence, coherence and constructiveness. The approach due to C. Wright \cite{wright} proposes that truth must not be a single discourse-invariant analog of identity, and that there are the only certain principles of application the truth predicate to a sentence, i.e. some platitudes about true sentences. Wright emphasizes the crucial role of the context, and def\/ines a truth predicate as superassertible if and only if it is assertible in a certain state of information. He did not proposed any mechanism for improving or growing of such a state of information. Because arbitrary standards, norms, and habits question the discourse, he gives the fundamental role for assertiveness. The approach proposed by M.P. Lynch \cite{lynch} claims that truth is multiply functional property. In \emph{Truth in Context} he proposed a path where metaphysical pluralism is consistent with robust realism about truth. His studies on investigated so called \emph{relativistic Kantianism}, i.e. taking of facts and propositions as relative without implications about relativity of an ordinary truth. According to Lynch truths are relative, but individual concepts of truth must not be. In \emph{True to Life} Lynch discussed basic truisms about truth: objectivity, goodness, and arising by worthiness of requesting. He considered mental origins of cynicism, and presented inadequacy of numerous theories of truth. Lynch defends caring about truth, and argues that truth has real value for a happy life.

H.N. Goodman \cite{goodman} gave far from modal realism contributions to the Multiverse hypothesis. He exalted artistry in human-world cognitive relationship, and argued artworks as symbols referring and constructing diverse worlds. Because any human activity is an artistry the Goodman approach is general. According to this the interpretation is fundamentally unif\/ied to the world via the symbols. The worlds demand interpretation of the symbols they contribute to construct, and with no interpretation of the symbols the worlds do exist. Perception, understanding, experience, and discovering use symbols. The interest in symbols is cognitive, what Goodman advocates as cognitivism.

\subsubsection{Many-Worlds Interpretation}
The straightforward philosophical implication of the Multiverse hypothesis to theoretical physics is the metatheory of quantum theory called \emph{relative state} (RS) formulation of quantum mechanics. The foundations of this theory were presented by H. Everett \cite{everett} in his doctoral dissertation supervised by B.S. DeWitt. On pp. 8-9, within the introductory part of the Everett thesis, one can f\/ind the statement\\

\noindent\textit{Since the universal validity of the state function description is asserted, one can regard the state functions themselves as the fundamental entities, and one can even consider the state function of the entire universe. In this sense this theory can be called the theory of the ''universal wave function,'' since all of physics is presumed to follow from this function alone.}\\

\noindent By direct application of the classical mechanical procedure for def\/ining probability, Everett derived the Born rule describing probabilities in quantum mechanics and proved its universality. DeWitt \cite{mwid} reincarnated relative state formulation as Many-Worlds Interpretation (MWI), and together with his another PhD student R.N. Graham alternatively derived the Born rule showing that for inf\/inite number of worlds, i.e. in the situation for which the statistical laws of quantum theory are inadequate, their norm becomes inf\/inite. Everett's thesis, M. Born paper and several another papers on MWI are collected in the book \cite{mwi}.

In addition A.M. Gleason \cite{gleason} and J.B. Hartle \cite{hartle} independently obtained the results of Everett's thesis. J.B. Hartle and S.W. Hawking \cite{harthawk} used MWI results to description of initial conditions for Big Bang cosmology by the solution of Wheeler--DeWitt equation. Applying cold chaotic/eternal inf\/lation A. Linde \cite{alin} proposed the first Multiverse cosmology, where randomly emerging events have independent initial conditions, and partially nucleate in space-time foam as bubbles. J.S. Bell in \cite{bell} straightforwardly supported Many-Worlds Interpretation. The direct cosmological results following from MWI context so called is Anthropic Cosmological Principle, discussed by in the book of J.D. Barrow \& F.J. Tipler \cite{bartip}, and in the book of Tipler \cite{tipp}. In the book edited by Penrose \& Isham one f\/inds topical Tipler article in which he expresses the following opinion about Many-Worlds Interpretation\\

\noindent\textit{I then asked, 'Who does not believe in the many-worlds interpretation?'. About 30 hands went up (including those of Roger Penrose and Bob Wald); clearly Bryce [DeWitt], David [Deutsch] (and I) were in a minority at this meeting. Finally I asked, 'Who is neutral on the many-worlds interpretation?'. The remaining 20 hands went up. I shall do my best in this short paper to persuade both the sceptics and those have not yet formed an opinion as to the validity of the many-worlds interpretation (MWI) that this interpretation is philosophically more beautiful than competing interpretations, and that it can be used in quantum cosmology as a powerful tool not only to interpret the wave function of the universe, but also to give us some information about the equation which this wave function obeys. (\ldots) Most sceptics, I've found, have a mistaken idea of what the MWI really means, so it behoves me to review it brief\/ly. The MWI is a theory of measurement, so it is concerned with describing how the universe looks to us qua human beings. At some stage during any measurement, the information is digitalized, and this is true even for the measurement of continuous variables, for example position or momentum. A typical position measurement — of an a-particle nucleus, say — would be carried out by letting the a-particle pass through an array of atoms such as those of a photographic plate. The array cannot make position measurements of unlimited accuracy; at best, the accuracy would be limited by the size of the atom. Even if we were to improve the accuracy of the position measurement at this level of the measuring process, the position measurement would in the end be digitalized when it is transmitted to human beings, for the data corresponding to an arbitrarily precise position measurement would in general exceed the storage capacity of a human brain. Hence we can model any measurement by a measurement of a discrete variable.}\\

Many-Worlds Interpretation possesses numerous applications and references. J.R. Gribbin \cite{gribbin} discussed Schr\"odinger's cat paradox and Multiverse. M. Lockwood \cite{lockwood}, M. Gell-Mann \& J.B. Hartle \cite{gellhart}, D. Albert \cite{albert}, R. Penrose \cite{penrose}, D.J. Chalmers \cite{chalmers}, and D.E. Deutsch \cite{deutsch} attempted to construct theory of evolution directly based on MWI. M. Kaku \cite{kaku} applied the idea of parallel worlds, i.e. worlds within the Multiverse, to speculations within String Theory. R. Plaga \cite{plaga} proposed the empirical test of Many-Worlds Interpretation. J.A. Barrett \cite{barrett} discussed details of both the Everett and ''no collapse'' interpretations of quantum mechanics. Deutsch \cite{deutsch1} also suggested to examine MWI by using of quantum computer. Applying MWI he derived the information-theoretic Born rule, and determined Multiverse by information f\/low. D. Page \cite{dpage} sees the essential support of MWI in cosmological observations. L. Polley \cite{polley} derived the Born rule by symmetry arguments instead of Deutsch's assumptions. Derivation of the Born rule by W.H. Zurek \cite{zurek} involved envariance, while he deduced probabilities from entanglement. In another considerations \cite{zurek1} Zurek discussed also the problem of causality, interaction with environment, and what he calls quantum darwinism describing proliferation, in the environment, of multiple records of selected states of a quantum system. The arguments due to Deutsch were improved on by D. Wallace \cite{wallace} and S. Saunders \cite{saunders}. J.A. Wheeler \cite{whebio} also expressed his views on Everettian relative state. L. Smolin \cite{smolin1} and B. Greene \cite{greene} supported Multiverse in the context of String Theory. M. Gardner \cite{gardner} did critical analysis of Many-Worlds Interpretation. C. Bruce \cite{bruce} focused on Schr\"odinger's cat paradox, and L. Randall \cite{randall} studied the context of brane worlds. Various experienced scientists and scholars expressed their opinions about Multiverse in the book edited by B. Carr \cite{carr}. M. Tegmark \cite{mtegm} provided the classif\/ication of multiple universes. P. Byrne \cite{byrne} performed a detailed analysis of the Everett heritage. Interesting looking studies which look like topically have been presented recently by V. Allori et al \cite{allori} and by S. Osnaghi et al \cite{osnaghi}. A. Jenkins \& G. Perez \cite{jenper}, and J. Feng \& M. Trodden \cite{fentro} have discussed the observational context of MWI. Also recently I have supported a certain particular context of the Multiverse hypothesis \cite{glin}.

There are also another various intriguing contexts of Many-World Interpretation and Multiverse. Interestingly recently, A. Kent \cite{kent} and N.P. Landsman \cite{landsman} have criticized the foundations of the Born rule applied in the context of Many-Worlds Interpretation. The constructive application of MWI was performed by D. Parf\/it \cite{parfit}, who discussed the concept of personal identity. His conclusions and deductions are essentially intriguing and are questioning the fundamental status quo of mental health. Namely, Parf\/it presented certain sample situations in which a unif\/ied person splits into several copies, and justif\/ied ambiguousness in f\/ixation of the state of personality. He concluded that dividing of ''I'' does leads to inadequacy of the concept of personal identity, which is the most celebrated and well-established concept of psychology and psychiatry. However, productive introduction of medical norms and social standards based on MWI is rather far perspective.

\subsubsection{String Theory and Anthropic Principle}
\noindent The Anthropic Principle leads to straightforward various implications of Many-World Interpretation and Multiverse hypothesis within String Theory. The contributions based on or related to Multiverse and MWI were presented by S. Weinberg \cite{weinanth}, G. 't Hooft \cite{thooft}, S.W. Hawking \cite{hawki}, M.J. Rees \cite{rees}, J.D. Bekenstein \cite{beken}, and L. Susskind \cite{susskind}. Weinberg, in his famous and intriguing book \emph{Dreams of a Final Theory} \cite{finalth}, unambiguously and manifestly expressed the beliefs which are strictly related to the Many-Worlds Interpretation\\

\noindent\textit{The f\/inal approach is to take the Schrodinger equation seriously (\ldots) In this way, a measurement causes the history of the universe for practical purposes to diverge into different non-interacting tracks, one for each possible value of the measured quantity. (\ldots) I prefer this last approach.}\\

\noindent There are also critical standpoints about the String Theory context of Multiverse hypothesis. Smolin in his another book \cite{smolin} is too critical\\

\noindent\textit{The search for quantum gravity is a true quest. The pioneers were explorers in a new landscape of ideas and possible worlds. (\ldots) The scenario of many unobserved universes plays the same logical role as the scenario of an intelligent designer. Each provides an untestable hypothesis that, if true, makes something improbable seem quite probable. (\ldots) The anthropic principle that Susskind refers to is an old idea proposed and explored by cosmologists since the 1970s, dealing with the fact that life can arise only in an extremely narrow range of all possible physical parameters and yet, oddly enough, here we are, as though the universe had been designed to accommodate us (hence the term "anthropic"). The specif\/ic version that Susskind invokes is a cosmological scenario that has been advocated by Andrei Linde for some time, called eternal inf\/lation. According to this scenario, the rapidly inf\/lating phase of the early universe gave rise not to one but to an inf\/inite population of universes. You can think of the primordial state of the universe as a phase that is exponentially expanding and never stops. Bubbles appear in it, and in these places the expansion slows dramatically. Our world is one of those bubbles, but there are an inf\/inite number of others. To this scenario, Susskind adds the idea that when a bubble forms, one of the vast number of string theories is chosen by some natural process to govern that universe. The result is a vast population of universes, each of which is governed by a string theory randomly chosen from the landscape of theories. Somewhere in the so-called multiverse is every possible theory in the landscape. (\ldots) I f\/ind it unfortunate that Susskind and others have embraced the anthropic principle, because it has been understood for some time that it is a very poor basis for doing science. Since every possible theory governs some part of the multiverse, we can make very few predictions. (\ldots) It is not hard to see why. To make a prediction in a theory that posits a vast population of universes satisfying randomly chosen laws, we would f\/irst have to write down all the things we know about our own universe. These things would apply to some number of other universes as well, and we can refer to the subset of universes where these facts are true as \emph{possibly true universes}.}\\

\noindent P. Woit in his book \emph{Not Even Wrong} \cite{woit} is critical too, but less radical\\

\noindent\textit{The anthropic principle comes in various versions, but they all involve the fact that the laws of physics must be of a nature that allows the development of intelligent beings such a ourselves. Many scientists believe that this is nothing more than a tautology, which while true, can never be used to create a falsif\/iable prediction, and thus can not be part of scientif\/ic reasoning. Controversy has arisen as a signif\/icant group of superstring theorists have begun to argue that superstring theory's inability to make predictions is not a problem with the theory, but a ref\/lection of the true nature of the universe. (\ldots) Weinberg suggested that perhaps the explanation of the problem of the small size of the cosmological constant was the anthropic principle. The idea is that there are huge number of consistent possible universes, and that our universe is part of some larger multiverse or megaverse. Quite naturally, we f\/ind ourselves in a part of this multiverse in which galaxies can be produced and thus intelligent life can evolve. If this is the case, there is no hope of ever predicting the value of the cosmological constant, since all one can do is note the tautology that it has a value consistent with one's existence.}\\

Anyway, in theoretical physics the crucial question is whether arbitrary mathematical creations can be applied to effective and constructive description of the physical Reality. In this book I shall present straightforwardly that certain selected ideas lead to such a constructive approach. It must be emphasized that in general it is clear what is the best argument for applicability of any mathematics to making of a constructive physical scenarios. Namely, this is a problem of choice which always should be verif\/ied empirically. A mathematics is physical if and only if it leads to a physical truth, which is independent on various diplomatic operations. Always, however, constructive failures are much more valuable then nonconstructive successes. Regarding the opinion due to Wolfgang Pauli: \emph{not even wrong}.

\mainmatter

\part{Lorentz Symmetry Violation}

\chapter{Deformed Special Relativity}

Special Relativity can be formulated basing on the momentum space, in which the Einstein energy-momentum relation \index{Einstein energy-momentum relation} holds
\begin{equation}\label{ec}
  E^2=m^2c^4+p^2c^2,
\end{equation}
where $c$ is speed of light in vacuum, $p$ the momentum value of a relativistic particle possessing mass $m$. Factually, the relation (\ref{ec}) can be rewritten in the more conventional form
\begin{equation}
E^2-p^2c^2-m^2c^4=0,
\end{equation}
which def\/ines the Einstein Hamiltonian constraint of Special Relativity\index{Einstein Hamiltonian constraint}. Solving of this constraint with respect to a particle energy $E$ leads to f\/ixation of the energy as the approximation. For Special Relativity the Lorentz symmetry\index{Lorentz symmetry} holds and (\ref{ec}), as a quadratic form on the Minkowski energy-momentum space \index{Minkowski energy-momentum space} of a particle $p^{\mu}=[E,p^ic]$, is Lorentz invariant.

From the modal realism point of view, the Lorentz invariance\index{Lorentz invariance} of the constraint (\ref{ec}) leads to the fundamental problem. Namely, the equation (\ref{ec}) is not the only one possible such a quadratic form. Naturally, the constructive generalization of Special Relativity is
\begin{equation}
E^2=m^2c^4+p^2c^2+\Delta(E,p),\label{dep}
\end{equation}
where the deformation $\Delta(E,p)$ contains also a set of free parameters, which I shall call deformation parameters. Such an extension, however, determines the Multiverse \index{Multiverse}in which the possible worlds are diverse {\AE}ther theories, and such an {\AE}ther \index{{\AE}ther} theory def\/ines new physical beings and effects. For Lorentz invariance\index{Lorentz invariance} a whole expression (\ref{dep}) must be a quadratic form on the Minkowski space. This chapter presents the updated results of the author paper \cite{laglinka1}. Certain part of these results is removed and replaced by more adequate ideas.

\section{The linear deformation}
Let us consider f\/irst the deformation of Special Relativity due to a simple linear term in a particle momentum $p$
\begin{equation}\label{ecx}
  E^2=m^2c^4+p^2c^2+\mathcal{P}^ip_ic^2,
\end{equation}
where $\mathcal{P}^i$, $i=1,2,3$ is a three-vector of deformation parameters, which in fact is certain constant reference momentum 3-vector distinguishing axes related to its direction. Let us introduce also the length of the 3-momentum vector $\mathcal{P}^i$ as
\begin{equation}
\mathcal{P}=\sqrt{\mathcal{P}^i\mathcal{P}_i}.\label{plength}
\end{equation}
It is not diff\/icult to deduce that the deformation momentum three-vector $\mathcal{P}^i$ can be physically understood as the {\AE}ther momentum vector\index{{\AE}ther momentum}. Therefore, the linear deformation of Special Relativity (\ref{ecx}) describes the simplest type interaction between a particle and the {\AE}ther in the Minkowski energy-momentum space \index{Minkowski energy-momentum space}. In such a situation the {\AE}ther is characterized by a constant momentum vector f\/ield dealt by the direction of the three momentum $\mathcal{P}^i$.

From the Minkowski space point of view the modif\/ied Special Relativity (\ref{ecx}) corresponds to nontrivially deformed invariant hyperboloid. The theory (\ref{ecx}) can be described by a quadratic form which by elementary algebraic manipulation can be obtained from (\ref{ecx}) as its canonical form
\begin{equation}\label{ec2}
  E^2+\dfrac{1}{4}\mathcal{P}^2c^2=\left(\dfrac{\mathcal{P}^i}{\mathcal{P}}p_ic+\dfrac{\mathcal{P}c}{2}\right)^2+m^2c^4,
\end{equation}
which can be worked out by multiple ways, i.e. in itself determines the Multiverse \index{Multiverse}of possible physical theories. By the identif\/ication method \index{identif\/ication method} there is a lot of routes of possible interpretations between all the parts of the equation (\ref{ec2}). Let us consider here the three basic interpretations.
\subsection{The Dirac equation and the new algebra}
The f\/irst case is the following identif\/ication
\begin{equation}
  \left\{\begin{array}{l}m^2c^4=\dfrac{\mathcal{P}^2c^2}{4}\\
E^2=\left(\dfrac{\mathcal{P}^i}{\mathcal{P}}p_ic+\dfrac{\mathcal{P}c}{2}\right)^2\end{array}\right.,
\end{equation}
which leads to the mass values
\begin{equation}
  \pm mc=\dfrac{\mathcal{P}}{2},\label{casedir}
\end{equation}
where we have included negative sign mass as physical. The second equation is not diff\/icult to solve. The solution determines energy as
\begin{equation}
\gamma^0E=\gamma^ip_ic\pm mc^2,\label{dirrac}
\end{equation}
where $\gamma$'s are def\/ined by the following relations
\begin{eqnarray}
{\gamma^0}^2&=&1,\label{gam0}\\
\gamma^i&=&\dfrac{\mathcal{P}^i}{\mathcal{P}}.\label{gami}
\end{eqnarray}
where the equality (\ref{gami}) for the present case (\ref{casedir}) takes the form
\begin{equation}
\gamma^i=\dfrac{\mathcal{P}^i}{2mc}.\label{gamix}
\end{equation}
In other words from Eq. (\ref{gami}) one has $\mathcal{P}^i=\mathcal{P}\gamma^i$ what, after application of the usual raising the lower index $\gamma_i=\delta_{ij}\gamma^j$, allows to write
\begin{equation}
\mathcal{P}^i\mathcal{P}_i=\mathcal{P}^2\gamma^i\gamma_i,\label{pipigam}
\end{equation}
i.e. for correctness with the def\/inition (\ref{plength}) the identity $\gamma^i\gamma_i\equiv1$ must hold in arbitrary case. It is, albeit blatantly incorrect if one treats $\gamma^i$ as the Dirac gamma matrices obeying the Clifford algebra \index{Clifford algebra}
\begin{equation}
\left\{\gamma^i,\gamma^j\right\}_C=2\delta^{ij},\label{alclifford}
\end{equation}
where $\delta_{ij}$ is $D\times D$ unit matrix, for $\gamma^i\gamma_i=\delta^i_i=D$ and therefore in 3-dimensional space $\gamma^i\gamma_i=3$. Factually such an algebraical treatment (\ref{alclifford}) was investigated by W. Pauli \cite{paulidirac}, and is commonly used in both relativistic quantum mechanics as well as quantum f\/ield theory \cite{standardth}. Moreover, the Clifford algebra \index{Clifford algebra}is the fundamental computational rule in particle physics, for instance for cross sections for reactions. Alternatively, the relations (\ref{plength}) and (\ref{pipigam}) can be understood as the suggestion that the linear deformation is true for the only $D=1$ dimensional space, i.e. that the {\AE}ther \index{{\AE}ther}exists only in one-dimensional space. It looks like, however, that such an explanation is logically inconsistent and the Clifford algebra \index{Clifford algebra}(\ref{alclifford}) must be exchanged for the more adequate one.

Because, however, the def\/inition (\ref{gami}) imparts the {\AE}ther momentum \index{{\AE}ther momentum} noncommutative nature there is difference between expressions $\mathcal{P}^i\mathcal{P}_i$ and $\mathcal{P}_i\mathcal{P}^i$ which in the Abelian case are identical. By this reason let us introduce the def\/inition of the {\AE}ther momentum square including the noncommutative nature of $\mathcal{P}^i$
\begin{equation}
\mathcal{P}^2=\dfrac{1}{2}\left(\mathcal{P}^i\mathcal{P}_i+\mathcal{P}_i\mathcal{P}^i\right),
\end{equation}
which in the commutative situation leads to (\ref{plength}). Thus one obtains
\begin{equation}
\mathcal{P}^2=\dfrac{1}{2}\left(\mathcal{P}^2\gamma^i\gamma_i+\mathcal{P}^2\gamma_i\gamma^i\right),
\end{equation}
what after using the identity $\gamma_i=\delta_{ij}\gamma^j$ gives
\begin{equation}
\mathcal{P}^2=\dfrac{\mathcal{P}^2}{2}\delta_{ij}\left(\gamma^i\gamma^j+\gamma^j\gamma^i\right),
\end{equation}
and results in the basic relation
\begin{equation}
1=\dfrac{1}{2}\delta_{ij}\left\{\gamma^i,\gamma^j\right\},\label{gammas}
\end{equation}
which can be expressed in more conventionally
\begin{equation}
\left\{\gamma^i,\gamma^j\right\}=\dfrac{2}{\delta_{ij}}.\label{gammasy}
\end{equation}
The Clifford algebra \index{Clifford algebra}(\ref{alclifford}) can be reconstructed from (\ref{gammas}) if and only if one put by hands the identity $\delta^{ij}=\dfrac{1}{\delta_{ij}}$, i.e. $\delta^{ij}\delta_{ij}=\delta^i_i=1$. However, such an algebraical strategy is in general incorrect because of in the $D$ dimensional space case $\dfrac{1}{\delta_{ij}}=\dfrac{\delta^{ij}}{\delta_{ij}\delta^{ij}}=
\dfrac{\delta^{ij}}{\delta^i_i}=\dfrac{1}{D}\delta^{ij}$. Strictly speaking it means that the spatial gamma matrices (\ref{gami}) do not belong to the Clifford algebra \index{Clifford algebra}for $D\neq1$.

In itself, however, the obtained problem is solved by introduction of the new algebra, which can be established straightforwardly and rather easy. Namely, application of the relation $\dfrac{1}{\delta_{ij}}=\dfrac{1}{D}\delta^{ij}$ within the equation (\ref{gammasy}) results in the following anticommutator
\begin{equation}
\left\{\gamma^i,\gamma^j\right\}=\dfrac{2}{D}\delta^{ij},\label{anticom}
\end{equation}
which for 3-dimensional case gives the rule
\begin{equation}
\left\{\gamma^i,\gamma^j\right\}=\dfrac{2}{3}\delta^{ij}.\label{anticomi}
\end{equation}

It is not diff\/icult to see straightforwardly that the extension of spatial gamma matrices to the space-time version
\begin{equation}\label{q1}
\gamma^i\rightarrow\gamma^\mu=\left(-\gamma^0,\gamma^i\right),
\end{equation}
leads to simple generalization of the basic formula (\ref{gammas})
\begin{equation}
1=\dfrac{1}{2}\eta_{\mu\nu}\left\{\gamma^\mu,\gamma^\nu\right\},\label{gammasa}
\end{equation}
where $\eta_{\mu\nu}=\mathrm{diag}(-1,1,1,1)$ is metric the Minkowski space-time. Therefore, one obtains the new four-dimensional gamma matrix algebra
\begin{equation}
\left\{\gamma^\mu,\gamma^\nu\right\}=\dfrac{2}{4}\eta^{\mu\nu}=\dfrac{1}{2}\eta^{\mu\nu},\label{new4alg}
\end{equation}
which is distinguishable from the space-time Clifford algebra\index{Clifford algebra}
\begin{equation}
\left\{\gamma^\mu,\gamma^\nu\right\}_C=2\eta^{\mu\nu}.\label{4cliff}
\end{equation}
It is not diff\/icult to proof that for $D+1$ dimensional space-time, where $D$ is the spatial dimension, the new algebra is
\begin{equation}
\left\{\gamma^\mu,\gamma^\nu\right\}=\dfrac{2}{D+1}\eta^{\mu\nu}.\label{newD}
\end{equation}
Let us consider now the constraint (\ref{dirrac}), and apply to this the canonical relativistic quantization\index{quantization!relativistic canonical} procedure
\begin{equation}\label{q2}
  (E,p_ic)\rightarrow i\hslash\partial_{\mu}=i\hslash(-\partial_0,c\partial_i).
\end{equation}
The resulting equation
\begin{equation}
-i\hslash\gamma^0\partial_0\psi=(ic\hslash \gamma^i\partial_i\pm mc^2)\psi,
\end{equation}
in fact is the Dirac equation \index{Dirac equation}
\begin{equation}
(i\hslash\gamma^\mu\partial_\mu\pm mc^2)\psi=0,\label{diracequation}
\end{equation}
for which the Lorentz symmetry\index{Lorentz symmetry} is fully validate. In other words we have obtained the Dirac equation\index{Dirac equation}, where however, the gamma matrices do not belong to the Clifford algebra \index{Clifford algebra}but obey the new algebra (\ref{new4alg}). Interestingly, factually we have generated the Dirac equation \index{Dirac equation}independently on the gamma matrix algebra, what suggests that the Dirac equation \index{Dirac equation}must not be obtained as ''square-root taking'' of the Klein--Gordon equation\index{Klein--Gordon equation}, like Dirac originally deduced and applied \cite{diraceqn}. Dirac's computations did not involve relations between gamma matrices manifestly, and therefore his deductions are true.

Interestingly, the general relation between the new algebra (\ref{anticom}) and the Clifford algebra \index{Clifford algebra} (\ref{alclifford}) can be established straightforwardly
\begin{equation}
\left\{\gamma^i,\gamma^j\right\}=\dfrac{1}{D}\left\{\gamma^i,\gamma^j\right\}_C.\label{relnc}
\end{equation}
In other words, because of the Clifford algebra \index{Clifford algebra}is here a unit $D\times D$ matrix, the limit $D\rightarrow\infty$ gives trivially
\begin{equation}
\lim_{D\rightarrow\infty}\left\{\gamma^i,\gamma^j\right\}=0,
\end{equation}
while the Clifford algebra \index{Clifford algebra}is conserved in such a limit situation. Albeit, in the light of the general relation (\ref{gammas}) this case generates the blatantly incorrect equality $1=0$. It suggests that $D\rightarrow\infty$ is a nonphysical situation when the spatial metric is a constant $D\times D$ unit matrix $\delta_{ij}$. Possibly, such a inf\/inite limit has a sense if and only if the space/space-time metric depends on the number of dimensions $D$. However, we shall not discuss such examples in this book. It is easy to see that similar situation is present for the space-time new algebra (\ref{newD}) and the space-time Clifford algebra (\ref{4cliff})\index{Clifford algebra}
\begin{eqnarray}
\left\{\gamma^\mu,\gamma^\nu\right\}&=&\dfrac{1}{D+1}\left\{\gamma^\mu,\gamma^\nu\right\}_C,\label{relnc1}\\
\lim_{D\rightarrow\infty}\left\{\gamma^\mu,\gamma^\nu\right\}&=&0.
\end{eqnarray}

Let us establish certain identity important for this chapter. Namely, let us compute $\gamma^i\gamma_i$ in the light of the basic relation (\ref{gammas})
\begin{eqnarray}
&&\gamma^i\gamma_i=\gamma^i\delta_{ij}\gamma^j=\delta_{ij}\gamma^i\gamma^j=
\dfrac{1}{2}\left(\delta_{ij}+\delta_{ji}\right)\gamma^i\gamma^j=
\dfrac{1}{2}\left(\delta_{ij}\gamma^i\gamma^j+\delta_{ji}\gamma^i\gamma^j\right)=\nonumber\\
&&=\dfrac{1}{2}\left(\delta_{ij}\gamma^i\gamma^j+\delta_{ij}\gamma^j\gamma^i\right)=
\dfrac{1}{2}\delta_{ij}\left(\gamma^i\gamma^j+\gamma^j\gamma^i\right)=
\dfrac{1}{2}\delta_{ij}\left\{\gamma^i,\gamma^j\right\}=1,\label{gigi}
\end{eqnarray}
what differs from the $D$-dimensional Clifford algebra \index{Clifford algebra}result $\gamma^i\gamma_i=D$. The result (\ref{gigi}) holds for any $D$, and f\/irst of all also for the space-time version of the new algebra $\gamma^\mu\gamma_\mu=1$, and is independent on the spatial dimension. This is an important elucidation.

Because of the presented approach, based on \emph{identif\/ication method} \index{identif\/ication method}, is essentially new and evidently changes deductions and explanations related to the Clifford algebra\index{Clifford algebra}, we shall call the new algebra \emph{the {\AE}ther algebra} \index{{\AE}ther algebra}. The presented way of reasoning shows that the Dirac equation \index{Dirac equation}is not related to the Clifford algebra \index{Clifford algebra} only, but factually can be deduced by a way very far from the Dirac ''square-root'' technique\index{Dirac ''square root'' technique} and produce other algebras of gamma matrices. We showed here that in general gamma matrices can be deduced by techniques different from the methods of relativistic physics, propagated by Dirac in his contributions to quantum mechanics. Albeit, we regard Dirac's results, and particularly their diverse consequences for particle physics, as the inspiration. The {\AE}ther algebra proposed above can be related to other, possibly unknown, particles and forces. Possibly, the Dirac equation \index{Dirac equation}with non-Dirac gamma matrices def\/ines an effective theory.

\subsection{Another Identif\/ications}
The second possible identif\/ication is
\begin{equation}\label{lcase1}
  \left\{\begin{array}{l}m^2c^4=E^2\\
  \dfrac{\mathcal{P}^2c^2}{4}=\left(\dfrac{\mathcal{P}^i}{\mathcal{P}}p_ic+\dfrac{\mathcal{P}c}{2}\right)^2\end{array}\right..
\end{equation}
In such a situation the f\/irst equality leads to the relation
\begin{equation}\label{gamener}
  \gamma^0E=\pm mc^2,
\end{equation}
where $\gamma^0$ is def\/ined by the relation (\ref{gam0}). The formula (\ref{gamener})looks like the Einstein mass-energy relation \index{Einstein mass-energy relation}. Similarly the second equality in (\ref{lcase1}) leads to the following nontrivial and manifestly distinguishable physical situations
\begin{eqnarray}
  \dfrac{\mathcal{P}^i}{\mathcal{P}}p_ic&=&0,\label{coi1}\\
  \dfrac{\mathcal{P}^i}{\mathcal{P}}p_ic&=&-\mathcal{P}c.\label{coi2}
\end{eqnarray}
Similarly as in the previous case one can introduce the Clifford algebra \index{Clifford algebra}of spatial gamma matrices given by the relations (\ref{gam0}) and (\ref{gami}), and applying the canonical relativistic quantization \index{quantization!relativistic canonical}procedure (\ref{q2}) one obtains the appropriate projections conditions. The f\/irst such a condition follows from (\ref{gamener}) and has a form
 \begin{equation}
 \left(i\hslash\gamma^0\partial_0\mp mc^2\right)\psi=0,\label{timecomp}
 \end{equation}
 while the second one, following from (\ref{coi1}) and (\ref{coi2}), has the form
 \begin{eqnarray}
 ic\hslash\gamma^i\partial_i\psi&=&0,\label{fdr1}\\ \left(ic\hslash\gamma^i\partial_i+\mathcal{P}c\right)\psi&=&0.\label{fdr2}
 \end{eqnarray}
Interestingly, the equation (\ref{timecomp}) added to the condition (\ref{fdr1}) leads to the usual Dirac equation \index{Dirac equation}(\ref{diracequation}), while addition of the condition (\ref{fdr1}) to the equation (\ref{timecomp}) allows to establish the new quantum relativistic equation jointing a Dirac particle and the {\AE}ther \index{{\AE}ther}
\begin{equation}
\left(i\hslash\gamma^\mu\partial_\mu+Mc^2\right)\psi=0,\label{neweqn1}
\end{equation}
where $M$ is the effective mass of the particle-{\AE}ther system
\begin{equation}
M=\mp m+\dfrac{1}{c}\mathcal{P},\label{masss}
\end{equation}
i.e. for $[\mathcal{P}]\sim[c]=3\cdot10^8$ the correction due to the {\AE}ther plays an essential physical role. Interestingly, for the positive sign near particle mass $m$ in (\ref{masss}) the effective mass $M$ is always positive, while for the negative sign the effective mass $M$ is positive for $mc<\mathcal{P}$, negative for $mc>\mathcal{P}$, and vanishes when $mc=\mathcal{P}$.

On the one hand the effective mass (\ref{masss}) manifestly contains the correction to a particle mass due to the {\AE}ther momentum value $\mathcal{P}$ but not due to $\mathcal{P}_i$. Such a property involves a situation when the {\AE}ther momentum vector $\mathcal{P}_i$ is nontrivial but its length $\mathcal{P}$ vanishes. In such a case an arbitrary component of the {\AE}ther momentum \index{{\AE}ther momentum} vector is determined by the two remained components which are still arbitrary. It can be seen that then a classical theory is a deformed Special Relativity (\ref{ecx}) while, because $M=\pm m$ by (\ref{masss}), quantum theory (\ref{neweqn1}) is the Dirac relativistic quantum mechanics. In other words the Dirac theory is related not only to the Einstein theory, but possesses wider sense.

On the other hand, however, the quantum theory given by the projections (\ref{timecomp}), and (\ref{fdr1}) and (\ref{fdr2}) carries different content than the usual Dirac relativistic quantum mechanics. In the Dirac theory there is the only one cumulative projection condition (\ref{diracequation}), while in the our theory the conditions (\ref{timecomp}), and (\ref{fdr1}) and (\ref{fdr2}) in general are not cumulative. We mean that we have obtained the condition for time evolution (\ref{timecomp}) and two alternative conditions for spatial evolution (\ref{fdr1}) and (\ref{fdr2}), while in the Dirac theory there is unif\/ied space-time evolution (\ref{diracequation}). By this reason the our situation is physically distinguished from the theory based on four-dimensional Dirac equation \index{Dirac equation}. However, the unif\/ication obtained by simple algebraic sum of the time projection and the spatial projection led us to the usual Dirac theory and the Dirac theory with the effective mass (\ref{masss}). This particular case is within the general theory given by the projection conditions (\ref{timecomp}), and (\ref{fdr1}) and (\ref{fdr2}).

Interestingly, also the classical physics context of the conditions (\ref{coi1}) and (\ref{coi2}) is nontrivial. Namely, these relations establish two possible constraints for a particle momentum components, what allows to express an arbitrary one component of a particle momentum via the {\AE}ther momentum\index{{\AE}ther momentum}. In other words the constraints (\ref{coi1}) and (\ref{coi2}) joint a classical particle with the {\AE}ther\index{{\AE}ther}. Factually, the f\/irst relation is
\begin{equation}\label{cm1}
\mathcal{P}^1p_1+\mathcal{P}^2p_2+\mathcal{P}^3p_3=0,
\end{equation}
and the latter one is
\begin{equation}\label{cm2}
\mathcal{P}^1(p_1+\mathcal{P}_1)+\mathcal{P}^2(p_2+\mathcal{P}_2)+\mathcal{P}^3(p_3+\mathcal{P}_3)=0.
\end{equation}
There are in general three types of solutions for each of these constraints, in which a one component of a particle momentum is dependent on two other (arbitrary) components of particle momentum and all components of the reference momentum 3-vector. The constraint (\ref{cm1}) can be solved by
\begin{eqnarray}
p_i&=&\left(-\dfrac{\mathcal{P}^2}{\mathcal{P}^1}p_2-\dfrac{\mathcal{P}^3}{\mathcal{P}^1}p_3,p_2,p_3\right),\label{st1}\\
p_i&=&\left(p_1,-\dfrac{\mathcal{P}^1}{\mathcal{P}^2}p_1-\dfrac{\mathcal{P}^3}{\mathcal{P}^2}p_3,p_3\right),\label{st2}\\
p_i&=&\left(p_1,p_2,-\dfrac{\mathcal{P}^1}{\mathcal{P}^3}p_1-\dfrac{\mathcal{P}^2}{\mathcal{P}^3}p_2\right).\label{st3}
\end{eqnarray}
Similarly, the constraint given by (\ref{cm2}) possesses in general three possible solutions
\begin{eqnarray}
p_i&=&\left(-\mathcal{P}_1-\dfrac{\mathcal{P}^2}{\mathcal{P}^1}(p_2+\mathcal{P}_2)-\dfrac{\mathcal{P}^3}{\mathcal{P}^1}(p_3+\mathcal{P}_3),p_2,p_3\right),\label{ste1}\\
p_i&=&\left(p_1,-\mathcal{P}_2-\dfrac{\mathcal{P}^1}{\mathcal{P}^2}(p_1+\mathcal{P}_1)-\dfrac{\mathcal{P}^3}{\mathcal{P}^2}(p_3+\mathcal{P}_3),p_3\right),\label{ste2}\\
p_i&=&\left(p_1,p_2,-\mathcal{P}_3-\dfrac{\mathcal{P}^1}{\mathcal{P}^3}(p_1+\mathcal{P}_1)-\dfrac{\mathcal{P}^2}{\mathcal{P}^3}(p_2+\mathcal{P}_2)\right).\label{ste3}
\end{eqnarray}
Particularly, the solution of the f\/irst constraint can be trivial, i.e. $p_i=0$, and the second constraint can be solved simply by $p_i=-\mathcal{P}_i$. Both these cases have a physical interpretation of an inertial reference frame of a particle: either rest frame or motion of a particle under the constant momentum opposite to the {\AE}ther momentum\index{{\AE}ther momentum}. For both these particular solutions the Lorentz symmetry\index{Lorentz symmetry} also holds.

The third interesting identif\/ication is
\begin{equation}\label{3rdid}
  \left\{\begin{array}{l}-E^2=\dfrac{\mathcal{P}^2c^2}{4}\\-m^2c^4=\left(\dfrac{\mathcal{P}^i}{\mathcal{P}}p_ic+\dfrac{\mathcal{P}c}{2}\right)^2\end{array}\right..
\end{equation}
The f\/irst constraint can be resolved straightforwardly with the result
\begin{equation}\label{cm3}
  \pm i\gamma^0E=\dfrac{\mathcal{P}c}{2}.
\end{equation}
The solution of the second equation also can be easy established
\begin{equation}\label{cm3a}
\pm imc^2=\dfrac{\mathcal{P}^i}{\mathcal{P}}p_ic+\dfrac{\mathcal{P}c}{2}.
\end{equation}
Employing (\ref{cm3}) this solution can be written as
\begin{equation}
\pm imc^2=\gamma^ip_ic\pm i\gamma^0E.\label{stton}
\end{equation}
After the canonical relativistic quantization \index{quantization!relativistic canonical}the solution (\ref{stton}) leads to the equation
\begin{equation}
\left(ic\hslash\gamma^i\partial_i\mp\hslash\gamma^0\partial_0\mp imc^2\right)\psi=0,\label{stton1}
\end{equation}
having blatantly real and imaginary parts which are
\begin{eqnarray}
\mp\hslash\gamma^0\partial_0\psi&=&0,\label{stton1a}\\
\left(c\hslash\gamma^i\partial_i\mp mc^2\right)\psi&=&0,\label{stton1b}
\end{eqnarray}
and must be treated as the system of equations.

With using of (\ref{cm3}) the solution (\ref{cm3a}), however, can be rewritten in another form and understood alternatively. Namely, because one has determined the {\AE}ther momentum \index{{\AE}ther momentum}value via a particle energy and non-determined the {\AE}ther momentum vector, the equation (\ref{cm3a}) can be presented in an equivalent form
\begin{equation}\label{cm3ab}
\pm imc^2=\dfrac{\mathcal{P}^ic}{\pm 2i\gamma^0E}p_ic\pm i\gamma^0E.
\end{equation}
One sees straightforwardly, however, that the equation (\ref{cm3ab}) after elementary algebraic manipulations can be presented in a form of a quadratic equation
\begin{equation}\label{cm4}
(\pm\gamma^0E)^2\mp mc^2(\pm\gamma^0E)-\dfrac{\mathcal{P}^ic}{2}p_ic=0.
\end{equation}
Now it is easy to conclude that essentially for arbitrary sign of $m$ this equation can be rewritten as
\begin{equation}\label{convfrom}
(\gamma^0E)^2\pm mc^2(\gamma^0E)-\dfrac{\mathcal{P}^ic}{2}p_ic=0,
\end{equation}
while one can treat the positive mass case in the constraint (\ref{cm3ab}) as the physical situation. The canonical relativistic quantization\index{quantization!relativistic canonical} applied to the constraint (\ref{convfrom}) results in the following projection condition
\begin{equation}
\left(-\hslash^2\partial_0^2\pm i\hslash mc^2\gamma^0\partial_0- ic\hslash\dfrac{\mathcal{P}^ic}{2}\partial_i\right)\psi=0,
\end{equation}
which after multiplication by $-1/mc^2$ and taking into account \emph{ad hoc} the following identif\/ication of the spatial gamma matrices
\begin{equation}
\gamma^i=\dfrac{\mathcal{P}^i}{2mc}.\label{spatgam}
\end{equation}
can be easily led to more convenient form
\begin{equation}\label{projec}
\left(\dfrac{\hslash^2}{mc^2}\partial_0^2\mp i\hslash\gamma^0\partial_0+ic\hslash\gamma^i\partial_i\right)\psi=0.
\end{equation}
Interestingly, when one takes the plus sign in (\ref{projec}) as the physical case than this condition can be written as
\begin{equation}
\left(i\hslash\gamma^\mu\partial_\mu+\dfrac{\hslash^2}{mc^2}\partial_0^2\right)\psi=0,
\end{equation}
and then the Dirac equation \index{Dirac equation}is recovered
\begin{equation}\label{direqu}
(i\hslash\gamma^\mu\partial_\mu+Mc^2)\psi=0,
\end{equation}
where $M$ is an effective mass term
\begin{equation}
Mc^2\psi\equiv\dfrac{\hslash^2}{mc^2}\partial_0^2\psi.
\end{equation}
It is easy to see that the Lorentz symmetry\index{Lorentz symmetry} is validate in such a mass generation mechanism\index{mass generation mechanism}.

The case of the minus sign in the equation (\ref{projec}) can be also considered in terms of the Dirac equation \index{Dirac equation}(\ref{direqu}), but then the mass term is determined via one of two equivalent conditions
\begin{eqnarray}
Mc^2\psi&\equiv&\left(-\dfrac{\hslash^2}{mc^2}\partial_0^2-2ic\hslash\gamma^i\partial_i\right)\psi,\label{mt1}\\
Mc^2\psi&\equiv&\left(\dfrac{\hslash^2}{mc^2}\partial_0^2-2i\hslash\gamma^0\partial_0\right)\psi.\label{mt2}
\end{eqnarray}

From the classical physics point of view, however, the mass term conditions (\ref{mt1}) and (\ref{mt2}) are respectively
\begin{eqnarray}
Mc^2&=&-\dfrac{E^2}{mc^2}-2\dfrac{\mathcal{P}^i}{2mc}p_ic,\\ Mc^2&=&\dfrac{E^2}{mc^2}-2\gamma^0E,
\end{eqnarray}
where we used ${\gamma_0}^2=1$, and their equivalence leads to the constraint
\begin{equation}
\dfrac{E^2}{mc^2}-\gamma^0E+\dfrac{\mathcal{P}^i}{2mc}p_ic=0,
\end{equation}
which can be rewritten in more conventional form
\begin{equation}\label{conny}
\gamma^0E=\dfrac{E^2}{mc^2}+\dfrac{\mathcal{P}^i}{2mc}p_ic.
\end{equation}
Applying the basic constraint (\ref{ecx}) to the right hand side of (\ref{conny}) one receives the relation
\begin{equation}\label{fv1}
\gamma^0E=mc^2+\dfrac{3\mathcal{P}^icp_ic}{2mc^2}+\dfrac{p^icp_ic}{mc^2},
\end{equation}
that def\/ines the resolution of the constraint (\ref{ecx}) in this case. Employing once again the canonical relativistic quantization \index{quantization!relativistic canonical}to (\ref{fv1}) and identifying the spatial gamma matrices as (\ref{spatgam}) one obtains the following equation
\begin{equation}
  i\hslash\gamma^0\partial_0\psi=
  \left(mc^2+3ic\hslash\gamma^i\partial_i-\dfrac{\hslash^2}{mc^2}\triangle\right)\psi,
\end{equation}
where $\triangle=\gamma^i\gamma_i\partial^i\partial_i=\partial^i\partial_i$ is the Laplace operator, which can be presented as the Dirac equation \index{Dirac equation}(\ref{direqu}) with the effective mass given by two equivalent mass generation rules
\begin{eqnarray}
Mc^2\psi=\left(-mc^2-4ic\hslash\gamma^i\partial_i+\dfrac{\hslash^2}{mc^2}\triangle\right)\psi,\\
Mc^2\psi=\dfrac{1}{3}\left(mc^2-4ic\hslash\gamma^0\partial_0-\dfrac{\hslash^2}{mc^2}\triangle\right)\psi.
\end{eqnarray}
It must be emphasized that the Laplace operator $\triangle=\gamma^i\gamma_i\partial^i\partial_i$ has the form $\triangle=\partial^i\partial_i$ for the new algebra, while for the Clifford algebra \index{Clifford algebra! and Laplace operator}it is manifestly different $\triangle=D\partial^i\partial_i$. It is the conf\/irmation of the correctness of the new algebra.

Another route to a new equation in the case under considering, can be obtained e.g. by application of the f\/irst relation (\ref{3rdid}) within the constraint (\ref{conny}). First one obtains
\begin{equation}
\gamma^0E=-\dfrac{\mathcal{P}^i}{2mc}\dfrac{\mathcal{P}_i}{2mc}mc^2+\dfrac{\mathcal{P}^i}{2mc}p_ic,
\end{equation}
then by using of (\ref{spatgam}) and the identity $\mathcal{P}_i=\delta_{ij}\mathcal{P}^j$ one has
\begin{equation}
\gamma^0E=-\delta_{ij}\gamma^i\gamma^jmc^2+\gamma^ip_ic,
\end{equation}
and by the new algebra result (\ref{gigi}) one receives f\/inally
\begin{equation}
\gamma^0E-\gamma^ip_ic+mc^2=0.
\end{equation}
In this manner, applying once again the canonical relativistic quantization \index{quantization!relativistic canonical}one obtains the equation
\begin{equation}\label{cuy1}
\left(i\hslash\gamma^0\partial_0-ic\hslash\gamma^i\partial_i+mc^2\right)\psi=0,
\end{equation}
which differs from the Dirac equation \index{Dirac equation}for particle with mass $m$ by the minus sign presence near spatial derivative. However, one sees straightforwardly that the equation (\ref{cuy1}) can be interpreted as the Dirac equation \index{Dirac equation}(\ref{direqu}) with the effective mass $M$, if and only if the effective mass term is given by one of the two equivalent relations
\begin{eqnarray}
M c^2\psi&\equiv& \left(-2ic\hslash\gamma^i\partial_i+mc^2\right)\psi,\\
M c^2\psi&\equiv& \left(-2i\hslash\gamma^0\partial_0-mc^2\right)\psi.
\end{eqnarray}

The obtained Dirac-like quantum theories can be always presented in the form of the Schr\"odinger equation\index{Schr\"odinger equation}, i.e.
\begin{equation}
i\hslash\partial_0\psi=H\psi,\label{schrodform}
\end{equation}
where $H=H(\partial_0,\partial_i)$ is the Hamilton operator describing a  quantum system. In comparison to the standard Dirac theory in the our case the operator $H$, however, in general must not be always a hermitean operator. The equation (\ref{schrodform}) should be solved with the usual spatial normalization condition for the wave function $\psi$
\begin{equation}
\int d^Dx|\psi(x)|^2=1,
\end{equation}
where $1$ is the D-dimensional unit matrix. The question of solvability of the systems deduced above, however,  is not the main theme of this book. It is good exercise for a reader.

As we proposed initially, the linear deformation of Special Relativity constraint in (\ref{ecx}) couples a particle and the {\AE}ther\index{{\AE}ther} in the Minkowski energy-momentum space \index{Minkowski energy-momentum space}. The reasoning done in the spirit of Dirac's relativistic quantum mechanics led us to the linkage (\ref{gami}) between the new algebra of the spatial gamma matrices, and the {\AE}ther momentum vector\index{{\AE}ther momentum}, and its particular case (\ref{spatgam}) was also discussed. Straightforward application of the {\AE}ther algebra \index{{\AE}ther algebra}(\ref{anticom}) results in the following noncommutative algebra of the {\AE}ther momentum
\begin{equation}\label{algae}
\left\{\dfrac{\mathcal{P}^i}{\mathcal{P}},\dfrac{\mathcal{P}^j}{\mathcal{P}}\right\}=\dfrac{2}{D}\delta^{ij}.
\end{equation}

Let us generalize this algebra to the Minkowski space-time \index{space-time!Minkowski}case
\begin{equation}\label{algae}
\left\{\dfrac{\mathcal{P}^\mu}{\mathfrak{P}},\dfrac{\mathcal{P}^\nu}{\mathfrak{P}}\right\}=\dfrac{2}{D+1}\eta^{\mu\nu},
\end{equation}
where $\mathfrak{P}$ is the length of the {\AE}ther four-momentum
\begin{equation}\label{aemom4}
\mathfrak{P}^2=\dfrac{1}{2}\left(\mathcal{P}^\mu\mathcal{P}_\mu+\mathcal{P}_\mu\mathcal{P}^\mu\right),
\end{equation}
which can be initially postulated as
\begin{equation}
\mathcal{P}^\mu=\left(\mathcal{P}^0,\mathcal{P}\gamma^i\right),
\end{equation}
where $\mathcal{P}^0$ is a time component of $\mathcal{P}^\mu$, and $\mathcal{P}$ is length of the {\AE}ther three-momentum. Let us compute (\ref{aemom4}) straightforwardly
\begin{equation}\label{aemom4a}
\mathfrak{P}^2=\dfrac{1}{2}\left(\mathcal{P}^0\mathcal{P}_0+\mathcal{P}_0\mathcal{P}^0+\mathcal{P}^2\delta_{ij}\left\{\gamma^i,\gamma^j\right\}\right),
\end{equation}
and postulate $\mathcal{P}^0=-\alpha\gamma^0$, where $\alpha$ is an unknown multiplier. Than, because of by definition one has $\mathcal{P}_0=\eta_{00}\mathcal{P}^0=\alpha\gamma_0$, one receives
\begin{equation}\label{aemom4b}
\mathfrak{P}^2-\dfrac{1}{2}\mathcal{P}^2\delta_{ij}\left\{\gamma^i,\gamma^j\right\}=-\alpha^2\left(1-\dfrac{1}{2}\delta_{ij}\left\{\gamma^i,\gamma^j\right\}\right),
\end{equation}
what in the light of the spatial algebra (\ref{gammas}) is satisfied if and only if
\begin{equation}
\mathfrak{P}^2=\mathcal{P}^2\quad,\quad\mathfrak{P}^2=-\alpha^2,
\end{equation}
and consequently time component of the {\AE}ther four-momentum \index{{\AE}ther momentum}is
\begin{equation}
\mathcal{P}^0=\pm i\mathfrak{P}\gamma^0=\pm i\mathcal{P}\gamma^0.
\end{equation}
Therefore, $\mathcal{P}^0=0$ if and only if $\mathfrak{P}=\mathcal{P}=0$.

For the particular situation $\gamma^i=\dfrac{\mathcal{P}^i}{2mc}$ a particle mass can be expressed via the {\AE}ther momentum\index{{\AE}ther momentum} vector
\begin{equation}
m=\dfrac{1}{2}\dfrac{\mathcal{P}}{c},
\end{equation}
so in fact the modif\/ied Einstein Hamiltonian constraint \index{Einstein Hamiltonian constraint}(\ref{ecx}) can be expressed via the {\AE}ther\index{{\AE}ther momentum} and a particle momenta only
\begin{equation}\label{enedue}
E^2=\left(\dfrac{1}{4}\mathcal{P}^i\mathcal{P}_i+p^ip_i+\mathcal{P}^ip_i\right)c^2.
\end{equation}
The f\/ixed value of square of the {\AE}ther momentum vector allows to establish three equivalent forms of the the momentum three-vector
\begin{eqnarray}
\mathcal{P}^i&=&\left[\sqrt{(2mc)^2-\mathcal{P}^2\mathcal{P}_2-\mathcal{P}^3\mathcal{P}_3},\mathcal{P}^2,\mathcal{P}^3\right]=\\
&=&\left[\mathcal{P}^1,\sqrt{(2mc)^2-\mathcal{P}^1\mathcal{P}_1-\mathcal{P}^3\mathcal{P}_3},\mathcal{P}^3\right]=\\
&=&\left[\mathcal{P}^1,\mathcal{P}^2,\sqrt{(2mc)^2-\mathcal{P}^1\mathcal{P}_1-\mathcal{P}^2\mathcal{P}_2}\right],
\end{eqnarray}
what allows also to derive easily the normalized {\AE}ther momentum vector $\dfrac{\mathcal{P}^i}{\mathcal{P}}$. Moreover, one can reconsider the situations def\/ined by the constraints  (\ref{coi1}) and (\ref{coi2}). In the f\/irst case the energetic constraint (\ref{enedue}) takes the following form
\begin{equation}\label{enedue1}
E^2=\left(\dfrac{1}{4}\mathcal{P}^i\mathcal{P}_i+p^ip_i\right)c^2,
\end{equation}
with the particle momentum def\/ined by the solutions (\ref{st1}), (\ref{st2}) or (\ref{st3}), while in the second situation one obtains
\begin{equation}\label{enedue2}
E^2=\left(-\dfrac{3}{4}\mathcal{P}^i\mathcal{P}_i+p^ip_i\right)c^2.
\end{equation}
where the particle momentum is given by (\ref{ste1}), (\ref{ste2}) or (\ref{ste3}). It is visible that in the case (\ref{enedue1}) the energy is always nonzero. However, in the case (\ref{enedue2}) the energy can be trivially vanishing if and only if the particle momentum is constrained by
\begin{equation}
p_i=\sqrt{\dfrac{3}{4}}\mathcal{P}_i.\label{11/12}
\end{equation}
It is easy to see for (\ref{ste1}), (\ref{ste2}) or (\ref{ste3}) that in the case (\ref{11/12}) holds
\begin{equation}
\mathcal{P}^i\mathcal{P}_i=0.\label{pipi}
\end{equation}
Particularly, the constraint (\ref{pipi}) is satisf\/ied when the {\AE}ther momentum \index{{\AE}ther momentum}vanishes identically $\mathcal{P}^i=0$, but by (\ref{11/12}) such a situation implies
\begin{equation}\label{1/11}
m=\dfrac{1}{\sqrt{3}}\dfrac{p}{c},
\end{equation}
where $p=\sqrt{p^ip_i}$ is the momentum value of a particle. Moreover, the vanishing {\AE}ther momentum implies that the deformation vanishes, i.e. Special Relativity should be reconstructed. Using of both $\mathcal{P}^i=0$ and (\ref{1/11}) in the light of the Hamiltonian constraint\index{Hamiltonian constraint} (\ref{ecx}) one obtains
\begin{equation}
E^2=4m^2c^4,
\end{equation}
but by (\ref{11/12}) the energy square (\ref{enedue2}) vanishes, so reconstruction of Special Relativity gives f\/inally $m^2=0$. Application of the constraint (\ref{pipi}) within the solutions (\ref{ste1})-(\ref{ste3}) reconstructs the case (\ref{st1})-(\ref{st3}), so (\ref{pipi}) is equivalent to $\mathcal{P}^ip_i=0$ in this case.

Interestingly, one can also consider other type linear deformations of Special Relativity. For example
\begin{eqnarray}
  \Delta_1&=&\mathcal{P}_ip^ic^2,\label{ecxa}\\
  \Delta_2&=&\dfrac{1}{2}\left(\mathcal{P}_ip^ic^2+\mathcal{P}^ip_ic^2\right),\label{ecxb}\\
  \Delta_3&=&\dfrac{1}{2}\left(\mathcal{P}_ip^ic^2+\mathcal{P}^ip_ic^2\right)\pm\mathcal{P}^2c^2,
\end{eqnarray}
etc., and perform analogous considerations. The deformations (\ref{ecxa}) and (\ref{ecxb}) are equivalent to (\ref{ecx}) one if and only if $\mathcal{P}_ip^i=\mathcal{P}^ip_i$, i.e.
\begin{equation}
\delta_{ik}\gamma^kp^i=\gamma^jp_j
\end{equation}
what, after multiplication of both sides by $\delta^{ik}\gamma_k=\gamma^i$ and taking into account the identity $\delta_{ik}\delta^{ik}=D$, takes the form
\begin{equation}
p^i=\dfrac{1}{D}\gamma^i\gamma^jp_j.
\end{equation}

In this section we have presented the approach based on the linear deformation of the Einstein Hamiltonian constraint \index{Einstein Hamiltonian constraint}of Special Relativity which generates new equations in frames of relativistic quantum mechanics. The linear deformation can be generalized for other deformations, and similar strategy can be used. All these new models can be treated as Dark Matter and/or Dark Energy models.\index{dark matter}\index{dark energy}

\section{The Snyder--Sidharth Hamiltonian}

Let us consider now more complex situation. Namely, we shall focus our attention on the following deformation of the Einstein Hamiltonian constraint \index{Einstein Hamiltonian constraint}(\ref{ec}) of Special Relativity
\begin{equation}\label{sss}
  E^2=m^2c^4+c^2p^2+\alpha\left(\dfrac{\ell}{\hslash}\right)^2c^2p^4,
\end{equation}
where $\ell$ is any minimal physical scale. This deformation was investigated by H. Snyder \cite{snyder} in the context of the infrared catastrophe\index{infrared catastrophe} of soft photons in the Compton scattering\index{Compton scattering}, and in general to renormalize quantum f\/ield theory by application of the noncommutative quantum space-time, as it is widely studied by numerous authors and scholars \cite{nspac}. In fact such a modif\/ication follows from the nontrivial manipulation in phase space of any special relativistic particle
\begin{eqnarray}\label{ndx}
  \dfrac{i}{\hslash}[p,x]&=&1+\alpha\dfrac{\ell^2}{\hslash^2}p^2\quad,\quad\alpha\sim1,\\
  \left[x,y\right]&=&O(\ell^2),
\end{eqnarray}
and therefore one has to deal with the structure of a non-differentiable manifold, or lattice model of space-time. We shall discuss wider mathematical details of the Snyder space-time\index{space-time!Snyder} (\ref{ndx}) in next chapters of this part. It must be emphasized that the deformation (\ref{ndx}) reveals Lorentz invariance\index{Lorentz invariance}. Factually B.G. Sidharth (Refs. \cite{sidbooks}) f\/irst accepted the Snyder noncommutative geometry\index{Snyder noncommutative geometry} as the serious argument for physics, has been studied the modif\/ied Einstein Hamiltonian constraint \index{Einstein Hamiltonian constraint!modif\/ied}(\ref{sss}) in the astroparticle physics context. He proposed taking into account the Hamiltonian constraint (\ref{sss}) and treating this deformation in generalized sense as a type of perturbational series in the minimum scale $\ell$, that can be e.g. the Planck scale\index{Planck scale} or the Compton scale\index{Compton scale}. By this reason we shall call the constraint (\ref{sss}) \emph{the Snyder--Sidharth Hamiltonian constraint}\index{Snyder--Sidharth Hamiltonian constraint} or brief\/ly the Snyder--Sidharth constraint/Hamiltonian.

A number of distinguished scholars like S. Glashow, S. Coleman, and others have considered diverse schemes which manifestly depart from the Einstein Special Relativity. It must be stressed here that these all schemes are purely \emph{ad hoc}. Recently observations of ultra-high-energy cosmic rays\index{ultra-high-energy cosmic rays} and rays from gamma bursts\index{gamma bursts} seem to suggest Lorentz symmetry\index{Lorentz symmetry!violation} violation \cite{lorvia}, and the Hamiltonian constraint\index{Hamiltonian constraint} (\ref{sss}) violates Lorentz symmetry. In this particular context the author \cite{laglinka1} has proposed to call the Hamiltonian constraint\index{Hamiltonian constraint} (\ref{sss}) the Snyder--Sidharth Hamiltonian constraint\index{Snyder--Sidharth Hamiltonian constraint} what is also preferred in this book. It is important to emphasize here, that in general situation the Hamiltonian constraint\index{Hamiltonian constraint} is not Hamiltonian, i.e. energy. In general an energy can be obtained by resolving a Hamiltonian constraint\index{Hamiltonian constraint}, as it was shown in the previous section, and shall be continued in next chapters of this book. Interestingly, the Snyder space-time\index{space-time!Snyder} has certain context in string theory, where is often referred as stringy-like space-time.

Similarly as in the case of the linear deformation analyzed in the previous section, the Snyder--Sidharth Hamiltonian constraint \index{Snyder--Sidharth Hamiltonian constraint}(\ref{sss}) as a quadratic form can be seen easily lead to the canonical form
\begin{equation}\label{qfss}
  E^2+\dfrac{\hslash^2c^2}{4\alpha\ell^2}=
  \alpha\left(\dfrac{c\ell}{\hslash}\right)^2\left(p^2+
  \dfrac{\hslash^2}{2\alpha\ell^2}\right)^2+m^2c^4,
\end{equation}
and as previously there are in general three possible mathematical interpretations of this relation. However, in the case of the fourth-order deformation (\ref{sss}) one has no linear terms in particle momentum $p$, but there are factually powers of $p^2$. In the light of the principles of relativistic quantum theory (See e.g. the Refs. \cite{greiner1} and \cite{peskin}) it means that in such a situation the structure of Clifford algebra \index{Clifford algebra}must be at least hidden if no hidden manifestly. Let us consider these identif\/ications to the case of fermions and the case of bosons.
\subsection{The Case of Fermions}
\begin{enumerate}
\item First, we can interpret the constraint equation (\ref{qfss}) as system of two equations
\begin{equation}
  \left\{\begin{array}{l}m^2c^4=\dfrac{\hslash^2c^2}{4\alpha\ell^2}\\
  E^2=\alpha\left(\dfrac{c\ell}{\hslash}\right)^2
  \left(p^2+\dfrac{\hslash^2}{2\alpha\ell^2}\right)^2\end{array}\right.
\end{equation}
The f\/irst equation leads to solution that looks like formally as the bosonic string tension
\begin{equation}\label{s1}
  m=\dfrac{\hslash}{2\sqrt{{\alpha}}c\ell}.
\end{equation}
Expressing $\ell$ via $m$ and constants, one can write the solution of the second equality in the following way
\begin{equation}\label{s2}
E=\dfrac{p^2}{2m}+mc^2,
\end{equation}
where we have omitted the solution with minus sign as non physical. This is the Hamiltonian of a free point particle in semi-classical mechanics, i.e. the Newtonian kinetic energy corrected by the Einstein--Poincar\`{e} rest energy term. Interestingly (\ref{s1}) and (\ref{s2}) are consistent if $m$ is the Planck mass\index{Planck mass}, i.e. $m=M_P=\sqrt{\dfrac{\hslash c}{G}}$, and $\ell$ is the Planck length\index{Planck length}, i.e. $\ell_P=\sqrt{\dfrac{\hslash G}{c^3}}$. Factually the Planck mass determines a unifying scale where the classical and the quantum meet and collaborate. The Schwarzschild radius\index{Schwarzschild radius} $r_S(m)=\dfrac{Gm}{c^2}$ evaluated on the Planck mass equals to the Compton wavelength \index{Compton wavelength}of a hypothetical particle possessing such a value of mass \cite{kiefer}. In comparison with Special Relativity, one has no here higher relativistic corrections to the semi-classical case. After the primary canonical quantization \index{quantization!primary canonical}one obtains exactly the Schr\"odinger equation \index{Schr\"odinger equation}of free quarks\index{free quarks} in Quantum Chromodynamics (QCD) because quarks are non-relativistic and massive \cite{lee}. For the case of vanishing scale and nonzero $\alpha$ as well as for vanishing $\alpha$ and f\/ixed nonzero scale $\ell$, formally $m\equiv\infty$ and energy is also inf\/inite, therefore it is a nonphysical black-hole type singularity. For the large scale limit and non vanishing $\alpha$, the mass spectrum is point-like $m=0$. For nonzero momentum energy is also inf\/inite, i.e. it is a nonphysical case. Anyway, it shows manifestly that (\ref{s2}) is compatible with (\ref{ndx}).

\item The second case changes the role of energy and mass
\begin{equation}
  \left\{\begin{array}{l}-m^2c^4=\alpha\left(\dfrac{c\ell}{\hslash}\right)^2\left(p^2+\dfrac{\hslash^2}{2\alpha\ell^2}\right)^2\\
  -E^2=\dfrac{\hslash^2c^2}{4\alpha\ell^2}\end{array}\right.,
\end{equation}
and leads to discrete energy spectrum for f\/ixed value of scale $\ell$. Because (See Sidharth's paper in \cite{lorvia}), however, the constraint (\ref{sss}) with positive $\alpha$ is true for fermions, and with negative $\alpha$ for bosons, for the case of fermions one has here
\begin{equation}\label{ene}
iE=\dfrac{\hslash c}{2\sqrt{{\alpha}}\ell},
\end{equation}
as well as the mass one
\begin{equation}\label{m1}
imc^2=\sqrt{\alpha}\dfrac{c\ell}{\hslash} p^2+\dfrac{\hslash c}{2\sqrt{\alpha}\ell},
\end{equation}
i.e. rejecting the negative value from. However, one can eliminate scale via using energy (\ref{ene}) with the result
\begin{equation}
  mc^2=-\dfrac{p^2c^2}{2E}+E,
\end{equation}
what can be rewritten in the form of the quadratic equation
\begin{equation}
E^2=mc^2E+\dfrac{p^2c^2}{2},
\end{equation}
and by using of the deformed constraint (\ref{sss}) it yields
\begin{equation}
 m^2c^4+c^2p^2+\alpha\left(\dfrac{\ell}{\hslash}\right)^2c^2p^4=mc^2E+\dfrac{p^2c^2}{2}.
\end{equation}
One can f\/ind now the energy (not square of energy!), that is $4th$-power in momentum
\begin{equation}\label{pauli1}
  E=mc^2+\dfrac{p^2}{2m}+\alpha\left(\dfrac{\ell}{\hslash}\right)^2\dfrac{p^4}{m}.
\end{equation}
So, again one can apply factorization of a quadratic form
\begin{equation}
  E+\dfrac{\hslash^2}{16\alpha\ell^2}\dfrac{1}{m}=
  \alpha\left(\dfrac{\ell}{\hslash}\right)^2\dfrac{1}{m}\left(p^2+\dfrac{\hslash^2}{4\alpha\ell^2}\right)^2+mc^2,
\end{equation}
and consider the identif\/ication method \index{identif\/ication method}for the relation mass-energy.
\begin{enumerate}
  \item The f\/irst obvious interpretation yields
\begin{equation}
  \left\{\begin{array}{l}mc^2=\dfrac{\hslash^2}{16\alpha\ell^2}\dfrac{1}{m}\vspace*{5pt}\\
  E=\alpha\left(\dfrac{\ell}{\hslash}\right)^2\dfrac{1}{m}\left(p^2+\dfrac{\hslash^2}{4\alpha\ell^2}\right)^2\end{array}\right.
\end{equation}
and again one can extract trivially the solution of the f\/irst equation
\begin{equation}\label{mmm1}
  m=\dfrac{\hslash}{4\sqrt{\alpha}c\ell},
\end{equation}
and solution of the latter equality can be written in the form of the Pauli Hamiltonian constraint\index{Pauli Hamiltonian constraint}
\begin{equation}
  E=mc^2+\dfrac{p^2}{2m}+\dfrac{p^4}{16m^3c^2},
\end{equation}
or after elimination of mass via using (\ref{mmm1}) one receives
\begin{equation}
E=\dfrac{\hslash c}{4\sqrt{\alpha}\ell}+2c\sqrt{\alpha}\dfrac{\ell}{\hslash}p^2+4c\left(\sqrt{\alpha}\dfrac{\ell}{\hslash}\right)^3p^4.
\end{equation}
However, it can be easily seen that the relation (\ref{ene}) jointed with (\ref{mmm1}) determines energy as
\begin{equation}
  iE=2mc^2,
\end{equation}
or equivalently the mass square value
\begin{equation}
m^2=-\dfrac{E^2}{2c^4}<0,
\end{equation}
what means that momentum values are non hermitian, so one has to deal with tachyon\index{tachyon}. Moreover, by straightforward application of the formula (\ref{mmm1}) together with the relation (\ref{m1}) one establishes that
\begin{equation}
p^2=\dfrac{\hslash^2}{2\alpha\ell^2}\left(-1+\dfrac{1}{2}i\right).
\end{equation}
Using the polar form of the complex number in brackets
\begin{equation}
  -1+\dfrac{1}{2}i=\dfrac{\sqrt{5}}{2}\exp\left(-i\arctan\dfrac{1}{2}+2ni\pi\right),
\end{equation}
where $n\in\mathbb{Z}$, one can take its square root
\begin{equation}
\sqrt{-1+\dfrac{1}{2}i}=\dfrac{1}{\sqrt{2}}\left(\sqrt{\dfrac{\sqrt{5}}{2}-1}+i\sqrt{\dfrac{\sqrt{5}}{2}+1}\right)e^{ni\pi},
\end{equation}
and obtains the momentum spectrum in dependence on $\ell$
\begin{equation}\label{pnplus}
p_n=\pm\dfrac{(-1)^n}{2}\left(\sqrt{\dfrac{\sqrt{5}}{2}-1}+i\sqrt{\dfrac{\sqrt{5}}{2}+1}\right)\dfrac{\hslash}{\sqrt{\alpha}\ell}.
\end{equation}
\item The latter subcase is
  \begin{equation}
  \left\{\begin{array}{l}mc^2=E\\
  \dfrac{\hslash^2}{16\alpha\ell^2}\dfrac{1}{m}=
  \alpha\left(\dfrac{\ell}{\hslash}\right)^2\dfrac{1}{m}\left(p^2+\dfrac{\hslash^2}{4\alpha\ell^2}\right)^2\end{array}\right.
\end{equation}
Again, solution of the f\/irst equation via using (\ref{ene}) is rather simple
\begin{equation}
  m=-i\dfrac{\hslash }{2\sqrt{{\alpha}}c\ell}\quad,\quad m^2<0,
\end{equation}
and once again the tachyon\index{tachyon} is obtained - there are particles with momentum spectrum
\begin{equation}
  p=\left\{0,\pm\dfrac{\hslash}{\sqrt{{2\alpha}}\ell}\right\}.
\end{equation}
This is discrete momenta spectrum for f\/ixed scale $\ell$. For running scale this is non compact spectrum, but compactif\/ication to the point is done in the large scale limit
\begin{equation}
  \lim_{\ell\rightarrow\infty} p = 0,
\end{equation}
and it is the rest. For $\alpha=0$ and f\/ixed scale $\ell$ there are two singular values of the momentum $p$. For all $\ell\neq0$ and $\alpha\neq0$, the case of nonzero $p$ is related to the existence of tachyon\index{tachyon}.
\item The third interpretation yields
\begin{equation}
  \left\{\begin{array}{l}-E=\dfrac{\hslash^2}{16\alpha\ell^2}\dfrac{1}{m}\vspace*{5pt}\\
  -mc^2=\alpha\left(\dfrac{\ell}{\hslash}\right)^2\dfrac{1}{m}\left(p^2+\dfrac{\hslash^2}{4\alpha\ell^2}\right)^2\end{array}\right..
\end{equation}
Similarly as in the previous subcase, employing the relation (\ref{ene}) one obtains from the f\/irst equation the mass
\begin{equation}
  m=-i\dfrac{\hslash}{8\sqrt{\alpha}c\ell},
\end{equation}
and again one has tachyon\index{tachyon}, i.e. $m^2<0$. Consequently the momentum spectrum can be established as
\begin{equation}
  p=\left\{\pm i\sqrt{\strut{\dfrac{1}{8\alpha}}}\dfrac{\hslash}{\ell},\pm i\sqrt{\strut{\dfrac{3}{8\alpha}}}\dfrac{\hslash}{\ell}\right\},
\end{equation}
and is non hermitian, as one has expected. By this reason this particular case, that is the Pauli Hamiltonian constraint \index{Pauli Hamiltonian constraint}with mass related to minimum scale, describes tachyon\index{tachyon}, the hypothetical particles moving with velocity faster then light.
\end{enumerate}
\item The third possible solution of the Snyder--Sidharth Hamiltonian constraint \index{Snyder--Sidharth Hamiltonian constraint}can be constructed by the system of equations
 \begin{equation}
  \left\{\begin{array}{l}E^2=m^2c^4\vspace*{5pt}\\
  \dfrac{\hslash^2c^2}{4\alpha\ell^2}=\alpha\left(\dfrac{c\ell}{\hslash}\right)^2
  \left(p^2+\dfrac{\hslash^2}{2\alpha\ell^2}\right)^2\end{array}\right..
\end{equation}
First equation gives the standard Einstein mass-energy relation\index{Einstein mass-energy relation}
\begin{equation}
  E=mc^2,
\end{equation}
and the latter equality results in the discrete momentum spectrum
\begin{equation}
  p=\left\{0,\pm\dfrac{1}{\sqrt{{\alpha}}}\dfrac{\hslash}{\ell}\right\},
\end{equation}
for f\/ixed scale $\ell$ value. For running scale this spectrum is non compact, but in the large scale limit the spectrum is manifestly compactif\/ied to the point
\begin{equation}
  \lim_{\ell\rightarrow\infty} p = 0,
\end{equation}
and it is the rest. For $\alpha=0$ and f\/ixed scale $\ell$ there are two singular values of the momentum $p$. For all $\ell\neq0$ and $\alpha\neq0$, the case of nonzero $p$ is related to the existence of a relativistic particle.
\end{enumerate}

The other situation is when particle mass vanishes identically, i.e. $m\equiv0$. In such a case one sees that the constraint (\ref{qfss}) takes the form
\begin{equation}\label{qfss0}
  E^2+\dfrac{\hslash^2c^2}{4\alpha\ell^2}=
  \alpha\left(\dfrac{c\ell}{\hslash}\right)^2\left(p^2+
  \dfrac{\hslash^2}{2\alpha\ell^2}\right)^2,
\end{equation}
and by the non-trivial identif\/ication
\begin{eqnarray}
E^2+\dfrac{\hslash^2c^2}{4\alpha\ell^2}&=&0,\\
p^2+\dfrac{\hslash^2}{2\alpha\ell^2}&=&0,
\end{eqnarray}
can be solved as the tachyonic case
\begin{eqnarray}
E^2&=&-\dfrac{\hslash^2c^2}{4\alpha\ell^2},\label{eny1}\\
p^2&=&-\dfrac{\hslash^2}{2\alpha\ell^2},\label{eny2}
\end{eqnarray}
which can be combined into a one constraint
\begin{equation}
E^2=\dfrac{1}{2}p^2c^2,\label{consy}
\end{equation}
which can be solved immediately
\begin{equation}
E=\pm\dfrac{1}{\sqrt{2}}pc,\label{mass0}
\end{equation}
The equation (\ref{mass0}) differs from the usual Special Relativity condition for massless particle, i.e. $E=pc$. The difference ref\/lects the fact that the Snyder--Sidharth deformation \index{Snyder--Sidharth deformation}is an algebraic deformation in the Minkowski energy-momentum space \index{Minkowski energy-momentum space}. It is easy to see from the relations (\ref{eny1}) and (\ref{eny2}) that energy and momentum of the massless particles are purely imaginary quantities, and therefore such a situation corresponds to massless fermionic tachyon\index{tachyon!fermionic}.

\subsection{The Case of Bosons}
Fermions obey the Dirac equation \index{Dirac equation}which is a square root of the Klein--Gordon equation \index{Klein--Gordon equation}ruling bosons. Therefore fermions are approximation and bosons are fundamental. Previous section results work for fermions when $\alpha>0$ and for $\alpha<0$ are true for boson, i.e. the case of bosons arises by the exchange
 \begin{equation}\label{cha}
   \alpha\longrightarrow-|\alpha|,
 \end{equation}
within mass and energy formulas. The basic relation (\ref{qfss}) then is
\begin{equation}\label{qfss1}
E^2-\dfrac{\hslash^2c^2}{4|\alpha|\ell^2}=
-|\alpha|\left(\dfrac{c\ell}{\hslash}\right)^2\left(p^2-\dfrac{\hslash^2}{2|\alpha|\ell^2}\right)^2+m^2c^4.
\end{equation}

\begin{enumerate}
\item In the f\/irst case one has the system
\begin{equation}
  \left\{\begin{array}{l}m^2c^4=-\dfrac{\hslash^2c^2}{4|\alpha|\ell^2}\\
  E^2=-|\alpha|\left(\dfrac{c\ell}{\hslash}\right)^2
  \left(p^2-\dfrac{\hslash^2}{2|\alpha|\ell^2}\right)^2\end{array}\right.
\end{equation}
which def\/ines the tachyon\index{tachyon} with mass and energy
\begin{eqnarray}
  m&=&i\dfrac{\hslash}{2\sqrt{{|\alpha|}}c\ell},\label{s1a}\\
  E&=&\dfrac{p^2}{2m}+mc^2.\label{s2a}
\end{eqnarray}
\item The second case is the system
\begin{equation}\label{ixiu}
  \left\{\begin{array}{l}m^2c^4=|\alpha|\left(\dfrac{c\ell}{\hslash}\right)^2
  \left(p^2-\dfrac{\hslash^2}{2|\alpha|\ell^2}\right)^2\\
  E^2=\dfrac{\hslash^2c^2}{4|\alpha|\ell^2}\end{array}\right.,
\end{equation}
which can be solved by
\begin{eqnarray}
mc^2&=&\dfrac{p^2c^2}{2E}-E,\label{masi}\\
E&=&\dfrac{\hslash c}{2\sqrt{{|\alpha|}}\ell}.\label{ene1}
\end{eqnarray}
However, the relation (\ref{masi}) is in itself the quadratic equation
\begin{equation}
E^2=-mc^2E+\dfrac{p^2c^2}{2},
\end{equation}
which, by using (\ref{sss}) and (\ref{cha}), is the constraint
\begin{equation}\label{pauli1a}
E=-mc^2-\dfrac{p^2}{2m}+|\alpha|\left(\dfrac{\ell}{\hslash}\right)^2\dfrac{p^4}{m},
\end{equation}
similar to the Pauli Hamiltonian constraint\index{Pauli Hamiltonian constraint}, and can be easy factorized
\begin{equation}
  E-\dfrac{\hslash^2}{16|\alpha|\ell^2}\dfrac{1}{m}=
  -|\alpha|\left(\dfrac{\ell}{\hslash}\right)^2\dfrac{1}{m}\left(p^2-\dfrac{\hslash^2}{4|\alpha|\ell^2}\right)^2+mc^2,
\end{equation}
and considered by three way.
\begin{enumerate}
  \item In the f\/irst one has the system
  \begin{equation}
  \left\{\begin{array}{l}mc^2=-\dfrac{\hslash^2}{16|\alpha|\ell^2}\dfrac{1}{m}\vspace*{5pt}\\
  E=-|\alpha|\left(\dfrac{\ell}{\hslash}\right)^2
  \dfrac{1}{m}\left(p^2-\dfrac{\hslash^2}{4|\alpha|\ell^2}\right)^2\end{array}\right.
\end{equation}
which gives
\begin{eqnarray}
  m&=&i\dfrac{\hslash}{4\sqrt{|\alpha|}c\ell},\label{mmx2}\\
  E&=&mc^2+\dfrac{p^2}{2m}+\dfrac{p^4}{16m^3c^2}.\label{enew}
\end{eqnarray}
Via using the mass (\ref{mmx2}) the energy (\ref{enew}) reads
\begin{equation}
E=i\dfrac{\hslash c}{4\sqrt{|\alpha|}\ell}-i2c\sqrt{|\alpha|}\dfrac{\ell}{\hslash}p^2+
i4c\left(\sqrt{|\alpha|}\dfrac{\ell}{\hslash}\right)^3p^4.
\end{equation}
Jointing of the energy (\ref{ene1}) and the mass (\ref{mmx2}) leads to
\begin{equation}
  iE=2mc^2,
\end{equation}
what means that
\begin{equation}
m^2=-\dfrac{E^2}{2c^4}<0,
\end{equation}
i.e. this case describes tachyon\index{tachyon}. Using of the f\/irst equation in (\ref{ixiu}) together with the mass (\ref{mmx2}) allows to establish
\begin{equation}
p^2=\dfrac{\hslash^2}{2|\alpha|\ell^2}\left(1+\dfrac{1}{2}i\right),
\end{equation}
what after using the fact that
\begin{equation}
\sqrt{-1+\dfrac{1}{2}i}=
\dfrac{1}{\sqrt{2}}\left(\sqrt{\dfrac{\sqrt{5}}{2}+1}+i\sqrt{\dfrac{\sqrt{5}}{2}-1}\right)e^{ni\pi},
\end{equation}
where $n\in\mathbb{Z}$ leads to the momentum spectrum
\begin{equation}\label{pnminus}
p_n=\pm\dfrac{(-1)^n}{2}\left(\sqrt{\dfrac{\sqrt{5}}{2}+1}+i\sqrt{\dfrac{\sqrt{5}}{2}-1}\right)
\dfrac{\hslash}{\sqrt{|\alpha|}\ell}.
\end{equation}
Interestingly, writing (\ref{pnplus}) as $p_n(|\alpha|)$ and (\ref{pnminus}) as $p_n(-|\alpha|)$ one can def\/ine the spectral mean
\begin{equation}
  \langle p_n(|\alpha|)p_n(-|\alpha|)\rangle=\dfrac{1}{|\alpha|-|\alpha_0|}\int_{|\alpha_0|}^{|\alpha|}p_n(x)p_n(-x)dx,
\end{equation}
which is easy to derive
\begin{equation}\label{spin1}
  \langle p_n(|\alpha|)p_n(-|\alpha|)\rangle=\pm i\dfrac{\sqrt{5}}{4}\dfrac{\hslash^2}{\ell^2}
  \dfrac{\ln\left|\dfrac{\alpha}{\alpha_0}\right|}{|\alpha|-|\alpha_0|},
\end{equation}
and it is not diff\/icult to see that
\begin{equation}
  \lim_{|\alpha|\rightarrow|\alpha_0|}\langle p_n(|\alpha|)p_n(-|\alpha|)\rangle=\pm\sqrt{5}\dfrac{\hslash}{2\sqrt{|\alpha_0|}\ell}
  \dfrac{\hslash}{2\sqrt{-|\alpha_0|}\ell}.
\end{equation}
Now it is not diff\/icult to establish also the spectral means
\begin{eqnarray}
  \langle |p_n(|\alpha|)|^2\rangle&=&\dfrac{1}{|\alpha|-|\alpha_0|}\int_{|\alpha_0|}^{|\alpha|}|p_n(x)|^2dx,\\
  \langle |p_n(-|\alpha|)|^2\rangle&=&\dfrac{1}{|\alpha|-|\alpha_0|}\int_{|\alpha_0|}^{|\alpha|}|p_n(-x)|^2dx,
\end{eqnarray}
which are equal to
\begin{eqnarray}
  \langle |p_n(|\alpha|)|^2\rangle&=&\dfrac{\sqrt{5}}{4}\dfrac{\hslash^2}{\ell^2}
  \dfrac{\ln\left|\dfrac{\alpha}{\alpha_0}\right|}{|\alpha|-|\alpha_0|},\label{gyp1}\\
  \langle |p_n(-|\alpha|)|^2\rangle&=&\dfrac{\sqrt{5}}{4}\dfrac{\hslash^2}{\ell^2}
  \dfrac{\ln\left|\dfrac{\alpha}{\alpha_0}\right|}{|\alpha|-|\alpha_0|},\label{gyp2}
\end{eqnarray}
and have the limiting values
\begin{eqnarray}
  \lim_{|\alpha|\rightarrow|\alpha_0|}\langle |p_n(|\alpha|)|^2\rangle&=&\dfrac{\sqrt{5}}{4}\dfrac{\hslash^2}{|\alpha_0|\ell^2},\\
  \lim_{|\alpha|\rightarrow|\alpha_0|}\langle |p_n(-|\alpha|)|^2\rangle&=&\dfrac{\sqrt{5}}{4}\dfrac{\hslash^2}{|\alpha_0|\ell^2}.
\end{eqnarray}
The formulas (\ref{gyp1})-(\ref{gyp2}) and (\ref{spin1}) allows to construct the variance
\begin{equation}
\sigma^2=\langle |p_n(|\alpha|)|^2\rangle\langle |p_n(-|\alpha|)|^2\rangle-|\langle p_n(|\alpha|)p_n(-|\alpha|)\rangle|^2,
\end{equation}
which vanishes in general
\begin{equation}
\sigma^2=0,
\end{equation}
as well as in the limiting case
\begin{equation}
\lim_{|\alpha|\rightarrow|\alpha_0|}\sigma^2=0.
\end{equation}

One can compute also the following means
\begin{eqnarray}
  \langle p_n(|\alpha|)\rangle&=&\dfrac{1}{|\alpha|-|\alpha_0|}\int_{|\alpha_0|}^{|\alpha|}p_n(x)dx,\\
  \langle p_n(-|\alpha|)\rangle&=&\dfrac{1}{|\alpha|-|\alpha_0|}\int_{|\alpha_0|}^{|\alpha|}p_n(-x)dx,
\end{eqnarray}
and obtain the quantity
\begin{eqnarray}
  \langle p_n(|\alpha|)\rangle\langle p_n(-|\alpha|)\rangle=\pm i\sqrt{5}\dfrac{\hslash^2}{\ell^2}\dfrac{1}{\left(\sqrt{|\alpha|}+\sqrt{|\alpha_0|}\right)^2},
\end{eqnarray}
which, together with (\ref{spin1}), allows to establish the variance
\begin{equation}
\sigma^2=\langle p_n(|\alpha|)\rangle\langle p_n(-|\alpha|)\rangle-\langle p_n(|\alpha|)p_n(-|\alpha|)\rangle,
\end{equation}
with the f\/inal result
\begin{equation}
\sigma^2=\pm i\dfrac{\sqrt{5}}{4}\dfrac{\hslash^2}{\ell^2}\left[\dfrac{4}{\left(\sqrt{|\alpha|}+\sqrt{|\alpha_0|}\right)^2}-
\dfrac{\ln\left|\dfrac{\alpha}{\alpha_0}\right|}{|\alpha|-|\alpha_0|}\right].\label{sigmasde}
\end{equation}
It is easy to derive now the limiting case of (\ref{sigmasde})
\begin{equation}
\lim_{|\alpha|\rightarrow|\alpha_0|}\sigma^2=i0.\label{sigmasde}
\end{equation}
\item The latter subcase is
  \begin{equation}
  \left\{\begin{array}{l}mc^2=E\\
  \dfrac{\hslash^2}{16\alpha\ell^2}\dfrac{1}{m}=
  \alpha\left(\dfrac{\ell}{\hslash}\right)^2\dfrac{1}{m}\left(p^2+\dfrac{\hslash^2}{4\alpha\ell^2}\right)^2\end{array}\right.
\end{equation}
Again, solution of the f\/irst equation via using (\ref{ene}) is rather simple
\begin{equation}
  m=-i\dfrac{\hslash }{2\sqrt{{\alpha}}c\ell}\quad,\quad m^2<0,
\end{equation}
and once again the tachyon\index{tachyon} is obtained - there are particles with momentum spectrum
\begin{equation}
  p=\left\{0,\pm\dfrac{\hslash}{\sqrt{{2\alpha}}\ell}\right\}.
\end{equation}
This is discrete momenta spectrum for f\/ixed scale $\ell$. For running scale this is non compact spectrum, but compactif\/ication to the point is done in the large scale limit
\begin{equation}
  \lim_{\ell\rightarrow\infty} p = 0,
\end{equation}
and it is the rest. For $\alpha=0$ and f\/ixed scale $\ell$ there are two singular values of the
momentum $p$. For all $\ell\neq0$ and $\alpha\neq0$, the case of nonzero $p$ is related to the existence of tachyon\index{tachyon}.
\item The third interpretation yields
\begin{equation}
  \left\{\begin{array}{l}-E=\dfrac{\hslash^2}{16\alpha\ell^2}\dfrac{1}{m}\vspace*{5pt}\\
  -mc^2=\alpha\left(\dfrac{\ell}{\hslash}\right)^2\dfrac{1}{m}\left(p^2+\dfrac{\hslash^2}{4\alpha\ell^2}\right)^2\end{array}\right..
\end{equation}
Similarly as in the previous subcase, employing the relation (\ref{ene}) one obtains from the f\/irst equation the mass
\begin{equation}
  m=-i\dfrac{\hslash}{8\sqrt{\alpha}c\ell},
\end{equation}
and again one has tachyon\index{tachyon}, i.e. $m^2<0$. Consequently the momentum spectrum can be established as
\begin{equation}
  p=\left\{\pm i\sqrt{\strut{\dfrac{1}{8\alpha}}}\dfrac{\hslash}{\ell},\pm i\sqrt{\strut{\dfrac{3}{8\alpha}}}\dfrac{\hslash}{\ell}\right\},
\end{equation}
and is non hermitian, as one has expected. By this reason this particular case, that is the Pauli Hamiltonian constraint \index{Pauli Hamiltonian constraint}with mass related to minimum scale, describes tachyon\index{tachyon}, the hypothetical particles with velocity faster then light.
\end{enumerate}
\item The third possible solution of the Snyder--Sidharth Hamiltonian constraint \index{Snyder--Sidharth Hamiltonian constraint}can be constructed by the system of equations
 \begin{equation}
  \left\{\begin{array}{l}E^2=m^2c^4\vspace*{5pt}\\
  \dfrac{\hslash^2c^2}{4\alpha\ell^2}=\alpha\left(\dfrac{c\ell}{\hslash}\right)^2
  \left(p^2+\dfrac{\hslash^2}{2\alpha\ell^2}\right)^2\end{array}\right..
\end{equation}
First equation gives the standard Einstein mass-energy relation\index{Einstein mass-energy relation}
\begin{equation}
  E=mc^2,
\end{equation}
and the latter equality results in the discrete momentum spectrum
\begin{equation}
  p=\left\{0,\pm\dfrac{1}{\sqrt{{\alpha}}}\dfrac{\hslash}{\ell}\right\},
\end{equation}
for f\/ixed scale $\ell$ value. For running scale this spectrum is non compact, but in the large scale limit the spectrum is manifestly compactif\/ied to the point
\begin{equation}
  \lim_{\ell\rightarrow\infty} p = 0,
\end{equation}
and it is the rest. For $\alpha=0$ and f\/ixed scale $\ell$ there are two singular values of the momentum $p$. For all $\ell\neq0$ and $\alpha\neq0$, the case of nonzero $p$ is related to the existence of a relativistic particle.
\end{enumerate}

\section{The Modif\/ied Compton Effect}

Particle astrophysics has a great interest in diverse situations between light and particles, and factually a lot of its conclusions arise via analysis of this type phenomena from both the theoretical and the experimental points of view (See e.g. general books in Ref. \cite{astpar}), which give a physical information about Cosmos. One of such phenomena is the Compton scattering\index{Compton scattering}, discovered by A.H. Compton in 1923 \cite{ahc}, which is a mid-energy interaction of light and matter. The scattering is realized via the electromagnetic radiation, X-rays\index{X-rays} and gamma ($\gamma$) rays, undergo in matter, i.e. electrons. A decrease in photon energy/wavelength due to the inelastic scattering is the point called the Compton effect\index{Compton effect}. We shall not discuss the detailed classical analysis of the Compton scattering\index{Compton scattering}, because of factually the analysis is based on a framework involving law of conservation of energy, law of conservation of momentum, and concepts of Einstein Special Relativity. Let us consider the case of the Compton effect \index{Compton effect}within the framework employing the Snyder--Sidharth Hamiltonian constraint \index{Snyder--Sidharth Hamiltonian constraint}(\ref{ss}). This can be done, however, in three different routs minimally. In the case of lack of deformation all the ways lead to the same result, called the Compton equation\index{Compton equation}. However, in presence of the deformation due to the Snyder geometry one obtains three various results. It is not clear which of the formulations is correct, and also in general it is not known are there other alternatives. Let us consider the three approaches step by step.

\subsection{The Relativistic Approach}
The our approach is based on the standard Special Relativity formulation (See e.g. Ref. \cite{tipler}). Let us consider a photon and an electron having the energy-momentum four-vectors
\begin{eqnarray}
p^\mu_\gamma&=&[E_\gamma~,~p_\gamma^ic],\\
p^\mu_e&=&[E_0~,~0],
\end{eqnarray}
where $E_0=m_ec^2$ is the electron rest energy. An electron at rest is scattered by an incoming photon, and an outgoing photon is observed under the scattering angle $\theta$ relatively to the incident direction of an incoming photon. The f\/inal energy-momentum four-vectors are
\begin{eqnarray}
p^\mu_{\gamma'}&=&[E_{\gamma'}~,~p_{\gamma'}^ic],\\
p^\mu_{e'}&=&[E_e~,~p_e^ic].
\end{eqnarray}
Let us preserve unchanged the Planck--Einstein relations \index{Planck--Einstein relations}of the wave-particle duality\index{wave-particle duality}. Thus for a photon one has
\begin{eqnarray}
E_\gamma&=&\hslash\omega,\label{plein1}\\
p_\gamma^i&=&\hslash k^i,\label{plein2}
\end{eqnarray}
where $\omega$ and $k_\gamma$ are angular frequency and wave vector, and the value of wave vector of a photon is
\begin{equation}
k=\sqrt{k_ik^i}=\dfrac{2\pi}{\lambda}=\dfrac{1}{\slashed{\lambda}},\label{wavevector}
\end{equation}
where $\slashed{\lambda}$ is reduced wavelength of a photon. Similarly for a matter particle, including electrons, one has
\begin{eqnarray}
E_e&=&\hslash\omega_e,\label{plein1}\\
p_e^i&=&\hslash k_e^i,\label{plein2}
\end{eqnarray}
where $\omega_e$ and $k_e$ are angular frequency and wave vector of an electron, and the value of wave vector of an electron is
\begin{equation}
k_e=\sqrt{{k_e}_ik_e^i}=\dfrac{2\pi}{\lambda_e}=\dfrac{1}{\slashed{\lambda}_e},
\end{equation}
where $\slashed{\lambda}_e$ is reduced wavelength of an electron.

In addition we assume that Special Relativity is deformed due to the Snyder noncommutative geometry\index{Snyder noncommutative geometry}, i.e. that for any photon $\gamma$ and any electron $e$ are satisf\/ied the Snyder--Sidharth Hamiltonian constraints\index{Snyder--Sidharth Hamiltonian constraint}
\begin{eqnarray}
E^2_\gamma&=&p_\gamma^2 c^2+\dfrac{1}{\epsilon^2}\left(p_\gamma^2 c^2\right)^2,\label{ssgam}\\
E^2_e&=&E_0^2+p_e^2 c^2+\dfrac{1}{\epsilon^2}\left(p_e^2 c^2\right)^2,\label{ssen}
\end{eqnarray}
where $p_e=\sqrt{{p_e}_ip_e^i}$ and $p_\gamma=\sqrt{{p_\gamma}_i{p_\gamma}^i}$ are values of momenta of an electron and a photon, respectively, and for shortened notation we have introduced the following energy parameter
\begin{equation}\label{enparam}
\epsilon=\dfrac{\hslash c}{\sqrt{\alpha}\ell}=\dfrac{\hslash c}{\ell_P}\dfrac{1}{\sqrt{\alpha}}\dfrac{\ell_P}{\ell}=E_P\dfrac{1}{\sqrt{\alpha}}\dfrac{\ell_P}{\ell},
\end{equation}
where $E_P=\sqrt{\dfrac{\hslash c^5}{G}}$ is the Planck energy. By application of the wave-particle duality\index{wave-particle duality} the Snyder--Sidharth Hamiltonian constraints for a photon and an electron can be presented as
\begin{eqnarray}
\dfrac{\omega^2}{c^2}&=&k^2+\dfrac{k^4}{\kappa^2},\label{ssgam1}\\
\dfrac{\omega^2_e}{c^2}&=&k_C^2+k_e^2+\dfrac{k_e^4}{\kappa^2},\label{ssen1}
\end{eqnarray}
where $k_C$ is the wave vector
\begin{equation}\label{kace}
k_C=\dfrac{2\pi}{\lambda_C}=\dfrac{1}{\slashed{\lambda}_C},
\end{equation}
related to the Compton wavelength\index{Compton wavelength!of electron}\index{Compton wavelength!reduced of electron} $\lambda_C$ of an electron, and its reduced form
\begin{equation}\label{lambdace}
\slashed{\lambda}_C=\dfrac{\hslash c}{E_0}=\dfrac{\hslash}{m_ec},
\end{equation}
and $\kappa$ is the parameter
\begin{equation}\label{kappa}
\kappa=\dfrac{\epsilon}{\hslash c}=\dfrac{1}{\sqrt{\alpha}\ell},
\end{equation}
which can be interpreted as a wave vector value associated to the (non-reduced!) wavelength $\lambda_{\kappa}=2\pi\sqrt{\alpha}\ell$, which for
\begin{equation}
\alpha=\dfrac{1}{(2\pi)^2},\label{alpha}
\end{equation}
becomes $\lambda_{\kappa}\equiv\ell$. The parameter (\ref{enparam}) can be physically interpreted as the {\AE}ther energy\index{{\AE}ther energy}, so $\kappa$ is the wave vector value of the {\AE}ther wave-particle.

In our interesting is the relation for the angle $\theta$ of scattered photon, i.e. the angle between the wave vectors $k_i$ and $k'_i$. Let us apply f\/irst the law of conservation of energy-momentum, i.e.
\begin{equation}\label{emcons}
p^\mu_\gamma+p^\mu_e=p^\mu_{\gamma'}+p^\mu_{e'}.
\end{equation}
For the momentum this principle gives
\begin{eqnarray}
p_\gamma^ic=p_{\gamma'}^ic+p_e^ic,
\end{eqnarray}
or in terms of the wave vectors
\begin{eqnarray}
k^i=k'^i+k_e^i,
\end{eqnarray}
and by elementary algebraic operations leads to the result
\begin{equation}\label{momres}
k_e^2=k^2+k'^2-2kk'\cos\theta.
\end{equation}
Next one can use the law of conservation of energy which, in the light of (\ref{emcons}), for the present case has the following form
\begin{equation}\label{encon}
E_\gamma+E_0=E_{\gamma'}+E_e,
\end{equation}
and by this reason one obtains
\begin{equation}\label{encon1}
\dfrac{\omega_e}{c}=k_C+\dfrac{\omega}{c}-\dfrac{\omega'}{c}.
\end{equation}
Taking square of both sides of the equation (\ref{encon1})
\begin{equation}\label{encon2}
\dfrac{\omega_e^2}{c^2}=k_C^2+\dfrac{\omega^2}{c^2}+\dfrac{\omega'^2}{c^2}-
2\dfrac{\omega}{c}\dfrac{\omega'}{c}+2k_C\left(\dfrac{\omega}{c}-\dfrac{\omega'}{c}\right),
\end{equation}
and applying the Snyder--Sidharth Hamiltonian constraint \index{Snyder--Sidharth Hamiltonian constraint}for photons and an electron, one receives the relation
\begin{eqnarray}
k_C^2+k_e^2+\dfrac{k_e^4}{\kappa^2}&=&k_C^2+k^2+\dfrac{k^4}{\kappa^2}+k'^2+\dfrac{k'^4}{\kappa^2}-2\sqrt{k^2+\dfrac{k^4}{\kappa^2}}\sqrt{k'^2+\dfrac{k'^4}{\kappa^2}}+\nonumber\\
&&2k_C\left(\sqrt{k^2+\dfrac{k^4}{\kappa^2}}-\sqrt{k'^2+\dfrac{k'^4}{\kappa^2}}\right),
\end{eqnarray}
or after small reduction
\begin{eqnarray}\label{fret}
k_e^2+\dfrac{k_e^4}{\kappa^2}&=&k^2+k'^2+\dfrac{k^4}{\kappa^2}+\dfrac{k'^4}{\kappa^2}-2\sqrt{k^2+\dfrac{k^4}{\kappa^2}}\sqrt{k'^2+\dfrac{k'^4}{\kappa^2}}+\nonumber\\
&&2k_C\left(\sqrt{k^2+\dfrac{k^4}{\kappa^2}}-\sqrt{k'^2+\dfrac{k'^4}{\kappa^2}}\right),
\end{eqnarray}
Applying the relation (\ref{momres}) within LHS of the equation (\ref{fret}), some algebraic identities, and factorization one obtains
\begin{eqnarray}\label{fret1}
\!\!\!\!\!\!\!\!\!\!\!\!\!\!\!\!\!\!\!\!\!\!\!\!\!\!\!\!\!\!\!\!\!\!\!\!\!\!\!\!\!\!\!\!\!\!\!\!\!\!
&&-\cos\theta+
\dfrac{\kappa^2}{2kk'}\dfrac{\left(k^2+k'^2\right)^2}{\kappa^4}-
2\dfrac{k^2+k'^2}{\kappa^2}\cos\theta+
\dfrac{2kk'}{\kappa^2}\cos^2\theta=\nonumber\\
\!\!\!\!\!\!\!\!\!\!\!\!\!\!\!\!\!\!\!\!\!\!\!\!\!\!\!\!\!\!\!\!\!\!\!\!\!\!\!\!\!\!\!\!\!\!\!\!\!\!
&&\dfrac{\kappa^2}{2kk'}\dfrac{k^4+k'^4}{\kappa^4}-
\sqrt{1+\dfrac{k^2}{\kappa^2}}\sqrt{1+\dfrac{k'^2}{\kappa^2}}+
k_C\left(\dfrac{1}{k'}\sqrt{1+\dfrac{k^2}{\kappa^2}}-\dfrac{1}{k}\sqrt{1+\dfrac{k'^2}{\kappa^2}}\right).
\end{eqnarray}
Now by ordering in $\cos\theta$ powers one receives the quadratic equation
\begin{eqnarray}\label{coseqn}
\beta_\kappa\cos^2\theta-\delta_\kappa\cos\theta+\eta_\kappa=0,
\end{eqnarray}
where we have introduced the notation
\begin{eqnarray}
\!\!\!\!\!\!\!\!\!\!\!\!\!\!\!\!\!\!\!\!\!\!\!\!\!\!\!\!\!\!\!\!\!\!\!
&&\beta_\kappa=\dfrac{2kk'}{\kappa^2},\label{cof1}\\
\!\!\!\!\!\!\!\!\!\!\!\!\!\!\!\!\!\!\!\!\!\!\!\!\!\!\!\!\!\!\!\!\!\!\!
&&\delta_\kappa=1+2\dfrac{k^2+k'^2}{\kappa^2},\label{cof2}\\
\!\!\!\!\!\!\!\!\!\!\!\!\!\!\!\!\!\!\!\!\!\!\!\!\!\!\!\!\!\!\!\!\!\!\!
&&\eta_\kappa=\dfrac{kk'}{\kappa^2}+
\sqrt{1+\dfrac{k^2}{\kappa^2}}\sqrt{1+\dfrac{k'^2}{\kappa^2}}-
k_C\left(\dfrac{1}{k'}\sqrt{1+\dfrac{k^2}{\kappa^2}}-
\dfrac{1}{k}\sqrt{1+\dfrac{k'^2}{\kappa^2}}\right).\label{cof3}
\end{eqnarray}

First of all, one sees straightforwardly that in the Special Relativity limit $\ell\rightarrow0$, i.e. $\kappa\rightarrow\infty$, the coeff\/icients (\ref{cof1}), (\ref{cof2}), and (\ref{cof3}) reads
\begin{eqnarray}
\!\!\!\!\!\!\!\!\!\!\!\!\!\!\!\!\!\!\!\!\!\!\!\!\!\!\!\!\!\!\!\!\!\!\!
&&\beta_\infty=0,\label{cof1a}\\
\!\!\!\!\!\!\!\!\!\!\!\!\!\!\!\!\!\!\!\!\!\!\!\!\!\!\!\!\!\!\!\!\!\!\!
&&\delta_\infty=1,\label{cof2a}\\
\!\!\!\!\!\!\!\!\!\!\!\!\!\!\!\!\!\!\!\!\!\!\!\!\!\!\!\!\!\!\!\!\!\!\!
&&\eta_\infty=1-k_C\left(\dfrac{1}{k'}-\dfrac{1}{k}\right),\label{cof3a}
\end{eqnarray}
so that the modif\/ied Compton equation \index{Compton equation!modif\/ied}(\ref{coseqn}) becomes linear in the cosinus
\begin{eqnarray}\label{coseqn1}
-\cos\theta+1-k_C\left(\dfrac{1}{k'}-\dfrac{1}{k}\right)=0,
\end{eqnarray}
and its solution can be presented in the form
\begin{equation}\label{comptoneqn}
  \slashed{\lambda}_C(1-\cos\theta)=\dfrac{1}{k'}-\dfrac{1}{k},
\end{equation}
called the Compton equation\index{Compton equation}, describing the photon-electron scattering in frames of Special Relativity as originally deduced by A.H. Compton. Because the boundaries $-1\leq\cos\theta\leq1$ hold, one obtains that in the Compton effect\index{Compton effect} the nontrivial restriction
\begin{equation}
\lambda\leq\lambda'\leq\lambda+2\lambda_C,
\end{equation}
for the outgoing photon wavelength, holds.
\subsection{The Relativistic Limit. The Lensing Hypothesis.}
However, in general the Snyder--Sidharth deformation \index{Snyder--Sidharth deformation}results in the modif\/ied Compton equation \index{Compton equation}(\ref{coseqn}) which has solutions different from the Compton equation (\ref{comptoneqn}). By straightforward easy computation one obtains formally two mathematically and physically distinguish solutions of the equation (\ref{coseqn})
\begin{equation}\label{cosdef}
\cos\theta=\dfrac{1}{2\beta_\kappa}\left(\delta_\kappa\pm
\sqrt{\delta^2_\kappa-4\beta_\kappa\eta_\kappa}\right),
\end{equation}
and in this way one can determine the exact formulas for the cosinus
\begin{eqnarray}\label{mycos}
\!\!\!\!\!\!\!\!\!\!\!\!\!\!\!
\cos\theta=\left[1\pm\sqrt{g^2\left(\dfrac{k}{\kappa},\dfrac{k'}{\kappa}\right)+
\dfrac{2k_C}{\kappa}g\left(\dfrac{k}{\kappa},\dfrac{k'}{\kappa}\right)+1}\right]\dfrac{\kappa^2}{4kk'}+
\dfrac{1}{2}\left(\dfrac{k}{k'}+\dfrac{k'}{k}\right),
\end{eqnarray}
where we have introduced the dimensionless quantity
\begin{equation}\label{gfun}
g\left(\dfrac{k}{\kappa},\dfrac{k'}{\kappa}\right)=\dfrac{2E_\gamma}{\kappa}-\dfrac{2E_{\gamma'}}{\kappa},
\end{equation}
where we have used the Snyder--Sidharth Hamiltonian constraint \index{Snyder--Sidharth Hamiltonian constraint}for photons (\ref{ssgam}). In the light of the law of conservation of energy (\ref{encon}) one sees that the function (\ref{gfun}) in fact can be completely expressed via energies of rest and scattered electron
\begin{equation}
g\left(\dfrac{k}{\kappa},\dfrac{k'}{\kappa}\right)=\dfrac{2(E_\gamma-E_{\gamma'})}{\kappa}=\dfrac{2(E_e-E_0)}{\kappa},
\end{equation}
where $E_e$ is given by the Snyder--Sidharth constraint \index{Snyder--Sidharth Hamiltonian constraint}for electrons (\ref{ssen}). It means that the function (\ref{gfun}) is always positively def\/ined $g\left(\dfrac{k}{\kappa},\dfrac{k'}{\kappa}\right)\geq0$. It can be seen straightforwardly that if one exchanges momenta of photons $k'\leftrightarrow k$ then the function (\ref{gfun}) changes the sign $g\left(\dfrac{k}{\kappa},\dfrac{k'}{\kappa}\right)\rightarrow-g\left(\dfrac{k}{\kappa},\dfrac{k'}{\kappa}\right)$, i.e. $g\left(\dfrac{k}{\kappa},\dfrac{k'}{\kappa}\right)$ is an odd function with respect to such an exchange. Therefore, the presence of the square root term in the cosinus formula (\ref{mycos}) breaks crossing symmetry\index{crossing symmetry} manifestly. This fact contradicts to the nature of cosinus which is an even function. It results in the conclusion that presence of a minimal scale breaks crossing symmetry\index{crossing symmetry}. Moreover, in general the cosinus (\ref{mycos}) does not belong to the range of cosinus $[-1,1]$, but we shall discuss and solve this problematic point later.

In the relativistic limit $\kappa\rightarrow\infty$ the Eqs. (\ref{mycos}) behave like
\begin{eqnarray}\label{mycosw}
\cos\theta=(1\pm1)\dfrac{\kappa^2}{4kk'}+\dfrac{1}{2}\left(\dfrac{k}{k'}+\dfrac{k'}{k}\right).
\end{eqnarray}
For the minus sign case the second term in (\ref{mycosw}) vanishes as $\kappa\rightarrow\infty$, while the plus sign case leads to non renormalizable divergence. In other words the physical solution is the solution with the minus sign, and when one applies the limit $\kappa\rightarrow\infty$ it gives the f\/inite result
\begin{equation}\label{rellim}
\slashed{\lambda}_C(1-\cos\theta)=\dfrac{1}{k_C}\left(1-\dfrac{1}{2}\left(\dfrac{k}{k'}+\dfrac{k'}{k}\right)\right),
\end{equation}
which blatantly differs from the Compton equation \index{Compton equation}(\ref{comptoneqn}). It means that despite using of the relativistic limit to the deformed scattering the received result does not reconstruct the relativistic scattering.

Anyway, for conceptual correctness one can compare \emph{ad hoc} the equation (\ref{rellim}) with the Compton equation \index{Compton equation}(\ref{comptoneqn})
\begin{equation}
\dfrac{1}{k'}-\dfrac{1}{k}=
\dfrac{1}{k_C}\left(1-\dfrac{1}{2}\left(\dfrac{k}{k'}+\dfrac{k'}{k}\right)\right),
\end{equation}
and easy obtain the condition for which the relativistic limit agrees with the Compton equation\index{Compton equation}
\begin{equation}
k'^2-2(k_C+k)k'+k^2+2k_Ck=0.
\end{equation}
This condition is satisf\/ied for two cases $k'=k_C+k\pm k_C$, i.e.
\begin{equation}\label{kkpr}
k'-k=2k_C\qquad\textrm{or}\qquad k'-k=0,
\end{equation}
which can be expressed by incoming and outgoing photons wavelengths
\begin{equation}\label{solus}
\dfrac{1}{\lambda'}-\dfrac{1}{\lambda}=\dfrac{2}{\lambda_C}\qquad\textrm{or}\qquad \dfrac{1}{\lambda'}-\dfrac{1}{\lambda}=0.
\end{equation}
The f\/irst solution looks like special case of the lensmaker's equation\index{lensmaker's equation},
\begin{equation}
\dfrac{1}{r_1}-\dfrac{1}{r_2}+\left(1-\dfrac{1}{n}\right)\dfrac{\delta}{r_1r_2}=\dfrac{1}{(n-1)f},
\end{equation}
for the refractive index $n=3/2=c/v$, focal length of the lens $f=\lambda_C$, and lens radii of curvature $r_1=\lambda'$ and $r_2=\lambda$, and the thickness of the lens $\delta=0$, i.e. the thin lens creating by the medium in which speed of light is $v=2c/3$. The second one corresponds with the telescopic lens case $f=\infty$ in this medium, but because photon wavelength does not change in this case we shall consider this solution as nonphysical.

In this way one can take into account the \emph{ad hoc} generalization
\begin{equation}\label{lensing}
\dfrac{\omega'}{c}-\dfrac{\omega}{c}+\delta\dfrac{n-1}{n}\dfrac{\omega}{c}\dfrac{\omega'}{c}=\dfrac{k_C}{n-1},
\end{equation}
which in the relativistic limit $\kappa\rightarrow\infty$ leads to
\begin{equation}
  k'-k+\delta\dfrac{n-1}{n}kk'=\dfrac{k_C}{n-1},
\end{equation}
i.e. with $k=1/\slashed{\lambda}=\dfrac{2\pi}{\lambda}$, $\delta=0$, and $n=3/2$ reconstructs the f\/irst solution (\ref{solus}). The problem is to establish the linkage between the thickness $d$ and a minimal scale $\ell$, and in general $n>1$ as a numerical coeff\/icient. We shall call (\ref{lensing}) \emph{the lensing hypothesis}\index{lensing hypothesis}.

\subsection{Bounds on the Modif\/ied Compton Equation}
Let us construct straightforwardly the general solution of the modif\/ied Compton equation (\ref{coseqn})\index{Compton equation!modif\/ied}. Because $\delta_\kappa\neq0$ always holds, let us extract $\cos\theta$ manifestly from the Eq. (\ref{coseqn})
\begin{equation}\label{cosik}
  \cos\theta=\dfrac{\beta_\kappa}{\delta_\kappa}\cos^2\theta+\dfrac{\eta_\kappa}{\delta_\kappa}.
\end{equation}
Then however, one must involve the fact that $\cos\theta$ is bounded function, i.e. $-1\leq\cos\theta\leq1$. Application of these boundaries to (\ref{cosik}) leads to the following restrictions for $\cos^2\theta$
\begin{equation}\label{cosik1}
-\dfrac{\delta_\kappa+\eta_\kappa}{\beta_\kappa}\leq\cos^2\theta\leq\dfrac{\delta_\kappa-\eta_\kappa}{\beta_\kappa}.
\end{equation}
The equation (\ref{cosik1}), however, must be considered with taking into account the fact $\cos^2\theta$ is naturally bounded to $0\leq\cos^2\theta\leq1$. Employment of these boundary values results in the system of constraints
\begin{eqnarray}
-\dfrac{\delta_\kappa+\eta_\kappa}{\beta_\kappa}&\equiv&0,\\
\dfrac{\delta_\kappa-\eta_\kappa}{\beta_\kappa}&\equiv&1,
\end{eqnarray}
which factually reduces momentum space of outgoing and incoming photons. In other words, we consider the equation (\ref{coseqn}) as the basic relation for the scattering, but no its solution which is secondary result. Let us see the consequences of such an approach.

For $\beta_\kappa\neq0$, i.e. factually for nonzero values of incoming and outgoing photon momenta, the reduction is given by the equivalent conditions
\begin{eqnarray}
\delta_\kappa+\eta_\kappa&=&0,\\
\delta_\kappa-\eta_\kappa&=&\beta_\kappa,
\end{eqnarray}
which are not diff\/icult to solve. The results are
\begin{eqnarray}
\delta_\kappa&=&\dfrac{1}{2}\beta_\kappa,\label{conny1}\\
\eta_\kappa&=&-\dfrac{1}{2}\beta_\kappa.\label{conny2}
\end{eqnarray}
It is easy to see that the f\/irst condition says that
\begin{equation}\label{se1}
2\dfrac{k'^2+k^2}{\kappa^2}-\dfrac{kk'}{\kappa^2}+1=0,
\end{equation}
while the second constraint can be written as
\begin{equation}\label{se2}
2\dfrac{kk'}{\kappa^2}+\sqrt{1+\dfrac{k^2}{\kappa^2}}\sqrt{1+\dfrac{k'^2}{\kappa^2}}-
k_C\left(\dfrac{1}{k'}\sqrt{1+\dfrac{k^2}{\kappa^2}}-
\dfrac{1}{k}\sqrt{1+\dfrac{k'^2}{\kappa^2}}\right)=0.
\end{equation}
Application of (\ref{se1}) and (\ref{se2}) allows to reduce the equation (\ref{coseqn})
\begin{equation}\label{cosen1}
\cos^2\theta-\dfrac{1}{2}\cos\theta-\dfrac{1}{2}=0.
\end{equation}
Because of (\ref{coseqn1}) does not contain any wave vectors therefore its solutions are constant and independent on photons wavelengths. In the usual Compton effect \index{Compton effect}the cosinus depends on the photon wavelengths. In the modif\/ied case the cosinus (\ref{mycos}) in general does not satisfy the cosinus variability range $[-1,1]$, and application of these limits to the equation (\ref{coseqn}) led us to the bounds (\ref{se1})-(\ref{se2}), what resulted in the equation (\ref{cosen1}). In this way the Compton effect modif\/ied due to the Snyder--Sidharth Hamiltonian constraint \index{Snyder--Sidharth Hamiltonian constraint}is solved by a constant scattering angle which solves the equation (\ref{cosen1}), and the values of the wave vectors $k$ and $k'$ following from the bounds (\ref{se1})-(\ref{se2}). The constant angle values are easy to derive
\begin{eqnarray}
\cos\theta&=&1\longrightarrow\theta=2n\pi,\label{1stsol}\\
\cos\theta&=&-\dfrac{1}{2}\longrightarrow\theta=\pm\dfrac{\pi}{3}+2n\pi\label{2ndsol},
\end{eqnarray}
where $n\in\textbf{Z}$. The f\/irst solution (\ref{1stsol}) means no scattering or backward scattering. Solutions of the system (\ref{se1})-(\ref{se2}) are not easy to extract. With no additional constraint(s) between $k$ and $k'$ this system leads to a polynomial equation of more than 40 degree which must be treated by complicated methods of the Galois group\index{Galois group}.

However, one can use the lensing hypothesis\index{lensing hypothesis} (\ref{lensing}) as the additional constraint which written out explicitly is
\begin{equation}
  \sqrt{k'^2+\dfrac{k'^4}{\kappa^2}}-\sqrt{k^2+\dfrac{k^4}{\kappa^2}}+
  \delta\dfrac{n-1}{n}\sqrt{k^2+\dfrac{k^4}{\kappa^2}}\sqrt{k'^2+\dfrac{k'^4}{\kappa^2}}=\dfrac{k_C}{n-1},
\end{equation}
what after elementary algebraic manipulations takes the form
\begin{equation}\label{addc}
k_C\left(\dfrac{1}{k'}\sqrt{1+\dfrac{k^2}{\kappa^2}}-\dfrac{1}{k}\sqrt{1+\dfrac{k'^2}{\kappa^2}}\right)=
k_C\delta\dfrac{n-1}{n}\sqrt{1+\dfrac{k^2}{\kappa^2}}\sqrt{1+\dfrac{k'^2}{\kappa^2}}-\dfrac{\kappa^2}{kk'}\dfrac{1}{n-1}\dfrac{k_C^2}{\kappa^2}.
\end{equation}
In other words, using of the additional constraint (\ref{addc}) within the constraint (\ref{se2}) gives the result
\begin{equation}
2\dfrac{kk'}{\kappa^2}+
\left(1+k_C\delta\dfrac{n-1}{n}\right)\sqrt{1+\dfrac{k^2}{\kappa^2}}\sqrt{1+\dfrac{k'^2}{\kappa^2}}-
\dfrac{\kappa^2}{kk'}\dfrac{1}{n-1}\dfrac{k_C^2}{\kappa^2}=0,
\end{equation}
which can be presented as
\begin{equation}\label{se2a}
\sqrt{1+\dfrac{k^2+k'^2}{\kappa^2}+\left(\dfrac{kk'}{\kappa^2}\right)^2}=
\dfrac{n}{(1+k_C\delta)n-k_C\delta}\left(\dfrac{1}{n-1}\dfrac{k_C^2}{\kappa^2}\dfrac{\kappa^2}{kk'}-2\dfrac{kk'}{\kappa^2}\right).
\end{equation}
Now one can apply the constraint (\ref{se1}) to the LHS of the equation (\ref{se2a}). It gives
\begin{equation}\label{se2b}
\sqrt{\dfrac{1}{2}+\dfrac{1}{2}\dfrac{kk'}{\kappa^2}+\left(\dfrac{kk'}{\kappa^2}\right)^2}=
\dfrac{n}{(1+k_C\delta)n-k_C\delta}\left(\dfrac{1}{n-1}\dfrac{k_C^2}{\kappa^2}\dfrac{\kappa^2}{kk'}-2\dfrac{kk'}{\kappa^2}\right).
\end{equation}
By squaring both sides of the equation (\ref{se2b}), doing very few elementary algebraic manipulations, and grouping the result with respect to powers of $x=\dfrac{kk'}{\kappa^2}$ one obtains f\/inally the condition
\begin{equation}\label{se2c}
Ax^4+Bx^3+Cx^2-D=0
\end{equation}
where we have introduced the coeff\/icients
\begin{eqnarray}
A&=&2\dfrac{(n-1)^3}{n}k_C\delta\left(2+\dfrac{n-1}{n}k_C\delta\right),\label{big1}\\
B&=&(n-1)^2\left(1+\dfrac{n-1}{n}k_C\delta\right)^2,\label{big2}\\
C&=&(n-1)^2\left(1+\dfrac{n-1}{n}k_C\delta\right)^2+4(n-1)\dfrac{k_C^2}{\kappa^2},\label{big3}\\
D&=&2\dfrac{k_C^4}{\kappa^4},\label{big4}
\end{eqnarray}
which are always positive and nonzero. Naturally, the equation (\ref{se2c}) in general can be solved and possesses solutions
\begin{equation}
k'=x\dfrac{\kappa^2}{k},\label{kprimk}
\end{equation}
where $x$ is a coeff\/icient. Because of $k'\geqslant0$ and $k>0$, the real and positive values of $x$ are physical. In this way there are two physical $x$
\begin{equation}\label{iks}
x=\dfrac{1}{4A}\left(\sqrt{2a-b-c\pm\dfrac{d}{\sqrt{a+b+c}}}-\sqrt{a+b+c}-B\right),
\end{equation}
with the conditions
\begin{eqnarray}
a+b+c&>&0,\label{condy1}\\
2a\pm\dfrac{d}{\sqrt{a+b+c}}&\geqslant&b+c,\label{condy2}\\
\sqrt{2a-b-c\pm\dfrac{d}{\sqrt{a+b+c}}}&\geqslant&B+\sqrt{a+b+c},\label{condy3}
\end{eqnarray}
where we have introduced the shortened notation
\begin{eqnarray}
a&=&B^2-\frac{8}{3}CA,\label{small1}\\
b&=&\dfrac{4}{3}2^{1/3}A\beta\left(\alpha+\sqrt{\alpha^2-4\beta^3}\right)^{-\frac{1}{3}},\label{small2}\\
c&=&2^{1/3}A\left(\alpha+\sqrt{\alpha^2-4\beta^3}\right)^{\frac{1}{3}},\label{small3}\\
d&=&\dfrac{2B}{A}\left(-B^2+4CA\right),\label{small4}
\end{eqnarray}
and $\alpha=2C^3-27B^2D+72ACD$, $\beta=C^2-12AD$.

Interestingly, one can see easy that the case $n=1$ applied to (\ref{se2c}) leads to $D=0$, what is true for the only relativistic limit $\kappa\rightarrow\infty$. However, the result of the relativistic limit applied to the equation (\ref{se2c}) has also different countenance. Before taking the limit one must reduce this equation maximally with respect to powers of $1/\kappa\rightarrow0$, i.e.
\begin{equation}
A\dfrac{\left(kk'\right)^4}{\kappa^4}+B\dfrac{\left(kk'\right)^3}{\kappa^2}+
C\left(kk'\right)^2-2k_C^4=0,
\end{equation}
and by this reason in such a limit one obtains the condition
\begin{equation}\label{se2d}
(n-1)^2\left(1+\dfrac{n-1}{n}k_C\delta\right)^2\left(kk'\right)^2-2k_C^4=0,
\end{equation}
which is easy to solve
\begin{equation}\label{se2e}
k'=\dfrac{\dfrac{\sqrt{2}}{n-1}}{1+\dfrac{n-1}{n}k_C\delta}\dfrac{k_C^2}{k}\sim\dfrac{1}{k},
\end{equation}
where we neglected negative solution as non physical. It is seen that $n=1$ is not appropriate for such a solution.

\section{The Dispersional Generalization}

Let us construct f\/inally the other approach to the modif\/ied Compton effect \index{Compton effect!modif\/ied}argued by the generalization due to the specif\/ic form of the dispersion relations. For this let us consider f\/irst the deduction due to the unmodif\/ied case, i.e. Special Relativity. In such a situation the dispersion relations for a photon and an electron have the form
\begin{eqnarray}
\dfrac{\omega^2}{c^2}&=&k^2,\label{disrel1}\\
\dfrac{\omega^2_e}{c^2}&=&k_C^2+k_e^2.\label{disrel2}
\end{eqnarray}
By this reason the laws of conservation of momentum (\ref{momres}) and energy (\ref{encon1}) expressed via the dispersion relations (\ref{disrel1}) and (\ref{disrel2}) have the following form
\begin{eqnarray}
\dfrac{\omega^2_e}{c^2}-k_C^2&=&\dfrac{\omega^2}{c^2}+\dfrac{\omega'^2}{c^2}-\dfrac{\omega}{c}\dfrac{\omega'}{c}\cos\theta,\label{dismom}\\
\dfrac{\omega_e}{c}&=&k_C+\dfrac{\omega}{c}-\dfrac{\omega'}{c}.\label{disen}
\end{eqnarray}
and after elementary algebraic manipulations lead to the result
\begin{equation}
  \slashed{\lambda}_C(1-\cos\theta)=\dfrac{c}{\omega'}-\dfrac{c}{\omega}, \label{discomp}
\end{equation}
which after application of the explicit form of the dispersion relation (\ref{disrel1}) gives the Compton equation \index{Compton equation}(\ref{comptoneqn}). In this manner we shall call this equation \emph{the generalized Compton equation}.\index{Compton equation!generalized}

Similar reasoning can be performed to the modif\/ied dispersion relations (\ref{ssgam1} and (\ref{ssen1}). Then the laws of conservation of momentum and energy expressed via the dispersion relations are preserved in the form (\ref{dismom})-(\ref{disen}) obtained by Special Relativity. Therefore, also the Compton equation \index{Compton equation} (\ref{discomp}) is preserved. However, the change is due to the specif\/ic dispersion relations (\ref{ssgam1}) and (\ref{ssen1}), so that the modif\/ied Compton effect \index{Compton effect!modif\/ied}is described by the equation
\begin{equation}
  \slashed{\lambda}_C(1-\cos\theta)=\dfrac{1}{\sqrt{k'^2+\dfrac{k'^4}{\kappa^2}}}-
  \dfrac{1}{\sqrt{k^2+\dfrac{k^4}{\kappa^2}}}. \label{discomp1}
\end{equation}
Interestingly, in the relativistic limit $\kappa\rightarrow\infty$ the modif\/ied Compton equation\index{Compton equation!modif\/ied} (\ref{discomp1}) coincides with the usual Compton equation (\ref{discomp}) with no additional presumptions. We shall call such an approach \emph{the dispersional generalization}\index{dispersional generalization}. It is clear that this generalization should be working also for elementary processes other than the Compton scattering\index{Compton scattering}. By this reason let us express the proposition
\begin{proposition}[The Dispersional Generalization]
Let us presume that there is an initial theory of elementary precesses having established results, which is characterized by certain dispersion relations and the laws of conservation. Let us consider the theory due to a modif\/ication of an initial theory. The consistent analysis of an arbitrary elementary process within the modif\/ied theory is based on the three-step procedure:
\begin{enumerate}
\item Derivation of the dispersion relations due of the modif\/ied theory,
\item Application of the dispersion relations to the laws of conservation of the modif\/ied theory,
\item Using of the explicit form of the dispersion relations within the results obtained due to the laws of conservation in frames of the modif\/ied theory.
\end{enumerate}
The analysis is consistent if and only if the results of the modif\/ied theory coincide with the results of an initial theory for lack of the modif\/ication.
\end{proposition}
Because of $-1\leqslant\cos\theta\leqslant1$, the relation (\ref{discomp1}) allows to establish the bounds for wave vector of an outgoing photon $k'_{min}\leqslant k\leqslant k'_{max}$, where
\begin{eqnarray}
k'_{min}&=&\dfrac{\kappa}{\sqrt{2}}\sqrt{\sqrt{1+\dfrac{4}{\kappa^2}\left(k^2+\dfrac{k^4}{\kappa^2}\right)}-1},\\
k'_{max}&=&\dfrac{\kappa}{\sqrt{2}}\sqrt{\sqrt{1+\dfrac{4}{\kappa^2}\dfrac{k^2+\dfrac{k^4}{\kappa^2}}{\left(1+2\slashed{\lambda}_C\sqrt{k^2+\dfrac{k^4}{\kappa^2}}\right)^2}}-1}.
\end{eqnarray}

Let us apply the lensing hypothesis\index{lensing hypothesis} to the obtained general result (\ref{discomp}). The formula (\ref{lensing}) can be rewritten in the form
\begin{equation}\label{lensing1}
\dfrac{c}{\omega'}-\dfrac{c}{\omega}=\delta\dfrac{n-1}{n}-\dfrac{c}{\omega}\dfrac{c}{\omega'}\dfrac{k_C}{n-1},
\end{equation}
what means that the generalized Compton equation \index{Compton equation!generalized}(\ref{discomp}) becomes
\begin{equation}
  \slashed{\lambda}_C(1-\cos\theta)=\delta\dfrac{n-1}{n}-\dfrac{c}{\omega}\dfrac{c}{\omega'}\dfrac{k_C}{n-1}. \label{discomp2}
\end{equation}
In this manner, by application of the identif\/ication method \index{identif\/ication method}, one can establish the following relations
\begin{eqnarray}
  \delta&=&\dfrac{n}{n-1}\slashed{\lambda}_C,\label{dee}\\
  \cos\theta&=&\dfrac{c}{\omega}\dfrac{c}{\omega'}\dfrac{k_C^2}{n-1},\label{lenscos}
\end{eqnarray}
what allows to derive
\begin{equation}\label{coslens}
  \cos\theta=\dfrac{c}{\omega}\dfrac{c}{\omega'}\dfrac{\delta-\slashed{\lambda}_C}{\slashed{\lambda}_C^3}.
\end{equation}
Because, however, $-1\leq\cos\theta\leq1$ one has the bound
\begin{equation}\label{ombound}
\dfrac{\omega'}{c}\geqslant\dfrac{\delta-\slashed{\lambda}_C}{\slashed{\lambda}_C^3}\dfrac{c}{\omega},
\end{equation}
which for the Compton effect \index{Compton effect}means that
\begin{equation}
k'\geqslant\dfrac{\delta-\slashed{\lambda}_C}{\slashed{\lambda}_C^3}\dfrac{1}{k},
\end{equation}
while for the modif\/ied Compton effect \index{Compton effect!modif\/ied}gives
\begin{equation}
k'\geqslant\dfrac{\kappa}{\sqrt{2}}\left[\sqrt{1+
\dfrac{4}{\kappa^2}\left(\dfrac{\delta-\slashed{\lambda}_C}{\slashed{\lambda}_C^3}\right)^2\dfrac{1}{k^2+\dfrac{k^4}{\kappa^2}}}-1\right].
\end{equation}
In the case of the usual Compton effect \index{Compton effect}the formula (\ref{coslens}) gives
\begin{equation}
  \cos\theta=\dfrac{1}{k}\dfrac{1}{k'}\dfrac{\delta-\slashed{\lambda}_C}{\slashed{\lambda}_C^3},
\end{equation}
while in the modif\/ied case one receives
\begin{equation}
  \cos\theta=\dfrac{1}{\sqrt{k'^2+\dfrac{k'^4}{\kappa^2}}}
  \dfrac{1}{\sqrt{k^2+\dfrac{k^4}{\kappa^2}}}\dfrac{\delta-\slashed{\lambda}_C}{\slashed{\lambda}_C^3}.
\end{equation}

The lensing hypothesis\index{lensing hypothesis} (\ref{lensing1}) together with (\ref{dee}) leads to
\begin{equation}\label{omprimlens}
  \dfrac{\omega'}{c}=\dfrac{1+\dfrac{c}{\omega}\dfrac{\delta-\slashed{\lambda}_C}{\slashed{\lambda}_C^2}}{\slashed{\lambda}_C+\dfrac{c}{\omega}},
\end{equation}
what together with (\ref{ombound}) leads to the another bound
\begin{equation}
  \dfrac{\omega}{c}\geqslant\dfrac{1}{\slashed{\lambda}_C}\sqrt{\dfrac{\delta}{\slashed{\lambda}_C}-1}=\dfrac{\sqrt{n-1}}{\slashed{\lambda}_C},
\end{equation}
which for the usual Compton effect \index{Compton effect}means that
\begin{equation}
  k\geqslant\dfrac{1}{\slashed{\lambda}_C}\sqrt{\dfrac{\delta}{\slashed{\lambda}_C}-1},
\end{equation}
while for the modif\/ied Compton effect leads to
\begin{equation}
k\geqslant\dfrac{\kappa}{\sqrt{2}}\sqrt{\sqrt{1+\dfrac{4}{\kappa^2}\dfrac{1}{\slashed{\lambda}_C}\sqrt{\dfrac{\delta}{\slashed{\lambda}_C}-1}}-1}.
\end{equation}

The formula (\ref{omprimlens}) can be rewritten in the form
\begin{equation}\label{omapprox}
\dfrac{\omega'}{c}=\dfrac{1}{\slashed{\lambda}_C}+
\left(\dfrac{1}{\slashed{\lambda}_C}\right)^2\dfrac{\delta-2\slashed{\lambda}_C}{1+\slashed{\lambda}_C\dfrac{\omega}{c}},
\end{equation}
and studied approximatively with respect to the point $\dfrac{\omega}{c}=\dfrac{1}{\slashed{\lambda}_C}$, which def\/ines the following value of wave vector of incoming photon.
\begin{equation}
k=\dfrac{\kappa}{\sqrt{2}}\sqrt{\sqrt{1+4\dfrac{k_C^2}{\kappa^2}}-1}.
\end{equation}
It is easy to see that, because of the formula (\ref{omprimlens}), in such a situation
\begin{equation}\label{omprimlens1}
  \dfrac{\omega'}{c}=\dfrac{1+\dfrac{\delta-\slashed{\lambda}_C}{\slashed{\lambda}_C}}{2\slashed{\lambda}_C}=\dfrac{\delta}{2\slashed{\lambda}_C^2},
\end{equation}
and by the this reason the cosinus formula (\ref{coslens}) takes the form
\begin{equation}\label{coslens1}
  \cos\theta=2\left(1-\dfrac{\slashed{\lambda}_C}{\delta}\right)=\dfrac{2}{n},
\end{equation}
i.e. the scattering is possible when the thickness of the lens is
\begin{equation}
  \dfrac{2}{3}\slashed{\lambda}_C\leqslant\delta\leqslant2\slashed{\lambda}_C,
\end{equation}
or equivalently the refraction index of the medium is
\begin{equation}
  n\geqslant2.
\end{equation}

For $\dfrac{\omega}{c}<\dfrac{1}{\slashed{\lambda}_C}$ the correct expansion of the formula (\ref{omapprox} is given by
\begin{equation}\label{app1}
\dfrac{\omega'}{c}=\dfrac{1}{\slashed{\lambda}_C}+
(\delta-2\slashed{\lambda}_C)\left(\dfrac{1}{\slashed{\lambda}_C}\right)^2\sum_{n=0}^{\infty}(-1)^n\slashed{\lambda}_C^n\left(\dfrac{\omega}{c}\right)^n,
\end{equation}
while for $\dfrac{\omega}{c}>\dfrac{1}{\slashed{\lambda}_C}$ the appropriate expansion has the following form
\begin{equation}\label{app2}
\dfrac{\omega'}{c}=\dfrac{1}{\slashed{\lambda}_C}+
(\delta-2\slashed{\lambda}_C)\left(\dfrac{1}{\slashed{\lambda}_C}\right)^2\sum_{n=1}^{\infty}(-1)^n\left(\dfrac{1}{\slashed{\lambda}_C}\right)^n\left(\dfrac{c}{\omega}\right)^n.
\end{equation}
In the neighborhood of the point $\dfrac{\omega}{c}=\dfrac{1}{\slashed{\lambda}_C}$ the approximations (\ref{app1}) and (\ref{app2})
\begin{eqnarray}
\dfrac{\omega'}{c}&\approx&\dfrac{\delta-2\slashed{\lambda}_C}{\slashed{\lambda}_C^2}-\dfrac{\delta-2\slashed{\lambda}_C}{\slashed{\lambda}_C}\dfrac{\omega}{c},\\
\dfrac{\omega'}{c}&\approx&\dfrac{1}{\slashed{\lambda}_C}-\dfrac{\delta-2\slashed{\lambda}_C}{\slashed{\lambda}_C^3}\dfrac{c}{\omega},
\end{eqnarray}
must coincide. By this reason in such a situation one obtains
\begin{equation}\label{3lambda}
\delta=3\slashed{\lambda}_C,
\end{equation}
i.e. the refraction index $n$ near $\dfrac{\omega}{c}=\dfrac{1}{\slashed{\lambda}_C}$ has the value
\begin{equation}
  n=1.5.
\end{equation}
In the light of the relation (\ref{3lambda}) the cosinus (\ref{coslens1}) becomes
\begin{equation}\label{coslens2}
  \cos\theta=\dfrac{4}{3},
\end{equation}
i.e. is non-physical. It means that near the point $\dfrac{\omega}{c}=\dfrac{1}{\slashed{\lambda}_C}$ the modif\/ied Compton effect\index{Compton effect!modif\/ied}, considered in frames of the lensing hypothesis\index{lensing hypothesis}, has no place because of lack of scattering.

If $\theta$, $\omega$, and $\omega'$ are established, e.g. via experimental data, then the thickness of the lens $d$ and the refractive index $n$ are
\begin{eqnarray}
  \delta&=&\slashed{\lambda}_C\left(1+\slashed{\lambda}_C^2 \dfrac{\omega}{c}\dfrac{\omega'}{c}\cos\theta\right),\label{thickness}\\
  n&=&1+\dfrac{1}{\slashed{\lambda}_C^2}\dfrac{c}{\omega}\dfrac{c}{\omega'}\sec\theta.\label{refraction}
\end{eqnarray}
Because, however, $n=c/v$ where $v$ is velocity of light in the medium in which the refraction has a place, one can see that
\begin{equation}
v=c\dfrac{\slashed{\lambda}_C^2 \dfrac{\omega}{c}\dfrac{\omega'}{c}\cos\theta}{1+\slashed{\lambda}_C^2 \dfrac{\omega}{c}\dfrac{\omega'}{c}\cos\theta}.
\end{equation}

In this section we presented the approach to the modif\/ied Compton effect\index{Compton effect!modif\/ied}. First, we constructed the relativistic analysis which produced the modif\/ied Compton equation \index{Compton equation!modif\/ied}coinciding to the Compton equation for lack of the modif\/ication due to the Snyder--Sidharth deformation\index{Snyder--Sidharth deformation}. Then, we solved the modif\/ied Compton equation and showed that the result of lack of modif\/ication differs from deductions of Special Relativity. By this reason we called out the lensing hypothesis\index{lensing hypothesis}. Then, we showed that the scattered angle values are independent on particle energies, and derived the equation jointing wave vectors of incoming and outgoing photon. Finally, we called out the dispersional generalization which led us to the correct result in lack of modif\/ication. Application of the lensing hypothesis\index{lensing hypothesis} and the identif\/ication method \index{identif\/ication method}together with the dispersional generalization produced a number of new results due to the modif\/ied Compton effect\index{Compton effect!modif\/ied}.

\chapter{The Neutrinos: Masses \& Chiral Condensate}

\section{Outlook on Noncommutative Geometry}

In 1947 an American physicist H.S. Snyder, for elimination of the infrared catastrophe\index{infrared catastrophe} in the Compton effect \index{Compton effect}and effectively resolving the problem of ultraviolet divergences in quantum f\/ield theory, proposed employing the model of space-time based on the commutators \cite{snyder}
\begin{eqnarray}
\dfrac{i}{\hslash}[x,p]=1+\alpha\left(\dfrac{\ell}{\hslash}\right)^2p^2,\label{ndA}\\ \dfrac{i}{\hslash}[x,y]=O(\ell^2)\quad,\label{ndB}
\end{eqnarray}
where $p$ is three momentum of a particle, $x$ and $y$ are two different points of space, $\ell$ is a fundamental length scale, $\hslash$ is the Planck constant, $\alpha\sim1$ is a dimensionless constant, $[\cdot,\cdot]$ is an appropriate Lie bracket. For the Lorentz and Poincar\'{e} invariance\index{Lorentz invariance}\index{Poincar\'{e} invariance} modif\/ied due to $\ell$, Snyder considered a momentum space of constant curvature isometry group, \emph{i.e.} the Poincar\'{e} algebra deformation\index{Poincar\'{e} algebra deformation} into the De Sitter space\index{De Sitter space}.

The Snyder space-time\index{space-time!Snyder} (\ref{ndA})--(\ref{ndB}) def\/ine a noncommutative geometry\index{noncommutative geometry} and a deformation (for basics of the theory and applications see e.g. the bibliography in Ref. \cite{ng}). Let us this in some detail. First, for better insight, let us sketch the rules of noncommutative geometry\index{noncommutative geometry} in a certain general outlook. Let us consider an associative Lie algebra $A$ for which $\tilde{A}=A[[\lambda]]$ is the module due to the ring of formal series $\mathbb{K}[[\lambda]]$ in a parameter $\lambda$. Let us call $\tilde{A}$ a deformation of $A$ i.e. $\mathbb{K}[[\lambda]]$-algebra such that $\tilde{A}/\lambda\tilde{A}\approx A$. If $A$ is endowed with a locally convex topology with continuous laws, \emph{i.e.} is a topological algebra, then $\tilde{A}$ is called topologically free. We presume that in the Lie algebra $A$ the law of composition is determined via an ordinary product and the related bracket is $[\cdot, \cdot]$. In such a situation $\tilde{A}$ is an associative Lie algebra if and only if for arbitrary two elements of the algebra $f,g\in A$ a new product $\star$ and the related bracket $[\cdot, \cdot]_\star$ are
def\/ined as follows
\begin{eqnarray}
f\star g&=&fg+\sum_{n=1}^\infty\lambda^nC_n(f,g),\label{star}\\
\left[f,g\right]_\star&\equiv&f\star g-g\star f=\left[f,g\right]+\sum_{n=1}^\infty\lambda^nB_n(f,g),\label{starb}
\end{eqnarray}
where $C_n$ and $B_n$ are the Hochschild and the Chevalley 2-cochains, and for arbitrary three elements of the algebra $f,g,h\in A$ are satisf\/ied two conditions: the Jacobi identity\index{Jacobi identity}
\begin{equation}
[[f,g]_\star,h]_\star+[[h,f]_\star,g]_\star+[[g,h]_\star,f]_\star=0
\end{equation}
and the law of associativity
\begin{equation}
(f\star g)\star h = f\star(g\star h).
\end{equation}
If $b$ and $\partial$ are the Hochschild and the Chevalley coboundary operators, i.e. such that
$b^2=0$ and $\partial^2=0$, then for each $n$ and $j,k\geqslant1$ such that $j+k=n$ the following relations hold
\begin{eqnarray}
\!\!\!\!\!\!\!\!bC_n(f,g,h)\!\!\!&=&\!\!\!\!\sum_{j,k}\left[C_j\left(C_k(f,g),h\right)-C_j\left(f,C_k(g,h)\right)\right],\\
\!\!\!\!\!\!\!\!\partial B_n(f,g,h)\!\!\!&=&\!\!\!\!\sum_{j,k}\left[B_j\left(B_k(f,g),h\right)+B_j\left(B_k(h,f),g\right)+B_j\left(B_k(g,h),f\right)\right],
\end{eqnarray}
Let $C^\infty(M)$ be an algebra of smooth functions on a differentiable manifold $M$. The law of associativity yields the Hochschild cohomologies\index{cohomology}. An antisymmetric contravariant 2-tensor $\theta$, which trivialize the Schouten–-Nijenhuis bracket $[\theta,\theta]_{SN}=0$ on $M$, determines the Poisson brackets\index{Poisson brackets} $\{f,g\}=i\theta df\wedge dg$ satisfying the Jacobi identity and the Leibniz rule. Than $(M,\{\cdot,\cdot\})$ is called a Poisson manifold.

In 1997 a Russian mathematician M.L. Kontsevich \cite{kontsevich} def\/ined deformation quantization\index{quantization!deformation} of a general Poisson differentiable manifold. Let $\mathbb{R}^d$ be endowed with a Poisson brackets\index{Poisson brackets} \begin{equation}
\alpha(f,g)=\sum_{1\leqslant i,j\leqslant n}\alpha^{ij}\dfrac{\partial f}{\partial x^i}\dfrac{\partial g}{\partial x^j},
\end{equation}
where $1\leqslant k\leqslant d$. For $\star$-product and $n\geqslant0$, exists a family $G_{n,2}$ of $(n(n+1))^n$ oriented graphs $\Gamma$. Let $V_\Gamma$ be the set of vertices of $\Gamma$. This set has $n+2$ elements collected in two subsets: the f\/irst type $\{1,\ldots,n\}$ and the second type $\{\bar{1},\bar{2}\}$. Let $E_\Gamma$ denotes the set of oriented edges of $\Gamma$, having $2n$ elements. The rule is that there is no edge starting at a second type vertex. Let $Star(k)$ denotes the set of oriented edges starting at a f\/irst type vertex $k$ with cardinality $\sharp k=2$, $\sum_{1\leqslant k \leqslant n}\sharp k = 2n$. Than $\{e^1_k,\ldots,e^{\sharp k}_k\}$ are the edges of $\Gamma$ starting at vertex $k$. Vortices starting and ending in the edge $v$ are $v = (s(v),e(v))$ where $s(v)\in\{1,\ldots,n\}$ and $e(v)\in\{1,\ldots,n;\bar{1},\bar{2}\}$. $\Gamma$ has no loop and no parallel multiple edges. For arbitrary two elements $f,g\in C^\infty(\mathbb{R}^d)$ a bidifferential operator $(f,g)\mapsto B_\Gamma(f,g)$ is associated to $\Gamma$. The symbols $\alpha^{e^1_ke^2_k}$ are associated to each f\/irst type vertex $k$ from where the edges $\{e^1_k,e^2_k\}$ start; $f$ is the vertex $1$, and $g$ is the vertex $\bar{2}$. An edge $e^1_k$ acts like differentiation operator $\partial/\partial x^{e^1_k}$ on its ending vertex. Than $B_\Gamma$ is a sum over all maps $I:E_\Gamma\rightarrow\{1,\ldots,d\}$
\begin{equation}
\!\!\!\!\!\!B_\Gamma(f,g)=\sum_I\left(\prod_{k=1}^n\prod_{k'=1}^n\partial_{I(k',k)}\alpha^{I(e^1_k)I(e^2_k)}\right)\!\!\left(\prod_{k_1=1}^n\partial_{I(k_1,\bar{1})}f\right)\!\!\left(\prod_{k_2=1}^n\partial_{I(k_2,\bar{2})}g\right).\!\!\!
\end{equation}
Let us denote by $\mathcal{H}_n$ the conf\/iguration space of $n$ distinct points in upper half-plane $\mathcal{H}=\{z\in\mathbb{C}|\Im(z)>0\}$ with the Lobachevsky hyperbolic metric, which is an open submanifold of $\mathbb{C}^n$. Let for the vertex $k$ such that $1\leqslant k \leqslant n$, $z_k\in\mathcal{H}$ denotes a variable associated to $\Gamma$. The vertices $1$ and $\bar{2}$ are associated to $0\in\mathbb{R}$ and $1\in\mathbb{R}$, respectively. If $\tilde{\phi}_v=\phi(s(v),e(v))$ is a function on $\mathcal{H}_n$, associated to $v$, determined by the angle function $\phi:\mathcal{H}_2\rightarrow\mathbb{R}/2\pi\mathbb{Z}$ having the following form
\begin{equation}
  \phi(z_1,z_2)=\mathrm{Arg}\dfrac{z_2-z_1}{z_2-\bar{z}_1}=\dfrac{1}{2i}\mathrm{Log}\dfrac{\bar{z}_2-z_1}{z_2-\bar{z}_1}\dfrac{z_2-z_1}{\bar{z}_2-\bar{z}_1},
\end{equation}
then the integral of $2n$-form $w(\Gamma)\in\mathbb{R}$ is a weight associated to $\Gamma\in G_{n,2}$
\begin{eqnarray}\label{weight}
w(\Gamma)=\dfrac{1}{n!(2\pi)^{2n}}\int_{\mathcal{H}_n}\bigwedge_{1\leqslant k \leqslant n}\left(d\tilde{\phi}_{e^1_k}\wedge d\tilde{\phi}_{e^2_k}\right), \end{eqnarray}
which does not depend on the Poisson structure or the dimension $d$. On $(\mathbb{R}^d,\alpha)$ the Kontsevich $\star$-product\index{Kontsevich star-product} maps $C^\infty(\mathbb{R})\times C^\infty(\mathbb{R})\rightarrow C^\infty(\mathbb{R})[[\lambda]]$
\begin{equation}\label{konts}
  (f,g)\mapsto f\star g=\sum_{n\geqslant 0}\lambda^nC_n(f,g),
\end{equation}
where $C_0(f,g)=fg$, $C_1(f,g)=\{f,g\}_\alpha=\alpha df\wedge dg$, and in general
\begin{equation}
C_n(f,g)=\sum_{\Gamma\in G_{n,2}}w(\Gamma)B_\Gamma(f,g).
\end{equation}
Equivalence classes of (\ref{konts}) are bijective to the equivalence classes of the Poisson brackets\index{Poisson brackets} $\alpha_\lambda=\sum_{k\geqslant0}\lambda^k\alpha_k$. For linear Poisson structures, \emph{i.e.} on coalgebra $A^\star$, the weight (\ref{weight}) of all even wheel graphs vanishes, and the Kontsevich star product\index{Kontsevich star-product} (\ref{konts}) coincides with the $\star$-product determined by the Duf\/lo isomorphism. This case allows to quantize the class of quadratic Poisson brackets\index{Poisson brackets} belonging to the image of the Drinfeld map which associates a quadratic to a linear bracket.

Let us consider the deformations of phase-space and space given by the parameters $\lambda_{ph}$, $\lambda_s$ being
\begin{eqnarray}
\lambda_{ph}=\dfrac{i\alpha}{2},\\
\lambda_s=\dfrac{i\beta}{2},
\end{eqnarray}
where $\alpha\sim1$ and $\beta\sim 1$ are dimensionless constants, and resulting in the appropriate star product (\ref{star}), or equivalently to the Kontsevich star-product\index{Kontsevich star-product} (\ref{konts}), on the phase space $(x,p)$ and between two distinct space points $x$ and $y$
\begin{eqnarray}
x\star p&=&px+\sum_{n=1}^\infty \left(\dfrac{i\alpha}{2}\right)^nC_n(x,p),\label{star1x}\\
x\star y&=&xy+\sum_{n=1}^\infty \left(\dfrac{i\beta}{2}\right)^nC_n(x,y),\label{star2x}
\end{eqnarray}
where $C_n(x,p)$ and $C_n(x,y)$ are the Hochschild cochains in the Kontsevich formula (\ref{konts}) related to the phase space and the space deformations, respectively. The Lie brackets arising from the star products (\ref{star1x}) and (\ref{star2x}) has the following form
\begin{eqnarray}
\left[x,p\right]_\star&=&\left[x,p\right]+\sum_{n=1}^\infty \left(\dfrac{i\alpha}{2}\right)^nB_n(x,p),\label{starb1x}\\
\left[x,y\right]_\star&=&\left[x,y\right]+\sum_{n=1}^\infty \left(\dfrac{i\beta}{2}\right)^nB_n(x,y),\label{starb2x}
\end{eqnarray}
where $B_n(x,p)$ and $B_n(x,y)$ are the Chevalley cochains related to phase-space and space, respectively. Taking the f\/irst approximation in the formulas (\ref{starb1x}) and (\ref{starb2x}), and application of the non-deformed commutators $[x,p]=-i\hslash$ and $[x,y]=0$ leads to
\begin{eqnarray}
\left[x,p\right]_\star&=&-i\hslash+\dfrac{i\alpha}{2}B_1(x,p),\label{starb1a}\\
\left[x,y\right]_\star&=&\dfrac{i\beta}{2}B_1(x,y).\label{starb1b}
\end{eqnarray}
or after using of the Dirac ''method of classical analogy''\index{Dirac ''method of classical analogy''} \cite{pamd}
\begin{eqnarray}
\dfrac{1}{i\hslash}\left[p,x\right]_\star&=&1-\dfrac{\alpha}{2\hslash}B_1(x,p),\label{starb2a1}\\
\dfrac{1}{i\hslash}\left[x,y\right]_\star&=&\dfrac{\beta}{2\hslash}B_1(x,y).\label{starb2a2}
\end{eqnarray}
Because of, however, for two arbitrary elements $f,g\in C^\infty(M)$ one has $B_1(f,g)=2\theta(df\wedge dg)$, therefore the formulas (\ref{starb2a1}) and (\ref{starb2a2}) can be rewritten in the following form
\begin{eqnarray}
\dfrac{1}{i\hslash}\left[p,x\right]_\star&=&1-\dfrac{\alpha}{\hslash}(dx\wedge dp),\label{starb2bx}\\
\dfrac{1}{i\hslash}\left[x,y\right]_\star&=&\dfrac{\beta}{\hslash} dx\wedge dy.\label{starb2b1}
\end{eqnarray}

Let us consider the space lattice characterized by an inf\/initesimal growth of a space coordinate identif\/ied with the fundamental scale $\ell$
\begin{equation}
dx=\ell,\label{dxell}
\end{equation}
and presume that the momentum of a particle is related to the coordinate of a particle via the De Broglie wave-particle duality\index{wave-particle duality} formula
\begin{equation}
  p=\dfrac{\hslash}{x}. \label{phx}
\end{equation}
Application of the model (\ref{dxell})-(\ref{phx}) allows to derive straightforwardly an inf\/initesimal growth of the momentum
\begin{equation}
 dp=-\dfrac{\hslash}{x^2}dx=-\dfrac{p^2}{\hslash}\ell.\label{phx1}
\end{equation}
Therefore, the inf\/initesimal growths (\ref{dxell}) and (\ref{phx1}) applied to the deformed brackets (\ref{starb2bx}) and (\ref{starb2b1}) allows to establish f\/inally
\begin{eqnarray}
\dfrac{i}{\hslash}\left[x,p\right]_\star&=&1+\dfrac{\alpha}{\hslash^2}\ell^2p^2,\label{nd1a}\\
\dfrac{i}{\hslash}\left[x,y\right]_\star&=&-\dfrac{\beta}{\hslash}\ell^2.\label{nd1b}
\end{eqnarray}
The relations (\ref{nd1a}) and (\ref{nd1b}) prove that the Snyder space-time\index{space-time!Snyder} (\ref{ndA})-(\ref{ndB}) is a noncommutative geometry\index{noncommutative geometry} obtained via the f\/irst approximation of the Kontsevich deformation quantization\index{quantization!deformation}.

In the 1960s a Soviet physicist M.A. Markov \cite{markov} proposed to take into account a fundamental length scale as the minimal scale identif\/ied with the Planck length\index{Planck length}, i.e. $\ell=\ell_{P}=\sqrt{\strut{\dfrac{\hslash c}{G}}}$, and expressed the hypothesis\index{Markov hypothesis} that a mass $m$ of any elementary particle is bounded by the maximal mass identif\/ied with the Planck mass, i.e. $m\leqslant M_{P}=\dfrac{\hslash}{c\ell_{P}}=\sqrt{\strut{\dfrac{G\hslash}{c^3}}}$. Applying this crucial idea, since 1978 a Soviet-Russian theoretician V.G. Kadyshevsky and his collaborators \cite{kadyshevsky} have studied widely certain aspects of the Snyder model of noncommutative geometry\index{noncommutative geometry} strictly related to particle physics. Recently also V.N. Rodionov has developed independently the stream of Kadyshevsky \cite{radionov}. The problems discussed in this chapter seem to be more related to a general current \cite{book}, where particularly the Snyder space-time\index{space-time!Snyder} (\ref{ndA})-(\ref{ndB}) has been found a number of applications.

Beginning 2000 an Indian scholar and philosopher B.G. Sidharth \cite{sidharth} showed that in spite of the self-evident Lorentz invariance \index{Lorentz invariance}of the deformation (\ref{ndA})-(\ref{ndB}), in general the Snyder noncommutative geometry\index{Snyder noncommutative geometry} breaks the two fundamental paradigms celebrated in relativistic physics: the Einstein energy-momentum relation\index{Einstein energy-momentum relation} as well as the Lorentz symmetry\index{Lorentz symmetry}. Sidharth (Cf. Ref. \cite{bgs2008foop1}) concluded that in such a situation the Hamiltonian constraint of Special Relativity is deformed due to the additional term proportional to the fourth power of three-momentum of a relativistic particle and the second power of a minimal scale $\ell$, which Sidharth has been identif\/ied with a minimal scale\index{minimal scale}, i.e. the Planck length\index{Planck length} or the Compton wavelength \index{Compton wavelength}of an electron
\begin{equation}\label{ss}
  E^2=m^2c^4+c^2p^2+\alpha\left(\dfrac{c}{\hslash}\right)^2\ell^2p^4.
\end{equation}
Neglecting negative mass states as nonphysical, Sidharth established a number of intriguing new facts \cite{bgs2005ijmp1}. Particularly, by straightforward application of Dirac ''square-root'' technique\index{Dirac ''square root'' technique} to the Hamiltonian constraint of the modif\/ied Special Relativity (\ref{ss}) he concluded the corresponding modif\/ied Dirac equation\index{Dirac equation}
\begin{equation}\label{dsit}
  \left(\gamma^\mu \hat{p}_\mu+ mc^2+\sqrt{\alpha}\dfrac{c}{\hslash}\ell\gamma^5 \hat{p}^2\right)\psi=0.
\end{equation}
which differs from the conventional Dirac relativistic quantum mechanics by a correction due to the $\gamma^5$-term proportional to the second power of the three momentum of a particle and to a minimal scale $\ell$.

However, it looks like that Sidharth has been neglected the fact that the modif\/ied Special Relativity (\ref{ss}) leads to a one more additional possibility which is physically nonequivalent to the modif\/ied Dirac equation\index{Dirac equation} (\ref{dsit}) considered by him as the physical quantum theory. Namely, he omitted the Dirac Hamiltonian constraint\index{Dirac Hamiltonian constraint} with the negative $\gamma^5$-term
\begin{equation}\label{ds1}
  \left(\gamma^\mu \hat{p}_\mu+ mc^2-\sqrt{\alpha}\dfrac{c}{\hslash}\ell\gamma^5 \hat{p}^2\right)\psi=0.
\end{equation}
Fortunately, however, such an issue seems to be easy to solve because of the possible physical results of the quantum theory (\ref{ds1}) can be straightforwardly concluded from the results following from the modif\/ied Dirac equation \index{Dirac equation}possessing the positive $\gamma^5$-term (\ref{dsit}) by application of the mirror ref\/lection in a minimal scale $\ell\rightarrow -\ell$. We are not going to neglect also the negative mass states in the modif\/ied Dirac theory as nonphysical, because this situation strictly corresponds with the results obtained from the equation (\ref{dsit}) transformed via a mirror ref\/lection in mass of a relativistic particle $m\rightarrow-m$. Moreover, it must be emphasized that in the standard relativistic quantum mechanics the negative sign corresponds to the Antimatter, which recently has been considered as the element of Reality (See e.g. Ref. \cite{antimatter}).

Therefore we propose to consider the result of the canonical relativistic quantization\index{quantization!relativistic canonical}
\begin{equation}
  p_{\mu}=(E,p_ic)\rightarrow \hat{p}_{\mu}=i\hslash(\partial_0,c\partial_i)
\end{equation}
applied to the deformed Special Relativity (\ref{ss}) linearized by the Dirac ''square-root'' technique\index{Dirac ''square root'' technique}. In general such a procedure leads to the four possible physically nonequivalent quantum theories which can be presented in a form of one compact equation
\begin{equation}\label{ds2}
  \left(\gamma^\mu \hat{p}_\mu\pm mc^2\pm\sqrt{\alpha}\dfrac{c}{\hslash}\ell\gamma^5 \hat{p}^2\right)\psi=0.
\end{equation}
We shall presume that on the analogy of the standard Dirac relativistic quantum mechanics, a wave function $\psi$ of the modif\/ied Dirac equation \index{Dirac equation}(\ref{ds2}) is a four-component spinor $\psi=\left[\phi_0,\phi_1,\phi_2,\phi_3\right]^\mathrm{T}$, and that the Dirac gamma matrices satisfy the four-dimensional {\AE}ther algebra \index{{\AE}ther algebra}$\left\{\gamma^\mu,\gamma^\nu\right\}=\dfrac{1}{2}\eta_{\mu\nu}$ introduced in the previous chapter. It must be emphasized that a presence of the $\gamma^5$-term in the modif\/ied Dirac equation \index{Dirac equation} (\ref{ds2}) results in manifest violation of parity symmetry, and therefore also the Lorentz symmetry \index{Lorentz symmetry!violation}is violated due to such a correction. For simplicity, however, we shall consider one of the four cases (\ref{ds2}) given by the Sidharth's Dirac equation (\ref{dsit})\index{Dirac equation!Sidharth's}. The results due to the three remained situations can be described by straightforward application of the mentioned mirror transformations in the mass of a particle and a minimal scale to the results due to the generic theory (\ref{dsit}).

Both this chapter and the next one are strictly based on the recent results of the author \cite{laglinka} enriched by necessary minor updates. The research value of these two chapters is justif\/ied by the fact that the approach based on deformations of Special Relativity, including the Snyder--Sidharth deformation \index{Snyder--Sidharth deformation} (\ref{ss}), recently has became one of the most intensively developing and fruitful research direction in astrophysics of gamma rays, especially in the context of CP violation \cite{maccione}. Interestingly, the modif\/ied Dirac equation \index{Dirac equation}(\ref{dsit}) was originally proposed \cite{bgs2005ijmp1} as the idea for ultra-high energy physics. Sidharth, however, has not presented computations based on this idea, which could result in experimentally verif\/iable physical predictions. In communication with the author \cite{sidpc}, Sidharth has presented a number of intriguing and interesting looking speculations about the extra mass terms deforming Special Relativity and the corresponding Dirac equations\index{Dirac equation}. Also we discussed a lot of philosophical issues and suggestions about the foundational role of noncommutative geometry\index{noncommutative geometry} for new physics based on the Lorentz symmetry violation\index{Lorentz symmetry!violation}. The discussions did not established a physical truth, and therefore derivation of the modif\/ied Dirac equation \index{Dirac equation} and the reasoning performed by Sidharth in general possesses philosophical countenance. Albeit, the physical role of deformation theory and noncommutative geometry\index{noncommutative geometry} is still a great riddle to the same degree as it is an amazing hope, and factually nobody established real meaning of such an abstractive mathematics for theoretical physics. In the author opinion the most hopeful observational research region to verif\/ication of the theories (\ref{ds2}) is high-energy and ultra-high-energy astrophysics. The astrophysical phenomena are probably the best test for the Planck scale\index{Planck scale}. Particularly, ultra-high-energy cosmic rays\index{ultra-high-energy cosmic rays} coming from gamma bursts\index{gamma bursts} sources, neutrinos coming from supernovas, and other effects observed in this energy region, are the most fruitful research material for tests of the modif\/ied theories. This cognitive aspect of the thing is both the most logical and rational justif\/ication for considering the equation (\ref{dsit}), arising due to the Snyder noncommutative geometry\index{Snyder noncommutative geometry} (\ref{ndA})-(\ref{ndB}), and trying pull out possibly novel valuable extensions of the well-grounded physical knowledge. It must be emphasized that an arbitrary abstractive mathematics creates potentially new physical theories, but extraction of physics is usually a heroic work. By its simplicity the Snyder noncommutative geometry\index{Snyder noncommutative geometry} is one of themes of this book, but another models are not forbidden. Possibly, however, the only Snyder noncommutative geometry\index{Snyder noncommutative geometry} possesses clear physical meaning. Such a hypothesis is also the good point for experimental verif\/ication.

\section{Massive neutrinos}

In fact the Sidharth $\gamma^5$-term, emerging from the Snyder noncommutative geometry\index{Snyder noncommutative geometry} of phase space (\ref{ndA}), is the shift of the conventional Dirac relativistic quantum mechanics. Let us presume that new physics arises from the physical picture in which the modif\/ied Dirac equation \index{Dirac equation}holds, but Special Relativity stays \emph{non-modif\/ied}. In other words, we shall preserve Einstein's Special Relativity unchanged, but change the Dirac relativistic quantum mechanics by the Snyder noncommutative geometry\index{Snyder noncommutative geometry}. Such a modif\/ication is an algebra deformation. It is easy to deduce that such a deformation can be realized by preservation of the hyperbolic geometry of both the Minkowski energy-momentum space \index{Minkowski energy-momentum space} as well as the space-time. While the physical foundation of the Einstein theory is dynamics of a relativistic particle, the physical foundations of an algebra deformation are based on a non-dynamical justif\/ication. For example a deformation can be due to f\/inite sizes of a particle. In this manner, in our view while the Snyder--Sidharth deformation \index{Snyder--Sidharth deformation}of Special Relativity (\ref{ss}) can be interpreted as a dynamical result, the corresponding modif\/ication of the Dirac equation \index{Dirac equation}(\ref{dsit}) is due to the non-dynamical $\gamma^5$-term despite this term is explicitly dependent on a particle three-momentum. To make such a constructive strategy evident we propose to apply the formalism of the Minkowski space, despite a presence of the $\gamma^5$-term, straightforwardly to both the modif\/ied Special Relativity (\ref{ss}) and the modif\/ied Dirac equation (\ref{dsit}).

The standard identity of the Minkowski energy-momentum space
\begin{equation}\label{minein}
p_\mu p^\mu = \left(\gamma^\mu p_\mu\right)^2 = E^2-c^2p^2,
\end{equation}
allows to extract square of three momentum which, together with the mass shell condition $p_\mu p^\mu=mc^2$, applied to the modif\/ied Special Relativity (\ref{ss}) results in the equation
\begin{equation}
\alpha\left(\dfrac{c}{\hslash}\right)^2\ell^2p^4=0,
\end{equation}
which for nonzero momentum has the unique and unambiguous solution $\ell=0$. In other words, the hyperbolic identity (\ref{minein}) on the mass shell leads to direct reconstruction of Special Relativity.

Let us now apply the identity (\ref{minein}) to the modif\/ied Dirac equation (\ref{dsit}). Square of three-momentum can be extracted via the identity (\ref{minein}) and applied within the equation (\ref{dsit})
\begin{equation}\label{hc}
\left[\gamma^\mu \hat{p}_\mu+mc^2+\dfrac{\sqrt{\alpha}}{\hslash c}\ell\gamma^5 \left[E^2-\left(\gamma^\mu \hat{p}_\mu\right)^2\right]\right]\psi=0,
\end{equation}
results in the quadratic equation
\begin{equation}
  \left[-\dfrac{\sqrt{\alpha}}{\hslash c}\ell\gamma^5\left(\gamma^\mu \hat{p}_\mu\right)^2+\gamma^\mu \hat{p}_\mu+mc^2+\dfrac{\sqrt{\alpha}}{\hslash c}\ell\gamma^5E^2\right]\psi=0,
\end{equation}
which after multiplication of both sides by $\gamma^5$ and using of the combination $\gamma^5\gamma^\mu p_\mu$ can be rewritten in the following form
\begin{equation}\label{qe}
 \left[ \left(\gamma^5\gamma^\mu \hat{p}_\mu\right)^2-\epsilon\left(\gamma^5\gamma^\mu \hat{p}_\mu\right)+E^2-\epsilon mc^2\gamma^5\right]\psi=0,
\end{equation}
where $\epsilon$ is a maximal energy due to a minimal scale $\ell$
\begin{equation}
  \epsilon = \dfrac{\hslash c}{\sqrt{\alpha}\ell}.\label{eps}
\end{equation}

The equation (\ref{qe}) expresses projection of the operator
\begin{equation}\label{oper}
  \left(\gamma^5\gamma^\mu \hat{p}_\mu\right)^2-\epsilon\left(\gamma^5\gamma^\mu \hat{p}_\mu\right)+E^2-\epsilon mc^2\gamma^5,
\end{equation}
on the spinor $\psi$. By application of elementary algebraic manipulations, however, the quadratic operator (\ref{oper}) can be factorized straightforwardly to the form
\begin{equation}\label{sol}
  (\gamma^5\gamma^\mu \hat{p}_\mu-\epsilon_+)(\gamma^5\gamma^\mu \hat{p}_\mu-\epsilon_-),
\end{equation}
where $\epsilon_{\pm}$ are the manifestly non-hermitian energies
\begin{equation}\label{mpm}
  \epsilon_{\pm}=\dfrac{\epsilon}{2}\left(1\pm\sqrt{\strut{1-\dfrac{4E^2}{\epsilon^2}}}\sqrt{\strut{1+\dfrac{4\epsilon mc^2}{\epsilon^2-4E^2}\gamma^5}}\right).
\end{equation}
Principally the quantities (\ref{mpm}) are due to the order reduction, and also cause the Dirac-like linearization.

Let us treat a particle energy $E$, a particle mass $m$, and a maximal energy $\epsilon$ (or equivalently a minimal scale $\ell$) in the formula (\ref{mpm}) as free parameters. It can be observed straightforwardly that the modif\/ied Dirac equation \index{Dirac equation}(\ref{dsit}) is equivalent to the two nonequivalent relativistic quantum theories
\begin{eqnarray}
  \left(\gamma^\mu \hat{p}_\mu-M_+c^2\right)\psi=0,\label{rhc1}\\
  \left(\gamma^\mu \hat{p}_\mu-M_-c^2\right)\psi=0,\label{rhc2}
\end{eqnarray}
where $M_{\pm}$ are the generated mass matrices
\begin{equation}\label{mmat}
  M_{\pm}=\dfrac{\epsilon}{2c^2}\left(-1\mp\sqrt{\strut{1-\dfrac{4E^2}{\epsilon^2}+\dfrac{4mc^2}{\epsilon}\gamma^5}}\right)\gamma^5.
\end{equation}

The result (\ref{mmat}) is in itself nontrivial. Factually, by application of the Minkowski energy-momentum space \index{Minkowski energy-momentum space}the Dirac equation \index{Dirac equation}modif\/ied due to the Snyder noncommutative geometry\index{Snyder noncommutative geometry} has been reduced to two distinguishable standard Dirac theories describing a kind of effective particles characterized by manifestly non-hermitian mass matrices $M_{\pm}$. Both these theories are all the more so intriguing because the total information about a minimal scale $\ell$, and therefore about the Snyder noncommutative geometry\index{Snyder noncommutative geometry}, was placed in the mass matrices $M_{\pm}$ only, while the relativistic space-time and energy-momentum formalisms are exactly the same as in the standard Dirac quantum theory. Note that such a procedure is also correct from the methodological point of view. We have applied the tensor formalism of the hyperbolic 4-dimensional energy-momentum geometry within the Dirac equation \index{Dirac equation}modif\/ied due to the $\gamma^5$-term. This point has not been noticed or has been omitted in the analysis made by Sidharth. In this manner we have constructed new type \emph{mass generation mechanism}\index{mass generation mechanism} which deduction is impossible to perform within the frames of Special Relativity only, \emph{i.e.} for in the situation when a minimal scale is vanishing $\ell=0$ or equivalently a maximal energy is inf\/inite $\epsilon=\infty$. In fact, application of the noncommutative geometry\index{noncommutative geometry} results in a f\/inite value of a maximal energy, what is a kind of renormalization of Special Relativity. Therefore the mass generation mechanism\index{mass generation mechanism} presented above results in the effect of such a nontrivial renormalization, possesses purely kinetic nature and, above all, is due to the factorization applied to the operator (\ref{oper}) projecting onto the spinor wave function $\psi$. It must be emphasized that this kinetic effect in the result due to the abstractive mathematics of noncommutative geometry\index{noncommutative geometry} and algebra deformation. The key problem is generalization of this mechanism to more general situations described in frames of the Kontsevich deformation quantization\index{quantization!deformation}. In other words the crucial unsolved issue is a reply to the question: \emph{does a physical contribution from noncommutative geometry\index{noncommutative geometry} in general contains in a mass generation mechanism only?}\index{mass generation mechanism}. The necessity of reply to this question is argued above all by the problem of experimental verif\/ication of the theoretical results due to noncommutative geometry\index{noncommutative geometry}, and possesses fundamental meaning for philosophical foundations of the new physics obtained via constructive applications of the abstractive mathematics. The reply to this question is, however, far from this book content.

Let us present now the mass matrices $M_{\pm}$ in more convenient form employing a linear dependence of the $\gamma^5$ matrix. Fist, let us apply the Taylor series\index{Taylor series} expansion to the square root part of the mass matrices (\ref{mmat}). The expansion can be performed in the following way
\begin{eqnarray}
\sqrt{\strut{1-\dfrac{4E^2}{\epsilon^2}+\dfrac{4mc^2}{\epsilon}\gamma^5}}&=&\sqrt{\strut{1-\dfrac{4E^2}{\epsilon^2}}}\sqrt{\strut{1+\dfrac{\dfrac{4mc^2}{\epsilon}}{1-\dfrac{4E^2}{\epsilon^2}}\gamma^5}}=\nonumber\\
&=&\sqrt{\strut{1-\dfrac{4E^2}{\epsilon^2}}}\sum_{n=0}^\infty \binom{1/2}{n}\left(\dfrac{\dfrac{4mc^2}{\epsilon}}{1-\dfrac{4E^2}{\epsilon^2}}\gamma^5\right)^n,\label{tay}
\end{eqnarray}
where the generalized Newton binomial symbol\index{Newton binomial symbol!generalized} was used $$\binom{n}{k}=\dfrac{\Gamma(n+1)}{\Gamma(k+1)\Gamma(n+1-k)}.$$ By application of the basic properties of $\gamma^5$ matrix, \emph{i.e.} $\left(\gamma^5\right)^{2n}=-1$ and $\left(\gamma^5\right)^{2n+1}=-\gamma^5$, one can decompose of the sum present in the last term of the formula (\ref{tay}) onto the two components related to the odd and even powers of $\gamma^5$ matrix
\begin{eqnarray}
  \!\!\!\!\!\!\!\!\!\!\!\!\!\!\!\!\!\!\!\!&&\sum_{n=0}^\infty \binom{1/2}{n}\left(\dfrac{\dfrac{4mc^2}{\epsilon}}{1-\dfrac{4E^2}{\epsilon^2}}\gamma^5\right)^n=\nonumber\\
  \!\!\!\!\!\!\!\!\!\!\!\!\!\!\!\!\!\!\!\!&&=-\sum_{n=0}^\infty \binom{1/2}{2n}\left(\dfrac{\dfrac{4mc^2}{\epsilon}}{1-\dfrac{4E^2}{\epsilon^2}}\right)^{2n}-\sum_{n=0}^\infty \binom{1/2}{2n+1}\left(\dfrac{\dfrac{4mc^2}{\epsilon}}{1-\dfrac{4E^2}{\epsilon^2}}\right)^{2n+1}\gamma^5.\label{comp}
\end{eqnarray}
Straightforward application of the standard summation procedure allows to establish the sums presented in both the components of the decomposition (\ref{comp})
\begin{eqnarray}
  \!\!\!\!\!\!\!\!\!\!\!\!\!\!\!\!\!\!\!\!\sum_{n=0}^\infty\binom{1/2}{2n}\left(\dfrac{\dfrac{4mc^2}{\epsilon}}{1-\dfrac{4E^2}{\epsilon^2}}\right)^{\!\!\!2n}&=&\sqrt{\strut{1+\dfrac{\dfrac{4mc^2}{\epsilon}}{1-\dfrac{4E^2}{\epsilon^2}}}}+\sqrt{\strut{1-\dfrac{\dfrac{4mc^2}{\epsilon}}{1-\dfrac{4E^2}{\epsilon^2}}}},\vspace*{10pt}\\
  \!\!\!\!\!\!\!\!\!\!\!\!\!\!\!\!\!\!\!\!\sum_{n=0}^\infty\binom{1/2}{2n+1}\left(\dfrac{\dfrac{4mc^2}{\epsilon}}{1-\dfrac{4E^2}{\epsilon^2}}\right)^{\!\!\!2n+1}&=&\sqrt{\strut{1+\dfrac{\dfrac{4mc^2}{\epsilon}}{1-\dfrac{4E^2}{\epsilon^2}}}}-\sqrt{\strut{1-\dfrac{\dfrac{4mc^2}{\epsilon}}{1-\dfrac{4E^2}{\epsilon^2}}}}.
\end{eqnarray}
In this manner one sees easily that both the mass matrices $M_{\pm}$ possess the following decomposition onto two components: the hermitian $\mathfrak{H}(M_{\pm})$ and the antihermitian $\mathfrak{A}(M_{\pm})$
\begin{equation}\label{mm1}
  M_{\pm}=\mathfrak{H}(M_{\pm})+\mathfrak{A}(M_{\pm}),
\end{equation}
where both the parts can be presented in a compact form
\begin{eqnarray}
\!\!\!\!\!\!\!\!\!\!\!\!\!\!\!\!\!\!\!\!\!\!\!\mathfrak{H}(M_{\pm})\!\!\!\!\!\!\!&=&\!\!\!\!\!\!\!\pm\dfrac{\epsilon}{2c^2}\left[\sqrt{\strut{1-\dfrac{4E^2}{\epsilon^2}}}\left(\sqrt{\strut{1+\dfrac{\dfrac{4mc^2}{\epsilon}}{1-\dfrac{4E^2}{\epsilon^2}}}}-\sqrt{\strut{1-\dfrac{\dfrac{4mc^2}{\epsilon}}{1-\dfrac{4E^2}{\epsilon^2}}}}\right)\right],\\
\!\!\!\!\!\!\!\!\!\!\!\!\!\!\!\!\!\!\!\!\!\!\!\mathfrak{A}(M_{\pm})\!\!\!\!\!\!\!&=&\!\!\!\!\!\!\!-\dfrac{\epsilon}{2c^2}\left[1\pm\sqrt{\strut{1-\dfrac{4E^2}{\epsilon^2}}}\left(\sqrt{\strut{1+\dfrac{\dfrac{4mc^2}{\epsilon}}{1-\dfrac{4E^2}{\epsilon^2}}}}+\sqrt{\strut{1-\dfrac{\dfrac{4mc^2}{\epsilon}}{1-\dfrac{4E^2}{\epsilon^2}}}}\right)\right]\gamma^5.
\end{eqnarray}

By application of elementary algebraic manipulations one obtains equivalent decomposition of the mass matrices $M_{\pm}$ into the basis of the commutating projectors $\left\{\Pi_i:\dfrac{1+\gamma^5}{2},\dfrac{1-\gamma^5}{2}\right\}$,
\begin{equation}\label{mm2}
  M_{\pm}=\sum_i\mu_i^\pm\Pi_i=\mu_R^\pm\dfrac{1+\gamma^5}{2}+\mu_L^\pm\dfrac{1-\gamma^5}{2},
\end{equation}
where $\mu_R^\pm$ and $\mu_L^\pm$ are the projected masses related to the Dirac theories with signs $\pm$ in agreement with the mass matrix signs. The values of the projected masses can be established easy
\begin{eqnarray}
  \mu_R^\pm&=&-\dfrac{1}{c^2}\left(\dfrac{\epsilon}{2}\pm\sqrt{\strut{\epsilon^2-4\epsilon mc^2-4E^2}}\right),\label{mu1x}\\
  \mu_L^\pm&=&\dfrac{1}{c^2}\left(\dfrac{\epsilon}{2}\pm\sqrt{\strut{\epsilon^2+4\epsilon mc^2-4E^2}}\right).\label{mu2x}
\end{eqnarray}
For physical correctness we shall presume that the masses of right-handed neutrinos (\ref{mu1x}) and the left-handed neutrinos (\ref{mu2x}) are real numbers. It is nontrivial conditions, which allows to establish the range of a maximal energy $\epsilon$ via the energy and the mass of an original quantum state obeying the Dirac equation\index{Dirac equation}. In the case of massive states one obtains
\begin{eqnarray}
  \epsilon&\in&\left(-\infty,-2mc^2\left(1+\sqrt{\strut{1+\left(\dfrac{E}{mc^2}\right)^2}}\right)\right]\cup\nonumber\\
  &\cup&\left[-2mc^2\left(1-\sqrt{\strut{1+\left(\dfrac{E}{mc^2}\right)^2}}\right),2mc^2\left(1+\sqrt{\strut{1+\left(\dfrac{E}{mc^2}\right)^2}}\right)\right]\cup\nonumber\\
  &\cup&\left[2mc^2\left(1+\sqrt{\strut{1+\left(\dfrac{E}{mc^2}\right)^2}}\right),\infty\right),
\end{eqnarray}
while in the case of massless quantum states possessing an energy $E$
\begin{eqnarray}
  \epsilon\in\left(-\infty,-2|E|\right]\cup\left[2|E|,\infty\right).
\end{eqnarray}

The properties of the projectors $\Pi_i^\dag\Pi_i=\mathbf{1}_4$, $\Pi_1\Pi_2=\dfrac{1}{2}\mathbf{1}_4$, $\Pi_1^\dag=\Pi_2$ and $\Pi_1+\Pi_2=\mathbf{1}_4$ allows to derive the relation
\begin{equation}
  M_\pm M_\pm^\dag=\dfrac{(\mu_R^\pm)^2+(\mu_L^\pm)^2}{2}\mathbf{1}_4.
\end{equation}
By introducing the right- and left-handed chiral Weyl f\/ields\index{chiral Weyl f\/ields}
\begin{eqnarray}
\psi_R&=&\dfrac{1+\gamma^5}{2}\psi,\label{rl1}\\
\psi_L&=&\dfrac{1-\gamma^5}{2}\psi,\label{rl2}
\end{eqnarray}
where a wave function $\psi$ is a solution of the appropriate Dirac equations \index{Dirac equation}(\ref{rhc1}) and (\ref{rhc2}), both the theories can be rewritten as the system of two equations
\begin{eqnarray}
  \left(\gamma^\mu \hat{p}_\mu+\mu^+ c^2\right)\left[\begin{array}{c}\psi_R^+\\ \psi_L^+\end{array}\right]=0,\label{neu1}\\
  \left(\gamma^\mu \hat{p}_\mu+\mu^- c^2\right)\left[\begin{array}{c}\psi_R^-\\ \psi_L^-\end{array}\right]=0,\label{neu2}
\end{eqnarray}
where now the mass matrices $\mu^\pm$, related to the chiral f\/ields $\psi_{R,L}^\pm$, are manifestly hermitian quantities
\begin{equation}\label{mm3}
  \mu^\pm=\left[\begin{array}{cc}\mu_R^\pm&0\\0&\mu_L^\pm\end{array}\right]=\left[\begin{array}{cc}\mu_R^\pm&0\\0&\mu_L^\pm\end{array}\right]^\dag.
\end{equation}
Note that the masses (\ref{mu1x}) and (\ref{mu2x}) are invariant with respect to a choice of representation of the Dirac matrices $\gamma^\mu$. By this reason these quantities have physical character. It is interesting that for the mirror ref\/lection in a minimal scale $\ell\rightarrow-\ell$, or equivalently for the mirror ref\/lection $\epsilon\rightarrow-\epsilon$, one has the exchange between the masses $\mu_R^\pm\leftrightarrow\mu_L^\pm$ while the chiral Weyl f\/ields\index{chiral Weyl f\/ields} stay unchanged. Similarly, in the situation of the mirror ref\/lection in the mass of an original quantum state $m\rightarrow-m$ one has the exchange between the projected masses $\mu_R^\pm\leftrightarrow-\mu_L^\pm$. The case of originally massless quantum states $m=0$ is also intriguing from a theoretical point of view. From the mass formulas (\ref{mu1x}) and (\ref{mu2x}) one sees straightforwardly that in such a case the equality $\mu_R=-\mu_L$ holds. In the limiting case of generic Einstein Special Relativity $\ell=0$, the projected masses have interesting properties
\begin{eqnarray}
\mu_{R}^\pm&=&\left\{\begin{array}{cc}-\infty&~\mathrm{for}~+\\ \infty&~\mathrm{for}~-\end{array}\right.,\\ \mu_{L}^\pm&=&\left\{\begin{array}{cc}\infty&~\mathrm{for}~+\\ -\infty&~\mathrm{for}~-\end{array}\right. .
\end{eqnarray}
It means that in the case of Special Relativity the massive Weyl theories are undetermined, i.e. the particles described by the Weyl equations (\ref{neu1}) and (\ref{neu2}) do not exist. In general, however, for correctness of the projection splitting (\ref{mm2}) both the projected masses (\ref{mu1x}) and (\ref{mu2x}) must be real numbers. Strictly speaking when the projected masses are complex numbers the decomposition (\ref{mm2}) does not yield hermitian mass matrices (\ref{mm3}). Therefore, in such a case the presented construction has no sense, and by this reason should be replaced by another theory based on different arguments.

In the approach based on the Standard Model\index{Standard Model} neutrinos are massless quantum particles. In this manner, it is evident that application of the Snyder noncommutative geometry\index{Snyder noncommutative geometry} generated the non-triviality which is \emph{the kinetic mass generation mechanism of the neutrino\index{neutrino} mass}\index{mass generation mechanism!kinetic}. It must be emphasized that in all the cited contributions Sidharth mentioned about a possibility of neutrino\index{neutrino} mass ''due to the mass term'', where by the mass term he understands the dynamical $\gamma^5$-term in the modif\/ied Dirac equation \index{Dirac equation}(\ref{dsit}). Albeit, regarding the Standard Model\index{Standard Model} which is the theory of elementary particles and fundamental interactions, a mass term can not be dynamical, and such a nomenclature is misleading in further analysis of the philosophical ideas. We have been received the massive neutrinos due to the two-step mass generation mechanism\index{mass generation mechanism}. The f\/irst one is factorization of the modif\/ied Dirac equation\index{Dirac equation} (\ref{dsit}). The second one is the decomposition of the mass matrices (\ref{mmat}) into the projectors basis and introducing the chiral Weyl f\/ields\index{chiral Weyl f\/ields} (\ref{rl1}) and (\ref{rl2}). Calling such an unique procedure as the result ''due to the mass term'' is at least inaccurate, creates a number of inequivalent interpretations, and in itself is escaping from physics to philosophy. It must be emphasized that any mass generation mechanism\index{mass generation mechanism} is manifestly absent in Sidharth's contributions and the reasoning presented there completely differs from our analysis, omits certain important physical and mathematical details (Cf. e.g. Ref. \cite{bgs2005ijmp2}). The procedure proposed above, i.e. by the unique treatment of the $\gamma^5$-term in the modif\/ied Dirac equation \index{Dirac equation}(\ref{dsit}) and preservation the hyperbolic geometry of Minkowski energy-momentum space \index{Minkowski energy-momentum space}(\ref{minein}) of the Einstein Special Relativity, resulted in generation of the Weyl equations (\ref{neu1})-(\ref{neu2}) describing two left- $\psi_L^\pm$ and two right- $\psi_R^\pm$ chiral f\/ields possessing nonzero masses. In other words we have established the constructive theory of massive neutrinos, related to both originally massive $m\neq0$ as well as massless $m=0$ quantum states described by relativistic quantum mechanics. By this reason the our approach changes the physical essence of the concepts \emph{neutrino\index{neutrino}} and \emph{neutrino\index{neutrino} mass}. Our neutrino\index{neutrino} is a chiral f\/ield due to any originally massive and massless quantum particle, which in itself is also a quantum particle. Our neutrino\index{neutrino} mass is generated due to a mass of originally quantum particle. Furthermore, it is easy to see that the Weyl theories (\ref{neu1}) and (\ref{neu2}) associate 4 massive neutrinos with a one original quantum state. Because of there are 4 possible theories (\ref{ds2}) the maximal number of the neutrinos due to the Snyder noncommutative geometry\index{Snyder noncommutative geometry} is 16. However, this number can be reduced due to experimental verif\/ication of the results of the theories.

\section{The Compton--Planck Scale}

The Planck scale\index{Planck scale}, def\/ined by taking into account a minimal scale $\ell$ identical with the Planck length\index{Planck length} $\ell_P=\sqrt{\strut{\dfrac{\hslash G}{c^3}}}$, in modern cosmology and particle physics is the energetic region at which quantum physics meets classical physics. At this scale the standard methodology of particle physics, i.e. the Standard Model\index{Standard Model}, is manifestly inadequate tool for constructive description because of necessity of a theory of quantum gravity. There is few approaches to the adequate formalism of the Planck scale\index{Planck scale} physics, including string theory\index{string theory}, M-theory\index{M-theory}, loop quantum gravity\index{loop quantum gravity}, and noncommutative geometry\index{noncommutative geometry} (See e.g. the Ref. \cite{planckscale}). A theory of quantum gravity, however, is the main requirement for the consistent description of the physics at the Planck scale\index{Planck scale}. We shall present certain proposal for such a theory in the second part of this book, while this part is related to noncommutative geometry\index{noncommutative geometry}.

Let us look on few coincidences at the Planck scale\index{Planck scale}, which are informative for general understanding. In such a situation a maximal energy (\ref{eps}) coincides with the Planck energy divided by $\sqrt{\alpha}$
\begin{equation}
  \epsilon(\ell_P)=\dfrac{1}{\sqrt{\alpha}}\sqrt{\strut{\dfrac{\hslash c^5}{G}}}=\dfrac{1}{\sqrt{\alpha}}M_{P}c^2=\dfrac{E_P}{\sqrt{\alpha}}.
\end{equation}

Let us take into account the Compton wavelength \index{Compton wavelength}of a particle possessing the rest mass $m$, i.e. $\lambda_C(m)=2\pi\dfrac{\hslash}{m c}$. A maximal energy (\ref{eps}) computed at the scale identical to the Compton wavelength $\lambda_C(m)$ is proportional to rest energy of a particle
\begin{equation}
  \epsilon(\lambda_C(m))=\dfrac{1}{2\pi\sqrt{\alpha}}m c^2,
\end{equation}
and, if $\alpha$ has the proposed value (\ref{alpha}), becomes simply the rest energy of a particle $\epsilon=mc^2$.

Interestingly, in the situation when the rest mass of a particle equals the Planck mass $m\equiv M_{P}=\sqrt{\dfrac{\hslash c}{G}}$ or equivalently the rest energy of a particle equals to the Planck energy $mc^2=E_P$, there is a number of non-trivialities. Let us denote the Compton wavelength of such a Planckian particle\index{Planckian particle} by $\ell_C=\lambda_C(M_P)$. In fact, this wavelength def\/ines the mixed scale which we propose to call \emph{the Compton--Planck (CP) scale}\index{Compton--Planck scale}\index{Planck scale}. It can be seen straightforwardly that in the CP scale holds
\begin{equation}
\epsilon(\ell_C)=\dfrac{\epsilon(\ell_P)}{2\pi}.
\end{equation}
Moreover, when one regards (\ref{alpha}), i.e. $\alpha=1/(2\pi)^2$, than also
\begin{equation}\label{alpha1}
\alpha=\left(\dfrac{\epsilon(\ell_C)}{\epsilon(\ell_P)}\right)^2=\left(\dfrac{\ell_P}{\ell_C}\right)^2.
\end{equation}
Moreover, in the CP scale the doubled Compton wavelength \index{Compton wavelength}equals to a circumference of a circle with a radius of the Schwarzschild radius $r_S(m)=\dfrac{2Gm}{c^2}$ of the Planck mass (Cf. Ref. \cite{kiefer})
\begin{equation}
  2\ell_C= 2\pi r_S\left(M_P\right).
\end{equation}
Straightforward and easy computation shows that the Compton wavelength of the Planck mass is identif\/ied with a circumference of a circle with a radius of the Planck length\index{Planck length}, i.e.
\begin{equation}
\ell_C=2\pi\ell_P,
\end{equation}
and by this reason the doubled Planck length\index{Planck length} equals to the Schwarzschild radius of the Planck mass
\begin{equation}
  2\ell_P=r_S(M_P).
\end{equation}

In general the ratio of the Planck length\index{Planck length} and the Compton wavelength \index{Compton wavelength}of a particle with mass $m$ is
\begin{equation}
\dfrac{\ell_P}{\lambda_C(m)}=\dfrac{1}{2\pi}\dfrac{m}{M_P}.
\end{equation}
Let us generalize the last relation in the equation (\ref{alpha1}) as follows
\begin{equation}
  \alpha\equiv\left(\dfrac{1}{2\pi}\dfrac{m}{M_P}\right)^2.\label{alpha2}
\end{equation}
Taking $\alpha$ as (\ref{alpha}) together with the presumption $\alpha\sim1$, which guarantees correctness of the Kontsevich deformation quantization\index{quantization!deformation}. Such a reasoning establishes the mass of a particle for which the Snyder noncommutative geometry\index{Snyder noncommutative geometry} is adequate. It is not diff\/icult to see that the mass of a particle must be of an order of the Planck mass, i.e.
\begin{equation}\label{kontmass}
m\sim M_P.
\end{equation}
If, however, one wishes to neglect (\ref{alpha}) but preserve (\ref{alpha2}) together with the condition $\alpha\sim1$ then
\begin{equation}
m\sim2\pi M_P,
\end{equation}
i.e. the mass of a particle described by the noncommutative geometry\index{noncommutative geometry} is of the order $m\sim (10^{22}-10^{23})\mathrm{MeV/c^2}$.

\section{The Global Effective Chiral Condensate}

Let us consider the meaning of the massive Weyl equations (\ref{neu1})-(\ref{neu1}) in the spirit of the gauge f\/ield theories \cite{gft}, which are the base of the Standard Model\index{Standard Model}. The problem is to to construct the Lagrangian $\mathcal{L}$, revealing Lorentz invariance\index{Lorentz invariance}, of the gauge f\/ield theory characterized by the massive Weyl equations treated as the Euler-Lagrange equations of motion for the chiral f\/ields $\psi_L^\pm$ and $\psi_R^\pm$. In general, such a construction is not easy to perform, but because in fact the massive Weyl equations are the Dirac theories, it can be seen by straightforward computations that the following four Dirac-like Lagrangians \index{Dirac-like Lagrangian}
\begin{eqnarray}
  \mathcal{L}^\pm_R&=&\bar{\psi}_R^\pm\left(\gamma^\mu \hat{p}_\mu +\mu_R^\pm c^2\right)\psi_R^\pm,\label{lagau1}\\
  \mathcal{L}^\pm_L&=&\bar{\psi}_L^\pm\left(\gamma^\mu \hat{p}_\mu +\mu_L^\pm c^2\right)\psi_L^\pm,\label{lagau2}
\end{eqnarray}
where $\bar{\psi}_{R,L}^\pm=\left(\psi_{R,L}^\pm\right)^\dag\gamma^0$ are the Dirac adjoint of $\psi_{R,L}^\pm$, lead to the massive Weyl equations by the appropriate principle of the least action. In other words, the Euler-Lagrange equations of motion
\begin{equation}\label{eulag}
\dfrac{\partial\mathcal{L}^\pm_{R,L}}{\partial\psi_{R,L}^\pm}-
\partial_{\mu}\dfrac{\partial\mathcal{L}^\pm_{R,L}}{\partial\left(\partial_\mu\psi_{R,L}^\pm\right)}=0,
\end{equation}
coincide with the massive Weyl equations (\ref{neu1})-(\ref{neu1}). In this manner, the appropriate full gauge f\/ield theory of massive neutrinos can be directly constructed by using of the Lagrangian which is an algebraical sum of the four gauge f\/ield theories (\ref{lagau1})-(\ref{lagau2})
\begin{equation}\label{tlag}
\mathcal{L}=\mathcal{L}^+_R +\mathcal{L}^-_R+\mathcal{L}^+_L +\mathcal{L}^-_L,
\end{equation}
i.e. the massive Weyl equations are obtained from the Euler-Lagrange equations of motion (\ref{eulag}) with the exchange $\mathcal{L}^\pm_{R,L}\rightarrow\mathcal{L}$. It must be emphasized that the choice of the Lagrangians in the form (\ref{lagau1})-(\ref{lagau2}) is due to straightforward analogy between the massive Weyl equation \index{Weyl equation!massive}and the Dirac equation\index{Dirac equation}. The essential difference between these theories is the only number of spinor components what, however, has no inf\/luence on the form of Lagrangian and the principle of the least action. The Lagrangians (\ref{lagau1}) and (\ref{lagau2}) describe the two components of the Weyl spinor. In other words, this choice is both the most intuitive and the simplest. However, it does not mean that there is no another, possibly more complicated, choice dictated by another justif\/ication.

One can see easy that the gauge f\/ield theories (\ref{lagau1}) and (\ref{lagau2}) exhibit several well-known gauge symmetries. Namely, the (local) chiral symmetry\index{symmetry!local chiral} $SU(2)_R^\pm\otimes SU(2)_L^\pm$ expressed via one of the transformations
\begin{eqnarray}
  &&\left\{\begin{array}{c}\psi_R^\pm\rightarrow \exp \left\{i\theta_R^\pm\right\}\psi_R^\pm\vspace*{5pt}\\
  \psi_L^\pm\rightarrow\psi_L^\pm\end{array}\right.,\vspace*{5pt}\\ &&\left\{\begin{array}{c}\psi_R^\pm\rightarrow\psi_R^\pm\vspace*{5pt}\\
  \psi_L^\pm\rightarrow \exp\left\{i\theta_L^\pm\right\}\psi_L^\pm\end{array}\right.,
\end{eqnarray}
the vector symmetry\index{symmetry!vector} $U(1)_V^\pm$
\begin{equation}
  \left\{\begin{array}{c}\psi_R^\pm\rightarrow \exp\left\{i\theta^\pm\right\}\psi_R^\pm\vspace*{5pt}\\
  \psi_L^\pm\rightarrow \exp\left\{i\theta^\pm\right\}\psi_L^\pm\end{array}\right. ,
\end{equation}
and the axial symmetry\index{symmetry!axial} $U(1)_A^\pm$
\begin{equation}
  \left\{\begin{array}{c}\psi_R^\pm\rightarrow \exp\left\{-i\theta^\pm\right\}\psi_R^\pm\vspace*{5pt}\\
  \psi_L^\pm\rightarrow \exp\left\{i\theta^\pm\right\}\psi_L^\pm\end{array}\right. .
\end{equation}
In this manner the total symmetry group is
\begin{equation}
  SU(3)_C^+\oplus SU(3)_C^-,\label{composite1}
\end{equation}
where $SU(3)_C^\pm$ are the global (chiral) 3-f\/lavor gauge symmetries related to each of the gauge theories (\ref{lagau1}) and (\ref{lagau2}), i.e.
\begin{eqnarray}\label{gs}
\!\!\!\!\!\!\!\!\!\!&SU(2)_R^+\otimes SU(2)_L^+\otimes U(1)_V^+\otimes U(1)_A^+\equiv SU(3)^+\otimes SU(3)^+=SU(3)_C^+,&\\
\!\!\!\!\!\!\!\!\!\!&SU(2)_R^-\otimes SU(2)_L^-\otimes U(1)_V^-\otimes U(1)_A^-\equiv SU(3)^-\otimes SU(3)^-=SU(3)_C^-,&
\end{eqnarray}
describing 2-f\/lavor massive free quarks - \emph{the neutrinos} in our proposition. Because of, the group (\ref{composite1}) does not possess a name in literature, we shall call the group \emph{the composite symmetry} $SU(3)_C^{TOT}$\index{symmetry!composite}.

By application of the def\/initions for the chiral Weyl f\/ields\index{chiral Weyl f\/ields} (\ref{rl1}) and (\ref{rl2}) one obtains
\begin{equation}
\bar{\psi}_{R,L}^\pm\gamma^\mu p_\mu \psi_{R,L}^\pm=\bar{\psi}^\pm\dfrac{1\pm\gamma^5}{2}\gamma^\mu p_\mu \dfrac{1\pm\gamma^5}{2}\psi^\pm=\bar{\psi}^\pm\left(\dfrac{1\pm\gamma^5}{2}\gamma^\mu \dfrac{1\pm\gamma^5}{2}\right)p_\mu \psi^\pm,
\end{equation}
where $\bar{\psi}^\pm=\left(\psi^\pm\right)^\dag\gamma^0$ is the Dirac adjoint of the Dirac f\/ields $\psi^\pm$ related to the chiral Weyl f\/ields\index{chiral Weyl f\/ields} by the transformations (\ref{rl1}) and (\ref{rl2}). Because of the identity
\begin{equation}
\dfrac{1\pm\gamma^5}{2}\gamma^\mu \dfrac{1\pm\gamma^5}{2}=
\dfrac{\gamma^\mu\pm\left\{\gamma^\mu,\gamma^5\right\}+\gamma^5\gamma^\mu\gamma^5}{4}=
\dfrac{1-(\gamma^5)^2}{4}\gamma^\mu=\dfrac{1}{2}\gamma^\mu,
\end{equation}
where we have applied the properties of $\gamma^5$ matrix $\left\{\gamma^\mu,\gamma^5\right\}=0$ and $(\gamma^5)^2=-1$, one obtains f\/inally
\begin{equation}
\bar{\psi}_{R,L}^\pm\gamma^\mu p_\mu \psi_{R,L}^\pm=\dfrac{1}{2}\bar{\psi}^\pm\gamma^\mu p_\mu\psi^\pm.\label{d1}
\end{equation}
Similarly, applying the identity
\begin{eqnarray}
\dfrac{1\pm\gamma^5}{2}\dfrac{1\pm\gamma^5}{2}&=&
\dfrac{1}{4}\left(1\pm2\gamma^5+(\gamma^5)^2\right)=\pm\dfrac{1}{2}\gamma^5,
\end{eqnarray}
one can establish the quantity
\begin{equation}
\mu_{R,L}^\pm c^2\bar{\psi}^\pm_{R,L}\psi_{R,L}^\pm=
\mu_{R,L}^\pm c^2\bar{\psi}^\pm\dfrac{1\pm\gamma^5}{2}\dfrac{1\pm\gamma^5}{2}\psi^\pm,
\end{equation}
with the result
\begin{equation}
\mu_{R,L}^\pm c^2\bar{\psi}^\pm_{R,L}\psi_{R,L}^\pm=
\dfrac{1}{2}\bar{\psi}^\pm\left(\pm\mu_{R,L}^\pm c^2\gamma^5\right)\psi^\pm,\label{d2}
\end{equation}
where the plus sign of the mass is appropriate for the right-handed neutrinos, while the minus sign is appropriate for the left-handed neutrinos. In this way one obtains f\/inally the partial Lagrangians
\begin{eqnarray}
\mathcal{L}_{R,L}^\pm&=&\bar{\psi}_{R,L}^\pm\left(\gamma^\mu p_\mu+\mu_{R,L}^\pm c^2\right)\psi_{R,L}^\pm=\nonumber\\
&=&\bar{\psi}_{R,L}^\pm\gamma^\mu p_\mu \psi_{R,L}^\pm+\mu_{R,L}^\pm c^2\bar{\psi}^\pm_{R,L}\psi_{R,L}^\pm=\nonumber\\
&=&\dfrac{1}{2}\bar{\psi}^\pm\gamma^\mu p_\mu\psi^\pm+\dfrac{1}{2}\bar{\psi}^\pm\left(\pm\mu_{R,L}^\pm c^2\gamma^5\right)\psi^\pm=\nonumber\\
&=&\dfrac{1}{2}\bar{\psi}^\pm\left(\gamma^\mu p_\mu\pm\mu_{R,L}^\pm c^2\gamma^5\right)\psi^\pm.\label{rllag}
\end{eqnarray}
which reveal the Lorentz invariance\index{Lorentz invariance}. In this manner it can be seen straightforwardly that with using of the partial Lagrangians (\ref{rllag}) the global chiral Lagrangian (\ref{tlag}) takes the following form
\begin{eqnarray}
\mathcal{L}&=&\dfrac{1}{2}\bar{\psi}^{+}\left(\gamma^\mu p_\mu+\mu_{R}^{+}c^2\gamma^5\right)\psi^{+}+\dfrac{1}{2}\bar{\psi}^{+}\left(\gamma^\mu p_\mu-\mu_{L}^{+}c^2\gamma^5\right)\psi^{+}+\nonumber\\
&+&\dfrac{1}{2}\bar{\psi}^{-}\left(\gamma^\mu p_\mu+\mu_{R}^{-}c^2\gamma^5\right)\psi^{-}+\dfrac{1}{2}\bar{\psi}^{-}\left(\gamma^\mu p_\mu-\mu_{L}^{-}c^2\gamma^5\right)\psi^{-},\label{tlag1}
\end{eqnarray}
or after summation
\begin{equation}
\mathcal{L}=\bar{\psi}^+\left(\gamma^\mu \hat{p}_\mu+\mu_{eff}^+ c^2\right)\psi^++\bar{\psi}^-\left(\gamma^\mu \hat{p}_\mu+\mu_{eff}^-c^2\right)\psi^-,\label{tlag1a}
\end{equation}
where $\mu_{eff}^{\pm}$ are the effective mass matrices of the gauge f\/ields $\psi^\pm$,
\begin{equation}
\mu_{eff}^\pm=\dfrac{\mu_R^\pm-\mu_L^\pm}{2}\gamma^5.
\end{equation}
After introduction of the global effective 8-component f\/ield
\begin{equation}
\Psi=\left[\begin{array}{c}{\psi^+}\\ {\psi^-}\end{array}\right],
\end{equation}
the theory (\ref{tlag1a}) becomes
\begin{equation}
  \mathcal{L}=\bar{\Psi}\left(\gamma^\mu\hat{p}_{\mu}+M_{eff}c^2\right)\Psi,\label{tlag1b}
\end{equation}
where $M_{eff}$ is the mass matrix given by
\begin{equation}
M_{eff}=\left[\begin{array}{cc}{\mu^+_{eff}}&0\\0&{\mu^-_{eff}}\end{array}\right],
\end{equation}
$\gamma^\mu$ are $8\times8$ matrices, and
\begin{equation}
\hat{p}_\mu=i\hslash\left[\begin{array}{c}\partial_\mu\\
\partial_\mu\end{array}\right],
\end{equation}
is 8-component momentum operator. Therefore (\ref{tlag1b}) can be associated with octonions. Hermiticity of both the effective mass matrices $\mu^\pm_{eff}$, and therefore also of the global effective mass matrix $M_{eff}$, depends on a choice of representation of the $\gamma^5$ matrix. For consistency the preferred representation of $\gamma^5$ matrix must be hermitian. It means that the effective global gauge f\/ield theory (\ref{tlag1b}) is physical for the only such a representation.

Obviously, the global effective gauge f\/ield theory (\ref{tlag1b}) demonstrates an invariance with respect to action of the composite gauge vector symmetry $SU(2)_V^{TOT}$\index{symmetry!composite vector}
\begin{equation}
SU(2)_V^{TOT}=SU(2)_V^+\oplus SU(2)_V^-,
\end{equation}
where $SU(2)_V^\pm$ are the $SU(2)\otimes SU(2)$ group transformations applied separately to each of the gauge f\/ields $\psi^\pm$
\begin{equation}
  \left\{\begin{array}{c}\psi^\pm\rightarrow \exp\left\{i\theta^\pm\right\}\psi^\pm\vspace*{5pt}\\ \bar{\psi}^\pm\rightarrow \bar{\psi}^\pm\exp\left\{-i\theta^\pm\right\}\end{array}\right..
\end{equation}
Such a situation means that there is realized the mechanism of spontaneous symmetry breakdown\index{spontaneous symmetry breakdown} for the global effective gauge f\/ield theory (\ref{tlag1b}). The broken symmetry is the composite global chiral symmetry $SU(3)_C^{TOT}$, and the result of the symmetry breakdown is its subgroup the composite isospin symmetry group $SU(2)_V^{TOT}$\index{symmetry!composite isospin}
\begin{equation}
SU(3)_C^{TOT}\longrightarrow SU(2)_V^{TOT}.
\end{equation}

Such a situation possesses an unambiguous physical interpretation. Namely, it is the syndrome of an existence of the global effective chiral condensate\index{chiral condensate} of the massive neutrinos, being a composition of two independent chiral condensates, which is the global effective gauge f\/ield theory invariant under action of the gauge symmetry (See, e.g. the book of S. Weinberg in Ref. \cite{gft})
\begin{equation}
SU(2)_V^{TOT}=(SU(2)^+\otimes SU(2)^+)\oplus(SU(2)^-\otimes SU(2)^-).
\end{equation}
However, because of action of the composite global chiral gauge symmetry $SU(3)_C^{TOT}$ (\ref{composite1}), the gauge f\/ield theories (\ref{lagau1}) and (\ref{lagau2}) looks like formally as the theories of non-interacting massive free quarks\index{free quarks}. Such a situation is very similar to the formalism of Quantum Chromodynamics (QCD)\index{quantum chromodynamics} \cite{greiner}, but factually in the presented physical scenario one has to deal with a composition of two independent copies of QCD. For each of these theories the space of f\/ields is distinguishable for the space of f\/ields of the single QCD. The difference is contained, namely, in the fact that there are only two massive chiral f\/ields - the left- and the right-handed Weyl f\/ields, which are the neutrinos\index{neutrino} in our proposition. The global effective chiral condensate\index{chiral condensate} of massive neutrinos (\ref{tlag1b}) is the result manifestly beyond the Standard Model\index{Standard Model}, but essentially it can be included into the fundamental theory of particle physics as the new contribution due to noncommutative geometry\index{noncommutative geometry}. Usually the spontaneous symmetry breakdown\index{spontaneous symmetry breakdown} results in the related Goldstone bosons generated via the mechanism. However, in the situation presented above this mechanism results in the chiral condensate, i.e. does not generate new particles. In the context of the massive neutrinos, the result of the mechanism of the spontaneous symmetry breakdown is the global effective chiral condensate\index{chiral condensate} of the massive neutrinos. By this reason such a situation is manifestly distinguishable, and is beyond methods of the Standard Model\index{Standard Model}.

\section{Conclusion}
The deformed Special Relativity given by the Snyder--Sidharth Hamiltonian constraint\index{Snyder--Sidharth Hamiltonian constraint} (\ref{ss}) obtained due to the Snyder geometry of noncommutative space-time\index{Snyder noncommutative geometry} (\ref{ndA})-(\ref{ndA}) manifestly and essentially differs from the usual Einstein energy-momentum relation\index{Einstein energy-momentum relation} well-known from Special Relativity. In particular as is self-evident from the form of the Snyder--Sidharth Hamiltonian constraint\index{Snyder--Sidharth Hamiltonian constraint}(\ref{ss}), the Snyder noncommutative geometry\index{Snyder noncommutative geometry} produces the extra contribution to the Einstein energy-momentum relation\index{Einstein energy-momentum relation} due to the additional $\ell^2$-term. As we have shown, this contribution can be neglected as the result of the algebra deformation. This is brought out very clearly in the Dirac equations \index{Dirac equation}(\ref{rhc1})-(\ref{rhc2}) which are manifestly non-hermitian, as well as in the massive Weyl equations \index{Weyl equation!massive}(\ref{neu1})-(\ref{neu2}) which are blatantly hermitian and are responsible for description of the neutrinos\index{neutrino} in our proposition.

A massless neutrino\index{neutrino}, characteristic for both the conventional Weyl theory as well as the Standard Model\index{Standard Model}, is now seen to argue as mass, and further, this mass has a two left-handed components and a two right-handed components, as it is straightforwardly noticeable from the formulas (\ref{mm1}) and (\ref{mm2}). Once this is recognized, the mass matrix which otherwise appears non-hermitian, turns out to be actually hermitian, as seen in the formula (\ref{mm3}), but if and only if when the masses (\ref{mm1}) and (\ref{mm2}) of the neutrinos are real numbers. There is no any restrictions, however, for their sign, i.e. the masses can be positive as well as negative. In other words, the underlying Snyder noncommutative geometry\index{Snyder noncommutative geometry} (\ref{ndA})-(\ref{ndA}) is ref\/lected in the modif\/ied Dirac equation \index{Dirac equation}(\ref{dsit}) and naturally and nontrivially gives rise to the mass of the neutrino\index{neutrino}.

As we have mentioned in partial discussions within this chapter, in analogy with the Standard Model Sidharth \cite{bgs2005ijmp2} suggested that such a situation is a possible result ''due to mass term'', however, with no any concrete calculations and proposals for the mass generation mechanism\index{mass generation mechanism}. The mass generation mechanism\index{mass generation mechanism}, proposed in this chapter for such a constructive and consistent formulation of this Sidharth idea, has purely kinetic nature, and moreover it is formally the result of the f\/irst approximation of more general noncommutative geometry\index{noncommutative geometry} determined by the Kontsevich deformation quantization\index{quantization!deformation}. In this manner we have shown that the mass generation mechanism ''due to mass term'' can be elegantly formulated in frames of noncommutative geometry\index{noncommutative geometry}, particularly in frames of the Snyder space-time\index{space-time!Snyder}.

We have shown also that the model of massive neutrinos can be understood and consistently described from the point of view of gauge f\/ield theories, which naturally includes Lorentz invariance\index{Lorentz invariance}. Such a formulation leads to interesting construction involving two independent copies of Quantum Chromodynamics and non-interacting massive free quarks\index{free quarks}, which is also employing effective composite isospin group\index{symmetry!composite isospin} resulting in the global effective chiral condensate\index{chiral condensate} of the massive neutrinos\label{neutrino}. The mechanism of spontaneous symmetry breakdown\index{spontaneous symmetry breakdown} presented above, which is the tool to receiving the composite isospin group, does not require existence of related Goldstone bosons\index{Goldstone bosons}, but the role of Goldstone bosons plays the chiral condensate\index{chiral condensate} of massive neutrinos. In itself this is new type of mass generation mechanism.

It must be remembered that in the Standard Model\index{Standard Model} the neutrino\index{neutrino} is massless, but the Super-Kamiokande experiments\index{Super-Kamiokande} in the late nineties showed that the neutrino\index{neutrino} does indeed have a mass and this is the leading motivation to an exploration of models beyond the Standard Model\index{Standard Model}. The model presented above is the b est example of such a situation. In this connection it is also relevant to mention that currently the Standard Model\index{Standard Model} requires the Higgs mechanism \index{Higgs mechanism}for the generation of mass in general\index{mass generation mechanism}, though the Higgs particle has been undetected for forty f\/ive years and it is hoped will be detected by researchers of Fermi National Accelerator Laboratory\index{Fermi National Accelerator Laboratory} or the Large Hadron Collider\index{Large Hadron Collider}. We hope for next development within the proposed here model of massive neutrinos.

\chapter{The Neutrinos: Energy Renormalization \& Integrability}

\section{Introduction}

In the previous chapter we have established that the modif\/ied Dirac equation \index{Dirac equation}arising due to the Snyder noncommutative geometry\index{Snyder noncommutative geometry}, yields the conventional Dirac theory with non-hermitian mass, or equivalently to the Weyl equation\index{Weyl equation} with a diagonal and hermitian mass matrices which describes the massive neutrinos. The obtained model of massive neutrinos involves 4 massive chiral f\/ields related to any originally massive or massless quantum state obeying the usual Dirac equation\index{Dirac equation}. By application of the mechanism of spontaneous symmetry breakdown\index{spontaneous symmetry breakdown} with respect to the global chiral symmetry the model was converted into the form of the isospin-symmetric global effective gauge f\/ield theory of the 8-component f\/ield $\Psi$ which is associated with the composed chiral condensate\index{chiral condensate} of massive neutrinos.

All these results violate the Lorentz symmetry\index{Lorentz symmetry!violation} manifestly, albeit their possible physical application can be considered in a diverse way. On the one hand the global effective gauge f\/ield theory is beyond the Standard Model\index{Standard Model}, yet can be considered as its contributory part due to the Snyder noncommutative geometry\index{Snyder noncommutative geometry}. On the other hand, in the model of massive neutrinos the masses of the two left-handed and two right-handed chiral Weyl f\/ields\index{chiral Weyl f\/ields} arise due to mass and energy of an original state and a minimal scale, \emph{e.g.} the Planck scale\index{Planck scale}. Therefore, its quantum mechanical countenance becomes almost a mystical riddle. In fact, possible existence of the massive neutrinos would be the logically consistent justif\/ication of physical correctness of the Snyder noncommutative geometry\index{Snyder noncommutative geometry}.

This chapter is mostly focused on the quantum mechanical aspect of the model of massive neutrinos. We shall present manifestly that the model in itself yields consistent physical explanation of the Snyder noncommutative geometry\index{Snyder noncommutative geometry} and consequently leads to energy renormalization of an original quantum relativistic particle. We shall perform computations arising directly from the Schr\"odinger equation\index{Schr\"odinger equation} formulation of both the modif\/ied Dirac equation\index{Dirac equation} and the massive Weyl equation\index{Weyl equation!massive}.

The f\/irst issue for discussion is the manifestly non-hermitian modif\/ied Dirac Hamiltonian. Its integrability is formulated by straightforward application of the Zassenhaus formula for exponentiation of sum of two non commuting operators. It is shown directly, however, that this approach does not lead to well-def\/ined solutions, because of for such a formulation the Zassenhaus exponents\index{Zassenhaus exponents} are still sums of two non commuting operators. Therefore such a integrability procedure possesses a cyclic problem which can not be removed, and by this reason is not algorithm. In this case the only approximations can be studied, but extraction of full solution is an extremely diff\/icult problem.

For solving the problem we shall change the integrability strategy, i.e. instead of the modif\/ied Dirac equation \index{Dirac equation}we shall employ the Schr\"odinger equation \index{Schr\"odinger equation}form of the Weyl equation \index{Weyl equation}with pure hermitian mass matrix. Integration of this equation is straightforward, elementary, and analogous to integration of the Dirac equation\index{Dirac equation}. Computations shall be presented in both the Dirac and the Weyl representations of the Dirac gamma matrices.

We perform calculations in the Clifford algebra \index{Clifford algebra}because of the representations of gamma matrices obeying the {\AE}ther algebra \index{{\AE}ther algebra}are not established and are very good problem for future research. This is caused by the fact, that the {\AE}ther algebra was proposed f\/irst in this book, and was not considered in earlier literature.

\newpage
\section{Energy renormalization}

Let us focus our attention on the masses of left-handed and right-handed chiral Weyl f\/ields\index{chiral Weyl f\/ields} (\ref{mu1x}) and (\ref{mu2x}). By straightforward elementary algebraic manipulations these two relations can be rewritten as the following system of equations
\begin{equation}\label{mus}
\left\{\begin{array}{c}
\left(\mu_R^{\pm}c^2+\dfrac{\epsilon}{2}\right)^2=\epsilon^2-4\epsilon mc^2 -4E^2\\
\left(\mu_L^{\pm}c^2-\dfrac{\epsilon}{2}\right)^2=\epsilon^2+4\epsilon mc^2 -4E^2
\end{array}\right.
\end{equation}
which allows to study dependence between the deformation parameter, i.e. a maximal energy $\epsilon$, and energy $E$ and mass $m$ of a particle and the masses $\mu_R^{\pm}$ and $\mu_L^{\pm}$ of neutrinos treated as physically measurable quantities which can be established via experimental data. Subtraction of the second equation from the f\/irst one in the system of equations (\ref{mus}), allows to obtain
\begin{equation}
 \left(\mu_L^{\pm}c^2-\dfrac{\epsilon}{2}\right)^2-\left(\mu_R^{\pm}c^2+\dfrac{\epsilon}{2}\right)^2=8\epsilon mc^2,
\end{equation}
or applying the difference of two squares $a^2-b^2=(a-b)(a+b)$
\begin{equation}
 \left[\left(\mu_L^{\pm}-\mu_R^{\pm}\right)c^2-\epsilon\right]\left(\mu_L^{\pm}+\mu_R^{\pm}\right)c^2=8\epsilon mc^2,
\end{equation}
what allows to derive a maximal energy as
\begin{equation}\label{def}
  \epsilon=\dfrac{\left(\mu_L^{\pm}-\mu_R^{\pm}\right)c^2}{1+\dfrac{8m}{\mu_L^{\pm}+\mu_R^{\pm}}}.
\end{equation}

Because of $\epsilon\geqslant0$ one has the condition of masses of neutrinos
\begin{equation}\label{neumascon}
  \mu_L^{\pm}\geqslant\mu_R^{\pm},
\end{equation}
what after using of the explicit formulas (\ref{mu1x}) and (\ref{mu2x}) leads to two alternative conditions for $\epsilon$
\begin{equation}
  \epsilon\geqslant\mp\left(\sqrt{\epsilon^2+4\epsilon mc^2-4E^2}+\sqrt{\epsilon^2-4\epsilon mc^2-4E^2}\right).
\end{equation}
The f\/irst condition leads to
\begin{equation}
  \epsilon\in\left(-\infty,-\dfrac{8}{3}mc^2-\dfrac{4}{3}\sqrt{3E^2+4m^2c^4}\right]\cup\left[-2mc^2+2\sqrt{E^2+m^2c^4},\infty\right),
\end{equation}
while the second one gives
\begin{equation}
  \epsilon\in\left[-2mc^2+2\sqrt{E^2+m^2c^4},-\dfrac{8}{3}mc^2+\dfrac{4}{3}\sqrt{3E^2+4m^2c^4}\right],
\end{equation}
and by taking these results together one obtains f\/inally
\begin{equation}\label{epsbound}
  -2mc^2+2\sqrt{E^2+m^2c^4}\leqslant\epsilon\leqslant -\dfrac{8}{3}mc^2+\dfrac{4}{3}\sqrt{3E^2+4m^2c^4}.
\end{equation}
Because, however, $\epsilon\geqslant0$ one has the conditions for mass and energy of a particle
\begin{eqnarray}
-2mc^2+2\sqrt{E^2+m^2c^4}&\geqslant&0,\\
-\dfrac{8}{3}mc^2+\dfrac{4}{3}\sqrt{3E^2+4m^2c^4}&\geqslant&0,\\
-\dfrac{8}{3}mc^2+\dfrac{4}{3}\sqrt{3E^2+4m^2c^4}&\geqslant&-2mc^2+2\sqrt{E^2+m^2c^4}.
\end{eqnarray}
The f\/irst and the second conditions leads to the trivial relation $E^2\geqslant0$, while the third one states that
\begin{equation}
\dfrac{9}{32}E^4+m^2c^4E^2+m^4c^8\geqslant0,
\end{equation}
what is satisf\/ied if and only if mass $m$ and energy $E$ of a particle are real numbers. By application of the bounds (\ref{epsbound}) to the relation (\ref{def}) one obtains the condition for masses
\begin{equation}\label{masbound}
m\leqslant\dfrac{\mu_L^{\pm}+\mu_R^{\pm}}{8},
\end{equation}
or equivalently the inequality for the ratio
\begin{equation}\label{ineqmass}
\dfrac{8m}{\mu_L^{\pm}+\mu_R^{\pm}}\leqslant1.
\end{equation}
By the def\/inition (\ref{def}) one has
\begin{equation}\label{def1}
  \dfrac{8m}{\mu_L^{\pm}+\mu_R^{\pm}}=\dfrac{\left(\mu_L^{\pm}-\mu_R^{\pm}\right)c^2}{\epsilon}-1,
\end{equation}
what in the light of the inequality (\ref{ineqmass}) leads to the bound
\begin{equation}
  \epsilon\geqslant\dfrac{\mu_L^{\pm}-\mu_R^{\pm}}{2}c^2,
\end{equation}
which can be translated into the language of a minimal scale
\begin{equation}\label{ellbound}
  \ell\leqslant\dfrac{\hslash}{\sqrt{\alpha}c}\dfrac{2}{\mu_L^{\pm}-\mu_R^{\pm}}.
\end{equation}
By using of the Compton wavelength of a neutrino\index{neutrino}\index{Compton wavelength!of neutrino\index{neutrino}}
\begin{equation}
\lambda_C(\mu_{R,L}^{\pm})=2\pi\dfrac{\hslash}{\mu_{R,L}^{\pm}c},
\end{equation}
the bound (\ref{ellbound}) can be written in the form
\begin{equation}\label{ellbound1}
  \ell\leqslant\dfrac{1}{2\pi\sqrt{\alpha}}\dfrac{2\lambda_C(\mu_{L}^{\pm})\lambda_C(\mu_{R}^{\pm})}{\lambda_C(\mu_{R}^{\pm})-\lambda_C(\mu_{L}^{\pm})},
\end{equation}
which, if one wishes to apply $\alpha=\dfrac{1}{2\pi}$ established by (\ref{alpha}), becomes
\begin{equation}
  \ell\leqslant\dfrac{2\lambda_C(\mu_{L}^{\pm})\lambda_C(\mu_{R}^{\pm})}{\lambda_C(\mu_{R}^{\pm})-\lambda_C(\mu_{L}^{\pm})}.
\end{equation}
Equivalently, the bound (\ref{ellbound}) with the condition (\ref{alpha}) can be presented in most conventional form
\begin{equation}
  \ell\leqslant\lambda_C\left(\dfrac{\mu_{R}^{\pm}-\mu_{L}^{\pm}}{2}\right).
\end{equation}
If one wishes to do not preserve (\ref{alpha}) then the bound (\ref{ellbound1}) for f\/ixed scale value establishes the following inequality for $\alpha$
\begin{equation}\label{ellbound1a}
  \alpha\leqslant\left(\dfrac{1}{2\pi}\dfrac{1}{\ell}\dfrac{2\lambda_C(\mu_{L}^{\pm})\lambda_C(\mu_{R}^{\pm})}{\lambda_C(\mu_{R}^{\pm})-\lambda_C(\mu_{L}^{\pm})}\right)^2,
\end{equation}
which in the light of the generalization (\ref{alpha2}) leads to
\begin{equation}\label{ellbound2}
  \dfrac{m}{M_P}\leqslant\dfrac{1}{\ell}\dfrac{2\lambda_C(\mu_{L}^{\pm})\lambda_C(\mu_{R}^{\pm})}{\lambda_C(\mu_{R}^{\pm})-\lambda_C(\mu_{L}^{\pm})},
\end{equation}
or with using of (\ref{ellbound})
\begin{equation}\label{ellbound}
  m\leqslant\left(\dfrac{\hslash}{c}\right)^2\dfrac{2\pi}{\ell\ell_P}\dfrac{2}{\mu_L^{\pm}-\mu_R^{\pm}}.
\end{equation}
This result in the light of the bound (\ref{masbound}), however, leads to
\begin{equation}
\left(\dfrac{\hslash}{c}\right)^2\dfrac{2\pi}{\ell\ell_P}\dfrac{2}{\mu_L^{\pm}-\mu_R^{\pm}}=\dfrac{\mu_L^{\pm}+\mu_R^{\pm}}{8},
\end{equation}
what results in the squared-mass difference\index{squared-mass difference}
\begin{equation}\label{boom}
\Delta\mu^2_{LR}=\left(\mu_L^{\pm}\right)^2-\left(\mu_R^{\pm}\right)^2=\left(\dfrac{\hslash}{c}\right)^2\dfrac{32\pi}{\ell\ell_P}=
32\pi\dfrac{\ell_P}{\ell}M_P^2.
\end{equation}
In this manner if $\Delta\mu^2_{LR}$ is f\/ixed by experimental data, then by the equation (\ref{boom}) establishes the minimal scale
\begin{equation}\label{boom1}
\ell=32\pi\dfrac{M_P^2}{\Delta\mu^2_{LR}}\ell_P,
\end{equation}
or approximatively
\begin{equation}
\ell\approx 2.4220\cdot10^{23}m\dfrac{1\dfrac{eV^2}{c^2}}{\Delta\mu^2_{LR}}.
\end{equation}
Interestingly, at the Planck scale\index{Planck scale} $\ell=\ell_P$, then (\ref{boom1}) generates
\begin{equation}
\Delta\mu^2_{LR}=32\pi M_P^2\approx1.4985\cdot10^{10}\dfrac{YeV^2}{c^2},
\end{equation}
where $1YeV=10^{24}eV$, while at the Compton scale \index{Compton scale}$\ell=\lambda_C(m_e)$
\begin{equation}
  \Delta\mu^2_{LR}\approx10^{5}\dfrac{PeV^2}{c^2},
\end{equation}
where $1PeV=10^{15}eV$. Similarly, at the Compton--Planck scale\index{Compton--Planck scale}$\ell=\lambda_C(M_P)$
\begin{equation}
  \Delta\mu^2_{LR}=16M_P^2\approx2.3850\cdot10^{9}\dfrac{YeV^2}{c^2}.
\end{equation}

In the light of the mass formulas (\ref{mu1x}) and (\ref{mu2x}) one can deduce the squared-mass difference\index{squared-mass difference}
\begin{equation}\label{boomx}
  \Delta\mu^2_{LR}=16\pi\dfrac{\ell_P}{\ell}M_P^2\left[1\pm
  \dfrac{\sqrt{\epsilon^2-4E^2+8\pi\dfrac{\ell_P}{\ell}M_P^2c^4}-\sqrt{\epsilon^2-4E^2-8\pi\dfrac{\ell_P}{\ell}M_P^2c^4}}{8mc^2}\right],
\end{equation}
where we have applied the generalization (\ref{alpha2}). Comparison of the equations (\ref{boom}) and (\ref{boomx}) gives
\begin{equation}
  1=\pm\dfrac{\sqrt{\epsilon^2-4E^2+8\pi\dfrac{\ell_P}{\ell}M_P^2c^4}-\sqrt{\epsilon^2-4E^2-8\pi\dfrac{\ell_P}{\ell}M_P^2c^4}}{8mc^2},
\end{equation}
what is the equation for a minimal scale $\ell$ as a function of mass $m$ and energy $E$ of a particle. The solution of this equation is easy to establish
\begin{equation}\label{alell0}
\alpha\ell=\dfrac{\sqrt{3}}{16\pi}\dfrac{\ell_P}{\sqrt{1+\left(\dfrac{E}{2mc^2}\right)^2}},
\end{equation}
what can be equivalently treated as the formula for the energy of a particle
\begin{equation}
\dfrac{E^2}{(2mc^2)^2}=\dfrac{3}{256\pi^2}\dfrac{\ell_P^2}{\alpha^2\ell^2}-1,
\end{equation}
and application of the generalization (\ref{alpha2}) leads to
\begin{equation}
E^2=\dfrac{3}{16}\dfrac{\ell_P^2}{\ell^2}E_P^2-(2mc^2)^2=\dfrac{3}{16}\epsilon^2-(2mc^2)^2,
\end{equation}
where $E_P=M_Pc^2$ is the Planck energy\index{Planck energy}. Because, however, $E^2\geqslant0$ one obtains the inequality
\begin{equation}
m\ell \leqslant\dfrac{\sqrt{3}}{8}M_P\ell_P=\dfrac{\sqrt{3}}{8}\dfrac{\hslash}{c},
\end{equation}
which for f\/ixed mass of a particle gives the upper bound for a minimal scale
\begin{equation}
\ell\leqslant\dfrac{2\pi\sqrt{3}}{8}\lambda_C(m),
\end{equation}
or the lower bound for maximal energy
\begin{equation}\label{epsboun}
  \epsilon\geqslant\dfrac{8}{\sqrt{3}}E_P,
\end{equation}
while for f\/ixed minimal scale leads to the upper bound for mass of a particle.

Applying a minimal scale (\ref{alell0}) within the def\/inition (\ref{eps}) allows to eliminate a minimal scale dependence
\begin{equation}
  \epsilon=\dfrac{16\pi}{\sqrt{3}}E_P\sqrt{\alpha}\sqrt{1+\left(\dfrac{E}{2mc^2}\right)^2},
\end{equation}
what after application of the generalization (\ref{alpha2}) takes the form
\begin{equation}
  \epsilon=\dfrac{4}{\sqrt{3}}\sqrt{E^2+(2mc^2)^2},
\end{equation}
and in the light of (\ref{epsboun}) leads to the bound for energy of a particle
\begin{equation}
E^2\geqslant4\left(E_P^2-m^2c^4\right),
\end{equation}
which by $E^2\geqslant0$ gives the bound for mass of a particle
\begin{equation}\label{massinequality}
 m\leqslant M_P,
\end{equation}
proving the Markov hypothesis\index{Markov hypothesis!proof} expressing supposition that the upper bound for mass of a particle is given by the Planck mass. Recall that we have deduced (\ref{kontmass}) that the only case $m\sim M_P$ coincides with the Kontsevich deformation quantization\index{quantization!deformation}. In the light of the inequality (\ref{massinequality}) such a situation suggests that a particle described by the Snyder noncommutative geometry\index{Snyder noncommutative geometry} is the Planckian particle\index{Planckian particle}, i.e. the particle equipped with the mass identical to the Planck mass $m=M_P$. In other words possible existence of the Planckian particle\index{Planckian particle} will be establishing the physical sense of both the Snyder noncommutative geometry\index{Snyder noncommutative geometry} and the Kontsevich deformation quantization\index{quantization!deformation}. It suggests also that physics at the Planck scale\index{Planck scale} is the physics of the Planckian particle\index{Planckian particle}.

A maximal energy (\ref{def}) does not vanish for all $\mu_L^{\pm}\neq\mu_R^{\pm}\neq0$, and is f\/inite if and only if $\mu_{L,R}^{\pm}<\infty$.                                                                    The mass $m$ of an original quantum state as well as the masses of neutrinos $\mu_R^{\pm}$ and $\mu_L^{\pm}$ are presumed to be physical quantities, which can be established by experimental data. In the case, when an original quantum state is massless, a maximal energy has the maximal value which equals to
\begin{equation}
  \epsilon(m=0)=\left(\mu_L^{\pm}-\mu_R^{\pm}\right)c^2\equiv\epsilon_0,
\end{equation}
that is f\/inite and non vanishing for f\/inite $\mu_R^{\pm}\neq0$ and $\mu_L^{\pm}\neq0$. In this manner one can study approximation around such def\/ined state. For $|\mu_R^{\pm}+\mu_L^{\pm}|>8m$ the appropriate Taylor series\index{Taylor series} expansion is
\begin{equation}\label{expan1}
  \epsilon=\epsilon_0\left[1-\dfrac{8m}{\mu_R^{\pm}+\mu_L^{\pm}}+O\left(\left(\dfrac{8m}{\mu_R^{\pm}+\mu_L^{\pm}}\right)^2\right)\right],
\end{equation}
while for $|\mu_R^{\pm}+\mu_L^{\pm}|<8m$ one has the following expansion
\begin{equation}
  \epsilon=\epsilon_0\left[\dfrac{\mu_R^{\pm}+\mu_L^{\pm}}{8m}+O\left(\left(\dfrac{\mu_R^{\pm}+\mu_L^{\pm}}{8m}\right)^2\right)\right].
\end{equation}
For the case $|\mu_R^{\pm}+\mu_L^{\pm}|=8m$ both these series coincide and
\begin{equation}
  \epsilon=\dfrac{\epsilon_0}{2}.
\end{equation}
The established inequality (\ref{ineqmass}), however, allows to remove from considerations the case $|\mu_R^{\pm}+\mu_L^{\pm}|>8m$ given by the Taylor series\index{Taylor series} expansion (\ref{expan1}).

On the other hand, however, addition of the second equation to the f\/irst one in the system of equations (\ref{mus}) gives the relation
\begin{equation}\label{c2}
 \left(\mu_L^{\pm}c^2-\dfrac{\epsilon}{2}\right)^2+\left(\mu_R^{\pm}c^2+\dfrac{\epsilon}{2}\right)^2=2\left(\epsilon^2-4E^2\right).
\end{equation}
The LHS of the equation (\ref{c2}) is always positive as a sum of two squares of real numbers, and therefore the RHS of this equation is always positive also. In this manner, one obtains the renormalization of energy of a particle vie a maximal energy
\begin{equation}\label{reg}
  -\dfrac{\epsilon}{2}\leqslant E\leqslant\dfrac{\epsilon}{2}.
\end{equation}
Naturally, for the generic case of Special Relativity we have $\epsilon\equiv\infty$ and by this reason values of energy $E$ of a particle are not bounded. Therefore, by the relation (\ref{reg}) it is evident that the Snyder noncommutative geometry\index{Snyder noncommutative geometry} results in renormalization of energy of a particle.

The relation (\ref{c2}) can be treated as the constraint for the energy $E$ of a particle, and immediately solved with respect to $E$. The solution is a quadratic form which can be presented in the canonical form with respect to a maximal energy $\epsilon$
\begin{equation}\label{er}
 E^2=\dfrac{3}{16}\left\{\left[\epsilon+\dfrac{\mu_L^{\pm}-\mu_R^{\pm}}{3}c^2\right]^2-\left[\dfrac{\mu_L^{\pm}-\mu_R^{\pm}}{3}c^2\right]^2\left[7+\dfrac{12\mu_L^{\pm}\mu_R^{\pm}}{\left(\mu_L^{\pm}-\mu_R^{\pm}\right)^2}\right]\right\}.
\end{equation}
By explicit application of a maximal energy (\ref{def}) within the energetic constraint (\ref{er}) one can present the particle energy via the only masses of the neutrinos related to this particle
\begin{equation}\label{er1}
 E^2=\dfrac{\left[\left(\mu_L^{\pm}-\mu_R^{\pm}\right)c^2\right]^2}{48}\left\{\left(\dfrac{4+\dfrac{8m}{\mu_L^{\pm}+\mu_R^{\pm}}}{1+\dfrac{8m}{\mu_L^{\pm}+\mu_R^{\pm}}}\right)^2-\left[7+\dfrac{12\mu_L^{\pm}\mu_R^{\pm}}{\left(\mu_L^{\pm}-\mu_R^{\pm}\right)^2}\right]\right\},
\end{equation}
which for the case of originally massless state has the value
\begin{equation}\label{e0}
  E^2(m=0)=\dfrac{1}{16}\left[\left(\mu_L^{\pm}-\mu_R^{\pm}\right)c^2\right]^{2}\left[3-4\dfrac{\mu_L^{\pm}\mu_R^{\pm}}{\left(\mu_L^{\pm}-\mu_R^{\pm}\right)^2}\right]\equiv E^2_0,
\end{equation}
and by this reason energy of a particle is
\begin{equation}
  E^2=E_0^2+\Delta E^2,
\end{equation}
where $\Delta E^2$ is the correction generated due to nonzero mass of an original quantum state
\begin{equation}
  \Delta E^2=-\dfrac{5}{16}\left[\left(\mu_L^{\pm}-\mu_R^{\pm}\right)c^2\right]^2\dfrac{\dfrac{8m}{\mu_L^{\pm}+\mu_R^{\pm}}}{1+\dfrac{8m}{\mu_L^{\pm}+\mu_R^{\pm}}}\dfrac{\dfrac{8}{5}+\dfrac{8m}{\mu_L^{\pm}+\mu_R^{\pm}}}{1+\dfrac{8m}{\mu_L^{\pm}+\mu_R^{\pm}}},
\end{equation}
which can be expanded into the Taylor series\index{Taylor series} around the massless state
\begin{equation}
  \Delta E^2=\dfrac{1}{16}\left[\left(\mu_L^{\pm}-\mu_R^{\pm}\right)c^2\right]^2\sum_{n=1}^{\infty}(-1)^n(3n+5)\left(\dfrac{8m}{\mu_L^{\pm}+\mu_R^{\pm}}\right)^n.
\end{equation}
Because of the relation (\ref{ineqmass}) it is more convenient to see the Taylor series\index{Taylor series} expansion of $E^2$ around the point $\mu_L^{\pm}+\mu_R^{\pm}=8m$, which we shall call the $8m$ point, when $|\mu_L^{\pm}+\mu_R^{\pm}|>8m$. In such a situation the decomposition, which we shall call \emph{the $8m$ expansion}\index{$8m$ expansion}, has somewhat different form
\begin{equation}
  E^2=E_{8m}^2+\Delta E^2_{8m},
\end{equation}
where
\begin{eqnarray}
  \!\!\!\!\!\!\!\!\!\!E_{8m}^2&=&E_0^2+\dfrac{13}{64}\left[\left(\mu_L^{\pm}-\mu_R^{\pm}\right)c^2\right]^2,\label{E8}\\
  \!\!\!\!\!\!\!\!\!\!\Delta E^2_{8m}&=&\dfrac{\left[\left(\mu_L^{\pm}-\mu_R^{\pm}\right)c^2\right]^2}{64}
  \sum_{n=1}^{\infty}\left(-\dfrac{1}{2}\right)^n(3n+7)\left(\dfrac{8m}{\mu_L^{\pm}+\mu_R^{\pm}}-1\right)^n.
\end{eqnarray}
In this way, in the $8m$ expansion the leading correction to square of energy of a particle is
\begin{equation}
{\Delta E^2_{8m}}^{(1)}=-\dfrac{5}{64}\left[\left(\mu_L^{\pm}-\mu_R^{\pm}\right)c^2\right]^2\left(\dfrac{8m}{\mu_L^{\pm}+\mu_R^{\pm}}-1\right).
\end{equation}
For def\/ined $E_{8m}^2$ the equation (\ref{E8}) establishes the relation between the masses of the neutrinos
\begin{equation}
\mu_L^{\pm}=\left(\dfrac{33}{25}\pm\sqrt{\dfrac{464}{625}+\left(\dfrac{8}{5}\dfrac{E_{8m}}{\mu_R^{\pm}c^2}\right)^2}\right)\mu_R^{\pm},
\end{equation}
and because of the condition $\mu_L^{\pm}-\mu_R^{\pm}\geqslant0$
\begin{equation}
\mu_L^{\pm}-\mu_R^{\pm}=\left(\dfrac{8}{25}\pm\sqrt{\dfrac{464}{625}+\left(\dfrac{8}{5}\dfrac{E_{8m}}{\mu_R^{\pm}c^2}\right)^2}\right)\mu_R^{\pm}\geqslant0.
\end{equation}
the case of minus sign leads to the inequality
\begin{equation}
\left(\dfrac{E_{8m}}{\mu_R^{\pm}c^2}\right)^2+\dfrac{1}{4}\leqslant0,
\end{equation}
which does not possess solutions for real values of $E_{8m}$ and $\mu_R^{\pm}$. Therefore, the physical solution is
\begin{equation}
\mu_L^{\pm}=\left(\dfrac{33}{25}+\dfrac{4}{25}\sqrt{29+4\left(\dfrac{E_{8m}}{\mu_R^{\pm}c^2}\right)^2}\right)\mu_R^{\pm},
\end{equation}
and has minimal value for $E_{8m}=0$ with the value
\begin{equation}
\mu_L^{\pm}=\left(\dfrac{33}{25}+\dfrac{4}{25}\sqrt{29}\right)\mu_R^{\pm}\approx2.1816\mu_R^{\pm}.
\end{equation}

In fact, for given value of energy of massless state $E_0$ the equation (\ref{e0}) can be used for establishment of the relation between masses of the neutrinos. In result one receives two possible solutions
\begin{equation}
  \mu_L^{\pm}=\left(\dfrac{5}{3}\pm\dfrac{4}{3}\sqrt{\strut{1+\dfrac{1}{3}\left(\dfrac{E_0}{\mu_L^{\pm}c^2}\right)^2}}\right)\mu_R^{\pm},
\end{equation}
and the physical solution is established by the condition (\ref{neumascon})
\begin{equation}\label{differ}
 \mu_L^{\pm}-\mu_R^{\pm}=\left(\dfrac{2}{3}\pm\dfrac{4}{3}\sqrt{\strut{1+\dfrac{1}{3}\left(\dfrac{E_0}{\mu_L^{\pm}c^2}\right)^2}}\right)\mu_R^{\pm}\geqslant0,
\end{equation}
which in the case of the minus sign states that
\begin{equation}
  \left(\dfrac{E_0}{\mu_L^{\pm}c^2}\right)^2+\dfrac{9}{4}\leqslant0,
\end{equation}
what is not satisf\/ied for real $E_0$ and $\mu_L^{\pm}$. This argument allows to generate the physical solution
\begin{equation}
  \mu_L^{\pm}=\left(\dfrac{5}{3}+\dfrac{4}{3}\sqrt{\strut{1+\dfrac{1}{3}\left(\dfrac{E_0}{\mu_L^{\pm}c^2}\right)^2}}\right)\mu_R^{\pm}.
\end{equation}
which is minimized by $E_0=0$ with the value
\begin{equation}
  \mu_L^{\pm}=3\mu_R^{\pm}.
\end{equation}

The relation (\ref{er1}) for squared energy $E^2$ of a particle, i.e. factually the constraint, is useful for analysis of certain situations. Albeit, in general if one knows approximative value of $E^2$ the value of energy $E$ of a particle can not be established by taking a square root of $E^2$. Square root taking is in itself an approximation, and if one wishes to study approximations of energy $E$ then this energy should be determined separately via the appropriate Taylor series\index{Taylor series} expansion. The problem is that the obtained series for $E^2$ and for $E$ shall be in general different, and taking a square root of arbitrary f\/ixed order of approximation of $E^2$ does not coincide with the same order of approximation of $E$.

In this manner, one can perform the deductions for energy $E$ of a particle which are analogous to the deductions for its square. Let us write out the formula for energy explicitly
\begin{equation}\label{er2}
 E=\dfrac{\left(\mu_L^{\pm}-\mu_R^{\pm}\right)c^2}{4}\sqrt{\dfrac{1}{3}\left(\dfrac{4+\dfrac{8m}{\mu_L^{\pm}+\mu_R^{\pm}}}{1+\dfrac{8m}{\mu_L^{\pm}+\mu_R^{\pm}}}\right)^2-\dfrac{7}{3}-\dfrac{4\mu_L^{\pm}\mu_R^{\pm}}{\left(\mu_L^{\pm}-\mu_R^{\pm}\right)^2}}.
\end{equation}
It is easy to see that for the originally massless case the value of energy is $E=E_0$ where
\begin{equation}
E_0=\dfrac{\left(\mu_L^{\pm}-\mu_R^{\pm}\right)c^2}{4}\sqrt{3-\dfrac{4\mu_L^{\pm}\mu_R^{\pm}}{\left(\mu_L^{\pm}-\mu_R^{\pm}\right)^2}},
\end{equation}
and in this way
\begin{eqnarray}\label{er2a}
 E-E_0&=&\dfrac{\left(\mu_L^{\pm}-\mu_R^{\pm}\right)c^2}{4}\Bigg\{\sqrt{\dfrac{1}{3}\left(\dfrac{4+\dfrac{8m}{\mu_L^{\pm}+\mu_R^{\pm}}}{1+\dfrac{8m}{\mu_L^{\pm}+\mu_R^{\pm}}}\right)^2-\left[\dfrac{7}{3}+\dfrac{4\mu_L^{\pm}\mu_R^{\pm}}{\left(\mu_L^{\pm}-\mu_R^{\pm}\right)^2}\right]}-\nonumber\\
 &&\sqrt{3-\dfrac{4\mu_L^{\pm}\mu_R^{\pm}}{\left(\mu_L^{\pm}-\mu_R^{\pm}\right)^2}}\Bigg\}.
\end{eqnarray}
Now one can apply the $8m$ expansion\index{$8m$ expansion} to the relation (\ref{er2a}). In this case, i.e. $E$ not $E^2$, the expansion is more diff\/icult to apply, because one has to deal with square roots. The expansion has the form
\begin{equation}
  E=\dfrac{\left(\mu_L^{\pm}-\mu_R^{\pm}\right)c^2}{4}\sum_{n=0}^\infty A_n\left(\dfrac{8m}{\mu_L^{\pm}+\mu_R^{\pm}}-1\right)^n.
\end{equation}
The explicit formula for the coeff\/icients $A_n$ is not easy to extract, because for $n\geqslant0$ they satisfy the recurrence equation
\begin{eqnarray}
&&(66+22n+8(3+n)a^2)A_{n+3}+(51+23n+12(2+n)a^2)A_{n+2}+\nonumber\\
&&+(9+8n+6(1+n)a^2)A_{n+1}+n(1+a^2)A_n=0,
\end{eqnarray}
with the initial conditions
\begin{eqnarray}
  A_0&=&\sqrt{\dfrac{11}{4}+a^2},\\
  A_1&=&-\dfrac{5}{8\sqrt{\dfrac{11}{4}+a^2}},\\
  A_2&=&\dfrac{59+26a^2}{64\left(\dfrac{11}{4}+a^2\right)^{3/2}},
\end{eqnarray}
where we have introduced the parameter
\begin{equation}
  a=\dfrac{4E_0/c^2}{\mu_L^{\pm}-\mu_R^{\pm}}.
\end{equation}

Similar situation has a place for expansion around massless state. In this case the expansion has the form
\begin{equation}
  E=\dfrac{\left(\mu_L^{\pm}-\mu_R^{\pm}\right)c^2}{4}\sum_{n=0}^\infty B_n\left(\dfrac{8m}{\mu_L^{\pm}+\mu_R^{\pm}}\right)^n,
\end{equation}
where in general for $n\geqslant0$ the expansion coeff\/icients $B_n$ satisfy the following recurrence equation
\begin{eqnarray}
&&(18+6n+(3+n)a^2)B_{n+3}+(24+10n+3(2+n)a^2)B_{n+2}+\nonumber\\
&&+(6+5n+3(1+n)a^2)B_{n+1}+n(1+a^2)B_n=0,
\end{eqnarray}
having the following initial conditions
\begin{eqnarray}
  B_0&=&\sqrt{6+a^2},\\
  B_1&=&-\dfrac{4}{\sqrt{6+a^2}},\\
  B_2&=&-\dfrac{50+11a^2}{2(6+a^2)^{3/2}}.
\end{eqnarray}
\section{The Integrability Problem for the Dirac equations}

The received Dirac equations\index{Dirac equation}
\begin{equation}\label{diraqeq}
  \left(i\hslash\gamma^\mu \partial_\mu-M_{\pm}\right)\psi=0,
\end{equation}
where the mass matrices $M_{\pm}$ is given by the formula (\ref{mm2}), can be rewritten in the form of the Schr\"odinger equation\index{Schr\"odinger equation}
\begin{equation}\label{scd}
  i\hslash\partial_0\psi^\pm=\hat{H}\psi^\pm,
\end{equation}
where in the present case the Hamilton operator $\hat{H}$ has the form
\begin{equation}\label{ham}
  \hat{H}=-i\hslash c\gamma^0\gamma^i\partial_i-\dfrac{\mu_L^{\pm}+\mu_R^{\pm}}{2}c^2\gamma^0+\dfrac{\mu_L^{\pm}-\mu_R^{\pm}}{2}c^2\gamma^0\gamma^5,
\end{equation}
which can be splitted into the hermitian $\mathfrak{H}(\hat{H})$ and the antihermitian $\mathfrak{A}(\hat{H})$ components
\begin{eqnarray}
\hat{H}&=&\mathfrak{H}(\hat{H})+\mathfrak{A}(\hat{H}),\label{hamm}\\
\mathfrak{H}(\hat{H})&=&-i\hslash c\gamma^0\gamma^i\partial_i-\dfrac{\mu_L^{\pm}+\mu_R^{\pm}}{2}c^2\gamma^0,\label{hp}\\
\mathfrak{A}(\hat{H})&=&\dfrac{\mu_L^{\pm}-\mu_R^{\pm}}{2}c^2\gamma^0\gamma^5,\label{ahp}
\end{eqnarray}
with (anti)hermiticity def\/ined standardly
\begin{eqnarray}
  \int d^3x \bar{\psi}^\pm\mathfrak{H}(\hat{H})\psi^\pm &=&\int d^3x \overline{\mathfrak{H}(\hat{H})\psi^\pm}\psi^\pm,\\
  \int d^3x \bar{\psi}^\pm\mathfrak{A}(\hat{H})\psi^\pm &=&-\int d^3x \overline{\mathfrak{A}(\hat{H})\psi^\pm}\psi^\pm.
\end{eqnarray}

Let us consider the situation when the masses of the neutrinos has the same value, which we shall call $\mu$
\begin{equation}\label{muuu}
\mu_R^{\pm}=\mu_L^{\pm}\equiv\mu.
\end{equation}
It is easy to see that then the antihermitian component (\ref{ahp}) vanishes identically. However, the hermitian component (\ref{hp}) is still nontrivial. Consequently, the Hamilton operator (\ref{hamm}) takes the form of the conventional Dirac Hamiltonian\index{Dirac Hamiltonian}
\begin{equation}
  \hat{H}_{D}=-\gamma^0\left(i\hslash c\gamma^i\partial_i+\mu c^2\right).\label{diru}
\end{equation}
In such a situation, however, by the relation (\ref{er1}) energy of a particle vanishes identically
\begin{equation}
E=0.
\end{equation}
Taking into account the bounds (\ref{reg}) one obtains that also a maximal energy trivializes identically
\begin{equation}
  \epsilon\equiv0,
\end{equation}
and therefore a minimal scale is inf\/inite $\ell=\infty$. In the light of the formulas (\ref{mu1x})-(\ref{mu2x}) for the masses of the neutrinos, one obtains
\begin{equation}
\mu_R^{\pm}=\mu_L^{\pm}\equiv0,
\end{equation}
i.e. by the def\/inition (\ref{muuu})
\begin{equation}
  \mu\equiv0.
\end{equation}
Therefore, the Dirac Hamiltonian\index{Dirac Hamiltonian!massless} (\ref{diru}) becomes massless
\begin{equation}
  \hat{H}_{D}=-i\hslash c\gamma^0\gamma^i\partial_i.\label{diraq}
\end{equation}
and consequently the Dirac equations \index{Dirac equation}(\ref{diraqeq}) becomes the Weyl equation\index{Weyl equation}
\begin{equation}\label{weyleq}
  i\hslash\gamma^\mu \partial_\mu\psi=0,
\end{equation}
describing massless particle - the Weyl neutrino\index{Weyl neutrino}. It means that equality between the masses of the neutrinos (\ref{muuu}) def\/ines the massless particle obeying the Weyl equation. However, in the light of the Super-Kamiokande\index{Super-Kamiokande} results neutrino\index{neutrino} is equipped with nonzero mass. In this manner, such an embarrassing situation created by the Weyl equation\index{Weyl equation} (\ref{weyleq}), and laying in the foundations of the Standard Model\index{Standard Model}, is manifestly non physical. In other words, in the light of the Snyder noncommutative geometry\index{Snyder noncommutative geometry} the theory of massive neutrinos is consistent if and only if the difference between masses of the left-handed and the right-handed neutrinos is nonzero and positive.

The full modif\/ied Hamiltonian (\ref{ham}) possesses non-hermitian nature evidently. Therefore consequently the Schr\"odinger equation \index{Schr\"odinger equation}form time evolution (\ref{scd}) is non unitary manifestly. Its formal integration, however, can be carried out by the standard method of quantum mechanics
\begin{equation}
  \psi^\pm(x,t)=G(t,t_0)\psi^\pm(x,t_0),
\end{equation}
involving the following time evolution operator
\begin{equation}
G(t,t_0)\equiv\exp\left\{-\dfrac{i}{\hslash}\int_{t_0}^td\tau\hat{H}(\tau)\right\}.\label{integ}
\end{equation}
By this reason, the integrability problem for the wave equation (\ref{scd}) with the Hamilton operator (\ref{ham}) can be formulated in terms of the appropriate Zassenhaus exponents\index{Zassenhaus exponents}
\begin{eqnarray}
\exp\left\{A+B\right\}&=&\exp(A)\exp(B)\prod_{n=2}^\infty \exp{C_n},\label{zassen}\\
C_2&=&-\frac{1}{2}C,\\
C_3&=&-\frac{1}{6}(2[C,B]+[C,A]),\\
C_4&=&-\frac{1}{24}([[C,A],A]+3[[C,A],B]+3[[C,B],B]),\\
&\ldots&\nonumber
\end{eqnarray}
where $C=[A,B]$. In the light of the def\/inition (\ref{integ}) one can establish the following identif\/ication
\begin{eqnarray}
  A&\equiv&A(t)=-\frac{i}{\hslash}\int_{t_0}^t d\tau\mathfrak{H}(\hat{H})(\tau),\\
  B&\equiv&B(t)=-\frac{i}{\hslash}\int_{t_0}^t d\tau \mathfrak{A}(\hat{H})(\tau),
\end{eqnarray}
and therefore the commutator $C$ can be derived straightforwardly and rather easy. The result is
\begin{equation}
  C=-\dfrac{1}{\hslash^2}\int_{t_0}^t d\tau' \int_{t_0}^t d\tau''\mathfrak{C}\left(\tau',\tau''\right),
\end{equation}
where we have introduced the quantity
\begin{equation}
  \mathfrak{C}\left(\tau',\tau''\right)\equiv\left[\mathfrak{H}(\hat{H})(\tau'),\mathfrak{A}(\hat{H})(\tau'')\right],
\end{equation}
which can be computed with using of elementary algebra
\begin{eqnarray}
\!\!\!\!\!\!\!\!\!\!\!\!\!\!\!\!\!\!\!\!\!\!\!\!\!
\mathfrak{C}&=&\left(i\hslash\dfrac{\mu_R^{\pm}-\mu_L^{\pm}}{2}c^3\partial_i\right)\gamma^0\gamma^i\gamma^0\gamma^5+\left(\dfrac{(\mu_R^{\pm})^2-(\mu_L^{\pm})^2}{4}c^4\right)\gamma^0\gamma^0\gamma^5-\nonumber\\
\!\!\!\!\!\!\!\!\!\!\!\!\!\!\!\!\!\!\!\!\!\!\!\!\!
&-&\left(i\hslash\dfrac{\mu_R^{\pm}-\mu_L^{\pm}}{2}c^3\partial_i\right)\gamma^0\gamma^5\gamma^0\gamma^i-\left(\dfrac{(\mu_R^{\pm})^2-(\mu_L^{\pm})^2}{4}c^4\right)\gamma^0\gamma^5\gamma^0=\nonumber\\
\!\!\!\!\!\!\!\!\!\!\!\!\!\!\!\!\!\!\!\!\!\!\!\!\!
&=&2\left(i\hslash\dfrac{\mu_R^{\pm}-\mu_L^{\pm}}{2}c^3\partial_i\right)\gamma^0\gamma^i\gamma^0\gamma^5+2\left(\dfrac{(\mu_R^{\pm})^2-(\mu_L^{\pm})^2}{4}c^4\right)\gamma^0\gamma^0\gamma^5,
\end{eqnarray}
where we have applied the relations
\begin{eqnarray}
\gamma^0\gamma^5\gamma^0\gamma^i&=&-\gamma^0\gamma^i\gamma^0\gamma^5,\\ \gamma^0\gamma^5\gamma^0&=&-\gamma^0\gamma^0\gamma^5,
\end{eqnarray}
arising from the property of the $\gamma^5$ matrix $\left\{\gamma^5, \gamma^\mu\right\}=0$. Therefore, consequently one obtains f\/inally the result
\begin{equation}
  \mathfrak{C}(\tau',\tau'')=2\mathfrak{H}(\hat{H})(\tau')\mathfrak{A}(\hat{H})(\tau''),
\end{equation}
that leads to the equivalent statement - for arbitrary two times $\tau'$ and $\tau''$ the Poisson brackets\index{Poisson brackets} of the hermitian $\mathfrak{H}(\hat{H})(\tau')$ and the antihermitian $\mathfrak{A}(\hat{H})(\tau'')$ components of the full Hamiltonian (\ref{hamm}) is trivial
\begin{equation}
  \left\{\mathfrak{H}(\hat{H})(\tau'),\mathfrak{A}(\hat{H})(\tau'')\right\}=0.
\end{equation}
Naturally, by simple factorization one obtains also
\begin{equation}
  C=2AB\quad,\quad\{A,B\}=0,
\end{equation}
and consequently
\begin{eqnarray}
  \left[C,A\right]&=&CA,\\
  \left[C,B\right]&=&CB,\\
  \left[\left[C,A\right],A\right]&=&2\left[C,A\right]A,\\
  \left[\left[C,A\right],B\right]&=&2\left[C,A\right]B,\\
  \left[\left[C,B\right],A\right]&=&2\left[C,B\right]A,
\end{eqnarray}
and so on. In this manner the 4th order approximation of the formula (\ref{zassen}) in the present case has the form
\begin{eqnarray}
\exp\left\{A+B\right\}&\approx&\exp(A)\exp(B)\exp{C_2}\exp{C_3}\exp{C_4},\\
C_2&=&-\frac{1}{2}C,\\
C_3&=&-\frac{1}{6}(CA+2CB),\\
C_4&=&-\frac{1}{12}\left(CA^2+3CB^2+\dfrac{3}{2}C^2\right).
\end{eqnarray}
For the case of constant in time masses $\mu_R^{\pm}$ and $\mu_L^{\pm}$ one can determine the relations
\begin{eqnarray}
  A&=&\dfrac{i}{\hslash}(t-t_0)\left(-i\hslash c \gamma^i\partial_i+\dfrac{\mu_L^{\pm}+\mu_R^{\pm}}{2}c^2\right)\gamma^0,\\
  B&=&\dfrac{i(\mu_L^{\pm}-\mu_R^{\pm})c^2}{2\hslash}(t-t_0)\gamma^5\gamma^0,\\
  C&=&\dfrac{(\mu_L^{\pm}-\mu_R^{\pm})c^2}{\hslash^2}(t-t_0)^2\left(-i\hslash c\gamma^i\partial_i+\dfrac{\mu_L^{\pm}+\mu_R^{\pm}}{2}c^2\right)\gamma^5,
\end{eqnarray}
and consequently by elementary algebraic manipulations one establishes the Zassenhaus exponents\index{Zassenhaus exponents} as
\begin{eqnarray}
C_2&=&-\dfrac{(\mu_L^{\pm}-\mu_R^{\pm})c^2}{2\hslash^2}(t-t_0)^2\left(-i\hslash c\gamma^i\partial_i+\dfrac{\mu_L^{\pm}+\mu_R^{\pm}}{2}c^2\right)\gamma^5,\\
C_3&=&-\dfrac{i}{6\hslash^3}(\mu_L^{\pm}-\mu_R^{\pm})c^2(t-t_0)^3\left(-i\hslash c\gamma^i\partial_i+\dfrac{\mu_L^{\pm}+\mu_R^{\pm}}{2}c^2\right)\times\nonumber\\
&\times&\left[\left(-i\hslash c\gamma^i\partial_i+\dfrac{\mu_L^{\pm}+\mu_R^{\pm}}{2}c^2\right)\gamma^5+(\mu_L^{\pm}-\mu_R^{\pm})c^2\right]\gamma^0,\\
C_4&=&\dfrac{(\mu_L^{\pm}-\mu_R^{\pm})c^2}{12\hslash^4}(t-t_0)^4\left(-i\hslash c\gamma^i\partial_i+\dfrac{\mu_L^{\pm}+\mu_R^{\pm}}{2}c^2\right)\times\nonumber\\
&\times&\Bigg\{\left[\left(-i\hslash c\gamma^i\partial_i+\dfrac{\mu_L^{\pm}+\mu_R^{\pm}}{2}c^2\right)^2+3\left(\dfrac{\mu_L^{\pm}-\mu_R^{\pm}}{2}c^2\right)^2\right]\gamma^5+\nonumber\\
&+&3\dfrac{\mu_L^{\pm}-\mu_R^{\pm}}{2}c^2\left(-i\hslash c\gamma^i\partial_i+\dfrac{\mu_L^{\pm}+\mu_R^{\pm}}{2}c^2\right)\Bigg\}.
\end{eqnarray}

This approximation is suff\/icient to conclude the general properties of the procedure and the conclusions following from these features. The explicit form of the Zassenhaus exponents\index{Zassenhaus exponents} $C_n$ shows manifestly that the integrability problem formulated in terms of the Zassenhaus formula is not well def\/ined. Namely, the problem is that when the Hamilton operator is a sum of non-commuting antihermitian and hermitian components then also the Zassenhaus exponents\index{Zassenhaus exponents} $C_n$ obtain analogous legacy, i.e. are sums of two non-commuting operators. The fundamental stage, \emph{i.e.} the exponentiation procedure, must be applied once again, and therefore consequently in the next step of the procedure one meets the same property, i.e. sums of two non-commuting operators. In this manner the problem is cyclic and can not be solved in any approximation, while computation of full integration formula (\ref{zassen}) becomes the tremendous computational problem and its convergence is unclear. Therefore for the case the Schr\"odinger equation \index{Schr\"odinger equation}(\ref{scd}) with the Hamilton operator (\ref{ham}), such a recurrence integrability procedure based on the Zassenhaus exponents\index{Zassenhaus exponents} is not an algorithm what results in the conclusion that the quantum system is non integrable. By this reason one must construct any different integrability procedure having f\/inite number of steps and being an algorithm. For realization of such a construction let us formulate the integrability problem in a certain different form.

\section{The Integrability Problem for the massive Weyl equations}

For constructive solving the problem, let us consider another integrability procedure. Instead of the Dirac  equation leads us focus on the massive Weyl equations (\ref{neu1})-(\ref{neu2})\index{Weyl equation!massive}, which def\/ine the model of massive neutrinos\index{neutrino}. These two equations can be straightforwardly rewritten in the form of the effective two-component time evolution described by the Schr\"odinger equation \index{Schr\"odinger equation!two-component}
\begin{equation}\label{qd}
  i\hslash\partial_0\left[\begin{array}{c}\psi^\pm_R(x,t)\\\psi^\pm_L(x,t)\end{array}\right]=\hat{H}\left(\partial_i\right)\left[\begin{array}{c}\psi^\pm_R(x,t)\\\psi^\pm_L(x,t)\end{array}\right],
\end{equation}
where the Hamilton operator $\hat{H}$
\begin{equation}\label{ham2}
  \hat{H}=-\gamma^0\left(i\hslash c\gamma^i\partial_i+\left[\begin{array}{cc}\mu_R^{\pm}c^2&0\\0&\mu_L^{\pm}c^2\end{array}\right]\right),
\end{equation}
is manifestly hermitian and therefore the Schr\"odinger time evolution (\ref{qd}) is unitary. In this manner the integration procedure can be performed in the usual way well known from quantum mechanics.

Integrability of (\ref{qd}) is well def\/ined. The solutions are
\begin{equation}\label{solutionn}
  \left[\begin{array}{c}\psi^\pm_R(x,t)\\\psi^\pm_L(x,t)\end{array}\right]=U(t,t_0)\left[\begin{array}{c}\psi^\pm_R(x,t_0)\\\psi^\pm_L(x,t_0)\end{array}\right],
\end{equation}
where $U(t,t_0)$ is the unitary time-evolution operator, that for the constant masses is explicitly given by
\begin{equation}\label{uexp}
  U(t,t_0)=\exp\left\{-\dfrac{i}{\hslash}(t-t_0)\hat{H}\right\},
\end{equation}
and $\psi^\pm_{R,L}(x,t_0)$ are the initial time $t_0$ eigenstates with def\/ined momenta
\begin{equation}\label{peig}
  i\hslash \sigma^i\partial_i\psi^\pm_{R,L}(x,t_0)=p_{R,L}^{\pm~0}\psi^\pm_{R,L}(x,t_0),
\end{equation}
where the initial momenta ${p_{R}^\pm}^0$ and ${p_{L}^\pm}^0$ are related to the right-handed $\psi^\pm_{R}(x,t_0)$ and the left-handed $\psi^\pm_{L}(x,t_0)$ chiral f\/ields, respectively. The eigenequation (\ref{peig}), however, can be straightforwardly integrated and the result will be determining the spatial part of the evolution. The result can be presented in the symbolic form
\begin{equation}\label{psix}
\psi^\pm_{R,L}(x,t_0)=\exp\left\{-\dfrac{i}{\hslash}p_{R,L}^{\pm~0}(x-x_0)_i\sigma^i\right\}\psi^\pm_{R,L}(x_0,t_0),
\end{equation}
which after direct exponentiation leads to
\begin{eqnarray}\label{psix1}
\psi^\pm_{R,L}(x,t_0)=\left(\mathbf{1}_2\cos\eta-i\eta_i\sigma^i\dfrac{\sin\eta}{\eta}\right)\psi^\pm_{R,L}(x_0,t_0),
\end{eqnarray}
where $\eta=|\eta_i|$ and $\eta_i$ is the three dimensionless vector
\begin{equation}
  \eta_i=\dfrac{p_{R,L}^{\pm~0}}{\hslash}(x-x_0)_i.
\end{equation}

In the present situation the embarrassing problem which emerges in the integration procedure for the Dirac equation\index{Dirac equation}, discussed in the previous section, is absent. Now the Zassenhaus exponents\index{Zassenhaus exponents} are not troublesome because of, by def\/inition, the $\gamma^5$ matrix is included into the chiral Weyl f\/ields\index{chiral Weyl f\/ields}. Therefore the Hamilton operator (\ref{ham2}) is manifestly hermitian, and consequently the exponentiation (\ref{uexp}) can be straightforwardly performed by the standard method of quantum mechanics. At f\/irst glance, however, the mass matrix presence in the Hamilton operator (\ref{ham2}) causes that one can choose between at least two nonequivalent representations of the Dirac $\gamma$ matrices. On the one hand, the straightforward analogy to the Weyl equation\index{Weyl equation} suggests that the appropriate choice is the Weyl basis\index{Weyl basis}. Albeit, on the other hand, the Hamilton operator (\ref{ham2}) can be treated as the usual hermitian Dirac Hamiltonian\index{Dirac Hamiltonian}, and therefore consequently the Dirac basis\index{Dirac basis} would be the right representation for the Dirac $\gamma$ matrices. Other choices can be also applied, but they have no unambiguous justif\/ication because of the mass matrix presence in the Weyl equation\index{Weyl equation}. For example the Majorana basis\index{Majorana basis}, which is an adequate choice for the case of massless neutrino\index{neutrino}, is not an adequate choice for the case of massive neutrino. In this manner, in fact, one should not prefer the representation but rather consider both the chiral f\/ields and the time evolution operator (\ref{uexp}) in both the Weyl and the Dirac representations. Let us denote by superscript $r$ the chosen representation. By this reason the adequate labeling is
 \begin{eqnarray}
   U(t,t_0)&\rightarrow&U^{r}(t,t_0),\\
   \psi^\pm_{R,L}(x,t_0)&\rightarrow&(\psi^\pm_{R,L})^{r}(x,t_0),\\
   \psi^\pm_{R,L}(x_0,t_0)&\rightarrow&(\psi^\pm_{R,L})^{r}(x_0,t_0)
 \end{eqnarray}
where the superscript $r=D,W$ means that the quantities are taken in the Dirac and the Weyl basis\index{Dirac basis}\index{Weyl basis}, respectively. Interestingly, the eigenequation (\ref{peig}) is independent on the representation choice, and therefore the initial momenta $p_{R,L}^{\pm~0}$ of the chiral Weyl f\/ields\index{chiral Weyl f\/ields} are measurable. For full consistency, let us test both the representations.

\subsection{The Dirac basis}

The Dirac basis\index{Dirac basis} of the gamma matrices is def\/ined as
\begin{equation}\label{dirrep}
  \gamma^0=\left[\begin{array}{cc}I&0\\0&-I\end{array}\right]\quad,\quad\gamma^i=\left[\begin{array}{cc}0&\sigma^i\\-\sigma^i&0\end{array}\right]\quad,\quad\gamma^5=\left[\begin{array}{cc}0&I\\I&0\end{array}\right],
\end{equation}
where $I$ is the $2\times2$ unit matrix, and $\sigma^i=[\sigma_x,\sigma_y,\sigma_z]$ is a vector of the $2\times2$ Pauli matrices
\begin{equation}\label{pauli}
\sigma_x=\left[\begin{array}{cc}0&1\\1&0\end{array}\right]\quad,\quad\sigma_y=\left[\begin{array}{cc}0&-i\\i&0\end{array}\right]\quad,\quad\sigma_z=\left[\begin{array}{cc}1&0\\0&-1\end{array}\right].
\end{equation}
Application of the Dirac basis\index{Dirac basis} (\ref{dirrep}) allows to express the Hamilton operator (\ref{ham2}) as follows
\begin{equation}\label{ham2a}
  \hat{H}=\left[\begin{array}{cc}\mu_R^{\pm}&i\dfrac{\hslash}{c}\sigma^i\partial_i\\
  i\dfrac{\hslash}{c}\sigma^i\partial_i&-\mu_L^{\pm}\end{array}\right]c^2,
\end{equation}
and for the case of constant in time neutrinos masses yields a solution (\ref{solutionn}) with the unitary time evolution operator $U$
\begin{equation}
  U^D=\exp\left\{-i\dfrac{c^2}{\hslash}(t-t_0)\left[\begin{array}{cc}\mu_R^{\pm}&i\dfrac{\hslash}{c}\sigma^i\partial_i\\
  i\dfrac{\hslash}{c}\sigma^i\partial_i&-\mu_L^{\pm}\end{array}\right]\right\}.\label{U}
\end{equation}
Straightforward exponentiation in (\ref{U}) leads to the result
\begin{eqnarray}\label{dteo}
U^D&=&\Bigg\{\left[\begin{array}{cc}I&0\\0&I\end{array}\right]
  \cos\left[\dfrac{t-t_0}{\hslash}c^2\sqrt{\strut{\left(\dfrac{\mu_R^{\pm}+\mu_L^{\pm}}{2}\right)^2}+\left(i\dfrac{\hslash}{c}\sigma^i\partial_i\right)^2}\right]-\nonumber\\
  &-&i\left[\begin{array}{cc}\dfrac{\mu_L^{\pm}+\mu_R^{\pm}}{2}&i\dfrac{\hslash}{c}\sigma^i\partial_i\\i\dfrac{\hslash}{c}\sigma^i\partial_i&-\dfrac{\mu_L^{\pm}+\mu_R^{\pm}}{2}\end{array}\right]\times\nonumber\\
  &\times&\dfrac{\sin\left[\dfrac{t-t_0}{\hslash}c^2\sqrt{\strut{\left(\dfrac{\mu_R^{\pm}+\mu_L^{\pm}}{2}\right)^2+\left(i\dfrac{\hslash}{c}\sigma^i\partial_i\right)^2}}\right]}{\sqrt{\strut{\left(\dfrac{\mu_R^{\pm}+\mu_L^{\pm}}{2}\right)^2+\left(i\dfrac{\hslash}{c}\sigma^i\partial_i\right)^2}}}\Bigg\}\times\nonumber\\
  &\times&\exp\left\{-i\dfrac{(\mu_R^{\pm}-\mu_L^{\pm})c^2}{2\hslash}(t-t_0)\right\},
\end{eqnarray}
where we understand that all the functions are treated by the appropriate Taylor series\index{Taylor series} expansions.

\subsection{The Weyl basis}

As we have mentioned, however, application of the Weyl representation\index{Weyl basis} of the Dirac $\gamma$ matrices is also justif\/ied by theoretical reasons. Such a basis is def\/ined as follows
\begin{equation}\label{wrep}
  \gamma^0=\left[\begin{array}{cc}0&I\\I&0\end{array}\right]\quad,\quad\gamma^i=\left[\begin{array}{cc}0&\sigma^i\\-\sigma^i&0\end{array}\right]\quad,\quad\gamma^5=\left[\begin{array}{cc}-I&0\\0&I\end{array}\right],
\end{equation}
and the Hamilton operator (\ref{ham2}) in this representation has the form
\begin{equation}\label{ham2b}
  \hat{H}=\left[\begin{array}{cc}i\dfrac{\hslash}{c}\sigma^i\partial_i&-\mu_L^{\pm}\\
  -\mu_R^{\pm}&-i\dfrac{\hslash}{c}\sigma^i\partial_i\end{array}\right]c^2.
\end{equation}
Consequently, for the case of constant in time neutrinos masses one establishes the unitary time evolution operator
\begin{equation}
  U^W=\exp\left\{-i\dfrac{c^2}{\hslash}(t-t_0)\left[\begin{array}{cc}i\dfrac{\hslash}{c}\sigma^i\partial_i&-\mu_L^{\pm}\\
  -\mu_R^{\pm}&-i\dfrac{\hslash}{c}\sigma^i\partial_i\end{array}\right]\right\},\label{U1}
\end{equation}
which after straightforward exponentiation becomes
\begin{eqnarray}\label{wteo}
\!\!\!\!\!\!\!\!\!\!\!\!\!\!\!\!\!\!\!\!\!\!\!\!\!\!
U^W&=&\left[\begin{array}{cc}I&0\\0&I\end{array}\right]
  \cos\left[\dfrac{t-t_0}{\hslash}c^2\sqrt{\strut{\mu_L^{\pm}\mu_R^{\pm}+\left(i\dfrac{\hslash}{c}\sigma^i\partial_i\right)^2}}\right]-\nonumber\\
\!\!\!\!\!\!\!\!\!\!\!\!\!\!\!\!\!\!\!\!\!\!\!\!\!\!
&-&i\left[\begin{array}{cc}i\dfrac{\hslash}{c}\sigma^i\partial_i&-\mu_L^{\pm}\\-\mu_R^{\pm}&-i\dfrac{\hslash}{c}\sigma^i\partial_i\end{array}\right]\dfrac{\sin\left[\dfrac{t-t_0}{\hslash}c^2\sqrt{\strut{\mu_L^{\pm}\mu_R^{\pm}+\left(i\dfrac{\hslash}{c}\sigma^i\partial_i\right)^2}}\right]}{\sqrt{\strut{\mu_L^{\pm}\mu_R^{\pm}+\left(i\dfrac{\hslash}{c}\sigma^i\partial_i\right)^2}}}.
\end{eqnarray}

Evidently, the time evolution operator evaluated in the Weyl representation (\ref{wteo}) has meaningfully simpler form then the result of the evaluation performed in the Dirac basis (\ref{dteo}). In this manner results of the choices distinguishable and therefore are not physically equivalent, \emph{i.e.} will yield different solutions of the same equation. Albeit, it is not the strangest property. Namely, both the choices can be related to physics in different energetic regions, and therefore it is useful to solve the massive Weyl equation\index{Weyl equation!massive} in both the representations.

\subsection{The space-time evolution}

It must be emphasized that the results obtained in both the previous subsections are strictly related to the massive Weyl equations\index{Weyl equation!massive} presented in the form of the Schr\"odinger equation \index{Schr\"odinger equation}(\ref{qd}). Presently, one can straightforwardly apply these key results, \emph{i.e.} the momentum eigenequations (\ref{peig}), the spatial evolutions (\ref{psix}), and the evaluations of the unitary time evolution operators (\ref{dteo}) and (\ref{wteo}), to exact determination of the corresponding wave functions of the massive Weyl equation (\ref{qd}) in both the Dirac and the Weyl bases\index{Weyl basis}\index{Dirac basis} of the Dirac gamma matrices.

\subsubsection{Dirac-like solutions}

Let us derive f\/irst the wave functions in the Dirac basis\index{Dirac basis}. Employing the shortened notation
\begin{equation}
E^D({p_R^\pm}^0)\equiv c^2\sqrt{\strut{\left(\mu_\pm^D\right)^2+\left(\dfrac{{p_{R}^\pm}^0}{c}\right)^2}},
\end{equation}
where $\mu_\pm^D$ is the arithmetic mean of the masses of the neutrinos
\begin{equation}
{\mu_\pm^D}=\dfrac{\mu_R^{\pm}+\mu_L^{\pm}}{2},
\end{equation}
by elementary algebraic manipulations one receives the right-handed chiral Weyl f\/ields\index{chiral Weyl f\/ields}
\begin{eqnarray}\label{dir1s}
&&(\psi^\pm_R)^D(x,t)=\Bigg\{\Bigg[\cos\left[\dfrac{t-t_0}{\hslash}E^D({p_R^\pm}^0)\right]-\nonumber\\
&-&i\mu_{\pm}^Dc^2\dfrac{\sin\left[\dfrac{t-t_0}{\hslash}E^D({p_R^\pm}^0)\right]}{E^D({p_R^\pm}^0)}\Bigg]\exp\left\{-\dfrac{i}{\hslash}{p_{R}^\pm}^0(x-x_0)_i\sigma^i\right\}(\psi^\pm_{R})^{D}_0-\nonumber\\
&-&i{p_{L}^\pm}^0c\dfrac{\sin\left[\dfrac{t-t_0}{\hslash}E^D({p_L^\pm}^0)\right]}{E^D({p_L^\pm}^0)}\exp\left\{-\dfrac{i}{\hslash}{p_{L}^\pm}^0(x-x_0)_i\sigma^i\right\}(\psi^\pm_{L})^{D}_0\Bigg\}\times\nonumber\\
&\times&\exp\left\{-i\dfrac{(\mu_R^{\pm}-\mu_L^{\pm})c^2}{2\hslash}(t-t_0)\right\},
\end{eqnarray}
where $(\psi^\pm_{R,L})^{D}_0=(\psi^\pm_{R,L})^{D}(x_0,t_0)$. Similarly, the left-handed chiral Weyl f\/ields\index{chiral Weyl f\/ields} also can be also established in an exact way
\begin{eqnarray}\label{dir2s}
&&(\psi^\pm_L)^D(x,t)=\Bigg\{\Bigg[\cos\left[\dfrac{t-t_0}{\hslash}E^D({p_L^\pm}^0)\right]+\nonumber\\
&+&i\mu_{\pm}^Dc^2\dfrac{\sin\left[\dfrac{t-t_0}{\hslash}E^D({p_L^\pm}^0)\right]}{E^D({p_L^\pm}^0)}\Bigg]\exp\left\{-\dfrac{i}{\hslash}{p_{L}^\pm}^0(x-x_0)_i\sigma^i\right\}(\psi^\pm_{L})^{D}_0-\nonumber\\
&-&i{p_{R}^\pm}^0c\dfrac{\sin\left[\dfrac{t-t_0}{\hslash}E^D({p_R^\pm}^0)\right]}{E^D({p_R^\pm}^0)}\exp\left\{-\dfrac{i}{\hslash}{p_{R}^\pm}^0(x-x_0)_i\sigma^i\right\}(\psi^\pm_{R})^{D}_0\Bigg\}\times\nonumber\\
&\times&\exp\left\{-i\dfrac{(\mu_R^{\pm}-\mu_L^{\pm})c^2}{2\hslash}(t-t_0)\right\}.
\end{eqnarray}

\subsubsection{Weyl-like solutions}

Similar line of reasoning can be carried out for derivation of the wave functions in the Weyl basis\index{Weyl basis}. Employing the following shortened notation
\begin{equation}
  E^W({p_R^\pm}^0)\equiv c^2\sqrt{\strut{\left(\mu_{\pm}^W\right)^2+\left(\dfrac{{p_R^\pm}^0}{c}\right)^2}},
\end{equation}
where $\mu_\pm^W$ is the geometric mean of the masses of the neutrinos
\begin{equation}
  {\mu_\pm^W}=\sqrt{\strut{\mu_R^{\pm}\mu_L^{\pm}}},
\end{equation}
and performing elementary calculation one can deduce the right-handed chiral Weyl f\/ields\index{chiral Weyl f\/ields}
\begin{eqnarray}\label{weyl1s}
&&(\psi^\pm_R)^W(x,t)=\Bigg\{\cos\left[\dfrac{t-t_0}{\hslash}E^W({p_R^\pm}^0)\right]-\nonumber\\
&-&i{p^\pm_R}^0c\dfrac{\sin\left[\dfrac{t-t_0}{\hslash}E^W({p_R^\pm}^0)\right]}{E^W({p_R^\pm}^0)}\Bigg\}\exp\left\{-\dfrac{i}{\hslash}{p_{R}^\pm}^0(x-x_0)_i\sigma^i\right\}(\psi^\pm_{R})^{W}_0+\nonumber\\
&+&i\mu_L^{\pm}c^2\dfrac{\sin\left[\dfrac{t-t_0}{\hslash}E^W({p_L^\pm}^0)\right]}{E^W({p_L^\pm}^0)}\exp\left\{-\dfrac{i}{\hslash}{p_{L}^\pm}^0(x-x_0)_i\sigma^i\right\}(\psi^\pm_{L})^{W}_0,
\end{eqnarray}
where similarly as in the case of the Dirac-like solutions we have introduced $(\psi^\pm_{R,L})^{W}_0=(\psi^\pm_{R,L})^{W}(x_0,t_0)$. For the left-handed chiral Weyl f\/ields\index{chiral Weyl f\/ields} one obtains the formula
\begin{eqnarray}\label{weyl2s}
&&(\psi^\pm_L)^W(x,t)=\Bigg\{\cos\left[\dfrac{t-t_0}{\hslash}E^W({p_L^\pm}^0)\right]-\nonumber\\
&+&i{p^\pm_L}^0c\dfrac{\sin\left[\dfrac{t-t_0}{\hslash}E^W({p_L^\pm}^0)\right]}{E^W({p_L^\pm}^0)}\Bigg\}\exp\left\{-\dfrac{i}{\hslash}{p_{L}^\pm}^0(x-x_0)_i\sigma^i\right\}(\psi^\pm_{L})^{W}_0+\nonumber\\
&+&i\mu_R^{\pm}c^2\dfrac{\sin\left[\dfrac{t-t_0}{\hslash}E^W({p_R^\pm}^0)\right]}{E^W({p_R^\pm}^0)}\exp\left\{-\dfrac{i}{\hslash}{p_{R}^\pm}^0(x-x_0)_i\sigma^i\right\}(\psi^\pm_{R})^{W}_0.
\end{eqnarray}

In this manner one sees that the difference between obtained wave functions is crucial. Straightforward comparison of the Weyl-like solutions (\ref{weyl1s}) and (\ref{weyl2s}) with the Dirac-like solutions (\ref{dir1s}) and (\ref{dir2s}) shows that in the case of the Dirac basis\index{Dirac basis} there are different coeff\/icients of cosinuses and sinuses, and there is an additional time-exponent. Moreover, the functions $M^D({p_R^\pm}^0)$ and $M^W({p_R^\pm}^0)$ having the basic status for both the received solutions also have different form which manifestly depends on the choice of the Dirac representation of the $\gamma$ matrices. As we have suggested earlier, the difference is not a problem, because the solutions can be related to different regions of energy. Anyway, however, the validation of both the representations, and also other ones, should be verif\/ied by experimental data.

\subsection{Probability density. Normalization}

If one knows explicit form of the chiral Weyl f\/ields\index{chiral Weyl f\/ields} then, applying the Dirac basis\index{Dirac basis}, one can derive the usual Dirac f\/ields by the following procedure
\begin{equation}\label{pdirac}
  (\psi^\pm)^D=\left[\begin{array}{cc}\dfrac{(\psi^\pm_R)^D+(\psi^\pm_L)^D}{2}\mathbf{1}_2&\dfrac{(\psi^\pm_R)^D-(\psi^\pm_L)^D}{2}\mathbf{1}_2\\
  \dfrac{(\psi^\pm_R)^D-(\psi^\pm_L)^D}{2}\mathbf{1}_2&\dfrac{(\psi^\pm_R)^D+(\psi^\pm_L)^D}{2}\mathbf{1}_2\end{array}\right],
\end{equation}
where we have used the shortened notation $(\psi^\pm)^D=(\psi^\pm)^D(x,t)$, and $(\psi^\pm_{R,L})^D=(\psi^\pm_{R,L})^D(x,t)$. Similarly, employing the Weyl basis\index{Weyl basis}, the Dirac f\/ields can be determined as follows
\begin{equation}\label{pweyl}
  (\psi^\pm)^W=\left[\begin{array}{cc}(\psi^\pm_L)^W\mathbf{1}_2&\mathbf{0}_2\\
  \mathbf{0}_2&(\psi^\pm_R)^W\mathbf{1}_2\end{array}\right],
\end{equation}
where like in the case of the Dirac basis\index{Dirac basis} we have applied the shortened notation $(\psi^\pm)^W=(\psi^\pm)^W(x,t)$, and $(\psi^\pm_{R,L})^W=(\psi^\pm_{R,L})^W(x,t)$. It is evident now, that in general these two cases are different from physical, mathematical, and computational points of view. In this manner, if we consider the quantum mechanical probability density and its normalization, we are forced to relate the probability density revealing Lorentz invariance \index{Lorentz invariance} to the chosen representation
\begin{equation}\label{pd}
  \Omega^{D,W}\equiv(\bar{\psi}^\pm)^{D,W}(\psi^\pm)^{D,W},
\end{equation}
\begin{equation}\label{normal}
  \int d^3x \Omega^{D,W}= \mathbf{1}_4.
\end{equation}
Applying the Dirac f\/ield in the Dirac basis\index{Dirac basis} (\ref{pdirac}) by elementary derivation one can obtain
\begin{equation}
  \Omega^D=\left[\begin{array}{cc}\dfrac{(\bar{\psi}^\pm_R)^D(\psi^\pm_R)^D+(\bar{\psi}^\pm_L)^D(\psi^\pm_L)^D}{2}\mathbf{1}_2&\dfrac{(\bar{\psi}^\pm_R)^D(\psi^\pm_R)^D-(\bar{\psi}^\pm_L)^D(\psi^\pm_L)^D}{2}\mathbf{1}_2\\
  \dfrac{(\bar{\psi}^\pm_R)^D(\psi^\pm_R)^D-(\bar{\psi}^\pm_L)^D(\psi^\pm_L)^D}{2}\mathbf{1}_2&\dfrac{(\bar{\psi}^\pm_R)^D(\psi^\pm_R)^D+(\bar{\psi}^\pm_L)^D(\psi^\pm_L)^D}{2}\mathbf{1}_2\end{array}\right],
\end{equation}
and similarly the probability density (\ref{pd}) computed for the Dirac f\/ield in the Weyl basis\index{Weyl basis} (\ref{pweyl}) has the form
\begin{equation}
  \Omega^W=\left[\begin{array}{cc}(\bar{\psi}^\pm_R)^W(\psi^\pm_R)^W\mathbf{1}_2&\mathbf{0}_2\\
  \mathbf{0}_2&(\bar{\psi}^\pm_L)^W(\psi^\pm_L)^W\mathbf{1}_2\end{array}\right].
\end{equation}
Employing the normalization condition (\ref{normal}) in the Dirac representation one obtains the system of equations
\begin{equation}
\dfrac{1}{2}\left(\int d^3x(\bar{\psi}^\pm_R)^D(\psi^\pm_R)^D+\int d^3x(\bar{\psi}^\pm_L)^D(\psi^\pm_L)^D\right)=1,
\end{equation}
\begin{equation}
\dfrac{1}{2}\left(\int d^3x(\bar{\psi}^\pm_R)^D(\psi^\pm_R)^D-\int d^3x(\bar{\psi}^\pm_L)^D(\psi^\pm_L)^D\right)=0,
\end{equation}
which leads to
\begin{eqnarray}
  \int d^3x(\bar{\psi}^\pm_R)^D(\psi^\pm_R)^D&=&1,\label{dirnorm1}\\
  \int d^3x(\bar{\psi}^\pm_L)^D(\psi^\pm_L)^D&=&1.\label{dirnorm2}
\end{eqnarray}
In the case of Weyl representation one receives
\begin{eqnarray}
  \int d^3x(\bar{\psi}^\pm_R)^W(\psi^\pm_R)^W&=&1,\label{weylnorm1}\\
  \int d^3x(\bar{\psi}^\pm_L)^W(\psi^\pm_L)^W&=&1.\label{weylnorm2}
\end{eqnarray}
In this manner one sees straightforwardly that the normalization conditions (\ref{dirnorm1}), (\ref{dirnorm2}) and (\ref{weylnorm1}), (\ref{weylnorm2})) are the same
\begin{equation}\label{norma}
  \int d^3x(\bar{\psi}^\pm_{R,L})^{D,W}(x,t)(\psi^\pm_{R,L})^{D,W}(x,t)=1,
\end{equation}
i.e. are invariant with respect to the choice of the gamma matrices representations, what means that they are physical conditions. Using of the fact that full space-time evolution is determined as
\begin{eqnarray}
  &&(\psi^\pm_{R,L})^{D,W}(x,t)=U^{D,W}(t,t_0)(\psi^\pm_{R,L})^{D,W}(x,t_0),\\
  &&\left[U^{D,W}(t,t_0)\right]^\dagger U^{D,W}(t,t_0)=\mathbf{1}_2,
\end{eqnarray}
one f\/inds easily that
\begin{equation}\label{norma1}
  \int d^3x(\bar{\psi}^\pm_{R,L})^{D,W}(x,t_0)(\psi^\pm_{R,L})^{D,W}(x,t_0)=1.
\end{equation}
By using of the spatial evolution (\ref{psix1}) one obtains the relation
\begin{equation}\label{norma1}
  \left|(\psi^\pm_{R,L})^{D,W}(x_0,t_0)\right|^2\int d^3x\left(\mathbf{1}_2+\dfrac{(x-x_0)_i}{|x-x_0|}\Im\sigma^i\sin\left|2\dfrac{p_{R,L}^{\pm~0}}{\hslash}(x-x_0)_i\right|\right)=1,
\end{equation}
where $\Im{\sigma^i}=\dfrac{\sigma^i-\sigma^{i\dagger}}{2i}$ is a imaginary part of the vector $\sigma^i$. The decomposition $\sigma_i=[\sigma_x,0,\sigma_z]+i[0,-i\sigma_y,0]$ yields $\Im\sigma^i=[0,-i\sigma_y,0]$, and the equation (\ref{norma1}) becomes
\begin{equation}\label{norma2}
  \left|(\psi^\pm_{R,L})^{D,W}(x_0,t_0)\right|^2\int d^3x\left(\mathbf{1}_2-i\dfrac{(x-x_0)_y}{|x-x_0|}\sigma_y\sin\left|2\dfrac{p_{R,L}^{\pm~0}}{\hslash}(x-x_0)_i\right|\right)=1.
\end{equation}
Introducing the change of variables $(x-x_0)_i\rightarrow {x'}_i$ in the following way
\begin{equation}
  {x'}_i\equiv 2\dfrac{p_{R,L}^{\pm~0}}{\hslash}(x-x_0)_i,
\end{equation}
and the effective volume $V'$ due to the vector ${x'}_i$
\begin{equation}\label{volum}
  V'\mathbf{1}_2=\int d^3x'\left\{\mathbf{1}_2-i\sigma_y{x'}_y\dfrac{\sin|x'|}{|x'|}\right\},
\end{equation}
the equation (\ref{norma2}) can be rewritten in the form
\begin{equation}\label{norma2a}
  \left|(\psi^\pm_{R,L})^{D,W}(x_0,t_0)\right|^2\left(\dfrac{2p_{R,L}^{\pm~0}}{\hslash}\right)^3V'\mathbf{1}_2=\mathbf{1}_2,
\end{equation}
and therefore one obtains f\/inally
\begin{equation}
  (\psi^\pm_{R,L})^{D,W}(x_0,t_0)=\left(\dfrac{\hslash}{2p_{R,L}^{\pm~0}}\right)^{3/2}\dfrac{1}{\sqrt{V'}}\exp{i\theta_{\pm}},
\end{equation}
where $\theta_{\pm}$ are arbitrary constant phases. The volume (\ref{volum}) differs from the standard one by the presence of the extra axial (y) volume $V_y$
\begin{equation}
  V_y=-i\sigma_y\int d^3x' {x'}_y\dfrac{\sin|x'|}{|x'|},
\end{equation}
which is the axial effect and has nontrivial feature, namely
\begin{equation}
  V_y=\left\{\begin{array}{cc}0&\mathrm{on}~\mathrm{f\/inite}~\mathrm{symmetrical}~\mathrm{spaces}\\
  \infty&\mathrm{on}~\mathrm{inf\/inite}~\mathrm{symmetrical}~\mathrm{spaces}\\
  <\infty&\mathrm{on}~\mathrm{sections}~\mathrm{of}~\mathrm{symmetrical}~\mathrm{spaces}\end{array}\right. .
\end{equation}

Now one sees straightforwardly that the normalization is strictly speaking dependent on the choice of an appropriate region of integrability. For inf\/inite symmetric spatial regions such a normalization procedure is not well def\/ined, because of the axial volume effect is inf\/inite. However, one can study certain reasonable situations which consider solutions of the quantum theory on f\/inite symmetric spatial regions. Moreover, the problem of integrability is def\/ined with respect to the choice of the initial momentum of the chiral Weyl f\/ields\index{chiral Weyl f\/ields} $p_{R,L}^{\pm~0}$. In fact, there are many possible nonequivalent physical situations connected with the choice of a concrete initial momentum eigenvalue. In the next section we are going to discuss the one of such situations related to a f\/inite symmetric spatial region, the concrete example of the model of massive neutrinos, which in general was solved in the present section.

\section{The Ultra-Relativistic Massive Neutrinos}

As the f\/inal piece of this chapter let us consider the concrete application of the general model presented in the previous section, which is based on the normalization in a f\/inite symmetrical box and putting \emph{ad hoc} the eigenvalue of the initial momenta of the chiral Weyl f\/ields\index{chiral Weyl f\/ields} according to the ultra-relativistic limit\index{ultra-relativistic limit} of Special Relativity
\begin{equation}\label{choice}
  p_{R,L}^{\pm~0}=\mu_{R,L}^\pm c.
\end{equation}
For such a simplif\/ied situation the normalization discussed in the previous section leads to the following initial data condition
\begin{equation}
  (\psi^\pm_{R,L})^{D,W}(x_0,t_0)=\sqrt{\left(\dfrac{\hslash}{2\mu_{R,L}^{\pm}c}\right)^3\dfrac{1}{V'}}\exp{i\theta_{\pm}}
  =\sqrt{\pi^3\dfrac{\lambda^3_C(\mu_{R,L}^{\pm})}{V'}}\exp{i\theta_{\pm}},
\end{equation}
where $\lambda_C(\mu_{R,L}^{\pm})$ is the Compton wavelength\index{Compton wavelength} of the right- or left-handed neutrino\index{neutrino}. Because of normalization in the symmetrical box gives
\begin{equation}
  V'=V=\int d^3x,
\end{equation}
one obtains f\/inally
\begin{equation}
  (\psi^\pm_{R,L})^{D,W}(x_0,t_0)=\sqrt{\pi^3\dfrac{\lambda^3_C(\mu_{R,L}^{\pm})}{V}}\exp{i\theta_{\pm}}.
\end{equation}
When the theory is normalized in the region of the volume
\begin{equation}
V=\pi^3\lambda^3_C(\mu_{R,L}^{\pm}),
\end{equation}
then initial values of the chiral Weyl f\/ields\index{chiral Weyl f\/ields} determine the phase
\begin{equation}
(\psi^\pm_{R,L})^{D,W}(x_0,t_0)=\exp{i\theta_{\pm}}.
\end{equation}
Interestingly, when one considers the normalization symmetrical spaces for the left- and right-handed neutrino\index{neutrino} as a spheres of radiuses $R_{R,L}^\pm$ then the normalization radiuses are
\begin{equation}
  R_{R,L}^\pm=\dfrac{\sqrt{3}}{2}\pi^{2/3}\lambda_C(\mu_{R,L}^{\pm})\approx1.858\lambda_C(\mu_{R,L}^{\pm}).
\end{equation}
In this manner, if one can measure the normalization radius $R_{R,L}^\pm$ then the masses of the neutrinos can be established as
\begin{equation}
\mu_{R,L}^{\pm}=\dfrac{\sqrt{3}}{4\pi^{1/3}}\dfrac{\hslash}{c}\dfrac{1}{R_{R,L}^\pm}=\dfrac{\sqrt{3}}{4\pi^{1/3}}M_P\dfrac{\ell_P}{R_{R,L}^\pm},
\end{equation}
what can be approximated as
\begin{equation}
\mu_{R,L}^{\pm}\approx1.04\cdot10^{-43}\dfrac{1 kg\cdot 1 m}{R_{R,L}^\pm}=0.583\dfrac{1 \dfrac{eV}{c^2}\cdot 1 nm}{R_{R,L}^\pm}.
\end{equation}
If one wishes to establish the squared-mass difference\index{squared-mass difference}
\begin{equation}
  \Delta\mu_{LR}^2=
  \dfrac{3}{16\pi^{2/3}}\left(\dfrac{\hslash}{c}\right)^2\left[\dfrac{1}{\left(R_{L}^\pm\right)^2}-\dfrac{1}{\left(R_{R}^\pm\right)^2}\right],
\end{equation}
or with using of the Planck mass\index{Planck mass} and the Planck length\index{Planck length}
\begin{equation}
  \Delta\mu_{LR}^2=
  \dfrac{3M_P^2}{16\pi^{2/3}}\left[\left(\dfrac{\ell_P}{R_{L}^\pm}\right)^2-\left(\dfrac{\ell_P}{R_{R}^\pm}\right)^2\right].
\end{equation}
Because of the squared-mass difference is positive $\Delta\mu^2_{LR}>0$ one has
\begin{equation}
  R_{R}^\pm>R_{L}^\pm.
\end{equation}
Moreover, application of the relation (\ref{boom}) leads to the conclusion
\begin{equation}\label{lensmaker}
\dfrac{3}{512\pi^{5/3}}\left[\dfrac{1}{\left(R_{L}^\pm\right)^2/\ell_P}-\dfrac{1}{\left(R_{R}^\pm\right)^2/\ell_P}\right]=\dfrac{1}{\ell},
\end{equation}
which can be treated as the lensmaker's equation\index{lensmaker's equation}
\begin{equation}
  (n-1)\left[\dfrac{1}{R_1}-\dfrac{1}{R_2}+\dfrac{n-1}{n}\dfrac{d}{R_1R_2}\right]=\dfrac{1}{f},
\end{equation}
for the convergent lens of thickness $d$ small compared with the radiuses of curvature $R_1$ and $R_2$. The lens has the focal length identical to a minimal scale
\begin{equation}
  f=\ell,
\end{equation}
the radiuses of curvature strictly related to \emph{the normalization radiuses}\index{normalization radiuses}
\begin{eqnarray}
  R_1&=&\dfrac{\left(R_{L}^\pm\right)^2}{\ell_P},\\
  R_2&=&\dfrac{\left(R_{R}^\pm\right)^2}{\ell_P},
\end{eqnarray}
and refractive index
\begin{equation}
  n=1+\dfrac{3}{512\pi^{5/3}}\approx1.00087.
\end{equation}
Let us call such a lens \emph{the neutrino\index{neutrino} lens}\index{neutrino lens}. The equation (\ref{lensmaker}) expresses a minimal scale via the normalization radiuses\index{normalization radiuses}, i.e. if and only if $R_{R,L}^{\pm}$ are established by experimental data then $\ell$ is also established.

Let us consider such a normalization, i.e. the particular case of the general solutions which describes the situation of the neutrino lens\index{neutrino lens}. First let us derive the appropriate wave functions in the Dirac representation. Introducing the function
\begin{equation}
  E^D(x,y)\equiv c^2\sqrt{\strut{\left(\dfrac{x+y}{2}\right)^2}+x^2},
\end{equation}
the right-handed chiral Weyl f\/ields\index{chiral Weyl f\/ields} are
\begin{eqnarray}
  &&(\psi^\pm_R)^D(x,t)=\Bigg\{\Bigg[\cos\left[\dfrac{t-t_0}{\hslash}E^D(\mu_R^{\pm},\mu_L^{\pm})\right]-\nonumber\\
  &-&i\dfrac{\mu_{\pm}^Dc^2}{E^D(\mu_R^{\pm},\mu_L^{\pm})}\sin\left[\dfrac{t-t_0}{\hslash}E^D(\mu_R^{\pm},\mu_L^{\pm})\right]\Bigg]\exp\left\{-\dfrac{ic}{\hslash}\mu_R^{\pm}(x-x_0)_i\sigma^i\right\}-\nonumber\\
  &-&i\dfrac{\mu_L^{\pm}c^2}{E^D(\mu_L^{\pm},\mu_R^{\pm})}\sin\left[\dfrac{t-t_0}{\hslash}E^D(\mu_L^{\pm},\mu_R^{\pm})\right]\exp\left\{-\dfrac{ic}{\hslash}\mu_L^{\pm}(x-x_0)_i\sigma^i\right\}\Bigg\}\times\nonumber\\
  &\times&\exp\left\{i\left[\theta_{\pm}-\dfrac{(\mu_R^{\pm}-\mu_L^{\pm})c^2}{2\hslash}(t-t_0)]\right]\right\}.
\end{eqnarray}
while the left-handed chiral Weyl f\/ields\index{chiral Weyl f\/ields} are
\begin{eqnarray}
&&(\psi^\pm_L)^D(x,t)=\Bigg\{\Bigg[\cos\left[\dfrac{t-t_0}{\hslash}E^D(\mu_L^{\pm},\mu_R^{\pm})\right]+\nonumber\\
    &+&i\dfrac{\mu_{\pm}^Dc^2}{E^D(\mu_L^{\pm},\mu_R^{\pm})}\sin\left[\dfrac{t-t_0}{\hslash}E^D(\mu_L^{\pm},\mu_R^{\pm})\right]\Bigg]\exp\left\{-\dfrac{ic}{\hslash}\mu_{L}^\pm(x-x_0)_i\sigma^i\right\}-\nonumber\\
  &-&i\dfrac{\mu_R^{\pm}c^2}{E^D(\mu_R^{\pm},\mu_L^{\pm})}\sin\left[\dfrac{t-t_0}{\hslash}E^D(\mu_R^{\pm},\mu_L^{\pm})\right]\exp\left\{-\dfrac{ic}{\hslash}\mu_{R}^\pm(x-x_0)_i\sigma^i\right\}\Bigg\}\times\nonumber\\
  &\times&\exp\left\{i\left[\theta_{\pm}-\dfrac{(\mu_R^{\pm}-\mu_L^{\pm})c^2}{2\hslash}(t-t_0)]\right]\right\}.
\end{eqnarray}
Similarly, one can derive the appropriate wave functions in the Weyl representation. Introducing the function
\begin{equation}
  E^W(x,y)\equiv c^2\sqrt{xy+x^2},
\end{equation}
for the right-handed chiral Weyl f\/ields\index{chiral Weyl f\/ields} are
\begin{eqnarray}
\!\!\!\!\!\!\!\!\!\!\!\!\!\!\!&&(\psi^\pm_R)^W(x,t)=\exp{i\theta_{\pm}}\Bigg\{\Bigg[\cos\left[\dfrac{t-t_0}{\hslash}E^W(\mu_R^{\pm},\mu_L^{\pm})\right]-\nonumber~~~~~~\\
\!\!\!\!\!\!\!\!\!\!\!\!\!\!\!&-&\dfrac{i\mu_R^{\pm}c^2}{E^W(\mu_R^{\pm},\mu_L^{\pm})}\sin\left[\dfrac{t-t_0}{\hslash}E^W(\mu_R^{\pm},\mu_L^{\pm})\right]\Bigg]\exp\left\{-\dfrac{ic}{\hslash}\mu_{R}^\pm(x-x_0)_i\sigma^i\right\}+\nonumber~~~~~~\\
\!\!\!\!\!\!\!\!\!\!\!\!\!\!\!&+&\dfrac{i\mu_L^{\pm}c^2}{E^W(\mu_L^{\pm},\mu_R^{\pm})}\sin\left[\dfrac{t-t_0}{\hslash}E^W(\mu_L^{\pm},\mu_R^{\pm})\right]\exp\left\{-\dfrac{ic}{\hslash}\mu_{L}^\pm(x-x_0)_i\sigma^i\right\}\Bigg\},~~~~~~
\end{eqnarray}
while the left-handed chiral Weyl f\/ields\index{chiral Weyl f\/ields} are
\begin{eqnarray}
\!\!\!\!\!\!\!\!\!\!\!\!\!\!\!&&(\psi^\pm_L)^W(x,t)=\exp{i\theta_{\pm}}\Bigg\{\Bigg[\cos\left[\dfrac{t-t_0}{\hslash}E^W(\mu_L^{\pm},\mu_R^{\pm})\right]-\nonumber~~~~~~\\
\!\!\!\!\!\!\!\!\!\!\!\!\!\!\!&-&\dfrac{i\mu_L^{\pm}c^2}{E^W(\mu_L^{\pm},\mu_R^{\pm})}\sin\left[\dfrac{t-t_0}{\hslash}E^W(\mu_L^{\pm},\mu_R^{\pm})\right]\Bigg]\exp\left\{-\dfrac{ic}{\hslash}\mu_{L}^\pm(x-x_0)_i\sigma^i\right\}+\nonumber~~~~~~\\
\!\!\!\!\!\!\!\!\!\!\!\!\!\!\!&+&\dfrac{i\mu_R^{\pm}c^2}{E^W(\mu_R^{\pm},\mu_L^{\pm})}\sin\left[\dfrac{t-t_0}{\hslash}E^W(\mu_R^{\pm},\mu_L^{\pm})\right]\exp\left\{-\dfrac{ic}{\hslash}\mu_{R}^\pm(x-x_0)_i\sigma^i\right\}\Bigg\}.~~~~~~
\end{eqnarray}

The situation considered above is only the example following from the model of massive neutrinos based on the massive Weyl equations\index{Weyl equation!massive} (\ref{neu1})-(\ref{neu2}) obtained via application of the Snyder noncommutative geometry\index{Snyder noncommutative geometry} (\ref{ndA})-(\ref{ndB}). Because of such a physical situation is related to the ultra-relativistic limit\index{ultra-relativistic limit}, therefore this type of massive neutrinos we shall call \emph{ultra-relativistic massive neutrinos}\index{ultra-relativistic massive neutrinos}. There are many other possibilities for determination of the relation between the initial values of eigenmomentum $p_{R,L}^{\pm~0}$ and the masses $\mu_{R,L}^{\pm}$ of the right- and left-handed chiral Weyl f\/ields\index{chiral Weyl f\/ields} $\psi_{R,L}^\pm$. However, the choice (\ref{choice}) tested above presents the crucial reasonability which is the straightforward consequence of its special-relativistic character. Such a situation expresses validation of special equivalence principle for the initial space-time evolution of the massive neutrinos\index{neutrino}, \emph{i.e.} $E_{R,L}^\pm=\mu^\pm_{R,L}c^2=p_{R,L}^{\pm~0}c$. This case, however, is also nontrivial from the high energy physics point of view \cite{uhep}. Namely, it is strictly related to the ultra-high energy region, widely studied in the modern astrophysics (See \emph{e.g.} the Refs. \cite{maccione} and \cite{uheap} and suitable references therein). The presented particular space-time evolution describes physics of the massive neutrinos in such an energetic region, and therefore its results should be verif\/ied by experimental data due to ultra-high energy astrophysics.

\part{Quantum General Relativity}
\chapter{The Quantum Cosmology}
\section{Introduction}
Quantum cosmology is one of the most important research tendencies within modern theoretical gravitational physics. Its necessity is def\/ined by problems of formulation of the physics of early and very early stages of our Universe. Its essence, however, is the problem of construction of the model of the Universe, in which quantum theory and statistical mechanics meet and work together. Such a strategy is well known as the crucial aspect of string theory\index{string theory}, however, and by this reason an arbitrary model of quantum cosmology which is the linkage between the quantum physics and the statistical physics will be possessing characteristic features and properties of string theory\index{string theory}. Moreover, it is possible even that any such a model will be def\/ining new kinds of string theories. Albeit, because of very abstractive mathematical form of string theory\index{string theory}, the problem is how to build the model of quantum cosmology which could be manifestly presenting the value for phenomenology. In other words, the crucial problem in formulation of quantum cosmology is an experimental verif\/ication of any its model.

One of the most essential steps in the history of quantum cosmology was quantum geometrodynamics (QGD\index{quantum geometrodynamics (QGD)})\index{quantum geometrodynamics (QGD)}, called also quantum General Relativity or quantum gravity, which had the beginning in the works due to J.A. Wheeler and B.S. DeWitt. Standardly, in such a strategy the primary canonical quantization \index{quantization!primary canonical}procedure is applied to arbitrary canonical formulation of classical theory of gravitation, General Relativity (GR). The most spread canonical formulation of General Relativity is the Hamiltonian approach which started by early works of P.A.M. Dirac and obtained a f\/inal and commonly accepted appearance in the works due to R. Arnowitt, S. Deser, and C.W. Misner. It is called the Arnowitt--Deser--Misner decomposition\index{Arnowitt--Deser--Misner decomposition} or $3+1$ splitting\index{$3+1$ splitting}.

In this part of this book we shall discuss certain both particular consequences as well as development of the strategy of quantum gravity based on the Wheeler--DeWitt equation \index{Wheeler--DeWitt equation}. The main purpose of this chapter is, however, the construction of such a model of quantum Universe, which will be leading to plausible theoretical predictions possessing both clearly def\/ined physical sense as well as the possibility of verif\/ication by comparison of its results with experimental data due to observations of the physical Universe. Applying the well-known epistemological justif\/ications of quantum theory, we propose to describe quantum cosmology as quantum f\/ield theory formulated in terms of the Fock space\index{Fock space} of creation and annihilation operators. The Fock space\index{Fock space} formalism has exceptionally well established phenomenological meaning for quantum physics. In this manner, the basic problem is the constructive application of the method of secondary quantization\index{quantization!secondary canonical}, which would be resulting in phenomenological deductions for quantum cosmology. The natural consequence of the formalism of the Fock rep\`{e}re\index{Fock rep\`{e}re} is the possibility for straightforward construction of statistical mechanics of quantum states of Universe. Moreover, the Fock space formalism applied in the context of quantum cosmology leads to natural def\/inition of Multiverse\index{Multiverse} as the collection of quantum universes.

By this reason we shall employ the quantum f\/ield theoretical reasoning for quantum cosmology. As the classical model of the Universe we shall consider the conformal-f\/lat metric f\/irst derived by A.A. Friedmann, and applied in the cosmological context by G. Lema\^{i}tre, H.P. Robertson, and A.G. Walker, which recently has been obtained the fundamental status for a number of models of observed Universe. We shall show that straightforward application of the Hamiltonian approach to General Relativity for such a concrete solution of the Einstein f\/ield equations \index{Einstein f\/ield equations}results in the Hamiltonian constraint\index{Hamiltonian constraint} corresponding to the fundamental state of a bosonic string.

This chapter is not too extensive. First, in the section B we shall sketch very shortly the classical model of Universe described by the conformal-f\/lat Friedmann--Lema\^{i}tre--Robertson--Walker metric \index{Friedmann--Lema\^{i}tre--Robertson--Walker metric}for the case of spatially f\/inite volume, which in this book is called the Einstein--Friedmann Universe\index{Einstein--Friedmann Universe}, possessing the cylindrical or toroidal topology corresponding to open and closed strings, respectively. We present the Hamiltonian approach which connects both the Dirac approach and the Arnowitt--Deser--Misner Hamiltonian formulation of General Relativity\index{Hamiltonian formulation of General Relativity!Arnowitt--Deser--Misner (ADM)}, in which there is splitting of evolution of four-dimensional space-time into the dynamics of three-dimensional space in one-dimensional time. There is received the Hamiltonian constraint\index{Hamiltonian constraint}, and the Hubble law\index{Hubble law} is obtained due to its resolution. Moreover, the dynamics of the model is identif\/ied with the dynamics of boson having negative squared-mass, i.e. with the tachyon\index{tachyon}. In the section C we perform the primary canonical quantization \index{quantization!primary canonical}procedure with respect to the classical model. By the primary quantization of the Hamiltonian constraint\index{Hamiltonian constraint} we receive the appropriate Wheeler--DeWitt equation\index{Wheeler--DeWitt equation}. Applying the separation of variables based on the Hamilton equations of motion we make reduction of order of such an evolution, and therefore the appropriate one-dimensional Dirac equation\index{Dirac equation!one-dimensional} for the Universe is deduced. Employing the method of secondary quantization \index{quantization!secondary canonical}to such a reduced evolution results in the quantum f\/ield theory formulated in the Fock space\index{Fock space}. Application of the appropriate Bogoliubov transformation\index{Bogoliubov transformation} and the Heisenberg equations of motion\index{Heisenberg equations of motion} leads to the diagonalization of the equations of motion to its canonical form. In this manner, we receive the static Fock rep\`{e}re\index{Fock rep\`{e}re} and quantization of the cosmological model is formulated in terms of the monodromy in the Fock space\index{Fock space}. In the section D we shall construct the statistical mechanics, especially thermodynamics, of quantum states of the Universe. Computations are performed according to the Bose--Einstein statistics \index{Bose--Einstein statistics}in frames of the method of density matrix\index{density matrix}. We apply the approximation in which the quantum Universe is a system with one degenerated state, i.e. one-particle approximation\index{one-particle approximation}. The section E is devoted to discussion of the particular thermodynamical situation of the system of many quantum Universe, i.e. the early light Multiverse \index{Multiverse!early light}corresponding to a cosmological radiation, which is characterized by minimal entropy. In the f\/inal section F we summarize brief\/ly the results of this chapter and present the perspective for further development of the proposed approach to quantum cosmology.

\section{The Classical Universe}

General Relativity based on the Einstein f\/ield equations \index{Einstein f\/ield equations}\cite{ein1}
\begin{equation}
	R_{\mu\nu}-\dfrac{1}{2}g_{\mu\nu}R=\kappa{T}_{\mu\nu},\label{efe1}
\end{equation}
where $\kappa=\dfrac{8\pi G}{c^4}$ is constant, $R_{\mu\nu}$ is the Ricci curvature tensor\index{Ricci curvature tensor}, $R$ is the Ricci scalar curvature\index{Ricci scalar curvature}, and $T_{\mu\nu}$ is the stress-energy tensor\index{stress-energy tensor}, is commonly accepted as the classical theory of gravitation describing the evolution of a metric tensor $g_{\mu\nu}$ of a four-dimensional Riemannian space-time manifold\index{space-time manifold} $M$ \cite{rie,car}. The f\/ield equations (\ref{efe1}) can be generated as the Euler--Lagrange\index{Euler--Lagrange equations of motion} equations of motion obtained via the Hilbert--Palatini action principle \cite{hil,pal} $\delta S_{EH}=0$ with respect to the fundamental f\/ield $g_{\mu\nu}$\index{Hilbert--Palatini action principle} applied to the Einstein--Hilbert action\index{Einstein--Hilbert action}
\begin{equation} \textit{S}_{\textrm{EH}}=\dfrac{1}{c}\int_Md^4x\sqrt{-g}\left(-\dfrac{1}{2\kappa\ell_P^2}R+\mathcal{L}_M\right),\label{eh}
\end{equation}
where $\ell_P$ is the Planck length\index{Planck length}, and the constant multiplier $\dfrac{1}{\ell_P^2}$ in the geometric action arises from dimensional correctness of the action, which should be $[E]\cdot[T]$. In the action (\ref{eh}) we denoted $g=\det g_{\mu\nu}$, and $\mathcal{L}_M$ denotes the Lagrangian density of Matter f\/ields\index{Matter f\/ields}. The variational principle allows to express the stress-energy tensor\index{stress-energy tensor} of Matter f\/ields\index{Matter f\/ields} via the Lagrangian density of Matter f\/ields\index{Matter f\/ields}
\begin{eqnarray}
T_{\mu\nu}=\dfrac{2}{\sqrt{-g}}\dfrac{\delta\left( \sqrt{-g}\mathcal{L}_M\right)}{\delta g^{\mu\nu}}\label{eh2}.
\end{eqnarray}

Let us consider the exact solution of the Einstein f\/ield equations\index{Einstein f\/ield equations} (\ref{efe1}) f\/irst derived by Friedmann, and studied in extensive cosmological context by Lema\^{i}tre, Robertson, and Walker \cite{frw}, for which the space-time interval has the following form
\begin{equation}\label{ef}
	ds^2=g_{\mu\nu}dx^{\mu}dx^{\nu}=-(dx^0)^2+a^2(x^0)\delta_{ij}dx^idx^j,
\end{equation}
where $a(x^0)$ is the cosmic scale factor parameter\index{cosmic scale factor parameter}, and $x^\mu$ $(\mu=0,1,2,3)$ is a Cartesian system of space-time coordinates in which the time coordinate $x^0$ is the object of the diffeomorphisms \cite{ein2,zel}
\begin{equation}
	x^0\rightarrow {x'}^0=x'(x^0).
\end{equation}
When volume of space is f\/inite
 \begin{equation}
 V = \int dx^1dx^2dx^3 < \infty,
 \end{equation}
then the interval (\ref{ef}) describes the Einstein--Friedmann Universe \index{Einstein--Friedmann Universe}possessing a topology related to f\/inite space. Such a cosmology is very far from triviality, because of the topological structure of f\/inite space relates the model to considerations of string theory\index{string theory}. Finite space is associated with cylindrical, toroidal, but also with more general stringy topologies and orbifolds \cite{orbifold}\index{orbifolds}. In other words, the Einstein--Friedmann Universe \index{Einstein--Friedmann Universe}having f\/inite space looks like to be a topological string\index{topological string}. However, such complicated theoretical situations shall not be developed in this book. Also cylindrical topology is far from our present considerations, because of presence of the boundaries in general can be resulting in very nontrivial physical consequences. The f\/inite space possessing toroidal topology has no boundary because of the ends of the cylinder are identif\/ied each other. By this reason, for simplicity of the cosmological model, in this book we shall study the case of the Einstein--Friedmann Universe \index{Einstein--Friedmann Universe!toroidal topology}equipped with the toroidal topology. Such a choice is in itself non trivial, because of the toroidal topology can be naturally identif\/ied with closed strings. In this manner, we shall consider the cosmological model possessing more general signif\/icance, which also can be extended on other topologies mentioned above.

Let us consider the Friedmann--Lema\^{i}tre--Robertson--Walker metric\index{Friedmann--Lema\^{i}tre--Robertson--Walker metric} (\ref{ef}) in frames of the Hamiltonian approach jointing the Dirac approach \cite{dir} and the Arnowitt--Deser--Misner Hamiltonian formulation of General Relativity\index{Hamiltonian formulation of General Relativity!Arnowitt--Deser--Misner (ADM)} \cite{adm}. In this chapter we shall not present the detailed analysis and explanations which can be found in more specialized student textbooks (See e.g. the Ref. \cite{poisson}), and shall be discussed in further part of this book. The Arnowitt--Deser--Misner decomposition\index{Arnowitt--Deser--Misner decomposition} of any metric satisfying the Einstein f\/ield equation, in a given coordinate system, has the form
\begin{equation}\label{adm}
	g_{\mu\nu}=\left[\begin{array}{cc}-N^2+N_iN^i&N^i\\
	N^j&h_{ij}\end{array}\right],
\end{equation}
where $N$ is the lapse function, $N_i$ is the shift vector, $h_{ij}$ is the metric of 3-dimensional space embedded in the 4-dimensional space-time, and $N^j=h^{ij}N_i$. The classical Universe described by the Friedmann--Lema\^{i}tre--Robertson--Walker metric\index{Friedmann--Lema\^{i}tre--Robertson--Walker metric} (\ref{ef}) can be parametrized by
\begin{eqnarray}
N^2&=&1,\\
N_i&=&[0,0,0],\\
h_{ij}&=&a^2(x^0)\delta_{ij}.
\end{eqnarray}
According to the strategy propagated by Dirac, the lapse function should be preserved explicitly but the shift vector, because of its trivialization, becomes absent. Moreover, at the end of calculations the lapse function should be putted explicitly. In this manner, the Hamiltonian approach involving the Dirac and the ADM formulations of General Relativity\index{Hamiltonian formulation of General Relativity!Arnowitt--Deser--Misner (ADM)} allows to present the Friedmann--Lema\^{i}tre--Robertson--Walker metric \index{Friedmann--Lema\^{i}tre--Robertson--Walker metric}in the following form
\begin{equation}\label{dadm}
	g_{\mu\nu}=\left[\begin{array}{cc}-N^2&0\\
	0&a^2(x^0)\delta_{ij}\end{array}\right],
\end{equation}
what is associated with the space-time interval
\begin{equation}\label{sti1}
	ds^2=-N^2(dx^0)^2+a^2(x^0)dx^idx^j.
\end{equation}
The diffeomorphism
\begin{equation}
ct=c\tau+x^0,\label{ct1}
\end{equation}
where $\tau$ is a reference constant, allows to transform the cosmological time $t$ to the conformal time\index{conformal time} $\eta$ on the level of the integral measure
\begin{equation}
d\eta=N(x^0)dx^0\equiv\dfrac{dt}{a(t)}\label{ct2},
\end{equation}
what allows to rewrite the interval (\ref{sti1}) in the following form
\begin{equation}
ds^2=a^2(\eta)\left[-c^2d\eta^2+\delta_{ij}dx^idx^j\right].
\end{equation}
It means that, in the coordinate system $(c\eta,x^i)$, the Universe is described by the scaled Minkowski space-time\index{space-time!Minkowski}
\begin{eqnarray}
g_{\mu\nu}&=&\Omega(\eta)\eta_{\mu\nu},
\end{eqnarray}
where the scaling function $\Omega(\eta)$ is
\begin{equation}
  \Omega(\eta)=a^2(\eta).\label{suncalif}
\end{equation}
 Derivation of the Christoffel symbols\index{Christoffel symbols} for the metric (\ref{dadm}),
\begin{equation}
	\Gamma^{\rho}_{\mu\nu}=\frac{1}{2}g^{\rho\sigma}
	\left(\partial_{\nu}g_{\mu\sigma}
	+\partial_{\mu}g_{\sigma\nu}-\partial_{\sigma}g_{\mu\nu}\right),
	\label{christ}
\end{equation}
leads to the nontrivial components
\begin{eqnarray}
	\Gamma^{0}_{00}&=&\frac{\dot{N}}{N},\\
	\Gamma^{0}_{ii}&=&\frac{a\dot{a}}{N^{2}},\\
	\Gamma^{i}_{i0}&=&\frac{\dot{a}}{a},
\end{eqnarray}
where dot means $x^0$-differentiation, and consequently the Ricci curvature tensor\index{Ricci curvature tensor}
\begin{eqnarray}
	R_{\mu\nu}=\partial_{\alpha}\Gamma^{\alpha}_{\mu\nu}
	-\partial_{\nu}\Gamma^{\alpha}_{\mu\alpha}
	+\Gamma^{\alpha}_{\beta\alpha}\Gamma^{\beta}_{\mu\nu}
	-\Gamma^{\alpha}_{\beta\nu}\Gamma^{\beta}_{\mu\alpha},\label{2.3}
\end{eqnarray}
takes the following form
\begin{equation}\label{3.6}
	R_{\mu\nu}\!=\!\left[\begin{array}{cccc}-3\left(\dfrac{\ddot{a}}{a}
	-\dfrac{\dot{a}}{a}\dfrac{\dot{N}}{N}\right)\!\!\!&\!\!\!
	\mathbf{0}^{\mathrm{T}}\\
	\mathbf{0}\!\!\!&\!\!\!\dfrac{a^2}{N^2}
	\left(\dfrac{\ddot{a}}{a}-2\left(\dfrac{\dot{a}}{a}\right)^{2}
	-\dfrac{\dot{a}}{a}\dfrac{\dot{N}}{N}\right)\delta_{ij}
	\end{array}\right]\!\!.
\end{equation}
 By straightforward calculation of the contravariant metric components and taking into account the Ricci curvature tensor\index{Ricci curvature tensor} (\ref{3.6}), one receives the Ricci scalar curvature\index{Ricci scalar curvature}
\begin{equation}
  	R=g^{\mu\nu}R_{\mu\nu}=\dfrac{6}{N^2}\left[\dfrac{\ddot{a}}{a}
  	-\left(\dfrac{\dot{a}}{a}\right)^2-\dfrac{\dot{a}}{a}
	\dfrac{\dot{N}}{N}\right],
\end{equation}
what allows to establish evaluation of the Einstein--Hilbert action\index{Einstein--Hilbert action} (\ref{eh}) for the classical Universe. Let us present the procedure in some detail. First, let us consider the geometric part of the Einstein--Hilbert action\index{Einstein--Hilbert action}, i.e.
 \begin{eqnarray}
 \!\!\!\!\!\!\!\!\!\!\!\!\!\!\!\!\!\!\!\!
\int_Md^4x\sqrt{-g}R&=&\int_Md^3x\int{dx^0}Na^3\dfrac{6}{N^2}\left[\dfrac{\ddot{a}}{a}-\left(\dfrac{\dot{a}}{a}\right)^2-\dfrac{\dot{a}}{a}\dfrac{\dot{N}}{N}\right]=\\
 \!\!\!\!\!\!\!\!\!\!\!\!\!\!\!\!\!\!\!\!
&=&6V\int{dx^0}\dfrac{\ddot{a}a^2N-\dot{a}^2aN-\dot{a}a^2\dot{N}}{N^2},
 \end{eqnarray}
where $V=\int{d^3x}$ is volume of space. Applying the total derivative
\begin{equation}
\dfrac{d}{dx^0}\left(\dfrac{\dot{a}a^2}{N}\right)=\dfrac{\ddot{a}a^2N+\dot{a}^2aN+\dot{a}a^2\dot{N}}{N^2},
\end{equation}
one obtains
 \begin{eqnarray}
\int_Md^4x\sqrt{-g}R&=&6V\int{dx^0}\left[\dfrac{d}{dx^0}\left(\dfrac{\dot{a}a^2}{N}\right)-\dfrac{3\dot{a}^2a}{N}\right]=\\
&=&-18V\int{dx^0}\dfrac{\dot{a}^2a}{N},
 \end{eqnarray}
where the boundary term was omitted as vanishing. Because of the relation holds
 \begin{equation}
   \dot{a}=\dfrac{1}{c}\dfrac{a'}{a},
 \end{equation}
where prime means $\eta$-differentiation, one has
\begin{equation}
\dot{a}^2a=\dfrac{1}{c^2}\dfrac{a'^2}{a},
\end{equation}
and because of $dx^0=cdt$ one obtains f\/inally
\begin{equation}
\int_Md^4x\sqrt{-g}R=-\dfrac{18V}{c}\int{dt}\dfrac{1}{N}\dfrac{a'^2}{a}=-\dfrac{18V}{c}\int{d\eta}\dfrac{a'^2}{N}.
\end{equation}

Let us consider now the action of Matter f\/ields\index{Matter f\/ields}
\begin{eqnarray}
\!\!\!\!\!\!\!\!\!\!\!\!\!\!\!\!\!\!\!\!\!\!\!\!\!\!\!\!\!\!
  \int_Md^4x\sqrt{-g}\mathcal{L}_M&=&\int{d^3x}\int{dx^0}Na^3\mathcal{L}_M=c\int{d^3x}\int{dt}Na^3\mathcal{L}_M=\\
\!\!\!\!\!\!\!\!\!\!\!\!\!\!\!\!\!\!\!\!\!\!\!\!\!\!\!\!\!\!
&=&c\int{d^3x}\int{d\eta}Na^4\mathcal{L}_M=c\int{d\eta}Na^4\int{d^3x}\mathcal{L}_M.
\end{eqnarray}
Let us take into account the Legendre transformation\index{Legendre transformation}
\begin{equation}
  \mathcal{L}_M=p_F\dot{F}-\mathcal{H}_M,
\end{equation}
where $F$ is a Matter f\/ield (for collection of Matter f\/ields\index{Matter f\/ields} one has a sum over f\/ields), $p_F=\dfrac{\partial\mathcal{L}}{\partial\dot{F}}$ is the conjugate momentum to $F$, and $\mathcal{H}_M$ is Hamiltonian density of Matter f\/ields\index{Matter f\/ields}. Taking into account the def\/inition of the conjugate momentum one can write
\begin{equation}
  p_F\dot{F}=\dot{F}\dfrac{\partial\mathcal{L}_M}{\partial\dot{F}}=\dfrac{\partial}{\partial\dot{F}}\left(\dot{F}\mathcal{L}_M\right)-\mathcal{L}_M,
\end{equation}
and therefore one receives the relation
\begin{equation}
  \mathcal{L}_M=p_F\dot{F}-\mathcal{H}_M=\dfrac{\partial}{\partial\dot{F}}\left(\dot{F}\mathcal{L}_M\right)-\mathcal{L}_M-\mathcal{H}_M,
\end{equation}
which allows to establish the Lagrangian density of Matter f\/ields\index{Matter f\/ields} in the form
\begin{equation}
  \mathcal{L}_M=\dfrac{1}{2}\dfrac{\partial}{\partial\dot{F}}\left(\dot{F}\mathcal{L}_M\right)-\dfrac{1}{2}\mathcal{H}_M.
\end{equation}
In this manner one can determine the volume integral of the Lagrangian density of Matter f\/ields\index{Matter f\/ields}
\begin{equation}
\int{d^3x}\mathcal{L}_M=\dfrac{1}{2}\int{d^3x}\dfrac{\partial}{\partial\dot{F}}\left(\dot{F}\mathcal{L}_M\right)-\dfrac{1}{2}\int{d^3x}\mathcal{H}_M.
\end{equation}
Let us consider the f\/irst term of this formula. It is evidently seen that the integrand is a total derivative, and applying the chain rule of differentiation\index{chain rule of differentiation} one can write the integrand as
\begin{equation}
\dfrac{\partial}{\partial\dot{F}}\left(\dot{F}\mathcal{L}_M\right)=\dfrac{\partial{V}}{\partial\dot{F}}\dfrac{\partial}{\partial{V}}\left(\dot{F}\mathcal{L}_M\right).
\end{equation}
However, in the light of the fact $dV=d^3x$ the derivative identically vanishes $\dfrac{\partial{V}}{\partial\dot{F}}=0$, so that in such a situation one has
\begin{equation}
  \int{d^3x}\dfrac{\partial}{\partial\dot{F}}\left(\dot{F}\mathcal{L}_M\right)=\dfrac{\partial{V}}{\partial\dot{F}}\int{dV}\dfrac{\partial}{\partial{V}}\left(\dot{F}\mathcal{L}_M\right)=\dfrac{\partial{V}}{\partial\dot{F}}\left(\dot{F}\mathcal{L}_M\right)=0,
\end{equation}
and therefore the volume integral of the Lagrangian density of Matter f\/ields\index{Matter f\/ields} is much more simplif\/ied
\begin{equation}
\int{d^3x}\mathcal{L}_M=-\dfrac{1}{2}\int{d^3x}\mathcal{H}_M.
\end{equation}
In this manner, f\/inally the action of Matter f\/ields\index{Matter f\/ields} can be entirely expressed via the only Hamiltonian density of Matter f\/ields\index{Matter f\/ields}
\begin{eqnarray}
  \int_Md^4x\sqrt{-g}\mathcal{L}_M&=&-\dfrac{c}{2}\int{d\eta}Na^4\int{d^3x}\mathcal{H}_M.
\end{eqnarray}
Collecting all these facts one can establish evaluation of the Einstein--Hilbert action on the Friedmann--Lema\^{i}tre--Robertson--Walker metric\index{Friedmann--Lema\^{i}tre--Robertson--Walker metric!Einstein--Hilbert action}. Interestingly, in such a situation the Einstein--Hilbert action has the Hamilton form in the conformal time\index{conformal time}, i.e.
\begin{equation}\label{action1}
S_{EH}=\int{d\eta}\mathrm{L}(a,a',\eta),
\end{equation}
and the Lagrangian $\mathrm{L}(a,a',\eta)$ of the cosmological model has the form
\begin{equation}\label{lagcosm}
  \mathrm{L}(a,a',\eta)=\dfrac{3}{2}M_P\ell_P^2\dfrac{V}{V_P}\dfrac{a'^2}{N}-\dfrac{1}{2}Na^4\mathrm{H}_{M}(\eta).
\end{equation}
where $V_P=\dfrac{4}{3}\pi\ell_P^3$ is the volume of the Planck sphere\index{Planck sphere}, and $\mathrm{H}_{M}(\eta)$ is the volume integral of the Hamiltonian density of Matter f\/ields\index{Matter f\/ields}
\begin{equation}
   \mathrm{H}_{M}(\eta)=\int d^{3}x\mathcal{H}_{M}(x,\eta),
\end{equation}
which has a natural interpretation of energy of Matter f\/ields\index{Matter f\/ields}.

The momentum conjugated to the cosmic scale factor parameter\index{cosmic scale factor parameter!momnetum conjugated to} can be derived straightforwardly from the reduced action (\ref{action1})
\begin{equation}\label{3.6}
\mathrm{P}_{a}=\dfrac{1}{\ell_P}\dfrac{\partial\mathrm{L}}{\partial(a')}=3M_P\ell_P\dfrac{V}{V_P}\dfrac{a'}{N},
\end{equation}
where the factor $\dfrac{1}{\ell_P}$ was putted \emph{ad hoc} for dimensional correctness of the momentum, which should be $[E]\cdot[T]\cdot[L]^{-1}$. It can be seen straightforwardly that by application of the Legendre transformation\index{Legendre transformation} to the reduced action (\ref{action1}) leads to
\begin{equation}\label{3.7}
S_{EH}=\int d\eta\left\{\ell_P\mathrm{P}_{a}a'-N\mathrm{H}(\eta)\right\},
\end{equation}
where $\mathrm{H}(\eta)$ is the Hamilton function of the cosmological model
\begin{equation}\label{hamcosm} \mathrm{H}(\eta)=\dfrac{V_P}{3V}\dfrac{\mathrm{P}_{a}^{2}}{2M_P}+\dfrac{1}{2}a^{4}\mathrm{H}_{M}(\eta).
\end{equation}
According to the Arnowitt--Deser--Misner Hamiltonian formulation of General Relativity\index{Hamiltonian formulation of General Relativity!Arnowitt--Deser--Misner (ADM)}, the Euler--Lagrange equations of motion\index{Euler--Lagrange equations of motion} obtained via vanishing of functional derivatives of the total action with respect to the parameters $N$, and $N_i$ are the constraints. In the present situation one has the only one constraint -- the Hamiltonian constraint\index{Hamiltonian constraint}
\begin{equation}
  \dfrac{\delta S_{EH}}{\delta N}=-\mathrm{H}(\eta)=0.
\end{equation}
In other words the Hamiltonian constraint to the case of the Friedmann--Lema\^{i}tre--Robertson--Walker metric\index{Friedmann--Lema\^{i}tre--Robertson--Walker metric!Hamiltonian constraint} is
\begin{equation}\label{3.8}
\dfrac{V_P}{3V}\dfrac{\mathrm{P}_{a}^{2}}{2M_P}+\dfrac{1}{2}a^{4}\mathrm{H}_{M}(\eta)\approx 0.
\end{equation}
Substitution of the explicit form of the conjugated momentum (\ref{3.6}) to the Hamiltonian constraint (\ref{3.8}) leads to the equation
\begin{equation}
	a'^2+\dfrac{N^2}{M_P\ell_P^2}\dfrac{V_P}{3V}a^4\mathrm{H}_{M}(\eta)=0.
\end{equation}
One can put now the value $N^2=1$ which is right for the present case. Using of the Hubble parameter\index{Hubble parameter}
\begin{eqnarray}\label{hubpar}
	 H(a)\equiv\dfrac{\dot{a}}{a}=\dfrac{a'}{a^2},
\end{eqnarray}
where in this context dot means $t$-differentiation, the Hamiltonian constraint\index{Hamiltonian constraint} can be presented as the equation for the energy of Matter f\/ields\index{Matter f\/ields}
\begin{equation}
   \mathrm{H}_{M}(\eta)=-3M_P\ell_P^2\dfrac{V}{V_P}H^2(a),\label{hmt}
\end{equation}
which suggests that energy of Matter f\/ields\index{Matter f\/ields} is negative. Therefore one can put
\begin{equation}
\mathrm{H}_{M}(\eta)=-\left|\mathrm{H}_{M}(\eta)\right|,
\end{equation}
so that the equation (\ref{hmt}) can be rewritten in the form
\begin{equation} H(a)=\pm\sqrt{\dfrac{1}{M_P\ell_P^2}\dfrac{V_P}{3V}\left|\mathrm{H}_M\right|},\label{hubble}
\end{equation}
where $\mathrm{H}_M=\mathrm{H}_M(\bullet)$, and $\bullet=t,\eta$. The solution (\ref{hubble}) is the Hubble law\index{Hubble law}.

The result (\ref{hmt}) is in itself nontrivial. Namely, it means that the energy of Matter f\/ields\index{Matter f\/ields} explicitly depends not on the conformal time\index{conformal time} $\eta$, but on the cosmic scale factor parameter\index{cosmic scale factor parameter}. In other words $\mathrm{H}_{M}(\eta)=\mathrm{H}_{M}(a)$. Applying the result (\ref{hmt}) within the Lagrangian (\ref{lagcosm}) and Hamiltonian (\ref{hamcosm}) of the cosmological model one sees that explicit dependence on the conformal time\index{conformal time} transits into explicit dependence on the cosmic scale factor parameter $a$, i.e.
\begin{eqnarray}
  \mathrm{L}(a,a')&=&\dfrac{M_P\ell_P^2}{2}\dfrac{3V}{V_P}\dfrac{1}{N}\left[a'^2+N^2a^4H^2(a)\right],\\
  \mathrm{H}(\mathrm{P}_a,a)&=&\dfrac{V_P}{3V}\left[\dfrac{\mathrm{P}_a^2}{2M_P}-\dfrac{M_P\ell_P^2}{2}\left(\dfrac{3V}{V_P}\right)^2a^4H^2(a)\right].\label{hamcosm1}
\end{eqnarray}
These formulas, however, show \emph{the tautology of quantum cosmology}\index{tautology of quantum cosmology}, because of in the light of the Dirac approach $\mathrm{H}(\mathrm{P}_a,a)\approx0$ generates the Hubble parameter $H(a)=\dfrac{1}{N}\dfrac{a'}{a^2}$, which after taking into account the value of $N=\pm1$ proper for the Friedmann--Lema\^{i}tre--Robertson--Walker metric\index{Friedmann--Lema\^{i}tre--Robertson--Walker metric} and the transformation to the cosmological time $t$, $a'=\dot{a}a$, becomes the usual Hubble parameter, i.e. $H(a)=\dfrac{\dot{a}}{a}$.

The Hamiltonian (\ref{hamcosm1}), however in itself in nontrivial because of after taking into account the fact $H(a)=\dfrac{a'}{a^2}$ up to the constant multiplier $\dfrac{V_P}{3V}$ it takes the form of the Hamiltonian of a one-dimensional Euclidean oscillator\index{one-dimensional Euclidean oscillator}
\begin{equation}
\mathrm{H}(\mathrm{P}_a,a)=\dfrac{V_P}{3V}\left[T(\mathrm{P}_a)-V(\mathrm{P}_a,a)\right].\label{hamcosm2}
\end{equation}
where the kinetic energy $T(\mathrm{P}_a)$ and the potential energy $V(\mathrm{P}_a,a)$ are
\begin{eqnarray}
T(\mathrm{P}_a)&=&\dfrac{\mathrm{P}_a^2}{2M_P},\\
V(\mathrm{P}_a,a)&=&\dfrac{P_a^2}{2M_P}a^2.
\end{eqnarray}
This oscillator is the one-dimensional Euclidean harmonic oscillator\index{one-dimensional Euclidean oscillator!harmonic} if one identif\/ies the \emph{cosmological coordinate}\index{cosmological coordinate} with
\begin{equation}
x_C=\ell_Pa
\end{equation}
and the potential energy with $V=\dfrac{kx_C^2}{2}$, then the constant $k$ is
\begin{equation}
  k=\dfrac{\mathrm{P}_a^2}{M_P\ell_P^2}.
\end{equation}
In other words the one-dimensional Euclidean harmonic oscillator\index{one-dimensional Euclidean oscillator!harmonic} is def\/ined by the conjugate momentum
\begin{equation}
\mathrm{P}_a=\pm\sqrt{kM_P}\ell_P,\label{paa}
\end{equation}
or equivalently
\begin{equation}
x_C'=\pm\sqrt{\dfrac{k}{M_P}}\ell_P\dfrac{V_P}{3V}.
\end{equation}
This equation can be solved immediately
\begin{equation}
  x_C(\eta)=\pm\sqrt{\dfrac{k}{M_P}}\ell_P\dfrac{V_P}{3V}\left(\eta-\eta_I\right).\label{solut}
\end{equation}
Applying the relation (\ref{paa}) and the solution (\ref{solut}) to the Hamiltonian (\ref{hamcosm2}) one obtains the Hamiltonian of the one-dimensional Euclidean harmonic oscillator\index{one-dimensional Euclidean oscillator!harmonic!Hamiltonian}
\begin{equation}\label{hamcosm3}
\mathrm{H}(\eta)=\dfrac{k\ell_P^2}{2}\dfrac{V_P}{3V}\left[1-\dfrac{k}{M_P}\left(\dfrac{V_P}{3V}\right)^2\left(\eta-\eta_I\right)^2\right].
\end{equation}
Interestingly, when one identif\/ies
\begin{equation}
  \eta_I=\sqrt{\dfrac{M_P}{k}}\dfrac{3V}{V_P},
\end{equation}
then the Hamiltonian of the one-dimensional Euclidean harmonic oscillator\index{one-dimensional Euclidean oscillator!harmonic!Hamiltonian} becomes
\begin{equation}\label{hamcosm4}
\mathrm{H}(\eta)=-\dfrac{k^2\ell_P^2}{2M_P}\left(\dfrac{V_P}{3V}\right)^3\left[\eta^2-2\sqrt{\dfrac{M_P}{k}}\dfrac{3V}{V_P}\eta\right].
\end{equation}
Its values are positive for $0<\eta<2\eta_I$, negative for $\eta>2\eta_I$, and zero for $\eta=2\eta_I$. Interestingly, when one puts \emph{ad hoc} $V=V_P$ and
\begin{equation}
  k=k_P=\dfrac{M_P}{t_P^2}=M_P\omega_P^2=\dfrac{E_P}{\ell_P^2}\approx7.4880571\cdot10^{78}\dfrac{\mathrm{kg}}{\mathrm{s}^2},
\end{equation}
then $\eta_I=3t_P$, and the Hamiltonian of one-dimensional Euclidean harmonic oscillatorr\index{one-dimensional Euclidean oscillator!harmonic!Hamiltonian} is simplif\/ied to
\begin{equation}\label{hamcosm5}
\mathrm{H}(\eta)=-\dfrac{E_P}{54}\left[\left(\dfrac{\eta}{t_P}\right)^2-6\dfrac{\eta}{t_P}\right],
\end{equation}
so that initially, i.e. for $\eta=3t_P$, one obtains
\begin{equation}\label{hamcosm6}
\mathrm{H}(\eta_I)=\dfrac{E_P}{6}.
\end{equation}
Let us call such a case \emph{the Planckian one-dimensional Euclidean harmonic oscillator}\index{one-dimensional Euclidean oscillator!harmonic!Planckian}.

In this manner, in the Hamiltonian approach presented above the Einstein--Friedmann Universe \index{Einstein--Friedmann Universe}is described by the Hamiltonian constraint\index{Einstein--Friedmann Universe!Hamiltonian constraint} (\ref{3.8}), and the Hubble law\index{Hubble law} is obtained due to straightforward integration of this constraint. The Hubble law can be expressed via both the cosmological as well as the conformal time\index{conformal time}
\begin{eqnarray}
\int_{a_I}^{a}\dfrac{da'}{a^{\prime 2}H(a')}&=&\eta_I-\eta,\label{hub1}\\
\int_{a_I}^{a}\dfrac{da'}{a' H(a')}&=&t_I-t,\label{hub2}
\end{eqnarray}
 where the subscript $I$ denotes the initial data values of given quantity. Let us consider the cosmological redshift\index{cosmological redshift} $z$ def\/ined by the formula \cite{wein},
\begin{equation}
	\dfrac{a}{a_I}\equiv\dfrac{1}{1+z},\label{redshift}
\end{equation}
where in terms of cosmological time $a=a(t)$, $a_I=a(t_I)$, $z=z(t_I,t)$, while in in terms of conformal time\index{conformal time} $a=a(\eta)$, $a_I=a(\eta_I)$, $z=z(\eta_I,\eta)$. Then the Hubble law (\ref{hub1})-(\ref{hub2}) can be expressed as
\begin{eqnarray}
\int_{z_I}^{z}\dfrac{dz'}{H(z')}&=&a(\eta_I)\left(\eta-\eta_I\right),\label{hubconf}\\
\int_{z_I}^{z}\dfrac{dz'}{(1+z')H(z')}&=&t-t_I,\label{hubcosm}
\end{eqnarray}
where $z_I=z(t_I,t_I)=z(\eta_I,\eta_I)$, and by the equation (\ref{hubble}) and the def\/inition (\ref{hubpar}) the cosmological redshift\index{cosmological redshift} can be expressed by two distinguishable ways
\begin{eqnarray}
\!\!\!\!\!\!\!\!\!\!\!\!\!z(\eta_I,\eta)&=&\pm a(\eta_I)\int_{\eta_I}^\eta d\eta'\sqrt{\dfrac{1}{M_P\ell_P^2}\dfrac{V_P}{3V}\left|\mathrm{H}_M(\eta')\right|}\label{redshift1}\\
\!\!\!\!\!\!\!\!\!\!\!\!\!z(t_I,t)&=&\exp\left\{\pm\dfrac{a(\eta_I)}
   {a(t_I)}\int_{t_I}^t dt'\sqrt{\dfrac{1}{M_P\ell_P^2}\dfrac{V_P}{3V}\left|\mathrm{H}_M(t')\right|}\right\}-1.\label{redshift2}
\end{eqnarray}
The relations (\ref{hub1}) and (\ref{hub2}), as well as, (\ref{hubconf}) and (\ref{hubcosm}) allow to establish the relation between conformal and cosmological time
\begin{eqnarray}
  \eta-\eta_I&=&t-t_I+\int_{a_I}^a\dfrac{a'-1}{a'^2}\dfrac{da'}{H(a')},\label{etacosm1}\\
  a(\eta_I)(\eta-\eta_I)&=&t-t_I+\int_{z_I}^z\dfrac{z'dz'}{(1+z')H(z')}.\label{etacosm2}
\end{eqnarray}

Interestingly, because of the universal def\/inition of the cosmological redshift\index{cosmological redshift} (\ref{redshift}) one can put \emph{ad hoc} that
\begin{equation}
z(\eta_I,\eta)=z(t_I,t),
\end{equation}
i.e. that cosmological redshift\index{cosmological redshift} is invariant with respect to the exchange between cosmological time and conformal time\index{conformal time}. In such a situation, for consistency one should apply the measure (\ref{ct2}) within the relation (\ref{redshift1}). It can seen by straightforward calculation that such an invariance is nontrivial, because of results in the Riccati equation\index{Riccati equation}
\begin{equation}
  \dot{g}=g+h,\label{riccati}
\end{equation}
where $\dot{g}=\dfrac{dg}{dt}$, and we have introduced the notation
\begin{eqnarray}
g&=&g(t)=\exp[f(t)]\ln[f(t)],\\
h&=&h(t)=\dfrac{a_I}{a(t)},
\end{eqnarray}
where $a_I=a(t_I)$ and the unknown function $f(t)$ has the form
\begin{equation}
f(t)=\pm\dfrac{a(\eta_I)}{a(t_I)}\sqrt{\dfrac{1}{M_P\ell_P^2}\dfrac{V_P}{3V}\left|\mathrm{H}_M(t')\right|}=H(t),
\end{equation}
where we have applied equality $a(\eta_I)=a(t_I)$. The Riccati equation\index{Riccati equation} (\ref{riccati}) can be solved immediately with the result
\begin{equation}\label{gfun}
  g(t)=e^{t-t_I}\left(1+\int_{t_I}^tdt'e^{-t'}h(t')\right).
\end{equation}
Differentiation of $g=\exp H\ln H$ leads to
\begin{equation}
\dot{g}=\dot{H}\exp H\ln H+\exp H\dfrac{\dot{H}}{H},\label{dotg}
\end{equation}
and applying of $\exp H=\dfrac{g}{\ln H}$ within (\ref{dotg}) one obtains
\begin{equation}
  \dfrac{\dot{g}}{g}=\dot{H}\left(1+\dfrac{1}{H\ln H}\right).\label{dgg}
\end{equation}
Differentiation of both the sides of the equation (\ref{dgg}) leads to
\begin{eqnarray}
  \dfrac{\ddot{g}}{g}-\dfrac{\dot{g}^2}{g^2}=\ddot{H}\left(1+\dfrac{1}{H\ln H}\right)-\dfrac{\dot{H}^2}{H^2\ln H}\left(1+\dfrac{1}{\ln H}\right),
\end{eqnarray}
or after using of the equation (\ref{dgg})
\begin{eqnarray}
  \dfrac{\ddot{g}}{g}-\dfrac{\dot{g}^2}{g^2}=
  \dfrac{\ddot{H}}{\dot{H}}\dfrac{\dot{g}}{g}-\dfrac{\dot{H}^2}{H^2\ln H}\left(1+\dfrac{1}{\ln H}\right).
\end{eqnarray}
By this reason, one can apply the following identif\/ication
\begin{eqnarray}
\dfrac{\ddot{g}}{g}&=&\dfrac{\ddot{H}}{\dot{H}}\dfrac{\dot{g}}{g},\label{1id}\\
\dfrac{\dot{g}^2}{g^2}&=&\dfrac{\dot{H}^2}{H^2\ln H}\left(1+\dfrac{1}{\ln H}\right).\label{2id}
\end{eqnarray}
The equation (\ref{1id}) can be rewritten as
\begin{equation}
\dfrac{\ddot{H}}{\dot{H}}=\dfrac{\ddot{g}}{\dot{g}}.
\end{equation}
Its integration can be performed straightforwardly
\begin{equation}
  \ln \dot{H}=\ln \dot{g}+\ln C_0,
\end{equation}
where $C_0$ is an integration constant, or equivalently
\begin{equation}
  \dot{H}=C_0\dot{g}.\label{hprim}
\end{equation}
Integration of the equation (\ref{hprim}) gives
\begin{equation}
  H=C_0g+C_1,\label{ha}
\end{equation}
where $C_1$ ia an integration constant. In this manner
\begin{equation}
\dfrac{\dot{H}^2}{H^2}=\left(\dfrac{C_0\dot{g}}{C_0g+C_1}\right)^2=\dfrac{\dot{g}^2}{\left(g+C_1'\right)^2},
\end{equation}
where $C_1'=\dfrac{C_1}{C_0}$, and the equation (\ref{2id}) can be rewritten as
\begin{equation}
\dfrac{\left(g+C_1'\right)^2}{g^2}=\dfrac{1}{\ln H}\left(1+\dfrac{1}{\ln H}\right),
\end{equation}
or in equivalent form of the algebraic equation for unknown $\ln H$
\begin{equation}
\left(g+C_1'\right)^2(\ln H)^2-g^2\ln H-g^2=0,\label{eqlnh}
\end{equation}
which has two solutions
\begin{equation}
\ln H=\dfrac{1}{2}\left(\dfrac{g}{g+C_1'}\right)^2\left[1\pm\sqrt{1+4\left(\dfrac{g+C_1'}{g}\right)^2}\right],
\end{equation}
and by $\ln H=\dfrac{g}{\exp H}>1$ the correct solution is
\begin{equation}
\ln H=\dfrac{1}{2}\left(\dfrac{g}{g+C_1'}\right)^2\left[1+\sqrt{1+4\left(\dfrac{g+C_1'}{g}\right)^2}\right].\label{logha}
\end{equation}
Jointing of the solutions (\ref{ha}) and (\ref{logha}) allows to establish the algebraic identity for the function $g$
\begin{equation}
\ln(C_0g+C_1)=\dfrac{1}{2}\left(\dfrac{C_0g}{C_0g+C_1}\right)^2\left[1+\sqrt{1+4\left(\dfrac{C_0g+C_1}{C_0g}\right)^2}\right],
\end{equation}
Equivalently, however, by using $\ln H=\dfrac{g}{\exp{H}}$ within the equation (\ref{eqlnh}) one obtains the algebraic equation for unknown function $\exp H$
\begin{equation}
(\exp H)^2+g\exp H-(g+C_1')^2=0,
\end{equation}
having also two possible solutions
\begin{equation}
  \exp H=\dfrac{g}{2}\left(1\pm\sqrt{1+4\left(\dfrac{g+C_1'}{g}\right)^2}\right),
\end{equation}
and because of $\exp H>1$, $g>1$ the correct solution is
\begin{equation}
  \exp H=\dfrac{g}{2}\left(1+\sqrt{1+4\left(\dfrac{g+C_1'}{g}\right)^2}\right),
\end{equation}
and by using of the solution (\ref{ha}) one obtains another identity for $g$
\begin{equation}
\exp\left(C_0g+C_1\right)=\dfrac{g}{2}\left(1+\sqrt{1+4\left(\dfrac{C_0g+C_1}{C_0g}\right)^2}\right).
\end{equation}
Anyway, however, employing the solution (\ref{ha}) within the def\/inition $g=\exp H\ln H$ allows to establish one more identity for $g$
\begin{equation}
  g=\exp\left(C_0g+C_1\right)\ln\left(C_0g+C_1\right).
\end{equation}

Another identity can be established via application of the equation (\ref{dgg}) and the relation (\ref{hprim}) which leads to
\begin{equation}
  \ln H=\dfrac{C_0g}{1-C_0g}\dfrac{1}{C_0g+C_1},
\end{equation}
what applied to the equation (\ref{eqlnh}) gives the algebraic equation for the function $g$
\begin{equation}\label{eqg3}
  C_0^2g^3+(C_0C_1-3C_0)g^2+(2-C_0^4-2C_1)g+C_0^3C_1-\dfrac{C_1}{C_0}=0,
\end{equation}
having in general three solutions. Taking into account the Riccati equation\index{Riccati equation} (\ref{riccati}), i.e. nonzero value of $\dot{g}=g+h$, one can differentiate both sides of the equation (\ref{eqg3}) and obtain
\begin{equation}\label{eqg2}
  [3C_0^2g^2+2(C_0C_1-3C_0)g+2-C_0^4-2C_1]\dot{g}=0,
\end{equation}
and conclude that the following constraint is fulf\/illed
\begin{equation}
  3C_0^2g^2+2(C_0C_1-3C_0)g+2-C_0^4-2C_1=0.
\end{equation}
It allows to establish the value of function $g$
\begin{equation}
  g=g(C_0,C_1)=-\dfrac{C_1-3}{3C_0}\pm\dfrac{1}{3C_0}\sqrt{C_1^2+3C_0^4+3}=constans.
\end{equation}
Taking the positive solution as the physical
\begin{equation}
g(C_0,C_1)=-\dfrac{C_1-3}{3C_0}+\dfrac{1}{3C_0}\sqrt{C_1^2+3C_0^4+3}.
\end{equation}
one can derive the value of the Hubble parameter
\begin{equation}
  H=1+\dfrac{2}{3}C_1+\dfrac{1}{3}\sqrt{C_1^2+3C_0^4+3}.
\end{equation}
Therefore, the problem is to establish the constants $C_0$ and $C_1$. It can be seen straightforwardly that
\begin{eqnarray}
  C_1&=&H_I-C_0g_I,\\
  C_0&=&\dfrac{\dot{H}_I}{\dot{g}_I},
\end{eqnarray}
where $g_I=g(t_I)$ and $\dot{g}_I=\dot{g}(t_I)$. Because of the function $g$ and its derivative are determined explicitly by the relation (\ref{gfun}) and the Riccati equation\index{Riccati equation} (\ref{riccati}), one has
\begin{eqnarray}
  g_I&=&1,\\
  \dot{g}_I&=&g_I+h(t_I)=2,
\end{eqnarray}
and by this reason one obtains f\/inally
\begin{eqnarray}
  C_1&=&H_I-\dfrac{\dot{H}_I}{2},\\
  C_0&=&\dfrac{\dot{H}_I}{2}.
\end{eqnarray}
In other words, in such a situation the Hubble parameter\index{Hubble parameter} has a form
\begin{equation}
  H=1+\dfrac{2}{3}\left(H_I-\dfrac{\dot{H}_I}{2}\right)+\dfrac{1}{3}\sqrt{H_I^2-H_I\dot{H}_I+\dfrac{3}{16}\dot{H}_I^4-\dfrac{1}{4}\dot{H}_I^2+3}.
\end{equation}
Particularly, in the case when $C_1=0$ one obtains
\begin{equation}
  H_I=H_0\exp\left\{2(t_I-t_0)\right\},
\end{equation}
and the actual value of the Hubble parameter\index{Hubble parameter} becomes
\begin{equation}
  H=1+\dfrac{1}{\sqrt{3}}\sqrt{1+H_0^4\exp\left\{8(t_I-t_0)\right\}}.
\end{equation}

It can be seen by straightforward computation that if one takes into account the usual def\/inition of the density of energy of Matter f\/ields\index{Matter f\/ields}
 \begin{equation}\label{endens}
 \mathcal{H}_M(x,\cdot)=\dfrac{d\epsilon_M(\cdot)}{dV(x)},
 \end{equation}
where $\cdot=t,\eta$, $dV(x)=d^3x$ is an inf\/initesimal volume, and $d\epsilon_M(\cdot)$ is an inf\/initesimal energy of Matter f\/ields\index{Matter f\/ields} contained in such a volume, then
\begin{equation}
  \epsilon_M=|\mathrm{H}_M|,
\end{equation}
and consequently the relations (\ref{redshift1}) and (\ref{redshift2}) allow to establish dependence of energy of Matter f\/ields\index{Matter f\/ields} from the redshift
\begin{eqnarray}
	\epsilon_M(t)&=&3M_P\ell_P^2\dfrac{V}{V_P}\left|\dfrac{a(t_I)}{a(\eta_I)}\right|^2
	\left|\dfrac{1}{1+z(t_I,t)}\dfrac{dz(t_I,t)}{dt}\right|^2,
	\label{ene1}\\
	\epsilon_M(\eta)&=&3M_P\ell_P^2\dfrac{V}{V_P}\left|\dfrac{1}{a(\eta_I)}\right|^2
	\left|\dfrac{dz(\eta_I,\eta)}{d\eta}\right|^2.\label{ene2}
\end{eqnarray}

 By application of the formulas (\ref{ene1}) and (\ref{ene2}), and def\/initions (\ref{ct2}),
 (\ref{hubble}) and (\ref{redshift}) it can be seen straightforwardly that $\mathcal{H}_M$,
 $\epsilon_M$, and $H$ are diffeoinvariants of the coordinate time
\begin{equation}
	\mathcal{H}_M=\mathrm{inv.},\qquad\epsilon_M=\mathrm{inv.},
	\qquad H(a)=\mathrm{inv.},
\end{equation}
and moreover the relation holds
\begin{equation}\label{eps} \epsilon_M(t)=3M_P\ell_P^2\dfrac{V}{V_P}\left|\dfrac{a(t_I)}{a(\eta_I)}\right|^2\left|\dfrac{\dot{a}(t)}{a(t)}\right|^2=3M_P\ell_P^2\dfrac{V}{V_P}\left|\dfrac{a(t_I)}{a(\eta_I)}\right|^2|H(a)|^2.
\end{equation}
Using of the relations (\ref{hubble}) and (\ref{eps}) one obtains
\begin{equation}
\left|\dfrac{a(t_I)}{a(\eta_I)}\right|^2=1.
\end{equation}
Integration of the equation (\ref{eps}) leads to the relation
\begin{equation} \dfrac{a(t)}{a(t_I)}=\exp\left\{\pm\dfrac{a(\eta_I)}{a(t_I)}\int_{t_I}^tdt'\sqrt{\dfrac{1}{M_P\ell_P^2}\dfrac{V_P}{3V}\epsilon_M(t')}\right\}.
\end{equation}
If one identif\/ies $a(t_I)=a(\eta_I)=a_I$, and $H(a)$ as real and positive then the energy of Matter f\/ields\index{Matter f\/ields} is simply
 \begin{equation}
   \dfrac{1}{M_P\ell_P^2}\dfrac{V_P}{3V}\epsilon_M=H^2(a),\label{emf}
 \end{equation}
and the cosmic scale factor parameter\index{cosmic scale factor parameter} takes the form
 \begin{equation}
a(t)=a_I\exp\left\{\pm\int_{t_I}^tdt'\sqrt{\dfrac{1}{M_P\ell_P^2}\dfrac{V_P}{3V}\epsilon_M(t')}\right\}.
\end{equation}
In general the energy of Matter f\/ields\index{Matter f\/ields} (\ref{emf}) def\/ines the evolution of the classical Universe. However, because of $\epsilon_M$ is invariant with respect to exchange between the cosmological time and the conformal time\index{conformal time}, it can be considered rather as function of the redshift $z$ or the cosmic scale factor parameter\index{cosmic scale factor parameter} $a$. Moreover, when there is several kinds of Matter f\/ields\index{Matter f\/ields} then $\epsilon_M$ can be considered as the total energy of Matter f\/ields\index{Matter f\/ields}, and by this reason is an algebraical sum of all contributions
\begin{equation}
\epsilon_M=\sum_{i}\epsilon_i,
\end{equation}
where the subscript $i$ is associated with a kind of Matter f\/ields\index{Matter f\/ields}. In this manner, the evolution (\ref{emf}) can be considered as the dependence of the redshift on the cosmological as well as the conformal time\index{conformal time}
\begin{eqnarray}
\sqrt{\sum_{i}\epsilon_i(z)}&=&\dfrac{1}{1+z}\dfrac{dz}{dt},\\
\sqrt{\sum_{i}\epsilon_i(z)}&=&\dfrac{dz}{d\eta},
\end{eqnarray}
as well as as the dependence of the cosmic scale factor on the cosmological as well as the conformal time\index{conformal time}
\begin{eqnarray}
\sqrt{\sum_{i}\epsilon_i(a)}&=&\dfrac{1}{a}\dfrac{da}{dt},\\
\sqrt{\sum_{i}\epsilon_i(a)}&=&\dfrac{1}{a^2}\dfrac{da}{d\eta}.
\end{eqnarray}

 Interestingly, the Hamiltonian constraint\index{Hamiltonian constraint} (\ref{3.8}), describing the classical toroidal Einstein--Friedmann Universe\index{Einstein--Friedmann Universe!}\index{toroidal topology}, can be parameterized in the way characteristic for a bosonic string, i.e. by using of the Einstein energy-momentum relation\index{Einstein energy-momentum relation}
\begin{equation}
	\mathrm{P}_a^2c^2+m^2(a)c^4=E^2(a),\label{pc}
\end{equation}
where the energy identically vanishes $E^2(a)=0$, and the squared mass is manifestly negative
\begin{equation}
m^2(a)=-\left(3M_Pt_P\dfrac{V}{V_P}\right)^2a^4H^2(a)<0,
\end{equation}
Hence, it is easy to deduce that the Universe can be understood as the tachyon\index{tachyon}, i.e. the groundstate of the classical bosonic string\index{bosonic string!groundstate} \cite{str}. The mass of the tachyon\index{tachyon} can be expressed via two ways
\begin{equation}
	\sqrt{|m^2(a)|}=3\dfrac{V}{V_P}M_Pt_Pa'=\dfrac{3}{2}\dfrac{V}{V_P}M_Pt_P\dot{\Omega},
\end{equation}
 where $\Omega=a^2(t)$ is the scaling function (\ref{suncalif}) and $S_P=4\pi\ell_P^2$ is the area of the Planck sphere\index{Planck sphere}. The Hubble law\index{Hubble law} (\ref{hub1})-(\ref{hub2}) can be interpreted as the relation between volume of space and the mass of the tachyon\index{tachyon}
\begin{eqnarray}
M_P\int_{a_I}^a\dfrac{da'}{\sqrt{|m^2(a')|}}&=&\dfrac{V_P}{3V}\dfrac{\eta-\eta_I}{t_P},\\
M_P\int_{a_I}^{a}\dfrac{a'da'}{\sqrt{|m^2(a')|}}&=&\dfrac{V_P}{3V}\dfrac{t-t_I}{t_P}.
\end{eqnarray}

In this manner, we have received the classical point of view on the Einstein--Friedmann Universe\index{Einstein--Friedmann Universe}. Let us construct the quantum theory of the Universe.

\section{Quantization of Hamiltonian Constraint}

Application of the primary canonical quantization\index{quantization!primary canonical} in the form
\begin{equation}\label{fq}
  	\left[\hat{\mathrm{P}}_{a},\hat{a}\right]=-i\dfrac{\hslash}{\ell_P},
\end{equation}
where the parameter of quantization\index{parameter of quantization} $\dfrac{\hslash}{\ell_P}=M_Pc$ is the Planck momentum\index{Planck momentum} having the value
\begin{equation}
  \dfrac{\hslash}{\ell_P}=M_Pc\approx6.524806271\dfrac{\mathrm{kg}\cdot\mathrm{m}}{\mathrm{s}}.
\end{equation}
In comparison to the standard quantum mechanics, in which the parameter of quantization is the reduced Planck constant $\hslash$, this value of the parameter of quantization is large, i.e. about $10^{34}$ times more. Such a procedure allows to generate the operator of the momentum conjugated to the cosmic scale factor parameter\index{cosmic scale factor parameter}
\begin{equation}
\hat{\mathrm{P}}_{a}=-i\dfrac{\hslash}{\ell_P}\dfrac{d}{da}.
\end{equation}
Applying such a primary canonical quantization\index{quantization!primary canonical} within the Hamiltonian constraint (\ref{pc}) of the Einstein--Friedmann Universe \index{Einstein--Friedmann Universe!Hamiltonian constraint}one receives the one-dimensional Klein--Gordon equation\index{Klein--Gordon equation!one-dimensional}
\begin{equation}\label{awdw}
	\left(\dfrac{d^2}{d a^2}+\omega^2(a)\right)\Psi(a)=0,
\end{equation}
where $\omega$ is the \emph{cosmological dimensionless frequency}\index{dimensionless frequency!cosmological} given by the formula
\begin{equation}
  \omega(a)=\dfrac{3V}{V_P}\dfrac{a^2H(a)}{\omega_P},
\end{equation}
where $V_P=\dfrac{4}{3}\pi\ell_P^3$ is volume of the Planck sphere\index{Planck sphere}, and $\omega_P=\dfrac{1}{t_P}$ is the Planck frequency\index{Planck frequency}. In the context of cosmology the Klein--Gordon equation (\ref{awdw}) is the analog of the Wheeler--DeWitt equation \index{Wheeler--DeWitt equation}\cite{whe,dew} of more general quantum geometrodynamics\index{quantum geometrodynamics (QGD)}. Usually this equation identif\/ied with non relativistic quantum mechanics, i.e. as the Schr\"odinger equation\index{Schr\"odinger equation}. Because of the cosmic scale factor parameter\index{cosmic scale factor parameter} $a$ is def\/ined as a function of cosmological or conformal time\index{conformal time}, in general the wave function $\Psi(a)$ is a functional, and the equation (\ref{awdw}) is a one-dimensional functional differential equation. In general by the spirit of the Wheeler superspace\index{Wheeler superspace}, the wave function is a wave functional on the space of three-dimensional metrics characterized by the parameter $a$. We shall discuss details of this approach in next chapters of this part. Albeit, it must be emphasized that the approach based on the Schr\"odinger equation, which has been worked out more than 50 years ago, since the mid-1980's does not lead to new constructive results.

A different possible interpretation of the quantum theory (\ref{awdw}) is the Klein--Gordon equation\index{Klein--Gordon equation}, i.e. the relativistic wave equation describing bosons. It is not diff\/icult to see by straightforward computations that the equation (\ref{awdw}) can be generated as the classical f\/ield theoretic Euler--Lagrange equations of motion\index{Euler--Lagrange equations of motion}
\begin{equation}
\left[\dfrac{\partial}{\partial\Psi}-\dfrac{\partial}{\partial a}\dfrac{\partial}{\partial \left(\dfrac{d\Psi}{da}\right)}\right]L\left(\Psi,\dfrac{d\Psi}{da}\right)=0,
\end{equation}
via the variational principle
\begin{equation}
\delta S[\Psi]=0,
\end{equation}
with respect to the action functional $S[\Psi]$ having the following form
\begin{equation}\label{ftact}
  	S[\Psi]=\int da L\left(\Psi,\dfrac{d\Psi}{da}\right),
\end{equation}
where the Lagrangian of the classical f\/ield, i.e. one-dimensional wave function $\Psi=\Psi(a)$, is
\begin{equation}
  L\left(\Psi,\dfrac{d\Psi}{da}\right)=\dfrac{1}{2}\left(\dfrac{d\Psi}{da}\right)^2-\dfrac{1}{2}\omega^2(a)\Psi^2.
\end{equation}
The momentum $\Pi_\Psi$ canonically conjugated to the wave function $\Psi$ can be established immediately from the action (\ref{ftact}) as
\begin{eqnarray}\label{eqnn1}
	\Pi_\Psi=\dfrac{\partial L\left(\Psi,\dfrac{d\Psi}{da}\right)}{\partial \left(\dfrac{d\Psi}{da}\right)}=\dfrac{d\Psi}{da},
\end{eqnarray}
and its straightforward application allows to rewrite the equation (\ref{awdw}) in the form
\begin{eqnarray}\label{eqnn2}
	\dfrac{d\Pi_\Psi}{da}+\omega^2(a)\Psi(a)=0.
\end{eqnarray}
The equations of motion (\ref{eqnn1}) and (\ref{eqnn2}) can be treated as the system of equations and employed for reduction of order of the Wheeler--DeWitt equation \index{Wheeler--DeWitt equation}(\ref{awdw}). Such a reduction can be done by introducing to the theory the two-component scalar f\/ield
\begin{equation}
  \Phi=\left[\begin{array}{c}\Pi_\Psi\\ \Psi\end{array}\right],
\end{equation}
which allows to express the quantum cosmology given by the Klein--Gordon equation(\ref{awdw}), which describes the Einstein--Friedmann Universe\index{Einstein--Friedmann Universe} as the Multiverse of quantum universes\index{Multiverse}, as the one-dimensional Dirac equation\index{Dirac equation!one-dimensional}
\begin{equation}\label{dir}
 	\left(-i\sigma_2\dfrac{d}{da}-M\right)\Phi=0,
\end{equation}
where $M$ is the mass matrix, and $\sigma_2$ is one of the Pauli matrices,
\begin{eqnarray}
     M&=&\left[\begin{array}{cc}
     -1&0\\0&-\omega^{2}\end{array}\right],\\
     \sigma_2&=&\left[\begin{array}{cc}0&-i\\
     i&0\end{array}\right].
\end{eqnarray}
On the other hand the one-dimensional Dirac equation \index{Dirac equation!one-dimensional}(\ref{dir}) describes two-component one-dimensional scalar f\/ield, i.e. axion\index{axion}, and as one-dimensional quantum mechanics is automatically supersymmetric\index{supersymmetry}. In this way the quantum cosmology based on spatially f\/inite and conformal-f\/lat Friedmann--Lema\^{i}tre--Robertson--Walker metric \index{Friedmann--Lema\^{i}tre--Robertson--Walker metric}, i.e. the Einstein--Friedmann Universe\index{Einstein--Friedmann Universe}, is strictly related to one-dimensional superstrings which we shall call the Fermi--Bose superstrings\index{Fermi--Bose superstrings}. This name arises from the fact that in the dimension $1$ there is no difference between Fermi--Dirac and Bose--Einstein statistics\index{Bose--Einstein statistics}\index{Fermi--Dirac statistics} because of the spin is zero. In this manner the quantum f\/ield theory which shall be constructed relying on the Fock rep\`{e}re\index{Fock rep\`{e}re} of creators and annihilators will be describing the Multiverse of Fermi--Bose superstrings, which we shall call the superstring Multiverse\index{Multiverse!superstring}, which in the quantum cosmology presented above is the system of multiple quantum Einstein--Friedmann Universes\index{Einstein--Friedmann Universe}.

Because of the quantum cosmology (\ref{awdw}) has manifestly bosonic character, let us perform the secondary canonical quantization \index{quantization!secondary canonical}of the one-dimensional Dirac equation \index{Dirac equation!one-dimensional}(\ref{dir}) appropriate to bosons \cite{bs}
\begin{equation}
  \left[\hat{\Pi}_{\Psi}[a],\hat{\Psi}[a']\right]=-i\delta\left(a-a'\right),
\end{equation}
and other canonical commutation relations\index{canonical commutation relations} are trivial. In the explicit form the bosonic f\/ield is
\begin{eqnarray}\label{sq}
	\left[\begin{array}{c}\hat{\Psi}\\ \hat{\Pi}_\Psi\end{array}\right]
	=\left[\begin{array}{cc}
	\dfrac{1}{\sqrt{2\omega(a)}}&\dfrac{1}{\sqrt{2\omega(a)}}\\
	-i\sqrt{\dfrac{\omega(a)}{2}}&i\sqrt{\dfrac{\omega(a)}{2}}\end{array}\right]
	\left[\begin{array}{c}{G}[a]\\ {G}^{\dagger}[a]\end{array}\right],
\end{eqnarray}
and is consistent with the algebraic approach to canonical commutation relations\index{canonical commutation relations} due to von Neumann, and Araki and Woods \cite{ccr}\index{von Neumann--Araki--Woods algebraic approach}. In result the Universe is described by the dynamical basis
\begin{equation}\label{db}
  {B}_{a}=\left\{\left[\begin{array}{c}{G}[a]\\{G}^{\dagger}[a]\!\!\end{array}\right]:\left[{G}[a],{G}^{\dagger}[a']\right]=\delta\left(a-a'\right),\left[{G}[a],{G}[a']\right]=0\right\},
\end{equation}
satisfying non-Heisenberg equations of motion\index{non-Heisenberg equations of motion}
\begin{equation}\label{nonh}
	\dfrac{d {B}_{a}}{d {a}}=\left[\begin{array}{cc}
	-i\omega(a)&\dfrac{1}{2\omega(a)}\dfrac{d\omega(a)}{da}\\
	\dfrac{1}{2\omega(a)}\dfrac{d\omega(a)}{da}&i\omega(a)\end{array}\right]{B}_{a},
\end{equation}
and via perturbation theory holds
 \begin{equation}\label{pertcon}
 \left|\dfrac{1}{2\omega(a)}\dfrac{d\omega(a)}{da}\right|\ll \dfrac{1}{a}.
 \end{equation}
The equations (\ref{nonh}) can be diagonalized via taking into account the new basis ${B}_{a}'$
\begin{equation}
{B}_{a}'=\left\{\left[\begin{array}{c}{G}'[a]\\{G}^{\prime\dagger}[a]\end{array}\right]:\left[{G}'[a],{G}^{\prime\dagger}[a']\right]=\delta\left(a-a'\right),\left[{G}'[a],{G}'[a']\right]=0\right\},
\end{equation}
obtained via taking together the Bogoliubov transformation\index{Bogoliubov transformation} and the Heisenberg equations of motion\index{Heisenberg equations of motion}
\begin{eqnarray}
{B}_{a}'&=&\left[\begin{array}{cc}u&v\\v^{\ast}&u^{\ast}\end{array}\right]{B}_{a},\\
\dfrac{d{B}_{a}'}{da}&=&\left[\begin{array}{cc}-i\omega'&0\\0&i\omega'\end{array}\right]{B}_{a}',
\end{eqnarray}
 where the hyperbolic constraint holds
 \begin{equation}
   |u|^2-|v|^2=1,
 \end{equation}
 and $\omega'$ is an unknown frequency. It can be seen by straightforward computation that such a procedure generates unambiguously $\omega'=0$, and therefore the new basis is the static Fock rep\`{e}re\index{Fock rep\`{e}re}
\begin{equation}\label{statb}
	{B}_{I}=\left\{\left[\begin{array}{c}\mathrm{w}_I\\
	\mathrm{w}^{\dagger}_I\end{array}\right]:
	\left[\mathrm{w}_I,\mathrm{w}^{\dagger}_I\right]=1,
	\left[\mathrm{w}_I,\mathrm{w}_I\right]=0\right\},
\end{equation}
and by this reason the static vacuum state\index{vacuum state} is obtained
\begin{equation}\label{vac}
	\lv=\left\{\lv:\mathrm{w}_I\lv=0\quad,\quad0=\rv\mathrm{w}_I^\dagger\right\}.
\end{equation}
The system of operator equations (\ref{nonh}) transits to the system of equations for the Bogoliubov coeff\/icients\index{Bogoliubov coeff\/icients}
\begin{equation}\label{uv}
	\dfrac{d}{da}\left[\begin{array}{c}v(a)\\
	u(a)\end{array}\right]=\left[\begin{array}{cc}
	-i\omega(a)&\dfrac{1}{2\omega(a)}\dfrac{\partial\omega(a)}{\partial a}\\
	\dfrac{1}{2\omega(a)}\dfrac{\partial\omega(a)}{\partial a}&i\omega(a)
	 \end{array}\right]\left[\begin{array}{c}v(a)\\u(a)\end{array}\right],
\end{equation}
 which is easy to solve in the superf\/luid parametrization \cite{bog}
\begin{eqnarray}
v(a)&=&\exp(i\theta(a))\sinh \phi(a),\\
u(a)&=&\exp(i\theta(a))\cosh \phi(a),
\end{eqnarray}
where $\theta$ and $\phi$ are the angles, which in the present situation express via the mass of the tachyon\index{tachyon} as follows
\begin{eqnarray}
	\theta(a)&=&\pm i\int_{a_I}^{a}\omega(a')da',\\
	\phi(a)&=&\ln{\sqrt{\dfrac{\omega_I}{\omega(a)}}},
\end{eqnarray}
where $\omega_I=\omega(a_I)$ is the initial data of $\omega(a)$.

In this manner, \emph{the quantum cosmology is completely determined via the monodromy matrix\index{monodromy matrix} $C$
between the dynamical and the static Fock rep\`{e}re\index{Fock rep\`{e}re} $B_a=CB_I$ which is explicitly given by}
\begin{equation}\label{mon}
C=\left[\begin{array}{cc}\left(\sqrt{\dfrac{\omega(a)}{\omega_I}}+\sqrt{\dfrac{\omega_I}{\omega(a)}}\right)\dfrac{e^{\lambda}}{2}&
\left(\sqrt{\dfrac{\omega(a)}{\omega_I}}-\sqrt{\dfrac{\omega_I}{\omega(a)}}\right)\dfrac{e^{-\lambda}}{2}\\
\left(\sqrt{\dfrac{\omega(a)}{\omega_I}}-\sqrt{\dfrac{\omega_I}{\omega(a)}}\right)\dfrac{e^{\lambda}}{2}&	 \left(\sqrt{\dfrac{\omega(a)}{\omega_I}}+\sqrt{\dfrac{\omega_I}{\omega(a)}}\right)\dfrac{e^{-\lambda}}{2}\end{array}\right],
\end{equation}
where $\lambda$ is integrated frequency
\begin{equation}
\lambda = \lambda(a)=i\theta(a)=\mp\int_{a_I}^{a}\omega(a')da'.
\end{equation}
Now the quantum f\/ield $\hat{\Psi}$ can be computed straightforwardly
\begin{eqnarray}\label{qfmuni}
\hat{\Psi}[a]=\dfrac{1}{\sqrt{2\omega_I}}\left(e^{\lambda(a)}\mathrm{w}_I+e^{-\lambda(a)}\mathrm{w}^{\dagger}_I\right),
\end{eqnarray}
and if one introduces the state
\begin{equation}\label{mpuni}
  |n\rangle\equiv\left(\hat{\Psi}[a]\right)^n\lv=\dfrac{e^{-n\lambda(a)}}{\left(2\omega_I\right)^{n/2}}\mathrm{w}^{\dagger~n}_I\rv,
\end{equation}
then the interesting relations can be derived
\begin{eqnarray}
\langle{m}|n\rangle&=&\dfrac{e^{(m-n)\lambda(a)}}{\left(2\omega_I\right)^{(m+n)/2}}\lv\mathrm{w}_I^m\mathrm{w}_I^{\dagger~n}\rv,\\
\langle{n}|n\rangle&=&\sum_{p=0}^{n}\dfrac{e^{i\pi(p+1)}}{\left(2\omega_I\right)^n}C^n_{n-p}\lv\left(\mathrm{w}_I^\dagger\mathrm{w}_I\right)^p\rv=\dfrac{e^{i\left(2n+1\right)\pi}}{\left(2\omega_I\right)^n},
\end{eqnarray}
where $C^n_{k}$ is the Newton binomial symbol\index{Newton binomial symbol}, which in the present case is
\begin{equation}
  C^n_{n-p}=\dfrac{n!}{p!(n-p)!},
\end{equation}
and we have applied the normalization of the stable vacuum state\index{vacuum state}
\begin{equation}
  \vv=1.
\end{equation}
Obviously the state $|n\rangle$ is a $n$-particle state describing the Multiverse\index{Multiverse}. Its physical sense arises via the straightforward analogy with the quantum theory of many body systems (See e.g. the Ref. \cite{ripka}). Namely, in the present case of the quantum cosmology one has to deal with the system of many quantum universes. Such a specif\/ic quantum many body system, i.e. the secondary-quantized Einstein--Friedmann Universe \index{Einstein--Friedmann Universe}given by the quantum f\/ield (\ref{qfmuni}) and the many particle states (\ref{mpuni}), is manifestly a realization of the Multiverse hypothesis\index{Multiverse}. Let us consider now the thermodynamics of the Multiverse.

\section{The Multiverse Thermodynamics}
Because of we have derived the static basis (\ref{statb}), formally also exists the thermal equilibrium for quantum states of the Einstein--Friedmann Universe \index{Einstein--Friedmann Universe}(\ref{ef}). In this section we shall formulate thermodynamics of many quantum Universe. We shall apply the simplest approximation, i.e. we shall consider the system with one degenerated state which is def\/ined by the density operator $\varrho_{{G}}$ having the form
\begin{eqnarray}
	\varrho_{{G}}&=&{G}^{\dagger}{G}=\\
    &=&{B}_{a}^{\dagger}\
    \left[\begin{array}{cc}1&0\\0&0\end{array}\right]{B}_{a}=\\
	&=&{B}_I^{\dagger}\left[\begin{array}{cc}|u|^2&-uv\\
	-u^{\ast}v^{\ast}&|v|^2\end{array}\right]{B}_I
	\equiv{B}_I^{\dagger}\rho_{\mathrm{eq}}{B}_I,
\end{eqnarray}
where $\rho_{\mathrm{eq}}$ is the density operator in the thermal equilibrium. By straightforward computation of the Boltzmann--von Neumann entropy\index{Boltzmann--von Neumann entropy} $\mathrm{S}$
\begin{eqnarray}
  	\mathrm{S}&=&\dfrac{\mathrm{tr}\left(\rho_{\mathrm{eq}}
  	\ln\rho_{\mathrm{eq}}\right)}{\mathrm{tr}\rho_{\mathrm{eq}}}=\\
  	 &=&\ln\left(2|u|^2-1\right)\equiv-\ln\Omega_{\mathrm{eq}},\label{bvn}
\end{eqnarray}
 it is easy to deduce the distribution function $\Omega_{\mathrm{eq}}$:
\begin{equation}
	\Omega_{\mathrm{eq}}=\dfrac{1}{2|u|^2-1}=\dfrac{1}{2|v|^2+1}.
\end{equation}
 On the other side, one can use the occupation number $n$
\begin{eqnarray}
	n&\equiv&\langle0|{G}^{\dagger}[a]{G}[a]|0\rangle=\\
    &=&\dfrac{1}{4}\left(\dfrac{\omega(a)}{\omega_I}+\dfrac{\omega_I}{\omega(a)}\right)-\dfrac{1}{2}=|v|^2,\label{s}
\end{eqnarray}
 where $m_I=m(a_I)$, to derivation of the entropy
\begin{equation}\label{ent}
  	\mathrm{S}=\ln\langle\mathrm{n}\rangle,
\end{equation}
where $\langle\mathrm{n}\rangle$ is averaged occupation number
 \begin{equation}
 \langle\mathrm{n}\rangle=2n+1=\dfrac{1}{2}
	\left(\dfrac{\omega(a)}{\omega_I}+\dfrac{\omega_I}{\omega(a)}\right).
 \end{equation}
 By combination of the formulas (\ref{bvn}), (\ref{s}), and (\ref{ent}), with the natural conditions $n\geq0$ and $\omega(a)\geq\omega_I$, one receives the mass spectrum of the system of many Einstein--Friedmann Universes\index{Einstein--Friedmann Universe}
\begin{equation}
  	\dfrac{\omega(a)}{\omega_I}=\langle\mathrm{n}\rangle
	+\sqrt{\langle\mathrm{n}\rangle^2-1}.
\end{equation}
 Similarly, by application of the averaging method one can derive the internal energy $\mathrm{U}$
\begin{eqnarray}\label{n}
	\mathrm{U}&\equiv&\dfrac{\mathrm{tr}(\rho_{\mathrm{eq}}
	 \mathrm{H}_{\mathrm{eq}})}{\mathrm{tr}\mathrm{\rho_{\mathrm{eq}}}}=\\
	&=&\left(\langle \mathrm{n}\rangle+\dfrac{1}{2}\right)
	\left(\langle\mathrm{n}\rangle+\sqrt{\langle\mathrm{n}\rangle^2-1}
	\right)\omega_I,\label{u}
\end{eqnarray}
and chemical potential\index{chemical potential} $\mu$
\begin{eqnarray}
	\mu&\equiv&\dfrac{\partial\mathrm{U}}{\partial\mathrm{n}}=\\
	&=&\left(\dfrac{\langle \mathrm{n}\rangle+\dfrac{1}{2}}{\sqrt{\langle \mathrm{n}\rangle^2-1}}+1\right)
	\left(\langle\mathrm{n}\rangle+\sqrt{\langle\mathrm{n}\rangle^2-1}
	\right)\omega_I,
\end{eqnarray}
where we have applied the Hamiltonian of the classical theory expressed in the static Fock rep\`{e}re\index{Fock rep\`{e}re}
\begin{equation}
  H=\left(G^\dagger[a]G[a]+G[a]G^\dagger[a]\right)\dfrac{\omega(a)}{2}=B_I^\dagger\mathrm{H}_{\mathrm{eq}}B_I,
\end{equation}
where $H_{\mathrm{eq}}$ is the Hamiltonian in equilibrium
\begin{eqnarray} H_{\mathrm{eq}}=\left[\begin{array}{cc}|u|^2+|v|^2&-2uv\\-2u^{\ast}v^{\ast}&|u|^2+|v|^2\end{array}\right]\dfrac{\omega(a)}{2}.
\end{eqnarray}

Let us establish temperature\index{temperature} of the system by using of the \emph{method of quantum statistics}\index{method of quantum statistics}. For this one must take into account the concrete form of the quantum statistics of the system. The bosonic character of the quantum cosmology (\ref{awdw}) suggests application of the Bose--Einstein statistics\index{Bose--Einstein statistics} which naturally describes bosonic systems. In this manner
\begin{equation}
	\Omega_{\mathrm{eq}}\equiv\dfrac{1}{\exp\left\{\dfrac{\mathrm{U}
	-\mu n}{\mathrm{T}}\right\}-1}.
\end{equation}
Because of the quantum statistics derived in this section has the form
\begin{equation}
\Omega_{\mathrm{eq}}=\dfrac{1}{|u|^2+|v|^2}=\dfrac{1}{2|u|^2-1},
\end{equation}
what after taking into account the fact $2|u|^2=\langle{n}\rangle+1$ results in the relation
\begin{equation}
\dfrac{\mathrm{U}-\mu n}{\mathrm{T}}=\ln\left(\langle{n}\rangle+1\right).
\end{equation}
Consequently such a procedure allows to determine the temperature\index{temperature} of the system of quantum states of the Universe as the f\/ixed parameter.
\begin{eqnarray}
\!\!\!\!\!\!\!\!\!\!\!\!\!\!\mathrm{T}&=&\omega_I\dfrac{\langle\mathrm{n}\rangle +\sqrt{\langle\mathrm{n}\rangle^2-1}}{\ln(\langle\mathrm{n}\rangle+1)}\times\\
\!\!\!\!\!\!\!\!\!\!\!\!\!\!&\times&\left[\langle\mathrm{n}\rangle+\dfrac{1}{2}- \dfrac{1}{2}\left(\langle\mathrm{n}\rangle+\sqrt{\langle\mathrm{n}\rangle^2-1}+ \dfrac{1}{2}\right)\sqrt{\dfrac{\langle\mathrm{n}\rangle-1}{\langle\mathrm{n}\rangle+1}}\right].\label{t}
\end{eqnarray}
\section{The Early Light Multiverse\index{Multiverse!early light}}
Interestingly, there is a certain particular thermodynamical situation of the system of many quantum Universes. Namely, these are the quantum Universes possessing minimal entropy
\begin{equation}
\mathrm{S}_{\mathrm{min}}=0.
\end{equation}
It can be deduced straightforwardly that such a Multiverse \index{Multiverse}is characterized by the following conditions
\begin{equation}
   	n=0\quad,\quad\langle\mathrm{n}\rangle=1\quad,\quad \omega(a)=\omega_I,
\end{equation}
and therefore its thermodynamical parameters have the values
\begin{eqnarray}
	\mathrm{U}_{\mathrm{min}}&=&\dfrac{3}{2}\omega_I,\label{max1}\\
	\mu_{\mathrm{min}}&=&\infty,\label{max2}\\
	 \mathrm{T}_{\mathrm{min}}&=&\dfrac{\mathrm{U}_{\mathrm{max}}}{\ln2}.\label{max3}
\end{eqnarray}
 The relations (\ref{max1}), (\ref{max2}), and (\ref{max3}) shows that such a specif\/ic collection of quantum Universes is determined by the only initial data, i.e. the parameter $\omega_I$. The relation (\ref{max3}) nontrivially connects temperature\index{temperature} and internal energy of these quantum states, and therefore in general spirit of statistical mechanics it is the equation of state of the collection of quantum Universes. Inf\/inite value of the chemical potential\index{chemical potential} is the only characteristic syndrome of openness of the system, and shows that in the point $n=0$ a phase transition happens. Such a point can be understood as the point in which quantum Universes start their existence. There is a question about the physical meaning of the quantum states of Universe characterized by the minimal entropy. It is not diff\/icult to see that for such Universes holds the relation
\begin{equation}
  	\omega_I = \pm\dfrac{3V}{V_P}\dfrac{a^2_IH(a_I)}{\omega_P},
\end{equation}
which def\/ines the initial value of the Hubble parameter\index{Hubble parameter!early light Multiverse}
\begin{equation}\label{maxi}
	H(a_I)=\dfrac{Q}{a^2_I},
\end{equation}
where $Q$ is a constant which depends on two free parameters: the initial data of the mass and the volume of space
\begin{equation}
	Q=\pm\dfrac{V_P}{3V}\omega_P\omega_I= constans.\label{qdef}
\end{equation}
Interestingly, when $\omega_I$ is f\/inite and nonzero and volume of space is inf\/inite then identically $Q\equiv0$, and therefore the initial value of the Hubble parameter\index{Hubble parameter} also vanishes. It proves that for consistency the volume of space must be f\/inite and nonzero. In such a situation the value of the Hubble parameter (\ref{maxi}) is associated for a radiation, and this is the physical sense of this collection of quantum Universes. In this manner, the Multiverse \index{Multiverse}of such a quantum Universes expresses a cosmological nature of light, and by this reason we propose to call this \emph{the light Multiverse}\index{Multiverse!early light}. It can be seen by straightforward derivation from the Hubble parameter (\ref{maxi}) and its def\/inition
\begin{equation}
H(a_I)=\dfrac{1}{a_I}\dfrac{da_I}{dt_I}=\dfrac{1}{a_I^2}\dfrac{da_I}{d\eta_I},
\end{equation}
that the light Multiverse is described by the following initial values of the cosmic scale factor parameter\index{cosmic scale factor parameter}
\begin{eqnarray}
	a_I(t_I,t_0)&=&\sqrt{a_0^2+2Q(t_I-t_0)},\label{ai1}\\
	a_I(\eta_I,\eta_0)&=&a_0+Q(\eta_I-\eta_0),\label{ai2}
\end{eqnarray}
where $t_0$, $\eta_0$, $a_0=a_I(t_0,t_0)=a_I(\eta_0,\eta_0)$ are the integration constants, and
\begin{eqnarray}
a_I(t_I,t_0)&\geqslant&a_0,\label{inequal1}\\
a_I(\eta_I,\eta_0)&\geqslant&a_0.\label{inequal2}
\end{eqnarray}
Because, however, both the scale factor parameters $a_I(t_I,t_0)$ and $a_I(\eta_I,\eta_0)$ are the only constants, one can suggest \emph{ad hoc} that they are equal
\begin{equation}
a_I(t_I,t_0)=a_I(\eta_I,\eta_0)\equiv a_I.
\end{equation}
Such a conjecture\index{conjecture} allows to establish the linkage between the cosmological and the conformal times for the early evolution of the Multiverse\index{Multiverse}
\begin{equation}\label{etat}
  \eta_I-\eta_0=\dfrac{a_0}{Q}\left(\sqrt{1+\dfrac{2Q}{a_0^2}(t_I-t_0)}-1\right).
\end{equation}
There can be also interesting the Taylor expansion\index{Taylor series} of this relation
\begin{equation}\label{etaseries}
  \eta_I-\eta_0=\sqrt{\pi}\sum_{n=0}^\infty\dfrac{(2Q/a_0)^n}{\Gamma(n+2)\Gamma\left(\dfrac{1}{2}-n\right)}(t_I-t_0)^{n+1}.
\end{equation}

In the sense of perturbation theory the proposed quantum cosmology is consistent if and only if the condition (\ref{pertcon}) is satisf\/ied. This inequality can be expressed in terms of the Hubble parameter\index{Hubble parameter}
 \begin{equation}\label{earuni}
 H(a)\ll H(a_I),
 \end{equation}
and straightforwardly integrated. Application of the cosmological time allows to rewrite the inequality (\ref{earuni}) as
\begin{equation}
  \dfrac{1}{a}\dfrac{da}{dt}\ll\dfrac{1}{a_I}\dfrac{da_I}{dt_I},
\end{equation}
and straightforward integration in the region $a_0\leqslant a_I\leqslant a$ gives
\begin{equation}
  \ln\left|\dfrac{a}{a_I}\right|\ll\ln\left|\dfrac{a_I}{a_0}\right|,
\end{equation}
what leads to the bound for cosmic scale factor parameter\index{cosmic scale factor parameter}
\begin{equation}\label{condit}
  a\ll\dfrac{a_I^2}{a_0}.
\end{equation}
In the light of the inequalities (\ref{inequal1}) and (\ref{inequal2}) one has
\begin{equation}
\dfrac{a_I^2}{a_0}=\dfrac{a_I}{a_0}a_I\geqslant a_I,
\end{equation}
results in the bound for cosmic scale factor parameter\index{cosmic scale factor parameter}
\begin{equation}
  a\geqslant a_I,
\end{equation}
which can be expressed equivalently as the bound for redshift
\begin{equation}
z(t_I,t)\leqslant0,
\end{equation}
and def\/ines the early Universe. Similarly, the inequality (\ref{earuni}) can be rewritten in terms of conformal time\index{conformal time}
\begin{equation}
\dfrac{1}{a^2}\dfrac{da}{d\eta}\ll\dfrac{1}{a^2_I}\dfrac{da_I}{d\eta_I}
\end{equation}
and straightforwardly integrated with the result
\begin{equation}
-\dfrac{1}{a}+\dfrac{1}{a_I}\ll-\dfrac{1}{a_I}+\dfrac{1}{a_0},
\end{equation}
which f\/inally gives the bound for cosmic scale factor parameter\index{cosmic scale factor parameter}
\begin{equation}\label{condit1}
a\ll\dfrac{a_0a_I}{2a_0-a_I}.
\end{equation}
Interestingly, the results (\ref{condit}) and (\ref{condit1}) coincide in the only one case $a_I=a_0$.

By this reason, with using of the formulas (\ref{ai1}) and (\ref{ai2}), the early light Multiverse \index{Multiverse}in itself expresses the applicability conditions for the model of quantum cosmology
\begin{eqnarray}
a&\ll&a_0+\dfrac{2Q}{a_0}(t_I-t_0),\\
a&\ll&a_0+\dfrac{Q(\eta_I-\eta_0)}{1-\dfrac{Q}{a_0}(\eta_I-\eta_0)}.
\end{eqnarray}
Because, however, one can suggest also that the cosmic scale factor parameter\index{cosmic scale factor parameter} expressed via cosmological and conformal time\index{conformal time} is the same, particularly in the case of early Universe, one obtains the equation
\begin{equation}\label{eqit}
 \dfrac{2Q}{a_0}(t_I-t_0)=\dfrac{Q(\eta_I-\eta_0)}{1-\dfrac{Q}{a_0}(\eta_I-\eta_0)},
\end{equation}
having the solution
\begin{equation}\label{solll}
\eta_I-\eta_0=\dfrac{2(t_I-t_0)}{a_0+2\dfrac{Q}{a_0}(t_I-t_0)},
\end{equation}
which compared with the relation (\ref{etat})
\begin{equation}
  \dfrac{2(t_I-t_0)}{a_0+2\dfrac{Q}{a_0}(t_I-t_0)}=\dfrac{a_0}{Q}\left(\sqrt{1+\dfrac{2Q}{a_0^2}(t_I-t_0)}-1\right),
\end{equation}
leads to the following equation
\begin{equation}\label{tit0}
2\dfrac{Q}{a_0}(t_I-t_0)\left[\left(2\dfrac{Q}{a_0}(t_I-t_0)\right)^2-2\dfrac{Q}{a_0}(t_I-t_0)-1\right]=0.
\end{equation}
Interestingly, despite the equation (\ref{tit0}) is satisf\/ied for $t_I=t_0$, i.e. if $a_I=a_0$, it also possesses two other solutions
\begin{equation}\label{golrat}
2\dfrac{Q}{a_0}(t_I-t_0)=\varphi_{\pm},
\end{equation}
where $\varphi_{\pm}$ are the irrational constants
\begin{equation}
  \varphi_{\pm}=\dfrac{1\pm\sqrt{5}}{2},
\end{equation}
which in the case $\varphi_+=\varphi\approx 1.6180339887$ is the Fibonacci golden ratio\index{Fibonacci golden ratio}, and in the case $\varphi_-=1-\varphi=\varphi-\sqrt{5}$. Employing the difference $t_I-t_0$ established via (\ref{golrat}) within the equation (\ref{solll}) one receives
\begin{equation}\label{solll1}
\eta_I-\eta_0=\dfrac{a_0}{Q}\dfrac{\varphi_{\pm}}{a_0+\varphi_{\pm}}.
\end{equation}
First let us consider the situation based on the cosmological time. When one knows the value of the difference
\begin{equation}
\tau=t_I-t_0,\label{cosmdif}
\end{equation}
the relation (\ref{golrat}) can be used for determination of the constant $Q$
\begin{equation}
Q=a_0\dfrac{\varphi_{\pm}}{2\tau}.
\end{equation}
Because of $Q=H_Ia_I^2$ one has
\begin{equation}\label{trew}
\dfrac{a_I^2}{a_0}=\dfrac{\varphi_{\pm}}{2H_I\tau},
\end{equation}
i.e. when one knows the initial value of the Hubble parameter\index{Hubble parameter} $H_I$ and the difference (\ref{cosmdif}) then by the condition (\ref{condit}) one obtains
\begin{equation}
  a\ll \dfrac{\varphi_{\pm}}{2H_I\tau}.
\end{equation}
It is easy to see from (\ref{trew}) that the initial data of cosmic scale factor parameter\index{cosmic scale factor parameter} is given by the beginning value of this parameter
\begin{equation}\label{ai}
  a_I=\sqrt{\dfrac{\varphi_{\pm}}{2H_I\tau}a_0},
\end{equation}
what means that for consistency must be $a_0\neq0$. The relations (\ref{trew}) and (\ref{ai}) allow to derive the ratio
\begin{equation}
  \dfrac{a_I}{a_0}=\dfrac{\varphi_{\pm}}{2H_I\tau}\dfrac{1}{a_I}=\sqrt{\dfrac{\varphi_{\pm}}{2H_I\tau}\dfrac{1}{a_0}}
\end{equation}
and consequently the initial data of redshift $z_I=z(t_0,t_I)$
\begin{equation}
  \dfrac{a_I}{a_0}=\dfrac{1}{1+z_I},
\end{equation}
where
\begin{equation}
z(t_0,t)=\exp\left\{\pm\dfrac{a(\eta_0)}{a(t_0)}\int_{t_0}^tdt'\sqrt{\dfrac{1}{M_P\ell_P^2}\dfrac{V_P}{3V}\left|\mathrm{H}_M(t')\right|}\right\}-1,\label{redshift2a}
\end{equation}
can be obtained straightforwardly
\begin{equation}
  z_I=\sqrt{\dfrac{2a_0}{\varphi_{\pm}}H_I\tau}-1,
\end{equation}
what after taking account that $a(\eta_0)=a(t_0)=a_0$ in the relation (\ref{redshift2a}) leads to the result
\begin{equation}\label{hm}
\pm\int_{t_0}^{t_I}dt'\sqrt{\dfrac{1}{M_P\ell_P^2}\dfrac{V_P}{3V}\left|\mathrm{H}_M(t')\right|}=\dfrac{1}{2}\ln\left|\dfrac{2a_0}{\varphi_{\pm}}H_I(t_I-t_0)\right|.
\end{equation}
After differentiating of both sides of the equation (\ref{hm}) with respect to $t_I$ one obtains
\begin{equation}
H_I=\sqrt{\dfrac{1}{M_P\ell_P^2}\dfrac{V_P}{3V}\left|\mathrm{H}_M(t_I)\right|}=\dfrac{1}{2(t_I-t_0)},
\end{equation}
what after straightforward integration
\begin{equation}
  \ln\dfrac{a_I}{a_0}=\dfrac{1}{2}\int_0^{t_I}\dfrac{dt'}{t'-t_0},
\end{equation}
on the one hand allows to establish the initial data of the cosmic scale factor parameter\index{cosmic scale factor parameter}
\begin{equation}
  a_I=a_0\sqrt{\dfrac{\tau}{t_0}},
\end{equation}
and on the other hand leads to the energy of Matter f\/ields\index{Matter f\/ields}
\begin{equation}
\epsilon_M(t_I)=\dfrac{3M_P\ell_P^2}{4\tau^2}\dfrac{V}{V_P}.
\end{equation}
Taking into account the Planck sphere\index{Planck sphere} $V=V_P$ and $\tau=t_P$ one receives
\begin{equation}
\epsilon_M(t_I)=\dfrac{3}{4}E_P.
\end{equation}
Similarly for the Planck cube\index{Planck cube} $V=\ell_P^3$ and $\tau=t_P$ one obtains
\begin{equation}
\epsilon_M(t_I)=\dfrac{9}{16\pi}E_P.
\end{equation}

Applying the formula (\ref{ai}) one receives noth the beginning value and the initial data of the cosmic scale factor parameter\index{cosmic scale factor parameter}
\begin{eqnarray}
  a_0&=&\dfrac{\varphi_{\pm}}{2H_I\tau}\dfrac{t_0}{\tau},\\
  a_I&=&\dfrac{\varphi_{\pm}}{2H_I\tau}\sqrt{\dfrac{t_0}{\tau}}.
\end{eqnarray}

Moreover, because of $Q=\pm\dfrac{V_P}{3V}\omega_I$ one can establish the initial data
\begin{equation}
\omega_I=\pm \dfrac{3}{2}a_0\varphi_\pm\dfrac{1}{\omega_P\tau}\dfrac{V}{V_P},
\end{equation}
which for the case of plus sign becomes
\begin{equation}
\omega_I^+=\dfrac{3}{2}a_0\varphi\dfrac{1}{\omega_P\tau}\dfrac{V}{V_P},
\end{equation}
while for the case of the minus sign is
\begin{equation}
\omega_I^-=\dfrac{3}{2}a_0(\varphi-1)\dfrac{1}{\omega_P\tau}\dfrac{V}{V_P}=\dfrac{\varphi-1}{\varphi}\omega_I^+.
\end{equation}
Interestingly, when one considers the Universe having volume of the Planck cube\index{Planck cube} $V=\ell_P^3$ and the time difference (\ref{cosmdif}) equal to the Planck time $\tau=t_P$
\begin{equation}
  t_P=\sqrt{\dfrac{\hslash G}{c^5}}\approx5.39124\cdot10^{-44}s,
\end{equation}
then one obtains approximatively
\begin{eqnarray}
 \omega_I^+&=&\dfrac{9\varphi}{8\pi}a_0,\\
 \omega_I^-&=&\dfrac{9}{8\pi}(\varphi-1)a_0.
\end{eqnarray}
Similarly, for the Planck sphere\index{Planck sphere} $V=\dfrac{4}{3}\pi\ell_P^3$ one receives
\begin{eqnarray}
  \omega_I^+&=&\dfrac{3}{2}\varphi a_0,\\
  \omega_I^-&=&\dfrac{3}{2}(\varphi-1)a_0.
\end{eqnarray}

Anyway, however, the relations (\ref{ai1}) and (\ref{ai2}) can be used to elimination of the parameter $Q$ by the following combination
\begin{equation}
 \dfrac{a_I^2(t_I,t_0)-a_0^2}{2(t_I-t_0)}=\dfrac{a_I(\eta_I,\eta_0)-a_0}{\eta_I-\eta_0},
\end{equation}
which after taking into account that $a_I(t_I,t_0)=a_I(\eta_I,\eta_0)=a_I$ becomes
\begin{equation}
  \dfrac{a_I+a_0}{2}=\dfrac{t_I-t_0}{\eta_I-\eta_0},
\end{equation}
and leads to another relation between cosmological and conformal times
\begin{equation}\label{aia0}
\eta_I-\eta_0=\dfrac{2}{a_I+a_0}(t_I-t_0),
\end{equation}
which in the case $a_0=0$ becomes
\begin{equation}
\eta_I-\eta_0=\dfrac{2}{a_I}(t_I-t_0).
\end{equation}
Applying the result (\ref{aia0}) to the series (\ref{etaseries}) one obtains the rule
\begin{equation}\label{etaseries1}
  \dfrac{1}{\sqrt{\pi}}=\dfrac{a_I+a_0}{2}\sum_{n=0}^\infty\dfrac{(2Q/a_0)^n}{\Gamma(n+2)\Gamma\left(\dfrac{1}{2}-n\right)}(t_I-t_0)^{n}.
\end{equation}

Let us consider the situation $\eta=\eta_0$ and $t=t_0$. Then the redshift (\ref{redshift1}) and (\ref{redshift2}) must be redef\/ined as follows
\begin{eqnarray}
z(\eta_0,\eta)=\pm a(\eta_0)\int_{\eta_0}^\eta d\eta'\sqrt{\dfrac{1}{M_P\ell_P^2}\dfrac{V_P}{3V}|\mathrm{H}_M(\eta')|},\label{redshift1a}
\end{eqnarray}
Let us denote by $z_0$ the value of the cosmological redshift\index{cosmological redshift}  at the beginning of evolution of the classical Universe. Then by using of the def\/inition (\ref{redshift1a}) one can establish
\begin{equation}
z_0=z(t_0,t_0)\equiv0,
\end{equation}
it is independent on a value of the cosmic scale factor parameter\index{cosmic scale factor parameter} $a_0$ at the beginning of evolution of the Universe. The relations (\ref{etacosm1}) and (\ref{etacosm2}) computed for $\eta=\eta_0$ and $t=t_0$ are
\begin{eqnarray}
  \eta_I-\eta_0&=&t_I-t_0+\int_{a_0}^{a_I}\dfrac{a'-1}{a'^2}\dfrac{da'}{H(a')},\\
  a_0(\eta_I-\eta_0)&=&t_I-t_0+\int_{0}^{z_I}\dfrac{z'dz'}{(1+z')H(z')},
\end{eqnarray}
and by application of the result (\ref{aia0}) become
\begin{eqnarray}
  \left(\dfrac{2}{a_I+a_0}-1\right)(t_I-t_0)&=&\int_{a_0}^{a_I}\dfrac{a'-1}{a'^2}\dfrac{da'}{H(a')},\label{krr1}\\
  \dfrac{a_I-a_0}{a_I+a_0}(t_I-t_0)&=&\int_{0}^{z_I}\dfrac{z'dz'}{(1+z')H(z')}.\label{krr2}
\end{eqnarray}

Interestingly, the LHS of the formula (\ref{krr1}) vanishes identically when the initial data $a_I$ and the beginning value $a_0$ of the cosmic scale factor parameter\index{cosmic scale factor parameter} are constrained by the equation
\begin{equation}
a_I+a_0=2.\label{cony}
\end{equation}
If one expresses the integral on the RHS of (\ref{krr1}) via time variables then this equation says that
\begin{equation}
  0=\int_{t_0}^{t_I}dt'-\int_{\eta_0}^{\eta_I}d\eta',
\end{equation}
what leads to the equality between the differences
\begin{equation}
  \eta_0-\eta_I=t_0-t_I,\label{etac}
\end{equation}
what is consistent with the relation (\ref{aia0}). Such an equality means that if the condition (\ref{cony}) holds then in such a region of evolution of the Universe the conformal time\index{conformal time} and the cosmological time f\/low in such a way that their difference is constant
\begin{equation}
\eta-t=constans,\label{times}
\end{equation}
and for convenience can be taken equal to zero. The equation (\ref{times}) expresses the law of conservation for the difference $\eta-t$. In such a situation the second relation (\ref{krr2}) becomes
\begin{equation}
  (1-a_0)(t_I-t_0)=\int_{0}^{z_I}\dfrac{z'dz'}{(1+z')H(z')},\label{krr2a}
\end{equation}
where the initial data of redshift is
\begin{equation}
z_I=2\dfrac{a_0-1}{2-a_0}=2\dfrac{1-a_I}{a_I}.
\end{equation}
Interestingly, when $a_0=1$ then $a_I=1$ the initial data of redshift is trivial $z_I=0$, and the LHS of the formula (\ref{krr2a}) identically vanishes. After expression of the integral on the RHS of (\ref{krr2a}) via time variables one obtains
\begin{equation}
  0=\int_{t_0}^{t_I}z(t_0,t')dt'=\int_{t_0}^{t_I}\left(\dfrac{a(t_0)}{a(t')}-1\right)dt'=a_0\int_{\eta_0}^{\eta_I}d\eta'-\int_{t_0}^{t_I}dt',
\end{equation}
what can be computed straightforwardly
\begin{equation}
  a_0(\eta_I-\eta_0)=t_I-t_0,
\end{equation}
and for $a_0=1$ leads once again to the law of conservation (\ref{etac}).

Let us see what happens in the particular situation for which $a_0=0$. First let us establish the initial data of redshift $z_I$
\begin{equation}
  z_I=-1.
\end{equation}
Because of the constraint (\ref{cony}) one has $a_I=2$ and therefore the relations (\ref{krr1}) and (\ref{krr2}) take the form
\begin{eqnarray}
  0&=&\int_{0}^{2}\dfrac{a'-1}{a'^2}\dfrac{da'}{H(a')},\label{zero0}\\
 t_0-t_I&=&\int_{-1}^0\dfrac{z'dz'}{(1+z')H(z')}.\label{tauu}
\end{eqnarray}
Interestingly, in such a case the equation (\ref{eqit}) is simplif\/ied
\begin{equation}\label{eqit1}
 \dfrac{Q}{2}(\eta_I-\eta_0)^2-(t_I-t_0)=0,
\end{equation}
and in the light of the relation (\ref{etac}) allows to establish the value of the parameter $Q$ crucial for the initial data Hubble law\index{Hubble law} (\ref{maxi})
\begin{equation}
Q=2\dfrac{t_I-t_0}{(\eta_I-\eta_0)^2}=\pm\dfrac{2}{\tau},
\end{equation}
where we have denoted $\tau=\eta_I-\eta_0=t_I-t_0$. In the light of the def\/inition (\ref{qdef}) one can determine the initial data $m_I$
\begin{equation}
\omega_I=\dfrac{6V}{V_P}\dfrac{1}{\omega_P\tau}.
\end{equation}
For f\/inite volume of space $V$ and f\/inite the time $\tau$ the initial data $\omega_I$ are established consistently. The time $\tau$ can be interpreted as the time between the beginning of the Universe and creation of initial data. If its value is taken \emph{ad hoc} as identical to the Planck time $\tau=t_P$ then
\begin{equation}
\omega_I=\dfrac{6V}{V_P},
\end{equation}
and therefore if value of $\omega_I$ is small, $\omega_I\sim1$ say, the volume of space is \begin{equation}
V\sim\dfrac{V_P}{6}.
\end{equation}
Such a volume can be treated as def\/inition of the early Multiverse\index{Multiverse}.

In the light of the result (\ref{tauu}) one obtains another def\/inition of the time $\tau$
\begin{equation}
 \tau=-\int_{-1}^0\dfrac{z'dz'}{(1+z')H(z')},\label{taur}
\end{equation}
which by using of the fact
\begin{equation}
H(z')=-\dfrac{1}{1+z'}\dfrac{dz'}{dt'},
\end{equation}
and $z'=z(t_0,t')$ becomes
\begin{equation}
 \tau=\int_{t_0}^{t_I}z(t_0,t')dt',\label{taur1}
\end{equation}
or after expressing via energy of Matter f\/ields\index{Matter f\/ields}
\begin{equation}
\tau=\int_{t_0}^{t_I}\left[\exp\left[\pm\int_{t_0}^{t'}dt''\sqrt{\dfrac{1}{M_P\ell_P^2}\dfrac{V_P}{3V}\epsilon_M(t'')}\right]-1\right]dt'.\label{taur2}
\end{equation}
f\/inally results in the relation
\begin{equation}\label{tauuu}
\tau=\dfrac{1}{2}\int_{t_0}^{t_I}\exp\left[\pm\int_{t_0}^{t'}dt''\sqrt{\dfrac{1}{M_P\ell_P^2}\dfrac{V_P}{3V}\epsilon_M(t'')}\right]dt'.
\end{equation}
The natural generalization of the equation (\ref{tauuu}) is
\begin{equation}\label{tauuu1}
t_2-t_1=\dfrac{1}{2}\int_{t_1}^{t_2}\exp\left[\pm\int_{t_1}^{t'}dt''\sqrt{\dfrac{1}{M_P\ell_P^2}\dfrac{V_P}{3V}\epsilon_M(t'')}\right]dt',
\end{equation}
which in the case $t_1=0$, and $t_2=t$ can be used for determination of the cosmological time
\begin{equation}\label{tauuu2}
t-t_0=\dfrac{1}{2}\int_{t_0}^{t}\exp\left[\pm\int_{t_0}^{t'}dt''\sqrt{\dfrac{1}{M_P\ell_P^2}\dfrac{V_P}{3V}\epsilon_M(t'')}\right]dt',
\end{equation}
In the light of the relations (\ref{hubble}) and (\ref{endens}) one has
\begin{equation}\label{hzem}
  H(z')=\sqrt{\dfrac{1}{M_P\ell_P^2}\dfrac{V_P}{3V}\epsilon_M(z')},
\end{equation}
where $\epsilon_M(z')$ is energy of Matter f\/ields\index{Matter f\/ields}. For given energy of Matter f\/ields\index{Matter f\/ields} dependence on redshift $z'$ the equation (\ref{hzem}) can be solved as a differential equation for the redshift as unknown function of cosmological or conformal time\index{conformal time}. Application of the explicit form of $\epsilon_M(z')$ to the formula (\ref{taur}) shall be resulting in the exact evaluation of the time $\tau$. Equivalently, the redshift derived from the evolution (\ref{hzem}) can be applied to the formula (\ref{taur1}) for determination of the time $\tau$.

Interestingly, one can apply the equation (\ref{hzem}) to the generalized relation (\ref{tauuu2})
\begin{eqnarray}
t-t_0&=&\dfrac{1}{2}\int_{t_0}^{t}\exp\left[\pm\int_{t_0}^{t'}dt''\dfrac{1}{a(t'')}\dfrac{da(t'')}{dt''}\right]dt'=\nonumber\\
&=&\dfrac{1}{2}\int_{t_0}^{t}\exp\left[\pm\int_{a_0}^{a(t')}\dfrac{da}{a}\right]dt'=\dfrac{1}{2}\int_{t_0}^{t}\exp\left[\pm\ln\dfrac{a(t')}{a_0}\right]dt'=\nonumber\\
&=&\dfrac{1}{2}\int_{t_0}^{t}\left(\dfrac{a(t')}{a_0}\right)^{\pm1}dt',\label{tauuu3}
\end{eqnarray}
and it is evident that the case of the plus sign, i.e. an expanding Universe, is distinguishable from the case of the minus sign, i.e. a collapsing Universe. The case of plus sign gives
\begin{eqnarray}
  (t-t_0)^+&=&\dfrac{1}{2a_0}\int_{t_0}^{t}a(t')dt'=\dfrac{1}{2a_0}\int_{t_0}^{t}a^2(t')\dfrac{dt'}{a(t')}=\nonumber\\
  &=&\dfrac{1}{2a_0}\int_{\eta_0}^{\eta}a^2(\eta')d\eta',\label{cosmexp}
\end{eqnarray}
what after application of the def\/inition of the Hubble parameter expressed in terms of conformal time\index{conformal time}
\begin{equation}
a^2(\eta)d\eta=\dfrac{da}{H(a)},
\end{equation}
allows to establish the result
\begin{equation}\label{cosmexp1}
(t-t_0)^+=\dfrac{1}{2a_0}\int_{a_0}^{a}\dfrac{da'}{H(a')}.
\end{equation}
The case of the minus sign leads to completely different result
\begin{equation}\label{cosmopp}
  (t-t_0)^-=\dfrac{a_0}{2}\int_{t_0}^{t}\dfrac{dt'}{a(t')}=\dfrac{a_0}{2}\int_{a_0}^{a}\dfrac{da'}{a'^2H(a')},
\end{equation}
or after straightforward computation
\begin{equation}\label{cosmopp1}
  (t-t_0)^-=\dfrac{a_0}{2}\int_{\eta_0}^{\eta}d\eta'=\dfrac{a_0}{2}(\eta-\eta_0),
\end{equation}
which allows to determine the nontrivial relation between the cosmological and the conformal time\index{conformal time}
\begin{equation}
  t=\dfrac{1}{2}a_0\eta.\label{teta}
\end{equation}
Because of the plus sign is related to expansion of the Multiverse \index{Multiverse}and the minus sign describes the opposite situation, the relations for the cosmological time (\ref{cosmexp}) and (\ref{cosmopp}) mean that the f\/low of cosmological time in an expanding Multiverse is distinguish then the f\/low in a collapsing Multiverse. Strictly speaking the equation is satisf\/ied
\begin{equation}
  a_0^2d(t-t_0)^+=a^2d(t-t_0)^-,\label{dift}
\end{equation}
and if $t_0^-=t_0^+$ then one has the condition
\begin{equation}
  a_0^2dt^+-a^2dt^-=0.\label{dift1}
\end{equation}
In other words the cosmological time f\/lows in the same way if and only if for arbitrary value of the conformal time\index{conformal time} or the cosmological time the relation $a^2\equiv a_0^2$ holds identically. In such a situation the relation (\ref{teta}) is automatically satisf\/ied, while the value of the Hubble parameter is trivial $H=\dfrac{1}{a^2}\dfrac{da}{d\eta}=0$. In this manner such an approach suggests that the Multiverse \index{Multiverse!non dynamical} has non dynamical nature or is dynamical but considered in a f\/ixed moment of time.

On the other side, however, it can be seen straightforwardly that application of the trivial value $H=0$ of the Hubble parameter within the def\/initions (\ref{cosmexp1}) and (\ref{cosmopp}) of the cosmological time leads to manifestly divergent integrands. Such a singular behavior suggests that the non dynamical Multiverse\index{Multiverse!non dynamical} has non physical nature or, in other words, that the f\/ixation of time has purely non physical nature. However, from the point of view of modern physics f\/ixation of time is the standard tool in the theoretical explanations of quantum f\/ield theory.

There is another possible interpretation of the relations (\ref{dift}) and (\ref{dift1}). Namely, one can say that during the expansion of the Multiverse\index{Multiverse!expansion} the beginning value of the cosmic scale factor parameter\index{cosmic scale factor parameter} is \emph{ad hoc} established to a certain value $a_0$, while during a collapse of the Multiverse\index{Multiverse!collapse} the beginning value is undetermined. In such a situation the identif\/ication $a^2=a_0^2$ means exactly that Multiverse is non expanding and is not collapsing, i.e. is non dynamical \index{Multiverse!non dynamical}.

\section{Summary}
The Multiverse, which we understand as the collection of multiple quantum universes, obtained via the quantization in the static Fock space\index{Fock space} of creators and annihilators applied to the classical Einstein--Friedmann Universe\index{Einstein--Friedmann Universe}, has showed us the general way for the constructive scenario of the physics of the observed Universe. By straightforward computations, involving mainly elementary mathematical analysis, we have received the thermodynamics of the system of many quantum universes, which in itself is the result possessing both the most natural and strong value for phenomenology and empirical verif\/ication of the theoretical results of the model. This elegant feature allows to conclude that the proposed programme of construction of the model of the quantum universe has been realized via using of the simple solution of the Einstein f\/ield equations\index{Einstein f\/ield equations}.

We have presented few important f\/indings of such a formulation. These are particularly:
\begin{enumerate}
\item Formulation of quantum cosmology via the models of theoretical physics having well-established value for phenomenology.
\item Rational and simple scheme of natural emergence of observed Universe as the system of multiple quantum universes - the Multiverse \index{Multiverse}in our understanding.
\item Ideological unif\/ication of quantum cosmology with the most fundamental natural sciences, like organic chemistry and evolutionary biology, in wider sense expressing neglecting of interference of supernatural forces in creation and development of Universe as misleading and groundless. Establishing of the Multiverse hypothesis\index{Multiverse} as the fundamental landscape for understanding of the observed Universe.
\item Description of (very) early Universe as the static Multiverse of the superf\/luid Fermi--Bose superstrings\index{Fermi--Bose superstrings}, which in general can be open or closed.
\item A conceptual way to understanding a physical role of quantum gravity, string theory\index{string theory}, and supersymmetry\index{supersymmetry} for birth and early evolution of the observed Universe.
\item Essential cosmological role of superf\/luidity and light for the Multiverse hypothesis.
\end{enumerate}

To see the sense of the proposed strategy, let us sketch brief\/ly the crucial elements of the constructed model of quantum cosmology
\begin{enumerate}
  \item Description of a certain selected solution of the Einstein f\/ield equations\index{Einstein f\/ield equations} in frames of the Hamiltonian approach jointing the Dirac approach and the Arnowitt--Deser--Misner Hamiltonian formulation of General Relativity\index{Hamiltonian formulation of General Relativity!Arnowitt--Deser--Misner (ADM)}.
  \item Application the received Hamiltonian constraint\index{Hamiltonian constraint} the methods of the primary\index{quantization!primary canonical} and the secondary quantization \index{quantization!secondary canonical}are applied, and the one-dimensional Dirac equation \index{Dirac equation!one-dimensional}is obtained.
  \item Elementary formulation of the thermodynamics of quantum states of the selected metric by application of the methods of statistical mechanics based on the static Fock rep\`{e}re\index{Fock rep\`{e}re}.
\end{enumerate}

There is an obvious conjecture\index{conjecture} following from such a programme of quantum cosmology. Namely, the proposed strategy of construction of quantum cosmology can be generalized for another solutions of the Einstein f\/ield equations\index{Einstein f\/ield equations}, and in result the appropriate models of quantum gravity can be straightforwardly obtained. In other words, the Multiverse hypothesis\index{Multiverse} based on the one-dimensional Dirac equation\index{Dirac equation!one-dimensional} can be straightforwardly generalized onto all solutions of the Einstein f\/ield equations\index{Einstein f\/ield equations} which can be parametrized by application of the Arnowitt--Deser--Misner decomposition\index{Arnowitt--Deser--Misner decomposition}. Such a strategy would be resulting in the elegant transition between quantum cosmology and quantum gravity. We shall present a certain idea for constructive realization of such a general strategy in the next chapters of this book.

\chapter{The Inf\/lationary Multiverse}
The quantum cosmology based on the one-dimensional Klein--Gordon equation can be applied straightforwardly. Let us consider an application of the one-dimensional quantum cosmology which leads to new results. It must be emphasized that this chapter is rather far from the main stream of this part. However, its content in itself is an essential linkage between the Multiverse\index{Multiverse} cosmology presented in this chapter and one of the most intriguing ideas of the modern theoretical cosmology, which is inf\/lation\index{inf\/lation}. In this section we shall the particular situation within the general idea of inf\/lation. This situation is the inf\/lation due to the Higgs inf\/laton\index{Higgs inf\/laton}.

\section{The Inf\/lationary Cosmology}
One of the main subjects of inf\/lationary cosmology is the theory of inf\/lationary cosmological perturbations of quantum-mechanical origin (For numerous details see e.g. the Ref. \cite{inflation}). The idea to go beyond the isotropic and homogeneous space-time given by the Friedmann--Lema\^{i}tre--Robertson--Walker metric\index{Friedmann--Lema\^{i}tre--Robertson--Walker metric}, for which the interval expressed via the conformal time\index{conformal time} $\eta$ has the form
\begin{equation}
ds^2=a^2(\eta)\left(-c^2d\eta^2+\delta_{ij}dx^idx^j\right).
\end{equation}
The question of is how small quantum perturbations around this solution of the Einstein f\/ield equations\index{Einstein f\/ield equations} behave during inf\/lation\index{inf\/lation}, the phase of accelerated expansion that took place in the early universe. It can be seen that the corresponding physics is similar to the Schwinger effect\index{Schwinger effect}\cite{martin}. In General Relativity, inf\/lation can be obtained by domination of a f\/luid which pressure is negative. Since, at very high energies, quantum f\/ield theory is the natural candidate to describe matter, it is natural and simple to postulate that a scalar f\/ield, called \emph{the inf\/laton}\index{inf\/laton} was responsible for the evolution of the universe in this regime. The action of Matter f\/ields\index{Matter f\/ields} which is considered in inf\/lationary cosmology has the form
\begin{equation}\label{scalaract}
\mathcal{S}=\dfrac{1}{c}\int_{M}d^4x\sqrt{-g}\left(\dfrac{E_P\ell_P}{2}g^{\mu\nu}\partial_\mu\varphi\partial_\nu\varphi+\dfrac{1}{\ell_P^3}V(\varphi)\right),
\end{equation}
where $\varphi$ is the inf\/laton f\/ield. It must be emphasized that the action (\ref{scalaract}) of scalar f\/ield was complemented by the Planck units for dimensional correctness of the action, such that the potential $V$ has a dimension of energy. In the most general situation the interval of the perturbed Friedmann--Lema\^{i}tre--Robertson--Walker metric\index{Friedmann--Lema\^{i}tre--Robertson--Walker metric} is \cite{mfb}
\begin{equation}
ds^2=a^2(\eta)\left(-c^2(1-2\phi)d\eta^2+2B_{,i}dx^id\eta+\left[(1-2\psi)\delta_{ij}+2E_{,ij}+h_{ij}\right]dx^idx^j\right),
\end{equation}
where the functions $\phi$, $B$, $\psi$ and $E$ represent the scalar sector whereas the tensor $h_{ij}$, satisfying $h_i^i=h_{ij}^{;j}=0$ describes the gravitational waves\index{gravitational waves}. There are no vector perturbations because a single scalar f\/ield cannot seed rotational perturbations. At the linear level, the two types of perturbations decouple and, therefore, can be treated separately. In the case of scalar perturbations of the geometry evoked above, by freedom to choose the coordinate system the four functions are in fact redundant, and the scalar f\/luctuations of the geometry can be characterized by the gauge-invariant Bardeen potential \cite{bardeen}
\begin{equation}
  \Phi_B=\phi+\dfrac{1}{a}[a(B-E')]',
\end{equation}
where prime means $\eta$-differentiation. The gauge-invariant perturbation which characterizes the f\/luctuations in the inf\/laton scalar f\/ield is
\begin{equation}
  \delta\varphi^g(\eta,x)=\delta\varphi+\varphi'(B-E').
\end{equation}
Because of the perturbed Einstein f\/ield equations\index{Einstein f\/ield equations} couple the Bardeen potential and the gauge-invariant perturbation one has one degree of freedom. Therefore the scalar sector formalism is reduced to study of the Mukhanov--Sasaki variable\index{Mukhanov--Sasaki variable}
\begin{equation}
  v(\eta,x)=a\sqrt{\dfrac{S_P}{\kappa c}}\left[\delta\varphi^g+\varphi'\dfrac{\Phi_B}{H}\right],
\end{equation}
where $S_P=4\pi\ell_P^2$ is the area of the Planck sphere\index{Planck sphere}, and $H=\dfrac{a'}{a}$ is the Hubble parameter\index{Hubble parameter}. Usually inf\/lationary perturbations are formulated in terms of the variable
\begin{equation}
  \mu_S(\eta,x)=-\sqrt{\kappa\hslash}v(\eta,x)=-2a\sqrt{\gamma}\zeta(\eta,x),
\end{equation}
where $\zeta(\eta,x)$ is the conserved quantity
\begin{equation}
  \zeta=\dfrac{\mathcal{H}^{-1}\Phi_B'+\Phi_B}{\dfrac{(\varphi')^2}{2a^2}-\mathrm{V}(\varphi)}+\Phi_B,
\end{equation}
and $\gamma$ is the background function
\begin{equation}
  \gamma=1-\dfrac{\mathcal{H}'}{\mathcal{H}^2}.
\end{equation}
Here $\mathcal{H}=\dfrac{a'}{a}$ is the conformal Hubble parameter\index{Hubble parameter!conformal}, which is connected with the Hubble parameter\index{Hubble parameter} $H=\dfrac{\dot{a}}{a}$ by the relation $\mathcal{H}=aH$. The automatically gauge-invariant tensor sector is described by the quantity $µ_T(\eta,x)$ def\/ined according by
\begin{equation}
  h_{ij}=\dfrac{\mu_T}{a}Q^{TT}_{ij},
\end{equation}
where $Q^{TT}_{ij}$ are the transverse and traceless eigentensors of the Laplace operator on the space-like sections. Usually the perturbations are studied mode by mode via using the Fourier transforms\index{Fourier transforms} $\widetilde{\mu}_{S}(\eta,k)$ and $\widetilde{\mu}_{T}(\eta,k)$
\begin{eqnarray}
\widetilde{\mu}_{S}(\eta,k)&=&\int d^3x\mu_S(\eta,x)e^{-ikx},\\
\widetilde{\mu}_{T}(\eta,k)&=&\int d^3x\mu_T(\eta,x)e^{-ikx},
\end{eqnarray}
obeying the following Euler--Lagrange equations of motion\index{Euler--Lagrange equations of motion}
\begin{eqnarray}
  \dfrac{d^2\widetilde{\mu}_S(\eta,k)}{d\eta^2}+\omega_S(k,\eta)\widetilde{\mu}_S(\eta,k)&=&0,\\
  \dfrac{d^2\widetilde{\mu}_T(\eta,k)}{d\eta^2}+\omega_T(k,\eta)\widetilde{\mu}_T(\eta,k)&=&0,
\end{eqnarray}
following from the second variation of the Einstein--Hilbert action\index{Einstein--Hilbert action}. Here $\omega_S(k,\eta)$ and $\omega_T(k,\eta)$ are the frequencies
\begin{eqnarray}
  \omega^2_S(k,\eta)&=&k^2c^2-\dfrac{(a\sqrt{\gamma})''}{a\sqrt{\gamma}},\\
  \omega^2_T(k,\eta)&=&k^2c^2-\dfrac{a''}{a}.
\end{eqnarray}

The cosmological perturbations obey exactly the same type of equation as a scalar f\/ield $\Phi(t,x)$ interacting with a classical electric f\/ield in the Schwinger effect\index{Schwinger effect}, i.e. the equation of a parametric oscillator
\begin{equation}
  \ddot{\widetilde{\Phi}}(t,k)+\omega^2(k,t)\widetilde{\Phi}(t,k)=0,
\end{equation}
where the frequency has the form
\begin{equation}
  \omega^2(k,t)=k^2c^2+\dfrac{m^2c^4}{\hslash^2}-2\dfrac{c^2}{\hslash}eEk_zt+\dfrac{1}{\hslash^2}e^2E^2t^2,
\end{equation}
where $e$ is the elementary charge, and $E$ is the electric f\/ield. The only difference is the classical source which, in the case of cosmological perturbations, is the background gravitational f\/ield. Also the time dependence of the frequencies is qualitatively different. The primary canonical quantization\index{quantization!primary canonical} of the theory proceeds as in the usual Schwinger effect\index{Schwinger effect}. The consequence of the interaction between the quantum cosmological perturbations and the classical background is creation of particles, which in the context of inf\/lation are gravitons. Classically, this corresponds to the amplif\/ication growing mode of the f\/luctuations.

Let us consider the slow-roll parameters\index{slow-roll parameters} \cite{martin1}
\begin{eqnarray}
  \epsilon&=&3\dfrac{\dfrac{M_P\ell_P}{2}\dot{\varphi}^2}{\dfrac{M_P\ell_P}{2}\dot{\varphi}^2+\dfrac{1}{\ell_P^3}\mathrm{V}(\varphi)}=-\dfrac{\dot{H}}{H^2},\\
  \delta&=&-\dfrac{\ddot{\varphi}}{H\dot{\varphi}}=\epsilon-\dfrac{1}{2H}\dfrac{\dot{\epsilon}}{\epsilon},\label{deldef1}\\
  \xi&=&\dfrac{\dot{\epsilon}-\dot{\delta}}{H},\label{xidef1}
\end{eqnarray}
obeying the equations of motion
\begin{eqnarray}
\dfrac{\dot{\epsilon}}{H}&=&2\epsilon(\epsilon-\delta),\\
\dfrac{\dot{\delta}}{H}&=&2\epsilon(\epsilon-\delta)-\xi.
\end{eqnarray}
It is convenient to express the slow-roll parameters\index{slow-roll parameters} via the horizon f\/low functions\index{horizon f\/low functions} $\epsilon_1$, $\epsilon_2$, and $\epsilon_3$
\begin{eqnarray}
  \epsilon&=&\epsilon_1\label{epsdef2},\\
  \delta&=&\epsilon_1-\dfrac{1}{2}\epsilon_2\label{deldef2},\\
  \xi&=&\dfrac{1}{2}\epsilon_2\epsilon_3.\label{xidef2}
\end{eqnarray}
It is easy to see that $\dfrac{\epsilon_1}{3}$ measures the ratio of of the kinetic energy to the total energy, whereas $\epsilon_2$ represents a model where the kinetic energy itself increases, when $\epsilon_2>0$, or decreases when $\epsilon_2<0$ with respect to the total energy. Provided the slow-roll conditions are valid $\epsilon_{1,2}\ll1$ one can express the slow-roll parameters\index{slow-roll parameters} via inf\/laton potential
\begin{eqnarray}
  \epsilon_1&\simeq&\dfrac{1}{4S_P}\left(\dfrac{\mathrm{V}'}{\mathrm{V}}\right)^2,\label{eps1def}\\
  \epsilon_2&\simeq&\dfrac{1}{S_P}\left[\left(\dfrac{\mathrm{V}'}{\mathrm{V}}\right)^2-\dfrac{\mathrm{V}''}{\mathrm{V}}\right],\label{eps2def}
\end{eqnarray}
where prime denotes $\varphi$-differentiation, and $S_P=4\pi\ell_P^2$ is the area of the Planck sphere\index{Planck sphere}. Derivation of the third horizon f\/low function $\epsilon_3$ is rather tedious. Straightforward application of the def\/initions (\ref{epsdef2}), (\ref{deldef2}), and (\ref{deldef1}) to the def\/inition (\ref{xidef1}), and taking into account the def\/inition (\ref{xidef2}) leads to the result
\begin{equation}
  \epsilon_3=2\epsilon_1\dfrac{\epsilon_2'}{\epsilon_1'}.
\end{equation}
while by using the def\/initions (\ref{eps1def}) and (\ref{eps2def}) one receives
\begin{equation}
  \dfrac{\epsilon_2'}{\epsilon_1'}=\left(1-\dfrac{1}{2}\dfrac{\dfrac{\mathrm{V}'''}{\mathrm{V}'}-\dfrac{\mathrm{V}''}{\mathrm{V}}}{\dfrac{\mathrm{V}''}{\mathrm{V}}-\left(\dfrac{\mathrm{V}'}{\mathrm{V}}\right)^2}\right)^{-1}.
\end{equation}
Therefore the f\/inal result can be presented in the compact form
\begin{equation}
\epsilon_3\simeq\dfrac{1}{2S_P}\left(\dfrac{\mathrm{V}'}{\mathrm{V}}\right)^2\left(1-\dfrac{1}{2}\dfrac{\dfrac{\mathrm{V}'''}{\mathrm{V}'}-\dfrac{\mathrm{V}''}{\mathrm{V}}}{\dfrac{\mathrm{V}''}{\mathrm{V}}-\left(\dfrac{\mathrm{V}'}{\mathrm{V}}\right)^2}\right)^{-1}.
\end{equation}
The frequencies of the gauge-invariant cosmological perturbations can be written as
\begin{eqnarray}
  \omega_S^2(k,\eta)&=&\omega_P^2\left[\ell_P^2k^2-\left(2+3\delta\right)\dfrac{t_P^2}{\eta^2}\right],\\
  \omega_T^2(k,\eta)&=&\omega_P^2\left[\ell_P^2k^2-\left(2+3\epsilon\right)\dfrac{t_P^2}{\eta^2}\right].
\end{eqnarray}
\section{The Power Law Inf\/laton}
Let us consider f\/irst the inf\/laton potential in the form of the power law\index{power law}
\begin{equation}
  \mathrm{V}(\varphi)=-\dfrac{\ell_P^p}{E_P}\dfrac{m^2c^4}{2}\varphi^p,
\end{equation}
where $m$ is the mass of the power law inf\/laton $\varphi$.\index{power law!inf\/laton}

In such a situation the horizon f\/low functions are easy to derive
\begin{eqnarray}
  \epsilon_1&\simeq&\dfrac{1}{4S_P}\dfrac{p^2}{\varphi^2},\label{eps1def}\\
  \epsilon_2&\simeq&\dfrac{1}{S_P}\dfrac{p}{\varphi^2},\label{eps2def}\\
  \epsilon_3&\simeq&-\dfrac{1}{2S_P}\dfrac{p^3/(p+1)}{\varphi^2},
\end{eqnarray}
so that the slow roll parameters have the form
\begin{eqnarray}
  \epsilon&=&\dfrac{1}{4S_P}\dfrac{p^2}{\varphi^2},\\
  \delta&=&\dfrac{1}{4S_P}\dfrac{p(p-2)}{\varphi^2},\\
  \xi&=&-\dfrac{1}{2S_P^2}\dfrac{p^4/(p+1)}{\varphi^4}.
\end{eqnarray}
In this manner one can establish straightforwardly the frequencies of the scalar and the tensor cosmological perturbations
\begin{eqnarray}
  \omega_S^2(k,\eta)&=&\omega_P^2\left[\ell_P^2k^2-\left(2+\dfrac{3}{4S_P}\dfrac{p(p-2)}{\varphi^2}\right)\dfrac{t_P^2}{\eta^2}\right],\\
  \omega_T^2(k,\eta)&=&\omega_P^2\left[\ell_P^2k^2-\left(2+\dfrac{3}{4S_P}\dfrac{p^2}{\varphi^2}\right)\dfrac{t_P^2}{\eta^2}\right].
\end{eqnarray}
The problem is to establish the total frequency of the inf\/lationary cosmological perturbations of the power law inf\/laton\index{power law!inf\/laton}. Let us postulate such an effective frequency by the Pythagorean theorem
\begin{equation}
  \omega_{\textrm{eff}}^2=\omega_S^2(k,\eta)+\omega_T^2(k,\eta),\label{omh}
\end{equation}
which we shall call the \emph{Pythagorean frequency}\index{Pythagorean frequency}, which has the following explicit form
\begin{equation}
  \omega_{\textrm{eff}}=\sqrt{2\omega_P^2\left[\ell_P^2k^2-\left(2+\dfrac{3}{4S_P}\dfrac{p(p-1)}{\varphi^2}\right)\dfrac{t_P^2}{\eta^2}\right]}.
\end{equation}
One can consider the effective energy of the cosmological perturbations of the power law inf\/laton\index{power law!inf\/laton}. Let us propose \emph{ad hoc} that such an energy of the inf\/lationary cosmological perturbations is simply given by the Planck wave-particle duality relation\index{Planck relation}\index{wave-particle duality}
\begin{equation}
  E_{\textrm{eff}}=\hslash\omega_{\textrm{eff}},
\end{equation}
so that applying the Pythagorean frequency\index{Pythagorean frequency} (\ref{omh}) one obtains
\begin{equation}\label{5}
  E_{\textrm{eff}}=\sqrt{2E_P^2\left[\ell_P^2k^2-\left(2+\dfrac{3}{4S_P}\dfrac{p(p-1)}{\varphi^2}\right)\dfrac{t_P^2}{\eta^2}\right]}.
\end{equation}
This chapter will be focused on discussion of several consequences for possible physical meaning of the Higgs inf\/laton following from the energy formula (\ref{5}).

\section{The Higgs--Hubble Inf\/laton}
Let us discuss f\/irst the nontrivial linkage of the power law inf\/laton\index{power law!inf\/laton} with the Multiverse \index{Multiverse}model presented in the previous chapter of this part. For this let us introduce the auxiliary f\/ield $\varphi=\varphi(\eta)$, which expressed in terms of the Planck units\index{Planck units} is
\begin{equation}\label{g}
\varphi(\eta)=\varphi_0a(\eta),
\end{equation}
where $a(\eta)$ is the cosmic scale factor parameter\index{cosmic scale factor parameter}, $\eta$ is the conformal time\index{conformal time} describing the classical general relativistic evolution of the Friedmann--Lema\^{i}tre--Robertson--Walker metric\index{Friedmann--Lema\^{i}tre--Robertson--Walker metric}, and $\varphi_0=\dfrac{\alpha}{\ell_P}$ is the initial datum of the auxiliary f\/ield $\varphi(\eta)$ (with $a(\eta_0)=1$, and dimensionless $\alpha$).

The Einstein--Hilbert action\index{Einstein--Hilbert action} of General Relativity evaluated on the Friedmann--Lema\^{i}tre--Robertson--Walker metric \index{Friedmann--Lema\^{i}tre--Robertson--Walker metric} $g_{\mu\nu}=\mathrm{diag}[-1,a^2(t)\delta_{ij}]$, which we established in the previous section as (\ref{action1}), can be presented in the following form
\begin{equation}\label{actino}
S[\varphi]=\int{d}\eta\left[M_P\ell_PV\dfrac{3}{2}\dfrac{\alpha^2\ell_P^3}{V_P}\varphi'^2+\dfrac{\varphi^4}{\varphi_0^4}\epsilon_M(\eta)\right],
\end{equation}
where $V=\int d^3x<\infty$ is the spatial volume, $\eta$ is the conformal time\index{conformal time}, $d\eta=\dfrac{dt}{a(t)}$, prime denotes $\eta$-differentiation, $\epsilon_M(\eta)=\int d^3x\mathcal{H}_{M}(x,\eta)$ is the energy of Matter f\/ields. In this manner the most convenient choice of the constant parameter $\alpha$ is
\begin{equation}
  \alpha=\dfrac{2}{3}\sqrt{\pi},
\end{equation}
so that the auxiliary f\/ield is
\begin{equation}\label{gl}
\varphi(\eta)=\varphi_0a(\eta)\quad,\quad\varphi_0=\dfrac{2\sqrt{\pi}}{3\ell_P}\approx7.3109596\cdot10^{34}\dfrac{1}{\mathrm{m}},
\end{equation}
and the Einstein--Hilbert action\index{Einstein--Hilbert action} becomes
\begin{equation}\label{actinox}
S[\varphi]=\int{d}\eta\left[\dfrac{M_P\ell_P}{2}V\varphi'^2+\dfrac{\varphi^4}{\varphi_0^4}\epsilon_M(\eta)\right].
\end{equation}

Application of the conjugate momentum
\begin{equation}
P_{\varphi}=\dfrac{1}{\ell_P^2}\dfrac{\delta S[\varphi]}{\delta\varphi'}=\dfrac{M_PV}{\ell_P}\varphi',
\end{equation}
allows to present the action (\ref{actinox}) in the Hamilton form by application of the Legendre transformation\index{Legendre transformation}
\begin{equation}
S[\varphi]=\int d\eta\left\{\ell_P^2P_\varphi\varphi'-H(\eta)\right\},
\end{equation}
where $H(\eta)$ is the Hamiltonian
\begin{equation}
H(\eta)=\dfrac{\ell_P^3}{V}\dfrac{P_\varphi^2}{2M_P}-\dfrac{\varphi^4}{\varphi^4_0}\epsilon_M(\eta)\approx0,
\end{equation}
which vanishes automatically due to the Dirac method of canonical primary quantization\index{quantization!primary canonical}. The Hamiltonian constraint\index{Hamiltonian constraint} can be straightforwardly resolved in the form
\begin{equation}
P_\varphi=\pm\sqrt{\dfrac{V}{\ell_P^3}}\dfrac{\varphi^2}{\varphi^2_0}\sqrt{2M_P\epsilon_M(\eta)},
\end{equation}
which generates the solution in the form of the Hubble law
\begin{equation}\label{ghub}
\frac{\varphi(\eta)}{\varphi_0}=\frac{1}{1+z(\eta_0,\eta)},
\end{equation}
where $z$ is the cosmological redshift\index{cosmological redshift}
\begin{equation}\label{redf}
z(\eta_0,\eta)=\pm\frac{1}{\varphi_0}\int_{\eta_0}^{\eta}\sqrt{\dfrac{2\epsilon_M(\eta')}{M_P\ell_PV}}d\eta'.
\end{equation}

Application of the Dirac method of canonical primary quantization\index{quantization!primary canonical}
\begin{equation}
  [\hat{P}_\varphi,\varphi]=-i\dfrac{\hslash}{\ell_P^2},
\end{equation}
leads to the momentum operator
\begin{equation}
\hat{P}_\varphi=-i\dfrac{\hslash}{\ell_P^2}\dfrac{d}{d\varphi},
\end{equation}
 which applied to the Hamiltonian constraint\index{Hamiltonian constraint}, leads to the Wheeler--DeWitt equation \index{Wheeler--DeWitt equation}-- the Klein--Gordon equation\index{Klein--Gordon equation} governing the Multiverse\index{Multiverse}
\begin{equation}
\left(\frac{d^2}{d\varphi^2}+\Omega_\varphi^2\right)\Psi(\varphi)=0,
\end{equation}
where $\Omega_\varphi$ is the frequency
\begin{equation}
\Omega_\varphi=\ell_P\sqrt{\dfrac{2V}{\ell_P^3}}\dfrac{\varphi^2}{\varphi^2_0}\sqrt{\dfrac{\epsilon_M(\eta)}{E_P}}.\label{frequen}
\end{equation}
The frequency (\ref{frequen}) can be presented in the equivalent form
\begin{equation}
\Omega_\varphi=\ell_P\sqrt{\dfrac{2V}{\ell_P^3}}\left(\dfrac{3E_P\mathrm{V}(\varphi)}{\sqrt{\pi}m^2c^4}\right)^{2/p}\sqrt{\dfrac{\epsilon_M(\eta)}{E_P}},\label{frequenn}
\end{equation}
which for the only $p=2$, i.e. for the case of the Higgs inf\/laton\index{Higgs inf\/laton}, becomes proportional to the inf\/laton potential $\mathrm{V}(\varphi)=\mathrm{V}_H(\varphi)=-\ell_P^2\dfrac{m^2c^4}{2}\varphi^2$. In such a situation the meaning of the frequency (\ref{frequenn}) becomes much more unambiguous. In other words such a Multiverse \index{Multiverse}emerges due to \emph{the Higgs--Hubble inf\/laton}\index{Higgs--Hubble inf\/laton}, and by this reason we shall call it \emph{the Higgs--Hubble Multiverse}\index{Higgs--Hubble Multiverse}. In general the auxiliary f\/ield (\ref{g}) satisf\/ies the Hubble law\index{Hubble law} (\ref{ghub}), and therefore we shall call it \emph{the Hubble auxiliary f\/ield}\index{Hubble auxiliary f\/ield}, what after identif\/ication with the power law inf\/laton becomes \emph{the Hubble inf\/laton}\index{Hubble inf\/laton}.

\section{The Chaotic Slow--Roll Inf\/lation}
Let us discuss in certain detail the Higgs--Hubble inf\/laton (\ref{g}) in the context of chaotic inf\/lation (See e.g. the Ref. \cite{mukha}). We shall work here in frames of the standard scalar perturbation theory of the inf\/lationary cosmology in which a scalar f\/ield $\varphi$ in a curved space given by the perturbed metric of the Friedmann--Lema\^{i}tre--Robertson--Walker space-time\index{space-time!Friedmann--Lema\^{i}tre--Robertson--Walker}.

The action of a scalar f\/ield $\sigma$ in an arbitrary curved space is
\begin{equation}
S=\dfrac{1}{c}\int{d^4x}\sqrt{-g}\left(\dfrac{E_P\ell_P}{2}g^{\mu\nu}\partial_\mu\sigma\partial_\nu\sigma+\dfrac{1}{\ell_P^3}\mathrm{V}(\sigma)\right),
\end{equation}
and therefore $\varphi$ satisf\/ies the Klein--Gordon equation\index{Klein--Gordon equation}
\begin{equation}
\dfrac{1}{\sqrt{-g}}\partial_\mu\left(\sqrt{-g}g^{\mu\nu}\partial_\nu\sigma\right)-\dfrac{2}{\ell_P^4E_P}\dfrac{d\mathrm{V}(\sigma)}{d\sigma}=0.
\end{equation}
If one performs the following perturbation
\begin{equation}
\sigma(x,\eta)=\varphi(\eta)+\delta\varphi(x,\eta),
\end{equation}
then the Klein--Gordon equation \index{Klein--Gordon equation}for the unperturbated homogeneous scalar f\/ield $\varphi=\varphi(\eta)$ takes the following form
\begin{equation}
\varphi''+2\mathcal{H}\varphi'+\dfrac{2a^2}{M_P\ell_P^4}\dfrac{d\mathrm{V}(\varphi)}{d\varphi}=0,\label{klegor}
\end{equation}
where $a=a(\eta)$ is the cosmic scale factor parameter\index{cosmic scale factor parameter}, $\mathcal{H}=\dfrac{a'(\eta)}{a(\eta)}$ is the conformal Hubble parameter\index{Hubble parameter!conformal}, and $\varphi_0=\dfrac{2\sqrt{\pi}}{3\ell_P}$ is the initial datum of the Higgs--Hubble inf\/laton. The perturbation scalar f\/ield $\delta\varphi(x,\eta)$ also possesses nontrivial dynamics which, however, we shall not discuss here.

Our proposal is to interpret the Higgs--Hubble inf\/laton\index{Higgs--Hubble inf\/laton} (\ref{g}) as the homogeneous unperturbated scalar f\/ield $\varphi$. In other words
\begin{equation}
  \varphi=\frac{\varphi_0}{1+z}=\varphi_0a.
\end{equation}
In such a situation one has
\begin{eqnarray}
  \varphi'&=&-\dfrac{z'}{\varphi_0}\varphi^2,\\
  \varphi''&=&2\left(\dfrac{z'}{\varphi_0}\right)^2\varphi^3,
\end{eqnarray}
\begin{eqnarray}
  a&=&\dfrac{\varphi}{\varphi_0},\\
  \mathcal{H}&=&\dfrac{\varphi'}{\varphi},\\
  \mathrm{V}(\varphi)&=&-\ell_P^2\dfrac{m^2c^4}{2E_P}\varphi^2.
\end{eqnarray}
Therefore the Klein--Gordon equation \index{Klein--Gordon equation}(\ref{klegor}) becomes
 \begin{equation}
\varphi''+2\dfrac{\varphi'^2}{\varphi}-\dfrac{2}{\varphi_0^2}\left(\dfrac{mc^2}{\hslash}\right)^2\varphi^3=0,
 \end{equation}
which after taking into account explicit form of the derivatives $\varphi'$ and $\varphi''$, and $\varphi\neq0$, becomes the differential equation for the cosmological redshift\index{cosmological redshift}
\begin{equation}\label{diffz}
  z'^2-\dfrac{1}{2}\left(\dfrac{mc^2}{\hslash}\right)^2=0.
\end{equation}
The equation (\ref{diffz}) can be solved straightforwardly
\begin{equation}
  z(\eta,\eta_0)=\pm\dfrac{1}{\sqrt{2}}\dfrac{mc^2}{\hslash}(\eta-\eta_0).
\end{equation}
In this manner in general case the Higgs--Hubble inf\/laton\index{Higgs--Hubble inf\/laton} has the form
\begin{equation}
  \varphi=\dfrac{\varphi_0}{1\pm\dfrac{1}{\sqrt{2}}\dfrac{mc^2}{\hslash}(\eta-\eta_0)}.
\end{equation}
Because of in the light of the def\/inition (\ref{redf}) one has
\begin{equation}
  z'=\pm\dfrac{1}{\varphi_0}\sqrt{\dfrac{2\epsilon_M(\eta)}{2M_P\ell_PV}},
\end{equation}
what compared to the result of the equation (\ref{diffz})
\begin{equation}
z'=\pm\dfrac{1}{\sqrt{2}}\dfrac{mc^2}{\hslash},
\end{equation}
allows to establish the energy of Matter f\/ields\index{Matter f\/ields}
\begin{equation}
\epsilon_M(\eta)=\dfrac{4}{27}\pi^2\dfrac{V}{V_P}\dfrac{m^2c^2}{M_P},\label{enmat}
\end{equation}
where $V_P=\dfrac{4}{3}\pi\ell_P^3$ is the volume of the Planck sphere\index{Planck sphere}. Therefore the frequency (\ref{frequen}) can be rewritten in the form
\begin{equation}
\Omega_\phi=\sqrt{\dfrac{9}{8\pi}}V\dfrac{m}{M_P}\varphi^2.\label{frequen1}
\end{equation}
If one takes into account \emph{ad hoc} the relation
\begin{equation}
  \Omega_\phi=\ell_P\dfrac{\ell_P^2}{E_P^2}\dfrac{m^2c^4}{2}\varphi^2,
\end{equation}
then one obtains the mass of the Higgs--Hubble inf\/laton\index{Higgs--Hubble inf\/laton!mass}
\begin{equation}
m=\sqrt{2\pi}\dfrac{V}{V_P}M_P.\label{mas1}
\end{equation}
Because of $\epsilon_M(\eta)$ is the energy of Matter f\/ields\index{Matter f\/ields} one can write \emph{ad hoc}
\begin{equation}
\epsilon_M(\eta)=\hslash\omega_{M}(\eta),
\end{equation}
where $\omega_{M}$ is the frequency f\/ield of the Matter f\/ields\index{Matter f\/ields}, which in the light of the formula (\ref{enmat}) is
\begin{equation}
\omega_{M}=\dfrac{4}{27}\pi^2\dfrac{V}{V_P}\left(\dfrac{mc^2}{\hslash}\right)^2\dfrac{1}{\omega_P},
\end{equation}
where $\omega_P=\dfrac{M_Pc^2}{\hslash}$ is the Planck frequency\index{Planck frequency}. One can suggest \emph{ad hoc} that $\omega_{M}=\omega_P$, and then the mass of the Higgs--Hubble inf\/laton\index{Higgs--Hubble inf\/laton!mass} is
\begin{equation}
m=\dfrac{3}{2\pi}\sqrt{\dfrac{3V_P}{V}}M_P.\label{mas2}
\end{equation}
In such a situation comparison of the formulas (\ref{mas1}) and (\ref{mas2}) leads to the following conclusion
\begin{equation}
  V=\dfrac{3}{2\pi}V_P=2\ell_P^3\approx8.4483\cdot10^{-105}m^3.
\end{equation}
In such a situation the Einstein--Hilbert action\index{Einstein--Hilbert action} of the Einstein--Friedmann Multiverse takes the form of the action of the one-dimensional $\varphi^4$-theory\index{$\varphi^4$-theory!one-dimensional} evolving in the conformal time\index{conformal time}
\begin{equation}\label{activ}
  S[\varphi]=\dfrac{3}{2\pi}\int{d\eta}\left(\dfrac{M_P\ell_PV_P}{2}\varphi'^2+\dfrac{3}{4}\ell_P^4\dfrac{m^2c^2}{M_P}\varphi^4\right).
\end{equation}
If one takes into account the usual action of a $\varphi^4$-theory\index{$\varphi^4$-theory} expressed in terms of the conformal time\index{conformal time} and complemented by the Planck units\index{Planck units}
\begin{equation}
  S[\varphi]=C\int{d\eta}\left(\dfrac{M_P\ell_PV_P}{2}\varphi'^2-\ell_P^4\dfrac{g}{4!}\varphi^4\right),
\end{equation}
where $g$ is the coupling parameter\index{coupling parameter} and $C$ is a constant which has not inf\/luence to the Euler--Lagrange equations of motion\index{Euler--Lagrange equations of motion}
\begin{equation}
M_PV_P\varphi''+\ell_P^3\dfrac{g}{3!}\varphi^3=0,
\end{equation}
then one obtains the coupling parameter\index{coupling parameter} in the form
\begin{equation}
  g=-3\cdot3!\dfrac{m^2c^2}{M_P}.\label{coupl}
\end{equation}
In such a situation one can establish the beta function\index{beta function}
\begin{equation}
  \beta(g)=\dfrac{dg}{d\ln m}
\end{equation}
which for the Higgs--Hubble inf\/laton\index{Higgs--Hubble inf\/laton!beta function} is
\begin{equation}
  \beta(g)=m\dfrac{dg}{dm}=2g.
\end{equation}
Interestingly, this value of the beta function coincides with the asymptotics $g\rightarrow\infty$ of $\beta(g)$ calculated from the duality relation for the two-dimensional Ising model\index{Ising model} \cite{ising}, where the dimension $D$ following from the asymptotic relation $\beta(g)=Dg$ is $D=2$.

However, the situation is rather strange, because one one has to deal with the dimension $1$. There is, however, different def\/inition of the beta function\index{beta function} (See e.g. the Ref. \cite{suslov})
\begin{equation}
  \beta(g)=\dfrac{dg}{d\ln m^2},
\end{equation}
which for the Higgs--Hubble inf\/laton is\index{Higgs--Hubble inf\/laton!beta function}
\begin{equation}
  \beta(g)=m^2\dfrac{dg}{dm^2}=g.\label{qedbeta}
\end{equation}
This value of $\beta(g)$ coincides with the asymptotics $g\rightarrow\infty$ of $\beta(g)$ in quantum electrodynamics\index{quantum electrodynamics}. If one takes into account the asymptotics $\beta(g)=Dg$, then $D=1$ coincides with the situation of the Higgs--Hubble inf\/laton\index{Higgs--Hubble inf\/laton}. In this manner the Einstein--Friedmann Universe\index{Einstein--Friedmann Universe} as well as the Higgs--Hubble inf\/laton\index{Higgs--Hubble inf\/laton} obtained nontrivial physical meaning.

In our situation, however, the potential can be straightforwardly deduced from the Einstein--Hilbert action\index{Einstein--Hilbert action} (\ref{activ}) as
\begin{equation}\label{activ1}
  V(\varphi,g,\ell_P)=\ell_P^4\dfrac{g(\ell_P)}{4!}\varphi^4,
\end{equation}
where the coupling parameter is
\begin{equation}
g(\ell_P)=3\cdot3!\dfrac{m^2c^3}{\hslash}\ell_P.\label{coupl1}
\end{equation}
Taking into account the scaling
\begin{equation}
  \varphi\rightarrow\varphi_\lambda=\dfrac{\varphi}{\lambda},
\end{equation}
for $\lambda=\ell_P$ one receives the property
\begin{equation}
V(\varphi,g(\lambda),\lambda)=\hat{V}\left(\varphi_\lambda,g(\lambda)\right),
\end{equation}
where the scaled potential is
\begin{equation}\label{activ1}
  \hat{V}\left(\varphi_\lambda,g(\lambda)\right)=\dfrac{g(\lambda)}{4!}\varphi^4.
\end{equation}
In general situation one can compute the Callan--Symanzik beta function\index{Callan--Symanzik beta function} by application of the Callan--Symanzik function\index{Callan--Symanzik $\psi(g)$-function} $\psi(g)$
\begin{equation}
\dfrac{d\ln g}{d\ln \lambda}=\psi(g)=\dfrac{\beta(g)}{g}.
\end{equation}
It is easy to see that in our case $\ln g=\ln\lambda+\ln C$, where $C$ is certain constant, and by this reason
\begin{equation}
\psi(g)=1.
\end{equation}
In this manner one can establish the Callan--Symanzik beta function of the Higgs--Hubble inf\/laton\index{Higgs--Hubble inf\/laton!Callan--Symanzik beta function} as
\begin{equation}
  \beta(g)=g.\label{hhbeta}
\end{equation}
It is manifestly seen that this beta function coincides with the beta function (\ref{qedbeta}) obtained from the asymptotics $g\rightarrow\infty$ of quantum electrodynamics\index{quantum electrodynamics}, i.e. corresponds with the dimension $1$. The problem is to construct the appropriate renormalization group equation\index{renormalization group equation}. Its construction can be performed by deformation of the usual renormalization group equation\index{renormalization group equation!deformation}
\begin{equation}
  \left(\lambda\dfrac{\partial}{\partial\lambda}-\beta(g)\dfrac{\partial}{\partial{g}}+\mu\right)V(\varphi,g,\lambda)=0,
\end{equation}
where $\mu$ is some deformation parameter which can be established by straightforward computation. Calculating the derivatives
\begin{eqnarray}
  \dfrac{\partial}{\partial\lambda}V(\varphi,g,\lambda)&=&\dfrac{4}{\lambda}V(\varphi,g,\lambda),\\
  \dfrac{\partial}{\partial{g}}V(\varphi,g,\lambda)&=&\dfrac{1}{g}V(\varphi,g,\lambda),
\end{eqnarray}
and taking into account the Callan--Symanzik beta function of the Higgs--Hubble inf\/laton\index{Higgs--Hubble inf\/laton!Callan--Symanzik beta function} (\ref{hhbeta}) established above, one obtains the following value of the deformation parameter
\begin{equation}
  \mu=3.
\end{equation}
Therefore, the renormalization group equation\index{renormalization group equation} of the Higgs--Hubble inf\/laton\index{Higgs--Hubble inf\/laton!renormalization group equation} has the form
\begin{equation}
  \left(\lambda\dfrac{\partial}{\partial\lambda}-\beta(g)\dfrac{\partial}{\partial{g}}+3\right)V(\varphi,g,\lambda)=0.
\end{equation}

\section{The Phononic Hubble Inf\/laton}
Let us see in some detail what happens in the Hubble Multiverse\index{Hubble Multiverse} and Higgs--Hubble Multiverse\index{Higgs--Hubble Multiverse}, i.e. the Multiverse generated by the Hubble inf\/laton and the Higgs--Hubble inf\/laton, respectively. Applying the explicit form of the inf\/laton (\ref{g}) to the inf\/laton energy (\ref{5}) one can straightforwardly express the inf\/laton energy in terms of the cosmic scale factor parameter\index{cosmic scale factor parameter}. The result is as follows
\begin{equation}\label{g1}
  E_H\equiv{E}_{\textrm{eff}}=\sqrt{2E_P^2\left[\ell_P^2k^2-\left(2+\dfrac{27}{64\pi^2}\dfrac{p(p-1)}{a^2}\right)\dfrac{t_P^2}{\eta^2}\right]}.
\end{equation}
First let us analyse this formula from the point of view of Special Relativity, i.e. the Einstein energy-momentum relation \index{Einstein energy-momentum relation}
\begin{equation}
E=\sqrt{p^2c^2+m^2c^4},
\end{equation}
and the wave-particle duality\index{wave-particle duality}, i.e. the Planck--Einstein relations\index{Planck--Einstein relations}
\begin{eqnarray}
E&=&\hslash\omega,\\
p&=&\hslash k.
\end{eqnarray}
It can be seen by direct computation from such a point of view the equation (\ref{g1}) describes the particle-universe possessing the following values of momentum and mass
\begin{eqnarray}
p&=&\sqrt{2}\hslash k,\\
m^2&=&-2M_P^2\left(2+\dfrac{27}{64\pi^2}\dfrac{p(p-1)}{a^2}\right)\dfrac{t_P^2}{\eta^2}\label{mpu}
\end{eqnarray}
Because of the squared mass is manifestly negative, one has to deal with \emph{tachyon}\index{tachyon} equipped with the mass
\begin{equation}
  m_t=im.
\end{equation}
Equivalently, the equation (\ref{mpu}) can be understood as the expression for the cosmic scale factor parameter\index{cosmic scale factor parameter} via the mass of a particle, i.e.
\begin{equation}
a^2(\eta)=\dfrac{27}{128\pi^2}\dfrac{p(p-1)}{\left(\dfrac{m_tc^2}{2\hslash}\eta\right)^2-1}.
\end{equation}
In such a context the Hubble inf\/laton becomes bosonic\index{Hubble inf\/laton!bosonic} equipped with negative squared mass. Applying the simple identif\/ication $E_H=\hslash\omega_H$
\begin{equation}
\omega_H=\sqrt{2\omega_P^2\left[\ell_P^2k^2-\left(2+\dfrac{27}{64\pi^2}\dfrac{p(p-1)}{a^2}\right)\dfrac{t_P^2}{\eta^2}\right]},
\end{equation}
one can derive straightforwardly the group velocity $v_g=\dfrac{d\omega_H}{dk}$ of the particle-universe
\begin{equation}
v_g=\frac{c}{\sqrt{1+\left(\dfrac{mc}{p}\right)^2}}=\frac{c}{\sqrt{1-\dfrac{1}{(k\ell_Pa)^2}\left(2a^2+\dfrac{27}{64\pi^2}p(p-1)\right)\dfrac{t_P^2}{\eta^2}}},
\end{equation}
and similarly the phase velocity $v_{ph}=\dfrac{\omega_H}{k}$ can be obtained
\begin{equation}
v_{ph}=c\sqrt{1+\left(\frac{mc}{p}\right)^2}=c\sqrt{1-\dfrac{1}{(k\ell_Pa)^2}\left(2a^2+\dfrac{27}{64\pi^2}p(p-1)\right)\dfrac{t_P^2}{\eta^2}}.
\end{equation}

On the other hand, however, the inf\/laton energy (\ref{g1}) can be analyzed by the point of view of phonons, i.e. quanta of sound in solids. In this context application of the dispersion relation for phonons
\begin{equation}\label{phonons}
\omega_k=\sqrt{2\omega^2(k)(1-\cos(k\ell_Pa))},
\end{equation}
leads to the identif\/ication
\begin{equation}
\omega=\omega_Pk\ell_P,
\end{equation}
and the cosmic scale factor parameter\index{cosmic scale factor parameter} $a$ becomes the lattice spacing. Then the Hubble inf\/laton becomes phononic\index{Hubble inf\/laton!phononic}, and straightforward comparison of the relations (\ref{g1}) and (\ref{phonons}) leads to the following non-algebraic equation
\begin{equation}\label{ka}
  (k\ell_Pa)^2\cos(k\ell_Pa)=\left(2a^2+\dfrac{27}{64\pi^2}p(p-1)\right)\dfrac{t_P^2}{\eta^2},
\end{equation}
where for known value of $\eta$ the unknown is the lattice spacing $a$. Solutions of the equation (\ref{ka}) can be found by the only numerical way. Let us apply the principles of quantum solid state physics (For basics, advances, and applications see e.g. the Ref. \cite{solids}).

According to the quantization rule for phonons, which we shall call \emph{the phononic quantization}\index{quantization!phononic}, the wave vector and the lattice spacing are nontrivially jointed by the relation
\begin{equation}\label{phq}
k\ell_Pa=\frac{n}{N}\pi,
\end{equation}
where $N$ is a number of identical atoms, $n=0,\pm1,\ldots,\pm N$, i.e. the product $k\ell_Pa$ takes integer values in the range $\left[-\dfrac{\pi}{N},\dfrac{\pi}{N}\right]$. These integers, however, can not be chosen arbitrary, because of they must solve the non algebraic equation (\ref{ka}). Possibly there is no any integer solution of this equation. In such a situation one must reinterpret the equation (\ref{ka}) as the equation for the conformal time\index{conformal time} $\eta$ while the product $k\ell_Pa$ is determined via the phononic quantization\index{quantization!phononic} (\ref{phq}).

Similarly as in the case of the particle-universe, one can derive straightforwardly the group velocity $v_g=\dfrac{d\omega_k}{dk}$ and the phase velocity $v_{ph}=\dfrac{\omega_k}{k}$ of the phonon-universe
\begin{eqnarray}
v_g&=&v_{ph}\left(2+k\ell_Pa\sqrt{1+\cos(k\ell_Pa)}\right),\label{phon1}\\
v_{ph}&=&c\sqrt{1-\cos(k\ell_Pa)},\label{phon2}
\end{eqnarray}
where
\begin{equation}
  \cos(k\ell_Pa)=\dfrac{1}{k^2\ell_P^2}\left(2+\dfrac{27}{64\pi^2}\dfrac{p(p-1)}{a^2}\right)\dfrac{t_P^2}{\eta^2}.
\end{equation}
Applying the constraint (\ref{ka}) within the formulas (\ref{phon1}) and (\ref{phon2}) one receives
\begin{eqnarray}
v_{ph}&=&c\sqrt{1-\dfrac{1}{(k\ell_Pa)^2}\left(2a^2+\dfrac{27}{64\pi^2}p(p-1)\right)\dfrac{t_P^2}{\eta^2}},\label{phon2a}\\
v_g&=&v_{ph}\left(2+\sqrt{(k\ell_Pa)^2+\left(2a^2+\dfrac{27}{64\pi^2}p(p-1)\right)\dfrac{t_P^2}{\eta^2}}\right).\label{phon1a}
\end{eqnarray}
Using the phononic quantization \index{quantization!phononic}(\ref{phq}) result in the quantization of the velocities
\begin{eqnarray}
v_{ph}&=&c\sqrt{1-\left(\dfrac{N}{\pi{n}}\right)^2\left(2a^2+\dfrac{27}{64\pi^2}p(p-1)\right)\dfrac{t_P^2}{\eta^2}},\label{phon2b}\\
v_g&=&v_{ph}\left(2+\sqrt{\left(\pi\dfrac{{n}}{N}\right)^2+\left(2a^2+\dfrac{27}{64\pi^2}p(p-1)\right)\dfrac{t_P^2}{\eta^2}}\right).\label{phon1a}
\end{eqnarray}
The phase velocity, however, is real if and only if
\begin{equation}
k\ell_Pa\geqslant\sqrt{2a^2+\dfrac{27}{64\pi^2}p(p-1)}\dfrac{t_P}{\eta},
\end{equation}
and therefore the quantization is not arbitrary, but restricted by the inequality
\begin{equation}
n\geqslant\dfrac{N}{\pi}\sqrt{2a^2+\dfrac{27}{64\pi^2}p(p-1)}\dfrac{t_P}{\eta}.
\end{equation}
It can be seen straightforwardly that also the following restriction holds
\begin{equation}
  n\geqslant\dfrac{N}{\sqrt{2}\pi}\dfrac{v_g-2v_{ph}}{v_{ph}}.
\end{equation}
In the limit situation $k\ell_Pa=\sqrt{2a^2+\dfrac{27}{64\pi^2}p(p-1)}\dfrac{t_P}{\eta}$ both the group velocity and the phase velocity vanish identically, and by the constraint (\ref{ka}) such a situation corresponds to the equation
\begin{equation}
\cos\left[\sqrt{2a^2+\dfrac{27}{64\pi^2}p(p-1)}\dfrac{t_P}{\eta}\right]=1,
\end{equation}
or in other words with the following quantization of the cosmic scale factor parameter\index{quantization!of cosmic scale factor parameter}\index{cosmic scale factor parameter}
\begin{equation}
a_n=\sqrt{\dfrac{1}{2}\left(2\pi n\omega_P\eta\right)^2-\dfrac{27}{128\pi^2}p(p-1)},
\end{equation}
where $n\in\mathbf{Z}$. Interestingly, $a_n\equiv0$, i.e. the Multiverse evolution starts, if and only if the conformal time\index{conformal time} is quantized as follows\index{quantization!of conformal time}
\begin{equation}
\eta_n=\dfrac{3t_P}{8\pi^2}\dfrac{\sqrt{3p(p-1)}}{2n}.
\end{equation}
This situation means that such a Multiverse is cyclic\index{Multiverse!cyclic}, i.e. its evolution begins few times.

It can be seen straightforwardly, however, that the phase velocities of the particle and the phonon are identical, while the group velocities are blatantly different. For full \emph{the bosonic-phononic duality}\index{bosonic-phononic duality} of the Hubble inf\/laton\index{Hubble inf\/laton}, the most natural way is to put \emph{ad hoc} the equality between the group velocities of the bosonic universe and the phononic universe. We shall call \emph{bonons}\index{bonons} the Hubble inf\/latons satisfying the bononic duality\index{bononic duality}. This type of duality is obviously nontrivial, because of in fact establishes the duality between sound (phonons) and matter (bosons). In this manner bonons\index{bonons} are the Hubble inf\/latons following from \emph{the matter-sound duality}\index{matter-sound duality}.

It is easy to see that such a duality condition can be presented as the non-algebraic equation
\begin{eqnarray}
\!\!\!\!\!\!\!\!\!\!\!\!\!\!\!\!\!\!\!\!\!\!\!\!\!\!\!\!\!\!
&&(k\ell_Pa)^2\cos^3(k\ell_Pa)-(k\ell_Pa)^2\cos^2(k\ell_Pa)-(k\ell_Pa)^2\cos(k\ell_Pa)-\nonumber\\
\!\!\!\!\!\!\!\!\!\!\!\!\!\!\!\!\!\!\!\!\!\!\!\!\!\!\!\!\!\!
&&2\cos(k\ell_Pa)+(k\ell_Pa)^2+1=0,
\end{eqnarray}
which with using of the constraint (\ref{ka}) can be presented as the algebraic equation of degree $3$
\begin{equation}
  x^3+(1-f)x^2-f(f+2)x+f^3=0,\label{xf}
\end{equation}
where we have introduced the notation
\begin{eqnarray}
  x&=&(k\ell_Pa)^2\geqslant0,\\
  f&=&\left(2a^2+\dfrac{27}{64\pi^2}p(p-1)\right)\dfrac{t_P^2}{\eta^2}.
\end{eqnarray}
If one wishes to use the phononic quantization\index{quantization!phononic} (\ref{phq}) then the equation (\ref{xf}) can be used to determination of the quantization of the cosmic scale factor parameter\index{quantization!of cosmic scale factor parameter}\index{cosmic scale factor parameter}. This quantization can be obtained as the solution of the equation
\begin{equation}
 f^3_n-x_nf^2_n-x_n(x_n+2)f_n+x_n^2+x_n^3=0,\label{fx}
\end{equation}
where
\begin{eqnarray}
  x_n&=&\left(\pi\dfrac{n}{N}\right)^2\geqslant0,\\
  f_n&=&\left(2a^2_n+\dfrac{27}{64\pi^2}p(p-1)\right)\dfrac{t_P^2}{\eta^2}.
\end{eqnarray}
The equation (\ref{fx}) does not possess real roots. It means the bononic duality has non physical nature.\index{bononic duality!non physical nature}.

Let us consider, however, seriously the phononic Hubble inf\/laton\index{Hubble inf\/laton!phononic}. Interestingly, after application of the phonon quantization\index{quantization!phononic}, the inf\/laton energy (\ref{g1}) in general is nontrivially quantized
\begin{equation}\label{rty}
E_{H}^{(n)}=\sqrt{2\left(\pi\dfrac{n}{N}\right)^2\left[1-\cos\left(\pi\dfrac{n}{N}\right)\right]}\dfrac{E_P}{a},
\end{equation}
where $E_P$ is the Planck energy\index{Planck energy}. Interestingly, in the most general situation this inf\/laton energy is even function with respect to $n$, and behaves as $E_{H}\sim\dfrac{1}{a}$. Therefore the total energy of the Hubble Multiverse\index{Hubble Multiverse!total energy}
\begin{equation}
E_H^{\mathrm{TOT}}(a)=\int_{a_I}^a da'E_H^{(n)}(a')=\sqrt{2\left(\pi\dfrac{n}{N}\right)^2\left[1-\cos\left(\pi\dfrac{n}{N}\right)\right]}E_P\ln\dfrac{a}{a_I},
\end{equation}
has divergent behavior in the limit $a\rightarrow\infty$
\begin{equation}
\lim_{a\rightarrow\infty}E_H^{\mathrm{TOT}}(a)=\infty.
\end{equation}

Another interesting quantity is the inf\/laton energy summarized with respect to the number excitations $n$. The general formula can be deduced as follows
\begin{equation}
\left\langle E_{H}\right\rangle_{N}=2\sum_{n=0}^{N}E_{H}^{(n)},
\end{equation}
where the multiplier $2$ follows from inclusion of the states with negative $n$. In this manner one can establish the mean inf\/laton energy in the Multiverse\index{Multiverse}. The result van be presented in the form
\begin{equation}
\left\langle E_{H}\right\rangle_{N}(a)=\Lambda_N\dfrac{E_P}{a},
\end{equation}
where $\Lambda_N$ is given by the formula
\begin{equation}\label{lambdan}
  \Lambda_N=2\sum_{n=0}^N\sqrt{2\left(\pi\dfrac{n}{N}\right)^2\left[1-\cos\left(\pi\dfrac{n}{N}\right)\right]}.
\end{equation}
For consistency one can also consider the mean inf\/laton energy\index{Hubble inf\/laton!mean energy} averaged over values of the cosmic scale factor parameter\index{cosmic scale factor parameter} $a$
\begin{equation}\label{aave}
\overline{\left\langle E_{H}\right\rangle_{N}}=\dfrac{1}{a-a_I}\int_{a_I}^ada'\left\langle E_{H}\right\rangle_{N}(a')=E_P\Lambda_N\dfrac{\ln\dfrac{a}{a_I}}{a-a_I},
\end{equation}
which gives the physical interpretation of the constant $\Lambda_N$
\begin{equation}
  \Lambda_N=\lim_{a\rightarrow a_I}\dfrac{\overline{\left\langle E_{H}\right\rangle_{N}}}{E_P}.
\end{equation}
Interestingly, when the Multiverse \index{Multiverse}becomes inf\/inite, i.e. $a\rightarrow\infty$, then the mean inf\/laton energy of the Hubble inf\/laton averaged over $a$ (\ref{aave}) tends to zero.

\section{The Inf\/laton Constant}
The question is, however, the convergence of quantity $\Lambda_N$ for the huge $N$ limit, $\Lambda_{\infty}$, i.e. when the Multiverse is full of the inf\/latons, for which
\begin{equation}
\left\langle E_H\right\rangle_{\infty}=\Lambda_\infty\dfrac{E_P}{a},\label{energit}
\end{equation}
where formally
\begin{equation}
  \Lambda_\infty=\lim_{N\rightarrow\infty}\Lambda_N.
\end{equation}
In the other words, the problem is the value of $\Lambda_\infty$ and whether $\Lambda_\infty$ is an universal constant. One can identify the identical atoms with spatial dimensions $N\equiv D$, and treat the Multiverse \index{Multiverse}model presented above as the multidimensional Universe. Another interpretation is that $N$ is a number of ''atoms of space'', and then the limit $N\rightarrow\infty$ defines the classical space, i.e. the space which is a solid-medium of the atoms - \emph{the phononic Hubble inf\/latons}\index{Hubble inf\/laton!phononic}. The most stable mean inf\/laton energy is obtained for infinite number of the identical atoms $N=\infty$, i.e. when the solid-medium is the \textbf{{\AE}ther model} \index{{\AE}ther}. Let us call $\Lambda_N$ \emph{the inf\/laton N-atomic constant}\index{inf\/laton constant!N-atomic}, and $\Lambda_\infty$ \emph{the inf\/laton constant}.\index{inf\/laton constant}

The inf\/laton constant is useful. One can consider the frequency $\omega_I$
\begin{equation}\label{omegai}
\omega_I=\Lambda_\infty\omega_P,
\end{equation}
where $\omega_P=E_P/\hslash$ is the Planck frequency\index{Planck frequency}, which can be interpreted as the Zero-Point Frequency f\/ield. This leads to the characteristic time of the inf\/lation
\begin{equation}\label{timei}
  t_I=\dfrac{2\pi}{\omega_I}=\dfrac{2\pi}{\Lambda_\infty}t_P,
\end{equation}
where $t_P$ is the Planck time\index{Planck time}. If one defines the cosmological potential energy $V_C(x_C)=\left\langle E_H\right\rangle_{\infty}$, where $x_C=\ell_Pa$ is the cosmological coordinate\index{cosmological coordinate} then one can determine the force
\begin{equation}
  F_H=-\dfrac{dV_C(x_C)}{dx_C}=-\dfrac{\Lambda_\infty\hslash{c}}{x_C^2}=-\dfrac{\Lambda_\infty}{4\pi}\dfrac{8}{\kappa}\left(\dfrac{p_Cc}{E_P}\right)^2=-\dfrac{G\Lambda_\infty{M}_P^2}{x_C^2},
\end{equation}
where $p_C=h/x_C$ is De Broglie cosmological momentum\index{De Broglie cosmological momentum} and $\kappa=8\pi\ell_P/E_P=8\pi G/c^4$ is the Einstein constant\index{Einstein constant}. This force defines the Newton law of universal gravitation\index{Newton law of universal gravitation} for $\Lambda_\infty M_P^2=m_1m_2$, where $m_{1,2}$ are masses of two interacting bodies.

Let us try to determine the value of the inf\/laton constant
\begin{equation}\label{lambdan1}
  \Lambda_N=2\sum_{n=0}^N\sqrt{2\left(\pi\dfrac{n}{N}\right)^2\left[1-\cos\left(\pi\dfrac{n}{N}\right)\right]}.
\end{equation}
One can change this sum by introduction of the index $k=\dfrac{n}{N}=\left\{0,1\right\}$. Then
\begin{equation}\label{lambdan1a}
  \Lambda_N=2\sum_{k=0}^1\sqrt{2\left(\pi k\right)^2\left[1-\cos\left(\pi k\right)\right]},
\end{equation}
what is easy to establish straightforwardly
\begin{equation}\label{lambdan1a}
  \Lambda_N=4\pi,
\end{equation}
and is independent on $N$. In this manner the N-atomic inf\/laton constant\index{inf\/laton constant!N-atomic} is the same as the inf\/laton constant\index{inf\/laton constant}. This result allows to write out the energy
\begin{equation}
\left\langle E_H\right\rangle_{N}=\left\langle E_H\right\rangle_{\infty}=4\pi\dfrac{E_P}{a}.
\end{equation}

By this reason one can evaluate the Zero-Point Frequency f\/ield (\ref{omegai}), the characteristic time of the inf\/lation (\ref{timei}) and the Newton law
\begin{eqnarray}
\omega_I&=&4\pi\omega_P\approx2.3308857\cdot10^{44}\mathrm{Hz},\\
t_I&=&t_P/2\approx2.9656213\cdot10^{-44}\mathrm{s},\\
F_H&=&-\dfrac{32\pi^2}{\kappa}\left(\dfrac{p_Cc}{E_P}\right)^2=-\dfrac{4\pi G{M}_P^2}{x_C^2}.
\end{eqnarray}

\chapter{Review of Quantum General Relativity}
In this chapter we shall present certain standard strategy having a basic status for quantum General Relativity. Namely, these are the $3+1$ Arnowitt--Deser--Misner Hamiltonian formulation of General Relativity \index{Hamiltonian formulation of General Relativity!Arnowitt--Deser--Misner (ADM)}and the Dirac method of canonical primary quantization \index{quantization!primary canonical}which leads to the Wheeler--DeWitt equation \index{Wheeler--DeWitt equation}and the concept of the Wheeler superspace\index{Wheeler superspace}.

\section{3+1 Splitting of General Relativity}
Let us consider a four-dimensional pseudo-Riemannian manifold $(M,g)$ (For differential geometric details see \emph{e.g.} Refs. \cite{rie,car,krie,pet}) equipped with the 4-volume form $g=\det{g_{\mu\nu}}$ related to a metric tensor $g_{\mu\nu}$ of signature $(1,3)$, the Christoffel symbols $\Gamma^\rho_{\mu\nu}$\index{Christoffel symbols}, the Riemann--Christoffel curvature tensor\index{Riemann--Christoffel curvature tensor} $R^\lambda_{\mu\alpha\nu}$, the Ricci curvature tensor $R_{\mu\nu}$\index{Ricci curvature tensor}, and the Ricci scalar curvature ${^{(4)}}\!R$\index{Ricci scalar curvature}
\begin{eqnarray}\label{eh2}
\Gamma^\rho_{\mu\nu}&=&\dfrac{1}{2}g^{\rho\sigma}\left(g_{\mu\sigma,\nu}+g_{\sigma\nu,\mu}-g_{\mu\nu,\sigma}\right),\label{chris}\\
R^\lambda_{\mu\alpha\nu}&=&\Gamma^\lambda_{\mu\nu,\alpha}-\Gamma^\lambda_{\mu\alpha,\nu}+\Gamma^\lambda_{\sigma\alpha}\Gamma^\sigma_{\mu\nu}-\Gamma^\lambda_{\sigma\nu}\Gamma^\sigma_{\mu\alpha},\\
R_{\mu\nu}&=&R^\lambda_{\mu\lambda\nu}=\Gamma^\lambda_{\mu\nu,\lambda}-\Gamma^\lambda_{\mu\lambda,\nu}+\Gamma^\lambda_{\sigma\lambda}\Gamma^\sigma_{\mu\nu}-\Gamma^\lambda_{\sigma\nu}\Gamma^\sigma_{\mu\lambda},\\
{^{(4)}}\!R&=&g^{\mu\nu}R_{\mu\nu},
\end{eqnarray}
where a holonomic basis \cite{mtw} was chosen. In General Relativity (For much more detailed books in its basics and applications see e.g. the Refs. \cite{mtw,poisson,wein}) the manifold $M$ is identif\/ied with space-time, and presence of Matter f\/ields\index{Matter f\/ields} ref\/lected by nonzero stress-energy tensor\index{stress-energy tensor}\index{energy-momentum tensor}\footnote{Some authors call $T_{\mu\nu}$ the energy-momentum tensor.} $T_{\mu\nu}$ is then studied. In such a situation the Einstein tensor\index{Einstein tensor}
\begin{equation}
  G_{\mu\nu}=R_{\mu\nu}-\frac{1}{2}g_{\mu\nu}{^{(4)}}\!R,\label{einten}
\end{equation}
allows to construct the Einstein f\/ield equations\index{Einstein f\/ield equations}
\begin{equation}\label{feq}
G_{\mu\nu}+\Lambda g_{\mu\nu}=\kappa\ell_P^2T_{\mu\nu},
\end{equation}
where $\kappa=\dfrac{8\pi G}{c^4}\approx 2.076\cdot10^{-43}~\mathrm{N}^{-1}$ is the Einstein constant\index{Einstein constant}, and $\Lambda$ is the cosmological constant\index{cosmological constant}. The constant $\kappa\ell_P^2=\dfrac{6V_P}{E_P}=\dfrac{6}{\varrho_Pc^2}$ up to the constant multiplier is reciprocal of the Planck energy\index{Planck energy} $E_P$ density $\varrho_P=\dfrac{E_P}{V_P}$ in the volume $V_P=\dfrac{4}{3}\pi\ell_P^3$ of the Planck sphere\index{Planck sphere}. This constant has the value \begin{equation}
\kappa\ell_P^2\approx5.424746\cdot10^{-129}\dfrac{\mathrm{m}^3}{\mathrm{J}},
\end{equation}
so that its reciprocal has the value
\begin{equation}
\dfrac{1}{\kappa\ell_P^2}=\dfrac{\varrho_Pc^2}{6}\approx1.843404\cdot10^{128}~\dfrac{\mathrm{J}}{\mathrm{m}^{3}}.
\end{equation}

To construct the Hamilton formulation of General Relativity it is necessary to foliate a space-time manifold $M$ with a family of space-like hypersurfaces, called also \emph{slices}\index{slices}. It is possible when $M$ is globally hyperbolic, i.e. pseudo-Riemannian, manifold what is the usual situation in General Relativity. Let $t(x^\mu)$ be a scalar f\/ield, an arbitrary single-valued function of coordinates $x^\mu$, such that the foliation $t=constans$ corresponds with a family of nonintersecting space-like hypersurfaces $\Sigma(t)$. Let us denote by $y^i$ the coordinates on all hypersurfaces $\Sigma(t)$. Let us choose a concrete hypersurface $\Sigma$ def\/ined by a parametric equations $x^\mu=x^\mu(y^i)$, where $i=1,\ldots,3$ indexes coordinates intrinsic to $\Sigma$. Equivalently, hypersurface $\Sigma$ can be selected by any restriction in the form $f(x^\mu)=0$. Then $\partial_\mu f(x^\mu)$ is a normal to $\Sigma$ which if is not null allows to def\/ine the unit normal vector f\/ield to $\Sigma$ as $n^\mu n_\mu=-1$. Then the normal vector f\/ield is given by the formula
\begin{equation}
n_\mu=-\dfrac{\partial_\mu f}{\sqrt{|\partial_\mu f\partial^\mu f|}}\quad,\quad n^\mu\partial_\mu f>0.
\end{equation}
In other words $\Sigma(t)$ are such that the unit normal to the hypersurfaces can be chosen to be future-directed time-like vector f\/ield $n_\mu\sim\partial_\mu t$ satisfying the condition $n^\mu n_\mu=-1$.

Let $\gamma$ be a congruence of curves intersecting the space-like hypersurfaces $\Sigma(t)$, which in general are not geodesics nor orthogonal to $\Sigma(t)$. Let $t$ be a parameter on the congruence $\gamma$, and let us denote by $t^\mu$ a tangent vector to $\gamma$. Then there is satisf\/ied the relation
\begin{equation}\label{basrel}
t^\mu\partial_\mu t=1.
\end{equation}
An arbitrary f\/ixed curve $\gamma_F$ is a mapping between points on all hypersurfaces $\Sigma(t)$
\begin{equation}
\gamma_F:P\in\Sigma(t)\mapsto P'\in\Sigma(t')\mapsto P''\in\Sigma(t'')\mapsto\ldots P^{(n)}\in\Sigma(t^{(n)}),
\end{equation}
where the index $n$ is an integer, and f\/ixing the coordinates on arbitrary two hypersurfaces leads to constant coordinates $y^i$ for arbitrary value of $n$. In this manner the coordinate system $(t,y^i)$ in $M$ is established. Assuming a transformation between this coordinate system and the another system $x^\mu$: $x^\mu=x^\mu(t,y^i)$ one can determine the tangent vector to the congruence $\gamma$
\begin{equation}
  t^\mu=(\partial_tx^\mu)_{y^i}=\delta^\mu_t,~\mathrm{in}~(t,y^i)
\end{equation}
as well as the tangent vectors on hypersurfaces $\Sigma(t)$
\begin{equation}
  e^\mu_i=(\partial_{y^i}x^\mu)_t=\delta^\mu_i,~\mathrm{in}~(t,y^i).
\end{equation}
In any coordinates the relation is satisf\/ied
\begin{equation}
  \mathcal{L}_te^\mu_i=0.
\end{equation}
Let us use the unit normal vector f\/ield to the hypersurfaces in the form
\begin{eqnarray}
  n_\mu&=&-N\partial_\mu t,\\
  n_\mu e^\mu_i&=&0,
\end{eqnarray}
where $N$ is called the lapse scalar, which is a function normalizing the vector f\/ield $n_\mu$. In general $t^\mu\nparallel n^\mu$, and therefore the tangent vector $t^\mu$ can be decomposed in the basis $(n^\mu, e^\mu_i)$
\begin{equation}\label{tmu}
 t^\mu=Nn^\mu+N^ie^\mu_i,
\end{equation}
where $N^i$ is a three-vector valued function called the shift vector. The coordinate transformation $x^\mu=x^\mu(t,y^i)$ allows to write in $(t,y^i)$
\begin{eqnarray}
 dx^\mu&=&t^\mu dt+e^\mu_idy^i=(Nn^\mu+N^ie^\mu_i)dt+e^\mu_idy^i\nonumber\\
 &=&(Ndt)n^\mu+(dy^i+N^idt)e^\mu_i,
\end{eqnarray}
and hence one can establish the evaluation of the space-time interval
\begin{eqnarray}
 ds^2&=&g_{\mu\nu}dx^\mu dx^\nu=dx^\mu dx_\mu\nonumber\\
&=&[(Ndt)n^\mu+(dy^i+N^idt)e^\mu_i][(Ndt)n_\mu+(dy^j+N^jdt)e_{\mu j}]\nonumber\\
&=&(n^\mu n_\mu)N^2dt^2+(dy^i+N^idt)(dy^j+N^jdt)e^\mu_ie_{\mu j}\nonumber\\
&=&-N^2dt^2+h_{ij}(dy^i+N^idt)(dy^j+N^jdt)\nonumber\\
&=&-\left(N^2-N_iN^i\right)dt^2+N_idx^idt+N_jdx^jdt+h_{ij}dx^idx^j,\label{adm}
\end{eqnarray}
where $h_{ij}$ is an induced metric on $\Sigma(t)$
\begin{equation}
h_{ij}=g_{\mu\nu}e^\mu_ie^\nu_j,
\end{equation}
which actually expresses the Pythagoras theorem\index{Pythagoras theorem} between two points lying on two distinguishable constant time hypersurfaces, and was investigated by R. Arnowitt, S. Deser and C.W. Misner \cite{adm}. By this reason, a space-time metric tensor $g_{\mu\nu}$ of a Lorentzian manifold satisfying the Einstein f\/ield equations\index{Einstein f\/ield equations} (\ref{feq}) obtains the following decomposition onto space and time
\begin{equation}\label{dec}
g_{\mu\nu}=\left[\begin{array}{cc}-N^2+N_iN^i&N_j\\N_i&h_{ij}\end{array}\right],
\end{equation}
where $N^j=h^{ij}N_i$ is the contravariant shift vector, and the spatial metric satisf\/ies the orthogonality condition \begin{equation}
h_{ik}h^{kj}=\delta_i^j.
\end{equation}
Completeness relations\index{completeness relations!metric} for the metric are
\begin{equation}\label{compmetric}
g_{\mu\nu}=-n_\mu n_\nu+h_{ij}e^i_\mu e^j_\nu.
\end{equation}
It can be verif\/ied straightforwardly that the transformation between the four-volume form and the three-volume form is
\begin{equation}
  \sqrt{-g}=N\sqrt{h},
\end{equation}
while the inverted metric has the form
\begin{equation}\label{indec}
g^{\mu\nu}=\left[\begin{array}{cc}-\dfrac{1}{N^2}&\dfrac{N^j}{N^2}\\{}\dfrac{N^i}{N^2}&h^{ij}-\dfrac{N^iN^j}{N^2}\end{array}\right].
\end{equation}
Completeness relations for the inverse metric\index{completeness relations!inverse metric} are
\begin{equation}\label{comptrans}
g^{\mu\nu}=-n^\mu n^\nu+h^{ij}e^\mu_ie^\nu_j.
\end{equation}
The second fundamental form of a slice is called the extrinsic curvature tensor\index{extrinsic curvature tensor} or induced curvature and has the form
\begin{equation}
  K_{ij}=n_{\mu;\nu}e^\mu_ie^\nu_j=-\nabla_{(i}n_{j)}-n_{(i}a_{j)},
\end{equation}
where $a_j$ is called the acceleration of the unit normal vector f\/ield
\begin{equation}
a_j=n^in_{j|i},
\end{equation}
and its trace, called the intrinsic curvature\index{intrinsic curvature}, has a form
\begin{equation}
K=K^i_i=h^{ij}K_{ij}=n^\mu_{;\mu}.
\end{equation}
The hypersurface is called convex when the congruence is diverging, i.e. $K>0$, and concave when the congruence is converging, i.e. $K<0$. The tangent vector satisf\/ies the Gauss--Weingarten equation\index{Gauss--Weingarten equation}
\begin{equation}
  e^\alpha_{i;\beta}e^\beta_j=\Gamma^k_{ij}e^\alpha_k+K_{ij}n^\mu,
\end{equation}
and the Gauss--Codazzi equations\index{Gauss--Codazzi equations} \cite{gau,cod,han} can be derived by straightforward computation
\begin{eqnarray}
  R_{\mu\nu\kappa\lambda}e^\mu_ie^\nu_je^\kappa_ke^\lambda_l&=&R_{ijkl}-K_{il}K_{jk}+K_{ik}K_{jl},\label{gc1}\\
  R_{\mu\nu\kappa\lambda}n^\mu e^\nu_ie^\kappa_je^\lambda_k&=&K_{ij|k}-K_{ik|j},\label{gc2}
\end{eqnarray}
which via using of the decomposition of the Ricci curvature tensor and Ricci scalar curvature\index{Ricci curvature tensor}\index{Ricci scalar curvature}
\begin{eqnarray}
  R_{\mu\nu}&=&-R_{\kappa\mu\lambda\nu}n^\kappa n^\lambda+h^{ij}R_{\kappa\mu\lambda\nu}e^\kappa_ie^\lambda_j,\label{ricdec}\\
  R&=&-2h^{kl}R_{\kappa\mu\lambda\nu}n^\kappa n^\lambda e^\mu_k e^\nu_l+h^{kl}h^{ij}R_{\kappa\mu\lambda\nu}e^\kappa_ie^\lambda_je^\mu_ke^\nu_l,\label{rscdec}
\end{eqnarray}
can be presented in terms of the Einstein tensor (\ref{einten})\index{Einstein tensor}
\begin{eqnarray}
  2G_{\mu\nu}n^\mu n^\nu&=&{^{(3)}}R+K^{ij}K_{ij}+K^2,\\
  G_{\mu\nu}e^\mu_jn^\nu&=&K^i_{j|i}-K_{,j}.
\end{eqnarray}
Another identity, called the Ricci equation\index{Ricci equation}
\begin{equation}
  \mathcal{L}_nK_{ij}=n^\mu n^\nu e^\kappa_ie^\lambda_j R_{\mu\nu\kappa\lambda}-\dfrac{1}{N}N_{|ij}-K_{ik}K^k_j,
\end{equation}
can be also derived by straightforward computation, which we omit here. By using of the completeness relations (\ref{comptrans}) and the fact
\begin{equation}
R_{\kappa\mu\lambda\nu}n^\kappa n^\mu n^\lambda n^\nu=0,
\end{equation}
one can see that the f\/irst term in (\ref{rscdec}) reduces to $-2R_{\mu\nu}n^\mu n^\nu$. By using of the relations
\begin{eqnarray}
R_{\mu\nu}n^\mu n^\nu&=&2\left(n^\mu_{;[\nu}n^\nu\right)_{;\mu]}+2n^\mu_{;[\mu}n^\nu_{;\nu]},\\
n^\mu_{;\nu}n^\nu_{;\mu}&=&K^{ij}K_{ij},
\end{eqnarray}
to the reduced f\/irst term in (\ref{rscdec}), and the Gauss--Codazzi equations\index{Gauss--Codazzi equations} (\ref{gc1}) to the second term in (\ref{rscdec})
\begin{eqnarray}
h^{kl}h^{ij}R_{\kappa\mu\lambda\nu}e^\kappa_ie^\lambda_je^\mu_ke^\nu_l&=&
h^{kl}h^{ij}\left(R_{ijkl}-K_{il}K_{jk}+K_{ik}K_{jl}\right)=\nonumber\\
&=&{^{(3)}}R+K^2-K^{ij}K_{ij},
\end{eqnarray}
one can obtains the three-dimensional evaluation of the four-dimensional Ricci scalar curvature\index{Ricci scalar curvature}
\begin{equation}\label{ric4}
{^{(4)}}R={^{(3)}}R+K^2-K^{ij}K_{ij}-2\left(n^\mu_{;\nu}n^\nu-n^\mu n^\nu_{;\nu}\right)_{;\mu}.
\end{equation}
We have denoted by stroke on the left of an index the intrinsic covariant differentiation with respect to a coordinate labeled by this index. Two indices before the stroke means taking two times the intrinsic covariant derivative with respect to each of the indices. For instance for a vector $V_i$ and a tensor $T_{ij}$ the intrinsic covariant differentiation is def\/ined as
\begin{eqnarray}
  V_{i|j}&=&\nabla_jV_i=\partial_jV_i-\Gamma_{ji}^kV_k,\\
  T_{ij|k}&=&\nabla_kT_{ij}=\partial_kT_{ij}-T_{lj}\Gamma^l_{ik}-T_{il}\Gamma^l_{jk},
\end{eqnarray}
where $\Gamma_{ij}^k$ are the spatial Christoffel symbols
\begin{equation}
  \Gamma^k_{ij}=\dfrac{1}{2}h^{kl}\left(h_{il,j}+h_{lj,i}-h_{ij,l}\right).
\end{equation}

The induced metric $h_{ij}$ and the extrinsic curvature $K_{ij}$ are the dynamical variables which describe the geometry of a submanifold $\partial M$ by the $3+1$ decomposed Einstein f\/ield equations\index{Einstein f\/ield equations}. The pair $(h_{ij},K_{ij})$ describes the local geometry of a single space-like (constant time) hypersurface $\partial M$, and then the evolution of the global four-dimensional geometry can be formulated in terms of the one-parameter family of the dynamical variables $(h_{ij}(t),K_{ij}(t))$ describing evolution of the local three-dimensional geometry of the space-like hypersurfaces $\partial M_t$. For consistency one must also specify the relation between the time evolution operator $\partial_t$ and the vector f\/ield $n$ normal to the $\partial M_t$
\begin{equation}
  \partial_t=Nn+N^i\partial_i,
\end{equation}
where $N$ and $N_i$ are called the lapse function and the shift vector, respectively. Albeit, in general stationarity of Matter f\/ields\index{Matter f\/ields}, i.e. $T_{\mu\nu}\equiv0$, results in existence of a global time-like Killing vector f\/ield\index{Killing vector f\/ield}  $\mathcal{K}_\mu$ for a metric tensor $g_{\mu\nu}$.

One can choose a coordinate system in such a way that the Killing vector f\/ield\index{Killing vector f\/ield} equals to $\dfrac{\partial}{\partial t}$ and the foliation $t=constans$ is space-like. In such a situation a metric tensor depends at most on a spatial coordinates $x^i$, and therefore the time $t$ can be treated as a global coordinate \cite{qft}. Let us introduce such a coordinate system chosen by this gauge condition in such a way that an induced three-dimensional boundary space is a constant time $t$ hypersurface. Then the space-time boundary $\partial M$ becomes an embedded space and satisf\/ies the Nash embedding theorem\index{Nash embedding theorem} (For detailed discussion of the theorem, its consequences and advanced development see \emph{e.g.} the Refs. \cite{nash,kowa,masa,gunt}).

\section{Geometrodynamics: Classical and Quantum}
Let the enveloping space-time manifold $M$ be compact and possesses a space-like boundary $(\partial M,h)$ equipped with the 3-volume form $h=\det{h_{ij}}$ related to the induced metric $h_{ij}$, and the second fundamental form $K_{ij}$. Let has the topology of space-time will be $\Sigma\times\mathbb{R}$ where $\Sigma$ is an unrestricted topology of the three-dimensional space. Then the Einstein f\/ield equations\index{Einstein f\/ield equations} (\ref{feq}) can be generated as the Euler-Lagrange equations of motion via using of the Hilbert--Palatini action principle\index{Hilbert--Palatini action principle} \cite{hil,pal} with respect to the fundamental f\/ield which for General Relativity is a metric tensor $g_{\mu\nu}$
\begin{equation}\label{eh1}
\dfrac{\delta S[g]}{\delta g_{\mu\nu}}=0,
\end{equation}
which must be complemented by the boundary condition
\begin{equation}
  \delta g_{\mu\nu}\left|_{\partial M}\right.=0,
\end{equation}
and applied to the Einstein--Hilbert action\index{Einstein--Hilbert action} complemented by the York--Gibbons--Hawking boundary action\index{York--Gibbons--Hawking boundary action} \cite{hawk01,york}, i.e. the action of a four-geometry with f\/ixed an induced three-geometry of a boundary
\begin{equation}\label{eh0}
S[g]=\dfrac{1}{2\kappa{c}\ell_P^2}\int_{M}d^4x\sqrt{-g}\left(-{^{(4)}}R+2\Lambda\right)+S_\phi[g]-\dfrac{1}{\kappa{c}\ell_P^2}\int_{\partial{M}}d^3x\sqrt{h}K,
\end{equation}
where $S_\phi[g]$ is the action of Matter f\/ields\index{Matter f\/ields}
\begin{equation}
S_\phi[g]=\dfrac{1}{c}\int_Md^4x\sqrt{-g}L_\phi.
\end{equation}
Einstein f\/ield equations\index{Einstein f\/ield equations} (\ref{feq}) can be obtained via straightforward computation of the variation $\delta S=\delta S_{EH}+\delta S_{YGH}+\delta S_\phi=0$ on $\partial M$, where
\begin{eqnarray}
\delta S_{G}&=&\dfrac{1}{2\kappa{c}\ell_P^2}\int_Md^{4}x\sqrt{-g}\left(G_{\mu\nu}+\Lambda g_{\mu\nu}\right)\delta g^{\mu\nu},\\
\delta S_{YGH}&=&\dfrac{1}{2\kappa{c}\ell_P^2}\int_{\partial M}d^3y\sqrt{|h|}h^{\mu\nu}n^\rho\delta g_{\mu\nu,\rho},\\
\delta S_\phi&=&-\dfrac{1}{2c}\int_Md^{4}x\sqrt{-g}T_{\mu\nu}\delta g^{\mu\nu},
\end{eqnarray}
where by $S_{G}=S_{EH}+S_{YGH}$ we have denoted the geometric part of the total Lagrangian (\ref{eh0}). Moreover, the variational principle allows to establish the relation between the stress-energy tensor\index{stress-energy tensor} and the Lagrangian of Matter f\/ields\index{Matter f\/ields}
\begin{equation}\label{varsten}
T_{\mu\nu}=-\dfrac{2}{\sqrt{-g}}\dfrac{\delta }{\delta g^{\mu\nu}}\left(\sqrt{-g}L_\phi\right).
\end{equation}

When the cosmological constant \index{cosmological constant}vanishes identically $\Lambda=0$, then a global time-like Killing vector f\/ield\index{Killing vector f\/ield} $\mathcal{K}_\mu$ on a space-time manifold $M$ exists. Recall (For more advanced and abstractive approach we suggest e.g. the books in the Ref. \cite{sachs}) that such a f\/ield follows from vanishing of the Lie derivative\index{Lie derivative} with respect to this f\/ield of the metric tensor $g_{\mu\nu}$
\begin{equation}
\mathcal{L}_{K}g_{\mu\nu}=\lim_{\epsilon\rightarrow0}\dfrac{g_{\mu\nu}(\tilde{x})-\tilde{g}_{\mu\nu}(\tilde{x})}{\epsilon}=0,
\end{equation}
where $\tilde{g}_{\mu\nu}(\tilde{x})$ is the metric tensor $g_{\mu\nu}(x)$ transformed under the inf\/initesimal transformation - an isometric mapping
\begin{equation}
  \tilde{x}^\mu=x^\mu+\epsilon \mathcal{K}^\mu,
\end{equation}
which is equivalent to the Killing equation\index{Killing equation}
\begin{equation}
\nabla_{(\mu}\mathcal{K}_{\nu)}(x)=0.
\end{equation}
In other words, the Killing vector f\/ields\index{Killing vector f\/ield} are the inf\/initesimal generators of isometries.

For positive value of the cosmological constant \index{cosmological constant}$\Lambda>0$ the Killing vector f\/ield\index{Killing vector f\/ield} $\mathcal{K}_\mu$ does not exist, and space-like boundary $\partial M$ only foliates an exterior to the horizons on geodesic lines. Therefore in such a situation the ADM decomposition (\ref{dec}) is a gauge of the f\/ield of metric. In all these situations, however, the total action (\ref{eh0}) evaluated for the $3+1$ decomposed metric tensor (\ref{dec}) takes the form of the Hamilton action functional
\begin{eqnarray}\label{gd}
  S[g]=\int dt L,
\end{eqnarray}
where $L$ is the total Lagrangian expressed via the $3+1$ splitting\index{$3+1$ splitting}. The most important contribution is the geometric part of the Einstein--Hilbert Lagrangian which is
\begin{eqnarray}
\!\!\!\!\!\!\!\!\!\!\!\!\!\!\!\!\!\!\!\!\sqrt{g}\left(-{^{(4)}}R+2\Lambda\right)&=&N\sqrt{h}\left(-K_{ij}K^{ij}+K^2-{^{(3)}}R+2\Lambda\right)\nonumber\\
\!\!\!\!\!\!\!\!\!\!\!\!\!\!\!\!\!\!\!\!&+&2\partial_0\left(\sqrt{h}K\right)-2\partial_i\left(\sqrt{h}(KN^i-h^{ij}N_{|j})\right),
\end{eqnarray}
and because the last two terms are total derivatives they can be dropped when performing a canonical formulation. The Lagrangian related to the York--Gibbons--Hawking boundary action\index{York--Gibbons--Hawking boundary action} in itself is total derivative, and therefore this term does not play a role here. Analysis of both the Lagrangian of Matter f\/ields\index{Matter f\/ields} and the cosmological constant \index{cosmological constant}term can be done easily, and in result one obtains the following Lagrangian of the total theory (\ref{gd})
\begin{equation}\label{totlag}
  L=\dfrac{1}{2\kappa\ell_P^2}\int_{\partial M} d^3x N\sqrt{h}\left(K^2-K_{ij}K^{ij}-{^{(3)}}R+2\Lambda+2\kappa\ell_P^2\rho\right).
\end{equation}
The Einstein f\/ield equations\index{Einstein f\/ield equations} can be decomposed in the $3+1$ splitting\index{$3+1$ splitting}\index{$3+1$ splitting}. In result one obtains the evolutionary equations for the induced metric $h_{ij}$ and the intrinsic curvature $K_{ij}$
\begin{eqnarray}
\!\!\!\!\!\!\!\!\!\!\!\!\!\!\!\!\!\!\!\!
\partial_t{h}_{ij}&=&N_{i|j}+N_{j|i}-2NK_{ij},\label{con0}\\
\!\!\!\!\!\!\!\!\!\!\!\!\!\!\!\!\!\!\!\!
\partial_t{K}_{ij}&=&-N_{|ij}+N(R_{ij}+KK_{ij}-2K_{ik}K^k_j)+N^kK_{ij|k}+K_{ik}N^k_{|j}+K_{jk}N^k_{|i}-\nonumber\\
\!\!\!\!\!\!\!\!\!\!\!\!\!\!\!\!\!\!\!\!
&-&\kappa\ell_P^2N\left[S_{ij}-\dfrac{1}{2}h_{ij}(S-\varrho)\right],
\end{eqnarray}
where the dot means differentiation with respect to the time coordinate. It can be seen by straightforward computation that the determinant of spatial metric $h=\det h_{ij}$ and the extrinsic curvature $K=K^i_i$ satisfy the equations
\begin{eqnarray}
\partial_t\ln\sqrt{h}&=&-NK+N^i_{|i},\\
\partial_t K&=&-h^{ij}N_{|ij}+N\left(K^{ij}K_{ij}+\dfrac{\kappa\ell_P^2}{2}(S+\varrho)\right)+N^iK_{|i}.
\end{eqnarray}
Here $\varrho$, called the energy density\index{energy density}, is double projection of the stress-energy tensor\index{stress-energy tensor} onto the normal vector f\/ield
\begin{equation}
  \varrho=T(n,n)={T}_{\mu\nu}n^\mu n^\nu,
\end{equation}
and $n^\mu$ is the normal vector f\/ield following from the $3+1$ splitting\index{$3+1$ splitting}
\begin{eqnarray}
  n^\mu&=&\left[\dfrac{1}{N},-\dfrac{N^i}{N}\right],\\
  n_{\mu}&=&\left[-N,0_i\right]^T,
\end{eqnarray}
where $0_i=[0,0,0]^T$ is the null three-vector. The tensor $S_{ij}$, called the spatial stress\index{spatial stress}, is double projection of the stress-energy tensor\index{stress-energy tensor} onto the spatial metric, and $S$ is its trace which we will call the spatial stress density\index{spatial stress density}
\begin{eqnarray}
  S_{ij}&=&T(h,h)=T_{\mu\nu}h^\mu_ih^\nu_j,\\
  S&=&h^{ij}S_{ij},
\end{eqnarray}
where $h^\mu_\nu=\delta^\mu_\nu+n^\mu n_\nu$. Straightforward calculation gives
\begin{equation}
  S-\varrho=T,\label{esrhodif}
\end{equation}
where $T=g^{\mu\nu}T_{\mu\nu}$ is the trace of the stress-energy tensor\index{stress-energy tensor}. In the light of the Einstein f\/ield equations\index{Einstein f\/ield equations} (\ref{feq}) one obtains
\begin{equation}
  T=\dfrac{G+4\Lambda}{\kappa\ell_P^2},
\end{equation}
where $G$ is the trace of the Einstein tensor\index{Einstein tensor} which can be computed straightforwardly
\begin{equation}
G=g^{\mu\nu}G_{\mu\nu}=g^{\mu\nu}\left(R_{\mu\nu}-\frac{1}{2}g_{\mu\nu}{^{(4)}}R\right)={^{(4)}}R-2{^{(4)}}R=-{^{(4)}}R.
\end{equation}
In this manner one obtains the constraint between the spatial stress\index{spatial stress}, the energy density\index{energy density}, the cosmological constant \index{cosmological constant}and the Ricci scalar curvature\index{Ricci scalar curvature} of an enveloping space-time manifold
\begin{equation}
S-\varrho=\dfrac{4\Lambda}{\kappa\ell_P^2}-\dfrac{{^{(4)}}R}{\kappa\ell_P^2}.\label{esrho}
\end{equation}

It can be seen by straightforward computation that the total Lagrangian (\ref{totlag}) leads to the Euler--Lagrange\index{Euler--Lagrange equations of motion} equations of motion. The f\/irst equation is
\begin{equation}\label{con}
{2c\kappa}h^{-1}\left(h_{ik}h_{jl}-\dfrac{1}{2}h_{ij}h_{kl}\right)\dfrac{\delta{S}}{\delta{h_{ij}}}\dfrac{\delta{S}}{\delta{h_{kl}}}-\dfrac{\ell_P^2}{2c\kappa}\left({^{(3)}}R-2\Lambda-2\kappa\ell_P^2\varrho\right)=0,
\end{equation}
while the second one has the form
\begin{equation}\label{eqmot1}
  \dfrac{c}{\ell_P^2}\pi^{ij}_{|j}+J^i=0,
\end{equation}
where $\pi^{ij}$ is the momentum conjugated to the induced metric
\begin{equation}
\pi^{ij}=\dfrac{1}{\ell_P}\dfrac{\delta{S[g]}}{\delta{h_{ij}}}=\dfrac{1}{\ell_P}\dfrac{\delta{L}}{\delta\left(\partial_t{h}_{ij}\right)}=-\dfrac{\ell_P}{2c\kappa}\sqrt{h}\left(K^{ij}-h^{ij}K\right),\label{mom}
\end{equation}
and $J^i$, called the momentum density, is  the stress-energy tensor\index{stress-energy tensor} projected onto the normal vector f\/ield and the spatial metric
\begin{eqnarray}
J^i=T(n,h)=T_{\mu\nu}n^\mu h^{\nu i}.
\end{eqnarray}
The resulting dynamical equation (\ref{con}) is the Hamilton--Jacobi equation\index{Hamilton--Jacobi equation} (For details of classical mechanics see \emph{e.g.} the Ref. \cite{gol}) to the case of General Relativity. Originally, the equation (\ref{con}) divided by $\sqrt{h}$ with $\Lambda=0$ and $\varrho=0$ was derived by A. Peres \cite{peres}, and by this reason we shall call it the Peres equation\index{Peres equation}. Interestingly, Wheeler \cite{whe} called the Hamilton--Jacobi equation\index{Hamilton--Jacobi equation} of General Relativity (\ref{con}) the Einstein--Hamilton--Jacobi equation\index{Einstein--Hamilton--Jacobi equation}. The Peres equation \index{Peres equation}def\/ines the classical geometrodynamics.

The total Lagrangian (\ref{totlag}) can be analyzed by the Hamiltonian approach. Let us determine the canonical momenta
\begin{eqnarray}
\pi_\phi&=&\dfrac{\beta}{\ell_P}\dfrac{\delta L}{\delta\left(\partial_t{\phi}\right)},\\
\pi&=&\dfrac{1}{\ell_P}\dfrac{\delta L}{\delta\left(\partial_t{N}\right)}=0,\\
\pi^i&=&\dfrac{1}{\ell_P}\dfrac{\delta L}{\delta\left(\partial_t{N_i}\right)}=0,
\end{eqnarray}
where $\beta$ is a constant of dimension of a Matter field $[\phi]$ constructed from the Planck units, conjugated to Matter f\/ields\index{Matter f\/ields}, lapse function, shift vector, respectively. Then the Legendre transformation\index{Legendre transformation} \cite{gol} allows to rewrite the total Lagrangian in the form
\begin{equation}
L=\int_{\partial{M}}d^3x\left[\dfrac{1}{2\kappa\ell_P}\left(\pi_\phi\partial_t{\phi}+\pi\partial_t{N}+\pi^i\partial_t{N_i}+\pi^{ij}\partial_t{h}_{ij}\right)-NH-N_iH^i\right],
\end{equation}
where the quantities $H$ and $H^i$ are def\/ined as
\begin{eqnarray}
\!\!\!\!\!\!\!\!\!\!\!\!\!\!\!\!\!\!\!\!
H&=&\dfrac{\sqrt{h}}{2\kappa\ell_P^2}\left(K^2-K_{ij}K^{ij}-{^{(3)}R}+2\Lambda+2\kappa\ell_P^2\varrho\right),\label{con1}\\ \!\!\!\!\!\!\!\!\!\!\!\!\!\!\!\!\!\!\!\!
H^i&=&-2\dfrac{c}{\ell_P^2}\pi^{ij}_{~|j}-2{J^i}=-2\dfrac{c}{\ell_P^2}\partial_j\pi^{ij}-\dfrac{c}{\ell_P^2}h^{il}\left(2h_{jl,k}-h_{jk,l}\right)\pi^{jk}-2{J^i},\label{con2}
\end{eqnarray}
where ${^{(3)}R}=h^{ij}R_{ij}$ is the Ricci scalar curvature\index{Ricci scalar curvature} of a three-dimensional embedded space. Application of the time-preservation \cite{dir} to the primary constraints
\begin{eqnarray}
\pi&\approx&0,\\
\pi^i&\approx&0,
\end{eqnarray}
leads to the secondary constraints
\begin{eqnarray}
H&\approx&0,\label{constA}\\
H^i&\approx&0,\label{constB}
\end{eqnarray}
called the Hamiltonian (scalar) constraint which yields the dynamics, and the diffeomorphism (vector) constraint which merely ref\/lects the spatial diffeoinvariance. B.S. DeWitt \cite{dew} showed that the quantities $H^i$ are generators of the spatial diffeomorphisms $\widetilde{x}^i=x^i+\xi^i$
\begin{eqnarray}
i\dfrac{\ell_P}{\hslash}\left[h_{ij},\int_{\partial{M}}H_{a}\xi^ad^3x\right]&=&c\ell_P\left(-h_{ij,k}\xi^k-h_{kj}\xi^{k}_{~,i}-h_{ik}\xi^{k}_{~,j}\right),\label{gene1}\\
i\dfrac{\ell_P}{\hslash}\left[\pi^{ij},\int_{\partial{M}}H_{a}\xi^ad^3x\right]&=&c\ell_P\left[-\left(\pi^{ij}\xi^k\right)_{,k}+\pi^{kj}\xi^{i}_{~,k}+\pi^{ik}\xi^{j}_{~,k}\right],\label{gene2}
\end{eqnarray}
where we have denoted the intrinsic covariant components $H_i=h_{ij}H^j$. Application of the structure constants of the diffeomorphism group\index{diffeomorphism group}, which can be presented in the most convenient compact form
\begin{equation}
c^a_{ij}=\delta^a_i\delta^b_j\delta^{(3)}_{,b}(x,z)\delta^{(3)}(y,z)-(x\rightarrow y).
\end{equation}
Applying the relations (\ref{gene1}) and (\ref{gene2}) one can derive the f\/irst-class constraints algebra
\begin{eqnarray}
  i\dfrac{\ell_P}{\hslash}\left[H_i(x),H_j(y)\right]&=&\dfrac{c}{\ell_P^5}\int_{\partial{M}}H_{a}c^a_{ij}d^3z,\label{com1}\\
  i\dfrac{\ell_P}{\hslash}\left[H(x),H_i(y)\right]&=&\dfrac{c}{\ell_P^5}H\delta^{(3)}_{,i}(x,y),\label{com2}
\end{eqnarray}
while involving of the elementary relation
\begin{equation}
\delta\left(\sqrt{h}{^{(3)}R}\right)=\sqrt{h}h^{ij}h^{kl}\left(\delta h_{ik,jl}-\delta h_{ij,kl}\right)-\sqrt{h}\left[R^{ij}-\dfrac{{^{(3)}R}}{2}h^{ij}\right]\delta h_{ij},
\end{equation}
allows to establish the third bracket
\begin{equation}
  i\dfrac{\ell_P}{\hslash}\left[\int_{\partial{M}}H\xi_1d^3x,\int_{\partial{M}}H\xi_2d^3x\right]=c\ell_P\int_{\partial{M}}H^a\left(\xi_{1,a}\xi_2-\xi_1\xi_{2,a}\right)d^3x.\label{com3}
\end{equation}
The constraints algebra (\ref{com1})-(\ref{com3}) was derived f\/irst by B.S. DeWitt, and by this reason we shall call it \emph{the DeWitt algebra}\index{DeWitt algebra}.

The method of canonical primary quantization \index{quantization!primary canonical}appropriate for constrained systems was investigated by Dirac \cite{dir} and developed for needs of quantum geometrodynamics\index{quantum geometrodynamics (QGD)} by L.D. Faddeev \cite{fad} (For general analysis and discussion see also the Refs. \cite{qg18,han}). In the light of the general method applied to the present situation the canonical commutation relations\index{canonical commutation relations} are
\begin{eqnarray}
i\dfrac{\ell_P}{\hslash}\left[\pi^{ij}(x),h_{kl}(y)\right]&=&\frac{1}{2}\left(\delta_{k}^{i}\delta_{l}^{j}+\delta_{l}^{i}\delta_{k}^{j}\right)\delta^{(3)}(x,y),\label{dpq}\\
i\dfrac{\ell_P}{\hslash}\left[\pi^i(x),N_j(y)\right]&=&\delta^i_j\delta^{(3)}(x,y),\label{dpq1}\\
i\dfrac{\ell_P}{\hslash}\left[\pi(x),N(y)\right]&=&\delta^{(3)}(x,y).\label{dpq2}
\end{eqnarray}
The solutions or rather representations of the momenta operators satisfying the commutators (\ref{dpq})-(\ref{dpq2}) is the question of choice. In quantum geometrodynamics\index{quantum geometrodynamics (QGD)} the Wheeler metric representation \index{Wheeler metric representation}is usually taken into account. In such a representation the momenta operators are analogous to the momentum operator in quantum mechanics
\begin{eqnarray}
\pi&=&-i\dfrac{\hslash}{\ell_P}\dfrac{\delta}{\delta N},\\
\pi^i&=&-i\dfrac{\hslash}{\ell_P}\dfrac{\delta}{\delta N_i},\\
\pi^{ij}&=&-i\dfrac{\hslash}{\ell_P}\dfrac{\delta}{\delta h_{ij}},
\end{eqnarray}
and applied to the Hamiltonian constraint\index{Hamiltonian constraint} (\ref{con1}) yields the Wheeler--DeWitt equation \index{Wheeler--DeWitt equation}\cite{dew,whe}
\begin{equation}\label{wdw}
\left\{2c\kappa\dfrac{\hslash^2}{\ell_P^2}G_{ijkl}\dfrac{\delta^2}{\delta h_{ij}\delta h_{kl}}+\dfrac{\ell_P^2}{2c\kappa} \sqrt{h}\left({^{(3)}R}-2\Lambda-2\kappa\varrho\right)\right\}\Psi[h_{ij},\phi]=0,
\end{equation}
where $G_{ijkl}$ is the DeWitt supermetric \index{DeWitt supermetric}on the conf\/igurational space of General Relativity called the Wheeler superspace \index{Wheeler superspace}$S(\partial M)$ (For more detailed analysis and discussion see \emph{e.g.} Refs. \cite{whe,dew,dewi,fish,pek,giul})
\begin{equation}\label{supmet}
G_{ijkl}=\dfrac{1}{2\sqrt{h}}\left(h_{ik}h_{jl}+h_{il}h_{jk}-h_{ij}h_{kl}\right).
\end{equation}
Other f\/irst-class constraints satisfy the canonical commutation relations\index{canonical commutation relations}
\begin{eqnarray}
\left[{\pi}(x),{\pi}^i(y)\right]&=&0,\\
\left[{\pi}(x),{H}^i(y)\right]&=&0,\\
\left[{\pi}^i(x),{H}^j(y)\right]&=&0,\\
\left[{\pi}^i(x),{H}(y)\right]&=&0,
\end{eqnarray}
and after the canonical primary quantization \index{quantization!primary canonical}are the supplementary conditions on a wave functional $\Psi[h_{ij},\phi]$. The primary constraints lead to the equations
\begin{eqnarray}
  -i\dfrac{\hslash}{\ell_P}\dfrac{\delta\Psi[h_{ij},\phi]}{\delta N}&=&0,\\
  -i\dfrac{\hslash}{\ell_P}\dfrac{\delta\Psi[h_{ij},\phi]}{\delta N_i}&=&0.
\end{eqnarray}
The diffeomorphism constraint\index{diffeomorphism constraint} also leads to such a condition
\begin{equation}
i\dfrac{E_P}{\ell_P^2}\left(\dfrac{\delta\Psi[h_{ij},\phi]}{\delta{h_{ij}}}\right)_{|j}={J^i}\Psi[h_{ij},\phi],
\end{equation}
which can be rewritten in explicit form
\begin{equation}
\left[i\dfrac{E_P}{\ell_P^2}\dfrac{\partial}{\partial{x^j}}\dfrac{\delta}{\delta{h_{ij}}}+i\dfrac{E_P}{\ell_P^2}{h}^{il}\left(h_{jl,k}-\dfrac{1}{2}h_{jk,l}\right)\dfrac{\delta}{\delta{h_{jk}}}-{J^i}\right]\Psi[h_{ij},\phi]=0.\label{momeq}
\end{equation}
The diffeomorphism constraint\index{diffeomorphism constraint} can be simply reduced
\begin{eqnarray}
  -2\dfrac{c}{\ell_P^2}\partial_j\pi^{ij}-\dfrac{c}{\ell_P^2}h^{il}\left(2h_{jl,k}-h_{jk,l}\right)\pi^{jk}-2{J^i}&=&\nonumber\\
-2\dfrac{c}{\ell_P^2}\partial_j\pi^{ij}-\dfrac{c}{\ell_P^2}h^{il}\left(2h_{jl,k}-h_{jk,l}\right)h^j_ih^k_j\pi^{ij}-2{J^i}&=&\nonumber\\
-2\dfrac{c}{\ell_P^2}\left[\partial_j+h^{il}h^j_ih^k_j\left(h_{jl,k}-\dfrac{1}{2}h_{jk,l}\right)\right]\pi^{ij}-2{J^i}&=&\nonumber\\
-2\dfrac{c}{\ell_P^2}\left[\partial_j+h^{lk}\left(h_{jl,k}-\dfrac{1}{2}h_{jk,l}\right)\right]\pi^{ij}-2{J^i}.
\end{eqnarray}
Application of the relations
\begin{eqnarray}
 h^{lk}h_{jl,k}&=&\left(h^{lk}h_{jl}\right)_{,k}-h^{lk}_{,k}h_{jl}=\partial_k\delta^k_j-h^{lk}_{,k}h_{jl}=-h^{lk}_{,k}h_{jl}\\
 h^{lk}h_{jk,l}&=&\left(h^{lk}h_{jk}\right)_{,l}-h^{lk}_{,l}h_{jk}=\partial_l\delta^l_j-h^{lk}_{,l}h_{jk}=-h^{lk}_{,l}h_{jk},
\end{eqnarray}
and manipulations in the indices
\begin{eqnarray}
 -h^{lk}_{,l}h_{jk}=-h^{kl}_{,l}h_{jk}=-h^{lk}_{,k}h_{jl},
\end{eqnarray}
allows to rewrite the diffeomorphism constraint\index{diffeomorphism constraint} in the form
\begin{equation}
  H^i=-2\dfrac{c}{\ell_P^2}\left(\partial_j-\dfrac{1}{2}h_{jl}h^{lk}_{,k}\right)\pi^{ij}-2\dfrac{\ell_P^2}{c\kappa}{J^i}\approx0,
\end{equation}
which via using of $h_{jl}h^{lk}_{,k}=\partial_k\delta^k_j-h_{jl,k}h^{kl}=-h_{jl,k}h^{kl}$ becomes
\begin{equation}
  H^i=-2\dfrac{c}{\ell_P^2}\left(\partial_j+\dfrac{1}{2}h_{jl,k}h^{kl}\right)\pi^{ij}-2{J^i}\approx0.
\end{equation}
By this reason the canonical primary quantization \index{quantization!primary canonical}of the diffeomorphism constraint\index{diffeomorphism constraint} leads to the following equation
\begin{equation}
\left[i\dfrac{E_P}{\ell_P^3}\left(\partial_j+\dfrac{1}{2}h_{jl,k}h^{kl}\right)\dfrac{\delta}{\delta{h_{ij}}}-{J^i}\right]\Psi[h_{ij},\phi]=0.
\end{equation}

The DeWitt supermetric\index{DeWitt supermetric} (\ref{supmet}) and the analogous metric following from the Peres equation \index{Peres equation}(\ref{con}) by multiplication of both the sides by $2\sqrt{h}$ are not the same despite the same procedure is applied. Factually, the relation between both the metrics follows from the change
\begin{equation}
  h_{ik}h_{jl}\longleftrightarrow \dfrac{h_{ik}h_{jl}+h_{il}h_{jk}}{2}=h_{i(k}h_{jl)},\label{condi}
\end{equation}
what means that DeWitt applied symmetrization $h_{i(k}h_{jl)}$ instead of $h_{ik}h_{jl}$, and by this reason included to the quantum geometrodynamics\index{quantum geometrodynamics (QGD)} one more stratum than Peres in derivation of the classical geometrodynamics. The equation (\ref{condi}) is in itself non trivial, because it generates
\begin{equation}
h_{ik}h_{jl}=h_{il}h_{jk},
\end{equation}
what after multiplication of both sides by $h^{ik}h^{jl}$ leads to
\begin{equation}
  D^2=h^k_lh^l_k=h^k_k=D,
\end{equation}
what is true if and only if the dimensionality of embedded space is $D=1$ or $D=0$. Naturally, it is not true in general. However, in the case when one uses the DeWitt method then one obtains the Peres equation\index{Peres equation} with the DeWitt supermetric\index{DeWitt supermetric}, i.e.
\begin{equation}\label{per}
  {2c\kappa}G_{ijkl}\dfrac{\delta{S}}{\delta{h_{ij}}}\dfrac{\delta{S}}{\delta{h_{kl}}}+\dfrac{\ell_P^2}{2c\kappa}\sqrt{h}\left({^{(3)}}R-2\Lambda-2\kappa\varrho\right)=0,
\end{equation}
and canonical primary quantization \index{quantization!primary canonical}of such a classical geometrodynamics leads to the Wheeler--DeWitt equation\index{Wheeler--DeWitt equation}. The classical geometrodynamics is argued by the fact that the only such a procedure establishes straightforward equivalence between quantization of the Hamiltonian constraint\index{Hamiltonian constraint} and quantization of the Euler--Lagrange\index{Euler--Lagrange equations of motion} equations of motion for geometrodynamics of an embedded space of arbitrary dimensionality $D$, including the situation $D=3$ which we are studying in this book. Factually, Wheeler \cite{whe} did not argued using of the symmetrization $h_{i(k}h_{jl)}$ and in this way he rather studied the quantum Peres equation \index{Peres equation!quantum}than the Wheeler--DeWitt equation\index{Wheeler--DeWitt equation}. DeWitt \cite{dew} established the supermetric\index{DeWitt supermetric} (\ref{supmet}) and derived the Wheeler--DeWitt equation\index{Wheeler--DeWitt equation}. The difference is crucial because in modern quantum geometrodynamics\index{quantum geometrodynamics (QGD)} the Wheeler superspace \index{Wheeler superspace}is def\/ined by the DeWitt supermetric \index{DeWitt supermetric}which possesses non trivial properties. The diffeomorphism constraint\index{diffeomorphism constraint} (\ref{constB}) with (\ref{con2}) and the equation of motion (\ref{eqmot1}) differs only by the constant factor $-2$, and by this reason lead to the same quantum and classical conditions independently on the DeWitt supermetric\index{DeWitt supermetric}. This is because of this condition merely ref\/lects diffeoinvariance, while the Wheeler--DeWitt equation \index{Wheeler--DeWitt equation}or the Peres equation \index{Peres equation}def\/ine quantum and classical dynamics, respectively. In other words, in quantum geometrodynamics\index{quantum geometrodynamics (QGD)} studying diffeomorphism constraint\index{diffeomorphism constraint} is worthless from the dynamical point of view.

DeWitt \cite{dew} argued that application of the Wheeler metric representation \index{Wheeler metric representation}for closed f\/inite worlds results in the feature: the wave functional $\Psi[h_{ij},\phi]$ depends only on components of three-metric, i.e. is $\Psi[h_{ij}]$. In such a situation the equations (\ref{momeq}) express the necessary and suff\/icient conditions for diffeoinvariance of the wave functional $\Psi[h_{ij}]$. For f\/inite worlds it means that the wave functional depends only on the geometry of an embedded space. He proposed to construct the related structure of $\Psi[h_{ij}]$ via the hypersurface integrals which can be constructed out of products of the Riemann-Christoffel tensor and its covariant derivatives, with the topology of three-space itself being separately specif\/ied. DeWitt discussed differences between inf\/inite and f\/inite worlds. He proposed that in the f\/inite case one can replace the wave functional $\Psi[h_{ij}]$ by $\Psi[{^{(3)}}\mathfrak{G}]$, where ${^{(3)}}\mathfrak{G}$ is the three-geometry. For description of the quantum gravity he introduced $\mathfrak{M}$ - the set of all possible three-geometries which a f\/inite world may possess, called today \emph{midisuperspace}\index{midisuperspace}, and asked for topological issues related to $\mathfrak{M}$.

However, such topological arguments are rather misleading in the light of the fact that the Wheeler--DeWitt equation \index{Wheeler--DeWitt equation}has never been solved in general. The only known solution is the Hartle--Hawking wave function\index{Hartle--Hawking wave function} \cite{hawk02}, called the wave function of the Universe, which is expressed via the Feynman path integral technique\index{Feynman path integral}
\begin{eqnarray}
\Psi[h_{ij}]&=&N\int_C\delta g(x)\exp(iS_E[g]),\label{harhaw1}\\
\Psi[h_{ij},\phi]&=&N\int_C\delta g\delta\phi\exp(iS[g,\phi]),\label{harhaw2}
\end{eqnarray}
where $N$ is normalization factor, and $S_E[g]$ and $S[g,\psi]$ are the geometric part of the total action and the total action, respectively. The functional integral is over all four-geometries with a space-like boundary on which the induced metric is $h_{ij}$. Albeit, the path integrals (\ref{harhaw1}) and (\ref{harhaw2}) are the only a kind tautology following from the fact that the Wheeler-DeWitt equation \index{Wheeler--DeWitt equation}can be treated as the non relativistic quantum mechanics, i.e. the Schr\"odinger equation\index{Schr\"odinger equation}. Such a strategy, however, has never been lead to a general evaluation of the Hartle--Hawking wave function\index{Hartle--Hawking wave function}. In other words, the Feynman path integrals of quantum geometrodynamics\index{quantum geometrodynamics (QGD)} (\ref{harhaw1})-(\ref{harhaw2}) can be established straightforwardly and easy in very few particular cases while its computation has never been performed for a general case. Such a situation is the legacy of the fact that in general the integration strategy based on functional integration carries a diff\/icult computational level, and above all in general functional integration is not well-def\/ined mathematical procedure. Such a wave functionals lead to the Wheeler--DeWitt equation\index{Wheeler--DeWitt equation}, and by this reason are its solutions. However, they def\/ine the only class of solutions of the Wheeler--DeWitt equation\index{Wheeler--DeWitt equation}.

\section{The Wheeler Superspace}
The mathematical structure of geometrodynamics is determined by the Wheeler superspace\index{Wheeler superspace}, i.e. the conf\/igurational space of General Relativity, which is a space of all equivalence class of metric f\/ields of General Relativity related by action of the diffeomorphism group\index{diffeomorphism group} of a compact, connected, orientable, Hausdorff, $C^\infty$ three-dimensional space-like manifold without boundary $\partial M$. Superspace is the factor space
\begin{equation}
  S(\partial M)=\dfrac{Riem(\partial M)}{Dif\!f(\partial M)},\label{supspace}
\end{equation}
where $Riem(\partial M)$ is a space of all $C^\infty$ Riemannian metrics on the boundary $\partial M$, while $Dif\!f(\partial M)$ is the group of all $C^\infty$ diffeomorphisms of $\partial M$ that preserve orientation. Superspace as a space of orbits of the diffeomorphism group\index{diffeomorphism group} is in itself a connected, second-countable, metrizeable space. From a topological point of view $Riem(\partial M)$ is an open positive convex cone in the inf\/inite dimensional vector space of all smooth $C^\infty$ symmetric second-rank tensor f\/ields over $\partial M$ having the point-set topology, i.e.
\begin{equation}
\forall \lambda\in\mathbb{R}_+ \cap\forall h_{ij}\in Riem(\partial M): \lambda h_{ij}\in Riem(\partial M).
\end{equation}
This vector space is a locally convex topological vector space possessing a translation-invariant metric $\bar{d}$ which induces its topology and def\/ines completeness of the space, i.e. is a Fr\'{e}chet space\index{Fr\'{e}chet space}. The metric can be chosen in such a way that $Dif\!f(\partial M)$ preserves distances, and $Riem(\partial M)$ inherits the metric and by this reason is metrizable topological space that is also paracompact and second countable. Factoring out $Dif\!f(\partial M)$ transits the topological information concerning $\partial M$ to the quotient space $S(\partial M)$. There are two problems. The f\/irst is the case of closed $\partial M$ equipped with metrics with non-trivial isometry group, for which $S(\partial M)$ is not manifold. As showed Fischer \cite{fish} in such a situation $Dif\!f(\partial M)$ does not act freely so $S(\partial M)$ is stratif\/ied manifold with nested sets of strata ordered according to the dimension of the isometry groups. However, then exists a way to resolving the singularities \cite{fimon} which involves the frame bundle\index{bundle!frame} $F(\partial M)$ over $\partial M$ (For details of the theory of bundles see e.g. the Ref. \cite{steen}) such that the quotient space
\begin{equation}
\dfrac{Riem(\partial M)\times F(\partial M)}{Dif\!f(\partial M)},
\end{equation}
is the ref\/inement of superspace. The action of the diffeomorphism group\index{diffeomorphism group} $Dif\!f(\partial M)$ is now free because non-trivial isometries f\/ixing a frame are removed. If $\phi$ is such an isometry, one can apply the exponential map and the relation valid for any isometry $\phi\circ\exp=\exp\circ\phi_\star$ to show that the subset of points in $\partial M$ f\/ixed by $\phi$ is open. Since this set is also closed and $\partial M$ is connected, $\phi$ must be the identity. The def\/inition (\ref{supspace}) can be more ref\/ined if one restricts the group of diffeomorphisms to the proper subgroup of those diffeomorphisms that f\/ix a preferred point, called $\infty\in\partial M$ and the tangent space at this point
\begin{equation}
Dif\!f_F(\partial M)=\left\{\phi\in Dif\!f(\partial M)|\phi(\infty)=\infty,\phi_\star(\infty)=id|_{T_\infty\partial M}\right\}.
\end{equation}
Then the quotient $Riem(\partial M)\times F(\partial M)/Dif\!f(\partial M)$ is isomorphic to
\begin{equation}
  S_F(\partial M)=\dfrac{Riem(\partial M)}{Dif\!f_F(\partial M)},\label{supspaceF}
\end{equation}
but a preferred point $\infty$ must be chosen arbitrary. $S_F(\partial M)$ is called \emph{the extended superspace}\index{superspace!extended}. The problematic situation is also asymptotic f\/latness. If one treats $\partial M$ as one-point compactif\/ication of a manifold with one end then diffeomorphisms have to respect the asymptotic geometry and by this reason extended superspace\index{superspace!extended} is right. The extended superspace\index{superspace!extended} would have been unnecessary in the closed case if one restricted attention to those manifolds $\partial M$ which do not allow for metrics with continuous symmetries, i.e. which degree of symmetry is zero. Recall that the degree of symmetry of a manifold $\partial M$ is def\/ined as
\begin{equation}
  \deg(\partial M)=\sup_{h_{ij}\in Riem(\partial M)}\dim J(\partial M,h_{ij}),
\end{equation}
where $J(\partial M,h_{ij})$ is the isometry group of $(\partial M,h_{ij})$
\begin{equation}
J(\partial M,h_{ij})=\left\{\phi\in Dif\!f(\partial M)|\phi^\star h_{ij}=h_{ij}\right\},
\end{equation}
and when the dimension of a manifold is $D=\dim\partial M$ then
\begin{equation}
\dim J(\partial M,h_{ij})\leqslant\dfrac{D(D+1)}{2}.
\end{equation}
$J(\partial M,h_{ij})$ is compact if $\partial M$ is compact. If $\partial M$ allows for an effective action of a compact group $G$ then it clearly allows for a metric $h_{ij}$ on which $G$ acts as isometries just average any Riemannian metric over $G$. For compact $\partial M$ the degree of symmetry is zero if and only if $\partial M$ cannot support an action of the circle group $SO(2)$. A list of three-manifolds
with $deg>0$ was done by A.E. Fischer \cite{fisher}, and with $\deg=0$ by A.E. Fischer and V.E. Moncrief \cite{fimon}. Because of the projection
\begin{equation}
Riem(\partial M)\rightarrow S_F(\partial M)
\end{equation}
is continuous, and $Dif\!f_F(\partial M)$ acts continuously on $Riem(\partial M)$, the topology of extended superspace\index{superspace!extended} is quotient and open. For arbitrary two geometries $x,y\in S_F(\partial M)$ a metric on $S_F(\partial M)$
\begin{equation}
  d(x,y)=\sup_{\phi_x,\phi_y\in Dif\!f_F(\partial M)}\bar{d}(\phi_x^\star x,\phi_y^\star y),
\end{equation}
where $\bar{d}$ is mentioned above a translation-invariant metric on $Riem(\partial M)$, turns $S_F(\partial M)$ into a connected, metrizable and second countable topological space. Hence $Riem(\partial M)$ and $S_F(\partial M)$ are perfectly decent connected topological spaces satisfying the axioms of strongest separability and countability. The basic geometric idea is to regard $Riem(\partial M)$ as principal f\/ibre bundle\index{bundle!principal f\/ibre} with structure group $Dif\!f_F(\partial M)$ and the extended superspace\index{superspace!extended} $S_F(\partial M)$
\[\begin{CD}
Dif\!f_F(\partial M)@>i>>Riem(\partial M)@>p>>S_F(\partial M)
\end{CD}\]
where the $i,p$ are the inclusion and projection maps, respectively. This is made possible by the so-called slice theorems and the fact that the group acts freely and properly. This bundle structure has two far-reaching consequences regarding the geometry and topology of $S_F(\partial M)$.

The topologically trivial space $Riem(\partial M)$ can be visualized as the box f\/ibred by the action of the diffeomorphism group\index{diffeomorphism group} $Dif\!f_F(\partial M)\in Dif\!f(\partial M)$ generated by the diffeomorphism constraint\index{diffeomorphism constraint}, where orbits of the diffeomorphism group\index{diffeomorphism group} $Dif\!f(\partial M)$ are represented by straight-lines in the box. The quotient space $S(\partial M)$ obtains non-trivial topology from $Diff(\partial M)$ with an orbit represented by one point only. The subgroup $Dif\!f_F(\partial M)$ acts as isometries on the DeWitt supermetric \index{DeWitt supermetric}on $Riem(\partial M)$. $S(\partial M)$ is the set of geometries of an embedding $\partial M$, which are equivalence classes of isometric Riemannian metrics. By the Metrization Theorem for Superspace\index{Metrization Theorem for Superspace}, $S(\partial M)$is a connected, second-countable, metrizeable space. In other words a countable basis of open sets exists for its topology, and there also exists a metric on $S(\partial M)$ inducting such a topology. The partially-ordered set of conjugacy classes of compact subgroups of $Dif\!f(\partial M)$ indexes a partition of $S(\partial M)$, i.e. is a set of nonempty subspaces $\{\Sigma_\alpha\}$ such that \begin{eqnarray}
S(\partial M)&=&\bigcup_{\alpha}\Sigma_\alpha,\\
\Sigma_\alpha\cap\Sigma_\beta\neq\emptyset&\Rightarrow&\alpha=\beta.
\end{eqnarray}
A partition is a manifold partition if each $\Sigma_\alpha$ is a manifold. All geometries with the same kind of symmetry have homeomorphic neighbourhoods, and therefore create a manifold in the Wheeler superspace \index{Wheeler superspace}$S(\partial M)$. The neighbourhoods of all symmetric geometries are not homeomorphic to the neighbourhoods of all nonsymmetric geometries, and therefore $S(\partial M)$ is not a manifold. According to the Decomposition Theorem of Superspace\index{Decomposition Theorem of Superspace}, $S(\partial M)$ can be decomposed by its subspaces $S_G(\partial M)$ on a countable, partially-ordered, $C^\infty$-Fr\'{e}chet manifold partition. By Stratif\/ication Theorem for Superspace\index{Stratif\/ication Theorem for Superspace}, $S_G(\partial M)$ is an inverted stratif\/ication indexed by the type of symmetry, i.e. geometries with a given symmetry are completely contained within the boundary of less symmetric geometries. There is a theorem due to D. Giulini which states that in a neighbourhood of the round three-sphere in $S(\partial M)$ the DeWitt supermetric\index{DeWitt supermetric} is an inf\/inite-dimensional Lorentzian metric, i.e. is of signature $(-1,\infty)$. However, at each point of space $\partial M$ the DeWitt supermetric \index{DeWitt supermetric}def\/ines a Lorentzian metric on the $1 + 5$ dimensional space of symmetric second-rank tensors at that point, which can be identif\/ied with the homogeneous quotient space
\begin{equation}
\dfrac{GL(3,\mathbb{R})}{SO(3)}\cong \dfrac{SL(3,\mathbb{R})}{SO(3)\times\mathbb{R}_+}.
\end{equation}
As showed Giulini and Kiefer \cite{giulkief} the Lorentzian signature of the DeWitt supermetric\index{DeWitt supermetric} has nothing to do with the Lorentzian signature of the space-time metric, i.e. it persists in Euclidean gravity. But it is related to the attractivity of gravity. Any two points of the Wheeler superspace \index{Wheeler superspace}which differ by an action of the diffeomorphism constraint\index{diffeomorphism constraint} are gauge equivalent and hence physically indistinguishable. However, the question of whether and when the diffeomorphism constraint\index{diffeomorphism constraint} actually generates all diffeomorphisms of $\partial M$ is unsolved. In fact, General Relativity is a dynamical system on the cotangent bundle\index{bundle!cotangent}, i.e. phase space, built over $S(\partial M)$. The topology of superspace is inherited from the topology of $\partial M$. The Hamiltonian evolution is varying embedding of space $\partial M$ into space-time $M$. Hence the images of an embedded space have the same topological type, what ref\/lects the fact that the classical geometrodynamics transitions of topology are impossible. This is not implied by the Einstein f\/ield equations\index{Einstein f\/ield equations}, but is a consequence of the restriction to space-times admitting a global space-like foliation. There is a number of solutions to the Einstein f\/ield equations\index{Einstein f\/ield equations} which do not satisfy such a requirement, i.e. such space-times cannot be constructed by integrating the Gauss--Codazzi equations\index{Gauss--Codazzi equations} with some reasonable initial data. From the Hartle--Hawking wave function\index{Hartle--Hawking wave function} point of view topology changing classical solutions should not be removed as possible contributors in the Feynman path integral\index{Feynman path integral}. In the evolutionary formulation of the Einstein f\/ield equations\index{Einstein f\/ield equations}, there is no space-time to start with. Only solutions of the dynamical equations construct the space-time. Then one can interpret the time dependence of the induced metric as being brought about by ‘wafting’ three-space through space-time via an embedding. Initially there is only a space-like submanifold of unrestricted topology.

The deformations of the space-like hypersurfaces, i.e. inf\/initesimal changes of embeddings $\mathcal{E}:\partial M\mapsto M$, possess nontrivial kinematics. The generators of normal and tangential deformations are
\begin{eqnarray}
  \mathcal{N}_{N}&=&\int_{\partial M}d^3x N(x) n^\mu[y(x)]\dfrac{\delta}{\delta y^\mu(x)},\label{genn1}\\
  \mathcal{T}_{N^i}&=&\int_{\partial M}d^3x N^i(x) y^\mu_{,i}(x)\dfrac{\delta}{\delta y^\mu(x)},\label{genn2}
\end{eqnarray}
where $y^\mu_{,i}=\partial_iy^\mu$, and $y^\mu$ and $x^i$ are local coordinates on $M$ and $\partial M$, respectively. The generators (\ref{genn1}) and (\ref{genn2}) can be understood as tangent vectors to the space of embeddings of $\partial M$ into M. An embedding is locally given by four functions $y^\mu(x)$, such that the $3\times4$ matrix $y^\mu_{,i}$ has its maximum rank 3. We have denoted by $n^\mu$ the components of the normal to the image $\mathcal{E}(\partial M)\in M$, which are functionals of $y^\mu(x)$, i.e. $n^\mu=n^\mu[y(x)]$. The generators (\ref{genn1}) and (\ref{genn2}) satisfy the following commutation relations
\begin{eqnarray}
  \left[\mathcal{T}_{N^i},\mathcal{T}_{{N'}^{i}}\right]&=&-\mathcal{T}_{\left[N^i,{N'}^{i}\right]},\label{al1}\\
  \left[\mathcal{T}_{N^i},\mathcal{N}_N\right]&=&-\mathcal{N}_{N^i(N)},\label{al2}\\
  \left[\mathcal{N}_N,\mathcal{N}_{N'}\right]&=&-\mathcal{T}_{N\nabla_hN'-N'\nabla_hN},\label{al3}
\end{eqnarray}
where $\nabla_hN=\left(h^{ab}\partial_bN\right)\partial_a$, and the Lie brackets (\ref{al2}) and (\ref{al3}) were obtained via taking variations of the basic identities $g_{\mu\nu}n^\mu n^{\nu}=-1$ and $g_{\mu\nu}y^\mu_{,i}n^\nu=0$. The space-time vector f\/ield
\begin{equation}
  V=Nn^\mu\partial_\mu+N^i\partial_i,
\end{equation}
induces the foliation-dependent decomposition of the tangent vector $\mathrm{T}(V)$ at $Y\in \textrm{Emb}(\partial M,M)$
\begin{equation}\label{tang}
  \mathrm{T}(V)=\int_{\partial M} d^3x V^\mu(y(x))\dfrac{\delta}{\delta y^\mu(x)},
\end{equation}
obeying the Lie algebra
\begin{equation}
  \left[\mathrm{T}(V),\mathrm{T}\left(V'\right)\right]=\mathrm{T}\left(\left[V,V'\right]\right),
\end{equation}
which means that $V\mapsto \mathrm{T}(V)$ is a Lie homomorphism from the tangent-vector f\/ields on $M$ to the tangent-vector f\/ields on $\textrm{Emb}(\partial M,M)$. In this sense, the Lie algebra of the four-dimensional diffeomorphism group\index{diffeomorphism group} is implemented on phase space of arbitrary generally covariant theory which phase space includes the embedding variables\index{embedding variables}, i.e. is so-called parametrized theory. Decomposing the vector (\ref{tang}) into normal and tangential components with respect to the leaves of the embedding at which the tangent-vector f\/ield to $Emb(\partial M,M)$ is evaluated, yields an embedding-dependent parametrization of $\mathrm{T}(V)$ via $(N,N^i)$
\begin{equation}\label{tang1}
  \mathrm{T}(N,N^i)=\int_{\partial M} d^3x \left[Nn^\mu[y(x)]+N^i(x)y^\mu_{,i}(x)\right]\dfrac{\delta}{\delta y^\mu(x)}.
\end{equation}
Computing the functional derivatives of $n$ with respect to $y$ one can establish the commutator of deformation generators
\begin{equation}
\left[\mathrm{T}\left(N,N^i\right),\mathrm{T}\left(N',{N'}^i\right)\right]=-\mathrm{T}\left(N'',{N''}^i\right),
\end{equation}
where
\begin{eqnarray}
N''&=&N^i(N)-{N'}^i(N'),\\
{N''}^i&=&[N^i,{N'}^i]+N\nabla_h N'-N'\nabla_hN.
\end{eqnarray}
The situation can be easy visualized. Composition of two an inf\/initesimal hypersurface deformations with parameters $(N,N^i)$ and $(N',{N'}^i)$ that maps $\partial M\mapsto\partial M_1$ and $\partial M_1\mapsto\partial M_{12}$ respectively, differs by the hypersurface deformation with parameters $(N'',{N''}^i)$ given by the composition that maps with the same parameters but in the opposite order.
\begin{equation}
\begin{tikzpicture}
\matrix(m)[matrix of math nodes, row sep=5em, column sep=5em, text height=1.5ex, text depth=0.25ex]
{
\partial{M_1}&\partial{M}&\partial{M_2}\\
\partial{M_{12}}& &\partial{M_{21}}\\
};
\path[->,>=angle 90,font=\normalsize]
(m-1-1) edge node[left]  {$(N',{N'}^i)$}   (m-2-1)
(m-1-2) edge node[above] {$(N',{N'}^i)$}   (m-1-3)
(m-1-3) edge node[right] {$(N,N^i)$}       (m-2-3)
(m-2-3) edge node[below] {$(N'',{N''}^i)$} (m-2-1)
(m-1-2) edge node[above] {$(N,N^i)$}       (m-1-1);
\end{tikzpicture}
\end{equation}

Hamiltonian General Relativity is a particular Lie-anti representation\index{Lie-anti representation} of the algebra (\ref{al1})-(\ref{al2}) as a Hamilton system on the phase space of physical f\/ields. When the latter merely depends on the induced metric $h_{ij}$ on $\partial M$, then the unconstrained phase space is the cotangent bundle\index{bundle!cotangent} $T^\star Riem(\partial M)$ over $Riem(\partial M)$, parameterized by the pair $(h_{ij}, \pi_{ij})$. In this simplest situation one can search for
\begin{equation}
  (N,N^i)\mapsto\left(\mathcal{H}(N,N^i):T^\star Riem(\partial M)\rightarrow\mathbb{R}\right)
\end{equation}
where $\mathcal{H}$ is a distribution,the test functions are $N$ and $N^i$, with values in real-valued
functions on $T^\star Riem(\partial M)$
\begin{equation}
  \mathcal{H}(N,N^i)[h_{ij},\pi_{ij}]=\int_{\partial M}d^3x\left(N(x)H(x)+N_i(x)H^i(x)\right).
\end{equation}
The fundamental condition is that the Poisson brackets\index{Poisson brackets} between two values of $\mathcal{H}(N,N^i)$ is
\begin{equation}\label{hambra}
  \left\{\mathcal{H}\left(N,N^i\right),\mathcal{H}\left(N',{N'}^i\right)\right\}=\mathcal{H}\left(N'',{N''}^i\right)
\end{equation}
The essential question is recovering the action of the Lie algebra of four-dimensional diffeomorphism on the extended phase space including embedding variables\index{embedding variables}.

The Hamiltonian ani-Lie representation\index{Lie-anti representation} of the algebra (\ref{al1})-(\ref{al2}) can be constructed via the identif\/ication
\begin{eqnarray}
\mathcal{N}_N\mapsto \mathcal{H}(N)&=&\int_{\partial M}d^3xN(x)H(x),\label{fam1}\\
\mathcal{T}_{N^i}\mapsto \mathcal{D}(N^i)&=&\int_{\partial M}d^3xN_i(x)H^i(x),\label{fam2}
\end{eqnarray}
and the resulting algebra is the Lie algebra of the diffeomorphism group\index{diffeomorphism group} $Dif\!f(\partial M)$
\begin{eqnarray}
  \left\{\mathcal{D}\left(N^i\right),\mathcal{D}\left({N'}^i\right)\right\}&=&\mathcal{D}\left(\left[N^i,{N'}^i\right]\right),\label{alg1}\\
  \left\{\mathcal{D}\left(N^i\right),\mathcal{H}\left(N\right)\right\}&=&\mathcal{H}\left(N^i\left(N\right)\right),\label{alg2}\\
  \left\{\mathcal{H}\left(N\right),\mathcal{H}\left(N'\right)\right\}&=&\mathcal{D}\left(N\nabla_hN'-N'\nabla_hN\right),\label{alg3}
\end{eqnarray}
which means that geometrodynamics does not def\/ine the extraordinary situation. Any four dimensional $Dif\!f(\partial M)$-invariant theory will gives rise to this same algebra. The Lie algebra of $Dif\!f(\partial M)$, namely, is universally satisf\/ied for a theory considered in an arbitrary $space+time$ decomposition, which is formed by hypersurface foliations and is space-time covariant. The universality of the diffeomorphism Lie algebra suggests searching for its another, possibly more general, representations on a given phase space. The theorem due to K. Kucha\v{r}, C. Teitelboim, and S.A. Hojman \cite{kth}, which states that for the the unique two-parameter family, given by $\kappa$ and $\Lambda$, of realizations for the algebra (\ref{fam1})-(\ref{fam2}) equipped with the constraints $H$ and $H^i$, in which the conjugated momentum can be expressed via the extrinsic curvature and the kinetic term is multiplied by an overall $n^\mu n_\mu$, is the most general Hamiltonian representation of the universal diffeomorphism Lie algebra (\ref{alg1})-(\ref{alg3}) on $T^\ast Riem(\partial M)$, i.e. the space tangent to all $C^\infty$ Riemannian metrics on the boundary $\partial M$, up to the residual canonical transformations\index{residual canonical transformations} having the following form
\begin{equation}
\pi^{ij}\mapsto\pi^{ij}+\dfrac{\delta F[h_{ij}]}{\delta h_{ij}},
\end{equation}
where $F$ is some function invariant with respect to action of the diffeomorphism group\index{diffeomorphism group} $Dif\!f(\partial M)$ determined on the space $Riem(\partial M)$.

Superspace possesses analogous ambiguity to the Aharonov--Bohm effect\index{Aharonov--Bohm effect} \cite{ahaboh} following from the lack of simple connectedness of topology. This depends on the topology of $\partial M$, i.e. contractibility of $Riem(\partial M)$ results in the relation for n-th homotopy group\index{homotopy group}
\begin{equation}\label{homotop}
  \pi_n\left(\dfrac{Riem(\partial M)}{Dif\!f_F(\partial M)}\right)\cong\pi_{n-1}\left(Dif\!f_F(\partial M)\right),
\end{equation}
where $n\geqslant1$. For $n=1$ one obtains the fundamental group of the extended superspace\index{superspace!extended!fundamental group}
\begin{equation}
  \pi_1\left(S_F(\partial M)\right)\cong\pi_0\left(Dif\!f_F(\partial M)\right)=\dfrac{Dif\!f_F(\partial M)}{Dif\!f_F^0(\partial M)}=MCG_F(\partial M),
\end{equation}
where $Dif\!f_F^0(\partial M)$ is the identity component of $Dif\!f_F(\partial M)$, and $MCG_F(\partial M)$ denotes abbreviation of the name mapping-class group for frame f\/ixing diffeomorphisms \cite{giul}. Uniqueness of representations of the Poisson brackets\index{Poisson brackets} (\ref{hambra}) has much more serious ambiguity. It is namely labeled by an additional $\mathbb{C}$-valued parameter, called the Barbero--Immirzi parameter\index{Barbero--Immirzi parameter}, obtained due to connection variables (See the Ref. \cite{barimi} and the books \cite{kiefer,qg45}). In such a situation the Poisson brackets\index{Poisson brackets} (\ref{hambra}) can not represented via a semi-direct product of it with the Lie algebra of $SU(2)$ gauge transformations, and by taking the quotient with respect to this algebra the Poisson brackets\index{Poisson brackets} (\ref{hambra}) are represented non locally. The Poisson brackets\index{Poisson brackets} (\ref{hambra}) do not seem to apply in case of connection variables, because of the connection variable does not admit an interpretation as a space-time gauge f\/ield restricted to space-like hypersurfaces and the dynamics generated by the constraints does not admit the interpretation of being induced by appropriately moving a hypersurface through a space-time with f\/ixed geometric structures on it. The homotopy groups\index{homotopy group} of the extended superspace\index{superspace!extended}, given by the formula (\ref{homotop}), encode much of its global topology. They were investigated by D. Giulini \cite{giul} and D. Witt \cite{witt}, and its quantum gravity context was investigated in the contributions due to J. Friedman and R. Sorkin \cite{frisor}, C.J. Isham \cite{isham}, and R. Sorkin \cite{sorkin}. Topological invariant of $S_F(\partial M)$ are also topological invariants of $\partial M$, while homotopy invariants\index{homotopy invariant} of $\partial M$ in general are not homotopy invariants\index{homotopy invariant} of $S_F(\partial M)$. Such a situation means that one can distinguish homotopy equivalent but non homeomorphic three-manifolds by looking at homotopy invariants\index{homotopy invariant} of their associated superspaces. When $\partial M$ is connected closed orientable three-manifold then the homology\index{homology groups} and cohomology\index{cohomology!groups} groups are determined by the fundamental group. Namely, the f\/irst four (zeroth to third, the only non-trivial ones) homology\index{homology groups} and cohomology groups are
\begin{eqnarray}
H_\star&=&(\mathbb{Z},A\pi_1,FA\pi_1,\mathbb{Z}),\label{homolog}\\
H^\star&=&(\mathbb{Z},FA\pi_1,A\pi_1,\mathbb{Z}),\label{cohom}
\end{eqnarray}
where $A$ and $F$ are the operation of abelianisation of a group and the operation of taking the free part of a f\/initely generated abelian group, respectively. Particularly interesting is the fundamental group of the extended superspace\index{superspace!extended}. The analogy with quantum mechanics already suggests that its classes of inequivalent irreducible unitary representations correspond to a superselection structure which here might serve as f\/ingerprint of the topology of $\partial M$ in the quantum theory. The sectors might, for instance, correspond to various statistics that preserve or violate a naively expected spin-statistics correlation (See e.g. papers in the Ref. \cite{spin}).

General three-manifolds can be understood by surgery\index{manifold surgery} (For theory see e.g. books in the Ref. \cite{surgery}, for excellent constructive application see e.g. the paper due to G. Perelman \cite{perel}). Particularly cutting along certain embedded two manifolds, such that the remaining pieces are simpler, is often applied technique. As the crucial and essential example, let us consider those simplif\/ications that are achieved by cutting along embedded two-spheres. An essential two-sphere is one which does not bound a three-ball and a splitting two-sphere is one which complement has two connected components. Let us consider a closed three-manifold $\partial M$, which is cutting along an essential splitting two-sphere, capping off the two-sphere boundary of each remaining component by a three-disk, and this process is repeating for each of the remaining closed three-manifolds. This process stops after a f\/inite number of steps where the resulting components are uniquely determined up to diffeomorphisms, orientation preserving if oriented manifolds are considered, and permutation \cite{milnor}. The process stops at that stage at which none of the remaining components, $\Pi_1,\ldots,\Pi_n$, allows for essential splitting two-spheres, i.e. at which each $\Pi_i$ is a prime manifold, i.e. such a manifold for which each embedded two-sphere either bounds a three-disc or does not split; it is called irreducible if each embedded two-sphere bounds a three-disc. In the latter case the second homotopy group\index{homotopy group!second}, $\pi_2$, must be trivial, since, if it were not, the so-called sphere theorem\index{sphere theorem}. The theorem states that if for connected three-manifold $\partial M$ $\pi_2 M=0$ then there is either an embedded $S^2$ in $M$ representing a nontrivial element in $\pi_2M$, or an embedded two-sided $\mathbb{R}\mathrm{P}^2$ in $\partial M$ such that the composition of the cover $S^2\rightarrow \mathbb{R}\mathrm{P}^2$ with the inclusion $\mathbb{R}\mathrm{P}^2 \hookrightarrow \partial M$ represents a nontrivial element of $\pi_2M$. This theorem ensures the existence of a non-trivial element of $\pi_2$ which could be represented by an embedded two-sphere. Conversely, it follows from the validity of the Poincar\'{e} conjecture\index{Poincar\'{e} conjecture} that a trivial $\pi_2$ implies irreducibility. Hence irreducibility is equivalent to a trivial $\pi_2$. There is precisely one non-irreducible prime three-manifold, and that is the handle $S^1\times S^2$. Hence a three-manifold is prime if and only if it is either a handle or if its $\pi_2$ is trivial. For a general three-manifold $\partial M$ given as connected sum of primes $\Pi_1,\ldots,\Pi_n$, there is a general method to establish $MCG_F(\partial M)$ in terms of $MCG_F(\Pi_i)$. The strategy is to look at the effect of elements in $MCG_F(\partial M)$ on the fundamental group of $\partial M$. As $\partial M$ is the connected sum of primes, and as connected sums in $D$ dimensions are taken along $D-1$ spheres which are simply-connected for $D\geqslant3$, the fundamental group of a connected sum is the free product of the fundamental groups of the primes for $D\geqslant3$. The group $MCG_F(\partial M)$ now naturally acts as automorphisms of $\pi(\partial M)$ by simply taking the image of a based loop that represents an element in $\pi(\partial M)$ by a based (same basepoint) diffeomorphism that represents the class in $MCG_F(\partial M)$. Hence there is a natural map of $MCG_F(\partial M)$ into group of automorphisms of the fundamental group of the D-dimensional manifold $\partial M$
\begin{equation}
  d_F:MCG_F(\partial M)\rightarrow\textrm{Aut}\left(\pi_1(\partial M)\right).\label{autom}
\end{equation}
The known f\/inite presentations of automorphism groups, i.e. its characterization in terms of a f\/inite number of generators and f\/inite number of relations, of free products in terms of presentations of the automorphisms of the individual factors and additional generators, basically exchanging isomorphic factors and conjugating whole factors by individual elements of others, can now be used to establish f\/inite presentations of $MCG_F(\partial M)$, provided f\/inite presentations for all prime factors are known. This presentation of the automorphism group of free products is due to D.I. Fouxe--Rabinovitch \cite{fouxe}, and its modern forms with corrections were performed by D. McCullough and A. Miller \cite{mcmil} and N.D. Gilbert \cite{gilbert}. This situation would be more complicated if $Dif\!f(\partial M)$ rather than $Dif\!f_F(\partial M)$, at least the diffeomorphisms f\/ixing a preferred point, had been considered. Only for $Dif\!f_F(\partial M)$, or the slightly larger group of diffeomorphisms f\/ixing the point, is it generally true that the mapping-class group of a prime factor injects into the mapping-class group of the connected sum in which it appears. For more on this, compare the discussion by D. Giulini \cite{giul1}. Clearly, one also needs to know which elements are in the kernel of the map (\ref{autom}).

The cotangent bundle\index{bundle!cotangent} over superspace is not the fully reduced phase space for matter-free General Relativity. It only takes account of the vector constraints and leaves the scalar constraint unreduced. However, under certain conditions, the scalar constraints can be solved by the ''conformal method'' which leaves only the conformal equivalence class of three-dimensional geometries as physical conf\/igurations. In those cases the fully reduced phase space is the cotangent bundle\index{bundle!cotangent} over conformal superspace, which is given by replacing $Dif\!f(\partial M)$ by the semi-direct product
\begin{equation}
Dif\!f^C(\partial M)=C(\partial M)\rtimes Dif\!f(\partial M),
\end{equation}
where $C(\partial M)$ is the abelian group of conformal re-scalings that acts on $Riem(\partial M)$ via pointwise multiplication $(f,h_{ij})\mapsto fh_{ij}$, where $f:\partial M\rightarrow\mathbb{R}_+$. The right action of $(f,\phi)\in Dif\!f_C(\partial M)$ on $h_{ij}\in Riem(\partial M)$ is then given by $R_{(f,\phi)}(h_{ij}) = f\phi^\star h_{ij}$, so that, using $R_{(f,\phi)}R_{(f',\phi')}=R_{(f,\phi)(f',\phi')}$, the semi-direct product structure is  $(f,\phi)(f',\phi')=(f'(f\circ\phi),\phi\circ\phi')$. Because of $(ff')\circ\phi=(f\circ\phi)(f'\circ\phi)$ $Dif\!f(\partial M)$ indeed acts as automorphisms of $C(\partial M)$. Conformal superspace\index{superspace!conformal} and extended conformal superspace\index{superspace!extended!conformal} would then be def\/ined as
\begin{eqnarray}
CS(\partial M)&=&\dfrac{Riem(\partial M)}{Dif\!f^C(\partial M)},\\
CS_F(\partial M)&=&\dfrac{Riem(\partial M)}{Dif\!f^C_F(\partial M)},
\end{eqnarray}
where
\begin{equation}
Dif\!f^C_F(\partial M)=C(\partial M)\rtimes Dif\!f_F(\partial M).
\end{equation}
Since $C(\partial M)$ is contractible, the topologies of $Dif\!f^C(\partial M)$ and $Dif\!f^C_F(\partial M)$ are those of $Dif\!f(\partial M)$ and $Dif\!f_F(\partial M)$ which also transcend to the quotient spaces whenever the groups act freely. In the f\/irst case this is essentially achieved by restricting to manifolds of vanishing degree of symmetry, whereas in the second case this follows almost as before, with the sole exception being $(S^3,h_{ij})$ with $h_{ij}$ conformal to the round metric. Let
\begin{equation}
CJ(\partial M, h_{ij}=\left\{\phi\in Dif\!f(\partial M) | \phi^\star h_{ij} = f h_{ij}, f: \partial M\rightarrow\mathbb{R}_+\right\}
\end{equation}
be the group of conformal isometries. For compact $\partial M$ it is known to be compact except if and only if $\partial M = S^3$ and $h_{ij}$ conformal to the round metric \cite{lelong}. Hence, for $\partial M\neq S^3$, we can average h over the compact group $CJ(\partial M, h_{ij})$ and obtain a new Riemannian metric $h'_{ij}$ in the conformal equivalence class of $h_{ij}$ for which $CJ(\partial M, h_{ij})$ acts as proper isometries. Therefore it cannot contain non-trivial elements f\/ixing a frame. Hence in contrast, the geometry for conformal superspace\index{superspace!conformal} differs insofar from that discussed above as the conformal modes that formed the negative directions of the DeWitt supermetric\index{DeWitt supermetric}. The horizontal subspaces, orthogonal to the orbits of $Dif\!f^C_F(\partial M)$, are now given by the transverse and traceless symmetric two-tensors. In that sense the geometry of conformal superspace\index{superspace!conformal}, if def\/ined as before by some ultralocal bilinear form on $Riem(\partial M)$, is manifestly positive due to the absence of trace terms and hence less pathological than the superspace metric discussed above. It might seem that its physical signif\/icance is less clear, as there is now no constraint left that may be said to induce this particular geometry. Whether it is a realistic hope to understand superspace\index{superspace} and conformal superspace\index{superspace!conformal}, its cotangent bundle\index{bundle!cotangent} being the space of solutions to the Einstein f\/ield equations\index{Einstein f\/ield equations}, well enough to actually gain a suff\/iciently complete understanding of its automorphism group is hard to say. An interesting strategy lies in the attempt to understand the solution space directly in a group-, or Lie algebra-, theoretic fashion in terms of a quotient $G^\infty/H^\infty$, where $G^\infty$ is an inf\/inite dimensional group (Lie algebra) that (locally) acts transitively on the space of solutions and $H^\infty$ is a suitable subgroup (algebra), usually the f\/ixed-point set of an involutive automorphism of $G$. The basis for the hope that this might work in general is the fact that it works for the subset of stationary and axially symmetric solutions of the Einstein f\/ield equations\index{Einstein f\/ield equations} for which $G^\infty$ is the Geroch group\index{Geroch group} \cite{geroch}.

One can distinguish homotopy equivalent but non homeomorphic three-dimensional $\partial M$ by looking at homotopy invariants\index{homotopy invariant} of their associated $S_F(\partial M)$. Good example are certain types of lens spaces $L(p,q)$\index{lens space} (For detailed discussion see e.g. the Ref. \cite{lens}), which are prime manifolds. Lens spaces were introduced by H. Tietze \cite{tietze} as the simplest possible examples of 3-manifolds obtained by identifying faces of a polyhedron. They were both f\/irst known three-manifolds not determined only by homology\index{homology groups} and fundamental group\index{fundamental group}, as well as the most simple closed manifolds for which the homotopy type does not determine the homeomorphism type. J.W. Alexander \cite{alex} proved that $L(5;1)$ and $L(5;2)$ are not homeomorphic despite their fundamental groups are isomorphic and their homology are identical, in spite that their homotopy type is not the same. Homotopy type of lens spaces\index{lens space} is the same, and therefore another lens spaces\index{lens space} have isomorphic fundamental groups and homology\index{homology}\index{fundamental group}. However, in general homeomorphism type is not the same for lens spaces\index{lens space}, and by this reason lens spaces\index{lens space} can are the birth of geometric topology of manifolds as distinct from algebraic topology. In dimension 3 are def\/ined as the quotient space of a 3-sphere $S^3$
\begin{equation}
  S^3=\left\{(z_1,z_2)\in\mathbb{C}\times\mathbb{C}||z_1|^2+|z_2|^2=1\right\},
\end{equation}
presented as the union of two solid tori
\begin{eqnarray}
  A_+&=&\left\{|z|=1-|w|^2,|w|\leqslant\dfrac{\sqrt{2}}{2}\right\},\\
  A_-&=&\left\{|z|\leqslant\dfrac{\sqrt{2}}{2},|w|=1-|z|^2\right\},
\end{eqnarray}
whose common boundary torus is given as the zero level of the function $f(z,w)=|z|^2-|w|^2$. These solid tori are invariant under isometric free action of the cyclic group $\mathbb{Z}_p$ of order $p$
\begin{equation}
\mathbb{Z}_p=\left\{1,\epsilon,\epsilon^2,\ldots,\epsilon^{p-1}\right\},
\end{equation}
where $\epsilon$ is a primitive $p$-th root of unity
\begin{equation}
\epsilon=\exp\dfrac{2\pi i}{p}.
\end{equation}
In other words the 3-dimensional lens space\index{lens space} is the orbit space\index{orbit space}
\begin{equation}
  L(p,q)=\dfrac{S^3}{\mathbb{Z}_p},
\end{equation}
where $(p,q)$ is the pair of relatively prime (coprime) integers with $p>1$. Moreover, after taking the quotient of the sphere, these solid tori are again transformed into solid tori into which the lens space\index{lens space} $L(p,q)$ splits. The function $f(z,w)$ def\/ined initially on the sphere generates a smooth function on $L(p,q)$. The levels of this function def\/ine some foliation on the lens space\index{lens space}. The three-manifold $L(p,q)$ can be visualized easy. Namely, the equator of a 3-ball, i.e. solid ball in $\mathbb{R}^3$, is divided into $p$ equal segments, so that the upper and lower hemispheres become $p$-sided polygons. These hemispherical faces are then identif\/ied by a rotation through $2\pi\dfrac{q}{p}$ about the vertical symmetry axis, where $0\leqslant q<p$ and $(p,q)=1$. If a corner is introduced along the equator of the 3-ball it assumes the lens-shaped appearance that gave these manifolds their name, the term lens space\index{lens space} being introduced by Seifert and Threlfall \cite{seth}. Tietze noted that $L(p,q)$ may also be described as the manifold with a genus $1$ Heegaard diagram consisting of a curve on the boundary of a solid torus which winds around p times latitudinally and $q$ times meridionally.

The action of $\mathbb{Z}_p$ on $S^3$ is
\begin{eqnarray}
  z_1'&=&\epsilon z_1,\\
  z_2'&=&\epsilon^qz_2.
\end{eqnarray}
In this way each set of $p$ equidistant points on the equator is identif\/ied to a single point. The fundamental group of $L(p,q)$ is independent on q
\begin{equation}
  \pi_1\left(L(p, q)\right)=\mathbb{Z}_p,
\end{equation}
and the higher homotopy groups\index{homotopy group!higher} are those of its universal covering space is $L(1,0) = S^3$ with $p$ sheets, i.e. $p$-fold 3-sphere. Therefore, also the standard invariants (\ref{homolog}) and (\ref{cohom}) taken for $L(p,q)$ are sensitive only to $p$. Two lens spaces\index{lens space} $L(p,q)$ and $L(p,q')$ are
\begin{enumerate}
  \item homotopy equivalent if and only if $qq'=\pm n^2 \mod p$ where $n\in \mathbb{Z}$ \cite{white}
  \item homeomorphic if and only if $q'=\pm q^{\pm 1} \mod p$ \cite{reid}
  \item orientation-preserving homeomorphic if and only if $q'=q^{\pm 1} \mod p$ \cite{reid}.
\end{enumerate}

The torsion linking form is the invariant allowing to perform the homotopy classif\/ication of three-dimensional lens spaces\index{lens space}, and the Reidemeister torsion\index{Reidemeister torsion} \cite{reid} allows to make the homeomorphism classif\/ication. This invariant was formalized and generalized to higher dimensions by Reidemeister's student Franz \cite{franz}. The latter was performed by K. Reidemeister as a classif\/ication up to piecewise linear homeomorphism (For PL topology see e.g. the Ref. \cite{piece}), and E.J. Brody \cite{brody} showed that the Reidemeister construction is also a homeomorphism classif\/ication. Lens spaces are determined by simple homotopy type, and there are no normal invariants, like e.g. characteristic classes, or surgery obstruction. J.H. Przytycki  and A. Yasuhara \cite{przyas} formulated the knot-theoretic classif\/ication: for a closed curve $C$ in the universal cover of the lens space\index{lens space} which lifts to a knot having a trivial Alexander polynomial, computation of the torsion linking form on the pair $(C,C)$ gives the homeomorphism classif\/ication. P. Salvatore and R. Longoni \cite{sallon} gave another showed that homotopy equivalent but not homeomorphic lens spaces\index{lens space} may have conf\/iguration spaces with different homotopy types, which can be detected by different Massey products.

The lens spaces\index{lens space} were also natural subjects for investigations of a more algebraic topological nature. In this vein, Rueff \cite{rueff} showed that there exists a degree 1 map $L(p, q)\rightarrow L(p,q')$ if and only if $qq'\equiv r^2 (\mod p)$, for some $r$. Lens space in the dimension 3 is the Seifert f\/iber space \cite{seif}, i.e. a $S^1$-bundle\index{bundle!circle} (circle bundle) over a 2-dimensional orbifold. In the dimension three the equivalence of the combinatorial and topological classif\/ications, called \emph{Hauptvermutung}\index{lens space!Hauptvermutung} and proved by E.E. Moise \cite{moise}, is validate (For detailed discussion see e.g. the Ref. \cite{haup}). For given $p$, we aim to attach a homotopy invariant\index{homotopy invariant} to $L(p,q)$ that depends on $q$. For this, one needs the Bockstein homomorphism\index{Bockstein homomorphism}. The short exact sequence of coeff\/icient groups
\begin{equation}
\begin{tikzpicture}
\matrix(m)[matrix of math nodes, row sep=3em, column sep=2.5em, text height=1.5ex, text depth=0.25ex]
{
0&\mathbb{Z}&\mathbb{Z}&\mathbb{Z}_p&0\\
};
\path[->,>=angle 90,font=\normalsize]
(m-1-1) edge (m-1-2)
(m-1-2) edge node[above] {$p$} (m-1-3)
(m-1-3) edge (m-1-4)
(m-1-4) edge (m-1-5);
\end{tikzpicture}
\end{equation}
induces the long exact sequence
\begin{equation}
\begin{tikzpicture}
\matrix(m)[matrix of math nodes, row sep=3em, column sep=2em, text height=1.5ex, text depth=0.25ex]
{
\ldots&H^n(X;\mathbb{Z})&H^n(X;\mathbb{Z})&H^n(X;\mathbb{Z}_p)&H^{n+1}(X;\mathbb{Z})&\ldots\\
};
\path[->,>=angle 90,font=\normalsize]
(m-1-1) edge (m-1-2)
(m-1-2) edge node[above] {$p$} (m-1-3)
(m-1-3) edge node[above] {$\rho$}       (m-1-4)
(m-1-4) edge node[above] {$\beta_0$}    (m-1-5)
(m-1-5) edge (m-1-6);
\end{tikzpicture}
\end{equation}
Let
\begin{equation}
  \beta=\rho\circ\beta_0:H^n(X;\mathbb{Z}_p)\rightarrow H^{n+1}(X;\mathbb{Z}_p).
\end{equation}
This, and $\beta^0$ itself, is called the Bockstein homomorphism\index{Bockstein homomorphism}. It is natural in a space $X$, and it is called a cohomology\index{cohomology} operation. The Hurewicz theorem\index{Hurewicz theorem}, the Universal Coeff\/icient Theorem\index{Universal Coeff\/icient Theorem}, and the Poincar\'{e} duality \index{Poincar\'{e} duality}allow to compute the cohomologies\index{lens space!cohomologies} ($L=L(p,q)$)
\begin{eqnarray}
  \!\!\!\!\!\!\!\!\!\!\!\!\!\!\!H_1(L;\mathbb{Z})\approx\mathbb{Z}_p\quad,\quad H^1(L;\mathbb{Z})\approx0\quad,\quad H_1(L;\mathbb{Z}_p)\approx H^1(L;\mathbb{Z}_p)\approx\mathbb{Z}_p,\\
  \!\!\!\!\!\!\!\!\!\!\!\!\!\!\!H_2(L;\mathbb{Z})\approx0\quad,\quad H^2(L;\mathbb{Z})\approx\mathbb{Z}_p\quad,\quad H_2(L;\mathbb{Z}_p)\approx H^2(L;\mathbb{Z}_p)\approx\mathbb{Z}_p,\\
  \!\!\!\!\!\!\!\!\!\!\!\!\!\!\!H_3(L;\mathbb{Z})\approx\mathbb{Z}\quad,\quad H^3(L;\mathbb{Z})\approx\mathbb{Z}\quad,\quad H_3(L;\mathbb{Z}_p)\approx H^3(L;\mathbb{Z}_p)\approx\mathbb{Z}_p.
\end{eqnarray}
and the exact sequence
\begin{equation}
\begin{tikzpicture}
\matrix(m)[matrix of math nodes, row sep=3em, column sep=2.5em, text height=1.5ex, text depth=0.25ex]
{
H^1(L;\mathbb{Z})&H^1(L;\mathbb{Z}_p)&H^2(L;\mathbb{Z})&H^2(L;\mathbb{Z})&H^2(L;\mathbb{Z}_p)\\
};
\path[->,>=angle 90,font=\normalsize]
(m-1-1) edge (m-1-2)
(m-1-2) edge node[above] {$\beta_0$} (m-1-3)
(m-1-3) edge node[above] {$0$}       (m-1-4)
(m-1-4) edge node[above] {$\rho$}    (m-1-5);
\end{tikzpicture}
\end{equation}
shows that $\beta_0$ and $\rho$ are isomorphisms, and therefore
\begin{equation}
  \beta:H^1(L;\mathbb{Z}_p)\rightarrow H^2(L;\mathbb{Z}),
\end{equation}
is an isomorphism.

\section{The Problems of Geometrodynamics}
The key problem of quantum geometrodynamics\index{quantum geometrodynamics (QGD)} are observables. The important notion is the space of solutions of all constraints $\mathcal{F}_0$ which is a subspace of a functional space $\mathcal{F}$ of quantum states represented in the Schr\"odinger picture by wave functionals of the three-metric $\Psi[h_{ij}]$. Physical states must belong to $\mathcal{F}_0$ in order to be invariant under the symmetries encoded in constraints. The Dirac observables\index{Dirac observables} must commute with all the f\/irst-class constraints generating gauge transformations, $[O,H]\Psi=0$, so the action of an observable on a physical state does not project the state out of the space of physical states $\mathcal{F}_0$. An inner product must be def\/ined on $\mathcal{F}_0$ in order to obtain an Hilbert space\index{Hilbert space} of physical normalized state vectors.

Kucha\v{r}\index{Kucha\v{r} observables} \cite{kuch} considered the problem of observables, and reached the conclusion that the observables are def\/ined by non vanishing Poisson brackets\index{Poisson brackets} with all of the constraints, but claimed that rightness of such a treatment is justif\/ied for the diffeomorphism constraint\index{diffeomorphism constraint}, while is manifestly wrong for the Hamiltonian constraint\index{Hamiltonian constraint} because of $H$ generates the dynamics between the hypersurfaces. Both a hypersurface as well as its points in itself are not directly observable. Albeit, the values of the canonical pair $(q_{ij}, \pi^{ij})$ are evidently distinguishable on an initial hypersurface and on an evolved hypersurface, and moreover the effects of the evolution are not possible to observe when there is a lack of difference between these values. For consistency Kucha\v{r} introduced two types of variable, i.e. observables and perennials\index{observables}\index{perennials}. In his approach observables are the diffeoinvariant dynamical variables which do not commute with the Hamiltonian constraint\index{Hamiltonian constraint}, meanwhile perennials are the observables commuting with the Hamiltonian constraint and can not be observed because of the Hamiltonian constraint\index{Hamiltonian constraint} should not be seen as a generator of the gauge transformations. By this reason the observables do not act on the space of physical solutions $\mathcal{F}_0$.

In spite of a number of details, in general the strategy proposed by Kucha\v{r} can be summarized concisely. First of all one must f\/ind the four kinematic variables $X^A:\partial M\mapsto M$ where $A=0,\ldots,3$ which represent a space-like embedding of a space-like hypersurface $\partial M$ into the space-time manifold $M$. These scalar f\/ields represent the space-time positions and observables evolving along $M$, which are the true gravitational degrees of freedom\index{degrees of freedom}, are the dynamical variables separated out from these f\/ields on the level of phase space. The second point is to interpret the constraints as conditions and identify the momenta $P_A$ conjugated to $X_A$, which determine the evolution of the degrees of freedom\index{degrees of freedom} between hypersurfaces, with the energy-momenta of the degrees of freedom\index{degrees of freedom}. Such a procedure involves solving the constraints on the classical level, necessity of the internal time, and quantization formulated in terms of the Tomonaga--Schwinger equation \index{Tomonaga--Schwinger equation}\cite{tomo,schw}
\begin{equation}
  i\dfrac{\delta\Psi[\phi^r(x)]}{\delta X^A(x)}=h_A\left(x;X^B,\phi^r,p^s\right)\Psi[\phi^r(x)],
\end{equation}
where $r,s=1,2$ and the variables $X^A$ are treated as classical, like time in quantum mechanics. There arise problems which include multiple-choice, no global time, problem in def\/inition of the Hamiltonian $h_A$ and many others. Brown and Kucha\v{r} \cite{brku} introduced matter variables, which label space-time points and are coupled to space-time geometry, instead of functionals of the gravitational variables. They proposed to take into account a dust f\/ield f\/illing all space and playing a role of time, what includes an internal time variable against which systems can evolve, and which can play a role of the f\/ixed background for the construction of quantum gravity. In the Brown--Kucha\v{r} formalism \index{Brown--Kucha\v{r} formalism}the Schr\"odinger equation \index{Schr\"odinger equation}can be written out and the emerging Hamiltonian does not depend on the dust variables.

Another version of the solution of the problem of time\index{problem of time}, called unimodular gravity\index{unimodular gravity} was proposed by W.A. Unruh \cite{unru}, who modif\/ied General Relativity such that the cosmological constant \index{cosmological constant}is a dynamical variable for which the conjugate is taken to be the cosmological time. The result is that the Hamiltonian constraint\index{Hamiltonian constraint} is augmented by a cosmological constant \index{cosmological constant}term giving the modif\/ied Hamiltonian constraint\index{Hamiltonian constraint} $\Lambda+\sqrt{h}H=0$. The presence of this extra term and the cosmological time $\tau$ unfreezes the dynamics and leas to the $\tau$-dependent Schr\"odinger equation\index{Schr\"odinger equation}.

Also DeWitt \cite{dew} tried to solve the problem of time\index{problem of time} in frames of quantum geometrodynamics\index{quantum geometrodynamics (QGD)}. It is the problem of extracting a notion of time from timeless dynamics described by the Wheeler--DeWitt equation\index{Wheeler--DeWitt equation}. A consequence of the timeless nature of this equation is the problematic implementation of an inner product for state vectors. In analogy to the inner product obtained from the Klein--Gordon equation\index{Klein--Gordon equation}, DeWitt proposed to def\/inition the inner product of two solutions of the Wheeler--DeWitt equation\index{Wheeler--DeWitt equation}
\begin{equation}\label{dewin}
  (\Psi_a,\Psi_b)=Z\int\Psi^\ast_a[{^{(3)}}\mathfrak{G}]\prod_{x}\left(d\Sigma^{ij}G_{ijkl}\dfrac{\overrightarrow{\delta}}{i\delta{h_{kl}}}-\dfrac{\overleftarrow{\delta}}{i\delta{h_{kl}}}G_{ijkl}d\Sigma^{ij}\right)\Psi_b[{^{(3)}}\mathfrak{G}],
\end{equation}
where the product is taken over all the points of a three-dimensional embedded space $\partial M$, the integration is over a $5\times\infty^3$-dimensional surface in $S(\partial M)$, $d\Sigma^{ij}(x)$ is the surface element of the topological product of a set of 5-dimensional hypersurfaces $\Sigma(x)$ one chosen at each point of $\partial M$, and $Z$ is normalization constant. The Klein--Gordon inner product\index{Klein--Gordon inner product} (\ref{dewin}) is invariant under the deformation of the $5\times\infty^3$ surface, but is not positively def\/ined and vanishes for real solutions of the Wheeler--DeWitt equation\index{Wheeler--DeWitt equation}. Moreover, by such a treatment, all problems related to the Klein--Gordon equation\index{Klein--Gordon equation}, like e.g. no separation into positive and negative frequencies and the negative probability, are available in general for the Wheeler-DeWitt equation\index{Wheeler--DeWitt equation}.

There are also another problems within quantum geometrodynamics\index{quantum geometrodynamics (QGD)} following from the Wheeler--DeWitt equation\index{Wheeler--DeWitt equation}. First of all, this is the initial data problem. Namely, by quantum geometrodynamics\index{quantum geometrodynamics (QGD)} the classical space-time is the history of space geometry governed by the deterministic evolution. There arises uncertainty relation between intrinsic and extrinsic geometry due to the Lie bracket of an induced metric and its conjugated momentum. By this reason interpretation of the properties of quantum space-time is unclear. The second important problem is that for consistency the standardly applied quantization of the constraints needs a choice of a regularization method. However, factor orderings lead to non-unique result and quantum anomalies, i.e. the most terrible ambiguities. The problem of indef\/initeness of measure in the Wheeler superspace \index{Wheeler superspace}follows from the def\/inition of the inner product. The canonical variables are present in the non polynomial way in the Wheeler--DeWitt equation\index{Wheeler--DeWitt equation}, what in itself gives rise to problematic analysis. As we have mentioned earlier, the Wheeler--DeWitt equation\index{Wheeler--DeWitt equation} has not been solved in general, the only simplest Feynman's path integral solutions are discussed and the general integrability problem seems to be omitted. Another question is the interpretation of both the wave functionals solving the Wheeler--DeWitt equation \index{Wheeler--DeWitt equation}as well as their normalization and superpositions. Factually, no Dirac observable \index{Dirac observables}of the quantum geometrodynamics\index{quantum geometrodynamics (QGD)} is known. Factually, both classical and quantum geometrodynamics\index{quantum geometrodynamics (QGD)} are time-independent evolutions, but both the problem of time\index{problem of time} and therefore also the quantum evolution are still unsolved.

The model of Quantum Cosmology presented in the previous chapter possesses the hidden structure of the Wheeler superspace\index{Wheeler superspace}. In such a situation the conf\/igurational space is the stratum of superspace called \emph{minisuperspace\index{minisuperspace}}. We shall continue studying of the $3+1$ decomposed metric f\/ields, that are all isotropic solutions of the Einstein f\/ield equations\index{Einstein f\/ield equations}, and in itself create another stratum of superspace called \emph{midisuperspace}\index{midisuperspace}. The midisuperspace\index{midisuperspace} models are not the most popular in the modern theoretical gravitational physics, and in general quantum geometrodynamics\index{quantum geometrodynamics (QGD)} by its functional nature has a status of rather not a very well-def\/ined mathematical theory than a theory of quantum gravity possessing physical signif\/icance. By this reason we shall present the new constructive analysis of such theories based on well established methods of quantum f\/ield theory, which leads to plausible sounding phenomenology. The plausibility is not a coincidence, but is the consequence of application of the models of quantum f\/ield theory having established meaning for physics. It must be emphasized that such a strategy is fully justif\/ied for one-dimensional quantum gravities. Albeit, its both applicability to and usefulness for another possible situations are the good question. One can suppose \emph{ad hoc} that the one-dimensionality of quantum gravity is its universal physical feature, and other situations are non physical. However, such a reasoning in itself is the attempt to preserve \emph{ad hoc} applicability of quantum f\/ield theory for theory of quantum gravity, while recently the legitimateness of quantum f\/ield theoretic methods applied to rather non usual situations in itself is a moot point. On the other hand, everything what is widely applied and developed in theoretical physics is primarily rooted in methodology of quantum f\/ield theory. The best example is string theory\index{string theory} which is a quantum f\/ield theory. By this reason, the necessity of doing the construction of the adequate quantum f\/ield theoretic formalism of quantum gravity, i.e. quantum theory of gravitational f\/ield or quantum f\/ield theory of gravity, is logically argued. The logical arguments, however, must not be satisf\/ied by Nature, and by this reason the results received via the adequate formalism must be empirically verif\/ied. Otherwise, the physical meaning of the theory will be unclear.

Factually the quantum geometrodynamics\index{quantum geometrodynamics (QGD)} based on the Wheeler--DeWitt equation\index{Wheeler--DeWitt equation} is the f\/irst constructive attempt to formulation of quantum General Relativity. Actually, however, QGD\index{quantum geometrodynamics (QGD)} has became the most inf\/luential motivation for development of both other theories of quantum gravity based on QGD\index{quantum geometrodynamics (QGD)} as well as building of completely different formulations. In the further part of this part we are going to present the model of quantum gravity strictly based on the Wheeler-DeWitt equation\index{Wheeler--DeWitt equation} (\ref{wdw}) presented above. Basics and applications of the ADM Hamiltonian approach to General Relativity\index{Hamiltonian formulation of General Relativity!Arnowitt--Deser--Misner (ADM)}, the classical and the quantum geometrodynamics\index{quantum geometrodynamics (QGD)}, and the Wheeler--DeWitt equation \index{Wheeler--DeWitt equation}have been studied intensively in the scientif\/ic and research literature since more than 50 years (See \emph{e.g.} the Refs. \cite{qgr0a}-\cite{ham1}). In the lack of other constructive competitors the theory still is the theory of quantum gravitational f\/ields, and factually the only one having real chances for predictions of constructive phenomenology.

\section{Other Approaches}
Another point of view on quantum gravity follows from application of the Ashtekar Hamiltonian formulation of General Relativity \cite{ash}, which applies the Einstein--Cartan theory\index{Einstein--Cartan theory} with a complex connection. Rovelli and Smolin \cite{rovsmol} used Ashtekar's new variables to investigation of the loop representation of quantum General Relativity. This direction was developed by Ashtekar, Rovelli, Smolin, Jacobson, and Lewandowski \cite{found} and in the quantum cosmological context by Bojowald \cite{bojo}. The resulting theories are called loop quantum gravity/cosmology\index{loop quantum gravity}\index{loop quantum cosmology} and take into account the fundamental role of diffeomorphisms, including the diffeomorphism constraint\index{diffeomorphism constraint} which does not play a crucial role for dynamics in the quantum geometrodynamics\index{quantum geometrodynamics (QGD)} formulated in terms of the Wheeler--DeWitt equation\index{Wheeler--DeWitt equation}. In loop quantum gravity important role plays the Ashtekar--Lewandowski group\index{Ashtekar--Lewandowski group}. Recently, this research direction has been received the well-established research status and is still under intensive development (See, \emph{e.g.} papers in the Ref. \cite{loop}).

The Arnowitt--Deser--Misner and Ashtekar Hamiltonian formulations of General Relativity \index{Hamiltonian formulation of General Relativity!Arnowitt--Deser--Misner (ADM)}\index{Hamiltonian formulation of General Relativity!Ashtekar} present different strategies. This heritage ref\/lects in evident differences between quantum geometrodynamics\index{quantum geometrodynamics (QGD)} and loop quantum gravity\index{loop quantum gravity}/cosmology\index{loop quantum cosmology}. Quantum gravity formulated by the Wheeler--DeWitt equation\index{Wheeler--DeWitt equation} is treated as established theory, while loop quantum gravity\index{loop quantum gravity} similarly to string theory\index{string theory} is presently intensively developed. The attempts of quantum geometrodynamics\index{quantum geometrodynamics (QGD)} were practically obscured by the alternative approach, while in itself Wheeler--DeWitt equation \index{Wheeler--DeWitt equation}still needs development and is a source of hidden constructive phenomenology.

There is a number of alternative evolution schemes, called numerical relativity (For modern analysis see \emph{e.g.} the Ref. \cite{nrel}), which is not taken into account in construction of quantum gravity. The privileged strong position of the ADM and the Ashtekar formulations\index{Hamiltonian formulation of General Relativity!Arnowitt--Deser--Misner (ADM)}\index{Hamiltonian formulation of General Relativity!Ashtekar} follows from their straightforward roots in the Hamiltonian analysis, because of the primary canonical quantization \index{quantization!primary canonical}procedure follows from the Hamiltonian analysis. Usually alternative evolution schemes are strictly based on these two canonical formulations, or are its particular cases. The crucial issue which connects all these schemes is a formulation of the Cauchy problem for the Einstein f\/ield equations\index{Einstein f\/ield equations}.

The pioneering approach to the Cauchy problem for General Relativity\index{Cauchy problem for General Relativity} in the case of analytic initial data was proposed by Darmois \cite{darm} in 1927 and Lichnerowicz \cite{lich39} in 1939. In 1944 Lichnerowicz \cite{lich44} proposed the f\/irst $3+1$ formalism based on the conformal decomposition of a spatial metric. In 1952 Four\`{e}s-Bruhat \cite{bruh} formulated the Cauchy problem for $C^5$ initial data via using of the local existence and uniqueness in harmonic coordinates, what in 1956 resulted in the $3+1$ formalism in moving frame. In 1962 Arnowitt, Deser, and Misner \cite{adm} proposed the $3+1$ formalism based on the Hamiltonian analysis of General Relativity. Soon after, in 1972, York \cite{york72} considered gravitational dynamical degrees of freedom\index{degrees of freedom} carried by the conformal spatial metric, and in 1974 \'{O} Murchadha and York \cite{omurd} introduced the conformal transverse-traceless (CTT) method for solving the constraint equations. In 1977 Smarr \cite{smar77} considered 2D axisymmetric head-on collision of two black holes and produced the f\/irst numerical solution beyond spherical symmetry of the Cauchy problem for asymptotically f\/lat spacetimes. In 1978 Smarr and York \cite{smaryork} proposed radiation gauge for numerical relativity what resulted in the elliptic-hyperbolic system with asymptotic TT behavior. In 1983 Bardeen and Piran \cite{barpir} considered 2D computations of partially constrained schemes. Nakamura \cite{naka} in 1983, and Stark and Piran \cite{stark} in 1985 applied 2D axisymmetric gravitational collapse to a black hole. In 1986 Ashtekar \cite{ash} proposed new variables. In 1987 Nakamura, Oohara, and Kojima \cite{nok} tested evolution of pure gravitational wave spacetimes in spherical coordinates. In 1989 Bona and Masso \cite{bona}, in 1995 Choquet-Bruhat and York \cite{chby}, in 2001 Kidder, Scheel and Teukolsky \cite{kst} considered the f\/irst-order symmetric hyperbolic formulations of the Einstein f\/ield equations\index{Einstein f\/ield equations} within the $3+1$ formalism. Shibata and Nakamura \cite{shinak} in 1995, and Baumgarte and Shapiro \cite{baum} in 1999 investigated so called BSSN formulation, i.e. conformal decomposition of the $3+1$ equations and promotion of some connection function as an independent variable. In 1999 York \cite{york99} introduced the conformal thin-sandwich (CTS) method for solving the constraint equations. In 2000 Shibata \cite{shib} performed 3D full computation of binary neutron star merger, what was the f\/irst full GR 3D solution of the Cauchy problem in the astrophysical context. In 2000 Hayward \cite{hayw} proposed a new scheme involving a dual-null decomposition of space-time and removing second-order terms from the Einstein f\/ield equations\index{Einstein f\/ield equations}, which would vanish in the case of spherically symmetric space-time. In 2004 Bonazzola, Gourgoulhon, Grandcl\'{e}ment and Novak \cite{bonaz} proposed the constrained scheme based on maximal slicing and Dirac's gauge.

\chapter{Global One-Dimensionality Conjecture}

\section{Introduction}
This chapter is devoted to the our proposition for theory of quantum gravity. The role of quantum gravity is a fundamental problem of modern theoretical physics. For instance, for lack of the consistent theory of quantum gravity we are not able to understand physics of our Universe at the Planck scale\index{Planck scale}. Factually, despite a number of signif\/icant efforts (For various approaches see \emph{e.g.} Refs. \cite{kiefer}, \cite{planckscale} and \cite{qg01}-\cite{qg48}), we are still very far of understanding the role of quantized gravitational f\/ields for physical phenomena at high and ultra-high energies. In this chapter we propose a very simple model of quantum gravity which can be useful for clarifying its some important aspects. However, the simplicity of the theory of quantum gravity presented here is far from triviality and is non obvious argument. In fact, the model is proposed \emph{ad hoc}, but is strictly based on the Wheeler--DeWitt equation\index{Wheeler--DeWitt equation}, and in itself is a certain particular realization of this rather general theory. Albeit, the our model is signif\/icantly simpler and by this reason is able to generate new facts and apply the straightforward analogy with the established phenomenological models of theoretical physics.

The f\/ield-theoretic formalism, so celebrated in modern physics, yields a plausible phenomenology for a number of experimental data coming from a rich spectrum of observations. In this chapter such a point of view is applied as the base for construction of a simple theory of quantum gravity. We shall perform the construction via the standard strategy resulting in the Wheeler--DeWitt equation\index{Wheeler--DeWitt equation} with, however, modif\/ied treatment of Matter f\/ields\index{Matter f\/ields} and the wave functional solving the Wheeler--DeWitt equation. The $3+1$ splitting\index{$3+1$ splitting} of a general relativistic metric tensor and the canonical primary quantization \index{quantization!primary canonical}of the appropriate Hamiltonian and diffeomorphism constraints are employed in the way well-grounded in numerous approaches to quantization of gravitation. The modif\/ication of the standard quantum geometrodynamics\index{quantum geometrodynamics (QGD)} is based on the \emph{global one-dimensionality conjecture\index{conjecture!global one-dimensionality}}, which in itself in not beyond the quantum geometrodynamics\index{quantum geometrodynamics (QGD)} and arises from the straightforward and strict analogy with the generic cosmological model \cite{gli} presented in the Chapter 4. The crucial idea of the model is the ansatz\index{ansatz} which can be summarized by the four brief phrases
\begin{enumerate}
  \item Investigation of the global one-dimensionality conjecture\index{conjecture!global one-dimensionality}, \emph{i.e.} taking into account a certain specif\/ic one-dimensional nature of Matter f\/ields\index{Matter f\/ields} and the wave functional,
  \item Reduction of quantum geometrodynamics\index{quantum geometrodynamics (QGD)}, resulting in the one-dimensional theory characteristic for bosonic f\/ields,
  \item Application of the Hamilton equations of motion, yielding the corresponding one-dimensional Dirac equation\index{Dirac equation!one-dimensional},
  \item Expression of the supposition that the quantum gravity is a one-dimensional f\/ield theory, and performing its secondary quantization\index{quantization!secondary canonical}.
\end{enumerate}
The Hamilton equations of motion allow to establish the appropriate one-dimensional Dirac equation and the corresponding Clifford algebra\index{Clifford algebra}. The secondary quantization\index{quantization!secondary canonical}, based on the Fock space\index{Fock space} and the diagonalization procedure consisting of the Bogoliubov transformation\index{Bogoliubov transformation} and the Heisenberg equations of motion\index{Heisenberg equations of motion}, yields correctly def\/ined quantum f\/ield theory formulated in terms of the static Fock rep\`{e}re\index{Fock rep\`{e}re} associated with initial data. We derive the 1D wave functional and discuss the corresponding 3-dimensional manifolds. Quantum correlations of the f\/ield are associated with physical scales. Mathematically, we employ the one-dimensional functional integrals, and therefore despite the model of quantum gravity corresponds to the trend initiated by S.W. Hawking and his collaborators \cite{hawk00}-\cite{hawk15} derivation of its solutions is signif\/icantly simplif\/ied.
\newpage
\section{The $\Gamma$-Scalar-Flat Space-times}\label{sec:1}
Let us consider f\/irst the relation (\ref{esrho}), i.e.
\begin{equation}
S-\varrho=\dfrac{4\Lambda}{\kappa\ell_P^2}-\dfrac{{^{(4)}}R}{\kappa\ell_P^2}.
\end{equation}
In the light of the rule (\ref{ric4}) one has
\begin{equation}
 {^{(4)}}R={^{(3)}}R+K^2-K^{ij}K_{ij},\label{rule}
\end{equation}
where we have omitted the total derivative, because of its vanishing in Hamiltonian analysis. Applying the Hamiltonian constraint\index{Hamiltonian constraint}
\begin{equation}
  {^{(3)}}R+K^2-K^{ij}K_{ij}-2\Lambda-2\kappa\ell_P^2\varrho\approx0,
\end{equation}
one obtains another relation between the energy density\index{energy density} and the spatial stress density\index{spatial stress density}
\begin{equation}
 \varrho=\dfrac{2\Lambda}{\kappa\ell_P^2}-S,
\end{equation}
which can be presented as the equation for the cosmological constant\index{cosmological constant}
\begin{equation}
  \Lambda=\dfrac{\kappa\ell_P^2}{2}(S+\varrho),\label{cosmconst}
\end{equation}
and gives the insight into the nature of the cosmological constant\index{cosmological constant}. In other words, the cosmological constant \index{cosmological constant}is an arithmetic mean of the spatial stress density\index{spatial stress density} and the energy density\index{energy density} multiplied by the Einstein constant\index{Einstein constant} $\kappa$. One can, however, also apply the difference (\ref{esrhodif}) $S-\varrho=T$ together with the equation (\ref{cosmconst}) and establish
\begin{eqnarray}
  \varrho&=&\dfrac{\Lambda}{\kappa\ell_P^2}-\dfrac{T}{2},\label{endensit}\\
  S&=&\dfrac{\Lambda}{\kappa\ell_P^2}+\dfrac{T}{2}.\label{stdensit}
\end{eqnarray}
By taking into account the fact $\varrho=T_{\mu\nu}n^\mu n^\nu$ the equation (\ref{endensit}) can be rewritten in the form
\begin{equation}
  T_{\mu\nu}\left(g^{\mu\nu}+2n^\mu n^\nu\right)=\dfrac{2\Lambda}{\kappa\ell_P^2},
\end{equation}
and solved immediately with respect to the stress-energy tensor\index{stress-energy tensor}
\begin{equation}
T_{\mu\nu}=\dfrac{2\Lambda}{\kappa\ell_P^2}\dfrac{1}{g^{\mu\nu}+2n^\mu n^\nu}.
\end{equation}
Let us consider the RHS of this expression. The tensor coeff\/icient multiplied by $\dfrac{2\Lambda}{\kappa}$ can be rewritten in the form
\begin{eqnarray}
  \!\!\!\!\!\!\!\!\!\!\!\!\!\!\!\dfrac{1}{g^{\mu\nu}+2n^\mu n^\nu}&=&\dfrac{(g_{\mu\nu}+2n_\mu n_\nu)}{(g^{\mu\nu}+2n^\mu n^\nu)(g_{\mu\nu}+2n_\mu n_\nu)}=\nonumber\\
  \!\!\!\!\!\!\!\!\!\!\!\!\!\!\!&=&\dfrac{(g_{\mu\nu}+2n_\mu n_\nu)}{g^{\mu\nu}g_{\mu\nu}+4(n^\mu n_\mu)^2+4n^\mu n_\mu}=\dfrac{1}{4}(g_{\mu\nu}+2n_\mu n_\nu),
\end{eqnarray}
where we have used the identities $g^{\mu\nu}n_\mu n_\nu=g_{\mu\nu}n^\mu n^\nu=n^\mu n_\mu$, $n_\mu n^\mu=-1$, and $g^{\mu\nu}g_{\mu\nu}=4$. By this reason one obtains the stress-energy tensor\index{stress-energy tensor}
\begin{equation}
  T_{\mu\nu}=\dfrac{\Lambda}{2\kappa\ell_P^2}g_{\mu\nu}+\dfrac{\Lambda}{\kappa\ell_P^2}n_\mu n_\nu,\label{seten}
\end{equation}
which covariant form also can be derived easy
\begin{equation}
  T^{\mu\nu}=\dfrac{\Lambda}{2\kappa\ell_P^2}g^{\mu\nu}+\dfrac{\Lambda}{\kappa\ell_P^2}n^\mu n^\nu.\label{seten1}
\end{equation}
and can be unambiguously recognized as the stress-energy tensor\index{stress-energy tensor} of the perfect f\/luid (See e.g. the Ref. \cite{stephani1})
\begin{equation}
T^{\mu\nu}=pg^{\mu\nu}+\left(\dfrac{p}{c^2}+\mu \right)u^\mu u^\nu,
\end{equation}
for which the four-velocity $u^{\mu}$ equals to the unit normal vector f\/ield multiplied by the speed of light $c$, i.e.
\begin{eqnarray}
u^\mu&=&cn^\mu,\\
u^\mu u_\mu&=&-c^2,
\end{eqnarray}
and the isotropic pressure $p$ and the mass density $\mu$ are as follows
\begin{eqnarray}
  p&=&\dfrac{\Lambda}{2\kappa\ell_P^2},\label{pressure}\\
  \mu&=&\dfrac{\Lambda}{2c^2\kappa\ell_P^2}.\label{massdens}
\end{eqnarray}
Now the trace of the stress-energy tensor\index{stress-energy tensor} can be established by straightforward easy computation
\begin{eqnarray}
  T&=&g^{\mu\nu}T_{\mu\nu}=\dfrac{\Lambda}{2\kappa\ell_P^2}g^{\mu\nu}g_{\mu\nu}+\dfrac{\Lambda}{\kappa\ell_P^2}g^{\mu\nu}n_\mu n_\nu=\dfrac{\Lambda}{2\kappa\ell_P^2}4+\dfrac{\Lambda}{\kappa\ell_P^2}n^\mu n_\mu=\nonumber\\
  &=&\dfrac{2\Lambda}{\kappa\ell_P^2}-\dfrac{\Lambda}{\kappa\ell_P^2}=\dfrac{\Lambda}{\kappa\ell_P^2},\label{cosmconst2}
\end{eqnarray}
and consequently the energy density\index{energy density} (\ref{endensit}) and the spatial stress density\index{spatial stress density} (\ref{stdensit}) have the values
\begin{eqnarray}
  \varrho&=&\dfrac{\Lambda}{2\kappa\ell_P^2},\label{endensit1}\\
  S&=&\dfrac{3\Lambda}{2\kappa\ell_P^2}.\label{stdensit1}
\end{eqnarray}
The momentum density related to such a situation can be established straightforwardly
\begin{eqnarray}
J^i&=&T_{\mu\nu}n^\mu h^{\nu i}=\left(\dfrac{\Lambda}{2\kappa\ell_P^2}g_{\mu\nu}+\dfrac{\Lambda}{\kappa\ell_P^2}n_\mu n_\nu\right)n^\mu h^{\nu i}=\dfrac{\Lambda}{2\kappa\ell_P^2}n_\nu h^{\nu i}-\dfrac{\Lambda}{\kappa\ell_P^2}n_\nu h^{\nu i}=\nonumber\\
&=&-\dfrac{\Lambda}{2\kappa\ell_P^2}n_\nu h^{\nu i}=-\dfrac{\Lambda}{2\kappa\ell_P^2}n_\nu\left(g^{\nu i}+n^\nu n^i\right)=-\dfrac{\Lambda}{2\kappa\ell_P^2}\left(n^i-n^i\right)=0.
\end{eqnarray}
By this reason the classical geometrodynamics becomes
\begin{equation}\label{perA}
  {2c\kappa}G_{ijkl}\dfrac{\delta{S[g]}}{\delta{h_{ij}}}\dfrac{\delta{S[g]}}{\delta{h_{kl}}}+\dfrac{\ell_P^2}{2c\kappa}\sqrt{h}\left({^{(3)}}R-3\Lambda\right)=0,
\end{equation}
while the Wheeler--DeWitt equation \index{Wheeler--DeWitt equation}is
\begin{equation}\label{wdwA}
\left\{2c\kappa\dfrac{\hslash^2}{\ell_P^2}G_{ijkl}\dfrac{\delta^2}{\delta h_{ij}\delta h_{kl}}+\dfrac{\ell_P^2}{2c\kappa}\sqrt{h}\left({^{(3)}R}-3\Lambda\right)\right\}\Psi[h_{ij},\phi]=0,
\end{equation}
and the quantized diffeomorphism constraint\index{diffeomorphism constraint} is
\begin{equation}
i\dfrac{E_P}{\ell_P^2}\left(\partial_j+\dfrac{1}{2}h_{jl,k}h^{kl}\right)\dfrac{\delta{\Psi}[h_{ij},\phi]}{\delta{h_{ij}}}=0.
\end{equation}
In other words, in such a situation both the classical and quantum geometrodynamics\index{quantum geometrodynamics (QGD)} become purely geometrical.

Good question is what is the Lagrangian of Matter f\/ields\index{Matter f\/ields} describing such a situation. Rewriting the stress-energy tensor\index{stress-energy tensor} (\ref{seten}) in the form
\begin{equation}
 T_{\mu\nu}=\dfrac{1}{2}\left(3p+\mu c^2\right)g_{\mu\nu},
\end{equation}
and using of the def\/inition (\ref{varsten}) of $T_{\mu\nu}$ following from the Hilbert--Palatini action principle\index{Hilbert--Palatini action principle} one obtains the equation
\begin{equation}
  -\dfrac{2}{\sqrt{-g}}\dfrac{\delta}{\delta g^{\mu\nu}}\left(\sqrt{-g}L_\phi\right)=\dfrac{1}{2}\left(3p+\mu c^2\right)g_{\mu\nu},
\end{equation}
which can be presented in equivalent form
\begin{equation}
  \delta\left(\sqrt{-g}L_\phi\right)=-\dfrac{1}{4}\left(3p+\mu c^2\right)\sqrt{-g}g_{\mu\nu}\delta g^{\mu\nu}.\label{lageq}
\end{equation}
Using of the Jacobi formula\index{Jacobi formula} for differentiating a determinant
\begin{equation}
  \delta g=g g^{\mu\nu}\delta g_{\mu\nu}=-g g_{\mu\nu}\delta g^{\mu\nu},
\end{equation}
allows to write the equation (\ref{lageq}) as
\begin{equation}
  L_\phi\delta\sqrt{-g}+\sqrt{-g}\delta{L_\phi}=\dfrac{1}{2}\left(3p+\mu c^2\right)\delta\sqrt{-g}.\label{lageq1}
\end{equation}
Because, however, the parameters $p$ and $\mu$ are constant one has uniquely
\begin{eqnarray}
  L_\phi&=&\dfrac{1}{2}\left(3p+\mu c^2\right),\\
  \delta{L_\phi}&=&0.
\end{eqnarray}
Applying the relations (\ref{pressure}) and (\ref{massdens}) one receives f\/inally
\begin{equation}
  L_\phi=\dfrac{\Lambda}{\kappa\ell_P^2}=-\varrho.
\end{equation}

The spatial stress density\index{spatial stress density} $S=h^{ij}S_{ij}$ together with the formula (\ref{stdensit1}) can be used for derivation of the spatial stress tensor\index{spatial stress tensor}
\begin{equation}\label{strten}
  S_{ij}=\dfrac{S}{3}h_{ij}=\dfrac{\Lambda}{2\kappa\ell_P^2}h_{ij},
\end{equation}
and together with the difference
\begin{equation}
  S-\varrho=\dfrac{\Lambda}{\kappa\ell_P^2},
\end{equation}
allow to establish the tensor
\begin{equation}
-\kappa\ell_P^2\left[S_{ij}-\dfrac{1}{2}h_{ij}(S-\varrho)\right]=0.
\end{equation}
Hence in such a situation the evolutionary equations for the extrinsic curvature tensor and the intrinsic curvature are given by
\begin{eqnarray}
\!\!\!\!\!\!\!\!\!\!\!\!\!\!\!\!\!\!\!\!
\partial_t{K}_{ij}&=&-N_{|ij}+N\left(R_{ij}+KK_{ij}-2K_{ik}K^k_j\right)+N^kK_{ij|k}+K_{ik}N^k_{|j}+K_{jk}N^k_{|i},\label{kijev}\\
\!\!\!\!\!\!\!\!\!\!\!\!\!\!\!\!\!\!\!\!
\partial_t K&=&-h^{ij}N_{|ij}+N\left(K^{ij}K_{ij}+\Lambda\right)+N^iK_{|i},\label{kev}
\end{eqnarray}
while the evolutionary equations for induced metrics and its determinant remain unchanged, i.e.
\begin{eqnarray}
\partial_t{h}_{ij}&=&N_{i|j}+N_{j|i}-2NK_{ij},\label{hijev}\\
\partial_t\ln\sqrt{h}&=&-NK+N^i_{|i}.\label{hev}
\end{eqnarray}
However, the spatial stress tensor\index{spatial stress tensor} (\ref{strten}) is not unique. Because of $h^{ij}n_in_j=n^in_i=1$, there are another inequivalent choices
\begin{eqnarray}
  S_{ij}&=&Sn_in_j,\\
  S_{ij}&=&\dfrac{S}{4}\left(h_{ij}+n_in_j\right),\\
  S_{ij}&=&\dfrac{S}{2}\left(h_{ij}-n_in_j\right),
\end{eqnarray}
and also much more general form
\begin{equation}
    S_{ij}=\dfrac{S}{3\alpha+\beta}\left(\alpha h_{ij}+\beta n_in_j\right)=S_{ij}(\alpha,\beta),\label{strgen}
\end{equation}
where $\alpha$ and $\beta$ are any numbers. All these forms of the spatial stress tensor\index{spatial stress tensor} have the same value of trace, but in the basis $(h_{ij},n_i)$ the two-parameter family (\ref{strgen}) is the most general solution. In other words all relations between the spatial stress density\index{spatial stress density} $S$ and the energy density\index{energy density} $\varrho$ are validate when the spatial stress tensor\index{spatial stress tensor} has a form (\ref{strgen}). It is easy to see that the following Poisson algebra is satisf\/ied
\begin{equation}
\left\{S_{ij}(\alpha,\beta),S_{kl}(\alpha',\beta')\right\}=
S\dfrac{\alpha\alpha'\chi_{ijkl}+\beta\beta'\lambda_{ijkl}+\alpha\beta'\zeta_{ijkl}
+\alpha'\beta\zeta_{klij}}{9\alpha\alpha'+\beta\beta'+3\alpha\beta'+3\alpha'\beta},\label{stralg}
\end{equation}
where we have introduced the tensors
\begin{eqnarray}
  \chi_{ijkl}&=&\left\{h_{ij},h_{kl}\right\},\\
  \lambda_{ijkl}&=&\left\{n_in_j,n_kn_l\right\},\\
  \zeta_{ijkl}&=&\left\{h_{ij},n_kn_l\right\}.
\end{eqnarray}
By the context we shall call (\ref{stralg} \emph{the stress algebra}\index{stress algebra}. Let us consider the RHS of the equation (\ref{stralg}). When the bracketed quantities are classical (C), i.e. are not operators corresponding to the classical quantities, the Poisson brackets\index{Poisson brackets} are easy to establish. Let us denote such a classical Poisson brackets\index{Poisson brackets} as
\begin{equation}
\sigma_{ijkl}(\alpha,\beta,\alpha',\beta')=\left\{S_{ij}(\alpha,\beta),S_{kl}(\alpha',\beta')\right\}_{C},
\end{equation}
or in explicit form
\begin{equation}
\sigma_{ijkl}(\alpha,\beta,\alpha',\beta')=2S\dfrac{\alpha\alpha'h_{ij}h_{kl}+\beta\beta'n_in_jn_kn_l+\alpha\beta'h_{ij}n_kn_l
+\alpha'\beta n_in_jh_{kl}}{9\alpha\alpha'+\beta\beta'+3\alpha\beta'+3\alpha'\beta}.\label{stralg1}
\end{equation}
It can be seen by straightforward computation that
\begin{eqnarray}
\!\!\!\!\!\!\!\!\!\!\!\!\!\!\!\!\!\!\!\!\!\!\!\!\!\!\!\!\!\!\!\!\!\!\!\!\!\!\!\!\!\!\!\!\!\!\!\!\!\!
&&\sigma_{jl}(\alpha,\beta,\alpha',\beta')=h^{ik}\sigma_{ijkl}=2S\dfrac{\alpha\alpha'h_{jl}+\left(\beta\beta'+\alpha\beta'+\alpha'\beta\right)n_jn_l}{9\alpha\alpha'+\beta\beta'+3\alpha\beta'+3\alpha'\beta},\label{stralg1a}\\
\!\!\!\!\!\!\!\!\!\!\!\!\!\!\!\!\!\!\!\!\!\!\!\!\!\!\!\!\!\!\!\!\!\!\!\!\!\!\!\!\!\!\!\!\!\!\!\!\!\!
&&\sigma(\alpha,\beta,\alpha',\beta')=h^{jl}\sigma_{jl}=2S\dfrac{3\alpha\alpha'+\beta\beta'+\alpha\beta'+\alpha'\beta}{9\alpha\alpha'+\beta\beta'+3\alpha\beta'+3\alpha'\beta}.\label{stralg1b}
\end{eqnarray}

By this reason for generality let us consider the spatial stress tensor\index{spatial stress tensor} (\ref{strgen}). Then one has
\begin{equation}
-\kappa\ell_P^2\left[S_{ij}-\dfrac{1}{2}h_{ij}(S-\varrho)\right]=\dfrac{\Lambda}{2}\dfrac{\beta}{3\alpha+\beta}\left(h_{ij}-3n_in_j\right).
\end{equation}
In this manner the evolutionary equations (\ref{kev}), (\ref{hijev}), and (\ref{hev}) remain unchanged, but the evolution of extrinsic curvature tensor\index{extrinsic curvature tensor} is
\begin{eqnarray}
\partial_t{K}_{ij}&=&-N_{|ij}+N\left[R_{ij}+KK_{ij}-2K_{ik}K^k_j+\dfrac{\Lambda}{2}\dfrac{\beta}{3\alpha+\beta}\left(h_{ij}-3n_in_j\right)\right]+\nonumber\\
&+&N^kK_{ij|k}+K_{ik}N^k_{|j}+K_{jk}N^k_{|i}
\end{eqnarray}

However, in the light of the Einstein f\/ield equations\index{Einstein f\/ield equations} one can express the trace of stress-energy tensor\index{stress-energy tensor} via the Ricci scalar curvature\index{Ricci scalar curvature} and the cosmological constant\index{cosmological constant}, i.e.
\begin{equation}
T=-\dfrac{{^{(4)}}R}{\kappa\ell_P^2}+\dfrac{4\Lambda}{\kappa\ell_P^2}
\end{equation}
Applying this fact to the equation (\ref{cosmconst2}) one obtains
\begin{equation}
 {^{(4)}}R\equiv 3\Lambda,\label{ric0}
\end{equation}
what means that the space-time manifold is four-dimensional pseudo-Riemannian manifold of constant Ricci scalar curvature\index{Ricci scalar curvature} given by the cosmological constant \index{cosmological constant}up to constant multiplier equal to the dimensionality of an embedded space $D=3$. In the light of the relation (\ref{seten}) the RHS of the Einstein f\/ield equations\index{Einstein f\/ield equations} (\ref{feq}) is
\begin{equation}
\kappa\ell_P^2T_{\mu\nu}=\dfrac{1}{2}\Lambda g_{\mu\nu}+\Lambda n_\mu n_\nu,
\end{equation}
while the LHS of the Einstein f\/ield equations\index{Einstein f\/ield equations} is
\begin{equation}
  R_{\mu\nu}-\dfrac{1}{2}{^{(4)}}Rg_{\mu\nu}+\Lambda g_{\mu\nu}=R_{\mu\nu}-\dfrac{1}{2}\Lambda g_{\mu\nu},\label{vacfeq}
\end{equation}
and by this reason the Einstein f\/ield equations\index{Einstein f\/ield equations} for such a situation are
\begin{equation}
  R_{\mu\nu}=\Lambda\left(g_{\mu\nu}+n_\mu n_\nu\right)=\Lambda h_{\mu\nu},\label{ricur}
\end{equation}
where we have applied the completeness relations (\ref{compmetric}), in which
\begin{equation}
  h_{\mu\nu}=h_{ij}e^i_\mu e^j_\nu.
\end{equation}
Let us consider the contravariant form of $h_{\mu\nu}$
\begin{equation}
  h^{\mu\nu}=g^{\mu\kappa}g^{\nu\lambda}h_{\kappa\lambda}=h_{ij}g^{\mu\kappa}e^i_\kappa g^{\nu\lambda}e^j_\lambda=h^{kl}(h_{ik}g^{\mu\kappa}e^i_\kappa)(h_{jl}g^{\nu\lambda}e^j_\lambda)=h^{kl}e^\mu_k e^\nu_l,
\end{equation}
where we have applied the notation
\begin{equation}
e^\mu_k=h_{ik}g^{\mu\kappa}e^i_\mu.
\end{equation}
Applying the inverted way one obtains
\begin{equation}
  h_{\mu\nu}=g_{\mu\kappa}g_{\nu\lambda}h^{\kappa\lambda}=g_{\mu\kappa}g_{\nu\lambda}h^{kl}e^\kappa_k e^\lambda_l=h^{kl}(g_{\mu\kappa}e^\kappa_k)(g_{\nu\lambda}e^\lambda_l)=h^{kl}e_{\mu k}e_{\nu l},
\end{equation}
where we have applied the notation
\begin{equation}
e_{\nu k}=g_{\mu\nu}e^\mu_k,
\end{equation}
following from transition between the completeness relations for metric (\ref{compmetric}) and the completeness relations for inverse metric (\ref{comptrans}). In this manner the equation (\ref{ricur}) bacomes
\begin{equation}
  R_{\mu\nu}=\left(\Lambda g_{\mu\kappa}g_{\nu\lambda}\right)h^{ij}e^\kappa_ie^\lambda_j,\label{ricura}
\end{equation}
and application of the Ricci curvature tensor\index{Ricci curvature tensor} evaluated on the three-boundary (\ref{ricdec})
\begin{equation}
  R_{\mu\nu}=-R_{\kappa\mu\lambda\nu}n^\kappa n^\lambda+R_{\kappa\mu\lambda\nu}h^{ij}e^\kappa_ie^\lambda_j,
\end{equation}
to the equation (\ref{ricura}) leads to the system of equations
\begin{eqnarray}
 \left\{\begin{array}{cc}R_{\kappa\mu\lambda\nu}n^\kappa n^\lambda=0\vspace*{10pt}\\
 \left(R_{\kappa\mu\lambda\nu}-\Lambda g_{\kappa\mu}g_{\lambda\nu}\right)h^{\kappa\lambda}=0\end{array}\right..\label{equations}
\end{eqnarray}
The f\/irst equation in (\ref{equations}) expresses the property that the double projection of the Riemann--Christoffel curvature tensor\index{Riemann--Christoffel curvature tensor} onto the unit normal vector f\/ield vanishes, while the second one expresses
the fact that the projection onto the metric $h^{\kappa\lambda}$ of the tensor
\begin{equation}
A_{\kappa\mu\lambda\nu}:=R_{\kappa\mu\lambda\nu}-\Lambda g_{\kappa\mu} g_{\lambda\nu},
\end{equation}
vanishes. In other words, in the space-time is characterized by the Riemann--Christoffel curvature tensor\index{Riemann--Christoffel curvature tensor}
\begin{equation}
  R_{\kappa\mu\lambda\nu}=\Lambda g_{\kappa\mu} g_{\lambda\nu}+A_{\kappa\mu\lambda\nu},
\end{equation}
where $A_{\kappa\mu\lambda\nu}$ is the tensor satisfying the equations
\begin{eqnarray}
  \Lambda n_\mu n_\nu+A_{\kappa\mu\lambda\nu}n^\kappa n^\lambda&=&0,\label{tena2}\\
  A_{\kappa\mu\lambda\nu}h^{\kappa\lambda}&=&0,\label{tena1}
\end{eqnarray}
where the f\/irst equation follows from the f\/irst equation of the system (\ref{equations}). The equation (\ref{tena2}) projected onto $n_\kappa n_\lambda$ and leads to
\begin{equation}\label{tena}
  A_{\kappa\mu\lambda\nu}=-\Lambda n_\mu n_\nu n_\kappa n_\lambda+\Gamma_{\kappa\mu\lambda\nu},
\end{equation}
where the tensor $\Gamma_{\kappa\mu\lambda\nu}$ satisfying the equations
\begin{eqnarray}
  \Gamma_{\kappa\mu\lambda\nu}n^\kappa n^\lambda&=&0,\label{tenb1}\\
  \Gamma_{\kappa\mu\lambda\nu}h^{\kappa\lambda}&=&0,\label{tenb2}
\end{eqnarray}
and the second equation was deduced from application of the tensor (\ref{tena}) to the equation (\ref{tena1}). In this manner, by application of the completeness relations for the metric $g_{\mu\nu}$, the Riemann--Christoffel curvature tensor\index{Riemann--Christoffel curvature tensor} describing the considered space-time has a form
\begin{equation}
  R_{\kappa\mu\lambda\nu}=\Lambda\left(g_{\kappa\mu} g_{\lambda\nu}-n_\mu n_\nu n_\kappa n_\lambda\right)+\Gamma_{\kappa\mu\lambda\nu},\label{riechr}
\end{equation}
with the tensor $\Gamma_{\kappa\mu\lambda\nu}$ being a solution of the equations (\ref{tenb1}) and (\ref{tenb2}). The problem is to solve the system (\ref{tenb1})--(\ref{tenb2}) in general, but we shall not perform this procedure in this book. The Riemann--Christoffel curvature tensor\index{Riemann--Christoffel curvature tensor} (\ref{riechr}) in general describes all four-dimensional space-times for which the Ricci scalar curvature\index{Ricci scalar curvature} is ${^{(4)}}R=3\Lambda$ and the Ricci curvature tensor\index{Ricci curvature tensor} is $R_{\mu\nu}=\Lambda(g_{\mu\nu}+n_\mu n_\nu)$. Interestingly, one can compute these curvatures immediately with using of (\ref{riechr})
\begin{eqnarray}
  R_{\mu\nu}&=&g^{\kappa\lambda}R_{\kappa\mu\lambda\nu}=\Lambda\left(g_{\mu\nu}+n_\mu n_\nu\right)+\Gamma_{\mu\nu},\label{ricgam}\\
  {^{(4)}}R&=&g^{\mu\nu}R_{\mu\nu}=3\Lambda+{^{(4)}}\Gamma,\label{scagam}
\end{eqnarray}
what gives equations for the contractions of the tensor $\Gamma_{\kappa\mu\lambda\nu}$
\begin{eqnarray}
 \Gamma_{\mu\nu}&:=&g^{\kappa\lambda}\Gamma_{\kappa\mu\lambda\nu}=0,\\
 {^{(4)}}\Gamma&:=&g^{\kappa\lambda}\Gamma_{\mu\nu}=0.
\end{eqnarray}
Moreover, double projection of the Ricci curvature tensor\index{Ricci curvature tensor} (\ref{ricgam}) onto the unit normal vector f\/ield leads to
\begin{equation}
R_{\mu\nu}n^\mu n^\nu=\Gamma_{\mu\nu}n^\mu n^\nu,
\end{equation}
i.e. in this projective sense the $\Gamma_{\mu\nu}$ curvature tensor carries the same information as the Ricci curvature tensor\index{Ricci curvature tensor}.

Interestingly, one can computed the Weyl curvature tensor
\begin{eqnarray}
  W_{\mu\kappa\nu\lambda}&=&R_{\mu\kappa\nu\lambda}+\dfrac{1}{2}\left(g_{\mu\lambda}R_{\nu\kappa}+g_{\kappa\nu}R_{\lambda\mu}-g_{\mu\nu}R_{\lambda\kappa}-g_{\kappa\lambda}R_{\nu\mu}\right)+\nonumber\\
  &+&\dfrac{1}{3}{^{(4)}}R\left(g_{\mu\nu}g_{\lambda\kappa}-g_{\mu\lambda}g_{\nu\kappa}\right),
\end{eqnarray}
which for the considered situation takes the form
\begin{equation}
 \!\!\!\!\!W_{\mu\kappa\nu\lambda}=\Gamma_{\kappa\mu\lambda\nu}+\Lambda\left(g_{\kappa\mu}g_{\lambda\nu}+2g_{\mu[\lambda}g_{\nu]\kappa}+4g_{[\mu(\lambda}n_{\nu)} n_{\kappa]}-n_\mu n_\nu n_\kappa n_\lambda\right).\label{weylten}
\end{equation}
It means that the curvature tensor $\Gamma_{\kappa\mu\lambda\nu}$ is not the Weyl tensor. The contractions of the Weyl curvature tensor are
\begin{eqnarray}
  W_{\kappa\lambda}&=&g^{\mu\nu}W_{\mu\kappa\nu\lambda}=\Gamma_{\kappa\lambda}-\Lambda\left(g_{\kappa\lambda}+n_\kappa n_\lambda\right)=2\Gamma_{\kappa\lambda}-R_{\kappa\lambda},\\
  {^{(4)}}W&=&g^{\kappa\lambda}W_{\kappa\lambda}=2{^{(4)}}\Gamma-{^{(4)}}R=\Gamma-3\Lambda,
\end{eqnarray}
what means that in the particular case considered in this section
\begin{eqnarray}
 \Gamma_{\kappa\lambda}&=&\dfrac{R_{\kappa\lambda}+W_{\kappa\lambda}}{2},\\
 {^{(4)}}\Gamma&=&\dfrac{{^{(4)}}R+{^{(4)}}W}{2}.
\end{eqnarray}

Let us call $\Gamma_{\mu\nu}$ \emph{the $\Gamma$ curvature tensor}\index{$\Gamma$ curvature tensor}, and ${^{(4)}}\Gamma$ \emph{the $\Gamma$ scalar curvature}\index{$\Gamma$ scalar curvature}. Then the space-times considered above is $\Gamma_{\mu\nu}$-f\/lat manifold\index{$\Gamma_{\mu\nu}$-f\/lat manifold} of zero $\Gamma$ scalar curvature\index{$\Gamma$ scalar curvature}, which we shall call \emph{the $\Gamma$-scalar-f\/lat manifolds}\index{$\Gamma$-scalar-f\/lat space-times}. The equations (\ref{ricgam}) and (\ref{scagam}) can be used for construction of the LHS of the Einstein f\/ield equations\index{Einstein f\/ield equations}
\begin{equation}
  G_{\mu\nu}+\Lambda g_{\mu\nu}=\dfrac{1}{2}\Lambda g_{\mu\nu}+\Lambda n_\mu n_\nu+\Gamma_{\mu\nu}+\dfrac{1}{2}{^{(4)}}\Gamma g_{\mu\nu},
\end{equation}
and because of the stress-energy tensor\index{stress-energy tensor} is given by (\ref{seten}) the RHS of the Einstein f\/ield equations\index{Einstein f\/ield equations} is
\begin{equation}
  \kappa\ell_P^2T_{\mu\nu}=\dfrac{1}{2}\Lambda g_{\mu\nu}+\Lambda n_\mu n_\nu,\label{setenx}
\end{equation}
and by this reason the Einstein f\/ield equations\index{Einstein f\/ield equations} expressed via the $\Gamma$ curvatures takes the form of the vacuum f\/ield equations
\begin{equation}
\Gamma_{\mu\nu}+\dfrac{1}{2}{^{(4)}}\Gamma g_{\mu\nu}=0.\label{vacgam}
\end{equation}
In other words the $\Gamma$ curvatures are the curvatures which for blatantly non stationary space-time given by the Ricci scalar curvature\index{Ricci scalar curvature} ${^{(4)}}R=3\Lambda$, the Ricci curvature tensor\index{Ricci curvature tensor} $R_{\mu\nu}=\Lambda(g_{\mu\nu}+n_\mu n_\nu)$, and the stress-energy tensor\index{stress-energy tensor} (\ref{setenx}) makes the non stationary solution of the Einstein f\/ield equations\index{Einstein f\/ield equations} the space-time obeying vacuum f\/ield equations (\ref{vacgam}). The constructive hypothesis is
\begin{hypothesis}[The $\Gamma$ Curvatures Hypothesis]\index{$\Gamma$ Curvatures Hypothesis}
In general the $\Gamma$ curvatures transforming non stationary four-dimensional Einstein f\/ield equations\index{Einstein f\/ield equations} to the vacuum f\/ield equations (\ref{vacgam}) can be constructed the only via using of the Riemann--Christoffel curvature tensor\index{Riemann--Christoffel curvature tensor}, its contractions with space-time metric, and combinations of all these quantities.
\end{hypothesis}

The Ricci curvature tensor\index{Ricci curvature tensor} (\ref{ricur}) can be presented in the form
\begin{equation}
  R_{\mu\nu}=\Lambda g_{\mu\nu}+\Lambda n_\mu n_\nu,\label{ricur1}
\end{equation}
and by this reason the second term on RHS of the equation (\ref{ricur}) can be interpreted as the correction to the four-dimensional Einstein manifold, i.e. the four-dimensional Riemannian manifold for which Ricci curvature tensor\index{Ricci curvature tensor} is proportional to metric $R_{\mu\nu}=\lambda g_{\mu\nu}$ and therefore the scalar curvature is constant ${^{(4)}}R=4\lambda$ (For advanced discussion of general Einstein manifolds\index{Einstein manifolds} e.g. the well-known Besse's book \cite{besse}), def\/ined by the sign identical to the cosmological constant\index{cosmological constant} $\lambda=\Lambda$. In other words the situation presented in this section corresponds to deformation of the four-dimensional Einstein manifolds\index{Einstein manifolds} of sign $\lambda=\Lambda$
\begin{equation}
  R_{\mu\nu}=\Lambda g_{\mu\nu}+\Delta_{\mu\nu},\label{ricur1}
\end{equation}
where $\Delta_{\mu\nu}=\Lambda n_\mu n_\nu$ is \emph{the deformation curvature tensor}\index{deformation curvature tensor}, for which
\begin{equation}
 {^{(4)}}R=4\Lambda+\Delta,\label{defric}
\end{equation}
where $\Delta=g^{\mu\nu}\Delta_{\mu\nu}=-\Lambda$ is \emph{the deformation scalar curvature}\index{deformation scalar curvature}. Comparison of the result (\ref{defric}) with the equation (\ref{scagam}) leads to expression of the $\Gamma$ scalar curvature\index{$\Gamma$ scalar curvature} via the deformation scalar curvature\index{deformation scalar curvature}
\begin{equation}
  \Gamma=\Lambda+\Delta.
\end{equation}

For vanishing cosmological constant\index{cosmological constant} $\Lambda\equiv 0$ one has to deal with the four-dimensional Ricci-f\/lat space-time manifold\index{Ricci-f\/lat manifold} of zero Ricci scalar curvature\index{Ricci scalar curvature}, which we shall call \emph{the Ricci-scalar-f\/lat manifold}\index{Ricci-scalar-f\/lat manifold}. This is however, the result of the fact that we have computed the value of the cosmological constant\index{cosmological constant} by using of (\ref{ric4}), i.e. the Ricci scalar curvature\index{Ricci scalar curvature} evaluated on the boundary $\partial M$. It means that in such a particular case the enveloping space-time is the Ricci-scalar-f\/lat manifold\index{Ricci-scalar-f\/lat manifold} from the point of view of an embedded space. It does not mean, however, that then space-time is f\/lat in general, because of its the Riemann--Christoffel curvature tensor\index{Riemann--Christoffel curvature tensor} must not be vanishing identically when both the Ricci curvature tensor\index{Ricci curvature tensor} and the Ricci scalar curvature\index{Ricci scalar curvature} are trivialized. This is in itself non trivial result because in such a situation both the cosmological constant\index{cosmological constant} and as well as the stress-energy tensor\index{stress-energy tensor} are in general non vanishing and arbitrary, what suggests that from the space point of view space-time looks like vacuum space-time. Such a situation, however, should be rather understood rather as a local property, i.e. related to quantum gravity given by the Wheeler--DeWitt equation\index{Wheeler--DeWitt equation}, than the classical space-time. Moreover, it must be emphasized that topology of such a space-time is still unrestricted, because of there is a lot of possible topologies of a four-dimensional Ricci-scalar-f\/lat manifold\index{Ricci-scalar-f\/lat manifold}. The Ricci-f\/lat manifolds\index{Ricci-f\/lat manifold} are the particular case of the Einstein manifolds\index{Einstein manifolds}, for which the sign is trivial $\lambda=0$. Such a class of the Einstein manifolds include e.g. the Calabi--Yau manifolds\index{Calabi--Yau manifolds} \cite{calyau} and the hyper-K\"ahler manifolds\index{hyper-K\"ahler manifolds} \cite{hypkah} which are in intensive interest of mathematical and theoretical physicists (For some particular applications see e.g. papers in the Ref. \cite{cyhk}), especially in context of string theory\index{string theory}. In a four-dimensional case every Calabi--Yau manifolds\index{Calabi--Yau manifolds} is hyper-K\"ahler manifold\index{hyper-K\"ahler manifolds}. There are also much more simpler solutions of the vacuum Einstein f\/ield equations\index{Einstein f\/ield equations}. For example the f\/lat Minkowski space-time\index{space-time!Minkowski} is the most simple vacuum solution, and nontrivial situations include the Schwarzschild space-time\index{space-time!Schwarzschild} and the Kerr space-time\index{space-time!Kerr} describing the geometry of space-time around a non-rotating spherical mass and a rotating massive body, respectively.

\section{The Ansatz for Wave Functionals}\label{sec:2}

According to the evaluation (\ref{rule}) the Ricci scalar curvature\index{Ricci scalar curvature} ${^{(4)}}R$ of the enveloping space-time manifold expresses via the Ricci scalar curvature\index{Ricci scalar curvature} ${^{(3)}}R$ of the embedded space, and its extrinsic $K_{ij}$ and intrinsic $K$ curvatures. All these embedding characterizations in general are functionals of an induced metric $h_{ij}$. Moreover, the trace of the spatial stress $S\equiv T(h,h)$ as double projection of the stress-energy tensor\index{stress-energy tensor} on an induced metric is also a functional of $h_{ij}$. The cosmological constant\index{cosmological constant} can be treated as a constant functional of $h_{ij}$. In this manner, the energy density\index{energy density} $\varrho$ is at the most a functional of $h_{ij}$, and by this reason in an arbitrary situation one has the functional dependence
\begin{equation}
 \varrho[h_{ij}]=-\dfrac{4\Lambda}{\kappa\ell_P^2}+S[h_{ij}]+\dfrac{1}{\kappa\ell_P^2}\left({^{(3)}}R+K^2-K^{ij}K_{ij}\right)=\dfrac{2\Lambda}{\kappa\ell_P^2}-S[h_{ij}],
\end{equation}
which allows to establish
\begin{eqnarray}
S[h_{ij}]&=&\dfrac{3\Lambda}{\kappa\ell_P^2}-\dfrac{1}{2\kappa\ell_P^2}\left({^{(3)}}R+K^2-K^{ij}K_{ij}\right),\label{esfun}\\
 \varrho[h_{ij}]&=&-\dfrac{\Lambda}{\kappa\ell_P^2}+\dfrac{1}{2\kappa\ell_P^2}\left({^{(3)}}R+K^2-K^{ij}K_{ij}\right).\label{varrhofun}
\end{eqnarray}
Factually, both the relations (\ref{esfun}) and (\ref{varrhofun}) are the results of application of the Hamiltonian constraint\index{Hamiltonian constraint}, i.e. strictly speaking they have a sense only for geometrodynamics. The functional nature of their LHS is a straightforward conclusion of the functional character of their RHS. Such a situation implies non trivial physical content. Namely, because of both the spatial stress density\index{spatial stress density} $S$ and the energy density\index{energy density} $\varrho$ are projections of the stress-energy tensor\index{stress-energy tensor} of Matter f\/ields\index{Matter f\/ields}, they depend on Matter f\/ields\index{Matter f\/ields} and their derivatives. In this manner by the functional nature of (\ref{esfun}) and (\ref{varrhofun}) one can conclude that such a situation is equivalent to the statement that Matter f\/ields\index{Matter f\/ields} are functionals of $h_{ij}$,
\begin{equation}
  \phi=\phi[h_{ij}],
\end{equation}
say. In this manner the DeWitt wave functional \index{DeWitt wave functional}
\begin{equation}
\Psi[h_{ij},\phi]=\Psi[h_{ij},\phi[h_{ij}]]\equiv\Psi[h_{ij}],
\end{equation}
is fully justif\/ied, and the Wheeler--DeWitt equation \index{Wheeler--DeWitt equation}becomes
\begin{equation}\label{wdwh}
\left\{2c\kappa\dfrac{\hslash^2}{\ell_P^2}G_{ijkl}\dfrac{\delta^2}{\delta h_{ij}\delta h_{kl}}+\dfrac{\ell_P^2}{2c\kappa}\sqrt{h}\left({^{(3)}R}[h_{ij}]-2\Lambda-2\kappa\ell_P^2\varrho[h_{ij}]\right)\right\}\Psi[h_{ij}]=0.
\end{equation}
Anyway, however, the crucial general problem is solving the Wheeler--DeWitt equation \index{Wheeler--DeWitt equation}in general. As we have mentioned earlier the Wheeler--DeWitt equation \index{Wheeler--DeWitt equation}has never been solved in general, and even taking into account the DeWitt wave functional does not simplify this general problem because of $\Psi[h_{ij}]$ is still a functional but not function. It means that it is not clear how to treat $\Psi[h_{ij}]$ mathematically. We shall present here the strategy for solution of the Wheeler--DeWitt equation \index{Wheeler--DeWitt equation}which is based on the DeWitt wave functional but reduces the functional $\Psi[h_{ij}]$ to a function. In itself such a reduction def\/ines a new model of quantum gravity within the quantum geometrodynamics\index{quantum geometrodynamics (QGD)} formulated in terms of the Wheeler--DeWitt equation\index{Wheeler--DeWitt equation}.

To start the deductions, we should rethink the quantum geometrodynamics\index{quantum geometrodynamics (QGD)} (\ref{wdwh}), particularly the structure of the DeWitt wave functional $\Psi[h_{ij}]$\index{DeWitt wave functional}. The fundamental interpretation of the Wheeler--DeWitt equation\index{Wheeler--DeWitt equation}, as the result of the primary canonical quantization\index{quantization!primary canonical}, is the Schr\"odinger equation \index{Schr\"odinger equation}or the Klein--Gordon equation\index{Klein--Gordon equation}. In both these situations, however, a wave function is always a scalar f\/ield. The operator acting on the wave functional in the quantum geometrodynamics\index{quantum geometrodynamics (QGD)} (\ref{wdwh}) is always scalar and is a functional on the conf\/igurational space, i.e. here the Wheeler superspace\index{Wheeler superspace}
\begin{equation}\label{wdwop}
  2c\kappa\dfrac{\hslash^2}{\ell_P^2}G_{ijkl}\dfrac{\delta^2}{\delta h_{ij}\delta h_{kl}}+\dfrac{\ell_P^2}{2c\kappa}\sqrt{h}\left({^{(3)}R}[h_{ij}]-2\Lambda-2\kappa\ell_P^2\varrho[h_{ij}]\right)=\mathcal{\hat{O}}[h_{ij}],
\end{equation}
what is similar to the case of the Schr\"odinger \index{Schr\"odinger equation}or the Klein--Gordon equation\index{Klein--Gordon equation}, in which the operator acting on the wave function is a functional on the conf\/igurational space, i.e. the product space $\mathbb{R}^4$. Moreover, the differential operator of the Wheeler--DeWitt equation\index{Wheeler--DeWitt equation} is the $Dif\!f(\partial M)$-invariant. It suggests clearly that the DeWitt wave functional $\Psi[h_{ij}]$ \index{DeWitt wave functional}must be a function invariant with respect to action of the diffeomorphism group\index{diffeomorphism group}, i.e. must be a function of another $Dif\!f(\partial M)$-invariant quantities. Furthermore, for full consistency these diffeoinvariant quantities must be constructed via using of the induced metric $h_{ij}$, $f=f(h_{ij})=inv$, say. Then, however, by the Kucha\v{r} formalism the wave functional $\Psi(f)$ inevitably will be becoming an observable or a perennial, and above all if one expresses the differential operator (\ref{wdwop}) via these invariant quantities then one can treat these invariants as solution of the problem of time\index{problem of time} in quantum geometrodynamics\index{quantum geometrodynamics (QGD)} by identif\/ication of the time $t$ with the invariant of an induced metric, i.e. $t\equiv f$. If a wave functional is an usual function then also one can perform straightforwardly and in extraordinary simply way the formalism of secondary quantization \index{quantization!secondary canonical}and product the theory of quantum gravity which is the quantum f\/ield theory of gravity. Let us apply such a strategy for quantum geometrodynamics\index{quantum geometrodynamics (QGD)}.

\subsubsection{Step 1: Global One-Dimensionality Conjecture}\index{conjecture!global one-dimensionality}
By the DeWitt construction based on the Wheeler metric representation \index{Wheeler metric representation}$\Psi[h_{ij}]$ is a functional of
the $3\times 3$ symmetric matrix of an induced metric. It suggests that the wave functional is a single functional
\begin{equation}
\Psi[h_{ij}]=\Psi\left[\left[\begin{array}{ccc}
h_{11}&h_{12}&h_{13}\\
h_{12}&h_{22}&h_{23}\\
h_{13}&h_{23}&h_{33}
\end{array}\right]\right].
\end{equation}
However, such a reasoning is not unique. The wave functional must not be a single functional but rather is a $3\times 3$ symmetric matrix which elements are dependent on a single element of an induced metric
\begin{equation}
\Psi[h_{ij}]=\left[\begin{array}{ccc}
\Psi[h_{11}]&\Psi[h_{12}]&\Psi[h_{13}]\\
\Psi[h_{12}]&\Psi[h_{22}]&\Psi[h_{23}]\\
\Psi[h_{13}]&\Psi[h_{23}]&\Psi[h_{33}]
\end{array}\right].
\end{equation}
The still unsolved problem of quantum gravity is the reduction procedure
\begin{equation}
\Psi\left[\left[\begin{array}{ccc}
h_{11}&h_{12}&h_{13}\\
h_{12}&h_{22}&h_{23}\\
h_{13}&h_{23}&h_{33}
\end{array}\right]\right]\rightarrow
\left[\begin{array}{ccc}
\Psi[h_{11}]&\Psi[h_{12}]&\Psi[h_{13}]\\
\Psi[h_{12}]&\Psi[h_{22}]&\Psi[h_{23}]\\
\Psi[h_{13}]&\Psi[h_{23}]&\Psi[h_{33}]
\end{array}\right].
\end{equation}
This is evidently perfectionist situation, because in general the wave functional can be considered as a $3\times 3$ symmetric matrix which elements are functional of several elements of an induced metric.

Albeit, the way of straightforward analogy with quantum mechanics suggests that the wave functional $\Psi[h_{ij}]$ is a classical scalar f\/ield like usual wave function in quantum mechanics based on the Schr\"odinger equation\index{Schr\"odinger equation}, i.e. in such a light $\Psi[h_{ij}]$ is a single functional. Let us accept such a state of things. For realization of this idea the wave functional should be dependent on a scalar function of an induced metric $h_{ij}$, which must be an invariant of the induced matrix as well as invariant with respect to action of the diffeomorphism group\index{diffeomorphism group}. The Cayley--Hamilton theorem\index{Cayley--Hamilton theorem} for any $3\times 3$ square matrix $\mathbf{h}$ states that the matrix obeys its characteristic equation
\begin{equation}
\mathbf{h}^3-I_{\mathbf{h}}\mathbf{h}^2+II_{\mathbf{h}}\mathbf{h}-III_{\mathbf{h}}\mathbf{I}_{3\times3}=0,
\end{equation}
where the coeff\/icients of the polynomial
\begin{eqnarray}
I_{\mathbf{h}}&=&\tr\mathbf{h},\\
II_{\mathbf{h}}&=&\dfrac{\left(\tr\mathbf{h}\right)^2-\tr\mathbf{h}^2}{2},\\
III_{\mathbf{h}}&=&\det\mathbf{h},
\end{eqnarray}
are the invariants of the matrix $\mathbf{h}$. A scalar valued matrix function $\Psi(h_{ij})$ that depends merely on the three invariants of a symmetric $3\times3$ matrix
\begin{equation}\label{oqg}
\Psi\left(h_{ij}\right)=\Psi\left(I_{\mathbf{h}},II_{\mathbf{h}},III_{\mathbf{h}}\right),
\end{equation}
is independent on rotations of the coordinate system, is called \emph{objective function}\index{objective function}. The invariants $I_{\mathbf{h}}$ and $II_{\mathbf{h}}$, however, are irrelevant because of do not carry full information about $I_{\mathbf{h}}$. The third invariant $III_{\mathbf{h}}$ as a function of all elements of a matrix carries full information about the matrix. In $3+1$ decomposition determinant is diffeoinvariant function of a $3\times 3$ induced metric $h_{ij}$. It suggests that the invariant dimension is $\det h_{ij}$. Then wave functional reduces to
\begin{equation}\label{GOD1}
\Psi\left(h_{ij}\right)=\Psi\left(III_{\mathbf{h}}\right)=\Psi(h).
\end{equation}
and the quantum geometrodynamics\index{quantum geometrodynamics (QGD)} becomes a one-dimensional quantum mechanics. We shall call $\det h_{ij}$ \emph{the global dimension}\index{global dimension}, because it is a function of local dimensions (coordinates) and some free parameters, and (\ref{GOD1}) \emph{the global one-dimensionality conjecture\index{conjecture!global one-dimensionality}}. We shall call \emph{generalized dimensions}\index{generalized dimensions} another, possibly more convenient, invariants constructed as $f(h)$. We shall call \emph{objective quantum gravity}\index{objective quantum gravity} a theory of quantum gravity related to wave functionals (\ref{oqg}), and \emph{global one-dimensional quantum gravity}\index{global one-dimensional quantum gravity} the theory of quantum gravity related to the wave functionals (\ref{GOD1}).

Such a global one-dimensional wave function can be constructed in the following way. Suppose that Matter f\/ields\index{Matter f\/ields} in general are functionals dependent on the one global variable
\begin{equation}
\phi=\phi[h],
\end{equation}
which is the determinant $h=\det h_{ij}$ of an induced metric on $\partial M$. Recall that in the dimension 3 one has
\begin{equation}
h=\dfrac{1}{3}\epsilon^{ijk}\epsilon^{lmn}h_{il}h_{jm}h_{kn},
\end{equation}
where $\epsilon^{abc}$ is the three-dimensional Levi-Civita symbol
\begin{equation}
\epsilon^{abc}=\dfrac{(a-b)(b-c)(c-a)}{2}.
\end{equation}
As the crucial point of the model let us assume that quantum gravity is globally one-dimensional. In result the DeWitt wave functional \index{DeWitt wave functional}becomes one-dimensional wave function
\begin{equation}\label{GOD}
  \Psi[h_{ij}]\rightarrow\Psi(h),
\end{equation}
and the Wheeler--DeWitt equation \index{Wheeler--DeWitt equation}is
\begin{equation}\label{wdw1}
\left\{-2c\kappa\dfrac{\hslash^2}{\ell_P^2}G_{ijkl}\dfrac{\delta^2}{\delta h_{ij}\delta h_{kl}}-\dfrac{\ell_P^2}{2c\kappa}h^{1/2}\left({^{(3)}R}-2\Lambda-2\kappa\ell_P^2\varrho[h]\right)\right\}\Psi(h)=0.
\end{equation}
In analogy to the generic cosmology \cite{gli} the conjecture\index{conjecture} (\ref{GOD}) describes isotropic spacetimes, and is related to the strata of the Wheeler superspace\index{Wheeler superspace}, called midisuperspace\index{midisuperspace}, in which wave functionals are functions of a one variable.

\subsubsection{Step 2: Reduction of Quantum Geometrodynamics}
Let us consider the Jacobi formula\index{Jacobi formula} for determinant of the space-time metric
\begin{equation}\label{dg}
  \delta g = gg^{\mu\nu}\delta g_{\mu\nu},
\end{equation}
which can be rewritten in components
\begin{equation}
\delta g = g\left(g^{00}\delta g_{00}+g^{ij}\delta g_{ij}+g^{0j}\delta g_{0j}+g^{i0}\delta g_{i0}\right).
\end{equation}
The $3+1$ splitting\index{$3+1$ splitting} (\ref{dec}) allows determine the partial variations
\begin{eqnarray}
  \delta g_{00}&=&-\delta N^2+N^iN^j\delta h_{ij}+h_{ij}N^i\delta N^j+h_{ij}N^j\delta N^i,\\
  \delta g_{ij}&=&\delta h_{ij},\\
  \delta g_{0j}&=&h_{ij}\delta N^i+N^i\delta h_{ij},\\
  \delta g_{i0}&=&h_{ij}\delta N^j+N^j\delta h_{ij},
\end{eqnarray}
as well as the total variation
\begin{eqnarray}
\delta g=N^2\delta h+h\delta N^2.
\end{eqnarray}
Collecting all one obtains the result relevant for an induced metric
\begin{equation}
N^2\delta h=N^2hh^{ij}\delta h_{ij},
\end{equation}
which allows to establish the Jacobian matrix for transformation of variables $h_{ij}\rightarrow h$
 \begin{eqnarray}\label{hij}
   \mathcal{J}\left(h_{ij},h\right)=\dfrac{\delta (h)}{\delta (h_{ij})}=\dfrac{\delta h}{\delta h_{ij}}\equiv hh^{ij}.
\end{eqnarray}
Because of the approximation (\ref{GOD}) the functional derivative $\dfrac{\delta}{\delta h_{ij}}$ acts on a wave functional depending only on $h$. It allows us to express the functional derivative with respect  $h_{ij}$ through the functional derivative $\dfrac{\delta}{\delta h}$. Therefore one has
\begin{equation}\label{hij1}
\dfrac{\delta \Psi[h]}{\delta h_{ij}} = h h^{ij}\dfrac{\delta\Psi[h]}{\delta h}.
\end{equation}
Consequently, application of (\ref{hij1}) within the differential operator of the Wheeler--DeWitt equation \index{Wheeler--DeWitt equation}(\ref{wdw1}) leads to
\begin{eqnarray}\label{gij1}
 G_{ijkl}\dfrac{\delta^2}{\delta h_{ij}\delta h_{kl}}=G_{ijkl}h^{ij}h^{kl}h^2\dfrac{\delta^2}{\delta h^2}.
\end{eqnarray}
So that the reduction is given by the double projection of the DeWitt supermetric \index{DeWitt supermetric}onto an induced metric
\begin{eqnarray}
  G_{ijkl}h^{ij}h^{kl}&=&\dfrac{1}{2\sqrt{h}}\left(h_{ik}h_{jl}+h_{il}h_{jk}-h_{ij}h_{kl}\right)h^{ij}h^{kl}=\nonumber\\
  &=&\dfrac{1}{2\sqrt{h}}\left(h_{ik}h^{kl}h^{ij}h_{jl}+h_{il}h^{ij}h_{jk}h^{kl}-h_{ij}h^{ij}h_{kl}h^{kl}\right)=\nonumber\\
  &=&\dfrac{1}{2\sqrt{h}}\left(\delta^l_i\delta^i_l+\delta^j_l\delta^l_j-\delta^i_i\delta^k_k\right)=\nonumber\\
  &=&\dfrac{1}{2\sqrt{h}}\left(\delta^i_i+\delta^j_j-(\delta^i_i)^2\right)=\nonumber\\
  &=&\dfrac{1}{2\sqrt{h}}\left(2\delta^i_i-(\delta^i_i)^2\right)=\nonumber\\
  &=&\dfrac{1}{2\sqrt{h}}\left(2\cdot3-(3)^2\right)=-\dfrac{3}{2}h^{-1/2},\label{gij2}
\end{eqnarray}
where we have used the relations for three-dimensional embedded space $h^{ab}h_{bc}=h^a_c$, $h^a_a=\delta^a_a=\tr h_{ab}=3$. Jointing (\ref{gij1}) and (\ref{gij2}) one obtains f\/inally the transformation
\begin{eqnarray}
 G_{ijkl}\dfrac{\delta^2}{\delta h_{ij}\delta h_{kl}}=-\dfrac{3}{2}h^{3/2}\dfrac{\delta^2}{\delta h^2},
\end{eqnarray}
which leads to the quantum geometrodynamics\index{quantum geometrodynamics (QGD)}
\begin{equation}\label{new}
  \left[2c\kappa\dfrac{\hslash^2}{\ell_P^2}\dfrac{3}{2}h^{3/2}\dfrac{\delta^2}{\delta h^2}-\dfrac{\ell_P^2}{2c\kappa}h^{1/2}\left({^{(3)}R}-2\Lambda-2\kappa\ell_P^2\varrho[h]\right)\right]\Psi(h)=0.
\end{equation}
Because the relation (\ref{hij}) arises due to $3+1$ approximation, so (\ref{gij2}) is an approximation within the ansatz\index{ansatz}.
\subsubsection{Step 3: Dimensional Reduction}
The quantum geometrodynamics\index{quantum geometrodynamics (QGD)} (\ref{new}) can be rewritten in the form of the Klein--Gordon equation\index{Klein--Gordon equation}
\begin{equation}\label{kgfeq}
\left(\dfrac{\delta^2}{\delta{h^2}}+\omega^2\right)\Psi=0,
\end{equation}
where $\omega^2$ is squared \emph{gravitational dimensionless frequency}\index{dimensionless frequency!gravitational} of the f\/ield $\Psi$
\begin{eqnarray}
\omega^2&=&-\dfrac{1}{6(8\pi)^2}\dfrac{1}{h}\left(^{(3)}R-2\Lambda-2\kappa\ell_P^2\varrho\right)=\label{masqr}\\
&=&-\dfrac{1}{6(8\pi)^2}\dfrac{1}{h}(K_{ij}K^{ij}-K^2),\label{masqr1}
\end{eqnarray}
where the Hamiltonian constraint\index{Hamiltonian constraint} was involved in the second line. In general the squared mass can be positive, negative or even vanishing identically. The equation (\ref{kgfeq}) can be treated as the classical-f\/ield-theoretical Euler--Lagrange\index{Euler--Lagrange equations of motion} equations of motion arising from stationarity of the action functional
\begin{equation}\label{act1}
  S[\Psi]=\int \delta h L\left(\Psi,\dfrac{\delta \Psi}{\delta h}\right),
\end{equation}
where $L=L\left(\Psi,\dfrac{\delta \Psi}{\delta h}\right)$ is the f\/ield-theoretic Lagrange function
\begin{eqnarray}
  L&=&\dfrac{1}{2}\left(\dfrac{\delta \Psi}{\delta h}\right)^2-\dfrac{\omega^2}{2}\Psi^2=\\
  &=&\dfrac{1}{2}\Pi_\Psi^2-\dfrac{\omega^2}{2}\Psi^2,\label{claslag}
\end{eqnarray}
where $\Pi_\Psi$ is the momentum conjugated to the classical scalar f\/ield $\Psi$
\begin{equation}
  \Pi_\Psi=\dfrac{\partial L}{\partial\left(\dfrac{\delta \Psi}{\delta h}\right)}=\dfrac{\delta \Psi}{\delta h}.\label{canmom}
\end{equation}
The action $S[\Psi]$ is a f\/ield-theoretic action functional in the classical f\/ield $\Psi$, and therefore arbitrary dependence on the variable $h$ of the mass $m=m[h]$ does not play a role, i.e. behaves as a coeff\/icient, in derivation of the Euler--Lagrange\index{Euler--Lagrange equations of motion} equations of motion
\begin{equation}
\delta S[\Psi]=\int\delta h\left[\dfrac{\partial L}{\partial\Psi}-\dfrac{\delta}{\delta h}\dfrac{\partial L}{\partial\left(\dfrac{\delta \Psi}{\delta h}\right)}\right]\delta\Psi+\int\delta h\dfrac{\delta}{\delta h}\left(\dfrac{\partial L}{\partial\Psi}\delta\Psi\right)=0,
\end{equation}
what gives the result
\begin{equation}
\dfrac{\partial L}{\partial\Psi}-\dfrac{\delta}{\delta h}\dfrac{\partial L}{\partial\left(\dfrac{\delta \Psi}{\delta h}\right)}=0,\label{eullag}
\end{equation}
where we have taken \emph{ad hoc} the f\/ield theoretical condition of vanishing of the boundary term
\begin{equation}
\int\delta h\dfrac{\delta}{\delta h}\left(\dfrac{\partial L}{\partial\Psi}\delta\Psi\right)=\int\delta\left(\dfrac{\partial L}{\partial\Psi}\delta\Psi\right)=\left.\dfrac{\partial L}{\partial\Psi}\delta\Psi\right|_{0}=0.
\end{equation}
It can be seen by straightforward computation that the equation (\ref{eullag}) coincides with (\ref{kgfeq}).

By application of the conjugate momentum $\Pi_\Psi$ one rewrites the equation (\ref{kgfeq}) in the following form
\begin{equation}\label{kgfeq1}
  \dfrac{\delta\Pi_\Psi}{\delta h}+\omega^2\Psi=0,
\end{equation}
and therefore the equations (\ref{canmom}) and (\ref{kgfeq1}) are the system of canonical Hamilton equations of motion\index{Hamilton canonical equations of motion}
\begin{eqnarray}
  \dfrac{\delta}{\delta h}\Psi&=&\dfrac{\delta}{\delta\Pi_\Psi}H\left(\Psi,\Pi_\Psi\right),\\
  \dfrac{\delta}{\delta h}\Pi_\Psi&=&-\dfrac{\delta}{\delta\Psi}H\left(\Psi,\Pi_\Psi\right),
\end{eqnarray}
where the Hamilton function $H\left(\Psi,\Pi_\Psi\right)$ is obtained from the Lagrange function (\ref{claslag}) via the Legendre transformation\index{Legendre transformation}
\begin{eqnarray}
H\left(\Psi,\Pi_\Psi\right)&=&\Pi_\Psi\dfrac{\delta \Psi}{\delta h}-L\left(\Psi,\dfrac{\delta \Psi}{\delta h}\right)=\\
&=&\dfrac{1}{2}\Pi_\Psi^2-\dfrac{\omega^2}{2}\Psi^2.\label{clasham}
\end{eqnarray}
If one recognizes the kinetic $T$ and the potential $V$ energies as
\begin{eqnarray}
  T&=&\dfrac{1}{2}\Pi_\Psi^2,\\
  V&=&\dfrac{1}{2}\omega^2\Psi^2,
\end{eqnarray}
then the Hamilton function (\ref{clasham}) is $H=T-V$ and the Lagrange function (\ref{claslag}) is $L=T+V$, what means that the f\/ield theory presented above is the Euclidean f\/ield theory\index{Euclidean f\/ield theory} of a simple harmonic oscillator of the mass $1$ and frequency $\omega$. In this context the classical scalar f\/ield - the wave function $\Psi$ - becomes the generalized coordinate.

Let us introduce the two-component f\/ield
\begin{equation}
  \Phi=\left[\begin{array}{c}\Pi_\Psi\\ \Psi\end{array}\right],
\end{equation}
which components obey the equations (\ref{canmom})-(\ref{kgfeq1}). The system of the Hamilton canonical equations of motion\index{Hamilton canonical equations of motion} (\ref{canmom})-(\ref{kgfeq1}) can be rewritten in the form of the vector equation
\begin{equation}
  \left(-i\left[\begin{array}{cc}0&-i\\i&0\end{array}\right]\dfrac{\delta}{\delta h}-\left[\begin{array}{cc}-\dfrac{1}{\Pi_\Psi}\dfrac{\delta}{\delta\Pi_\Psi}&0\\0&-\dfrac{1}{\Psi}\dfrac{\delta}{\delta\Psi}\end{array}\right]H\left(\Psi,\Pi_\Psi\right)\right)\Phi=0,
\end{equation}
which for the situation given by the Hamiltonian (\ref{clasham}) leads the appropriate one-dimensional Dirac equation \index{Dirac equation!one-dimensional}for the classical two-component f\/ield $\Phi$
\begin{equation}\label{diraceq}
 \left(-i\gamma\dfrac{\delta}{\delta h}-M\right)\Phi=0,
\end{equation}
where $M$ is the mass matrix of the f\/ield $\Phi$
\begin{equation}
  M=\left[\begin{array}{cc}-1&0\\0&-\omega^{2}\end{array}\right],
\end{equation}
and the $\gamma$ matrix is the Pauli matrix $\sigma_y$
\begin{equation}
  \gamma=\sigma_y=\left[\begin{array}{cc}0&-i\\i&0\end{array}\right],
\end{equation}
obeying the following algebra
\begin{equation}\label{cliff}
  \gamma^2=\mathbf{I}_2\quad,\quad\left\{\gamma,\gamma\right\}=2\mathbf{I}_2\quad,\quad\mathbf{I}_2=\left[\begin{array}{cc}1&0\\0&1\end{array}\right].
\end{equation}

The algebra (\ref{cliff}) is the four-dimensional Clifford algebra \index{Clifford algebra!$\mathcal{C}\ell_{2}(\mathbb{C})$}over the complex vector space $\mathbb{C}^2$ (For basics and advances in Clifford algebras see e.g. the Ref. \cite{clifford})
\begin{equation}
\mathcal{C}\ell_{2}(\mathbb{C})=\mathcal{C}\ell_0(\mathbb{C})\otimes\mathrm{M}_2(\mathbb{C})\cong\mathrm{M}_2(\mathbb{C})=\mathbb{C}\oplus\mathbb{C},
\end{equation}
where $\mathcal{C}\ell_{n}\equiv \mathcal{C}\ell_{n,0}$, and $\mathrm{M}_2(\mathbb{C})$ denotes algebra of all $2\times 2$ matrices over $\mathbb{C}$. The Clifford algebra\index{Clifford algebra!$\mathcal{C}\ell_{2}(\mathbb{C})$} $\mathcal{C}\ell_{2,0}(\mathbb{C})$ possesses a two-dimensional complex representation. Restriction to the pinor group $\textrm{Pin}_{2,0}(\mathbb{R})$ yields a complex representation of two-dimensional pinor group, i.e. the two-dimensional spinor representation, whereas restriction to the spinor group $\textrm{Spin}_{2,0}(\mathbb{R})$ splits $\mathcal{C}\ell_{1,1}(\mathbb{R})$ onto a sum of two half spin representations of dimension 1, i.e. the one dimensional Weyl representations. There is the isomorphism
\begin{equation}
\textrm{Spin}_{2,0}(\mathbb{R})\cong \mathrm{U}(1)\cong\mathrm{SO}(2),
\end{equation}
and the spinor group $\textrm{Spin}_{2,0}(\mathbb{R})$ acts on a 1-sphere $S^1$ in such a way that one has a f\/ibre bundle\index{bundle!f\/ibre} with f\/ibre $\textrm{Spin}_{1,0}(\mathbb{R})$
\begin{equation}
\textrm{Spin}_{1,0}(\mathbb{R})\longrightarrow\textrm{Spin}_{2,0}(\mathbb{R})\longrightarrow S^1,
\end{equation}
and the homotopy sequence\index{homotopy sequence} is
\begin{equation}
\pi_1\left(\textrm{Spin}_{1,0}(\mathbb{R})\right)\longrightarrow\pi_1\left(\textrm{Spin}_{2,0}(\mathbb{R})\right)\longrightarrow \pi_1\left(S^1\right).
\end{equation}

The Clifford algebra $\mathcal{C}\ell_{2}(\mathbb{C})$\index{Clifford algebra!$\mathcal{C}\ell_{2}(\mathbb{C})$} can be generated by complexif\/ication
\begin{equation}
\mathcal{C}\ell_{2}(\mathbb{C})\cong\mathcal{C}\ell_{1,1}(\mathbb{R})\otimes\mathcal{C}\ell_{0}(\mathbb{C}),
\end{equation}
where $\mathcal{C}\ell_{1,1}(\mathbb{R})$ is the four-dimensional Clifford algebra \index{Clifford algebra}over the real vector space $\mathbb{R}^{2,0}$
\begin{equation}
\mathcal{C}\ell_{1,1}(\mathbb{R})\cong\mathrm{M}_2(\mathbb{R})\otimes\mathcal{C}\ell_{0}(\mathbb{R})\cong{\mathrm{M}_2(\mathbb{R})},\label{cl11}
\end{equation}
with $\mathrm{M}_2(\mathbb{R})$ being algebra of $2\times 2$ matrices over $\mathbb{R}$, and
\begin{eqnarray}
\mathcal{C}\ell_{0}(\mathbb{R})=\mathbb{R},\\
\mathcal{C}\ell_{0}(\mathbb{C})=\mathbb{C}.
\end{eqnarray}
The Clifford algebra \index{Clifford algebra!$\mathcal{C}\ell_{2}(\mathbb{C})$}(\ref{cl11}) can be decomposed into a direct sum of central simple algebras isomorphic to matrix algebra over $\mathbb{R}$
\begin{eqnarray}
\mathcal{C}\ell_{1,1}(\mathbb{R})&=&\mathcal{C}\ell^+_{1,1}(\mathbb{R})\oplus\mathcal{C}\ell^-_{1,1}(\mathbb{R}),\\
\mathcal{C}\ell^{\pm}_{1,1}(\mathbb{R})&=&\dfrac{1\pm\gamma}{2}\mathcal{C}\ell_{1,1}(\mathbb{R})\cong\mathbb{R},
\end{eqnarray}
as well as into a tensor product
\begin{eqnarray}
  \mathcal{C}\ell_{1,1}(\mathbb{R})&=&\mathcal{C}\ell_{2,0}(\mathbb{R})\otimes\mathcal{C}\ell_{0,0}(\mathbb{R}),\\
  \mathcal{C}\ell_{2,0}(\mathbb{R})&=&\mathrm{M}_2(\mathbb{R})\otimes\mathcal{C}\ell_{0,0}(\mathbb{R})\cong{\mathrm{M}_2(\mathbb{R})}.
\end{eqnarray}
\section{Field Quantization in Static Fock Space\index{Fock space}}\label{sec:3}
The one-dimensional Dirac equation \index{Dirac equation!one-dimensional}(\ref{diraceq}) can be canonically quantized
\begin{equation}\label{qdiraceq}
 \left(-i\gamma\dfrac{\delta}{\delta h}-M\right)\hat{\Phi}=0,
\end{equation}
according to the canonical commutation relations\index{canonical commutation relations!bosonic} (CCR) characteristic for bosonic f\/ields
\begin{eqnarray}
i\left[\hat{\Pi}_{\Psi}[h'],\hat{\Psi}[h]\right]&=&\delta(h'-h),\\
i\left[\hat{\Pi}_{\Psi}[h'],\hat{\Pi}_{\Psi}[h]\right]&=&0,\\
i\left[\hat{\Psi}[h'],\hat{\Psi}[h]\right]&=&0,\label{c3}
\end{eqnarray}
where the choice of the bosonic CCR follows form the fact that one has the one-dimensional situation in which there is no difference between bosons and fermions. Particles obeying one-dimensional quantum evolutions are called \emph{axions}\index{axion}, and in this manner the second quantized one-dimensional Dirac equation \index{Dirac equation!one-dimensional}(\ref{qdiraceq}) describes axions\index{axion} obeying the Bose--Einstein statistics, which are \emph{gravitons}\index{graviton} in our understanding.

Let us apply the Fock space\index{Fock space} formalism, which allows to write out explicitly the decomposition of the solution
\begin{equation}\label{sqx}
  \hat{\Phi}=\mathbf{Q}\mathfrak{B},
\end{equation}\
where $\mathbf{Q}$ is the matrix of secondary quantization\index{quantization!secondary canonical}
\begin{equation}
  \mathbf{Q}=\left[\begin{array}{cc}\sqrt{\dfrac{1}{2\omega}}&\sqrt{\dfrac{1}{2\omega}}\\
-i\sqrt{\dfrac{\omega}{2}}&i\sqrt{\dfrac{\omega}{2}}\end{array}\right],
\end{equation}
and $\mathfrak{B}=\mathfrak{B}[h]$ is a dynamical rep\`{e}re
\begin{equation}\label{db}
  \mathfrak{B}=\left\{\left[\begin{array}{c}\textsf{G}[h]\\
\textsf{G}^{\dagger}[h]\end{array}\right]:\left[\textsf{G}[h'],\textsf{G}^{\dagger}[h]\right]=\delta\left(h'-h\right), \left[\textsf{G}[h'],\textsf{G}[h]\right]=0\right\},
\end{equation}
on the Fock space\index{Fock space} of creation and annihilation operators
\begin{equation}
  \mathcal{F}=\left(\textsf{G},\textsf{G}^{\dagger}\right).
\end{equation}
Application of the decomposition (\ref{sqx}) yields the Heisenberg equations of motion\index{Heisenberg equations of motion} modif\/ied by the non-diagonal components
\begin{equation}\label{df}
\dfrac{\delta\mathfrak{B}}{\delta h}=\mathbf{X}\mathfrak{B},
\end{equation}
where $\mathbf{X}$ is the matrix
\begin{equation}
  \mathbf{X}=\left[\begin{array}{cc}-i\omega&\dfrac{1}{2\omega}\dfrac{\delta \omega}{\delta h}\\ \dfrac{1}{2\omega}\dfrac{\delta \omega}{\delta h}&i\omega\end{array}\right].
\end{equation}
Let us suppose that there is another rep\`{e}re $\mathfrak{F}$ determined by the Bogoliubov transformation\index{Bogoliubov transformation}
\begin{eqnarray}
\mathfrak{F}&=&\left[\begin{array}{cc}u&v\\v^{\ast}&u^{\ast}\end{array}\right]\mathfrak{B},\label{pr2}
\end{eqnarray}
where the Bogoliubov coeff\/icients\index{Bogoliubov coeff\/icients} $u$ and $v$ forms the Gauss--Lobachevsky--Bolyai hyperbolic space\index{Gauss--Lobachevsky--Bolyai hyperbolic space} and obey the constraint
\begin{equation}
 |u|^2-|v|^2=1,
\end{equation}
and together with the frequency $\Omega$ are functionals of $h$. As the second requirement let us suppose also that dynamics of the rep\`{e}re is governed by the Heisenberg equations of motion\index{Heisenberg equations of motion}
\begin{equation}
  \dfrac{\delta\mathfrak{F}}{\delta h}=\left[\begin{array}{cc}-i\Omega&0\\0&i\Omega\end{array}\right]\mathfrak{F}.\label{pr3}
\end{equation}
Application of the system of equations (\ref{pr2})- (\ref{pr3}) to the equations (\ref{df}) leads to the equation for the vector of the Bogoliubov coeff\/icients\index{Bogoliubov coeff\/icients}
\begin{equation}
  \mathbf{b}=\left[\begin{array}{c}u\\v\end{array}\right],
\end{equation}
which is given by the following vector equation
\begin{equation}
  \dfrac{\delta\mathbf{b}}{\delta h}=\mathbf{X}\mathbf{b},\label{bcof}
\end{equation}
and gives trivial value of the unknown frequency
\begin{equation}
\Omega\equiv0.
\end{equation}
Therefore, the conjectured rep\`{e}re $\mathfrak{F}$ becomes the static Fock rep\`{e}re\index{Fock rep\`{e}re} with respect to initial data ($I$)
\begin{equation}\label{in}
\mathfrak{F}=\left\{\left[\begin{array}{c}\textsf{G}_I\\
\textsf{G}^{\dagger}_I\end{array}\right]: \left[\textsf{G}_I,\textsf{G}^{\dagger}_I\right]=1, \left[\textsf{G}_I,\textsf{G}_I\right]=0\right\},
 \end{equation}
and the vacuum state\index{vacuum state} $\rv$ is correctly def\/ined
\begin{eqnarray}
\textsf{G}_I\rv&=&0,\\
\lv \textsf{G}_I^\dagger&=&0.
\end{eqnarray}
Integrability of the system of equations (\ref{bcof}) is the crucial element of the scheme presented above. The Bogoliubov transformation\index{Bogoliubov transformation} (\ref{pr2}), however, suggests application of the superf\/luid parametrization
\begin{eqnarray}
u&=&e^{i\theta}\cosh \phi,\\
v&=&e^{i\theta}\sinh \phi,
\end{eqnarray}
where $\theta$ and $\phi$ are the angles which for the present situation are
\begin{eqnarray}
	\theta&=&\pm i\int_{h_I}^{h}\omega'\delta h',\label{sup}\\
	\phi&=&\ln{\sqrt{\left|\dfrac{\omega_I}{\omega}\right|}},
\end{eqnarray}
where $\omega'=\omega(h')$ and $\omega_I$ is the initial datum of gravitational dimensionless frequency\index{dimensionless frequency!gravitational!initial datum}
\begin{equation}
  \omega_I=-\dfrac{1}{8\pi\sqrt{6}},
\end{equation}
which yield the Bogoliubov coeff\/icients\index{Bogoliubov coeff\/icients}
\begin{eqnarray}
u&=&\dfrac{\mu+1}{2\sqrt{\mu}}\exp\left\{i\int_{h_I}^{h}\omega'\delta h'\right\},\\ v&=&\dfrac{\mu-1}{2\sqrt{\mu}}\exp\left\{-i\int_{h_I}^{h}\omega'\delta h'\right\},
\end{eqnarray}
where $\mu=\dfrac{\omega}{\omega_I}$ measures the relative gravitational dimensionless frequency\index{dimensionless frequency!gravitational!relative}. For convenience one can apply also the reciprocal of $\mu$, i.e. the parameter $\lambda=\dfrac{\omega_I}{\omega}=\dfrac{1}{\mu}$
\begin{equation}
\lambda=\sqrt{\left|\dfrac{h}{^{(3)}R-2\Lambda-2\kappa\ell_P^2\varrho}\right|}=\sqrt{\left|\dfrac{h}{K_{ij}K^{ij}-K^2}\right|},
\end{equation}
and we understand $\lambda\equiv\lambda[h]$, $\lambda'=\lambda[h']$.

Consequently, the integrability problem is solved by the equation
\begin{equation}\label{phi}
\hat{\Phi}=\mathbf{Q}\mathbf{G}\mathfrak{F},
\end{equation}
where $\mathbf{G}$ is the monodromy matrix\index{monodromy matrix}
\begin{equation}\label{mono}
\mathbf{G}=\left[\begin{array}{cc}
\dfrac{1+\mu}{2\sqrt{\mu}}\exp\left\{-i\int_{h_I}^{h}\omega'\delta h'\right\}\vspace*{10pt}&
\dfrac{1-\mu}{2\sqrt{\mu}}\exp\left\{i\int_{h_I}^{h}\omega'\delta h'\right\}\\
\dfrac{1-\mu}{2\sqrt{\mu}}\exp\left\{-i\int_{h_I}^{h}\omega'\delta h'\right\}&
\dfrac{1+\mu}{2\sqrt{\mu}}\exp\left\{i\int_{h_I}^{h}\omega'\delta h'\right\}\end{array}\right].
\end{equation}

Now it can be seen straightforwardly that the presented version of quantum geometrodynamics\index{quantum geometrodynamics (QGD)} formulates quantum gravity as a quantum f\/ield theory of gravity, where the quantum gravitational f\/ield is associated with conf\/iguration of embedded space and given by the decomposition (\ref{phi}) in the static Fock space\index{Fock space}. In this manner one can write out straightforwardly conclusions following form the global one-dimensional model of quantum gravity.

It must be noticed that the functional measure $\delta h$ in any integrals of the form $\int \delta h'f[h']$ for the case of a f\/ixed conf\/iguration of space, i.e. $h=constant$, becomes the Riemann--Lebesgue measure\index{Riemann--Lebesgue measure} $dh$. However, because of $h$ in general is a smooth function of space-time coordinates and free parameters, the measure $\delta h$ as a total variation over space-time coordinates is the Lebesgue--Stieltjes measure\index{Lebesgue--Stieltjes measure} which can be rewritten as the Riemann--Lebesgue measure\index{Riemann--Lebesgue measure} on space-time. In the most general case $h=h(x_0,x_1,x_2,x_3)$ one can use the transformation
\begin{equation}\label{measures}
\delta h = \dfrac{\partial^4 h(x_0,x_1,x_2,x_3)}{\partial x_0\partial x_1\partial x_2\partial x_3}d^4x,
\end{equation}
where $d^4x=dx_0dx_1dx_2dx_3$, and compute the integral
\begin{equation}
\int \delta h'f[h']=\int d^4x' \dfrac{\partial^4 h(x_0',x_1',x_2',x_3')}{\partial x_0'\partial x_1'\partial x_2'\partial x_3'}f(x_0',x_1',x_2',x_3').
\end{equation}
In this manner the transformation (\ref{measures}) establishes the relation between the Wheeler superspace\index{Wheeler superspace} and space-time.

The initial data condition $m=m_I$ generates the equation for the initial manifold
\begin{equation}
{^{(3)}}R^{(I)}-2\Lambda-2\kappa\ell_P^2\varrho_I=h_I,
\end{equation}
or equivalently
\begin{equation}
K^{(I)}_{ij}K^{(I)ij}-K^{(I)2}=h^{I},
\end{equation}
where the superscript $I$ means initial value of given quantity. The quantum evolution (\ref{kgfeq}) in such a situation takes the form
\begin{equation}
  \left(\dfrac{\delta^2}{\delta h_I^2}-\dfrac{1}{6(8\pi)^2}\right)\Psi(h_I)=0,\label{ideq}
\end{equation}
and after taking into account the suitable boundary conditions
\begin{eqnarray}
  \Psi(h_I=h_0)=\Psi_0,\\
  \left.\dfrac{\delta\Psi(h_I)}{\delta h_I}\right|_{h_I=h_0}=\Pi_\Psi^0,
\end{eqnarray}
can be solved straightforwardly
\begin{equation}
  \Psi(h_I)=\Psi_0\cosh\left\{\dfrac{h_I-h_0}{8\pi\sqrt{6}}\right\}+8\pi\sqrt{6}\ell_P^2\Pi_\Psi^0\sinh\left\{\dfrac{h_I-h_0}{8\pi\sqrt{6}}\right\}.
\end{equation}

\section{Several Implications}\label{sec:4}
The quantum f\/ield-theoretic geometrodynamics just was formulated. However, still we do not know what it the role of an one-dimensional wave function which solves the equation (\ref{wdw1}). The same problem is to def\/ine any geometric quantities related to the midisuperspace\index{midisuperspace} quantum geometrodynamics\index{quantum geometrodynamics (QGD)}. The quantum f\/ield theory of gravity (\ref{phi}) has also unprecise signif\/icance. Let us present now several conclusions arising from the previous section, which shall clarify our doubts in some detail.\\
\subsection{The Global 1D Wave Function}
The one-dimensional Dirac equation \index{Dirac equation!one-dimensional}(\ref{dira}) can be rewritten in the form of Schr\"odinger equation\index{Schr\"odinger equation}
\begin{equation}\label{evol}
  \dfrac{\delta\Phi}{\delta h}=H\Phi,
\end{equation}
where $H$ is the hermitian Hamiltonian
\begin{equation}
  H=i\gamma{M}=\left[\begin{array}{cc}0&-\omega^2\\1&0\end{array}\right]=\left[\begin{array}{cc}0&-\dfrac{\omega_I^2}{\lambda^2}\\1&0\end{array}\right],
\end{equation}
yielding the evolution operator
\begin{equation}
U[h,h_I]=\exp \int_{h_I}^h \delta h'H[h']=\exp\left[\begin{array}{cc}0&-\int_{h_I}^h\delta{h'}\omega^2[h']\\h-h_I&0\end{array}\right],
\end{equation}
where $h\geqslant h_I$, which is explicitly
\begin{equation}\label{evoloper}
U[h,h_I]=\left[\begin{array}{cc}\cos f[h,h_I]&-\left(\int_{h_I}^h\delta h'{\omega'}^2\right)\dfrac{\sin f[h,h_I]}{f[h,h_I]}\\(h-h_I)\dfrac{\sin f[h,h_I]}{f[h,h_I]}&\cos f[h,h_I]\end{array}\right],
\end{equation}
where $f[h,h_I]$ is the functional
\begin{equation}
f[h,h_I]=\sqrt{\strut{(h-h_I)\int_{h_I}^h{\omega'}^2\delta h'}},
\end{equation}
so that the solution of the equation (\ref{evol}) is
\begin{equation}
  \Phi[h,h_I]=U[h,h_I]\Phi[h_I].
\end{equation}
Straightforward elementary algebraic manipulations allow to determine the global one-dimensional wave function as
\begin{equation}\label{wfun}
  \Psi[h,h_I]]=\Psi^I\cos f[h,h_I]+\Pi_\Psi^I(h-h_I)\dfrac{\sin f[h,h_I]}{f[h,h_I]},
\end{equation}
and similarly the canonical conjugate momentum is
\begin{equation}\label{pfun}
  \Pi_\Psi[h,h_I]=\Pi_\Psi^I\cos f[h,h_I]-\Psi^I\left(\int_{h_I}^h\delta h'{\omega'}^2\right)\dfrac{\sin f[h,h_I]}{f[h,h_I]},
\end{equation}
where $\Psi^I$ and $\Pi_\Psi^I$ are initial data
\begin{eqnarray}
\Psi^I&=&\Psi[h_I],\\
\Pi_\Psi^I&=&\Pi_{\Psi}[h_I]=\left.\dfrac{\delta\Psi}{\delta h}\right|_{h=h_I}.
\end{eqnarray}
Because of $U^\dag[h,h_I]=U^T[h,h_I]$ one sees that
\begin{eqnarray}
\Psi^\star[h,h_I]&=&\Psi[h,h_I],\\
\Pi_\Psi^\star[h,h_I]&=&\Pi_\Psi[h,h_I],
\end{eqnarray}
and for consistency also must be $(\Psi^I)^\star=\Psi^I$, $(\Pi_\Psi^I)^\star=\Pi_\Psi^I$.
The probability density in the quantum mechanics is
\begin{equation}
\Omega[h,h_I]=\Phi^\dag[h,h_I]\Phi[h,h_I]=\Phi^\dag[h_I]U^\dag[h,h_I]U[h,h_I]\Phi[h_I].\label{denpro}
\end{equation}
Computing the matrix $U^\dag[h,h_I]U[h,h_I]$
\begin{eqnarray}
\!\!\!\!\!\!\!\!\!\!\!\!\!\!\!\!\!\!\!\!
U^\dag[h,h_I]U[h,h_I]&=&\left[\begin{array}{cc}\cos^2f[h,h_I]+\left[(h-h_I)\dfrac{\sin f[h,h_I]}{f[h,h_I]}\right]^2\\\left(h-h_I-\int_{h_I}^h\delta h'{\omega'}^2\right)\dfrac{\sin2f[h,h_I]}{2f[h,h_I]}\end{array}\right.\nonumber\\
\!\!\!\!\!\!\!\!\!\!\!\!\!\!\!\!\!\!\!\!
&&\left.\begin{array}{cc}\left(h-h_I-\int_{h_I}^h\delta h'{\omega'}^2\right)\dfrac{\sin 2f[h,h_I]}{2f[h,h_I]}\\ \cos^2f[h,h_I]+\left[\int_{h_I}^h\delta h'{\omega'}^2\dfrac{\sin f[h,h_I]}{f[h,h_I]}\right]^2\end{array}\right],
\end{eqnarray}
and taking into account that
\begin{equation}
\Phi^\dag[h_I]=\left[\begin{array}{c}\Pi_\Psi^I\\ \Psi^I\end{array}\right]^\dag=\left[\begin{array}{c}(\Pi_\Psi^I)^\star\\ (\Psi^I)^\star\end{array}\right]^T=\left[\begin{array}{c}\Pi_\Psi^I\\ \Psi^I\end{array}\right]^T=\left[\Pi_\Psi^I,\Psi^I\right],
\end{equation}
one obtains f\/inally
\begin{equation}
\Omega[h,h_I]=A[h,h_I]\left(\Psi^I\right)^2+2B[h,h_I]\Psi^I\Pi_\Psi^I+C[h,h_I]\left(\Pi_\Psi^I\right)^2,\label{proba}
\end{equation}
where the functional coeff\/icients $A[h,h_I]$, $B[h,h_I]$, $C[h,h_I]$ in (\ref{proba}) are
\begin{eqnarray}
A[h,h_I]&=&\cos^2f[h,h_I]+\left(\int_{h_I}^h\delta h'{\omega'}^2\right)^2\left(\dfrac{\sin f[h,h_I]}{f[h,h_I]}\right)^2,\\
B[h,h_I]&=&\left(h-h_I-\int_{h_I}^h\delta h'{\omega'}^2\right)\dfrac{\sin 2f[h,h_I]}{2f[h,h_I]},\\
C[h,h_I]&=&\cos^2f[h,h_I]+(h-h_I)^2\left(\dfrac{\sin f[h,h_I]}{f[h,h_I]}\right)^2,
\end{eqnarray}
The initial data $\Psi^I$ and $\Pi_\Psi^I$ are not arbitrary, but constrained by the normalization condition for the probability density
\begin{equation}
  \int_{h_I}^{h_F}\Omega[h']\delta h'=1,\label{normcond1}
\end{equation}
where $h_F$ is some maximal value of $h$, which in explicit form leads to the algebraic equation
\begin{equation}
  C[h_I](\Pi_\Psi^I)^2+2B[h_I]\Psi^I\Pi_\Psi^I+A[h_I](\Psi^I)^2-1=0,\label{abcX}
\end{equation}
with the coeff\/icients $A[h_I]$, $B[h_I]$, $C[h_I]$ given by the integrals
\begin{eqnarray}
  A[h_I]&=&\int_{h_I}^{h_F}\delta h'A[h',h_I],\\
  B[h_I]&=&\int_{h_I}^{h_F}\delta h'B[h',h_I],\\
  C[h_I]&=&\int_{h_I}^{h_F}\delta h'C[h',h_I].
\end{eqnarray}
The equation (\ref{abcX}) can be solved straightforwardly. In result one obtains
\begin{eqnarray}\label{pii}
\Pi_\Psi^I=-\dfrac{B[h_I]}{C[h_I]}\Psi^I\pm\sqrt{\strut{\dfrac{B^2[h_I]-A[h_I]C[h_I]}{C^2[h_I]}(\Psi^I)^2+\dfrac{1}{C[h_I]}}}.
\end{eqnarray}
Application of the explicit form of initial data of the conjugate momentum $\Pi_\Psi^I=\dfrac{\delta \Psi^I}{\delta h_I}$ to the equation (\ref{abcX}) yields the differential equation for initial data of the classical scalar f\/ield $\Psi^I$
\begin{equation}\label{pii2}
  C[h_I]\left(\dfrac{\delta\Psi^I}{\delta h_I}\right)^2+2B[h_I]\Psi^I\dfrac{\delta \Psi^I}{\delta h_I}+A[h_I](\Psi^I)^2-1=0,
\end{equation}
which in general is very hard to solve. However, there is the case def\/ined by the values of the coeff\/icients $A[h_I]=A$, $B[h_I]=B$, $C[h_I]=C$ which are independent on $h_I$. In such a situation the equation (\ref{pii2}) possesses solutions which are easy to extract
\begin{equation}\label{wf}
  \Psi^I_{\mp}=f_{\pm}^{(-1)}\left(\mp\dfrac{h_I}{C}+C_1\right),
\end{equation}
where $C_1$ is an integration constant, and $f_{\pm}(x)$ are the functions
\begin{eqnarray}\label{func}
\!\!\!\!\!\!\!\!\!\!\!\!\!\!\!\!\!\!\!\!\!\!\!\!\!\!\!\!\!\!\!\!\!\!\!\!\!\!\!
&&f_{\pm}(x)=\pm\dfrac{B}{AC}\Bigg\{\dfrac{1}{2}\ln\left|Ax^2-1\right|\pm\artanh\left[\dfrac{Bx}{\sqrt{C+\left(B^2-AC\right)x^2}}\right]\mp\nonumber\\
\!\!\!\!\!\!\!\!\!\!\!\!\!\!\!\!\!\!\!\!\!\!\!\!\!\!\!\!\!\!\!\!\!\!\!\!\!\!\!
&&\mp\dfrac{\sqrt{B^2-AC}}{B}\ln\left|2\left(B^2-AC\right)x+2\sqrt{B^2-AC}\sqrt{C+\left(B^2-AC\right)x^2}\right|\Bigg\},
\end{eqnarray}
The solution (\ref{wf}) has been received computationally, and by using it one can construct the analytical solutions straightforwardly. Let us introduce the following parameter
\begin{equation}
  x=\pm\dfrac{h_I-h_0}{C},
\end{equation}
where for consistency we have taken $C_1=\mp\dfrac{h_0}{C}$. Differentiating the solutions (\ref{wf}) with respect to $x$ one obtains
\begin{equation}\label{wf1}
  \dfrac{\delta\Psi^I_{\mp}}{\delta\left(\pm\dfrac{h_I}{C}+C_1\right)}=\pm C\dfrac{\delta\Psi^I_{\mp}}{\delta h_I}=-\dfrac{1}{f'_{\mp}(x)},
\end{equation}
where $f'(x)=\dfrac{df(x)}{dx}$, and by this reason
\begin{equation}\label{wf1a}
  \dfrac{\delta\Psi^I_{\mp}}{\delta h_I}=\mp\dfrac{1}{Cf'_{\mp}(x)}.
\end{equation}
Substitution of the derivative (\ref{wf1a}) to the equation (\ref{pii2}) gives
\begin{equation}\label{pii2a}
\dfrac{1}{C(f'_{\mp}(x))^2}\mp\dfrac{2B}{Cf'_{\mp}(x)}\Psi^I_{\mp}+A(\Psi^I_{\mp})^2-1=0,
\end{equation}
or equivalently one obtains quadratic equation for $\Psi^I$
\begin{equation}\label{pii2b}
AC(f'_{\mp}(x))^2(\Psi^I_{\mp})^2\mp2Bf'_{\mp}(x)\Psi^I_{\mp}+1-C(f'(x))^2=0.
\end{equation}
The equation (\ref{pii2b}) possesses two solutions
\begin{equation}\label{psii}
  \Psi^I_{\mp}=\pm\dfrac{B}{ACf'_{\mp}(x)}\left(1+\dfrac{1}{ABC}\sqrt{B^2-AC+4AC^2(f'_{\mp}(x))^2}\right),
\end{equation}
where the derivative $f'_{\mp}(x)$ can be established straightforwardly
\begin{equation}\label{funct1}
 f'_{\mp}(x)=\pm\dfrac{B}{C}\left[\dfrac{x}{Ax^2-1}\mp\dfrac{B}{\sqrt{C+(B^2-AC)x^2}}\left(\dfrac{x^2}{Ax^2-1}-\dfrac{C}{B^2}\right)\right].
\end{equation}
Substitution of the derivative (\ref{funct1}) to the formula (\ref{psii}) gives
\begin{eqnarray}
&&\Psi^I_{\mp}=\left(\dfrac{Ax}{Ax^2-1}\mp\dfrac{AB}{\sqrt{C+(B^2-AC)x^2}}\left(\dfrac{x^2}{Ax^2-1}-\dfrac{C}{B^2}\right)\right)^{-1/2}\times\nonumber\\
&&\Bigg\{1+\dfrac{1}{AC}\Bigg[1-\dfrac{AC}{B^2}+\dfrac{4Ax^2}{(Ax^2-1)^2}+\dfrac{4AB}{C+(B^2-AC)x^2}\left(\dfrac{x^2}{Ax^2-1}-\dfrac{C}{B^2}\right)^2\pm\nonumber\\
&&\dfrac{ABx}{(Ax^2-1)\sqrt{C+(B^2-AC)x^2}}\left(\dfrac{x^2}{Ax^2-1}-\dfrac{C}{B^2}\right)\Bigg]^{1/2}\Bigg\}.
\end{eqnarray}

\subsection{The Unitary Three-Manifolds}\index{unitary three-manifolds}
The evolution operator (\ref{evoloper}) is in general non unitary, i.e. it can be shown by straightforward calculation that the condition
\begin{equation}
  U^\dag U=UU^\dag=\mathbf{1}_2,
\end{equation}
is broken. However, there are situations within the theory for which the evolution operator (\ref{evoloper}) is unitary. It can be proved easy that for the unitarity of $U$ the necessary and suff\/icient conditions are
\begin{eqnarray}
  \left(h-h_I\int_{h_I}^h\delta h'{\omega'}^2\right)\dfrac{\sin 2f[h,h_I]}{2f[h,h_I]}&=&0,\\
  \cos^2f[h,h_I]+\left(\int_{h_I}^h\delta h'{\omega'}^2\right)^2\left(\dfrac{\sin f[h,h_I]}{f[h,h_I]}\right)^2&=&1,\\
  \left[\left(\int_{h_I}^h\delta h'{\omega'}^2\right)^2-\left(h-h_I\right)^2\right]\left(\dfrac{\sin f[h,h_I]}{f[h,h_I]}\right)^2&=&0.
\end{eqnarray}
These equations possess two solutions. The f\/irst solution is trivial
\begin{equation}
f[h,h_I]=0,
\end{equation}
and corresponds to the initial data point
\begin{equation}
  h=h_I.
\end{equation}
In this situation the evolution operator is trivially equal to the unit $2\times2$ matrix, what in fact means that there is no evolution. The second solution, however, is non trivial
\begin{equation}
  f^2[h,h_I]=-(h-h_I)^2,
\end{equation}
and corresponds to the equation
\begin{equation}
  \omega^2=-1,\label{tacheq}
\end{equation}
which is associated with purely imaginary frequency and therefore also energy, i.e. tachyon\index{tachyon}. With using of the def\/inition (\ref{masqr}) the condition (\ref{tacheq}) generates the equation for the embedded three-dimensional space
\begin{equation}
  {^{(3)}}R=6(8\pi)^2h+2\Lambda+2\kappa\ell_P^2\varrho,\label{useq1}
\end{equation}
or equivalently with using of the second def\/inition (\ref{masqr1})
\begin{equation}
  K_{ij}K^{ij}-K^2=6(8\pi)^2h.\label{useq2}
\end{equation}
In this case the unitary evolution operator has the form
\begin{equation}\label{uoper}
  U[h,h_I]=\left[\begin{array}{cc}\cosh(h-h_I)&\sinh(h-h_I)\\-\sinh(h-h_I)&\cosh(h-h_I)\end{array}\right],
\end{equation}
and is the rotation matrix of the unitary Lie group $U(1)\cong SO(2)$, where the angle of the rotation is $i(h-h_I)$.
Hence in such a situation the classical scalar f\/ield $\Psi$ and its conjugate momentum f\/ield $\Pi_\Psi$ are \begin{eqnarray}
  \Psi[h,h_I]&=&\Psi^I\cosh(h-h_I)-\Pi_\Psi^I\sinh(h-h_I),\\
  \Pi_\psi[h,h_I]&=&\Psi^I\sinh(h-h_I)+\Pi_\Psi^I\cosh(h-h_I),
\end{eqnarray}
and the corresponding probability density equals to
\begin{eqnarray}
  \Omega[h,h_I]&=&\left(\Psi^I\right)^2\cosh^2(h-h_I)+\left(\Pi_\Psi^I\right)^2\sinh^2(h-h_I)-\\
  &-&2\Psi^I\Pi^I\sinh(h-h_I)\cosh(h-h_I).
\end{eqnarray}
In this manner the normalization condition (\ref{normcond1}) generates the equation for the initial data
\begin{equation}
\alpha(h_F-h_I)\left(\Pi_\Psi^I\right)^2-2\beta(h_F-h_I)\Psi^I\Pi^I+\gamma(h_F-h_I)\left(\Psi^I\right)^2=\dfrac{1}{h_F-h_I},\label{idequat}
\end{equation}
where the coeff\/icients-functions of $h_I$ are
\begin{eqnarray}
  \alpha(h_F-h_I)&=&\dfrac{\sinh(2(h_F-h_I))-2(h_F-h_I)}{4},\\
  \beta(h_F-h_I)&=&\dfrac{1}{2}\sinh^2(2(h_F-h_I)),\\
  \gamma(h_F-h_I)&=&\dfrac{\sinh(2(h_F-h_I))+2(h_F-h_I)}{4}.
\end{eqnarray}
Application of the basic def\/inition of initial data of the conjugate momentum $\Pi_\Psi^I=\dfrac{\delta\Psi_I}{\delta h_I}$ to the equation (\ref{idequat}) gives the differential equation for the initial data do the scalar f\/ield $\Psi_I$
\begin{equation}
\alpha(h_F-h_I)\left(\dfrac{\delta\Psi_I}{\delta h_I}\right)^2-2\beta(h_F-h_I)\Psi^I\dfrac{\delta\Psi_I}{\delta h_I}+\gamma(h_F-h_I)\left(\Psi^I\right)^2-\dfrac{1}{h_F-h_I}=0.\label{idequat}
\end{equation}
for f\/inite values of $h_F$ this equation is diff\/icult to solve straightforwardly. When the upper limit is inf\/inite, i.e. $h_F\rightarrow\infty$, then the third term in (\ref{idequat}) vanishes identically whereas the coeff\/icients $\alpha$, $\beta$, and $\gamma$ tends to inf\/inity. Then however, one can divide both sides of the equation (\ref{idequat}) by $\dfrac{1}{2}\sinh(2(h_F-h_I))$ and obtain the f\/inite limit
\begin{equation}
\left(\dfrac{\delta\Psi_I}{\delta h_I}\right)^2-2\Psi^I\dfrac{\delta\Psi_I}{\delta h_I}+\left(\Psi^I\right)^2=\left(\dfrac{\delta\Psi_I}{\delta h_I}-\Psi^I\right)^2=0.\label{idequat1}
\end{equation}
With using of the boundary condition $\Psi_I(h_0)=\Psi_0$, this equation can be solved immediately
\begin{equation}
\Psi_I=\Psi_0\exp\left\{h_I-h_0\right\}.
\end{equation}
Because of the equations (\ref{useq1}) and (\ref{useq2}) are related to the unitary operator of evolution (\ref{uoper}) we shall call their solutions \emph{the unitary three-manifolds}\index{unitary three-manifolds}. Interestingly, the equation (\ref{useq1}) suggests that in general the unitary three-manifolds are deformations of the three-manifolds def\/ined by the Ricci scalar-curvature proportional to determinant of an induced metric, i.e. to the global dimension
\begin{equation}
{^{(3)}}R=6(8\pi)^2h,\label{einus}
\end{equation}
and deformation is due to the cosmological constant\index{cosmological constant} $\Lambda$ and the energy density\index{energy density} $\varrho$ of Matter f\/ields\index{Matter f\/ields}. Because of the three-dimensional manifolds def\/ined by (\ref{einus}) are also the unitary three-manifolds we shall call them \emph{the global unitary three-manifolds}\index{unitary three-manifolds!global}.

\subsection{The Fourier Analysis}
The quantum gravity given by one-dimensional the Klein--Gordon equation \index{Klein--Gordon equation}(\ref{kgfeq})
\begin{eqnarray}
  &&\left(\dfrac{\delta^2}{\delta h^2}+\omega^2[h]\right)\Psi(h)=0,\label{kleing}\\
  &&\omega^2[h]=-\dfrac{1}{6(8\pi)^2}\dfrac{1}{h}\left({^{(3)}}\!R-2\Lambda-2\kappa\ell_P^2\varrho\right),
\end{eqnarray}
can be considered as the equation for the 3-dimensional scalar curvature ${^{(3)}\!R}$
\begin{equation}\label{kgfq2}
^{(3)}\!R=2\left(\Lambda+\kappa\ell_P^2\varrho\right)+6(8\pi)^2\varphi(h)h,
\end{equation}
where we have introduced the function
\begin{equation}
  \varphi(h)=\dfrac{1}{\Psi(h)}\dfrac{\delta^2\Psi(h)}{\delta{h^2}}.
\end{equation}
In the stationary case
\begin{equation}
\varrho\equiv0 \cap \Lambda\equiv0\quad\textrm{or}\quad\varrho=-\dfrac{\Lambda}{\kappa\ell_P^2},
\end{equation}
one obtains from (\ref{kgfq2}) that
\begin{equation}\label{kgfq3}
^{(3)}\!R=6(8\pi)^2\varphi_nh,
\end{equation}
where $\varphi_n$ are eigenvalues determined by the equation
\begin{equation}
  \dfrac{\delta^2\Psi}{\delta{h^2}}=\varphi_n\Psi.\label{eigeneq}
\end{equation}
If one wishes to consider the non-stationary situation then it is easy to see that
\begin{equation}
  \varphi(h)=\varphi_n-\dfrac{2}{6(8\pi)^2}\dfrac{\Lambda+\kappa\ell_P^2\varrho}{h}.
\end{equation}
One can apply the analytical form of the classical scalar f\/ield $\Psi$
\begin{equation}
\Psi=\sum_{n=-\infty}^{\infty}a_n(h-h_I)^n,
\end{equation}
to the eigenequation (\ref{eigeneq})
\begin{equation}
  \sum_{n=-\infty}^\infty(n+1)(n+2)a_{n+2}(h-h_I)^n=\varphi_n\sum_{n=-\infty}^{\infty}a_n(h-h_I)^n,
\end{equation}
and express eigenvalues $\varphi_n$ by the series coeff\/icients $a_n$
\begin{equation}
\varphi_n=(n+1)(n+2)\dfrac{a_{n+2}}{a_n}.\label{eigan}
\end{equation}
Because, however, the series coeff\/icients are def\/ined as
\begin{equation}
  a_n=\left.\dfrac{\delta^n\Psi}{\delta h^n}\right|_{h=h_I},
\end{equation}
one obtains from (\ref{eigan})
\begin{equation}
\varphi_n=(n+1)(n+2)\left.\dfrac{\dfrac{\delta^{n+2}}{\delta h^{n+2}}\Psi(h)}{\dfrac{\delta^n}{\delta h^n}\Psi(h)}\right|_{h=h_I}.\label{eigan1}
\end{equation}
In the light of the fact
\begin{equation}
\dfrac{\delta^{n+2}\Psi}{\delta h^{n+2}}=\dfrac{\delta^{n}}{\delta h^{n}}\left(\dfrac{\delta^{2}\Psi}{\delta h^{2}}\right)=-\dfrac{\delta^{n}}{\delta h^{n}}\left(\omega^2[h]\Psi(h)\right),
\end{equation}
where we have applied the equation (\ref{kleing}), one has
\begin{equation}
\varphi_n=-(n+1)(n+2)\left.\dfrac{\dfrac{\delta^{n}}{\delta h^{n}}\left(\omega^2[h]\Psi(h)\right)}{\dfrac{\delta^n}{\delta h^n}\Psi(h)}\right|_{h=h_I}.\label{vph}
\end{equation}

Let us assume that there are generalized Fourier transforms\index{Fourier transforms}
\begin{eqnarray}
  \widetilde{\Psi}(s)&=&\int\delta h e^{-2i\pi sh}\Psi(h),\label{ft2b1}\\
  \widetilde{m^2}(s)&=&\int\delta h e^{-2i\pi sh}m^2[h],\label{ft2b2}
\end{eqnarray}
as well as the inverted Fourier transforms\index{Fourier transforms}
\begin{eqnarray}
  \Psi(h)&=&\int\delta s e^{2i\pi sh}\widetilde{\Psi}(s),\label{ft2a1}\\
  m^2[h]&=&\int\delta s e^{2i\pi sh}\widetilde{m^2}(s).\label{ft2a2}
\end{eqnarray}
Applying the generalized Leibniz product formula\index{Leibniz product formula}
\begin{equation}
  \dfrac{\delta^n}{\delta h^n}\left(\omega^2[h]\Psi(h)\right)=\sum_{r=0}^n\binom{n}{r}\left(\dfrac{\delta^r}{\delta h^r}\omega^2[h]\right)\left(\dfrac{\delta^{n-r}}{\delta h^{n-r}}\Psi(h)\right),\label{leib}
\end{equation}
one obtains
\begin{eqnarray}
\dfrac{\dfrac{\delta^{n}}{\delta h^{n}}\left(\omega^2[h]\Psi(h)\right)}{\dfrac{\delta^n}{\delta h^n}\Psi(h)}&=&
\dfrac{1}{\dfrac{\delta^n}{\delta h^n}\Psi(h)}\sum_{r=0}^n\binom{n}{r}\left(\dfrac{\delta^r}{\delta h^r}\omega^2[h]\right)\left(\dfrac{\delta^{n-r}}{\delta h^{n-r}}\Psi(h)\right)=\nonumber\\
&=&\dfrac{\dfrac{\delta^n}{\delta h^n}\Psi(h)}{\dfrac{\delta^n}{\delta h^n}\Psi(h)}\sum_{r=0}^n\binom{n}{r}\left(\dfrac{\delta^r}{\delta h^r}\omega^2[h]\right)\left(\dfrac{\delta^{-r}}{\delta h^{-r}}\Psi(h)\right)=\nonumber\\
&=&\sum_{r=0}^n\binom{n}{r}\left(\dfrac{\delta^r}{\delta h^r}\omega^2[h]\right)\left(\dfrac{\delta^{-r}}{\delta h^{-r}}\Psi(h)\right).\label{quot}
\end{eqnarray}
In this manner using of the inverted Fourier transforms\index{Fourier transforms} (\ref{ft2a1}) and (\ref{ft2a2}) gives
\begin{eqnarray}
\dfrac{\delta^r}{\delta h^r}\omega^2[h]&=&\int\delta s e^{2i\pi sh}(2i\pi s)^r\widetilde{\omega^2}(s),\\
\dfrac{\delta^{-r}}{\delta h^{-r}}\Psi(h)&=&\int\delta s e^{2i\pi sh}(2i\pi s)^{-r}\widetilde{\Psi}(s),
\end{eqnarray}
and by this reason the formula (\ref{quot}) becomes
\begin{equation}
  \dfrac{\dfrac{\delta^n}{\delta h^n}\left(\omega^2[h]\Psi(h)\right)}{\dfrac{\delta^n}{\delta h^n}\Psi(h)}=\int\int\delta s\delta s'e^{2i\pi('+s')h}\left[\sum_{r=0}^n\binom{n}{r}\left(\dfrac{s}{s'}\right)^r\right]\widetilde{\omega^2}(s)\widetilde{\Psi}(s').\label{quot1}
\end{equation}
Applying the standard summation procedure
\begin{equation}
  \sum_{r=0}^n\binom{n}{r}x^r=(1+x)^n,
\end{equation}
one obtains
\begin{equation}
  \dfrac{\dfrac{\delta^n}{\delta h^n}\left(\omega^2[h]\Psi(h)\right)}{{\dfrac{\delta^n}{\delta h^n}\Psi(h)}}=\int\int\delta{s}\delta{s'}e^{2i\pi(s'+s)h}\left(1+\dfrac{s}{s'}\right)^n\widetilde{\omega^2}(s)\widetilde{\Psi}(s').\label{quot2}
\end{equation}
By this reason the eigenvalues (\ref{vph}) are
\begin{equation}\label{vp1}
\varphi_n=-(n+1)(n+2)\left.\int\int\delta{s}\delta{s'}e^{2i\pi(s+s')h}\left(1+\dfrac{s}{s'}\right)^n\widetilde{\omega^2}(s)\widetilde{\Psi}(s')\right|_{h=h_I}.
\end{equation}
Applying the Fourier transforms\index{Fourier transforms} (\ref{ft2b1}) and (\ref{ft2b2}) one receives
\begin{equation}
\widetilde{\omega^2}(s)\widetilde{\Psi}(s')=\int\int\delta h\delta h'e^{-2i\pi(sh+s'h')}\omega^2[h]\Psi(h'),
\end{equation}
and by this reason one obtains f\/inally
\begin{eqnarray}
\varphi_n&=&-\left.\int\int\delta h\delta h'\mathcal{G}_n(h-h')\omega^2[h]\Psi(h')\right|_{h=h_I,h'=h_I}=\label{vp2}\\
&=&-\left.\int\int\delta h\delta h'\mathcal{G}_n(h-h')\dfrac{\delta^2\Psi(h)}{\delta h^2}\dfrac{\Psi(h')}{\Psi(h)}\right|_{h=h_I,h'=h_I}=\\
&=&-\left.\int\int\delta h\delta h'\mathcal{G}_n(h-h')\dfrac{\delta\Pi_\Psi(h)}{\delta h}\dfrac{\Psi(h')}{\Psi(h)}\right|_{h=h_I,h'=h_I},
\end{eqnarray}
where in the second line we have used equations of motion, and in the third line we have applied def\/inition of the conjugate momentum f\/ield. The kernel $\mathcal{G}(h-h')$ is given by the relation
\begin{equation}
  \mathcal{G}_n(h-h')=(n+1)(n+2)\int\int\delta s\delta s'e^{2i\pi s'(h-h')}\left(1+\dfrac{s'}{s}\right)^n.\label{kern1}
\end{equation}
Estimation of the kernel (\ref{kern1}) is the crucial element of the proposed analysis. To make it consistently let us determine the range of $s$ and $s'$. In quantum mechanics the Fourier analysis transforms theory to the momentum space representation. Therefore, $s$ and $s'$ are the momenta conjugated to $h$ and $h'$, respectively.

The double integral in the kernel (\ref{kern1}) can be transformed as
\begin{equation}
\int\delta s'e^{2i\pi s'(h-h')}\left[\int\delta s\left(1+\dfrac{s}{s'}\right)^n\right],\label{doubint}
\end{equation}
and therefore one sees that the variable $s$ is \emph{internal}, while the variable $s'$ is \emph{external}. Let us take \emph{ad hoc} the f\/inite range of the internal variable $s\in[0,S]$ and the inf\/inite range of the external variable $s'\in[0,\infty]$. The internal integral can be easy computed
\begin{eqnarray}
\int\delta s\left(1+\dfrac{s}{s'}\right)^n&=&s'\int\delta\left(1+\dfrac{s}{s'}\right)\left(1+\dfrac{s}{s'}\right)^n=\\
&=&s'\int_{t=1+\frac{s}{s'}}\delta t t^n=\dfrac{s'}{n+1}\left(1+\dfrac{S}{s'}\right)^{n+1},
\end{eqnarray}
and by this reason the double integral (\ref{doubint}) becomes
\begin{equation}
\dfrac{1}{n+1}\int\delta s'e^{2i\pi s'(h-h')}s'\left(1+\dfrac{S}{s'}\right)^{n+1}.\label{doubint1}
\end{equation}
In the range $s'\in[0,\infty]$ the integral in (\ref{doubint1}) converges for $n<1$
\begin{equation}
\int\delta s'e^{2i\pi s'(h-h')}s'\left(1+\dfrac{S}{s'}\right)^{n+1}=-\dfrac{\Gamma(1-n)}{4\pi^2(h-h')^2}U\left(-1-n,-1,-2i\pi S\left(h-h'\right)\right),\label{integrum}
\end{equation}
where $\Gamma(z)$ is the Euler gamma-function\index{Euler gamma-function}
\begin{equation}
  \Gamma(z)=\int_0^\infty t^{z-1}e^{-t} dt,
\end{equation}
and $U\left(a,b,z\right)$ is the Tricomi conf\/luent hypergeometric function\index{hypergeometric function!Tricomi conf\/luent} (See e.g. the Ref. \cite{spec} for basic knowledge about special functions)
\begin{equation}
U(a,b,z)=\dfrac{\Gamma(1-b)}{\Gamma(a-b+1)}M(a,b,z)+\dfrac{\Gamma(b-1)}{\Gamma(a)}z^{1-b}M(a-b+1,2-b,z),
\end{equation}
where $M(a,b,z)$ is the Kummer conf\/luent hypergeometric function\index{hypergeometric function!Kummer conf\/luent}
\begin{equation}
M(a,b,z)=\sum_{n=0}^\infty\dfrac{(a)_n}{(b)_n}\dfrac{z^n}{n!}=1+\dfrac{\Gamma(b)}{\Gamma(a)}\sum_{n=1}^\infty\dfrac{\Gamma(a+n)}{\Gamma(b+n)}\dfrac{z^n}{n!},
\end{equation}
where $(\alpha)_n$ are the Pochhammer symbols\index{Pochhammer symbols}
\begin{equation}
  (\alpha)_n=\dfrac{\Gamma(\alpha+n)}{\Gamma(\alpha)}\quad,\quad(\alpha)_0=1,
\end{equation}
and for $\Re a>0$ there is the integral relation
\begin{equation}
\dfrac{(a)_n}{(b)_n}=\dfrac{\Gamma(b)}{\Gamma(a)\Gamma(b-a)}\int_0^1t^{a-1+n}(1-t)^{b-a-1}dt.
\end{equation}
Applying the values $a=1-n>0$, $b=3$, $z=-2i\pi S(h-h')$ to the Kummer transformation\index{Kummer transformation}
\begin{equation}
  U(a,b,z)=z^{1-b}U(1+a-b,2-b,z),
\end{equation}
one receives
\begin{equation}
U\left(-1-n,-1,-2i\pi S\left(h-h'\right)\right)=-4\pi^2S^2 (h-h')^2U\left(1-n,3,-2i\pi S\left(h-h'\right)\right),
\end{equation}
and by this reason the integral (\ref{integrum}) becomes
\begin{equation}
\int\delta s'e^{2i\pi s'(h-h')}s'\left(1+\dfrac{S}{s'}\right)^{n+1}=S^2\Gamma(1-n)U\left(1-n,3,-2i\pi S\left(h-h'\right)\right).
\end{equation}
Because of now $1-n>0$ the Tricomi conf\/luent hypergeometric functions\index{hypergeometric function!Tricomi conf\/luent} can be computed in the integral representation
\begin{equation}
U(a,b,z)=\dfrac{1}{\Gamma(a)}\int_0^\infty e^{-zt}t^{a-1}(1+t)^{b-a-1}dt.
\end{equation}
In this manner the kernel (\ref{kern1}) (for $n<1$) is given by the formula
\begin{equation}
\mathcal{G}_n(h-h')=S^2(n+2)\Gamma(1-n)U\left(1-n,3,-2i\pi S\left(h-h'\right)\right),
\end{equation}
and consequently the eigenvalue (\ref{eigan}) can be evaluated as
\begin{equation}
  \varphi_n=-S^2(n+2)\Gamma(1-n)\left.\int\int\delta h\delta h'U\left(1-n,3,-2i\pi S\left(h-h'\right)\right)\Upsilon(h,h')\right|_{h=h_I,h'=h_I},
\end{equation}
where we have introduced the symbol $\Upsilon(h,h')$
\begin{equation}
  \Upsilon(h,h')=\dfrac{\Psi(h')}{\Psi(h)}\dfrac{\delta^2\Psi(h)}{\delta h^2}=\dfrac{\Psi(h')}{\Psi(h)}\dfrac{\delta\Pi_\Psi(h)}{\delta h},
\end{equation}
which can be straightforwardly established when both wave function $\Psi(h)$ and its second derivative $\dfrac{\delta^2\Psi(h)}{\delta h^2}$, or equivalently derivative of its conjugate momentum f\/ield $\dfrac{\delta\Pi_\Psi(h)}{\delta h}$, are given explicitly.

\subsection{Quantum Correlations}\index{quantum correlations}
With using of the matrices (\ref{mono}) and (\ref{sqx}), and the relation (\ref{phi}) one derives the quantum f\/ield
\begin{equation}\label{field}
  \hat{\Psi}(h)=\sqrt{\dfrac{\omega_I}{8}}\dfrac{1}{\omega}\left(\exp\left\{-i\int_{h_I}^h\omega'\delta h'\right\}\textsf{G}_I+\exp\left\{i\int_{h_I}^h \omega'\delta h'\right\}\textsf{G}_I^\dagger\right).
\end{equation}
Let us take into account the $n$-particle one-point quantum states determined as
\begin{equation}
|h,n\rangle\equiv\hat{\Psi}^n\rv=\left(\sqrt{\dfrac{\omega_I}{8}}\dfrac{1}{\omega}\exp\left\{\int_{h_I}^h\omega'\delta h'\right\}\right)^n\textsf{G}^{\dagger n}_I\rv,
\end{equation}
yields two-point correlators $\cor_{n'n}(h',h)\equiv\langle n',h'|h,n\rangle$ or explicitly
\begin{eqnarray}\label{gencor}  \cor_{n'n}(h',h)&=&\left(\dfrac{\omega_I}{8}\right)^{(n'+n)/2}\exp\left\{i\left(n'\int_{h'}^{h_I}+n\int_{h_I}^{h}\right)\omega''\delta h''\right\}\times\nonumber\\
  &\times&\dfrac{\lv\textsf{G}_I^{n'}\textsf{G}^{\dagger n}_I\rv}{{{\omega'}^{n'}\omega^n}}.\label{cor0}
\end{eqnarray}
Basically one obtains
\begin{eqnarray}
&&\cor_{00}(h,h)=\cor_{00}(h',h)=\cor_{00}(h_I,h_I)=\vv,\\
&&\cor_{11}(h_I,h_I)=\dfrac{1}{8\omega_I},\\
&&\dfrac{\cor_{n'n}(h_I,h_I)}{\left[\cor_{11}(h_I,h_I)\right]^{(n'+n)/2}}=\lv\textsf{G}_I^{n'}\textsf{G}^{\dagger n}_I\rv,
\end{eqnarray}
and by elementary algebraic manipulations one receives
\begin{eqnarray}
\!\!\!\!\!\!\!\!\!\!\!\!\!\!\!
&&\cor_{11}(h',h)=\dfrac{\sqrt{\strut{\cor_{11}(h',h')\cor_{11}(h,h)}}}{\cor_{11}(h_I,h_I)}\exp\left\{i\int_{h'}^{h}\omega''\delta h''\right\},\\
\!\!\!\!\!\!\!\!\!\!\!\!\!\!\!
&&\dfrac{\cor_{nn}(h',h)}{\cor_{00}(h_I,h_I)}=\left[\dfrac{\cor_{11}(h',h)}{\cor_{00}(h_I,h_I)}\right]^n,\\
\!\!\!\!\!\!\!\!\!\!\!\!\!\!\!
&&\dfrac{\cor_{11}(h,h)}{\cor_{00}(h_I,h_I)}=\left(\dfrac{m_I}{m}\right)^2\cor_{11}(h_I,h_I).\label{cor2}
\end{eqnarray}
Straightforwardly from (\ref{cor2}) one can relate a size scale with quantum correlations
\begin{eqnarray}
\lambda=\dfrac{\omega_I}{\omega}=\sqrt{\strut{\dfrac{\cor_{11}(h,h)}{\cor_{11}(h_I,h_I)\cor_{00}(h_I,h_I)}}},\label{lam}
\end{eqnarray}
and consequently one receives the formulas
\begin{eqnarray}
&&\dfrac{\cor_{n'n}(h,h)}{\cor_{n'n}(h_I,h_I)}=\lambda^{n'+n}\exp\left\{-i(n'-n)\int_{h_I}^h\omega''\delta h''\right\},\label{cor4}\\
&&\dfrac{\cor_{11}(h',h)}{\cor_{00}(h_I,h_I)\cor_{11}(h_I,h_I)}=\lambda'\lambda\exp\left\{i\int_{h'}^{h}\omega''\delta h''\right\},\\
&&\dfrac{\cor_{nn}(h',h)}{\cor_{00}(h_I,h_I)}=\lambda'^n\lambda^n[\cor_{11}(h_I,h_I)]^n\exp\left\{in\int_{h'}^{h}\omega''\delta h''\right\}.
\end{eqnarray}
A whole information about the quantum gravity is contained in the parameters of the theory, i.e. $m$, $\lambda$, and the initial data $m_I$. It is evident that the quantum correlations are strictly determined by these fundamental quantities only. In other words measurement of quantum correlations can be used for deduction of values of the fundamental parameters of the theory.

The conclusions presented in this section have purely formal character, however, they show manifestly a general feature of the proposed theory of quantum gravity. These conclusions are partial, but they show a non trivial both physical and mathematical implications following from the theory of quantum gravity. Let us see more consequences of the theory.

\chapter{The Invariant Global Dimension}
\section{The Invariant Global Quantum Gravity}
Let us included explicitly presence of Matter f\/ields\index{Matter f\/ields}, and extend the global one-dimensional wave function $\Psi(h)$ to a functional $\Psi[h,\phi]$, which we shall call \emph{extended global wave function}\index{extended global wave function}. Still we take the global dimension $h=\det h_{ij}$.

The theory of quantum gravity proposed in the previous chapter can be rewritten as
\begin{equation}\label{kgf}
\left(\dfrac{\delta^2}{\delta{h^2}}+V_{eff}[h, \phi]\right)\Psi[h,\phi]=0.
\end{equation}
where $V_{eff}$ is \emph{the (effective) gravitational potential}\index{gravitational potential!effective}
\begin{equation}\label{eff}
  V_{eff}\equiv \dfrac{1}{6(8\pi)^2}\left(-\dfrac{{^{(3)}\!R}}{h}+\dfrac{2\Lambda}{h}+\dfrac{2\kappa\ell_P^2}{h}\varrho[\phi]\right).
\end{equation}
The manifestly singular behavior $\sim1/h$ of the potential (\ref{eff}), however, can be regularized by the suitable change of variables
\begin{equation}
h\rightarrow \xi=\xi[h],\label{kgfrd}
\end{equation}
which generates the adequate Jacobi formula
\begin{equation}
  \delta\xi=\left(\dfrac{\delta \xi}{\delta h}\right)hh^{ij}\delta h_{ij}.
\end{equation}
The dimension $\xi[h]$, as a functional of the global dimension $h$, is also diffeoinvariant. Applying the change of variables (\ref{kgfrd}) within the equation (\ref{kgf}) one obtains
\begin{equation}\label{kgfr}
\left\{\left(\dfrac{\delta \xi}{\delta h}\right)^2\dfrac{\delta^2}{\delta{\xi^2}}+V_{eff}\left[\xi,\phi\right]\right\}\Psi\left[\xi,\phi\right]=0,
\end{equation}
and therefore for all nonsingular situations $\dfrac{\delta \xi}{\delta h}\neq0$ it can be rewritten in more convenient form
\begin{equation}\label{kgfr}
\left\{\dfrac{\delta^2}{\delta{\xi^2}}+V[\xi,\phi]\right\}\Psi\left[\xi,\phi\right]=0,
\end{equation}
where the potential $V[\xi,\phi]$ is
\begin{equation}
  V[\xi,\phi]=\left(\dfrac{\delta \xi}{\delta h}\right)^{-2}V_{eff}\left[\xi,\phi\right].
\end{equation}
We shall call it \emph{the generalized gravitational potential}\index{gravitational potential!generalized}. In fact, the choice of the invariant dimension $\xi$ is a kind of the choice of a gauge for the theory of quantum gravity.

Naturally, the generic gauge is the global dimension $\xi[h]\equiv h$. Another situations can be generated straightforwardly from this fundamental case. There is a lot of possible choices of gauge for quantum gravity. However, note that the particular choice
\begin{equation}
  \xi=\dfrac{1}{4\pi}\sqrt{\strut{\dfrac{h}{6}}},\label{xi1}
\end{equation}
which is associated to the measure
\begin{equation}
  \delta\xi=\dfrac{1}{8\pi}\sqrt{\strut{\dfrac{h}{6}}}h^{ij}\delta h_{ij}\label{xi2},
\end{equation}
removes the singularity $1/h$ present in the gravitational potential\index{gravitational potential!effective} $V_{eff}\left[h,\phi\right]$ (\ref{eff}), and consequently the equation (\ref{kgfr}) reads
\begin{equation}\label{eqn}
\left\{\dfrac{\delta^2}{\delta{\xi^2}}-\left({^{(3)}\!R[\xi]}-2\Lambda-2\kappa\ell_P^2\varrho[\phi]\right)\right\}\Psi\left[\xi,\phi\right]=0.
\end{equation}
The appropriate normalization condition should be chosen as
\begin{equation}\label{norm}
  \int\left|\Psi\left[\xi,\phi\right]\right|^2\delta\mu(\xi,\phi)=1,
\end{equation}
where $\delta\mu(\xi,\phi)=\delta\xi\delta\phi.$ is the invariant product functional measure. Note that similarly as $\delta h$ the measure $\delta \sqrt{h}$ is also the Lebesgue--Stieltjes\index{Lebesgue--Stieltjes integral} (Radon) type integral measure, which can be transformed to the Riemann--Lebesgue measure\index{Riemann--Lebesgue measure} over space-time. In the general case $h=h(x_0,x_1,x_2,x_3)$
\begin{equation}
\delta \sqrt{h} = \dfrac{\partial^4 \sqrt{h}}{\partial x_0\partial x_1\partial x_2\partial x_3}d^4x.
\end{equation}
By the special role of the change of variables (\ref{xi1}) we shall call this dimension \emph{invariant global dimension}\index{invariant global dimension}, and the theory of quantum gravity (\ref{eqn}) will be called \emph{the invariant global quantum gravity}\index{invariant global quantum gravity}.

\section{The One-Dimensional Dirac Equation}\label{sec:3}\index{Dirac equation!one-dimensional}
The equation (\ref{kgfr}) can be derived as the Euler--Lagrange equations of motion\index{Euler--Lagrange equations of motion} by the f\/ield theoretical variational principle $\delta S[\Psi]=0$ applied to the action functional
\begin{eqnarray}
S[\Psi]&=&-\dfrac{1}{2}\int \delta \xi\delta\phi\Psi[\xi,\phi]\left(\dfrac{\delta^2}{\delta{\xi^2}}+V[\xi,\phi]\right)\Psi[\xi,\phi]=\nonumber\\
&=&-\dfrac{1}{2}\int\delta\phi\Psi[\xi,\phi]\dfrac{\delta\Psi[\xi,\phi]}{\delta\xi}+\nonumber\\
&+&\dfrac{1}{2}\int\delta\xi\delta\phi \left\{\left(\dfrac{\delta \Psi[\xi,\phi]}{\delta \xi}\right)^2+V[\xi,\phi]\Psi^2[\xi,\phi]\right\},\label{actionf}
\end{eqnarray}
where the integration by parts was applied. One can choose the coordinate system in the Wheeler superspace \index{Wheeler superspace}by the condition of vanishing of the material surface term
\begin{equation}\label{choice}
-\dfrac{1}{2}\int\delta\phi\Psi[\xi,\phi]\dfrac{\delta\Psi[\xi,\phi]}{\delta\xi}=0,
\end{equation}
which we shall call \emph{the material coordinate system}\index{material coordinate system}. In the material coordinates the action functional (\ref{actionf}) is reduced to the action of the Euclidean f\/ield theory
\begin{equation}
S[\Psi]\equiv\int\delta \xi\delta\phi L\left[\Psi[\xi,\phi],\dfrac{\delta \Psi[\xi,\phi]}{\delta \xi}\right],
\end{equation}
where the Euclidean Lagrangian has the form
\begin{equation}
  L\left[\Psi[\xi,\phi],\dfrac{\delta \Psi[\xi,\phi]}{\delta \xi}\right]=\dfrac{1}{2}\left(\dfrac{\delta \Psi[\xi,\phi]}{\delta \xi}\right)^2+\dfrac{V[\xi,\phi]}{2}\Psi^2[\xi,\phi],
\end{equation}
and the corresponding canonical momentum conjugated to the classical scalar-valued f\/ield $\Psi^[\xi,\phi]$ is
\begin{equation}\label{1}
  \Pi_\Psi[\xi,\phi]=\dfrac{\partial L}{\partial \left(\dfrac{\delta \Psi[\xi,\phi]}{\delta \xi}\right)}=\dfrac{\delta \Psi[\xi,\phi]}{\delta \xi}.
\end{equation}
Therefore the choice of the coordinate system (\ref{choice}) actually means the choice of the orthogonal coordinates in the space f\/ield-theoretic phase space $(\Psi[\xi,\phi],\Pi_{\Psi}[\xi,\phi])$
\begin{equation}\label{coord}
  \forall \xi,\phi\in \Sigma(\xi,\phi): \int\delta\phi\Psi[\xi,\phi]\Pi_{\Psi}[\xi,\phi]=0,
\end{equation}
where by $\Sigma(\xi,\phi)$ we denoted the midisuperspace\index{midisuperspace} strata of the Wheeler superspace\index{Wheeler superspace}. With using of the conjugate momentum (\ref{1}) the equation (\ref{kgfr}) can be rewritten in the form
\begin{equation}\label{2}
  \dfrac{\delta\Pi_\Psi[\xi,\phi]}{\delta \xi}+V[\xi,\phi]\Psi[\xi,\phi]=0,
\end{equation}
and together with the equation (\ref{1}) creates the Hamilton canonical equations of motion\index{Hamilton canonical equations of motion}, yielding the appropriate one-dimensional Dirac equation\index{Dirac equation!one-dimensional}
\begin{equation}\label{dira}
 \left(-i\gamma\dfrac{\delta}{\delta \xi}-M[\xi,\phi]\right)\Phi[\xi,\phi]=0,
\end{equation}
where $\Phi[\xi,\phi]$ is the two-component classical f\/ield
\begin{equation}
  \Phi[\xi,\phi]=\left[\begin{array}{c}\Pi_\Psi[\xi,\phi]\\ \Psi[\xi,\phi]\end{array}\right],
\end{equation}
$M[\xi,\phi]$ is the mass matrix of this f\/ield
\begin{equation}
  M[\xi,\phi]=\left[\begin{array}{cc}1&0\\0&V[\xi,\phi]\end{array}\right],
\end{equation}
and the $\gamma$-matrices algebra
\begin{eqnarray}
  \gamma&=&\left[\begin{array}{cc}0&-i\\i&0\end{array}\right]\equiv\sigma_y,\\
  \gamma^2&=&\left[\begin{array}{cc}1&0\\0&1\end{array}\right]\equiv\mathbf{I}_2,
\end{eqnarray}
that in itself creates the Clifford algebra \index{Clifford algebra!$\mathcal{C}\ell_{2}(\mathbb{C})$}$\mathcal{C}\ell_{2}(\mathbb{C})$
\begin{eqnarray}
  \left\{\gamma,\gamma\right\}&=&2\mathbf{I}_2.
\end{eqnarray}

\section{The Cauchy-Like Wave Functionals}\label{sec:4}
The one-dimensional Dirac equation \index{Dirac equation!one-dimensional}(\ref{dira}) can be rewritten in the form of the Schr\"odinger equation\index{Schr\"odinger equation}
\begin{equation}\label{evol}
  i\dfrac{\delta\Phi[\xi,\phi]}{\delta \xi}=H[\xi,\phi]\Phi[\xi,\phi],
\end{equation}
where the Hamiltonian is
\begin{equation}
  H[\xi,\phi]=i\left[\begin{array}{cc}0&-V[\xi,\phi]\\1&0\end{array}\right].
\end{equation}
Solution of the $\xi$-evolution (\ref{evol}) can be written out straightforwardly
\begin{equation}
  \Phi[\xi,\phi]=U[\xi,\phi]\Phi[\xi^I,\phi],
\end{equation}
where $\Phi[\xi^I,\phi]$ is an initial data vector with respect to $\xi$ only, and $U[\xi,\phi]$ is the operator of $\xi$-evolution
\begin{eqnarray}
U=\exp\left\{-i\int_{\Sigma(\xi)}\delta\xi' H[\xi',\phi]\right\}=\exp\left\{-i\Omega(\xi,\phi) \langle H\rangle(\xi,\phi)\right\},
\end{eqnarray}
where $\Sigma(\xi)$ is the f\/inite integration region in the subset of midisuperspace\index{midisuperspace}, which we shall call \emph{$\xi$-space}\index{$\xi$-space}, $\Omega=V\left(\Sigma(\phi,\xi)\right)$ is the volume of full conf\/iguration space, and $\langle H\rangle(\phi)$ is the Hamiltonian averaged on midisuperspace\index{midisuperspace}
\begin{eqnarray}\label{formulas}
\Omega(\xi,\phi)&=&\int_{\Sigma(\xi,\phi)}\delta\xi'\delta\phi',\\
\langle H\rangle(\xi,\phi)&=&\dfrac{1}{\Omega(\xi,\phi)}\int_{\Sigma(\xi)}\delta\xi' H[\xi',\phi],
\end{eqnarray}
where $\Sigma(\xi,\phi)$ is the midisuperspace\index{midisuperspace}
\begin{equation}
\Sigma(\xi,\phi)=\Sigma(\xi)\times\Sigma(\phi),
\end{equation}
which is assumed to be f\/inite integration region. Explicitly one obtains
\begin{eqnarray}
  \!\!\!\!\!\!\!\!\!\!\!\!\!\!\!
  &&U[\xi,\phi]=\mathbf{I}_2\cos\left[\Omega(\xi,\phi)\sqrt{\strut{\langle V\rangle(\xi,\phi)}}\right]+\nonumber\\
  \!\!\!\!\!\!\!\!\!\!\!\!\!\!\!
  &&+\left[\begin{array}{cc}0&-\sqrt{\strut{\langle V\rangle(\xi,\phi)}}\\ \dfrac{1}{\sqrt{\strut{\langle V\rangle(\xi,\phi)}}}&0\end{array}\right]\sin\left[\Omega(\xi,\phi)\sqrt{\strut{\langle V\rangle(\xi,\phi)}}\right],
\end{eqnarray}
where we have introduced averaged generalized gravitational potential\index{gravitational potential!generalized}
\begin{equation}
  \langle V\rangle(\xi,\phi)=\dfrac{1}{\Omega(\xi,\phi)}\int_{\Sigma(\xi)}\delta\xi'V[\xi',\phi].
\end{equation}
Elementary algebraic manipulations yield the extended global wave functional
\begin{eqnarray}\label{wfun}
  \Psi[\xi,\phi]&=&\Psi[\xi^I,\phi]\cos\left[\Omega(\xi,\phi)\sqrt{\strut{\langle V\rangle(\xi,\phi)}}\right]+\nonumber\\
  &+&\Pi_\Psi[\xi^I,\phi]\dfrac{\sin\left[\Omega(\xi,\phi)\sqrt{\strut{\langle V\rangle(\xi,\phi)}}\right]}{\sqrt{\strut{\langle V\rangle(\xi,\phi)}}},
\end{eqnarray}
and the canonical conjugate momentum as the solution is
\begin{eqnarray}\label{pfun}
  \Pi_\Psi[\xi,\phi]&=&\Pi_\Psi[\xi^I,\phi]\cosh\left[\Omega(\xi,\phi)\sqrt{\strut{\langle V\rangle(\xi,\phi)}}\right]+\nonumber\\
  &-&\Psi[\xi^I,\phi]\sqrt{\strut{\langle V\rangle(\xi,\phi)}}\sinh\left[\Omega(\xi,\phi)\sqrt{\strut{\langle V\rangle(\xi,\phi)}}\right],
\end{eqnarray}
where $\Psi[\xi^I,\phi]$ and $\Pi_\Psi[\xi^I,\phi]$ are initial data with respect to $\xi$ only. Applying, however, the equation (\ref{1}) to (\ref{pfun}) one obtains the relation
\begin{eqnarray}
\!\!\!\!\!\!\!\!\!\!\!\!\!\!\!\!\!\!\!\!
&&\Pi_\Psi[\xi,\phi]=\dfrac{\Pi_\Psi[\xi^I,\phi]}{\sqrt{\langle V\rangle(\xi,\phi)}}\dfrac{\delta}{\delta\xi}\left[\Omega(\xi,\phi)\sqrt{\langle V\rangle(\xi,\phi)}\right]\cos\left[\Omega(\xi,\phi)\sqrt{\langle V\rangle(\xi,\phi)}\right]\nonumber\\
\!\!\!\!\!\!\!\!\!\!\!\!\!\!\!\!\!\!\!\!
&&-\Bigg\{\Psi[\xi^I,\phi]\dfrac{\delta}{\delta\xi}\left[\Omega(\xi,\phi)\sqrt{\langle V\rangle(\xi,\phi)}\right]-\Pi_\Psi[\xi^I,\phi]\dfrac{\delta}{\delta\xi}\left[\dfrac{1}{\sqrt{\langle V\rangle(\xi,\phi)}}\right]\Bigg\}\nonumber\\
\!\!\!\!\!\!\!\!\!\!\!\!\!\!\!\!\!\!\!\!
&&\times\sin\left[\Omega(\xi,\phi)\sqrt{\langle V\rangle(\xi,\phi)}\right],\label{pi1}
\end{eqnarray}
which after calculation of the functional derivatives
\begin{equation}
  \dfrac{\delta}{\delta\xi}\left[\Omega(\xi,\phi)\sqrt{\langle V\rangle(\xi,\phi)}\right]=\dfrac{\sqrt{\langle V\rangle(\xi,\phi)}}{2}\left(\dfrac{\delta\Omega(\xi,\phi)}{\delta\xi}+1\right),
\end{equation}
and
\begin{equation}
\dfrac{\delta}{\delta\xi}\left[\dfrac{1}{\sqrt{\langle V\rangle(\xi,\phi)}}\right]=\dfrac{1}{2}\dfrac{1}{\Omega(\xi,\phi)\sqrt{\langle V\rangle(\xi,\phi)}}\left(\dfrac{\delta\Omega(\xi,\phi)}{\delta\xi}-1\right)
\end{equation}
and using them within the formula (\ref{pi1}) yields
\begin{eqnarray}\label{pi2}
\!\!\!\!\!\!\!\!\!\!\!\!\!\!\!\!\!\!\!\!\!\!\!\!\!\!\!\!\!\!\!\!\!\!\!
&&\Pi_\Psi[\xi,\phi]=\Pi_\Psi[\xi^I,\phi]\dfrac{1}{2}\left(\dfrac{\delta\Omega(\xi,\phi)}{\delta\xi}+1\right)\cos\left[\Omega(\xi,\phi)\sqrt{\langle V\rangle(\xi,\phi)}\right]\nonumber\\
\!\!\!\!\!\!\!\!\!\!\!\!\!\!\!\!\!\!\!\!\!\!\!\!\!\!\!\!\!\!\!\!\!\!\!
&&-\Bigg[\Psi[\xi^I,\phi]\dfrac{\sqrt{\langle V\rangle(\xi,\phi)}}{2}\left(\dfrac{\delta\Omega(\xi,\phi)}{\delta\xi}+1\right)\nonumber\\
\!\!\!\!\!\!\!\!\!\!\!\!\!\!\!\!\!\!\!\!\!\!\!\!\!\!\!\!\!\!\!\!\!\!\!
&&-\dfrac{\Pi_\Psi[\xi^I,\phi]}{2\Omega(\xi,\phi)\sqrt{\langle V\rangle(\xi,\phi)}}\left(\dfrac{\delta\Omega(\xi,\phi)}{\delta\xi}-1\right)\Bigg]\sin\left[\Omega(\xi,\phi)\sqrt{\langle V\rangle(\xi,\phi)}\right].
\end{eqnarray}
After comparison with (\ref{pfun}) one obtains the equations
\begin{equation}
\dfrac{1}{2}\left(\dfrac{\delta\Omega(\xi,\phi)}{\delta\xi}+1\right)=1,\label{seq}
\end{equation}
and
\begin{equation}\label{seq1}
\Psi[\xi^I,\phi]\dfrac{1}{2}\left(\dfrac{\delta\Omega(\xi,\phi)}{\delta\xi}+1\right)-\dfrac{\Pi_\Psi[\xi^I,\phi]}{\Omega(\xi,\phi)\langle V\rangle(\xi,\phi)}\dfrac{1}{2}\left(\dfrac{\delta\Omega(\xi,\phi)}{\delta\xi}-1\right)=\Psi[\xi^I,\phi],
\end{equation}
The equation (\ref{seq}) yields the relation
\begin{equation}
\dfrac{\delta\Omega}{\delta\xi}=1=\int_{\Sigma(\phi)}\delta\phi',\label{intphi}
\end{equation}
where the last integral arises by the formula (\ref{formulas}). Application of the result (\ref{intphi}) to the equation (\ref{seq1}) leads to the self-consistent identity $\Psi[\xi^I,\phi]=\Psi[\xi^I,\phi]$. Such a situation means that the volume of midisuperspace\index{midisuperspace} $\Omega(\xi,\phi)$ is $\phi$-invariant
\begin{equation}
  \Omega(\xi,\phi)=\int_{\Sigma(\xi,\phi)}\delta\xi'\delta\phi'=\int_{\Sigma(\phi)}\delta\phi'\int_{\Sigma(\xi)}\delta\xi'=\int_{\Sigma(\xi)}\delta\xi'=\Omega(\xi),
\end{equation}
i.e. does not depend on Matter f\/ields\index{Matter f\/ields}.

Directly from (\ref{wfun}) the probability density can be deduced easily as
\begin{eqnarray}
  |\Psi[\xi,\phi]|^2&=&(\Psi[\xi^I,\phi])^2\cos^2\left[\Omega(\xi)\sqrt{\langle V\rangle(\xi,\phi)}\right]\nonumber\\
  &+&(\Pi_\Psi[\xi^I,\phi])^2\left(\dfrac{\sin\left[\Omega(\xi)\sqrt{\langle V\rangle(\xi,\phi)}\right]}{\sqrt{\langle V\rangle(\xi,\phi)}}\right)^2\nonumber\\
  &+&\Psi[\xi^I,\phi]\Pi_\Psi[\xi^I,\phi]\dfrac{\sin\left[2\Omega(\xi)\sqrt{\langle V\rangle(\xi,\phi)}\right]}{\sqrt{\langle V\rangle(\xi,\phi)}},
\end{eqnarray}
and in the light of the relation (\ref{coord}) it simplif\/ies to
\begin{eqnarray}
  |\Psi[\xi,\phi]|^2&=&(\Psi[\xi^I,\phi])^2\cos^2\left[\Omega(\xi)\sqrt{\langle V\rangle(\xi,\phi)}\right]\nonumber\\
  &+&(\Pi_\Psi[\xi^I,\phi])^2\left(\dfrac{\sin\left[\Omega(\xi)\sqrt{\langle V\rangle(\xi,\phi)}\right]}{\sqrt{\langle V\rangle(\xi,\phi)}}\right)^2.
\end{eqnarray}
Let us take into account \emph{ad hoc} the following separation conditions
\begin{eqnarray}
  \Psi[\xi^I,\phi]&=&\Psi[\xi^I]\Gamma_\Psi[\phi],\label{sep1}\\
  \Pi_\Psi[\xi^I,\phi]&=&\Pi_\Psi[\xi^I]\Gamma_\Pi[\phi],\label{sep2}
\end{eqnarray}
where $\Gamma_\Psi$ and $\Gamma_\Pi$ are functionals of $\phi$ only and $\Psi[\xi^I]$, and $\Pi_\Psi[\xi^I]$ are constant functionals. Applying the usual normalization condition
\begin{equation}\label{normcond}
\int_{\Sigma(\xi,\phi)}|\Psi[\xi',\phi']|^2\delta\xi'\delta\phi'=1,
\end{equation}
one obtains the simple constraint for the initial data
\begin{equation}\label{abc}
  A(\Pi_\Psi[\xi^I])^2+B(\Psi[\xi^I])^2-1=0,
\end{equation}
where the constants $A$ and $B$ are given by the integrals
\begin{eqnarray}
  A&=&\int_{\Sigma(\xi,\phi)}\Gamma_\Pi[\phi']\left(\dfrac{\sin\left[\Omega(\xi')\sqrt{\langle V\rangle(\xi',\phi')}\right]}{\sqrt{\langle V\rangle(\xi',\phi')}}\right)^2\delta\xi'\delta\phi',\\
  B&=&\int_{\Sigma(\xi,\phi)}\Gamma_\Psi[\phi']\cos^2\left[\Omega(\xi')\sqrt{\langle V(\xi',\phi')\rangle}\right]\delta\xi'\delta\phi',
\end{eqnarray}
which in our assumption are convergent, f\/inite, and independent on the initial data $\xi^I$. The equation (\ref{abc}), however, can be solved straightforwardly. In result one obtains
\begin{eqnarray}\label{pii}
\Pi_\Psi[\xi^I]=\pm\sqrt{\strut{\dfrac{1}{A}-\dfrac{B}{A}(\Psi[\xi^I])^2}},
\end{eqnarray}
which together with the def\/inition
\begin{equation}
\Pi_\Psi[\xi^I,\phi]=\dfrac{\delta \Psi[\xi^I,\phi]}{\delta \xi^I},
\end{equation}
and the separability conditions (\ref{sep1})-(\ref{sep2}) yields the differential equation for the initial data of the f\/ield $\Psi[\xi]$
\begin{equation}\label{de}
  \dfrac{1}{\Gamma[\phi]}\dfrac{\delta \Psi[\xi^I]}{\delta \xi^I}=\pm\sqrt{\strut{\dfrac{1}{A}-\dfrac{B}{A}(\Psi[\xi^I])^2}},
\end{equation}
where $\Gamma[\phi]$ is the coeff\/icient dependent on Matter f\/ields\index{Matter f\/ields} only
\begin{equation}
\Gamma[\phi]\equiv\dfrac{\Gamma_\Pi[\phi]}{\Gamma_\Psi[\phi]},
\end{equation}
which can be integrated straightforwardly
\begin{equation}
  \sqrt{A}\int\dfrac{\delta \Psi[\xi^I]}{\sqrt{\strut{1-B(\Psi[\xi^I])^2}}}=\pm\Gamma[\phi]\xi^I+C,
\end{equation}
where $C$ is a constant of integration, with the result
\begin{equation}
  \sqrt{\strut{\dfrac{A}{B}}}\arcsin\left\{\sqrt{\strut{\dfrac{B}{A}}}\Psi[\xi^I]\right\}=\pm\Gamma[\phi]\xi^I+C,
\end{equation}
so that after elementary algebraic manipulations one obtains
 \begin{equation}
 \Psi[\xi^I]=\sqrt{\strut{\dfrac{A}{B}}}\sin\theta(\xi^I,\phi),
 \end{equation}
 where $\theta(\xi^I,\phi)$ is the phase
 \begin{equation}
   \theta(\xi^I,\phi)=\sqrt{\strut{\dfrac{B}{A}}}\left(\pm\Gamma[\phi]\xi^I+C\right),
 \end{equation}
 Albeit, because of $\Psi[\xi^I]$ must be a functional of $\xi^I$ only, must hold $\Gamma[\phi]=\Gamma_0$, where $\Gamma_0$ is a constant independent on $\phi$ and $\xi^I$, for which the phase $\theta(\xi^I,\phi)$ is reduced to
 \begin{equation}
   \theta(\xi^I)=\sqrt{\strut{\dfrac{B}{A}}}\left(\pm\Gamma_0\xi^I+C\right).
 \end{equation}
 Taking into account the relation (\ref{pii}) one obtains f\/inally
 \begin{eqnarray}
\Psi[\xi^I]&=&\sqrt{\dfrac{A}{B}}\sin\theta(\xi^I),\\
\Pi_\Psi[\xi^I]&=&\pm\sqrt{\strut{\dfrac{1}{A}-\sin^2\theta(\xi^I)}}.
 \end{eqnarray}
 In the light of the equation (\ref{coord}), however, one of the relations
 \begin{eqnarray}
\sin\theta(\xi^I)&=&0,\label{1st1}\\
\sin\theta(\xi^I)&=&\pm\sqrt{\dfrac{1}{A}},\label{1st2}
 \end{eqnarray}
is always true. One sees that both these conditions def\/ine discrete values of the initial data of the invariant global dimension $\xi_I$. Namely, the f\/irst relation (\ref{1st1}) leads to the solution
 \begin{equation}
   \sqrt{\strut{\dfrac{B}{A}}}\left(\pm\Gamma_0\xi^I+C\right)=k\pi,
 \end{equation}
 where $k\in \mathbb{Z}$ is an integer, what leads to
 \begin{equation}
   \xi^I=\pm\dfrac{1}{\Gamma_0}\left(\sqrt{\strut{\dfrac{A}{B}}}k\pi-C\right).
 \end{equation}
 Similarly the second relation (\ref{1st2}) can be solved immediately
 \begin{equation}
 \xi^I=\pm\dfrac{1}{\Gamma_0}\left(\pm\sqrt{\strut{\dfrac{A}{B}}}\arcsin\sqrt{\strut{\dfrac{1}{A}}}-C\right).
 \end{equation}
 For the f\/irst case one has
 \begin{eqnarray}
   \Psi[\xi^I]&=&0,\\
   \Pi_\Psi[\xi^I]&=&\pm\sqrt{\strut{\dfrac{1}{A}}},
 \end{eqnarray}
 and for the second one hold
 \begin{eqnarray}
   \Psi[\xi^I]&=&\pm\sqrt{\strut{\dfrac{1}{B}}},\\
   \Pi_\Psi[\xi^I]&=&0.
 \end{eqnarray}
 Finally one sees that the invariant global wave functional (\ref{wfun}) is
 \begin{equation}\label{sol1}
 \Psi[\xi,\phi]=\pm\Gamma_\Psi[\phi]\Gamma_0\sqrt{\strut{\dfrac{1}{A}}}\dfrac{\sin\left[\Omega(\xi)\sqrt{\strut{\langle V\rangle(\xi,\phi)}}\right]}{\sqrt{\strut{\langle V\rangle(\xi,\phi)}}},
 \end{equation}
 in the f\/irst case (\ref{1st1}), and
 \begin{equation}\label{sol2}
\Psi[\xi,\phi]=\pm\Gamma_\Psi[\phi]\sqrt{\strut{\dfrac{1}{B}}}\cos\left[\Omega(\xi)\sqrt{\strut{\langle V\rangle(\xi,\phi)}}\right],
\end{equation}
for the second one (\ref{1st2}).

Applying the normalization condition (\ref{normcond}) to the solutions (\ref{sol1}) and (\ref{sol2}) one receives the equations
\begin{eqnarray}
  |\Gamma_\Psi[\phi]\Gamma_0|^2&=&1,\\
  \Gamma_\Psi[\phi]\Gamma_0&=&1,
\end{eqnarray}
which can be solved easy and lead to the relation
\begin{equation}
\Gamma_\Psi[\phi]=\dfrac{1}{\Gamma_0}.
\end{equation}
Therefore, one obtains f\/inally
\begin{eqnarray}
\Psi_1[\xi,\phi]&=&\pm\sqrt{\strut{\dfrac{1}{|A|}}}\dfrac{\sin\left[\Omega(\xi)\sqrt{\strut{\langle V\rangle(\xi,\phi)}}\right]}{\sqrt{\strut{\langle V\rangle(\xi,\phi)}}},\label{solmi1}\\
\Psi_2[\xi,\phi]&=&\pm\sqrt{\strut{\dfrac{1}{|B|}}}\cos\left[\Omega(\xi)\sqrt{\strut{\langle V\rangle(\xi,\phi)}}\right],~~~~~~~~\label{solmi2}
\end{eqnarray}
where now the constants $A$ and $B$ are equal to
\begin{eqnarray}
\!\!\!\!\!\!\!\!\!\!\!\!\!\!\!\!\!\!\!\!\!\!\!\!\!\!\!\!\!\!\!\!\!\!\!\!\!\!\!\!
&&A=\int_{\Sigma(\xi,\phi)}\left(\dfrac{\sin\left[\Omega(\xi')\sqrt{\langle V\rangle(\xi',\phi')}\right]}{\sqrt{\langle V\rangle(\xi',\phi')}}\right)^2\delta\xi'\delta\phi',\\
\!\!\!\!\!\!\!\!\!\!\!\!\!\!\!\!\!\!\!\!\!\!\!\!\!\!\!\!\!\!\!\!\!\!\!\!\!\!\!\!
&&B=\dfrac{1}{\Gamma_0}\int_{\Sigma(\xi,\phi)}\cos^2\left[\Omega(\xi')\sqrt{\langle V\rangle(\xi',\phi')}\right]\delta\xi'\delta\phi'.
\end{eqnarray}

In this manner the general solutions of the theory of quantum gravity can be now constructed straightforwardly by using of the solutions (\ref{sol1}) and (\ref{sol2}), in which the integrals must be putted
\begin{eqnarray}\label{ints}
\Omega(\xi)&=&\int_{\Sigma(\xi)}\delta\xi'=\xi,\\
\langle V\rangle(\xi,\phi)&=&\dfrac{1}{\Omega(\xi)}\int_{\Sigma(\xi)}\delta\xi'V[\xi',\phi],
\end{eqnarray}
where the $\xi$-measure is
\begin{eqnarray}
\delta\xi=\dfrac{1}{8\pi}\dfrac{\delta h}{\sqrt{6h}}.
\end{eqnarray}
Because, however, the generalized gravitational potential\index{gravitational potential!generalized} $V[\xi,\phi]$ has the form of an algebraic sum
\begin{equation}
V[\xi,\phi]=-{^{(3)}\!R}+2\Lambda+2\kappa\ell_P^2\varrho,
\end{equation}
one has very convenient separability
\begin{equation}
\langle{V}\rangle(\xi,\phi)=-\dfrac{1}{\Omega(\xi)}\int_{\Sigma(\xi)}\delta\xi'~{^{(3)}\!R}+2\Lambda+\dfrac{2\kappa\ell_P^2}{\Omega(\xi)}\int_{\Sigma(\phi)}\delta \phi'\int_{\Sigma(\xi)}\delta\xi'\rho.
\end{equation}
Therefore, for a concretely given geometry of a three-dimensional space-like embedded space one should estimate the functionally averaged three-dimensional Ricci scalar
\begin{eqnarray}
\langle{^{(3)}\!R}\rangle=\dfrac{1}{\Omega(\xi)}\int_{\Sigma(\xi)}\delta\xi'~{^{(3)}\!R},
\end{eqnarray}
and the functionally averaged energy density\index{energy density} of Matter f\/ields\index{Matter f\/ields}
\begin{equation}
\langle\rho\rangle=\dfrac{1}{\Omega(\xi)}\int_{\Sigma(\phi)}\delta \phi'\int_{\Sigma(\xi)}\delta\xi'\rho,
\end{equation}
and using these quantities construct the functionally averaged generalized gravitational potential\index{gravitational potential!generalized}
\begin{equation}\label{formula}
\langle V\rangle(\xi,\phi)=-\langle{^{(3)}\!R}\rangle+2\Lambda+2\kappa\ell_P^2\langle\rho\rangle.
\end{equation}
Applying this averaging method one can construct the solutions (\ref{sol1}) and (\ref{sol2}) straightforwardly. If the Ricci scalar curvature\index{Ricci scalar curvature} ${^{(3)}\!R}$ of an embedded space and/or the energy density\index{energy density} of Matter f\/ields\index{Matter f\/ields} $\varrho$ are functions of a space-time point $x^\mu=(x^0,x^1,x^2,x^3)$ or any one space-time coordinate the integration procedure is understood as the procedure of performing of the Lebesgue--Stieltjes integral. For instance in the case when the integrand is a function of $x^\mu$ then the Lebesgue--Stieltjes measure\index{Lebesgue--Stieltjes measure} has the following form
\begin{equation}
\delta \xi = \dfrac{\partial^4 \xi}{\partial x_0\partial x_1\partial x_2\partial x_3}d^4x,
\end{equation}
or in terms of determinant of induced metric $h$
\begin{equation}
\delta \xi = \dfrac{1}{8\pi}\dfrac{\partial^4 h}{\partial x_0\partial x_1\partial x_2\partial x_3}\dfrac{d^4x}{\sqrt{6h}}.
\end{equation}

\section{Problem I: Inverted Transformation}
There is the problem of inverted transformation
\begin{equation}
  \xi\rightarrow h_{ij},
\end{equation}
within the obtained solutions. In fact, this problem is strictly related to the problem of f\/inding of a induced metric if one knows its determinant. Such a procedure can not be performed analytically, but there are functional methods which enable performing of this step.

The solutions (\ref{sol1}) and (\ref{sol2}) expressed via $h_{ij}$ are
\begin{eqnarray}
\Psi_1[h_{ij},\phi]&=&\pm\sqrt{\strut{\dfrac{1}{|A|}}}\dfrac{\sin\left[\Omega(h_{ij})\sqrt{\strut{\langle V\rangle(h_{ij},\phi)}}\right]}{\sqrt{\strut{\langle V\rangle(h_{ij},\phi)}}},\label{solmi1}\\
\Psi_2[h_{ij},\phi]&=&\pm\sqrt{\strut{\dfrac{1}{|B|}}}\cos\left[\Omega(h_{ij})\sqrt{\strut{\langle V\rangle(h_{ij},\phi)}}\right],\label{solmi2}
\end{eqnarray}
where now the constants $A$ and $B$ are equal to
\begin{eqnarray}
\!\!\!\!\!\!\!\!\!\!\!\!\!\!\!\!\!\!\!\!\!\!\!\!\!\!\!\!\!\!\!\!\!\!\!\!\!\!\!\!
&&A=\dfrac{1}{8\pi}\int_{\Sigma(h_{ij},\phi)}\left(\dfrac{\sin\left[\Omega(h_{ij}')\sqrt{\langle V\rangle(h_{ij}',\phi')}\right]}{\sqrt{\langle V\rangle(h_{ij}',\phi')}}\right)^2\sqrt{\strut{\dfrac{h'}{6}}}{h^{ij}}'\delta h_{ij}'\delta\phi',\\
\!\!\!\!\!\!\!\!\!\!\!\!\!\!\!\!\!\!\!\!\!\!\!\!\!\!\!\!\!\!\!\!\!\!\!\!\!\!\!\!
&&B=\dfrac{1}{8\pi\Gamma_0}\int_{\Sigma(h_{ij},\phi)}\cos^2\left[\Omega(h_{ij}')\sqrt{\langle V\rangle(h_{ij}',\phi')}\right]\sqrt{\strut{\dfrac{h'}{6}}}{h^{ij}}'\delta h_{ij}'\delta\phi',
\end{eqnarray}
and assumed to be convergent and f\/inite. The solutions (\ref{solmi1}) and (\ref{solmi2}) are two independent states transformed to the standard quantum geometrodynamics\index{quantum geometrodynamics (QGD)}, but in general they are not solutions of the Wheeler--DeWitt equation\index{Wheeler--DeWitt equation}.

Interestingly, in such a situation the normalization condition (\ref{normcond}) becomes
\begin{equation}\label{normaliz}
  \dfrac{1}{8\pi}\int_{\Sigma(h_{ij},\phi)}|\Psi[h_{ij}',\phi']|^2\sqrt{\strut{\dfrac{h'}{6}}}{h^{ij}}'\delta h_{ij}'\delta\phi'=1,
\end{equation}
and can be used to construct the $\pi$ number def\/inition\index{$\pi$ number def\/inition}
\begin{equation}
\pi=\dfrac{1}{8}\int_{\Sigma(h_{ij},\phi)}|\Psi[h_{ij}',\phi']|^2\sqrt{\strut{\dfrac{h'}{6}}}{h^{ij}}'\delta h_{ij}'\delta\phi'.
\end{equation}
\newpage
\section{Problem II: The Hilbert Space and Superposition}
Because, however, both the equations of quantum geometrodynamics\index{quantum geometrodynamics (QGD)} (\ref{wdw}) and (\ref{eqn}) are linear in the f\/ield $\Psi$, in general the superposition\index{superposition principle}
\begin{equation}
\Psi=\sum_{i=1,2}\alpha_i\Psi_i,\label{solmi3}
\end{equation}
where $\alpha_i$ are arbitrary constants, and $\Psi_i$ are (\ref{solmi1}) and (\ref{solmi2}), should be also a solution. The normalization (\ref{normaliz}) of (\ref{solmi3}) gives
\begin{equation}\label{relc}
|\alpha_1|^2+|\alpha_2|^2+(\alpha^\star_1\alpha_2+\alpha_1\alpha^\star_2)I=1,
\end{equation}
where $I$ is the integral
\begin{equation}
  I=\dfrac{1}{\sqrt{|A||B|}}\dfrac{1}{8\pi}\int_{\Sigma(h_{ij},\phi)} \dfrac{\sin\left[2\Omega(h_{ij}')\sqrt{\strut{\langle V\rangle(h_{ij}',\phi')}}\right]}{2\sqrt{\strut{\langle V\rangle(h_{ij}',\phi')}}}\sqrt{\strut{\dfrac{h'}{6}}}{h^{ij}}'\delta h_{ij}'\delta\phi'.\label{Int}
\end{equation}
For vanishing $I=0$ one obtains form (\ref{relc}) simply
\begin{equation}
  |\alpha_2|=\sqrt{\strut{1-|\alpha_1|^2}}\quad,\quad|\alpha_1|\geqslant1.
\end{equation}
The case of $I\neq0$ is much more complicated. Note that the equation (\ref{relc}) can be rewritten in form
\begin{equation}
  (\alpha_1+\alpha_2I)\alpha^\star_1+(\alpha_2+\alpha_1I)\alpha_2^\star=0,
\end{equation}
what leads to the result
\begin{equation}
  \dfrac{\alpha^\star_1}{\alpha_2^\star}=\dfrac{-\alpha_1I+\alpha_2}{\alpha_1+\alpha_2I},
\end{equation}
or equivalently
\begin{eqnarray}
C\alpha^\star_1&=&-\alpha_1I+\alpha_2,\label{CC1}\\
C\alpha_2^\star&=&\alpha_1+\alpha_2 I,\label{CC2}
\end{eqnarray}
where $0\neq C\in\mathbb{R}$ is a constant. The relations (\ref{CC1})-(\ref{CC2}) lead to
\begin{eqnarray}
C|\alpha_1|^2&=&-\alpha^2_1 I+\alpha_2\alpha_1,\\
C|\alpha_2|^2&=&\alpha_1\alpha_2+\alpha_2^2 I.
\end{eqnarray}
Mutual addition and application of (\ref{relc}) yields
\begin{equation}
  CI[(\alpha^\star_1-\alpha_2)\alpha_2+(\alpha_2^\star+\alpha_1)\alpha_1]=\alpha_1\alpha_2+\alpha_2\alpha_1,
\end{equation}
which generates the equations
\begin{eqnarray}
CI(\alpha^\star_1-\alpha_2)&=&\alpha_1,\label{in1}\\
CI(\alpha_2^\star+\alpha_1)&=&\alpha_2.\label{in2}
\end{eqnarray}
Complex decomposition of (\ref{in1})-(\ref{in2}) leads to
\begin{eqnarray}
\Re\alpha_2&=&(CI-1)\Re\alpha_1,\\
\Im\alpha_2&=&(CI-1)\Im\alpha_1,
\end{eqnarray}
or equivalently
\begin{eqnarray}
\alpha_2&=&(CI-1)\alpha_1,\label{eqi1}\\
|\alpha_2|^2&=&(CI-1)^2|\alpha_1|^2.\label{eqi2}
\end{eqnarray}
Employing (\ref{eqi1})-(\ref{eqi2}) within the constraint (\ref{relc}) yields to
\begin{equation}\label{alp}
\dfrac{1}{|\alpha_1|^2}=IC^2+(I^2-2I)C-I+2.
\end{equation}
Because of the natural condition $\dfrac{1}{|\alpha_1|^2}>0$ one has the inequality for $C$
\begin{equation}
IC^2+I(I-2)C-(I-2)>0,\label{ineqp}
\end{equation}
where $I\geqslant0$. One sees that the case $I=0$ gives $2>0$ what is true. The inequality (\ref{ineqp}) has two different solutions which are dependent on $\sgn(I-2)$. When $0\leqslant I<2$ then the solution is
\begin{equation}
C\in\left(C_-,C_+\right).
\end{equation}
When $I>2$ the solution is somewhat different
\begin{equation}
C\in(-\infty,C_-)\cup(C_+,\infty).\label{ceee}
\end{equation}
Here are the constants
\begin{equation}
  C_\mp=\dfrac{2-I}{2}\mp\dfrac{1}{2}\sqrt{\dfrac{I^3-4I^2+8I-8}{I}}.
\end{equation}
For consistency the constants $C_\mp$ must be real numbers. For this it is necessary and suff\/icient to satisfy the condition
\begin{equation}
I^3-4I^2+8I-8\geqslant0,
\end{equation}
what is satisf\/ied for $I\geqslant2$ or $I<0$. Because of $I>0$ the f\/irst case is true, and therefore (\ref{ceee}) is true region of validity of $C$. Applying the def\/inition of $I$, $A$, and $B$ one receives the following inequality
\begin{eqnarray}
&&\int_{\Sigma(h_{ij},\phi)} \dfrac{\sin\left[2\Omega(h_{ij}')\sqrt{\strut{\langle V\rangle(h_{ij}',\phi')}}\right]}{2\sqrt{\strut{\langle V\rangle(h_{ij}',\phi')}}}\sqrt{h'}{h^{ij}}'\delta h_{ij}'\delta\phi'\geqslant\nonumber\\
&&\dfrac{2}{\sqrt{|\Gamma_0|}}\left(\int_{\Sigma(h_{ij},\phi)}\left(\dfrac{\sin\left[\Omega(h_{ij}')\sqrt{\langle V\rangle(h_{ij}',\phi')}\right]}{\sqrt{\langle V\rangle(h_{ij}',\phi')}}\right)^2\sqrt{h'}{h^{ij}}'\delta h_{ij}'\delta\phi'\right)^{1/2}\times\nonumber\\
&&\left(\int_{\Sigma(h_{ij},\phi)}\cos^2\left[\Omega(h_{ij}')\sqrt{\langle V\rangle(h_{ij}',\phi')}\right]\sqrt{h'}{h^{ij}}'\delta h_{ij}'\delta\phi'\right)^{1/2},
\end{eqnarray}
or in terms of the invariant global dimension
\begin{eqnarray}
&&\int_{\Sigma(\xi,\phi)} \dfrac{\sin\left[2\Omega(\xi')\sqrt{\strut{\langle V\rangle(\xi',\phi')}}\right]}{2\sqrt{\strut{\langle V\rangle(\xi',\phi')}}}\delta\xi'\delta\phi'\geqslant\nonumber\\
&&\dfrac{2}{\sqrt{|\Gamma_0|}}\left(\int_{\Sigma(\xi,\phi)}\left(\dfrac{\sin\left[\Omega(\xi')\sqrt{\langle V\rangle(\xi',\phi')}\right]}{\sqrt{\langle V\rangle(\xi',\phi')}}\right)^2\delta\xi'\delta\phi'\right)^{1/2}\times\nonumber\\
&&\left(\int_{\Sigma(\xi,\phi)}\cos^2\left[\Omega(\xi')\sqrt{\langle V\rangle(\xi',\phi')}\right]\delta\xi'\delta\phi'\right)^{1/2}.\label{idineq}
\end{eqnarray}

Let us consider the concept of an inner product space, called also pre-Hilbert space\index{pre-Hilbert space}. Such a space is a vector space equipped with an additional structure called an inner product\index{inner product}
\begin{equation}
  \left<\bullet,\bullet\right>:V\times V\rightarrow\mathbb{F},
\end{equation}
where $V$ is a vector space over the f\/ield of scalars $\mathbb{F}$. A pre-Hilbert space\index{pre-Hilbert space} is called a Hilbert space\index{Hilbert space} if is complete as a normed space under the induced norm $||\bullet||$. Every Hilbert space is the Banach space\index{Banach space}. The structure of an inner product allows to build intuitive geometrical notions such as length of a vector or the angle between two vectors, and deduce orthogonality between vectors by vanishing of an inner product. Pre-Hilbert spaces generalize Euclidean spaces to vector spaces of any, including inf\/inite, dimension and are the theme of functional analysis (For more detailed discussion of functional analysis see e.g. the Ref. \cite{funan}). The Lebesgue space $L^2\left(\Sigma,\mu\right)$\index{Lebesgue space $L^2$} where $\Sigma$ is the conf\/igurational space equipped with the Lebesgue measure $\mu$, called also the space of square-integrable functions, which is the special case of $L^p$-spaces (For more details see e.g. the Ref. \cite{lebe}), and moreover is a Hilbert space\index{Hilbert space}. This particular Hilbert space\index{Hilbert space} is the fundament of quantum mechanics, in which wave functions belong to $L^2$.

Let us construct the Lebesgue space $L^2$\index{Lebesgue space $L^2$} for the theory of quantum gravity \emph{in accordance} with the superposition principle. In such a situation the conf\/igurational space is the midisuperspace\index{midisuperspace} $\Sigma(\xi,\phi)$ with the measure $\mu=\delta\xi\delta\phi$. The Lebesgue space\index{Lebesgue space $L^2$} $L^2\left(\Sigma(\xi,\phi),\delta\xi\delta\phi\right)$ is the Hilbert space\index{Hilbert space} with the inner product
\begin{equation}
 \left\langle{f(\xi,\phi),g(\xi,\phi)}\right\rangle=\int_{\Sigma(\xi,\phi)}f^\star(\xi',\phi')g(\xi',\phi')\delta\xi'\delta\phi',
\end{equation}
satisfying the conditions
\begin{eqnarray}
\!\!\!\!\!\!\!\!\!\!\!\!\!\!\!\!\!\!\!\!\!\!\!\!\!\!\!\!\!\!
\left\langle{f(\xi,\phi),g(\xi,\phi)}\right\rangle&=&\overline{\left\langle{f(\xi,\phi),g(\xi,\phi)}\right\rangle},\\
\!\!\!\!\!\!\!\!\!\!\!\!\!\!\!\!\!\!\!\!\!\!\!\!\!\!\!\!\!\!
\left\langle{af(\xi,\phi),g(\xi,\phi)}\right\rangle&=&a\left\langle{f(\xi,\phi),g(\xi,\phi)}\right\rangle,\\
\!\!\!\!\!\!\!\!\!\!\!\!\!\!\!\!\!\!\!\!\!\!\!\!\!\!\!\!\!\!
\left\langle{f(\xi,\phi)+h(\xi,\phi),g(\xi,\phi)}\right\rangle&=&\left\langle{f(\xi,\phi),g(\xi,\phi)}\right\rangle+\left\langle{h(\xi,\phi),g(\xi,\phi)}\right\rangle.
\end{eqnarray}
The induced norm\index{norm}
\begin{equation}
  \left|\left|f(\xi,\phi)\right|\right|=\sqrt{\left\langle{f(\xi,\phi),f(\xi,\phi)}\right\rangle}=\left(\int_{\Sigma(\xi,\phi)}|f(\xi',\phi')|^2\delta\xi'\delta\phi'\right)^{1/2}<\infty,
\end{equation}
is homogeneous
\begin{equation}
\left|\left|af(\xi,\phi)\right|\right|=|a|\left|\left|f(\xi,\phi)\right|\right|,
\end{equation}
and satisf\/ies the triangle inequality\index{triangle inequality}
\begin{equation}
\left|\left|f(\xi,\phi)+g(\xi,\phi)\right|\right|\leqslant\left|\left|f(\xi,\phi)\right|\right|+\left|\left|g(\xi,\phi)\right|\right|.
\end{equation}
For the orthogonal situation
\begin{equation}
\left\langle{f(\xi,\phi),g(\xi,\phi)}\right\rangle=0,
\end{equation}
the Pythagoras theorem\index{Pythagoras theorem} holds
\begin{equation}
\left|\left|f(\xi,\phi)+g(\xi,\phi)\right|\right|^2=\left|\left|f(\xi,\phi)\right|\right|^2+\left|\left|g(\xi,\phi)\right|\right|^2.
\end{equation}
The parallelogram law\index{parallelogram law}
\begin{equation}
\left|\left|f(\xi,\phi)+g(\xi,\phi)\right|\right|^2+\left|\left|f(\xi,\phi)-g(\xi,\phi)\right|\right|^2=2\left|\left|f(\xi,\phi)\right|\right|^2+2\left|\left|g(\xi,\phi)\right|\right|^2,
\end{equation}
is a necessary and suff\/icient condition for the existence of a scalar product corresponding to a given norm. This scalar product\index{scalar product} follows from the polarization identity\index{polarization identity}
\begin{equation}
\left|\left|f(\xi,\phi)+g(\xi,\phi)\right|\right|^2=\left|\left|f(\xi,\phi)\right|\right|^2+\left|\left|g(\xi,\phi)\right|\right|^2+\Re\left\langle{f(\xi,\phi),g(\xi,\phi)}\right\rangle,
\end{equation}
called also the law of cosinuses, and equals to
\begin{eqnarray}
  \left(f(\xi,\phi),g(\xi,\phi)\right)&=&\dfrac{\left|\left|f(\xi,\phi)+g(\xi,\phi)\right|\right|^2-\left|\left|f(\xi,\phi)-g(\xi,\phi)\right|\right|^2}{2}=\nonumber\\
  &=&\Re\left\langle{f(\xi,\phi),g(\xi,\phi)}\right\rangle.
\end{eqnarray}
When $f_i(\xi,\phi)$, $i=1,\ldots,N$, are orthogonal vectors then holds the relation
\begin{equation}
  \sum_{i=1}^N\left|\left|f_i(\xi,\phi)\right|\right|=\left|\left|\sum_{i=1}^Nf_i(\xi,\phi)\right|\right|.\label{orth}
\end{equation}
When $V$ is a complete pre-Hilbert space\index{pre-Hilbert space}, i.e. is a Hilbert space\index{Hilbert space}, then the equation (\ref{orth}) becomes the Parseval identity\index{Parseval identity}
\begin{equation}
  \sum_{i=1}^\infty\left|\left|f_i(\xi,\phi)\right|\right|=\left|\left|\sum_{i=1}^\infty f_i(\xi,\phi)\right|\right|,
\end{equation}
provided that the inf\/inite series on the LHS is convergent. Recall that completeness of the space is necessary for convergence of the sequence of partial sums on the space
\begin{equation}
s_k=\sum_{i=1}^k\left|\left|f_i(\xi,\phi)\right|\right|,
\end{equation}
which is a Cauchy sequence.

The inner product and the norm can be used for def\/inition of the angle $\alpha$ between two vectors $f(\xi,\phi)$ and $g(\xi,\phi)$
\begin{equation}
\alpha\left(f(\xi,\phi),g(\xi,\phi)\right)=\arccos\dfrac{\left\langle{f(\xi,\phi),g(\xi,\phi)}\right\rangle}{\left|\left|f(\xi,\phi)\right|\right|\left|\left|g(\xi,\phi)\right|\right|},\label{angl}
\end{equation}
and it can be shown by straightforward computation that the Cauchy--Bunyakovsky--Schwarz inequality\index{Cauchy--Bunyakovsky--Schwarz inequality} holds
\begin{equation}\label{csin}
\left|\left\langle{f(\xi,\phi),g(\xi,\phi)}\right\rangle\right|\leqslant\left|\left|f(\xi,\phi)\right|\right|\left|\left|g(\xi,\phi)\right|\right|.
\end{equation}

Taking into account
\begin{eqnarray}
f(\xi,\phi)&=&\dfrac{\sin\left[\Omega(\xi)\sqrt{\langle V\rangle(\xi,\phi)}\right]}{\sqrt{\langle V\rangle(\xi,\phi)}},\\
g(\xi,\phi)&=&\cos\left[\Omega(\xi)\sqrt{\langle V\rangle(\xi,\phi)}\right],
\end{eqnarray}
one obtains
\begin{eqnarray}
\!\!\!\!\!\!\!\!\!\!\!\!\!\!\!\!\!\!\!\!\!\!\!\!\!\!\!\!\!\!
\left\langle{f(\xi,\phi),g(\xi,\phi)}\right\rangle&=&\int_{\Sigma(\xi,\phi)}\dfrac{\sin\left[2\Omega(\xi')\sqrt{\langle V\rangle(\xi',\phi')}\right]}{2\sqrt{\langle V\rangle(\xi',\phi')}}\delta\xi'\delta\phi',\\
\!\!\!\!\!\!\!\!\!\!\!\!\!\!\!\!\!\!\!\!\!\!\!\!\!\!\!\!\!\!
\left|\left|f(\xi,\phi)\right|\right|^2&=&\int_{\Sigma(\xi,\phi)}\left(\dfrac{\sin\left[\Omega(\xi')\sqrt{\langle V\rangle(\xi',\phi')}\right]}{\sqrt{\langle V\rangle(\xi',\phi')}}\right)^2\delta\xi'\delta\phi',\\
\!\!\!\!\!\!\!\!\!\!\!\!\!\!\!\!\!\!\!\!\!\!\!\!\!\!\!\!\!\!
\left|\left|g(\xi,\phi)\right|\right|^2&=&\int_{\Sigma(\xi,\phi)}\cos^2\left[\Omega(\xi)\sqrt{\langle V\rangle(\xi,\phi)}\right]\delta\xi'\delta\phi'.
\end{eqnarray}
By this reason one can rewrite the integral (\ref{Int}) as
\begin{equation}
  I=\dfrac{\left\langle{f(\xi,\phi),g(\xi,\phi)}\right\rangle}{\left|\left|f(\xi,\phi)\right|\right|\left|\left|g(\xi,\phi)\right|\right|},
\end{equation}
or with using of the def\/inition (\ref{angl})
\begin{equation}
  I=\cos\alpha\left(f(\xi,\phi),g(\xi,\phi)\right).\label{Intcos}
\end{equation}
In this manner the inequality (\ref{idineq}) can be rewritten as
\begin{equation}
\cos\alpha\left(f(\xi,\phi),g(\xi,\phi)\right)\geqslant\dfrac{2}{\sqrt{|\Gamma_0|}}
\end{equation}
whereas the Cauchy--Bunyakovsky--Schwarz inequality\index{Cauchy--Bunyakovsky--Schwarz inequality} (\ref{csin}) says that
\begin{equation}
\left|\cos\alpha\left(f(\xi,\phi),g(\xi,\phi)\right)\right|\leqslant1.
\end{equation}
Hence the superposition principle\index{superposition principle} requires
\begin{equation}
\dfrac{2}{\sqrt{|\Gamma_0|}}\leqslant\left|\cos\alpha\left(f(\xi,\phi),g(\xi,\phi)\right)\right|\leqslant1,
\end{equation}
or equivalently
\begin{equation}
\alpha\left(f(\xi,\phi),g(\xi,\phi)\right)\in\left[\arccos\dfrac{2}{\sqrt{|\Gamma_0|}}+2k\pi,2\pi+2k\pi\right],\quad k\in\mathbb{Z},
\end{equation}
what is consistent if and only if
\begin{equation}
  \sqrt{|\Gamma_0|}\geqslant2.\label{gamin}
\end{equation}

Let us  consider the wave functions (\ref{solmi1}) and (\ref{solmi2}) expressed via the invariant global dimension
\begin{eqnarray}
\Psi_1[\xi,\phi]&=&\dfrac{1}{\sqrt{|A|}}\dfrac{\sin\left[\Omega(\xi)\sqrt{\strut{\langle V\rangle(\xi,\phi)}}\right]}{\sqrt{\strut{\langle V\rangle(\xi,\phi)}}}=\dfrac{1}{\sqrt{|A|}}f(\xi,\phi),\label{solmit1}\\
\Psi_2[\xi,\phi]&=&\dfrac{1}{\sqrt{|B|}}\cos\left[\Omega(\xi)\sqrt{\strut{\langle V\rangle(\xi,\phi)}}\right]=\dfrac{1}{\sqrt{|B|}}g(\xi,\phi),\label{solmit2}
\end{eqnarray}
with the constants $A$ and $B$ given by
\begin{eqnarray}
\!\!\!\!\!\!\!\!\!\!\!\!\!\!\!\!\!\!\!\!\!\!\!\!\!\!\!\!\!\!
A&=&\int_{\Sigma(\xi,\phi)}\left(\dfrac{\sin\left[\Omega(\xi')\sqrt{\langle V\rangle(\xi',\phi')}\right]}{\sqrt{\langle V\rangle(\xi',\phi')}}\right)^2\delta\xi'\delta\phi'\equiv\left|\left|f(\xi,\phi)\right|\right|^2,\\
\!\!\!\!\!\!\!\!\!\!\!\!\!\!\!\!\!\!\!\!\!\!\!\!\!\!\!\!\!\!
B&=&\dfrac{1}{|\Gamma_0|}\int_{\Sigma(\xi,\phi)}\cos^2\left[\Omega(\xi')\sqrt{\langle V\rangle(\xi',\phi')}\right]\delta\xi'\delta\phi'\equiv\dfrac{1}{|\Gamma_0|}\left|\left|g(\xi,\phi)\right|\right|^2.
\end{eqnarray}
In other words
\begin{eqnarray}
\Psi_1[\xi,\phi]&=&\dfrac{f(\xi,\phi)}{\left|\left|f(\xi,\phi)\right|\right|},\label{solmita1}\\
\Psi_2[\xi,\phi]&=&\sqrt{|\Gamma_0|}\dfrac{g(\xi,\phi)}{\left|\left|g(\xi,\phi)\right|\right|},\label{solmita2}\\
I&=&\cos\alpha\left(\Psi_1[\xi,\phi],\Psi_2[\xi,\phi]\right).
\end{eqnarray}
The wave functions (\ref{solmita1}) and (\ref{solmita2}) are elements of the Lebesgue space $L^2\left(\Sigma(\xi,\phi),\delta\xi\delta\phi\right)$\index{Lebesgue space $L^2$} with the scalar product\index{scalar product}
\begin{equation}
\left({\Psi_1(\xi,\phi),\Psi_2(\xi,\phi)}\right)=\Re\left\langle{\Psi_1(\xi,\phi),\Psi_2(\xi,\phi)}\right\rangle,
\end{equation}
where the inner product is
\begin{equation}
\left\langle{\Psi_1(\xi,\phi),\Psi_2(\xi,\phi)}\right\rangle=\int_{\Sigma(\xi,\phi)}\Psi_1^\star(\xi',\phi')\Psi_2(\xi',\phi')\delta\xi'\delta\phi',
\end{equation}
and the induced norm
\begin{equation}
\left|\left|\Psi_{i}(\xi,\phi)\right|\right|=\left(\int_{\Sigma(\xi,\phi)}|\Psi_i(\xi',\phi')|^2\delta\xi'\delta\phi'\right)^{1/2},
\end{equation}
where $i=1,2$. It is easy to see that
\begin{equation}
\left({\Psi_1(\xi,\phi),\Psi_2(\xi,\phi)}\right)=\sqrt{|\Gamma_0|}I,\label{scalpro}
\end{equation}
and because of $I\geqslant2$ the scalar product satisf\/ies the inequality
\begin{equation}
\left({\Psi_1(\xi,\phi),\Psi_2(\xi,\phi)}\right)\geqslant2\sqrt{|\Gamma_0|}.
\end{equation}
In the light of the inequality (\ref{gamin}) one has
\begin{equation}
\left({\Psi_1(\xi,\phi),\Psi_2(\xi,\phi)}\right)\geqslant4.
\end{equation}
It can be seen also that
\begin{eqnarray}
  ||\Psi_1||&=&1,\\
  ||\Psi_2||&=&\sqrt{|\Gamma_0|}\geqslant2,
\end{eqnarray}
i.e. the state $\Psi_1$ is a ray whereas the state $\Psi_2$ is not a ray in the Hilbert space\index{Hilbert space}. Because, however the superposition is given by
\begin{equation}
  \Psi=\alpha_1\Psi_1+\alpha_2\Psi_2,
\end{equation}
one can selected the constants $\alpha_1$ and $\alpha_2$ in such a way that the wave function $\Psi_2$ will be normalized to unity. The choice is easy to deduce. Because of $\Psi_1$ is normalized to unity the coeff\/icient $\alpha_1$ is an arbitrary constant, while the coeff\/icient $\alpha_2$ should be exchanged on
\begin{equation}
  \alpha_2'=\dfrac{1}{\sqrt{|\Gamma_0|}}\alpha_2,\label{all2}
\end{equation}
where $\alpha_2$ is an arbitrary constant. Still, however, in general the problem of normalization of the state $\Psi_2$ is unsolved. Redef\/inition of the scalar product
\begin{equation}
  \left({\Psi_1(\xi,\phi),\Psi_2(\xi,\phi)}\right)=\dfrac{\left\langle{\Psi_1(\xi,\phi),\Psi_2(\xi,\phi)}\right\rangle}{\left|\left|\Psi_{1}(\xi,\phi)\right|\right|\left|\left|\Psi_{2}(\xi,\phi)\right|\right|}=I\geqslant2,
\end{equation}
does not work consistently in the light of (\ref{Intcos}). The only solution is to apply \emph{ad hoc} the scaling
\begin{equation}
  \Psi_2'=\dfrac{1}{\sqrt{|\Gamma_0|}}\Psi_2\leqslant\dfrac{1}{\sqrt{2}}\Psi_2,\label{scaledstate}
\end{equation}
which, however, is consistent with the choice (\ref{all2}).

Another alternative can be constructed as follows. Let us assume that the superposition state
\begin{equation}
  \Psi'=\alpha_1'\Psi_1'+\alpha_2'\Psi_2',\label{supstate}
\end{equation}
is such that the states $\Psi_1'$ and $\Psi_2'$ are orthonormal, i.e. are orthogonal rays in the Hilbert space\index{Hilbert space}
\begin{eqnarray}
||\Psi_1'||&=&1,\\
||\Psi_2'||&=&1,\\
\left(\Psi_1',\Psi_2'\right)&=&0.
\end{eqnarray}
The problem is to construct such constants $\alpha_1'$, $\alpha_2'$, and states $\Psi_1'$ and $\Psi_2'$ which satisfy these requirements. This problem can be solved with help of the Gram--Schmidt algorithm\index{Gram--Schmidt algorithm}, in which the orthonormal states $\Psi_i'$ (here $i=1,2$) are expressed via the non-orthonormal states $\Psi_i$ via using of the Gram determinants\index{Gram determinant}
\begin{eqnarray}
  G_0&=&1,\\
  G_1&=&\left|(\Psi_1,\Psi_1)\right|,\\
  G_2&=&\det\left[\begin{array}{cc}(\Psi_1,\Psi_1)&(\Psi_1,\Psi_2)\\
  (\Psi_1,\Psi_2)&(\Psi_2,\Psi_2)\end{array}\right],
\end{eqnarray}
as follows
\begin{eqnarray}
  \Psi_1'&=&\dfrac{1}{\sqrt{G_0G_1}}\Psi_1,\\
  \Psi_2'&=&\dfrac{1}{\sqrt{G_1G_2}}\det\left[\begin{array}{cc}(\Psi_1,\Psi_1)&(\Psi_1,\Psi_2)\\
  \Psi_1&\Psi_2\end{array}\right].
\end{eqnarray}
It can be seen by straightforward computation that
\begin{eqnarray}
  G_1&=&1,\\
  G_2&=&|\Gamma_0|\left(\dfrac{1}{\sqrt{|\Gamma_0|}}-I^2\right),
\end{eqnarray}
and consequently
\begin{eqnarray}
  \Psi_1'&=&\Psi_1,\label{psi1prim}\\
  \Psi_2'&=&\left(\dfrac{1}{\sqrt{|\Gamma_0|}}-I^2\right)^{-1/2}\left(-I\Psi_1+\dfrac{1}{\sqrt{|\Gamma_0|}}\Psi_2\right).\label{psi2prim}
\end{eqnarray}
One sees now that the second term in (\ref{psi2prim}) converges with the proposed scaled state (\ref{scaledstate}). In this manner the superposition state (\ref{supstate}) can be rewritten as
\begin{equation}
  \Psi'=\alpha_1'\Psi_1+\alpha_2'\Psi_2',
\end{equation}
where now $\Psi_2'=\dfrac{1}{\sqrt{|\Gamma_0|}}\Psi_2$ is the scaled state (\ref{scaledstate}) and
\begin{eqnarray}
  \alpha_1'&=&\alpha_1-\alpha_2I\left(\dfrac{1}{\sqrt{|\Gamma_0|}}-I^2\right)^{-1/2},\\
  \alpha_2'&=&\alpha_2\left(\dfrac{1}{\sqrt{|\Gamma_0|}}-I^2\right)^{-1/2}.
\end{eqnarray}
The states $\Psi_1$ and $\Psi_2$ can be expressed via the orthonormal states $\Psi_1'$ and $\Psi_2'$
\begin{eqnarray}
  \Psi_1&=&\Psi_1',\\
  \Psi_2&=&\sqrt{|\Gamma_0|}\left(\Psi_2'+I\left(\dfrac{1}{\sqrt{|\Gamma_0|}}-I^2\right)^{1/2}\Psi_1'\right),
\end{eqnarray}
and by this reason
\begin{eqnarray}
\!\!\!\!\!\!\!\!\!\!\!\!\!\!\!\!\!\!\!\!\!\!\!\!\!\!\!\!\!\!
  (\Psi_1,\Psi_2)&=&\left(\Psi_1',\sqrt{|\Gamma_0|}\left(\Psi_2'+I\left(\dfrac{1}{\sqrt{|\Gamma_0|}}-I^2\right)^{1/2}\Psi_1'\right)\right)=\nonumber\\
\!\!\!\!\!\!\!\!\!\!\!\!\!\!\!\!\!\!\!\!\!\!\!\!\!\!\!\!\!\!
&=&\sqrt{|\Gamma_0|}\left[(\Psi_1',\Psi_2')+I\left(\dfrac{1}{\sqrt{|\Gamma_0|}}-I^2\right)^{1/2}(\Psi_1',\Psi_1')\right]=\nonumber\\
\!\!\!\!\!\!\!\!\!\!\!\!\!\!\!\!\!\!\!\!\!\!\!\!\!\!\!\!\!\!
&=&\sqrt{|\Gamma_0|}I\left(\dfrac{1}{\sqrt{|\Gamma_0|}}-I^2\right)^{1/2}.
\end{eqnarray}
In the light of the relation (\ref{scalpro}) one has
\begin{equation}
 \sqrt{|\Gamma_0|}I\left(\dfrac{1}{\sqrt{|\Gamma_0|}}-I^2\right)^{1/2}=\sqrt{|\Gamma_0|}I,
\end{equation}
which, because of $I\geqslant2$, gives uniquely and unambiguously
\begin{equation}
  \left(\dfrac{1}{\sqrt{|\Gamma_0|}}-I^2\right)^{1/2}=1,
\end{equation}
or equivalently
\begin{equation}
  I^2=\dfrac{1}{\sqrt{|\Gamma_0|}}-1.
\end{equation}
In the light of the fact $I\geqslant2$ one has
\begin{equation}
  \sqrt{|\Gamma_0|}\leqslant\dfrac{1}{5},
\end{equation}
what however is inconsistent in the light of the fact (\ref{gamin}). In this manner the problem of superposition in the theory of quantum gravity can not be solved by application of the Gram--Schmidt algorithm\index{Gram--Schmidt algorithm} of orthonormalization to the Hilbert space\index{Hilbert space} constructed in this subsection.
\section{Problem III: The Problem of Time}

It is easy to see that within the invariant global one-dimensional quantum gravity the problem of time\index{problem of time} is naturally solved. Namely the role of physical time plays the invariant global dimension
\begin{equation}
\xi=\dfrac{1}{4\pi}\sqrt{\dfrac{h}{6}}\equiv \dfrac{t_{\textrm{physical}}}{\tau},
\end{equation}
where $\tau$ is a reference constant, which can be taken \emph{ad hoc} as the Planck time, i.e. $\tau=t_P$. Such an identification is justif\/ied by the fact that $\xi$ is invariant with respect to action of the spatial diffeomorphisms group, because of the volume form $\sqrt{h}$ is such a diffeoinvariant. In this manner, in the sense of Kucha\v{r} $\xi$ is an observable\index{Kucha\v{r} observable!time}.

Such a situation is analogous to the quantum cosmology presented in the chapter 5, in which the role of time plays the cosmic scale factor parameter $a$\index{cosmic scale factor parameter}. This parameter is a function of the conformal or cosmological time\index{conformal time}. By this reason the proposed model of quantum gravity can be rewritten in more fashionable form
\begin{equation}
  \left(\dfrac{d^2}{dt^2}+V[t]\right)\Psi(t,\phi)=0.
\end{equation}
This global one-dimensional evolutionary equation defines in itself non-trivial dynamics. Namely, this is the time-dependent $0+1$ Schr\"odinger wave equation\index{Schr\"odinger equation!$0+1$ time-dependent}.

\chapter{Examples of Invariant 1D Wave Functions}

Let us present with no detailed computations the wave functionals associated to the three classical solutions of the Einstein f\/ield equations\index{Einstein f\/ield equations}.

\section{The Minkowski Space-time}

Let us consider f\/irst empty space with no cosmological constant\index{cosmological constant}, i.e. the Minkowski space-time\index{space-time!Minkowski}. In such a case, in the Cartesian coordinates, the spatial metric coincides with the metric of the Euclidean space
\begin{equation}
  h_{ij}=\left[\begin{array}{ccc}1&0&0\\0&1&0\\
  0&0&1\end{array}\right],
\end{equation}
and therefore $h=1$ so that the invariant global dimension is
\begin{equation}
  \xi=\dfrac{1}{4\pi}\dfrac{1}{\sqrt{6}}.
\end{equation}
The Minkowski space-time is characterized by
\begin{equation}
  {^{(3)}}R=0\quad,\quad\varrho=0\quad,\quad\Lambda=0,
\end{equation}
and by this reason the averaged generalized gravitational potential\index{gravitational potential!generalized} is
\begin{equation}
  \langle{V}\rangle(\xi,\phi)=0.
\end{equation}
In this manner the wave functions for this case are
\begin{eqnarray}
  \Psi_1&=&\dfrac{1}{\sqrt{|A|}}\xi,\\
  \Psi_2&=&\dfrac{1}{\sqrt{|B|}},
\end{eqnarray}
where the integration constants $A$ and $B$ are
\begin{eqnarray}
A&=&\dfrac{1}{18(4\pi)^3\sqrt{6}},\\
B&=&\dfrac{1}{4\Gamma_0\pi\sqrt{6}}.
\end{eqnarray}
In other words the wave functions
\begin{eqnarray}
  \Psi_1&=&\sqrt{12\pi\sqrt{6}},\\
  \Psi_2&=&\sqrt{4|\Gamma_0|\pi\sqrt{6}},
\end{eqnarray}
are constant on the midisuperspace\index{midisuperspace}. This case is singular because of constant wave functions can not be normalized to unity separately. However, taking the superposed state
\begin{equation}
  \Psi=\alpha_1\Psi_1+\alpha_2\Psi_2,
\end{equation}
one obtains the normalization condition
\begin{equation}
  3|\alpha_1|^2+|\Gamma_0||\alpha_2|^2+\left(\alpha_1^\star\alpha_2+\alpha_1\alpha_2^\star\right)\sqrt{3|\Gamma_0|}=1,
\end{equation}
having the following solutions
\begin{eqnarray}
  \alpha_1&=&\dfrac{\beta_1}{\sqrt{3}}\exp(i\alpha),\\
  \alpha_2&=&\dfrac{\beta_2}{\sqrt{|\Gamma_0|}}\exp(i\alpha),
\end{eqnarray}
where $\alpha$ is an arbitrary real phase, and the constants $\beta_1$ and $\beta_2$ are constrained by
\begin{equation}
  \beta_2=\pm1-\beta_1.
\end{equation}
In other words the superposed states
\begin{equation}\label{minwf}
  \Psi_{\pm}^M=\exp(i\alpha)\left(\dfrac{\beta}{\sqrt{3}}\Psi_1+\dfrac{\pm1-\beta}{\sqrt{|\Gamma_0|}}\Psi_2\right)=\pm\exp(i\alpha)\sqrt{4\pi\sqrt{6}},
\end{equation}
are consistent wave functions of the Minkowski space-time. Because of the Minkowski vacuum is nonlinearly stable, the wave functions (\ref{minwf}) are the reference states for another $\pi$ number def\/inition \index{$\pi$ number def\/inition}
\begin{equation}
  \pi:=\dfrac{\left|\Psi_{\pm}^M\right|^2}{4\sqrt{6}}.
\end{equation}

\section{The Kasner Space-time}

Let us consider the simple solution of the Einstein f\/ield equations\index{Einstein f\/ield equations} describing an anisotropic universe without matter, called the Kasner metric. The spatial part of this metric is
\begin{equation}
  h_{ij}=\left[\begin{array}{ccc}t^{2p_1}&0&0\\0&t^{2p_2}&0\\0&0&t^{2p_3}\end{array}\right],
\end{equation}
where $t$ is time coordinate, and
\begin{eqnarray}
  \sum_ip_i&=&1,\\
  \sum_ip_i^2&=&1.
\end{eqnarray}
In this case the global dimension is
\begin{equation}
  h=t^2,
\end{equation}
while the invariant global dimension is
\begin{equation}
  \xi=\dfrac{t}{4\pi\sqrt{6}}.
\end{equation}
Because of the cosmological constant\index{cosmological constant} and the Matter f\/ields\index{Matter f\/ields} are absent here, and the three-dimensional Ricci scalar curvature\index{Ricci scalar curvature} vanishes one has $\langle{V}\rangle=0$. In this manner the wave functions of the Kasner space-time\index{space-time!Kasner} are
\begin{eqnarray}
  \Psi_1^{K}&=&\dfrac{1}{\sqrt{|A|}}\dfrac{t}{4\pi\sqrt{6}},\\
  \Psi_2^{K}&=&\dfrac{1}{\sqrt{|B|}},
\end{eqnarray}
where the constants of integration are
\begin{eqnarray}
  A&=&\dfrac{1}{3}\dfrac{T^3}{6(4\pi)^3\sqrt{6}},\\
  B&=&\dfrac{1}{\Gamma_0}\dfrac{T}{4\pi\sqrt{6}},
\end{eqnarray}
where $T$ is some reference value of time $t$.

\section{The Schwarzschild Space-time}

The second situation which we shall present here is the Schwarzschild space-time\index{space-time!Schwarzschild}. This metric is a spherically symmetric vacuum solution of the Einstein f\/ield equations\index{Einstein f\/ield equations} with no cosmological constant\index{cosmological constant}, i.e. a situation
${^{(3)}\!R[h]}=0$, $\Lambda=0$, $\varrho=0$. It means that in this case also $\langle V\rangle(h_{ij},\phi)=0$. The spatial part of the metric has the form
\begin{equation}
  h_{ij}=\left[\begin{array}{ccc}\left(1-\dfrac{r_S}{r}\right)^{-1}&0&0\\0&r^2&0\\
  0&0&r^2\sin^2\theta\end{array}\right],
\end{equation}
where $r_S$ is the Schwarzschild radius of the spherically symmetric non-rotating object of mass $M$
\begin{equation}
  r_S=\dfrac{2GM}{c^2}=\dfrac{\kappa}{4\pi}Mc^2.
\end{equation}
Therefore in this case the global dimension is
\begin{equation}
  h=\dfrac{r^4\sin^2\theta}{1-\dfrac{r_S}{r}},
\end{equation}
so that the invariant global dimension is
\begin{equation}
  \xi=\dfrac{1}{4\pi}\dfrac{r^2\sin\theta}{\sqrt{6\left(1-\dfrac{r_S}{r}\right)}},
\end{equation}
and the total volume of the Schwarzschild midisuperspace\index{midisuperspace} is $\Omega(\xi)=\xi$. The integration constants $A$ and $B$ can be established easy
\begin{eqnarray}
A&=&\int\Omega^2(\xi)\delta\xi=\dfrac{\Xi^3}{3}=\dfrac{1}{(4\pi)^3}\dfrac{R^6\sin^3\Theta}{\left(6\left(1-\dfrac{r_S}{R}\right)\right)^{3/2}},\\
B&=&\dfrac{1}{\Gamma_0}\int\delta\xi=\dfrac{\Xi}{\Gamma_0}=\dfrac{1}{4\Gamma_0\pi}\dfrac{R^2\sin\Theta}{\sqrt{6\left(1-\dfrac{r_S}{R}\right)}},
\end{eqnarray}
where $\Xi=\xi(R,\Theta)$ is
\begin{equation}
  \Xi=\dfrac{1}{4\pi}\dfrac{R^2\sin\Theta}{\sqrt{6\left(1-\dfrac{r_S}{R}\right)}}.
\end{equation}
In this manner the wave functions of the Schwarzschild space-time are
\begin{eqnarray}
  \Psi_1^{S}&=&\sqrt{4\pi\sqrt{6}}\sqrt{\dfrac{1}{R^2\sin\Theta}\sqrt{1-\dfrac{r_S}{R}}}\sqrt{\dfrac{1-\dfrac{r_S}{R}}{1-\dfrac{r_S}{r}}}\dfrac{r^2\sin\theta}{R^2\sin\Theta},\\
  \Psi_2^{S}&=&\sqrt{4\pi|\Gamma_0|\sqrt{6}}\sqrt{\dfrac{1}{R^2\sin\Theta}\sqrt{1-\dfrac{r_S}{R}}}.
\end{eqnarray}

\section{The (Anti-) De Sitter Space-time}

Let us consider now the spherically symmetric solution of vacuum Einstein's f\/ield equations in presence of the cosmological constant\index{cosmological constant}, i.e. lambdavacuum Einstein's f\/ield equations
\begin{equation}
R_{\mu\nu}=\Lambda g_{\mu\nu},\label{lamvac}
\end{equation}
called the (Anti-) De Sitter space-time\index{space-time!De Sitter}. Such a space-time is a submanifold of the Minkowski space-time of one higher dimension described by the hyperboloid of one sheet
\begin{equation}
  -x_0^2+x_1^2+x_2^2+x_3^2=\pm\alpha^2,
\end{equation}
where the sign plus describes De Sitter space-time\index{space-time!De Sitter}, whereas the sign minus is related to the Anti-De Sitter space\index{space-time!Anti-De Sitter}. The (Anti-) De Sitter space-time is an Einstein manifold
\begin{equation}
R_{\mu\nu}=\pm\dfrac{3}{\alpha^2}g_{\mu\nu},
\end{equation}
what in the light of the lambdavacuum Einstein f\/ield equations\index{Einstein f\/ield equations} (\ref{lamvac}) the cosmological constant\index{cosmological constant} is
\begin{equation}
  \Lambda=\pm\dfrac{3}{\alpha^2}.
\end{equation}
The three-dimensional Ricci scalar curvature\index{Ricci scalar curvature} of the (Anti-) De Sitter space-time\index{space-time!Anti-De Sitter}\index{space-time!De Sitter} can be established easy
\begin{equation}
  {^{(3)}}R=h^{ij}R_{ij}=3\Lambda,
\end{equation}
and because of there is no Matter f\/ields\index{Matter f\/ields} $\varrho=0$, so that the averaged generalized gravitational potential\index{gravitational potential!generalized!averaged} has the value
\begin{equation}
  \langle{V}\rangle(\xi,\phi)=-3\Lambda+2\Lambda=-\Lambda.
\end{equation}
In other words for the De Sitter space-time\index{space-time!De Sitter}
\begin{equation}
\langle{V}\rangle(\xi,\phi)=-\dfrac{3}{\alpha^2},
\end{equation}
while for the Anti-De Sitter space-time\index{space-time!Anti-De Sitter}
\begin{equation}
\langle{V}\rangle(\xi,\phi)=\dfrac{3}{\alpha^2}.
\end{equation}
The spatial part of the (Anti-) De Sitter metric is
\begin{equation}
  h_{ij}=\left[\begin{array}{ccc}\left(1-\dfrac{\Lambda}{3}r^2\right)^{-1}&0&0\\0&r^2&0\\
  0&0&r^2\sin^2\theta\end{array}\right],
\end{equation}
and by this reason the global dimension is
\begin{equation}
  h=\dfrac{r^4\sin^2\theta}{1-\dfrac{\Lambda}{3}r^2},
\end{equation}
whereas the invariant global dimension is
\begin{equation}
  \xi=\dfrac{1}{4\pi\sqrt{6}}\dfrac{r^2\sin\theta}{\sqrt{1-\dfrac{\Lambda}{3}r^2}},
\end{equation}
and of course the volume of the related midisuperspace\index{midisuperspace} is $\Omega(\xi)=\xi$.

Let us consider f\/irst the case of the De Sitter space-time\index{space-time!De Sitter}, i.e. $\Lambda=\dfrac{3}{\alpha^2}$. The constants of integration can be evaluated as follows
\begin{eqnarray}
A&=&\int\left(\dfrac{\sinh\left(\sqrt{\Lambda}\xi\right)}{\sqrt{\Lambda}}\right)^2\delta\xi=\dfrac{1}{\Lambda}\left(-\dfrac{\Xi}{2}+\dfrac{\sinh\left(2\sqrt{\Lambda}\Xi\right)}{4\sqrt{\Lambda}}\right),\\
B&=&\dfrac{1}{\Gamma_0}\int\cosh^2\left(\sqrt{\Lambda}\xi\right)\delta\xi=\dfrac{1}{\Gamma_0}\left(\dfrac{\Xi}{2}+\dfrac{\cosh\left(2\sqrt{\Lambda}\Xi\right)}{4\sqrt{\Lambda}}\right),
\end{eqnarray}
where $\Xi$ is the reference constant
\begin{equation}
\Xi=\dfrac{1}{4\pi\sqrt{6}}\dfrac{R^2\sin\Theta}{\sqrt{1-\dfrac{\Lambda}{3}R^2}}.
\end{equation}
By this reason the wave functions of the De Sitter space-time\index{space-time!De Sitter} are
\begin{eqnarray}
  \Psi_1^{DS}&=&\dfrac{1}{\sqrt{|A|}}\dfrac{1}{\sqrt{\Lambda}}\sinh\left(\dfrac{\sqrt{\Lambda}}{4\pi\sqrt{6}}\dfrac{r^2\sin\theta}{\sqrt{1-\dfrac{\Lambda}{3}r^2}}\right),\\
  \Psi_2^{DS}&=&\dfrac{1}{\sqrt{|B|}}\cosh\left(\dfrac{\sqrt{\Lambda}}{4\pi\sqrt{6}}\dfrac{r^2\sin\theta}{\sqrt{1-\dfrac{\Lambda}{3}r^2}}\right).
\end{eqnarray}

Similarly for the case of Anti-De Sitter space-time \index{space-time!Anti-De Sitter}one obtains the following wave functions
\begin{eqnarray}
  \Psi_1^{ADS}&=&\dfrac{1}{\sqrt{|A|}}\dfrac{1}{\sqrt{|\Lambda|}}\sin\left(\dfrac{\sqrt{|\Lambda|}}{4\pi\sqrt{6}}\dfrac{r^2\sin\theta}{\sqrt{1+\dfrac{|\Lambda|}{3}r^2}}\right),\\
  \Psi_2^{ADS}&=&\dfrac{1}{\sqrt{|B|}}\cos\left(\dfrac{\sqrt{|\Lambda|}}{4\pi\sqrt{6}}\dfrac{r^2\sin\theta}{\sqrt{1+\dfrac{|\Lambda|}{3}r^2}}\right),
\end{eqnarray}
where $|\Lambda|=\dfrac{3}{\alpha^2}$, and the constants of integration $A$ and $B$ are
\begin{eqnarray}
A&=&\int\left(\dfrac{\sin\left(\sqrt{|\Lambda|}\xi\right)}{\sqrt{|\Lambda|}}\right)^2\delta\xi=\dfrac{1}{|\Lambda|}\left(\dfrac{\Xi}{2}-\dfrac{\sin\left(2\sqrt{|\Lambda|}\Xi\right)}{4\sqrt{|\Lambda|}}\right),\\
B&=&\dfrac{1}{\Gamma_0}\int\cos^2\left(\sqrt{|\Lambda|}\xi\right)\delta\xi=\dfrac{1}{\Gamma_0}\left(\dfrac{\Xi}{2}+\dfrac{\cos\left(2\sqrt{|\Lambda|}\Xi\right)}{4\sqrt{|\Lambda|}}\right),
\end{eqnarray}
where $\Xi$ is the reference constant
\begin{equation}
\Xi=\dfrac{1}{4\pi\sqrt{6}}\dfrac{R^2\sin\Theta}{\sqrt{1+\dfrac{|\Lambda|}{3}R^2}}.
\end{equation}

\section{The (Anti-) De Sitter--Schwarzschild Space-time}
The (Anti-) De Sitter--Schwarzschild space-time\index{space-time!Anti-De Sitter--Schwarzschild}\index{space-time!De Sitter--Schwarzschild} is the case jointing the (Anti-) De Sitter and Schwarzschild space-times. This is the spherically symmetric solution of lambdavacuum Einstein's f\/ield equations with suitable boundary conditions. The spatial part of space-time metric has the form
\begin{equation}
  h_{ij}=\left[\begin{array}{ccc}\left(1-\dfrac{r_S}{r}-\dfrac{\Lambda}{3}r^2\right)^{-1}&0&0\\0&r^2&0\\
  0&0&r^2\sin^2\theta\end{array}\right].
\end{equation}
In this manner the global dimension is
\begin{equation}
  h=\dfrac{r^4\sin^2\theta}{1-\dfrac{r_S}{r}-\dfrac{\Lambda}{3}r^2},
\end{equation}
while the invariant global dimension is
\begin{equation}
  \xi=\dfrac{1}{4\pi\sqrt{6}}\dfrac{r^2\sin\theta}{\sqrt{1-\dfrac{r_S}{r}-\dfrac{\Lambda}{3}r^2}},
\end{equation}
and of course the volume of the related midisuperspace\index{midisuperspace} is $\Omega(\xi)=\xi$. The wave functions of the De Sitter--Schwarzschild space-time\index{space-time!De Sitter--Schwarzschild}, called also the Kottler space-time\index{space-time!Kottler}, can be written in the form
\begin{eqnarray}
  \Psi_1^{DS-S}&=&\dfrac{1}{\sqrt{|A|}}\dfrac{1}{\sqrt{\Lambda}}\sinh\left(\dfrac{\sqrt{\Lambda}}{4\pi\sqrt{6}}\dfrac{r^2\sin\theta}{\sqrt{1-\dfrac{r_S}{r}-\dfrac{\Lambda}{3}r^2}}\right),\\
  \Psi_2^{DS-S}&=&\dfrac{1}{\sqrt{|B|}}\cosh\left(\dfrac{\sqrt{\Lambda}}{4\pi\sqrt{6}}\dfrac{r^2\sin\theta}{\sqrt{1-\dfrac{r_S}{r}-\dfrac{\Lambda}{3}r^2}}\right).
\end{eqnarray}
where the constants of integration $A$ and $B$ are
\begin{eqnarray}
A&=&\int\left(\dfrac{\sinh\left(\sqrt{\Lambda}\xi\right)}{\sqrt{\Lambda}}\right)^2\delta\xi=\dfrac{1}{\Lambda}\left(-\dfrac{\Xi}{2}+\dfrac{\sinh\left(2\sqrt{\Lambda}\Xi\right)}{4\sqrt{\Lambda}}\right),\\
B&=&\dfrac{1}{\Gamma_0}\int\cosh^2\left(\sqrt{\Lambda}\xi\right)\delta\xi=\dfrac{1}{\Gamma_0}\left(\dfrac{\Xi}{2}+\dfrac{\cosh\left(2\sqrt{\Lambda}\Xi\right)}{4\sqrt{\Lambda}}\right),
\end{eqnarray}
where $\Xi$ is the reference constant
\begin{equation}
\Xi=\dfrac{1}{4\pi\sqrt{6}}\dfrac{R^2\sin\Theta}{\sqrt{1-\dfrac{r_S}{R}-\dfrac{\Lambda}{3}R^2}}.
\end{equation}

Similarly, the wave functions of the Anti-De Sitter--Schwarzschild space-time \index{space-time!Anti-De Sitter--Schwarzschild}are
\begin{eqnarray}
  \Psi_1^{ADS-S}&=&\dfrac{1}{\sqrt{|A|}}\dfrac{1}{\sqrt{|\Lambda|}}\sin\left(\dfrac{\sqrt{|\Lambda|}}{4\pi\sqrt{6}}\dfrac{r^2\sin\theta}{\sqrt{1-\dfrac{r_S}{r}+\dfrac{|\Lambda|}{3}r^2}}\right),\\
  \Psi_2^{ADS-S}&=&\dfrac{1}{\sqrt{|B|}}\cos\left(\dfrac{\sqrt{|\Lambda|}}{4\pi\sqrt{6}}\dfrac{r^2\sin\theta}{\sqrt{1-\dfrac{r_S}{r}+\dfrac{|\Lambda|}{3}r^2}}\right),
\end{eqnarray}
where $|\Lambda|=\dfrac{3}{\alpha^2}$, and the constants of integration $A$ and $B$ are
\begin{eqnarray}
A&=&\int\left(\dfrac{\sin\left(\sqrt{|\Lambda|}\xi\right)}{\sqrt{|\Lambda|}}\right)^2\delta\xi=\dfrac{1}{|\Lambda|}\left(\dfrac{\Xi}{2}-\dfrac{\sin\left(2\sqrt{|\Lambda|}\Xi\right)}{4\sqrt{|\Lambda|}}\right),\\
B&=&\dfrac{1}{\Gamma_0}\int\cos^2\left(\sqrt{|\Lambda|}\xi\right)\delta\xi=\dfrac{1}{\Gamma_0}\left(\dfrac{\Xi}{2}+\dfrac{\cos\left(2\sqrt{|\Lambda|}\Xi\right)}{4\sqrt{|\Lambda|}}\right),
\end{eqnarray}
where $\Xi$ is the reference constant
\begin{equation}
\Xi=\dfrac{1}{4\pi\sqrt{6}}\dfrac{R^2\sin\Theta}{\sqrt{1-\dfrac{r_S}{R}+\dfrac{|\Lambda|}{3}R^2}}.
\end{equation}

\section{The Kerr Space-Time}

Kerr's space-time\index{space-time!Kerr} is a solution of vacuum Einstein's f\/ield equations for spherical body rotating with the angular momentum $J$. Because of the spatial Ricci scalar curvature\index{Ricci scalar curvature} is zero, cosmological constant\index{cosmological constant} is not present, and Matter f\/ields\index{Matter f\/ields} are absent, in such a situation the averaged generalized gravitational potential\index{gravitational potential!generalized!averaged} identically vanishes. The spatial part of the Kerr metric has the form
\begin{equation}
  h_{ij}=\left[\begin{array}{ccc}\dfrac{r^2+\alpha^2\cos^2\theta}{r^2-r_sr+\alpha^2}&0&0\\0&r^2+\alpha^2\cos^2\theta&0\\
  0&0&r^2+\alpha^2+\dfrac{\alpha^2r_Sr\sin^2\theta}{r^2+\alpha^2\cos^2\theta}\end{array}\right],
\end{equation}
where $\alpha$ is the coeff\/icient related to the angular momentum and the mass of the rotating object
\begin{equation}
  \alpha=\dfrac{J}{Mc}.
\end{equation}
In this manner the global dimension is
\begin{equation}
  h=\dfrac{\left(r^2+\alpha^2\cos^2\theta\right)^2}{r^2-r_sr+\alpha^2}\left(r^2+\alpha^2+\dfrac{\alpha^2r_Sr\sin^2\theta}{r^2+\alpha^2\cos^2\theta}\right),
\end{equation}
while the invariant global dimension is
\begin{equation}
  \xi=\dfrac{1}{4\pi\sqrt{6}}\left(r^2+\alpha^2\cos^2\theta\right)\sqrt{\dfrac{\left(r^2+\alpha^2\right)^2-\alpha^2\left(r^2-r_Sr+\alpha^2\right)\sin^2\theta}{\left(r^2-r_sr+\alpha^2\right)\left(r^2+\alpha^2\cos^2\theta\right)}},
\end{equation}
and of course the volume of the Kerr midisuperspace\index{midisuperspace} is $\Omega(\xi)=\xi$. Introducing the reference parameter
\begin{equation}
  \Xi=\dfrac{1}{4\pi\sqrt{6}}\left(R^2+\alpha^2\cos^2\Theta\right)\sqrt{\dfrac{\left(R^2+\alpha^2\right)^2-\alpha^2\left(R^2-r_SR+\alpha^2\right)\sin^2\Theta}{\left(R^2-r_sR+\alpha^2\right)\left(R^2+\alpha^2\cos^2\Theta\right)}},
\end{equation}
one can write out the constants of integration $A$ and $B$
\begin{eqnarray}
A&=&\dfrac{\Xi^3}{3},\\
B&=&\dfrac{1}{\Gamma_0}\Xi,
\end{eqnarray}
so that the wave functions of the Kerr space-time \index{space-time!Kerr}have the form
\begin{eqnarray}
\!\!\!\!\!\!\!\!\!\!\!\!\!\!\!
\Psi_1^{Kerr}&=&\dfrac{r^2+\alpha^2\cos^2\theta}{4\pi\sqrt{|A|}\sqrt{6}}\sqrt{\dfrac{\left(r^2+\alpha^2\right)^2-\alpha^2\left(r^2-r_Sr+\alpha^2\right)\sin^2\theta}{\left(r^2-r_sr+\alpha^2\right)\left(r^2+\alpha^2\cos^2\theta\right)}},\\
\!\!\!\!\!\!\!\!\!\!\!\!\!\!\!
\Psi_2^{Kerr}&=&\dfrac{1}{\sqrt{|B|}}.
\end{eqnarray}

\section{The Kerr--Newman Space-time}

The Kerr--Newman space-time \index{space-time!Kerr--Newman}is a solution of the electrovacuum Einstein f\/ield equations\index{Einstein f\/ield equations} called also Einstein--Maxwell equations\index{Einstein--Maxwell equations}, which are combination of the Einstein f\/ield equations\index{Einstein f\/ield equations} with the stress-energy tensor of electromagnetic f\/ield\index{stress-energy tensor!electromagnetic f\/ield} and no cosmological constant\index{cosmological constant}, and the source-free Maxwell equations\index{Maxwell equations}
\begin{eqnarray}
  &&R_{\mu\nu}=\kappa\ell_P^2{T}_{\mu\nu}^{em},\label{emeq}\\
  &&F_{\mu\nu;\kappa}+F_{\nu\kappa;\mu}+F_{\kappa\mu;\nu}=0,\\
  &&F^{\mu\nu}_{;\nu}=0,
\end{eqnarray}
where $T_{\mu\nu}^{em}$ is the stress-energy tensor\index{stress-energy tensor} of electromagnetic f\/ield, which can be given \emph{ad hoc} in analogy with the case of f\/lat space-time
\begin{equation}\label{setem}
  T_{\mu\nu}^{em}=\dfrac{1}{\mu_0}\left(F^\alpha_\mu{F}_{\alpha\nu}-\dfrac{1}{4}g_{\mu\nu}F_{\alpha\beta}F^{\alpha\beta}\right),
\end{equation}
and $F_{\mu\nu}$ is the electromagnetic f\/ield tensor\index{electromagnetic f\/ield tensor}
\begin{eqnarray}
  F_{\mu\nu}&=&\left[\begin{array}{cccc}0&\dfrac{E_x}{c}&\dfrac{E_y}{c}&\dfrac{E_z}{c}\\
  -\dfrac{E_x}{c}&0&-B_z&-B_y\\
  -\dfrac{E_y}{c}&B_z&0&-B_x\\
  -\dfrac{E_z}{c}&-B_y&B_x&0\end{array}\right],\\
  F^{\mu\nu}&=&\left[\begin{array}{cccc}0&-\dfrac{E_x}{c}&-\dfrac{E_y}{c}&-\dfrac{E_z}{c}\\
  \dfrac{E_x}{c}&0&-B_z&-B_y\\
  \dfrac{E_y}{c}&B_z&0&-B_x\\
  \dfrac{E_z}{c}&-B_y&B_x&0\end{array}\right],
\end{eqnarray}
where $\vec{E}=[E_x,E_y,E_z]$ and $\vec{B}=[B_x,B_y,B_z]$ are three-vectors of the electric and the magnetic f\/ield, respectively. The speed of light $c$ can be expressed in terms of the vacuum permeability $\mu_0$ and the vacuum permittivity $\epsilon_0$ via the formula $\epsilon_0\mu_0=1/c^2$.

There is the question how to derive the stress-energy tensor\index{stress-energy tensor} of electromagnetic f\/ield (\ref{setem}) via using of the def\/inition
\begin{equation}\label{sethp}
  T_{\mu\nu}=-\dfrac{2}{\sqrt{-g}}\dfrac{\delta(\sqrt{-g}\mathcal{L})}{\delta g^{\mu\nu}},
\end{equation}
following from the Hilbert--Palatini action principle, and describing Matter f\/ields\index{Matter f\/ields} characterized by Lagrangian $\mathcal{L}$ in a four-dimensional Riemannian space-time with a metric $g^{\mu\nu}$. The Lagrangian of the Maxwell electromagnetic f\/ield has the form (For more general approach see e.g. the Ref. \cite{ibb})
\begin{equation}
  \mathcal{L}=-\dfrac{1}{4\mu_0}F_{\alpha\beta}F^{\alpha\beta}.
\end{equation}
With using of the Jacobi formula $\delta\sqrt{-g}=-\dfrac{1}{2}\sqrt{-g}g_{\mu\nu}\delta{g}^{\mu\nu}$ the stress-energy tensor\index{stress-energy tensor} (\ref{sethp}) can be rewritten in more convenient form
\begin{equation}
  T_{\mu\nu}=-2\dfrac{\delta\mathcal{L}}{\delta g^{\mu\nu}}+g_{\mu\nu}\mathcal{L},
\end{equation}
so that the problem is the establish the functional derivative
\begin{equation}
\dfrac{\delta\mathcal{L}}{\delta g^{\mu\nu}}=-\dfrac{1}{4\mu_0}\dfrac{\delta}{\delta g^{\mu\nu}}\left(F_{\alpha\beta}F^{\alpha\beta}\right)=-\dfrac{1}{4\mu_0}\left(\dfrac{\delta{F}_{\alpha\beta}}{\delta g^{\mu\nu}}F^{\alpha\beta}+F_{\alpha\beta}\dfrac{\delta{F}^{\alpha\beta}}{\delta g^{\mu\nu}}\right).
\end{equation}
Let us notice that the tensor $F_{\alpha\beta}$ can be presented in more complex form
\begin{eqnarray}
\!\!\!\!\!\!\!\!\!\!\!\!\!\!\!\!\!\!\!\!\!\!\!\!\!\!\!\!\!\!
F_{\alpha\beta}&=&g^\gamma_\alpha{g}^\delta_\beta{F}_{\gamma\delta}=\left(g^\gamma_\mu{g}^{\mu\nu}g_{\nu\alpha}\right)\left(g^\delta_\nu{g}^{\nu\mu}g_{\mu\beta}\right){F}_{\gamma\delta}=\nonumber\\
\!\!\!\!\!\!\!\!\!\!\!\!\!\!\!\!\!\!\!\!\!\!\!\!\!\!\!\!\!\!
&=&\left(g^\gamma_\mu{g}^{\mu\nu}g_{\nu\alpha}\right)\left(g^\delta_\nu{g}^{\mu\nu}g_{\mu\beta}\right){F}_{\gamma\delta}={g}^{\mu\nu}{g}^{\mu\nu}\left(g^\gamma_\mu{g}_{\nu\alpha}g^\delta_\nu{g}_{\mu\beta}\right){F}_{\gamma\delta}.
\end{eqnarray}
In this manner one has
\begin{eqnarray}
  \dfrac{\delta{F}_{\alpha\beta}}{\delta{g}^{\mu\nu}}&=&2{g}^{\mu\nu}\left(g^\gamma_\mu{g}_{\nu\alpha}g^\delta_\nu{g}_{\mu\beta}\right){F}_{\gamma\delta}=2g^\gamma_\mu{g}^\delta_\nu\left({g}^{\mu\nu}g_{\nu\alpha}\right){g}_{\mu\beta}{F}_{\gamma\delta}=\nonumber\\
  &=&2\left(g^\gamma_\mu{g}^\delta_\nu{g}^\mu_\alpha{g}_{\mu\beta}\right){F}_{\gamma\delta}=2\left(g^\gamma_\mu{g}^\mu_\alpha\right){g}^\delta_\nu{g}_{\mu\beta}{F}_{\gamma\delta}=\nonumber\\
  &=&2\left(g^\gamma_\alpha{g}^\delta_\nu{g}_{\mu\beta}\right){F}_{\gamma\delta}=2\left({g}^\delta_\nu{g}_{\mu\beta}\right)g^\gamma_\alpha{F}_{\gamma\delta}=2\left({g}^\delta_\nu{g}_{\mu\beta}\right){F}_{\alpha\delta}=\nonumber\\
  &=&-2{g}_{\mu\beta}\left({g}^\delta_\nu{F}_{\delta\alpha}\right)=-2{g}_{\mu\beta}F_{\nu\alpha},
\end{eqnarray}
and consequently
\begin{equation}
  \dfrac{\delta{F}_{\alpha\beta}}{\delta g^{\mu\nu}}F^{\alpha\beta}=-2{g}_{\mu\beta}F_{\nu\alpha}F^{\alpha\beta}=2F_{\nu\alpha}\left({g}_{\mu\beta}F^{\beta\alpha}\right)=2F_{\nu\alpha}F_\mu^\alpha=2F_\mu^\alpha{F}_{\alpha\nu}.\label{derf}
\end{equation}
Similarly, one can rewrite ${F}^{\alpha\beta}$ in the form
\begin{equation}
  F^{\alpha\beta}=g^{\alpha\kappa}g^{\beta\lambda}F_{\kappa\lambda},
\end{equation}
and consequently one has
\begin{eqnarray}
\dfrac{\delta{F}^{\alpha\beta}}{\delta{g}^{\mu\nu}}&=&\dfrac{\delta{g}^{\alpha\kappa}}{\delta{g}^{\mu\nu}}g^{\beta\lambda}F_{\kappa\lambda}+{g}^{\alpha\kappa}\dfrac{\delta{g}^{\beta\lambda}}{\delta{g}^{\mu\nu}}F_{\kappa\lambda}+{g}^{\alpha\kappa}{g}^{\beta\lambda}\dfrac{\delta{F}_{\kappa\lambda}}{\delta{g}^{\mu\nu}}=\nonumber\\
&=&g^\alpha_\mu{g}^\kappa_\nu{g}^{\beta\lambda}F_{\kappa\lambda}+{g}^{\alpha\kappa}g^\beta_\mu{g}^\lambda_\nu{F}_{\kappa\lambda}+{g}^{\alpha\kappa}{g}^{\beta\lambda}\dfrac{\delta{F}_{\kappa\lambda}}{\delta{g}^{\mu\nu}}=\nonumber\\
&=&-g^\alpha_\mu{g}^\kappa_\nu{g}^{\beta\lambda}{F}_{\lambda\kappa}-g^{\alpha\kappa}g^\beta_\mu{g}^\lambda_\nu{F}_{\lambda\kappa}+{g}^{\alpha\kappa}{g}^{\beta\lambda}\dfrac{\delta{F}_{\kappa\lambda}}{\delta{g}^{\mu\nu}}=\nonumber\\
&=&-g^\alpha_\mu{g}^\kappa_\nu{F}^\beta_\kappa-g^{\alpha\kappa}g^\beta_\mu{F}_{\nu\kappa}+{g}^{\alpha\kappa}{g}^{\beta\lambda}\dfrac{\delta{F}_{\kappa\lambda}}{\delta{g}^{\mu\nu}}=\nonumber\\
&=&-g^\alpha_\mu{F}^\beta_\nu-g^\beta_\mu{g}^{\alpha\kappa}{F}_{\nu\kappa}+{g}^{\alpha\kappa}{g}^{\beta\lambda}\dfrac{\delta{F}_{\kappa\lambda}}{\delta{g}^{\mu\nu}}=\nonumber\\
&=&g^\alpha_\mu{F}^\beta_\nu+g^\beta_\mu{g}^{\alpha\kappa}{F}_{\kappa\nu}+{g}^{\alpha\kappa}{g}^{\beta\lambda}\dfrac{\delta{F}_{\kappa\lambda}}{\delta{g}^{\mu\nu}}=\nonumber\\
&=&g^\alpha_\mu{F}^\beta_\nu+g^\beta_\mu{F}^\alpha_\nu+{g}^{\alpha\kappa}{g}^{\beta\lambda}\dfrac{\delta{F}_{\kappa\lambda}}{\delta{g}^{\mu\nu}}.
\end{eqnarray}
By this reason one obtains
\begin{eqnarray}
F_{\alpha\beta}\dfrac{\delta{F}^{\alpha\beta}}{\delta{g}^{\mu\nu}}&=&F_{\alpha\beta}\left[g^\alpha_\mu{F}^\beta_\nu+g^\beta_\mu{F}^\alpha_\nu+{g}^{\alpha\kappa}{g}^{\beta\lambda}\dfrac{\delta{F}_{\kappa\lambda}}{\delta{g}^{\mu\nu}}\right]=\nonumber\\
&=&\left(F_{\alpha\beta}g^\alpha_\mu\right){F}^\beta_\nu+\left(F_{\alpha\beta}g^\beta_\mu\right){F}^\alpha_\nu+\left(F_{\alpha\beta}{g}^{\alpha\kappa}{g}^{\beta\lambda}\right)\dfrac{\delta{F}_{\kappa\lambda}}{\delta{g}^{\mu\nu}}=\nonumber\\
&=&\left(-F_{\beta\alpha}g^\alpha_\mu\right){F}^\beta_\nu-F_{\alpha\mu}{F}^\alpha_\nu-\left(F_{\beta\alpha}{g}^{\alpha\kappa}{g}^{\beta\lambda}\right)\dfrac{\delta{F}_{\kappa\lambda}}{\delta{g}^{\mu\nu}}=\nonumber\\
&=&-F_{\beta\mu}{F}^\beta_\nu+{F}^\alpha_\nu{F}_{\alpha\mu}-\left(F_\beta^\kappa{g}^{\beta\lambda}\right)\dfrac{\delta{F}_{\kappa\lambda}}{\delta{g}^{\mu\nu}}=\nonumber\\
&=&{F}^\beta_\nu{F}_{\beta\mu}+{F}^\alpha_\nu{F}_{\alpha\mu}-F^{\kappa\lambda}\dfrac{\delta{F}_{\kappa\lambda}}{\delta{g}^{\mu\nu}}=\nonumber\\
&=&2{F}^\alpha_\nu{F}_{\alpha\mu}-2F_\mu^\alpha{F}_{\alpha\nu}=0,\label{derf1}
\end{eqnarray}
where we have applied the identity
\begin{equation}
{F}^\alpha_\nu{F}_{\alpha\mu}=g^{\alpha\beta}{F}_{\beta\nu}{F}_{\alpha\mu}=g^{\beta\alpha}{F}_{\alpha\mu}{F}_{\beta\nu}=F_\mu^\beta{F}_{\beta\nu}=F_\mu^\alpha{F}_{\alpha\nu}.
\end{equation}
Taking into account (\ref{derf}) and (\ref{derf1}) one receives
\begin{equation}
\dfrac{\delta{F}_{\alpha\beta}}{\delta{g}^{\mu\nu}}F^{\alpha\beta}+F_{\alpha\beta}\dfrac{\delta{F}^{\alpha\beta}}{\delta{g}^{\mu\nu}}=-2{F}^\alpha_\mu{F}_{\alpha\nu}.
\end{equation}
By this reason one obtains
\begin{equation}
\dfrac{\delta\mathcal{L}}{\delta g^{\mu\nu}}=-\dfrac{1}{4\mu_0}\left(-2{F}^\alpha_\mu{F}_{\alpha\nu}\right)=\dfrac{1}{2\mu_0}{F}^\alpha_\mu{F}_{\alpha\nu}=-\dfrac{1}{2\mu_0}{F}^\alpha_\mu{F}_{\alpha\nu},
\end{equation}
where in the last step we have changed order of indexes in ${F}^\alpha_\mu$, which is invisible by the only inconvenient notation. In this manner one receives f\/inally
\begin{equation}
  T_{\mu\nu}^{em}=-2\dfrac{\delta\mathcal{L}}{\delta g^{\mu\nu}}+g_{\mu\nu}\mathcal{L}=\dfrac{1}{\mu_0}\left({F}^\alpha_\mu{F}_{\alpha\nu}-\dfrac{1}{4}g_{\mu\nu}F_{\alpha\beta}F^{\alpha\beta}\right),
\end{equation}
what coincides with the stress-energy tensor\index{stress-energy tensor} of electromagnetic f\/ield (\ref{setem}) given \emph{ad hoc} in analogy with the case of f\/lat space-time.

For the case of the Kerr--Newman metric in general the spatial Ricci scalar curvature\index{Ricci scalar curvature} and energy density\index{energy density} of Matter f\/ields\index{Matter f\/ields} are manifestly non zero. Let us calculate them straightforwardly. The spatial Ricci curvature tensor\index{Ricci curvature tensor} can be taken from (\ref{emeq}) as $R_{ij}=\kappa{T}^{em}_{ij}$, where
\begin{equation}\label{sttem}
  T_{ij}^{em}=\dfrac{1}{\mu_0}\left(F^\alpha_i{F}_{\alpha{j}}-\dfrac{1}{4}h_{ij}F_{\alpha\beta}F^{\alpha\beta}\right),
\end{equation}
is the spatial part of the stress-energy tensor\index{stress-energy tensor} for electromagnetic f\/ield. When the space-time metric is the Minkowski metric $\eta_{\mu\nu}$ then the spatial part of the stress-energy tensor of electromagnetic f\/ield is
\begin{equation}
 \left.T_{ij}^{em}\right|_{g_{\mu\nu}=\eta_{\mu\nu}}=-\sigma^{f\/lat}_{ij},
\end{equation}
where $\sigma_{ij}^{f\/lat}$ is the Maxwell stress tensor of electromagnetic f\/ield\index{Maxwell stress tensor!electromagnetic f\/ield}
\begin{equation}
  \sigma_{ij}^{f\/lat}=\epsilon_0E_iE_j+\dfrac{1}{\mu_0}B_iB_j-\dfrac{1}{2}\left(\epsilon_0\vec{E}^2+\dfrac{1}{\mu_0}\vec{B}^2\right)\delta_{ij}.
\end{equation}
In this manner the natural generalization of the Maxwell stress tensor of electromagnetic f\/ield\index{Maxwell stress tensor!electromagnetic f\/ield} to the case of non-f\/lat space-time is the tensor
\begin{equation}
  \sigma_{ij}=-\dfrac{1}{\mu_0}\left(F^\alpha_i{F}_{\alpha{j}}-\dfrac{1}{4}h_{ij}F_{\alpha\beta}F^{\alpha\beta}\right),
\end{equation}
which we shall call the the curved-space Maxwell stress tensor of electromagnetic f\/ield\index{Maxwell stress tensor!curved-space!electromagnetic f\/ield}. Therefore the three-dimensional Ricci scalar curvature\index{Ricci scalar curvature} ${^{(3)}}R=h^{ij}R_{ij}$ up to the minus sign becomes the curved-space Maxwell stress tensor of electromagnetic f\/ield\index{Maxwell stress tensor!curved-space!electromagnetic f\/ield} projected onto the induced metric
\begin{eqnarray}
{^{(3)}}R&=&-\kappa\ell_P^2{h}^{ij}\sigma_{ij}=\kappa\ell_P^2{h}^{ij}\dfrac{1}{\mu_0}\left(F^\alpha_i{F}_{\alpha{j}}-\dfrac{1}{4}h_{ij}F_{\alpha\beta}F^{\alpha\beta}\right)=\nonumber\\
&=&\dfrac{\kappa\ell_P^2}{\mu_0}\left({h}^{ij}F^\alpha_i{F}_{\alpha{j}}-\dfrac{3}{4}F_{\alpha\beta}F^{\alpha\beta}\right).
\end{eqnarray}
where we have used the identity $h^{ij}h_{ij}=3$. Using of the transformation
\begin{equation}
{h}^{ij}F^\alpha_i{F}_{\alpha{j}}={h}^{ij}h_{ik}F^{k\alpha}{F}_{\alpha{j}}=F^{j\alpha}{F}_{\alpha{j}}=F^{\alpha{j}}{F}_{{j}\alpha},
\end{equation}
and the def\/inition
\begin{equation}
F_{\alpha\beta}F^{\alpha\beta}=F^{\alpha{j}}{F}_{\alpha{j}}+F^{\alpha{0}}{F}_{\alpha{0}}=-F^{\alpha{j}}{F}_{{j}\alpha}-F^{\alpha{0}}{F}_{{0}\alpha},
\end{equation}
together with the properties of the electromagnetic f\/ield tensor
\begin{eqnarray}
{F}_{\alpha\beta}F^{\alpha\beta}&=&2\left(\vec{B}^2-\dfrac{\vec{E}^2}{c^2}\right),\label{prop1}\\
{F}_{\alpha0}F^{0\alpha}&=&\dfrac{\vec{E}^2}{c^2},
\end{eqnarray}
one obtains
\begin{equation}
  {h}^{ij}F^\alpha_i{F}_{\alpha{j}}=-F_{\alpha\beta}F^{\alpha\beta}-F^{\alpha{0}}{F}_{{0}\alpha}=-2\left(\vec{B}^2-\dfrac{\vec{E}^2}{c^2}\right)-\dfrac{\vec{E}^2}{c^2}=2\vec{B}^2+\dfrac{\vec{E}^2}{c^2}.
\end{equation}
Collecting all together one receives f\/inally
\begin{equation}
{^{(3)}}R=\dfrac{\kappa\ell_P^2}{\mu_0}\left[2\vec{B}^2+\dfrac{\vec{E}^2}{c^2}-\dfrac{3}{2}\left(\vec{B}^2-\dfrac{\vec{E}^2}{c^2}\right)\right]=\dfrac{\kappa\ell_P^2}{2}\left(\dfrac{1}{\mu_0}\vec{B}^2+5\epsilon_0\vec{E}^2\right).
\end{equation}

Similarly one can calculate the energy density\index{energy density!electromagnetic f\/ield} of electromagnetic f\/ield. Applying the def\/inition $n^{\mu}={n}^\beta{g}^\mu_\beta$, and the identity $g_{\mu\nu}n^\mu{n}^\nu=n^\mu{n}_{\mu}=-1$ one obtains
\begin{eqnarray}
  \varrho&=&T_{\mu\nu}n^\mu{n}^\nu=\dfrac{1}{\mu_0}\left(F_{\mu\alpha}g^{\alpha\beta}F_{\nu\beta}-\dfrac{1}{4}g_{\mu\nu}F_{\alpha\beta}F^{\alpha\beta}\right)n^\mu{n}^\nu=\nonumber\\
  &=&\dfrac{1}{\mu_0}\left({F}_{\mu\alpha}({g}^{\alpha\beta}n^\mu{n}^\nu)F_{\nu\beta}-\dfrac{1}{4}g_{\mu\nu}n^\mu{n}^\nu{F}_{\alpha\beta}F^{\alpha\beta}\right).\label{endem}
\end{eqnarray}
The second term in the formula (\ref{endem}) can be easy transformed with using of the identity $g_{\mu\nu}n_\mu{n}_\nu=n^\mu{n}_{\mu}=-1$. The f\/irst term, however, is not so easy to transform. Let us notice that
\begin{equation}
{g}^{\alpha\beta}n^\mu{n}^\nu=g^{\alpha\mu}g^{\nu\beta}(g_{\mu\nu}n^\mu{n}^\nu)=-g^{\alpha\mu}g^{\nu\beta},
\end{equation}
where we have applied $g_{\mu\nu}n^\mu{n}^\nu=n_\nu n^\nu=-1$. In this manner
\begin{eqnarray}
{F}_{\mu\alpha}({g}^{\alpha\beta}n^\mu{n}^\nu)F_{\nu\beta}&=&-{F}_{\mu\alpha}(g^{\alpha\mu}g^{\nu\beta})F_{\nu\beta}={F}_{\mu\alpha}(g^{\alpha\mu}g^{\nu\beta})F_{\beta\nu}=\nonumber\\
&=&{F}^\mu_\mu{F}^\nu_\nu=F_{\mu\nu}(g^{\nu\mu}g_{\nu\mu})F^{\mu\nu}=\nonumber\\
&=&F_{\mu\nu}(g^{\nu\mu}g_{\mu\nu})F^{\mu\nu}=4F_{\mu\nu}F^{\mu\nu},
\end{eqnarray}
where we have applied the identities $F^\mu_\mu=F_{\mu\nu}g^{\nu\mu}$, $F^\nu_\nu=g_{\nu\mu}F^{\mu\nu}$, and $g^{\nu\mu}g_{\mu\nu}=4$. In this manner one receives f\/inally the energy density\index{energy density}
\begin{eqnarray}
\varrho&=&\dfrac{1}{\mu_0}\left(4{F}_{\alpha\beta}{F}^{\alpha\beta}+\dfrac{1}{4}{F}_{\alpha\beta}F^{\alpha\beta}\right)=\dfrac{17}{4\mu_0}{F}_{\alpha\beta}F^{\alpha\beta}=\nonumber\\
  &=&\dfrac{17}{2\mu_0}\left(\vec{B}^2-\dfrac{\vec{E}^2}{c^2}\right)=\dfrac{17}{2}\left(\dfrac{1}{\mu_0}\vec{B}^2-\epsilon_0\vec{E}^2\right),
\end{eqnarray}
where we have applied the property (\ref{prop1}) of electromagnetic f\/ield tensor. Consequently the generalized gravitational potential\index{gravitational potential!generalized} is
\begin{eqnarray}
\!\!\!\!\!\!\!\!\!\!\!\!\!\!\!\!\!\!\!\!
&&V=-{^{(3)}}R+2\kappa\ell_P^2\varrho=-\dfrac{\kappa\ell_P^2}{2}\left(\dfrac{1}{\mu_0}\vec{B}^2+5\epsilon_0\vec{E}^2\right)+17\kappa\ell_P^2\left(\dfrac{1}{\mu_0}\vec{B}^2-\epsilon_0\vec{E}^2\right)=\nonumber\\
\!\!\!\!\!\!\!\!\!\!\!\!\!\!\!\!\!\!\!\!
&&=\dfrac{33\kappa\ell_P^2}{2}\left(\dfrac{1}{\mu_0}\vec{B}^2-\dfrac{13}{11}\epsilon_0\vec{E}^2\right).
\end{eqnarray}
Now it is easy to see that $V$ averaged on midisuperspace\index{midisuperspace} is
\begin{equation}\label{kngpot}
  \langle{V}\rangle=\dfrac{33\kappa\ell_P^2}{2}\left(\dfrac{1}{\mu_0}\langle\vec{B}^2\rangle-\dfrac{13}{11}\epsilon_0\langle\vec{E}^2\rangle\right),
\end{equation}
where $\langle\vec{B}^2\rangle$ and $\langle\vec{E}^2\rangle$ are the midisuperspace means\index{midisuperspace!means} of the squared f\/ields $\vec{B}$ and $\vec{E}$, respectively. One has explicitly
\begin{eqnarray}
\langle\vec{B}^2\rangle&=&\dfrac{1}{\Omega(\xi)}\int\vec{B}^2\delta\xi,\label{ave1}\\
\langle\vec{E}^2\rangle&=&\dfrac{1}{\Omega(\xi)}\int\vec{E}^2\delta\xi.\label{ave2}
\end{eqnarray}
Because of the f\/ields $\vec{B}$ and $\vec{E}$ are in general functions on space-time, it is evident that both the averages (\ref{ave1}) and (\ref{ave2}) must be treated as functions on space-time, of course after performing in all the functional integrals the suitable transformation from midisuperspace\index{midisuperspace} to space-time $\xi\rightarrow\xi(x)$. In other words one has
\begin{equation}
  \langle{V}\rangle(\xi,\phi)=\langle{V}\rangle(x).
\end{equation}

Because of the considerations presented above are independent on the concrete form of the electrovacuum solution, they can be applied to any solution of the Einstein--Maxwell equations\index{Einstein--Maxwell equations}. Interestingly, one can \emph{ad hoc} add the cosmological constant\index{cosmological constant}
\begin{equation}\label{kngpot1}
  \langle{V}\rangle=\dfrac{33\kappa\ell_P^2}{2}\left(\dfrac{1}{\mu_0}\langle\vec{B}^2\rangle-\dfrac{13}{11}\epsilon_0\langle\vec{E}^2\rangle\right)+2\Lambda,
\end{equation}
so that the averaged generalized gravitational potential\index{gravitational potential!generalized!averaged} can be ordered by two equivalent ways
\begin{eqnarray}\label{kngpot2}
  \langle{V}\rangle&=&\dfrac{33\kappa\ell_P^2}{2}\left(\dfrac{1}{\mu_0}\langle\vec{B}^2\rangle-\epsilon_0\langle\vec{E}^2\rangle\right)+2\Lambda-3\kappa\ell_P^2\epsilon_0\langle\vec{E}^2\rangle=\\
  &=&\dfrac{39\kappa\ell_P^2}{2}\left(\dfrac{1}{\mu_0}\langle\vec{B}^2\rangle-\epsilon_0\langle\vec{E}^2\rangle\right)+2\Lambda-\dfrac{3\kappa\ell_P^2}{\mu_0}\langle\vec{B}^2\rangle.
\end{eqnarray}
If the averages $\langle\vec{B}^2\rangle=B_\Lambda^2$ and $\langle\vec{E}^2\rangle=E_\Lambda^2$ are constant then one can take
\begin{equation}
  2\Lambda=3\kappa\ell_P^2\epsilon_0E^2_\Lambda=\dfrac{3\kappa\ell_P^2}{\mu_0}B^2_\Lambda,\label{lambd}
\end{equation}
and one the averaged generalized gravitational potential\index{gravitational potential!generalized!averaged} vanishes
\begin{equation}
\langle{V}\rangle=0.
\end{equation}
The condition (\ref{lambd}, however, leads to the relation
\begin{equation}
E_\Lambda=cB_\Lambda,
\end{equation}
which can be used to def\/ine the speed of light
\begin{equation}
  c:=\dfrac{E_\Lambda}{B_\Lambda},
\end{equation}
if one knows $E_\Lambda$ and $B_\Lambda$.

Let us return to the Kerr--Newman space-time\index{space-time!Kerr}. The spatial part of the Kerr--Newman metric has the form
\begin{equation}
  h_{ij}=\left[\begin{array}{ccc}\dfrac{r^2+\alpha^2\cos^2\theta}{\Delta}&0&0\\
  0&r^2+\alpha^2\cos^2\theta&0\\
  0&0&\dfrac{(r^2+\alpha^2)^2-\alpha^2\Delta\sin^2\theta}{r^2+\alpha^2\cos^2\theta}\sin^2\theta
  \end{array}\right],
\end{equation}
where $\Delta=r^2\alpha^2-r_Sr+r_Q^2$, and by this reason the global dimension can be deduced easy
\begin{equation}
  h=\left(r^2+\alpha^2\cos^2\theta\right)\left[r^2+\alpha^2\cos^2\theta+\dfrac{(r^2+\alpha^2)(r_Sr-r_Q^2)}{r^2+\alpha^2-r_Sr+r_Q^2}\right]\sin^2\theta,
\end{equation}
so that the invariant global dimension is
\begin{equation}
  \xi=\dfrac{1}{4\pi\sqrt{6}}\sqrt{\left(r^2+\alpha^2\cos^2\theta\right)\left[r^2+\alpha^2\cos^2\theta+\dfrac{(r^2+\alpha^2)(r_Sr-r_Q^2)}{r^2+\alpha^2-r_Sr+r_Q^2}\right]}\sin\theta.
\end{equation}
The volume of the Kerr-Newman midisuperspace\index{midisuperspace} is $\Omega(\xi)=\xi$. Here $r_S$ is the Schwarzschild radius\index{Schwarzschild radius} and $r_Q$ is a length-scale corresponding to the electric charge Q of the mass
\begin{equation}
  r_Q^2=\dfrac{Q^2G}{4\pi\epsilon_0c^4}=\dfrac{Q^2}{4\pi\epsilon_0}\dfrac{\kappa}{8\pi}.
\end{equation}

There are three possible situations. Namely, when the averaged generalized gravitational potential\index{gravitational potential!generalized!averaged} (\ref{kngpot}) is $1^\circ$ vanishing, $2^\circ$ positive, $3^\circ$ negative. The f\/irst case is rather trivial. $\langle{V}\rangle$ vanishes if and only if
\begin{equation}
\langle\vec{B}^2\rangle=\dfrac{13}{11}\dfrac{\langle\vec{E}^2\rangle}{c^2},
\end{equation}
what in fact means that
\begin{equation}
|\vec{B}|=\sqrt{\dfrac{13}{11}}\dfrac{|\vec{E}|}{c}\approx1.087\dfrac{|\vec{E}|}{c}.
\end{equation}
In such a situation one can construct straightforwardly and easy the wave functions of the Kerr--Newman space-time\index{space-time!Kerr}
\begin{equation}
\Psi_1^{KN}=\dfrac{\sin\theta}{4\pi\sqrt{|A|}\sqrt{6}}\sqrt{\left(r^2+\alpha^2\cos^2\theta\right)\left[r^2+\alpha^2\cos^2\theta+\dfrac{(r^2+\alpha^2)(r_Sr-r_Q^2)}{r^2+\alpha^2-r_sr+r_Q^2}\right]},
\end{equation}
\begin{equation}
\Psi_2^{KN}=\dfrac{1}{\sqrt{|B|}},
\end{equation}
where the constants of integration $A$ and $B$ are
\begin{eqnarray}
  A&=&\dfrac{\Xi^3}{3},\\
  B&=&\dfrac{1}{\Gamma_0}\Xi,
\end{eqnarray}
where $\Xi$ is the reference constant
\begin{equation}
  \Xi=\dfrac{1}{4\pi\sqrt{6}}\sqrt{\left(R^2+\alpha^2\cos^2\Theta\right)\left[R^2+\alpha^2\cos^2\Theta+\dfrac{(R^2+\alpha^2)(r_SR-r_Q^2)}{R^2+\alpha^2-r_sR+r_Q^2}\right]}\sin\Theta.
\end{equation}

The second situation, i.e. positive $\langle{V}\rangle$, is def\/ined for
\begin{equation}
\langle\vec{B}^2\rangle>\dfrac{13}{11}\dfrac{\langle\vec{E}^2\rangle}{c^2},
\end{equation}
what can be presented in the form
\begin{equation}
|\vec{B}|>\sqrt{\dfrac{13}{11}}\dfrac{|\vec{E}|}{c}.
\end{equation}
In such a situation the constants of integration $A$ and $B$ are not easy to calculate for general f\/ields $\vec{B}$ and $\vec{E}$, but can be presented in the compact form
\begin{eqnarray}
  A&=&\int\left[\dfrac{\sin\left(\xi\sqrt{\dfrac{33\kappa\ell_P^2}{2}\left(\dfrac{1}{\mu_0}\langle\vec{B}^2\rangle-\dfrac{13}{11}\epsilon_0\langle\vec{E}^2\rangle\right)}\right)}{\sqrt{\dfrac{33\kappa\ell_P^2}{2}\left(\dfrac{1}{\mu_0}\langle\vec{B}^2\rangle-\dfrac{13}{11}\epsilon_0\langle\vec{E}^2\rangle\right)}}\right]^2\delta\xi,\\
  B&=&\int\cos^2\left(\xi\sqrt{\dfrac{33\kappa\ell_P^2}{2}\left(\dfrac{1}{\mu_0}\langle\vec{B}^2\rangle-\dfrac{13}{11}\epsilon_0\langle\vec{E}^2\rangle\right)}\right)\delta\xi,
\end{eqnarray} and the wave functions of the Kerr--Newman space-time\index{space-time!Kerr} are
\begin{eqnarray}
  \Psi_1^{KN}&=&\dfrac{1}{\sqrt{|A|}}\dfrac{\sin\left(\xi\sqrt{\dfrac{33\kappa\ell_P^2}{2}\left(\dfrac{1}{\mu_0}\langle\vec{B}^2\rangle-\dfrac{13}{11}\epsilon_0\langle\vec{E}^2\rangle\right)}\right)}{\sqrt{\dfrac{33\kappa\ell_P^2}{2}\left(\dfrac{1}{\mu_0}\langle\vec{B}^2\rangle-\dfrac{13}{11}\epsilon_0\langle\vec{E}^2\rangle\right)}},\\
  \Psi_2^{KN}&=&\dfrac{1}{\sqrt{|B|}}\cos\left(\xi\sqrt{\dfrac{33\kappa\ell_P^2}{2}\left(\dfrac{1}{\mu_0}\langle\vec{B}^2\rangle-\dfrac{13}{11}\epsilon_0\langle\vec{E}^2\rangle\right)}\right).
\end{eqnarray}
The case of negative averaged generalized gravitational potential\index{gravitational potential!generalized!averaged} can be considered analogously. In this case
\begin{equation}
|\vec{B}|<\sqrt{\dfrac{13}{11}}\dfrac{|\vec{E}|}{c}.
\end{equation}
So that the wave functions of the Kerr--Newman space-time\index{space-time!Kerr} are
\begin{eqnarray}
  \Psi_1^{KN}&=&\dfrac{1}{\sqrt{|A|}}\dfrac{\sinh\left(\xi\sqrt{\dfrac{33\kappa\ell_P^2}{2}\left|\dfrac{1}{\mu_0}\langle\vec{B}^2\rangle-\dfrac{13}{11}\epsilon_0\langle\vec{E}^2\rangle\right|}\right)}{\sqrt{\dfrac{33\kappa\ell_P^2}{2}\left|\dfrac{1}{\mu_0}\langle\vec{B}^2\rangle-\dfrac{13}{11}\epsilon_0\langle\vec{E}^2\rangle\right|}},\\
  \Psi_2^{KN}&=&\dfrac{1}{\sqrt{|B|}}\cosh\left(\xi\sqrt{\dfrac{33\kappa\ell_P^2}{2}\left|\dfrac{1}{\mu_0}\langle\vec{B}^2\rangle-\dfrac{13}{11}\epsilon_0\langle\vec{E}^2\rangle\right|}\right).
\end{eqnarray}
where the integration constants $A$ and $B$ are
\begin{eqnarray}
  A&=&\int\left[\dfrac{\sinh\left(\xi\sqrt{\dfrac{33\kappa\ell_P^2}{2}\left|\dfrac{1}{\mu_0}\langle\vec{B}^2\rangle-\dfrac{13}{11}\epsilon_0\langle\vec{E}^2\rangle\right|}\right)}{\sqrt{\dfrac{33\kappa\ell_P^2}{2}\left|\dfrac{1}{\mu_0}\langle\vec{B}^2\rangle-\dfrac{13}{11}\epsilon_0\langle\vec{E}^2\rangle\right|}}\right]^2\delta\xi,\\
  B&=&\int\cosh^2\left(\xi\sqrt{\dfrac{33\kappa\ell_P^2}{2}\left|\dfrac{1}{\mu_0}\langle\vec{B}^2\rangle-\dfrac{13}{11}\epsilon_0\langle\vec{E}^2\rangle\right|}\right)\delta\xi,
\end{eqnarray}

\section{The Reissner--Nordstr\"om Space-time}

Another solution of the electrovacuum Einstein f\/ield equations\index{Einstein f\/ield equations} is the Reissner--Nordstr\"om metric describing spherically symmetric static massive charged object. Therefore in this case one has also
\begin{equation}
  \langle{V}\rangle=\dfrac{33\kappa\ell_P^2}{2}\left(\dfrac{1}{\mu_0}\langle\vec{B}^2\rangle-\dfrac{13}{11}\epsilon_0\langle\vec{E}^2\rangle\right),
\end{equation}
and the only change is the metric. The spatial part of Reissner--Nordstr\"om space-time\index{space-time!Reissner--Nordstr\"om} metric is
\begin{equation}
  h_{ij}=\left[\begin{array}{ccc}\left(1-\dfrac{r_S}{r}+\dfrac{r_Q^2}{r^2}\right)^{-1}&0&0\\0&r^2&0\\
  0&0&r^2\sin^2\theta\end{array}\right],
\end{equation}
so that the global dimension is
\begin{equation}
  h=\dfrac{r^4\sin^2\theta}{1-\dfrac{r_S}{r}+\dfrac{r_Q^2}{r^2}}
\end{equation}
while the invariant global dimension has the form
\begin{equation}
  \xi=\dfrac{1}{4\pi\sqrt{6}}\dfrac{r^2\sin\theta}{\sqrt{1-\dfrac{r_S}{r}+\dfrac{r_Q^2}{r^2}}}
\end{equation}
and of the volume of the Reissner--Nordstr\"om midisuperspace\index{midisuperspace} is $\Omega(\xi)=\xi$. There are three possible situations def\/ined by absolute values of the electric and magnetic f\/ields. For the case of negative $\langle{V}\rangle$ the wave functions of the Reissner--Nordstr\"om space-time\index{space-time!Reissner--Nordstr\"om} are
\begin{eqnarray}
  \Psi_1^{RN}&=&\dfrac{1}{\sqrt{|A|}}\dfrac{\sinh\left(\xi\sqrt{\dfrac{33\kappa\ell_P^2}{2}\left|\dfrac{1}{\mu_0}\langle\vec{B}^2\rangle-\dfrac{13}{11}\epsilon_0\langle\vec{E}^2\rangle\right|}\right)}{\sqrt{\dfrac{33\kappa\ell_P^2}{2}\left|\dfrac{1}{\mu_0}\langle\vec{B}^2\rangle-\dfrac{13}{11}\epsilon_0\langle\vec{E}^2\rangle\right|}},\\
  \Psi_2^{RN}&=&\dfrac{1}{\sqrt{|B|}}\cosh\left(\xi\sqrt{\dfrac{33\kappa\ell_P^2}{2}\left|\dfrac{1}{\mu_0}\langle\vec{B}^2\rangle-\dfrac{13}{11}\epsilon_0\langle\vec{E}^2\rangle\right|}\right).
\end{eqnarray}
where the integration constants $A$ and $B$ are
\begin{eqnarray}
  A&=&\int\left[\dfrac{\sinh\left(\xi\sqrt{\dfrac{33\kappa\ell_P^2}{2}\left|\dfrac{1}{\mu_0}\langle\vec{B}^2\rangle-\dfrac{13}{11}\epsilon_0\langle\vec{E}^2\rangle\right|}\right)}{\sqrt{\dfrac{33\kappa\ell_P^2}{2}\left|\dfrac{1}{\mu_0}\langle\vec{B}^2\rangle-\dfrac{13}{11}\epsilon_0\langle\vec{E}^2\rangle\right|}}\right]^2\delta\xi,\\
  B&=&\int\cosh^2\left(\xi\sqrt{\dfrac{33\kappa\ell_P^2}{2}\left|\dfrac{1}{\mu_0}\langle\vec{B}^2\rangle-\dfrac{13}{11}\epsilon_0\langle\vec{E}^2\rangle\right|}\right)\delta\xi.
\end{eqnarray}
When $\langle{V}\rangle$ is positive then
\begin{eqnarray}
  A&=&\int\left[\dfrac{\sin\left(\xi\sqrt{\dfrac{33\kappa\ell_P^2}{2}\left(\dfrac{1}{\mu_0}\langle\vec{B}^2\rangle-\dfrac{13}{11}\epsilon_0\langle\vec{E}^2\rangle\right)}\right)}{\sqrt{\dfrac{33\kappa\ell_P^2}{2}\left(\dfrac{1}{\mu_0}\langle\vec{B}^2\rangle-\dfrac{13}{11}\epsilon_0\langle\vec{E}^2\rangle\right)}}\right]^2\delta\xi,\\
  B&=&\int\cos^2\left(\xi\sqrt{\dfrac{33\kappa\ell_P^2}{2}\left(\dfrac{1}{\mu_0}\langle\vec{B}^2\rangle-\dfrac{13}{11}\epsilon_0\langle\vec{E}^2\rangle\right)}\right)\delta\xi,
\end{eqnarray}
and the wave functions of the Reissner--Nordstr\"om space-time \index{space-time!Reissner--Nordstr\"om}are
\begin{eqnarray}
  \Psi_1^{RN}&=&\dfrac{1}{\sqrt{|A|}}\dfrac{\sin\left(\xi\sqrt{\dfrac{33\kappa\ell_P^2}{2}\left(\dfrac{1}{\mu_0}\langle\vec{B}^2\rangle-\dfrac{13}{11}\epsilon_0\langle\vec{E}^2\rangle\right)}\right)}{\sqrt{\dfrac{33\kappa\ell_P^2}{2}\left(\dfrac{1}{\mu_0}\langle\vec{B}^2\rangle-\dfrac{13}{11}\epsilon_0\langle\vec{E}^2\rangle\right)}},\\
  \Psi_2^{RN}&=&\dfrac{1}{\sqrt{|B|}}\cos\left(\xi\sqrt{\dfrac{33\kappa\ell_P^2}{2}\left(\dfrac{1}{\mu_0}\langle\vec{B}^2\rangle-\dfrac{13}{11}\epsilon_0\langle\vec{E}^2\rangle\right)}\right).
\end{eqnarray}
For vanishing $\langle{V}\rangle$ one has
\begin{eqnarray}
\Psi_1^{RN}&=&\dfrac{1}{4\pi\sqrt{|A|}\sqrt{6}}\dfrac{r^2\sin\theta}{\sqrt{1-\dfrac{r_S}{r}+\dfrac{r_Q^2}{r^2}}},\\
\Psi_2^{RN}&=&\dfrac{1}{\sqrt{|B|}},
\end{eqnarray}
where the constants of integration $A$ and $B$ are
\begin{eqnarray}
  A&=&\dfrac{\Xi^3}{3},\\
  B&=&\dfrac{1}{\Gamma_0}\Xi,
\end{eqnarray}
where $\Xi$ is the reference constant
\begin{equation}
  \Xi=\dfrac{1}{4\pi\sqrt{6}}\dfrac{R^2\sin\Theta}{\sqrt{1-\dfrac{r_S}{R}+\dfrac{r_Q^2}{R^2}}}.
\end{equation}

\section{The G\"odel Space-time}

Let us consider the solution of the Einstein f\/ield equations\index{Einstein f\/ield equations} with presence of the cosmological constant\index{cosmological constant} and the stress-energy tensor\index{stress-energy tensor} of dust
\begin{eqnarray}
  &&R_{\mu\nu}-\dfrac{1}{2}g_{\mu\nu}{^{(4)}}R+\Lambda{g}_{\mu\nu}=\kappa\ell_P^2{T}_{\mu\nu},\label{goddust1}\\
  &&T_{\mu\nu}=\varepsilon{u}_\mu{u}_\nu,\label{goddust2}
\end{eqnarray}
which is given by the G\"odel space-time\index{space-time!G\"odel}
\begin{eqnarray}
g_{\mu\nu}&=&\left[\begin{array}{cccc}-a^2&0&0&-a^2e^x\\0&a^2&0&0\\0&0&a^2&0\\-a^2e^x&0&0&-\dfrac{1}{2}a^2e^{2x}\end{array}\right],\\
g^{\mu\nu}&=&\left[\begin{array}{cccc}\dfrac{1}{a^2}&0&0&-\dfrac{2e^{-x}}{a^2}\\0&\dfrac{1}{a^2}&0&0\\0&0&\dfrac{1}{a^2}&0\\-\dfrac{2e^{-x}}{a^2}&0&0&\dfrac{2e^{-2x}}{a^2}\end{array}\right].
\end{eqnarray}
having the following spatial part
\begin{equation}
h_{ij}=\left[\begin{array}{ccc}a^2&0&0\\0&a^2&0\\0&0&-\dfrac{1}{2}a^2e^{2x}\end{array}\right]\quad,\quad h^{ij}=\left[\begin{array}{ccc}\dfrac{1}{a^2}&0&0\\0&\dfrac{1}{a^2}&0\\0&0&\dfrac{2e^{-2x}}{a^2}\end{array}\right],\label{godelspat}
\end{equation}
where $a=a(x,y,z)$. In this case the global dimension is
\begin{equation}
  |h|=\dfrac{a^6}{2}e^{2x},
\end{equation}
so the invariant global dimension has the form
\begin{equation}
  \xi=\dfrac{1}{8\pi\sqrt{3}}a^3e^x,
\end{equation}
and therefore the volume of the G\"odel midisuperspace\index{midisuperspace} is $\Omega(\xi)=\xi$. The G\"odel metric satisf\/ies the Einstein f\/ield equations\index{Einstein f\/ield equations} for which the cosmological constant\index{cosmological constant} is related to the parameter $a$ by the relation
\begin{equation}
  \Lambda=-\dfrac{1}{2a^2}.
\end{equation}
The energy density\index{energy density} $\epsilon$ of the dust is
\begin{equation}
  \epsilon=\dfrac{1}{\kappa\ell_P^2a^2}.
\end{equation}
The G\"odel metric can be decomposed in the ADM $3+1$ form. It is easy to see that such an \emph{ad hoc} decomposition generates the equations
\begin{eqnarray}
  -N^2+N_iN^i&=&-a^2,\\
  N_i&=&\left[0,0,-a^2e^x\right].
\end{eqnarray}
Taking into account the spatial metric (\ref{godelspat}) one receives
\begin{equation}
N^i=h^{ij}N_j=\left[0,0,-2e^{-x}\right],
\end{equation}
what gives $N_iN^i=2a^2$ and by this reason
\begin{equation}
N=\sqrt{3}a.
\end{equation}
In this manner one can establish the normal unit vector f\/ield
\begin{eqnarray}
n^\mu&=&\left[\dfrac{1}{N},-\dfrac{N^i}{N}\right]=\dfrac{1}{\sqrt{3}}\left[\dfrac{1}{a},0,0,\dfrac{2e^{-x}}{a}\right],\\
n_\mu&=&\left[-N,0_i\right]=\sqrt{3}\left[-a,0,0,0\right].
\end{eqnarray}
The problem is to choose the velocity vector $u_\mu$ according to the general rules
\begin{eqnarray}
  u_\mu{u}^\mu&=&-1,\label{ucon1}\\
  u_\mu{n}^\mu&=&-1,\label{ucon2}\\
  u_\mu{n}^\mu{u}_\nu{n}^\nu&=&1.\label{ucon3}
\end{eqnarray}
However, it can be seen by straightforward calculation that the following choice
\begin{eqnarray}
  u^\mu&=&\dfrac{1}{\sqrt{3}}\left[\dfrac{1}{a},0,0,0\right],\\
  u_\mu&=&g_{\mu\nu}u^\nu=\sqrt{3}\left[-a,0,0,-ae^x\right],
\end{eqnarray}
satisf\/ies the conditions (\ref{ucon1})-(\ref{ucon3}). The contraction of the Einstein f\/ield equations\index{Einstein f\/ield equations} (\ref{goddust1})-(\ref{goddust2}) with metric $g^{\mu\nu}$ leads to
\begin{equation}
  {^{(4)}}R=4\Lambda+\kappa\ell_P^2\epsilon,
\end{equation}
so that the equations (\ref{goddust1})-(\ref{goddust2}) can be presented in the form
\begin{equation}
  R_{\mu\nu}=\Lambda{g}_{\mu\nu}+\kappa\ell_P^2\epsilon\left({u}_\mu{u}_\nu+\dfrac{1}{2}g_{\mu\nu}\right),
\end{equation}
what after including the fact that for the G\"odel Universe one has
\begin{equation}
\kappa\ell_P^2\epsilon=-2\Lambda,
\end{equation}
one receives the four-dimensional Ricci curvature tensor\index{Ricci curvature tensor}
\begin{equation}
  R_{\mu\nu}=\kappa\ell_P^2\epsilon{u}_\mu{u}_\nu=-2\Lambda{u}_\mu{u}_\nu.
\end{equation}
In this manner the three-dimensional Ricci curvature tensor\index{Ricci curvature tensor} has the form
\begin{equation}
  R_{ij}=-2\Lambda{u}_i{u}_j,
\end{equation}
and consequently the three-dimensional Ricci scalar curvature\index{Ricci scalar curvature} is
\begin{equation}
  {^{(3)}}R=h^{ij}R_{ij}=-2\Lambda{u}_i{u}^i=0.
\end{equation}
Similarly one can establish the energy density\index{energy density} of Matter f\/ields\index{Matter f\/ields}
\begin{equation}
  \varrho=T_{\mu\nu}n^\mu{n}^\nu=\epsilon{u}_\mu{n}^\mu{u}_\nu{n}^\nu=\epsilon=-2\dfrac{\Lambda}{\kappa\ell_P^2}.
\end{equation}
In this manner the generalized gravitational potential\index{gravitational potential!generalized} for the G\"odel Universe has the following form
\begin{equation}
  V=-{^{(3)}}R+2\Lambda+2\kappa\ell_P^2\varrho=-2\Lambda=\kappa\ell_P^2\epsilon=\dfrac{1}{a^2}>0,
\end{equation}
so that
\begin{eqnarray}
  \langle{V}\rangle&=&\dfrac{1}{a^3e^x}\int\dfrac{d({a}^3e^x)}{{a}^2}=\dfrac{1}{a^3e^x}\int\dfrac{3a^2dae^x+a^3e^xdx}{{a}^2}=\\
  &=&\dfrac{1}{a^3e^x}\int\left(3e^xda+ae^xdx\right),
\end{eqnarray}
and by this reason
\begin{equation}
  \sqrt{\langle{V}\rangle}\left(\xi,\phi\right)=\dfrac{1}{a^{3/2}e^{x/2}}\sqrt{\int\left(3e^xda+ae^xdx\right)}.
\end{equation}
In this manner the wave functions of the G\"odel dust Universe are
\begin{eqnarray}
  \Psi_1^G&=&\dfrac{1}{\sqrt{|A|}}\dfrac{\sin\left(\dfrac{a^{3/2}e^{x/2}}{8\pi\sqrt{3}}\sqrt{\int\left(3e^xda+ae^xdx\right)}\right)}{\dfrac{1}{a^{3/2}e^{x/2}}\sqrt{\int\left(3e^xda+ae^xdx\right)}},\\
  \Psi_2^G&=&\dfrac{1}{\sqrt{|B|}}\cos\left(\dfrac{a^{3/2}e^{x/2}}{8\pi\sqrt{3}}\sqrt{\int\left(3e^xda+ae^xdx\right)}\right),
\end{eqnarray}
where $\sgn(a)\neq0$, and the constants of integration $A$ and $B$ are
\begin{eqnarray}
\!\!\!\!\!\!\!\!\!\!\!\!\!\!\!\!\!\!\!\!\!\!\!\!\!\!\!\!\!\!
A&=&\dfrac{1}{2\sqrt{3}S_P}\int\left(\dfrac{\sin\left(\dfrac{a^{3/2}e^{x/2}}{8\pi\sqrt{3}}\sqrt{\int\left(3e^xda+ae^xdx\right)}\right)}{\dfrac{1}{a^{3/2}e^{x/2}}\sqrt{\int\left(3e^xda+ae^xdx\right)}}\right)^2d(a^3e^x),\\
\!\!\!\!\!\!\!\!\!\!\!\!\!\!\!\!\!\!\!\!\!\!\!\!\!\!\!\!\!\!
B&=&\dfrac{1}{\Gamma_0}\dfrac{1}{2\sqrt{3}S_P}\int\cos^2\left(\dfrac{a^{3/2}e^{x/2}}{8\pi\sqrt{3}}\sqrt{\int\left(3e^xda+ae^xdx\right)}\right)d(a^3e^x).
\end{eqnarray}
Of course when $a=constans$ the situation is much more simpler, but we will not derive these particular results.

\section{The Einstein--Rosen Gravitational Waves}

The example of vacuum solution are also the Einstein--Rosen cylindrical gravitational waves\index{cylindrical gravitational waves}. The spatial part of the metric has the form
\begin{equation}
  h_{ij}=\left[\begin{array}{ccc}e^{2\gamma-2\psi}&0&0\\0&e^{2\psi}&0\\0&0&r^2e^{-2\psi}\end{array}\right],
\end{equation}
where $r$ is radial distance from the $z$ axis, and $\psi=\psi(t,r)$ and $\gamma=\gamma(t,r)$ are functions satisfying the equations
\begin{eqnarray}
  \ddot{\psi}&=&\dfrac{\psi'}{r}+\psi'',\label{psieq}\\
  \dfrac{\gamma'}{r}&=&\psi'^2+\dot{\psi}^2,\\
  \dfrac{\dot{\gamma}}{r}&=&2\dot{\psi}\psi'.
\end{eqnarray}
In other words when one solves the equation (\ref{psieq}) then the function $\gamma$ can be established by
\begin{equation}
  \gamma=\int\left[r\left(\psi'^2+\dot{\psi}^2\right)dr+2r\dot{\psi}\psi'dt\right].
\end{equation}
In this manner the global dimension for the Einstein--Rosen waves is
\begin{equation}
  h=r^2e^{2\gamma-2\psi},
\end{equation}
while the invariant global dimension has the form
\begin{equation}
  \xi=\dfrac{1}{\sqrt{6}S_P}re^{\gamma-\psi}.
\end{equation}
The generalized gravitational potential\index{gravitational potential!generalized} vanishes in this case, so that the wave functions are
\begin{eqnarray}
\Psi_1^{ER}&=&\dfrac{1}{\sqrt{|A|}}\dfrac{1}{4\pi\sqrt{6}}re^{\gamma-\psi},\\
\Psi_2^{ER}&=&\dfrac{1}{\sqrt{|B|}},
\end{eqnarray}
where the constants of integration $A$ and $B$ are determined by the relations
\begin{eqnarray}
  A&=&\dfrac{\Xi^3}{3},\\
  B&=&\dfrac{1}{\Gamma_0}\Xi,
\end{eqnarray}
where $\Xi$ is the reference constant
\begin{equation}
  \Xi=\dfrac{1}{4\pi\sqrt{6}}Re^{\gamma(T,R)-\psi(T,R)},
\end{equation}
where $T$ and $R$ are reference values of $t$ and $R$, respectively.

\section{The Taub--Newman--Unti--Tamburino Space-time}

Let us consider the generalized axisymmetric solution of the vacuum Einstein f\/ield equations\index{Einstein f\/ield equations} presented in the Weyl canonical coordinates
\begin{equation}
  h_{ij}=\left[\begin{array}{ccc}e^{2\gamma-2\psi}&0&0\\0&e^{2\gamma-2\psi}&0\\0&0&r^2e^{-2\psi}-A^2e^{2\gamma}\end{array}\right],
\end{equation}
for which the f\/ield equations are
\begin{eqnarray}
\bigtriangleup\psi&=&0,\\
\dfrac{\gamma'}{r}&=&\psi'^2+\dot{\psi}^2,\\
\dfrac{\gamma'_z}{r}&=&2\dot{\psi}\psi'.
\end{eqnarray}
Here $r$ is the radial distance from the axis of symmetry $z$. In such a situation the global dimension is
\begin{equation}
  h=e^{4\gamma-4\psi}\left(r^2e^{-2\psi}-A^2e^{2\gamma}\right),
\end{equation}
so that the invariant global dimension is
\begin{equation}
  \xi=\dfrac{1}{4\pi\sqrt{6}}e^{2\gamma-2\psi}\sqrt{r^2e^{-2\psi}-A^2e^{2\gamma}}.
\end{equation}
For this case the cosmological constant\index{cosmological constant} as well as Matter f\/ields\index{Matter f\/ields} are absent. Therefore the three-dimensional Ricci scalar curvature\index{Ricci scalar curvature}, and the generalized gravitational potential\index{gravitational potential!generalized} are trivial. The wave functions of any static axisymmetric vacuum solution can be established as
\begin{eqnarray}
\Psi_1^{SAVS}&=&\dfrac{1}{\sqrt{|A|}}\dfrac{1}{4\pi\sqrt{6}}e^{2\gamma-2\psi}\sqrt{r^2e^{-2\psi}-A^2e^{2\gamma}},\label{savs1}\\
\Psi_2^{SAVS}&=&\dfrac{1}{\sqrt{|B|}},\label{savs2}
\end{eqnarray}
where the constants of integration $A$ and $B$ are
\begin{eqnarray}
  A&=&\dfrac{\Xi^3}{3},\label{savsa}\\
  B&=&\dfrac{1}{\Gamma_0}\Xi,\label{savsb}
\end{eqnarray}
where $\Xi$ is the reference constant
\begin{equation}
  \Xi=\dfrac{1}{4\pi\sqrt{6}}e^{2\gamma(R,Z)-2\psi(R,Z)}\sqrt{R^2e^{-2\psi(R,Z)}-A^2(R,Z)e^{2\gamma(R,Z)}},\label{savsxi}
\end{equation}
where $R$ and $Z$ are reference values of $r$ and $z$. Let us see the wave functions of two particular cases.

The Taub--Newman--Unti--Tamburino space-time\index{space-time!Taub--Newman--Unti--Tamburino} is the particular case the static axisymmetric vacuum space-time characterized by
\begin{eqnarray}
  e^{2\psi}&=&\dfrac{(r_++r_-)^2-(r_S^2+4l^2)}{(r_++r_-+r_S)^2+4l^2},\\
  e^{2\gamma}&=&\dfrac{(r_++r_-)^2-(r_S^2+4l^2)}{4r_+r_-},\\
  A&=&\dfrac{2l(r_+-r_-)}{\sqrt{r_S^2+4l^2}},\\
  r_\pm^2&=&r^2+\left(z\pm\dfrac{1}{2}\sqrt{r_S^2+4l^2}\right)^2.
\end{eqnarray}

\chapter{The Functional Objective Geometry}

\section{Effective Scalar Curvature}
Let us assume that the concrete form of the gravitational potential\index{gravitational potential!effective} $V_{eff}$ (\ref{eff}) is f\/ixed \emph{ad hoc} as functional or function of the global dimension $h$. In such a situation one can express the Ricci scalar of a three-dimensional embedded space as follows
\begin{equation}\label{kgf0}
{^{(3)}}R=2\Lambda+2\kappa\ell_P^2\varrho-6(8\pi)^2hV_{eff},
\end{equation}
whereas the global one-dimensional quantum gravity is given by the evolutionary equation
\begin{equation}
  \left(\dfrac{\delta^2}{\delta h^2}+V_{eff}\right)\Psi[h]=0.\label{godeff}
\end{equation}
In this manner the quantum gravity is the system describing geometry (\ref{kgf0}), and quantum mechanics (\ref{godeff}) of an embedded space.

The quantum gravity (\ref{kgf0})-(\ref{godeff}) is in itself non trivial. In fact, this can be expressed in more general notation
\begin{eqnarray}
{^{(3)}}R&=&f[h_{ij}],\label{fog1}\\
\dfrac{\delta^2\Psi[h]}{\delta h^2}&=&-V_{eff}[h_{ij}]\Psi[h],\label{fog2}
\end{eqnarray}
where both $f[h_{ij}]$ as well as $V_{eff}[h_{ij}]$ are scalar-valued functionals of an induced three-dimensional metric $h_{ij}$, i.e. are \emph{objective functionals}\index{objective functionals}
\begin{eqnarray}
f[h_{ij}]&=&f[I_{\mathbf{h}},II_{\mathbf{h}},III_{\mathbf{h}}],\\
\Psi[h_{ij}]&=&\Psi[I_{\mathbf{h}},II_{\mathbf{h}},III_{\mathbf{h}}],
\end{eqnarray}
where $I_{\mathbf{h}}$, $II_{\mathbf{h}}$, and $III_{\mathbf{h}}$ are the $3\times3$ matrix invariants of an induced metric $h_{ij}$
\begin{equation}
I_{\mathbf{h}}=\tr\mathbf{h}\quad,\quad{II}_{\mathbf{h}}=\dfrac{\left(\tr\mathbf{h}\right)^2-\tr\mathbf{h}^2}{2}\quad,\quad{III}_{\mathbf{h}}=\det\mathbf{h},
\end{equation}
which according to the Cayley--Hamilton theorem\index{Cayley--Hamilton theorem} are the coeff\/icients of the characteristic polynomial of the matrix $h_{ij}$
\begin{equation}
\mathbf{h}^3-I_{\mathbf{h}}\mathbf{h}^2+II_{\mathbf{h}}\mathbf{h}-III_{\mathbf{h}}\mathbf{I}_{3\times3}=0.
\end{equation}
In this manner we shall call \emph{the functional objective geometry}\index{functional objective geometry} the quantum gravity given by system of equations (\ref{fog1})- (\ref{fog2}).

We shall call the Ricci scalar curvature\index{Ricci scalar curvature} (\ref{kgf0}) describing the three-geometry of an embedded space \emph{the effective scalar curvature}\index{effective scalar curvature} and study its meaning in this section. For convenience let us present $V_{eff}$ as an algebraic sum of three elementary energetic constituents
\begin{equation}
  V_{eff}=V_G+V_C+V_M,
\end{equation}
where $V_G$, $V_C$, and $V_M$ are the geometric and the cosmological, and the material contributions
\begin{eqnarray}
V_G&=&-\dfrac{1}{6(8\pi)^2}\dfrac{{^{(3)}\!R}}{h},\\
V_C&=&\dfrac{1}{6(8\pi)^2}\dfrac{2\Lambda}{h},\\
V_M&=&\dfrac{1}{6(8\pi)^2}\dfrac{2\kappa}{h}\varrho.
\end{eqnarray}
One can list several examples of physical scenarios within the global one-dimensional quantum gravity, with respect to the choice of the form of the potential $V_{eff}$.
\begin{enumerate}
\item The case of constant non vanishing effective gravitational potential\index{gravitational potential!effective} $V_{eff}=V_c\neq0$. In such a situation the Ricci scalar curvature\index{Ricci scalar curvature} of an embedded space and the global one-dimensional quantum gravity are
\begin{eqnarray}
&&{^{(3)}}R=2\Lambda+2\kappa\ell_P^2\varrho-6(8\pi)^2hV_c,\label{kgf0a}\\ &&\left(\dfrac{\delta^2}{\delta{h^2}}+V_c\right)\Psi_c[h]=0,
\end{eqnarray}
where $\Psi_c[h]$ is a wave functional related to $V_{eff}=V_c$.
\item The case of trivial effective gravitational potential\index{gravitational potential!effective} $V_{eff}=0$. In such a situation the three-dimensional Ricci scalar curvature\index{Ricci scalar curvature} and the global one-dimensional quantum gravity are
\begin{eqnarray}
&&{^{(3)}\!R}=2\Lambda+2\kappa\ell_P^2\varrho,\label{kgf1}\\
&&\dfrac{\delta^2}{\delta{h^2}}\Psi_0[h]=0,
\end{eqnarray}
where $\Psi_0$ is a ''free'' wave functional related to $V_{eff}=0$.
\item The case when a sum of geometric and cosmological contributions is trivial $V_G+V_C=0$, but the effective potential does not vanish identically $V_{eff}\neq0$. In such a situation the three-dimensional Ricci scalar curvature\index{Ricci scalar curvature} and the global one-dimensional quantum gravity are
\begin{eqnarray}
&&{^{(3)}}R=2\Lambda,\label{kgf2}\\
&&\left(\dfrac{\delta^2}{\delta{h^2}}-\dfrac{1}{6(8\pi)^2}\dfrac{2\kappa\ell_P^2}{h}\varrho[h]\right)\Psi_M[h]=0,
\end{eqnarray}
where $\Psi_M$ is a ''material'' wave functional related to $V_M\neq0$.
\item The case when a sum of geometric and material contributions vanishes $V_G+V_M=0$, but the gravitational potential\index{gravitational potential!effective} is in general non trivial $V_{eff}\neq0$. In such a situation the Ricci scalar curvature\index{Ricci scalar curvature} of an embedded space and the global one-dimensional quantum gravity are
\begin{eqnarray}
&&{^{(3)}\!R}=2\kappa\ell_P^2\varrho,\label{kgf4}\\ &&\left(\dfrac{\delta^2}{\delta{h^2}}+\dfrac{1}{6(8\pi)^2}\dfrac{2\Lambda}{h}\right)\Psi_C[h]=0.
\end{eqnarray}
Here $\Psi_C$ is the ''cosmological'' wave functional related to $V_C\neq0$.
\item The case when a sum of cosmological and material contributions is trivial $V_C+V_M=0$, but the effective gravitational potential\index{gravitational potential!effective} is non zero $V_{eff}\neq0$. In such a situation the energy density\index{energy density} of Matter f\/ields and the global one-dimensional quantum gravity are
\begin{eqnarray}
&&\varrho=-\dfrac{\Lambda}{\kappa\ell_P^2},\label{kgf3}\\ &&\left(\dfrac{\delta^2}{\delta{h^2}}-\dfrac{1}{6(8\pi)^2}\dfrac{{^{(3)}\!R}}{h}\right)\Psi_G[h]=0,
\end{eqnarray}
where $\Psi_G$ is the ''geometric'' wave functional related to $V_G\neq0$.
\item More general approach can be based on complex analysis (For basics advances see e.g. \cite{compan}). Let us pu \emph{ad hoc} the functional Laurent series\index{Laurent series} expansion in the global dimension $h$ of the effective gravitational potential\index{gravitational potential!effective} $V_{eff}[h]$ in an inf\/initesimal neighborhood, i.e. a 1-sphere (circle) of a radius $h_\epsilon$, of any f\/ixed initial value $h_0$
\begin{equation}\label{exp}
  V_{eff}[h]=\sum_{-\infty}^{\infty}a_{n}\left(h-h_0\right)^n\quad\mathrm{in}\quad C(h_\epsilon)=\left\{h:|h-h_0|<h_{\epsilon}\right\},
\end{equation}
where $a_n$ are the series coeff\/icients given by the classical functional integral
\begin{equation}\label{calka}
  a_n=\dfrac{1}{2\pi i}\int_{C(h_\epsilon)}\dfrac{V_{eff}[h]}{\left(h-h_0\right)^{n+1}}\delta h,
\end{equation}
which is the Cauchy integral with the Radon/Lebesgue--Stieltjes measure\index{Lebesgue--Stieltjes measure} $\delta h$. Let us take into considerations $h_0=0$. Then the Ricci scalar curvature\index{Ricci scalar curvature} of a three-dimensional embedded space is
\begin{equation}\label{eq2}
  {^{(3)}\!R}=2\Lambda+2\kappa\ell_P^2\varrho-6(8\pi)^2\sum_{-\infty}^{\infty}b_{n}(h-h_0)^{n},
\end{equation}
where $b_n$ is the series coeff\/icient
\begin{equation}
  b_n=a_{n-1}+h_0a_n=\dfrac{1}{2\pi i}\int_{C(h_\epsilon)}\dfrac{h}{\left(h-h_0\right)^{n+1}}V_{eff}[h]\delta h,
\end{equation}
and the global one-dimensional quantum gravity yields
\begin{equation}\label{eq1}
  \left(\dfrac{\delta^2}{\delta{h^2}}+\sum_{-\infty}^{\infty}a_{n}(h-h_0)^n\right)\Psi[h]=0.
\end{equation}
By the triangle inequality\index{triangle inequality} one has
\begin{equation}
  |b_n|\leqslant|a_{n-1}|+|h_0||a_n|
\end{equation}
so it is easy to see that
\begin{equation}\label{been}
  \dfrac{|b_n|}{|a_n|}\leqslant\dfrac{|a_{n-1}|}{|a_n|}+|h_0|.
\end{equation}
Applying the inequality
\begin{equation}
  \left|\int{f}\right|\leqslant\int|f|,
\end{equation}
where $f$ is considered as Riemann-integrable function and integral is def\/ined, to the coeff\/icients $a_n$ and $b_n$ one has
\begin{eqnarray}
|a_n|&\leqslant&\dfrac{1}{h_\epsilon^{n+1}}\dfrac{1}{2\pi}\int_{C(h_\epsilon)}\left|V_{eff}\right|\delta{h}\leqslant\dfrac{1}{h_\epsilon^{n+1}}|a_{-1}|,
\end{eqnarray}
where $a_{-1}$ is the residue of the effective gravitational potential\index{gravitational potential!effective} in the point $h=h_0$ given by the Cauchy integral formula\index{Cauchy integral formula}
\begin{equation}
  a_{-1}=\mathrm{Res}(V_{eff},h_0)=\dfrac{1}{2\pi i}\int_{C(h_\epsilon)}{V}_{eff}\delta{h},
\end{equation}
where $C(h_\epsilon)$ traces out a circle around $h_0$ in a counterclockwise manner on the punctured disk $D=\left\{z:0<|h-h_0|<R\right\}$. If the point $h=h_0$ is a pole of order $n$, then
\begin{equation}
\mathrm{Res}(V_{eff},h_0)=\dfrac{1}{\Gamma(n)}\lim_{h\rightarrow h_0}\dfrac{\delta^{n-1}}{\delta{h}^{n-1}}\left((h-h_0)V_{eff}\right).
\end{equation}
It can be seen straightforwardly that
\begin{equation}
  \dfrac{|a_{n-1}|}{|a_n|}\geqslant{h}_\epsilon,
\end{equation}
and hence the inequality (\ref{been}) gives
\begin{equation}
  \dfrac{|b_n|}{|a_n|}\geqslant{h}_\epsilon+|h_0|.\label{inequalit1}
\end{equation}
Because by the triangle inequality\index{triangle inequality}
\begin{equation}
  |b_{n+1}|=|a_n+h_0a_{n+1}|\leqslant|a_n|+|h_0||a_{n+1}|,
\end{equation}
and
\begin{equation}
\dfrac{|a_{n+1}|}{|a_n|}\leqslant\dfrac{1}{h_\epsilon},
\end{equation}
one obtains
\begin{equation}
\dfrac{|b_{n+1}|}{|a_n|}\leqslant1+\dfrac{|h_0|}{h_\epsilon}.
\end{equation}
applying the inequality (\ref{inequalit1}) in the equivalent form
\begin{equation}
  \dfrac{|a_n|}{|b_n|}\leqslant\dfrac{1}{{h}_\epsilon+|h_0|}.\label{inequalit2}
\end{equation}
one receives the upper bound
\begin{equation}\label{boundy1}
  \dfrac{|b_{n+1}|}{|a_n|}\dfrac{|a_n|}{|b_n|}=\dfrac{|b_{n+1}|}{|b_n|}\leqslant\dfrac{1}{{h}_\epsilon+|h_0|}\left(1+\dfrac{|h_0|}{h_\epsilon}\right).
\end{equation}

Another bound for $\dfrac{|b_{n+1}|}{|b_n|}$ can be obtained as follows. Because of
\begin{equation}
  a_n=\dfrac{b_n-a_{n-1}}{h_0},
\end{equation}
one has
\begin{equation}
  b_{n+1}=\dfrac{b_n-a_{n-1}}{h_0}+h_0a_{n+1},
\end{equation}
or after small algebraic manipulations
\begin{equation}
  h_0b_{n+1}+a_{n-1}=h_0a_{n+1}+b_n.
\end{equation}
This equation can be rewritten in the form
\begin{equation}
  1=\left|\dfrac{h_0b_{n+1}}{h_0a_{n+1}+b_n}+\dfrac{a_{n-1}}{h_0a_{n+1}+b_n}\right|,
\end{equation}
which after taking into account the triangle inequality\index{triangle inequality}
\begin{equation}
\left|\dfrac{h_0b_{n+1}}{h_0a_{n+1}+b_n}+\dfrac{a_{n-1}}{h_0a_{n+1}+b_n}\right|\leqslant\left|\dfrac{h_0b_{n+1}}{h_0a_{n+1}+b_n}\right|+\left|\dfrac{a_{n-1}}{h_0a_{n+1}+b_n}\right|,
\end{equation}
leads to
\begin{equation}
  |h_0a_{n+1}+b_n|\leqslant|h_0||b_{n+1}|+|a_{n-1}|.
\end{equation}
Applying again the triangle inequality\index{triangle inequality}
\begin{equation}
  |h_0a_{n+1}+b_n|\leqslant|h_0||a_{n+1}|+|b_n|,
\end{equation}
one receives
\begin{equation}
|h_0||b_{n+1}|-|b_n|\leqslant|h_0||a_{n+1}|-|a_{n-1}|.
\end{equation}
This inequality can be rewritten as
\begin{equation}
|h_0|\dfrac{|b_{n+1}|}{|a_n|}-\dfrac{|b_n|}{|a_n|}\leqslant|h_0|\dfrac{|a_{n+1}|}{|a_n|}-\dfrac{|a_{n-1}|}{|a_n|},
\end{equation}
or equivalently
\begin{equation}
|h_0|\dfrac{|b_{n+1}|}{|b_n|}-1\leqslant\dfrac{|a_n|}{|b_n|}\left(|h_0|\dfrac{|a_{n+1}|}{|a_n|}-\dfrac{|a_{n-1}|}{|a_n|}\right).
\end{equation}
In the light of the inequality (\ref{inequalit2}) and the relation
\begin{equation}
  |h_0|\dfrac{|a_{n+1}|}{|a_n|}-\dfrac{|a_{n-1}|}{|a_n|}\leqslant\dfrac{|h_0|}{{h}_\epsilon}-{h}_\epsilon,
\end{equation}
one obtains the bound
\begin{equation}
\dfrac{|b_{n+1}|}{|b_n|}\leqslant\dfrac{1}{|h_0|}\left(1+\dfrac{1}{h_\epsilon+|h_0|}\left(\dfrac{|h_0|}{h_\epsilon}-h_\epsilon\right)\right).
\end{equation}
Comparing this bound to the previous one (\ref{boundy1}) one receives
\begin{equation}
|h_0|(|h_0|-1)\geqslant0,
\end{equation}
what gives the condition for $h_0$
\begin{equation}
  |h_0|\in\{0\}\cup[1,\infty).\label{h0cond}
\end{equation}
\end{enumerate}
Naturally, there is many other opportunities for selection of a form of the effective gravitational potential\index{gravitational potential!effective} $V_{eff}[h]$. However, in this section we shall discuss the only a particular case.

\section{The Newton--Coulomb Potential\index{Newton--Coulomb potential}}\label{sec:2}
Let us consider \emph{the residual approximation}\index{residual approximation} in which the series coeff\/icient of the effective gravitational potential\index{gravitational potential!effective} are
\begin{equation}
  a_n=\left\{\begin{array}{cc}a_{-1}=const&\mathrm{for}~n=-1\\0&\mathrm{for}~n\neq-1\end{array}\right.,
\end{equation}
i.e. the effective gravitational potential\index{gravitational potential!effective} (\ref{eff}) has the form of the Newton--Coulomb potential\index{Newton--Coulomb potential}
\begin{equation}
  V_{eff}=\dfrac{a_{-1}}{h-h_0}.
\end{equation}
In such a situation the coeff\/icients $b_n$ are
\begin{equation}
b_n=\left\{\begin{array}{cc}
b_{-1}=h_0a_{-1}&\mathrm{for}~n=-1\\
b_{0}=a_{-1}&\mathrm{for}~n=0\\0&\mathrm{for}~n\neq-1,0\end{array}\right.,
\end{equation}
so that the Ricci scalar curvature\index{Ricci scalar curvature} of a three-dimensional space is
\begin{equation}\label{ric}
  {^{(3)}\!R}=2\Lambda+2\kappa\ell_P^2\varrho-6(8\pi)^2a_{-1}\left(1+\dfrac{h_0}{h-h_0}\right),
\end{equation}
whereas the Klein--Gordon equation\index{Klein--Gordon equation} (\ref{kgf}) is
\begin{equation}\label{eq1a}
\left(\dfrac{\delta^2}{\delta h^2}+\dfrac{a_{-1}}{h-h_0}\right)\Psi=0.
\end{equation}
The equation (\ref{ric}) def\/ines some three-geometries, but even in the vacuum situation, i.e. $\varrho=0$ and $\Lambda=0$, it is diff\/icult to establish an induced geometry which Ricci scalar curvature\index{Ricci scalar curvature} behaves like
\begin{equation}
{^{(3)}\!R}\sim1+\dfrac{h_0}{h-h_0}.\label{rich}
\end{equation}
Interestingly, in general the residue of the three-dimensional Ricci scalar curvature\index{Ricci scalar curvature} in the point $h_0$ is
\begin{equation}
  \mathrm{Res}({^{(3)}}R,h_0)=2\kappa\ell_P^2\mathrm{Res}(\varrho,h_0)-6(8\pi)^2a_{-1}h_0,
\end{equation}
i.e. it can be taken equal to zero if and only if the residue of energy density\index{energy density} of Matter f\/ields\index{Matter f\/ields} is
\begin{equation}
\mathrm{Res}(\varrho,h_0)=\dfrac{3(8\pi)^2}{\kappa\ell_P^2}a_{-1}h_0.
\end{equation}
If one takes \emph{ad hoc} the relation
\begin{equation}
a_{-1}=\dfrac{\Lambda}{3(8\pi)^2},
\end{equation}
then the Ricci scalar curvature\index{Ricci scalar curvature} of induced three-geometry has the form
\begin{equation}\label{ric}
  {^{(3)}\!R}=2\kappa\ell_P^2\varrho-\dfrac{2\Lambda{h_0}}{h-h_0},
\end{equation}
and its residue
\begin{equation}
  \mathrm{Res}({^{(3)}}R,h_0)=2\kappa\ell_P^2\mathrm{Res}(\varrho,h_0)-2\Lambda{h_0},
\end{equation}
vanishes if and only if
\begin{equation}
\mathrm{Res}(\varrho,h_0)=\dfrac{\Lambda}{\kappa\ell_P^2}h_0.
\end{equation}
Then also the geometry of an embedded three-manifold is Ricci-f\/lat\index{Ricci-f\/lat manifold} if and only if the energy density\index{energy density} of Matter f\/ields\index{Matter f\/ields} has the following form
\begin{equation}
  \varrho=\dfrac{\Lambda}{\kappa\ell_P^2}\dfrac{{h_0}}{h-h_0}.
\end{equation}
Another possible Ricci-f\/lat three-manifold\index{Ricci-f\/lat manifold} is obtained for $\Lambda=0$ and
\begin{eqnarray}
\varrho=\dfrac{3(8\pi)^2}{\kappa\ell_P^2}a_{-1}\left(1+\dfrac{h_0}{h-h_0}\right).
\end{eqnarray}

In general three-spaces having induced metrics satisfying the Ricci scalar curvature\index{Ricci scalar curvature} (\ref{rich}) are not known yet. However, it is evidently seen that in the particular case $h_0=0$, which is in full accordance with the general condition (\ref{h0cond}), the situation is much more simpler, i.e.
\begin{eqnarray}
&&{^{(3)}\!R}=2\Lambda+2\kappa\ell_P^2\varrho-6(8\pi)^2a_{-1},\label{ncqg1}\\
&&\left(\dfrac{\delta^2}{\delta h^2}+\dfrac{a_{-1}}{h}\right)\Psi=0.\label{ncqg2}
\end{eqnarray}
Let us consider this particular case as the basic situation. We shall call the global one-dimensional quantum gravity described by the system of equations (\ref{ncqg1})-(\ref{ncqg2}) \emph{the Newton--Coulomb quantum gravity}.\index{Newton--Coulomb quantum gravity}

As the example we shall consider vanishing of the energy density\index{energy density}
\begin{equation}\label{kgf3}
\varrho\equiv 0,
\end{equation}
i.e. stationarity of Matter f\/ields\index{Matter f\/ields}. We shall call this case \emph{the Newton--Coulomb stationary quantum gravity}\index{Newton--Coulomb quantum gravity!stationary}. In such a situation the Ricci scalar curvature\index{Ricci scalar curvature} of three-dimensional embedded space becomes
\begin{equation}
  {^{(3)}}R=2\Lambda-6(8\pi)^2a_{-1}=constant.
\end{equation}
Therefore the Ricci curvature tensor\index{Ricci curvature tensor} of the three-dimensional manifolds describes the three-dimensional Einstein manifolds\index{Einstein manifolds} \cite{besse}
\begin{equation}\label{eins}
  R_{ij}=\lambda h_{ij},
\end{equation}
where the sign $\lambda$ of the Einstein manifolds\index{Einstein manifolds} is def\/ined by the Newton--Coulomb stationary quantum gravity\index{Newton--Coulomb stationary quantum gravity} as follows
\begin{equation}\label{sign}
  \dfrac{2}{3}\Lambda-2(8\pi)^2a_{-1}=\lambda.
\end{equation}
Because, however, energy density\index{energy density} of Matter f\/ields  vanishes therefore the Einstein manifolds\index{Einstein manifolds} described by the sign (\ref{sign}) possess \emph{maximal symmetry}. By this reason the Newton--Coulomb stationary quantum gravity\index{Newton--Coulomb stationary quantum gravity} geometrically corresponds to three-dimensional \emph{the maximally symmetric Einstein manifolds}.\index{Einstein manifolds!maximally symmetric}

In this manner in general one can consider the classif\/ication of the three-dimensional spaces, which are maximally symmetric three-dimensional Einstein manifolds\index{Einstein manifolds} (\ref{eins}), with respect to the value of the sign $\lambda$ (\ref{sign}) of a manifold. The following conclusion can be deduced straightforwardly.

\begin{conclusion}
The Newton--Coulomb stationary quantum gravity\index{Newton--Coulomb stationary quantum gravity}, def\/ined by the effective gravitational potential\index{gravitational potential!effective} $V_{eff}[h]=\dfrac{a_{-1}}{h}$, determines the three-dimensional embedded spaces which are the maximally symmetric three-dimensional Einstein manifolds\index{Einstein manifolds}, characterized by the sign (\ref{sign}). There are particular situations:
\begin{enumerate}
\item When the sign is non zero $\lambda\neq0$ and the residue of the effective gravitational potential\index{gravitational potential!effective} is negative $a_{-1}=-|\alpha|$, then the effective potential $V_{eff}[h]$ becomes the Newtonian attractive potential energy
    \begin{equation}
    V_{eff}=-\dfrac{|\alpha|}{h}=-\dfrac{Gm_1m_2}{\ell_Ph}.\label{i1}
    \end{equation}
    When the cosmological constant\index{cosmological constant} is positive $\Lambda=+|\Lambda|$ then the maximally symmetric Einstein three-manifolds are characterized by positive Ricci scalar curvature\index{Ricci scalar curvature}
    \begin{equation}
    {^{(3)}}R=\dfrac{2}{3}|\Lambda|+2(8\pi)^2|\alpha|.
    \end{equation}
    When the cosmological constant\index{cosmological constant} is negative $\Lambda=-|\Lambda|$ the maximally symmetric Einstein three-manifolds are characterized by the Ricci scalar curvature\index{Ricci scalar curvature}
    \begin{equation}
    {^{(3)}}R=-\dfrac{2}{3}|\Lambda|+2(8\pi)^2|\alpha|,
    \end{equation}
    which is negative if $|\Lambda|>3(8\pi)^2|\alpha|$, and positive if $|\Lambda|<3(8\pi)^2|\alpha|$.
\item When the sign is non zero $\lambda\neq0$ and the residue of the effective gravitational potential\index{gravitational potential!effective} is positive $a_{-1}=+|\alpha|$, then the effective potential $V_{eff}[h]$ becomes the Coulomb repulsive potential energy
    \begin{equation}
    V_{eff}=\dfrac{|\alpha|}{h}=\dfrac{q_1q_2}{4\pi\epsilon_0\ell_Ph}.\label{i2}
    \end{equation}
    When the cosmological constant\index{cosmological constant} is negative $\Lambda=-|\Lambda|$ then the maximally symmetric Einstein three-manifolds are characterized by negative Ricci scalar curvature\index{Ricci scalar curvature}
    \begin{equation}
      {^{(3)}}R=-\dfrac{2}{3}|\Lambda|-2(8\pi)^2|\alpha|.
    \end{equation}
    When the cosmological constant\index{cosmological constant} is positive $\Lambda=+|\Lambda|$ then the maximally symmetric Einstein three-manifolds are characterized by the Ricci scalar curvature\index{Ricci scalar curvature}
    \begin{equation}
      {^{(3)}}R=\dfrac{2}{3}|\Lambda|-2(8\pi)^2|\alpha|,
    \end{equation}
    which is negative if $|\Lambda|<3(8\pi)^2|\alpha|$, an positive if $|\Lambda|>3(8\pi)^2|\alpha|$.
\item When the sign is vanishing $\lambda=0$, i.e. the maximally symmetric Einstein three-manifolds\index{Einstein manifolds!maximally symmetric} are Ricci-f\/lat\index{Ricci-f\/lat manifold}, one determines uniquely the value of the reside of the effective gravitational potential\index{gravitational potential!effective} as
    \begin{equation}
    a_{-1}=\pm\dfrac{|\Lambda|}{3(8\pi)^2},\label{i3}
    \end{equation}
In such a case one obtains the values of cosmological constant\index{cosmological constant}
\begin{equation}\label{lambda}
  |\Lambda|=\left\{\begin{array}{rl}\dfrac{3(8\pi)^3E_P}{4\ell_P^2}r_g(m_1)r_g(m_2)&\mathrm{for~the~Newton~law}\vspace*{10pt}\\
  \dfrac{3(8\pi)^3E_P}{\ell_P^2}r_e(q_1)r_e(q_2)&\mathrm{for~the~Coulomb~law}\end{array}\right.
\end{equation}
where $m$ is mass of a body generating Newtonian gravitational f\/ield in vacuum and $r_g(m)=\dfrac{2Gm}{c^2}=\kappa{c^2}\dfrac{m}{4\pi}$ is its gravitational radius, $q$ is charge generating Coulombic electrical f\/ield in vacuum and $r_e(q)=\dfrac{q}{\sqrt{4\pi\epsilon_0}}\sqrt{\dfrac{G}{c^4}}=\sqrt{\dfrac{\kappa}{2\epsilon_0}}\dfrac{q}{4\pi}$ is its electrical radius.
\end{enumerate}
\end{conclusion}

Note that in fact, by assuming the relation for the series coeff\/icients (\ref{calka}), the residue $a_{-1}$ is the Cauchy integral of the effective potential $V_{eff}$ in the f\/ixed point $h_0=0$
\begin{equation}
  a_{-1}=\textrm{Res}\left[\dfrac{1}{6(8\pi)^2}\dfrac{1}{h}\left(-{^{(3)}\!R}+2\Lambda+2\kappa\ell_P^2\varrho\right),h=0\right],
\end{equation}
and its value can be straightforwardly established as
\begin{equation}\label{resi}
  a_{-1}=\dfrac{1}{6(8\pi)^2}\left.\left(-{^{(3)}\!R}+2\Lambda+2\kappa\ell_P^2\varrho\right)\right|_{h=0}=-\dfrac{{^{(3)}}R_0}{6(8\pi)^2}+\dfrac{\Lambda}{3(8\pi)^2}+\dfrac{\kappa\ell_P^2\varrho_0}{3(8\pi)^2}\quad,
\end{equation}
where subscript ''$0$'' on the LHS means value of a quantity in $h=0$.

When one associates the effective gravitational potential\index{gravitational potential!effective} $V_{eff}=\dfrac{a_{-1}}{h}$ the Newton or the Coulomb potential energy, then the global dimension becomes a spatial distance $r=\sqrt{x^2+y^2+z^2}$
\begin{equation}\label{detr}
  h\equiv r\quad,
\end{equation}
so that the evolution (\ref{eq1a}) becomes radial wave equation
\begin{equation}\label{eq1ab}
  \left(\dfrac{d^2}{d{r^2}}+\dfrac{\mp|\alpha|}{r}\right)\Psi(r)=0,
\end{equation}
describing a quantum Kepler problem\index{Kepler problem}. There is a lot of metrics $h_{ij}$ possessing the same determinant, for instance
\begin{equation}
h_{ij}=r^{1/3}r_{ij},
\end{equation}
where $r_{ij}$ is $SO(3)$ group rotation matrix, which can be expressed via the Euler angles $(\theta,\varphi,\phi)$
\begin{equation}
  r_{ij}(\theta,\varphi,\phi)\equiv r_{il}^{(3)}(\theta)r_{lk}^{(2)}(\varphi)r_{kj}^{(3)}(\phi)\quad,
\end{equation}
where $r_{ij}^{(p)}(\vartheta)$ are rotation matrices around a $p$-axis
\begin{eqnarray}
r_{ij}^{(3)}(\vartheta)&=&\left[\begin{array}{ccc}\cos\vartheta&-\sin\vartheta&0\\\sin\vartheta&\cos\vartheta&0\\0&0&1\end{array}\right],\\ r_{ij}^{(2)}(\vartheta)&=&\left[\begin{array}{ccc}\cos\vartheta&0&\sin\vartheta\\0&1&0\\-\sin\vartheta&0&\cos\vartheta\end{array}\right].
\end{eqnarray}

Interestingly, one can connect the radial wave equation (\ref{eq1ab}) with the radial Schr\"odinger equation \index{Schr\"odinger equation!radial}for a particle equipped with energy $E$, mass $m$, and potential energy $V$ (For more detailed discussion of the radial Schr\"odinger equation see e.g. the Ref. \cite{radial}). Applying the substitution $\Psi(r)=rR(r)$ where $R(r)$ is the radial part of the wave function of a particle, one can present the radial Schr\"odinger equation in the following form
\begin{equation}
\left[\dfrac{d^2}{dr^2}-\left(\dfrac{2m}{\hslash^2}V+\dfrac{l(l+1)}{r^2}\right)\right]\Psi(r)=\dfrac{2mE}{\hslash^2}\psi(r),
\end{equation}
where $l$ is one of subscripts of the spherical harmonics\index{spherical harmonics} $Y_l^m(\theta,\phi)$ of degree $l$ and order $m$
\begin{equation}
Y_l^m(\theta,\phi)=\sqrt{\dfrac{2l+1}{4\pi}\dfrac{(l-m)!}{(l+m)!}}P_{lm}(\cos\theta)e^{im\phi},
\end{equation}
which are the angular part of the wave function of a particle. Here $P_{lm}(x)$ are associated Legendre polynomials\index{Legendre polynomials!associated}
\begin{equation}
  P_{lm}(x)=\dfrac{1}{2^ll!}(1-x^2)^{m/2}\dfrac{d^{l+m}}{dx^{l+m}}(x^2-1)^l.
\end{equation}
Putting \emph{ad hoc} the values $E=0$, $l=0$ and $-\dfrac{2m}{\hslash^2}V=V_{eff}$ one receives excellent coincidence with the equation (\ref{eq1ab}). This nice property allows to construct the physical interpretation of the wave functional $\Psi(r)$. Namely, it is strictly related to the wave function $\psi(r,\theta,\phi)=R(r)Y_0^m(\theta,\phi)$ of a particle equipped with mass $m$ and having the total energy $E=0$ and the potential energy $V=-\dfrac{\hslash^2}{2m}\dfrac{\pm|\alpha|}{r}$. The identif\/ication is as follows
\begin{equation}
  \Psi(r)=r\dfrac{\psi(r,\theta,\phi)}{Y_0^0(\theta,\phi)},
\end{equation}
where $Y_0^0(\theta,\phi)=\dfrac{1}{2\sqrt{\pi}}$ is the only non trivial value of $Y_0^m(\theta,\phi)$.

Another linkage to the radial Schr\"odinger equation \index{Schr\"odinger equation!radial}is also possible. Let the wave functional $\Psi(r)$ is the radial part $R(r)$ of the wave function of a particle, i.e. $\psi(r,\theta,\phi)=\Psi(r)Y_l^m(\theta,\phi)$, equipped with mass $m$ and total energy $E$, and moving in \emph{arbitrary} potential $V(r)$. Then $\Psi(r)$ satisf\/ies the radial Schr\"odinger equation
\begin{equation}
-\dfrac{1}{r^2}\dfrac{d}{dr}\left(r\dfrac{d\Psi}{dr}\right)+\left(\dfrac{2m}{\hslash^2}V+\dfrac{l(l+1)}{r^2}\right)\Psi=\dfrac{2mE}{\hslash^2}\Psi,
\end{equation}
or equivalently
\begin{equation}\label{radschr}
\dfrac{d^2\Psi}{dr^2}+\dfrac{1}{r}\dfrac{d\Psi}{dr}-\left(\dfrac{2m}{\hslash^2}rV+\dfrac{l(l+1)}{r}\right)\Psi=-\dfrac{2mE}{\hslash^2}r\Psi.
\end{equation}
This equations can be separated on the system of two equations
\begin{eqnarray}
\dfrac{d^2\Psi}{dr^2}-\dfrac{2m}{\hslash^2}rV\Psi&=&0,\label{schr1}\\
\dfrac{d\Psi}{dr}-l(l+1)\Psi+\dfrac{2mE}{\hslash^2}r^2\Psi&=&0.\label{schr2}
\end{eqnarray}
The equation (\ref{schr1}) is exactly the wave equation (\ref{eq1ab}) if and only if the equation holds
\begin{equation}
  V(r)=-\dfrac{\hslash^2}{2mr}V_{eff}.\label{potent1}
\end{equation}
The equation (\ref{schr2}) must be solved. It is easy to see that this equation can be integrated straightforwardly
\begin{equation}
  \int_{\Psi(r_0)}^{\Psi(r)}\dfrac{d\Psi}{\Psi}=\int_{r_0}^r\left(l(l+1)-\dfrac{2mE}{\hslash^2}r^2\right)dr,
\end{equation}
where $r_0$ is some reference value of $r$, what gives
\begin{equation}
  \ln\left|\dfrac{\Psi(r)}{\Psi(r_0)}\right|=l(l+1)(r-r_0)-\dfrac{2mE}{\hslash^2}\dfrac{r^3-r_0^3}{3},
\end{equation}
and results in the solution
\begin{equation}
\Psi(r)=\Psi(r_0)\exp\left\{-l(l+1)r_0+\dfrac{2mE}{3\hslash^2}r_0^3\right\}\exp\left\{l(l+1)r-\dfrac{2mE}{3\hslash^2}r^3\right\}.
\end{equation}
Now one can differentiate the equation (\ref{schr2}) with respect to $r$ and obtains
\begin{eqnarray}
\dfrac{d^2\Psi}{dr^2}&=&l(l+1)\dfrac{d\Psi}{dr}-\dfrac{2mE}{\hslash^2}\left(2r\Psi+r^2\dfrac{d\Psi}{dr}\right)=\nonumber\\
&=&\left[l(l+1)-\dfrac{2mE}{\hslash^2}r^2\right]\dfrac{d\Psi}{dr}-\dfrac{4mE}{\hslash^2}r\Psi=\nonumber\\
&=&\left\{\left[l(l+1)-\dfrac{2mE}{\hslash^2}r^2\right]^2-\dfrac{4mE}{\hslash^2}r\right\}\Psi.\label{derv2}
\end{eqnarray}
Application of the equation (\ref{schr1}) to the result (\ref{derv2}) leads to
\begin{equation}
  \dfrac{2m}{\hslash^2}rV=\left\{\left[l(l+1)-\dfrac{2mE}{\hslash^2}r^2\right]^2-\dfrac{4mE}{\hslash^2}r\right\},
\end{equation}
what allows to establish the potential $V$
\begin{equation}
V(r)=\dfrac{\hslash^2}{2m}\dfrac{l^2(l+1)^2}{r}-2E-2El(l+1)r+\dfrac{2mE^2}{\hslash^2}r^3.
\end{equation}
Using of the identif\/ication (\ref{potent1}) and the explicit form of $V_{eff}=\dfrac{\pm|\alpha|}{r}$, one receives the equation for $r$
\begin{equation}\label{5th}
\left(\dfrac{2mE}{\hslash^2}\right)^2r^5-\dfrac{4mE}{\hslash^2}l(l+1)r^3-\dfrac{4mE}{\hslash^2}r^2+l^2(l+1)^2r\pm|\alpha|=0,
\end{equation}
which is very diff\/icult to solve, but simplif\/ies for suggested value $l=0$
\begin{equation}
\left(\dfrac{2mE}{\hslash^2}\right)^2r^5-\dfrac{4mE}{\hslash^2}r^2\pm|\alpha|=0.\label{eqnr}
\end{equation}
The equation (\ref{eqnr}) can be solved easy. Let us focus on the real solution which for the Coulombic case is
\begin{eqnarray}
 r&=&\Bigg[\dfrac{\hslash^2}{2mE}\left(\dfrac{mE}{\hslash^2}|\alpha|\right)^{1/3}\left(\sqrt{1-\dfrac{32}{27}\dfrac{2mE}{\hslash^2|\alpha|^2}}-1\right)^{1/3}+\nonumber\\
 &+&\dfrac{2}{3}\left(\dfrac{mE}{\hslash^2}|\alpha|\right)^{-1/3}\left(\sqrt{1-\dfrac{32}{27}\dfrac{2mE}{\hslash^2|\alpha|^2}}-1\right)^{-1/3}\Bigg]^{1/2},\label{errr1}
\end{eqnarray}
while for the Newtonian case one obtains
\begin{eqnarray}
 r&=&\Bigg[\dfrac{\hslash^2}{2mE}\left(\dfrac{mE}{\hslash^2}|\alpha|\right)^{1/3}\left(\sqrt{1-\dfrac{32}{27}\dfrac{2mE}{\hslash^2|\alpha|^2}}+1\right)^{1/3}+\nonumber\\
 &+&\dfrac{2}{3}\left(\dfrac{mE}{\hslash^2}|\alpha|\right)^{-1/3}\left(\sqrt{1-\dfrac{32}{27}\dfrac{2mE}{\hslash^2|\alpha|^2}}+1\right)^{-1/3}\Bigg]^{1/2},\label{errr2}
\end{eqnarray}
Recall that in our theory $h=r$, i.e. in fact the results (\ref{errr1})-(\ref{errr2}) f\/ix value of $h$. The distance must be real number. For this must be
\begin{equation}
E\leqslant\dfrac{27}{32}\dfrac{\hslash^2}{2m}|\alpha|^2.
\end{equation}
In general solutions of the algebraic equation (\ref{5th}) can be found by methods of the Galois group\index{Galois group}. Despite the real and positive solution is diff\/icult to extract, some simpler solution can be constructed. Let us substitute to the equation (\ref{5th}) as the unknown $r=x+iy$, where the imaginary part $y$ will be $y\rightarrow0$ in a certain stage of the construction. In such a situation the equation (\ref{5th}) is equivalent to the statement that its both real and imaginary part vanish. Such a method generates the system of equations
\begin{eqnarray}
\!\!\!\!\!\!\!\!\!\!\!\!\!\!\!\!\!\!\!\!\!\!\!\!\!\!\!\!\!\!
&&\left(\dfrac{2mE}{\hslash^2}\right)^2y^4+2\left(\dfrac{2mE}{\hslash^2}\right)\left(l(l+1)-5\left(\dfrac{2mE}{\hslash^2}\right)x^2\right)y^2-\nonumber\\
\!\!\!\!\!\!\!\!\!\!\!\!\!\!\!\!\!\!\!\!\!\!\!\!\!\!\!\!\!\!
&&-\left(\dfrac{2mE}{\hslash^2}\right)x\left(4+6l(l+1)x-5\left(\dfrac{2mE}{\hslash^2}\right)x^3\right)+l^2(l+1)^2=0,
\end{eqnarray}
\begin{eqnarray}
\!\!\!\!\!\!\!\!\!\!\!\!\!\!\!\!\!\!\!\!\!\!\!\!\!\!\!\!\!\!
&&5\left(\dfrac{2mE}{\hslash^2}\right)^2xy^4+2\left(\dfrac{2mE}{\hslash^2}\right)\left(1+3l(l+1)x-5\left(\dfrac{2mE}{\hslash^2}\right)x^3\right)y^2+\nonumber\\
\!\!\!\!\!\!\!\!\!\!\!\!\!\!\!\!\!\!\!\!\!\!\!\!\!\!\!\!\!\!
&&+x\left(-2\left(\dfrac{2mE}{\hslash^2}\right)x+\left(l(l+1)-\left(\dfrac{2mE}{\hslash^2}\right)^2\right)^2\right)\pm|\alpha|=0,
\end{eqnarray}
where we introduced the shortened notation $L=l(l+1)$. Now one can put $y=0$, so that $x=r$, and obtain the system of equations
\begin{eqnarray}
\!\!\!\!\!\!\!\!\!\!\!\!\!\!\!\!\!\!\!\!\!\!\!\!\!\!\!\!\!\!
&&-\left(\dfrac{2mE}{\hslash^2}\right)r\left(4+6l(l+1)r-5\left(\dfrac{2mE}{\hslash^2}\right)r^3\right)+l^2(l+1)^2=0,\\
\!\!\!\!\!\!\!\!\!\!\!\!\!\!\!\!\!\!\!\!\!\!\!\!\!\!\!\!\!\!
&&l^2(l+1)^2r-\left(\dfrac{2mE}{\hslash^2}\right)r^2\left(2+2l(l+1)r-\left(\dfrac{2mE}{\hslash^2}\right)r^3\right)\pm|\alpha|=0,
\end{eqnarray}
which can be presented in the form
\begin{eqnarray}
\left(\dfrac{2mE}{\hslash^2}\right)r^2&=&\dfrac{l^2(l+1)^2r}{\left(4+6l(l+1)r-5\left(\dfrac{2mE}{\hslash^2}\right)r^3\right)},\\
\left(\dfrac{2mE}{\hslash^2}\right)r^2&=&\dfrac{l^2(l+1)^2r\pm|\alpha|}{\left(2+2l(l+1)r-\left(\dfrac{2mE}{\hslash^2}\right)r^3\right)},
\end{eqnarray}
and lead to the relation
\begin{equation}
\dfrac{l^2(l+1)^2r}{\left(4+6l(l+1)r-5\left(\dfrac{2mE}{\hslash^2}\right)r^3\right)}=\dfrac{l^2(l+1)^2r\pm|\alpha|}{\left(2+2l^2(l+1)^2r-\left(\dfrac{2mE}{\hslash^2}\right)r^3\right)}.\label{redux}
\end{equation}
The relation (\ref{redux}), however, can be rewritten in much more convenient form of the 4th order algebraic equation
\begin{eqnarray}\label{redux2}
&&4\left(\dfrac{2mE}{\hslash^2}\right)l^2(l+1)^2r^4\pm5\left(\dfrac{2mE}{\hslash^2}\right)|\alpha|r^3-4l^3(l+1)^3r^2-\nonumber\\
&&2l(l+1)\left(l(l+1)\pm3|\alpha|\right)r-(\pm4|\alpha|)=0.
\end{eqnarray}
In other words the 5th order algebraic equation very diff\/icult to solve has been reduced to the 4th order algebraic equation which in general can solved straightforwardly. In our interest is the real and positive solution of the equation (\ref{redux2}) which is given by the formula
\begin{eqnarray}
\!\!\!\!\!\!\!\!\!\!\!\!\!\!\!
&&r=-\dfrac{\pm5|\alpha|}{16l^2(l+1)^2}+\dfrac{1}{2}\Bigg(\dfrac{1}{2^{2/3}3A_l}\left(45\dfrac{|\alpha|^2}{l(l+1)}-81(\pm|\alpha|)+\dfrac{4\hslash^2l^4(l+1)^4}{mE}\right)+\nonumber\\
\!\!\!\!\!\!\!\!\!\!\!\!\!\!\!
&&\dfrac{25|\alpha|^2}{64l^4(l+1)^4}+\dfrac{\hslash^2l(l+1)}{3mE}+\dfrac{\hslash^2A_l}{{2}^{10/3}3mEl^2(l+1)^2}\Bigg)^{1/2}+\dfrac{1}{2}\Bigg(\dfrac{25|\alpha|^2}{32l^4(l+1)^4}+\nonumber\\
\!\!\!\!\!\!\!\!\!\!\!\!\!\!\!
&&\dfrac{2\hslash^2l(l+1)}{3mE}-\dfrac{1}{2^{2/3}3A_l}\left(45\dfrac{|\alpha|^2}{l(l+1)}-81(\pm|\alpha|)+\dfrac{4\hslash^2l^4(l+1)^4}{mE}\right)-\nonumber\\
\!\!\!\!\!\!\!\!\!\!\!\!\!\!\!
&&\dfrac{\hslash^2A_l}{2^{10/3}3mEl^2(l+1)^2}+\dfrac{1}{4l(l+1)}\left(\dfrac{2\hslash^2l(l+1)}{mE}-\dfrac{\pm125|\alpha|^3}{64l^5(l+1)^5}+\dfrac{\pm7\hslash^2|\alpha|}{2mE}\right)\times\nonumber\\
\!\!\!\!\!\!\!\!\!\!\!\!\!\!\!
&&\Bigg(\dfrac{25|\alpha|^2}{64l^4(l+1)^4}+\dfrac{\hslash^2l(l+1)}{3mE}+\dfrac{\hslash^2A_l}{{2}^{10/3}3mEl^2(l+1)^2}+\nonumber\\
\!\!\!\!\!\!\!\!\!\!\!\!\!\!\!
&&\dfrac{1}{2^{2/3}3A_l}\left(45\dfrac{|\alpha|^2}{l(l+1)}-81(\pm|\alpha|)+\dfrac{4\hslash^2l^4(l+1)^4}{mE}\right)\Bigg)^{-1/2}\Bigg)^{1/2},
\end{eqnarray}
where
\begin{eqnarray}
\!\!\!\!\!\!\!\!\!\!\!\!\!\!\!
&&A_l=\dfrac{2mE}{\hslash^2}\Bigg(-\dfrac{\pm675\hslash^2|\alpha|^3}{2mE}+\dfrac{\hslash^4l^6(l+1)^6}{2(mE)^2}\bigg(\dfrac{351|\alpha|^2}{l(l+1)}-\dfrac{\pm297|\alpha|}{l(l+1)}-\nonumber\\
\!\!\!\!\!\!\!\!\!\!\!\!\!\!\!
&&\dfrac{8\hslash^2l^3(l+1)^3}{mE}+54\bigg)+\dfrac{\hslash^2}{2mE}\Bigg(-\dfrac{\hslash^2l^6(l+1)^6}{mE}\bigg(\dfrac{45|\alpha|^2}{l(l+1)}+\dfrac{4\hslash^2l^4(l+1)^4}{mE}-\nonumber\\
\!\!\!\!\!\!\!\!\!\!\!\!\!\!\!
&&81(\pm|\alpha|)\bigg)^3+\bigg(\pm675|\alpha|^3-\dfrac{\hslash^2l^8(l+1)^8}{mE}\bigg(\dfrac{351|\alpha|^2}{l(l+1)}-\dfrac{\pm297|\alpha|}{l(l+1)}-\nonumber\\ \!\!\!\!\!\!\!\!\!\!\!\!\!\!\!
&&\dfrac{8\hslash^2l^3(l+1)^3}{mE}\bigg)\bigg)^2+54\Bigg)^{1/2}\Bigg)^{1/3}.
\end{eqnarray}

Another separation of the radial Schr\"odinger equation \index{Schr\"odinger equation!radial}(\ref{radschr}) is possible. Namely, one can rewrite this equation as the system
\begin{eqnarray}
\dfrac{d^2\Psi}{dr^2}-\left(\dfrac{2m}{\hslash^2}rV+\dfrac{l(l+1)}{r}\right)\Psi&=&0,\label{schr1a}\\
\dfrac{d\Psi}{dr}+\dfrac{2mE}{\hslash^2}r^2\Psi&=&0.\label{schr2a}
\end{eqnarray}
The procedure analogous to the previous separation gives the solution
\begin{equation}
\Psi(r)=\Psi(r_0)\exp\left\{\dfrac{2mE}{3\hslash^2}r_0^3\right\}\exp\left\{-\dfrac{2mE}{3\hslash^2}r^3\right\},
\end{equation}
the potential
\begin{equation}
  V(r)=-\dfrac{\hslash^2}{2m}\dfrac{l(l+1)}{r^2}-2E+\dfrac{2mE^2}{\hslash^2}r^3,
\end{equation}
and the equation for $r$
\begin{equation}
\left(\dfrac{2mE}{\hslash^2}\right)^2r^4-\dfrac{4mE}{\hslash^2}r-\dfrac{\hslash^2}{2m}l^2(l+1)^2\pm|\alpha|=0.
\end{equation}
This is equation can be solved for arbitrary $l$ with no problems. There are real and positive solutions if and only if
\begin{equation}
\pm|\alpha|\leqslant\dfrac{\hslash^2}{2m}l^2(l+1)^2.
\end{equation}
Then there is the only one real and positive solution given by
\begin{eqnarray}
 &&r=\dfrac{1}{2}\Bigg\{\dfrac{2\hslash^2}{mE}\Bigg[\dfrac{2^{-1/3}}{(mE/\hslash^2)^{2/3}}\beta_l^{1/3}+\dfrac{2^{2/3}\alpha_l}{3(mE/\hslash^2)^{4/3}}\beta_l^{-1/3}\Bigg]^{-1/2}\nonumber\\
 &&-\dfrac{2^{-1/3}}{(mE/\hslash^2)^{2/3}}\beta_l^{1/3}-\dfrac{2^{2/3}\alpha_l}{3(mE/\hslash^2)^{4/3}}\beta_l^{-1/3}\Bigg\}^{1/2}\nonumber\\
 &&+\dfrac{1}{2}\left(\dfrac{2^{-1/3}}{(mE/\hslash^2)^{2/3}}\beta_l^{1/3}+\dfrac{2^{2/3} \alpha_l}{3(mE/\hslash^2)^{4/3}}\beta_l^{-1/3}\right)^{1/2}
\end{eqnarray}
where we have introduced the shortened notation
\begin{eqnarray}
\alpha_l&=&-\dfrac{\hslash^2}{2m}l^2(l+1)^2\pm|\alpha|,\\
\beta_l&=&1+\sqrt{1-\dfrac{4}{27}\dfrac{\hslash^4}{m^2E^2}\alpha_l^3}.
\end{eqnarray}
It is easy to deduce the solution for $l=0$.

\section{Boundary Conditions for The Wave Functionals}\label{sec:3}
In this subsection we shall consider certain solutions of the global one-dimensional quantum gravity (\ref{kgf}) for the approximation of the effective gravitational potential\index{gravitational potential!effective} $V_{eff}$ discussed in the previous subsection. For the considered situation the $h$-evolution
\begin{equation}\label{kgf5}
\left(\dfrac{\delta^2}{\delta{h^2}}\mp\dfrac{|\alpha|}{h}\right)\Psi^{\mp}[h]=0,
\end{equation}
is solved by two type of wave functions $\Psi^{\mp}$ where the attractive wave functions $\Psi_G^{-}[h]$ are associated with the the Newton-like effective gravitational potential\index{gravitational potential!effective!Newton-like}, and the repulsive ones $\Psi^{+}[h]$ are associated with the Coulomb-like effective gravitational potential\index{gravitational potential!effective!Coulomb-like}. Because of manifest one-dimensionality of the functional evolutionary equation (\ref{kgf5}) one can solve this equation in frames of the theory of ordinary differential equations by treatment of the functional derivative as the ordinary one, i.e. $\dfrac{\delta}{\delta h} = \dfrac{d}{dh}$, and the functional and a function $\Psi[h]=\Psi(h)$ with no loss of the generality.

In this manner we shall consider here the equation
\begin{equation}\label{kgf6}
\left(\dfrac{d^2}{dh^2}\mp\dfrac{|\alpha|}{h}\right)\Psi^{\mp}(h)=0.
\end{equation}
The general solution of this differential equation can be constructed straightforwardly by application of the Bessel functions\index{Bessel functions!of the f\/irst kind}\index{Bessel functions!of the second kind} $J_n$ and $Y_n$ for the case of the attractive potential
\begin{equation}
  \Psi^{-}[h]=\sqrt{|\alpha| h}\left[C_1^-J_1\left(2\sqrt{|\alpha| h}\right) + 2iC_2^-Y_1\left(2\sqrt{|\alpha| h}\right)\right],\label{solp}
\end{equation}
and in terms of the modif\/ied Bessel\index{Bessel functions!modif\/ied!of the f\/irst kind}\index{Bessel functions!modif\/ied!of the second kind} $I_n$ and $K_n$ for the case of the repulsive potential
\begin{equation}
\Psi^{+}[h]=-\sqrt{|\alpha| h}\left[C_1^+I_1\left(2\sqrt{|\alpha| h}\right) + 2C_2^+K_1\left(2\sqrt{|\alpha| h}\right)\right],\label{solm}
\end{equation}
where $C_1^{\pm}$ and $C_2^{\pm}$ are constants of integration. In concrete calculations one can take the standard def\/initions of the Bessel functions of f\/irst and second kind \cite{spec}, $J_\alpha(x)$ and $Y_\alpha(x)$, which are
\begin{eqnarray}
   J_\alpha(x)&=&\dfrac{1}{\pi}\int_0^\pi dt\cos\left(x\cos t-\alpha t\right),\\
  Y_\alpha(x)&=&\dfrac{J_\alpha(x)\cos\left(\alpha\pi\right)-J_{-\alpha}(x)}{\sin\left(\alpha\pi\right)},
\end{eqnarray}
as well as the modif\/ied Bessel functions of the f\/irst and the second kind, $I_\alpha(x)$ and $K_\alpha(x)$, which are
\begin{eqnarray}
   I_\alpha(x)&=&\dfrac{1}{\pi}\int_0^\pi dt\exp\left(x\cos t\right)\cos\left(\alpha t\right),\\
  K_\alpha(x)&=&\dfrac{\pi}{2}\dfrac{I_{-\alpha}(x)-I_\alpha(x)}{\sin\left(\alpha \pi\right)}.
\end{eqnarray}
Recall that standardly the values of the Bessel functions of second kind and the modif\/ied Bessel functions of second kind for any integers $n$ can be received by application of the limiting procedure
\begin{eqnarray}
Y_n(x)&=&\lim_{\alpha \rightarrow n}Y_\alpha(x),\\
{K}_n(x)&=&\lim_{\alpha \rightarrow n}K_\alpha(x).
\end{eqnarray}

In further part of this section we shall present solutions to the one-dimensional quantum mechanics (\ref{kgf6}) with respect to several selected boundary conditions for the general solutions (\ref{solp}) and (\ref{solm}).

\subsubsection{Boundary Conditions I}\label{sub:4}
Let us consider the global one-dimensional quantum mechanics (\ref{kgf}) with the boundary conditions for some selected initial value of the dimension $h=h_I$:
\begin{eqnarray}
  \Psi[h_I]&=&\Psi_I,\label{bound1a}\\
  \dfrac{\delta\Psi}{\delta h}[h_I]&=&\Psi'_I.\label{bound1b}
\end{eqnarray}
For construction of the solution one can use the regularized hypergeometric functions\index{hypergeometric function!regularized} ${_p}\tilde{F}_q$
\begin{equation}
{_p}\tilde{F}_q\left(\begin{array}{c} a_1,\ldots,a_p\\b_1,\ldots,b_q \end{array};x\right)=\dfrac{{_p}F_q\left(\begin{array}{c} a_1,\ldots,a_p\\b_1,\ldots,b_q \end{array};x\right)}{\Gamma(b_1)\ldots\Gamma(b_q)},
\end{equation}
where ${_p}F_q$ is the conf\/luent hypergeometric function
\begin{equation}
{_p}F_q\left(\begin{array}{c} a_1,\ldots,a_p\\b_1,\ldots,b_q \end{array};x\right)=\sum_{r=0}^{\infty}\dfrac{(a_1)_r\ldots(a_p)_r}{(b_1)_r\ldots(b_q)_r}\dfrac{x^r}{r!},
\end{equation}
where $(a)_r=\dfrac{\Gamma(a+r)}{\Gamma(a)}$ are Pochhammer symbols\index{Pochhammer symbols}. The general solutions (\ref{solp}) and (\ref{solm}) with respect to the boundary conditions (\ref{bound1a})-(\ref{bound1b}) can be written out in the form
\begin{equation}\label{sol1}
\Psi^{-}=C^-_1\left(2\sqrt{\strut{|\alpha| h}}\right)K_1\left(2\sqrt{\strut{|\alpha| h}}\right)+C^-_2\left(2\sqrt{\strut{|\alpha| h}}\right)^2{_0}\tilde{F}_1\left(\begin{array}{c} -\\2\end{array};|\alpha| h\right),
\end{equation}
where constants of integration are
\begin{eqnarray}
  C^-_1&=&\Psi_I\,{_0}\tilde{F}_1\left(\begin{array}{c} -\\1\end{array};|\alpha| h_I\right)-\Psi'_Ih_I\,{_0}\tilde{F}_1\left(\begin{array}{c} -\\2\end{array};|\alpha| h_I\right),\\
  C^-_2&=&\dfrac{1}{2}\left(\Psi_IK_0\left(2\sqrt{\strut{|\alpha| h_I}}\right)+\Psi'_I\sqrt{\strut{\dfrac{h_I}{|\alpha|}}}K_1\left(2\sqrt{\strut{|\alpha| h_I}}\right)\right),
\end{eqnarray}
for the Newton-like case, and
\begin{equation}\label{sol2}
\Psi^{+}=C^+_1\left(2\sqrt{\strut{|\alpha| h}}\right)Y_1\left(2\sqrt{\strut{|\alpha| h}}\right)+C^+_2\left(2\sqrt{\strut{|\alpha| h}}\right)^2{_0}\tilde{F}_1\left(\begin{array}{c} -\\2\end{array};-|\alpha| h\right),
\end{equation}
where constants of integration are
\begin{eqnarray}
\!\!\!\!\!\!\!\!\!\!
C^+_1&=&\dfrac{\pi}{2}\left(\Psi'_Ih_I\,{_0}\tilde{F}_1\left(\begin{array}{c} -\\2\end{array};-|\alpha| h_I\right)-\Psi_I\,{_0}\tilde{F}_1\left(\begin{array}{c} -\\1\end{array};-|\alpha| h_I\right)\right),\\
\!\!\!\!\!\!\!\!\!\!
C^+_2&=&\dfrac{\pi}{2}\left(\Psi_IY_0\left(2\sqrt{\strut{|\alpha| h_I}}\right)-\Psi'_I\sqrt{\strut{\dfrac{h_I}{|\alpha|}}}Y_1\left(2\sqrt{\strut{|\alpha| h_I}}\right)\right),
\end{eqnarray}
for the Coulomb-like case.

\subsubsection{Boundary Conditions II}\label{sub:5}
The second case which we want to present in this paper, are the boundary conditions for 1st and 2nd functional derivatives
\begin{eqnarray}
  \dfrac{\delta\Psi}{\delta h}[h_I]&=&\Psi'_I,\label{bound2a}\\
  \dfrac{\delta^2\Psi}{\delta h^2}[h_I]&=&\Psi''_I.\label{bound2b}
\end{eqnarray}
By using of the hypergeometric functions, one can express the solution for attractive case as follows
\begin{equation}\label{sol1a}
\Psi^{-}=C^-_1\left(2\sqrt{\strut{|\alpha| h}}\right)K_1\left(2\sqrt{\strut{|\alpha| h}}\right)+C^-_2\left(2\sqrt{\strut{|\alpha| h}}\right)^2{_0}\tilde{F}_1\left(\begin{array}{c} -\\2\end{array};|\alpha| h\right),
\end{equation}
where $C^-_1$ and $C^-_2$ are constants def\/ined as
\begin{eqnarray}
\!\!\!\!\!\!\!\!\!\!\!\!\!\!\!\!\!\!\!\!
C^-_1&=&-h_I\left(\Psi'_I\,{_0}\tilde{F}_1\left(\begin{array}{c} -\\2\end{array};|\alpha| h_I\right)-\dfrac{\Psi''_I}{|\alpha|}\,{_0}\tilde{F}_1\left(\begin{array}{c} -\\1\end{array};|\alpha| h_I\right)\right),\\
\!\!\!\!\!\!\!\!\!\!\!\!\!\!\!\!\!\!\!\!
C^-_2&=&\dfrac{1}{2}\sqrt{\dfrac{h_I}{|\alpha|}}\left(\Psi''_I\sqrt{\dfrac{h_I}{|\alpha|}}K_0\left(2\sqrt{\strut{|\alpha| h_I}}\right)+\Psi'_IK_1\left(2\sqrt{\strut{|\alpha| h_I}}\right)\right).
\end{eqnarray}
Similarly for the repulsive case one obtains easily
\begin{equation}\label{sol2a}
\Psi^{+}=C^+_1\left(2\sqrt{\strut{|\alpha| h}}\right)Y_1\left(2\sqrt{\strut{|\alpha| h}}\right)+C^+_2\left(2\sqrt{\strut{|\alpha| h}}\right)^2{_0}\tilde{F}_1\left(\begin{array}{c} -\\2\end{array};-|\alpha| h\right),
\end{equation}
where the constants of integration are
\begin{eqnarray}
\!\!\!\!\!\!\!\!\!\!\!\!\!\!\!\!\!\!\!\!
  C^+_1&=&\dfrac{\pi h_I}{2}\left(\Psi'_I\,{_0}\tilde{F}_1\left(\begin{array}{c} -\\2\end{array};-|\alpha| h_I\right)+\dfrac{\Psi''_I}{|\alpha|}\,{_0}\tilde{F}_1\left(\begin{array}{c} -\\1\end{array};-|\alpha| h_I\right)\right),\\
\!\!\!\!\!\!\!\!\!\!\!\!\!\!\!\!\!\!\!\!
  C^+_2&=&\dfrac{\pi}{4}\sqrt{\dfrac{h_I}{|\alpha|}}\left(\Psi''_I\sqrt{\dfrac{h_I}{|\alpha|}}Y_0\left(2\sqrt{\strut{|\alpha| h_I}}\right)+\Psi'_IY_1\left(2\sqrt{\strut{|\alpha| h_I}}\right)\right).
\end{eqnarray}

\subsubsection{Boundary Conditions III}\label{sub:6}
The third possible choice of the boundary conditions for the considered problem has the following form
\begin{eqnarray}
  \Psi[h_I]&=&\Psi_I,\label{bound3a}\\
  \dfrac{\delta^2\Psi}{\delta h^2}[h_I]&=&\Psi''_I.\label{bound3b}
\end{eqnarray}
These conditions are formally improper for the problem, because of they lead to manifestly singular solutions. In such a situation, however, one can present the solutions in the form in which constants of integration are formally singular. For the case of the attractive Newton-like potential one has
\begin{equation}\label{sol1c}
  \Psi_G^-=C^-_1\left(2\sqrt{|\alpha|h}\right)K_1\left(2 \sqrt{\left|\alpha\right|h}\right)+C^-_2\left(2\sqrt{|\alpha|h}\right)^2{_0}\tilde{F}_1\left(\begin{array}{c} -\\2\end{array};|\alpha| h\right),
\end{equation}
where constants of integration are ($\epsilon\rightarrow0$)
\begin{eqnarray}
  C^-_1&=&\dfrac{2}{\epsilon}\sqrt{|\alpha|h_I}\left(\Psi_I-\dfrac{h_I}{\left|\alpha\right| }\Psi''_I \right){_0}\tilde{F}_1\left(\begin{array}{c} -\\2\end{array};|\alpha| h_I\right),\\
  C^-_2&=&\dfrac{1}{\epsilon}\left(\Psi_I-\dfrac{h_I}{|\alpha|}\Psi''_I\right)K_1\left(2\sqrt{|\alpha|h_I}\right).
\end{eqnarray}
Similarly for case of the repulsive Coulomb-like potential one obtains
\begin{equation}\label{sol2c}
  \Psi_G^+=C^+_1\left(2\sqrt{|\alpha|h}\right)Y_1\left(2 \sqrt{\left|\alpha\right|h}\right)+C^+_2\left(2\sqrt{|\alpha|h}\right)^2{_0}\tilde{F}_1\left(\begin{array}{c} -\\2\end{array};-|\alpha| h\right),
\end{equation}
where constants of integration are ($\epsilon\rightarrow0$)
\begin{eqnarray}
  C^+_1&=&\dfrac{2}{\epsilon}\sqrt{|\alpha|h_I}\left(\Psi_I+\dfrac{h_I}{\left|\alpha\right| }\Psi''_I \right){_0}\tilde{F}_1\left(\begin{array}{c} -\\2\end{array};-|\alpha| h_I\right),\\
  C^+_2&=&\dfrac{1}{\epsilon}\left(\Psi_I+\dfrac{h_I}{|\alpha|}\Psi''_I\right)Y_1\left(2\sqrt{|\alpha|h_I}\right).
\end{eqnarray}

There is the question how to regularize these solutions to obtain non-singular constants of integration. There is no general method for such an productive procedure. Let us propose some constructive regularization. For regularization of the solution let us take into account the following \emph{the ansatz for boundary conditions}\index{ansatz!for boundary conditions}
\begin{equation}\label{cond}
  \pm\dfrac{h_I}{\left|\alpha\right| }{\Psi^{\pm}_I}''+\Psi^{\pm}_I\equiv\epsilon f_{\pm}[h_I,|\alpha|],
\end{equation}
where $f_{\pm}[h_I,|\alpha|]\neq0$ are some nonsingular functionals of $h_I$ and $|\alpha|$, which are presently unknown and arbitrary. For formal correctness and consistency of the method we shall put the limiting procedure $\epsilon\rightarrow0$ in some step of the regularization process. The sign $+$ is related to the Newton-like case, and the sign $-$ to the Coulomb-like case. It can be seen by straightforward calculation that in such a situation the singularity of the solutions (\ref{sol1c}) and (\ref{sol2c}) can be removed. We are going to show now that the ansatz\index{ansatz} (\ref{cond}) allows to express the problem of choice of the initial data via integral equations for the functionals $f_{\pm}[h_I,|\alpha|]$. For basics and applications of the theory of integral equations we suggest to see e.g. the books in the Ref. \cite{inteq}.

For the attractive Newton-like case the initial value $\Psi_I$ can be obtained as follows
\begin{eqnarray}
\!\!\!\!\!\!\!\!\!\!\!\!\!\!\!
&&\Psi^-_I=-|\alpha|h_I\,{_0}{F}_1\left(\begin{array}{c} -\\2\end{array};|\alpha|h_I\right)\left[c^-_1+2\epsilon\sqrt{|\alpha|}\int_1^{h_I}\dfrac{dt}{\sqrt{t}}f_{-}[t,|\alpha|]K_1\left(2\sqrt{|\alpha|t}\right)\right]\nonumber\\
\!\!\!\!\!\!\!\!\!\!\!\!\!\!\!
&&+\,2\sqrt{|\alpha|h_I}K_1\left(2\sqrt{|\alpha|h_I}\right)\left[c^-_2+\epsilon|\alpha|\int_1^{h_I}dtf_{-}[t,|\alpha|]\,{_0}{F}_1\left(\begin{array}{c} -\\2\end{array};|\alpha|t\right)\right],\nonumber\\
\!\!\!\!\!\!\!\!\!\!\!\!\!\!\!
&&\label{eq1}
\end{eqnarray}
and similarly for the repulsive Coulomb-like  case one receives
\begin{eqnarray}
\!\!\!\!\!\!\!\!\!\!\!\!\!\!\!
&&\Psi^+_I=|\alpha|h_I\,{_0}{F}_1\left(\begin{array}{c} -\\2\end{array};-|\alpha|h_I\right)\left[c^+_1-\epsilon\pi\sqrt{|\alpha|}\int_1^{h_I}\dfrac{dt}{\sqrt{t}}f_{+}[t,|\alpha|]Y_1\left(2\sqrt{|\alpha|t}\right)\right]\nonumber\\
\!\!\!\!\!\!\!\!\!\!\!\!\!\!\!
&&+\,2i\sqrt{|\alpha|h_I}Y_1\left(2\sqrt{|\alpha|h_I}\right)\left[c^+_2-\epsilon\dfrac{i\pi}{2}|\alpha|\int_1^{h_I}dtf_{+}[t,|\alpha|]\,{_0}{F}_1\left(\begin{array}{c} -\\2\end{array};-|\alpha|t\right)\right],\nonumber\\
\!\!\!\!\!\!\!\!\!\!\!\!\!\!\!
&&\label{eq2}
\end{eqnarray}
where $c^{\pm}_{1,2}$ are now non-singular constants of integration. The functionals $f_{\pm}[h_I,|\alpha|]$ can be established by straightforward application of the ansatz\index{ansatz} (\ref{cond}) within the general solutions (\ref{sol1c}) and (\ref{sol2c}). Such an application yields the following results
\begin{eqnarray}
  \Psi_I^-&=&8|\alpha|h_IK_1\left(2\sqrt{|\alpha|h_I}\right)\,{_0}\tilde{F}_1\left(\begin{array}{c} -\\2\end{array};|\alpha|h_I\right)f_{-}[h_I,|\alpha|],\\
  \Psi_I^+&=&8|\alpha|h_IY_1\left(2\sqrt{|\alpha|h_I}\right)\,{_0}\tilde{F}_1\left(\begin{array}{c} -\\2\end{array};-|\alpha|h_I\right)f_{+}[h_I,|\alpha|].
\end{eqnarray}
Employing straightforwardly these results within the equations (\ref{eq1}) and (\ref{eq2}) one obtains the integral equations for the functionals $f_{\pm}$. For the Coulomb-like situation one receives the following Volterra integral equation of the second kind\index{Volterra integral equation of the second kind}
\begin{equation}
f_{-}[h_I,|\alpha|]=g^-[h_I,|\alpha|]+\epsilon\int_1^{h_I}dt\mathcal{K}^-(t,|\alpha|,h_I)f_{-}[t,|\alpha|],\label{93}
\end{equation}
where the function $g^-[h_I,|\alpha|]$ causing the non-homogeneity is
\begin{equation}
g^-[h_I,|\alpha|]=-\dfrac{c^-_1}{8K_1\left(2\sqrt{|\alpha|h_I}\right)}+\dfrac{c^-_2}{2\left(2\sqrt{|\alpha|h_I}\right)\,{_0}\tilde{F}_1\left(\begin{array}{c}-\\2\end{array};|\alpha|h_I\right)},
\end{equation}
and the kernel $\mathcal{K}^-(t,|\alpha|,h_I)$ has the form
\begin{eqnarray}
\mathcal{K}^-(t,|\alpha|,h_I)&=&\dfrac{|\alpha|}{\sqrt{t}}\Bigg[\dfrac{\,{_0}{F}_1\left(\begin{array}{c}-\\2\end{array};|\alpha|t\right)}{2\left(2\sqrt{|\alpha|h_I}\right)\,{_0}\tilde{F}_1\left(\begin{array}{c}-\\2\end{array};|\alpha|h_I\right)}-\nonumber\\
&-&\dfrac{\sqrt{h_I}K_1\left(2\sqrt{|\alpha|t}\right)}{2\left(2\sqrt{|\alpha|h_I}\right)K_1\left(2\sqrt{|\alpha|h_I}\right)}\Bigg].
\end{eqnarray}

Similar procedure can be performed for the Newton-like situation. In this case the Volterra integral equation of the second kind\index{Volterra integral equation of the second kind} is
\begin{equation}
f_{+}[h_I,|\alpha|]=g^+[h_I,|\alpha|]+\epsilon\int_1^{h_I}dt\mathcal{K}^+(t,|\alpha|,h_I)f_{+}[t,|\alpha|]\label{94}
\end{equation}
where the function $g^-[h_I,|\alpha|]$ causing the non-homogeneity is
\begin{equation}
g^+[h_I,|\alpha|]=\dfrac{c^+_1}{8Y_1\left(2\sqrt{|\alpha|h_I}\right)}+\dfrac{ic^+_2}{2\left(2\sqrt{|\alpha|h_I}\right)\,{_0}\tilde{F}_1\left(\begin{array}{c}-\\2\end{array};-|\alpha|h_I\right)},
\end{equation}
and the kernel $\mathcal{K}^+(t,|\alpha|,h_I)$ has the form
\begin{eqnarray}
\mathcal{K}^+(t,|\alpha|,h_I)&=&\dfrac{\pi|\alpha|}{2\sqrt{t}}\Bigg[-\dfrac{\sqrt{h_I}Y_1\left(2\sqrt{|\alpha|t}\right)}{2\left(2\sqrt{|\alpha|h_I}\right)Y_1\left(2\sqrt{|\alpha|h_I}\right)}-\nonumber\\
&-&\dfrac{\,i\,{_0}{F}_1\left(\begin{array}{c} -\\2\end{array};-|\alpha|t\right)}{2\left(2\sqrt{|\alpha|h_I}\right)\,{_0}\tilde{F}_1\left(\begin{array}{c} -\\2\end{array};-|\alpha|h_I\right)}\Bigg].
\end{eqnarray}

The number $\epsilon$ plays the role analogous to the eigenvalue in linear algebra. It is evidently seen that the integral operators acting on the functionals $f_{\pm}$ are non-singular when the limiting procedure $\epsilon\rightarrow0$ is performed. Such a property guarantees stability and effectiveness of the regularization method given by the ansatz\index{ansatz} (\ref{cond}). By this reason one can perform the limit $\epsilon\rightarrow0$ straightforwardly within the Volterra integral equations of the second kind\index{Volterra integral equation of the second kind} (\ref{93}) and (\ref{94}), and extract the unknown functionals
\begin{equation}
  f_{\pm}[h_I,|\alpha|]=g^{\pm}[h_I,|\alpha|].
\end{equation}
Taking into account elementary properties of the hypergeometric function the f\/inal resultscan be presented in the following form
\begin{eqnarray}
f_{-}[h_I,|\alpha|]&=&\dfrac{-c^-_1}{8K_1\left(2\sqrt{|\alpha|h_I}\right)}+\dfrac{c^-_2}{4I_1\left(2\sqrt{|\alpha|h_I}\right)},\\
f_{+}[h_I,|\alpha|]&=&\dfrac{c^+_1}{8Y_1\left(2\sqrt{|\alpha|h_I}\right)}+\dfrac{ic^+_2}{4J_1\left(2\sqrt{|\alpha|h_I}\right)}.
\end{eqnarray}

In this manner the conditions for the initial data in the considered situation given by the improper boundary conditions (\ref{bound3a})-(\ref{bound3b}) and in the light to the ansatz\index{ansatz} (\ref{cond}) can not be chosen arbitrary, but according to the following selection rules
\begin{eqnarray}
  \Psi_I^-&=&\sqrt{|\alpha|h_I}\left[-c^-_1I_1\left(2\sqrt{|\alpha|h_I}\right)+2c^-_2K_1\left(2\sqrt{|\alpha|h_I}\right)\right],\\
  \Psi_I^+&=&\sqrt{|\alpha|h_I}\left[c^+_1J_1\left(2\sqrt{|\alpha|h_I}\right)+2ic^+_2Y_1\left(2\sqrt{|\alpha|h_I}\right)\right].
\end{eqnarray}

It must be emphasized that the proposed ansatz\index{ansatz} for the boundary conditions (\ref{cond}) is not unique, and can be replaced by another proposal. However, as we have mentioned earlier, this type of regularization method assures stability and effectiveness of the limiting procedure $\epsilon\rightarrow0$, and therefore in general guarantees consistency of the method. The ansatz\index{ansatz} (\ref{cond}) allowed to formulate the regularization in terms of the Volterra integral equation of the second kind which, however, was not solved because of the limit $\epsilon\rightarrow0$ was performed. Albeit, it must be emphasized also that according to the theory of integral equations the obtained Volterra integral equations of the second kind\index{Volterra integral equation of the second kind} should be solved for arbitrary $\epsilon$ and then the limit $\epsilon\rightarrow0$ should be performed. Such a solution can be constructed straightforwardly by application of the Neumann series\index{Neumann series} called also the Born series\index{Born series}. If one rewrites the Volterra integral equations (\ref{93}) and (\ref{94}) in more convenient symbolic form
\begin{equation}
f_{\pm}[h_I,|\alpha|]=g^{\pm}[h_I,|\alpha|]+\epsilon\textsf{K}^{\pm}f_{\pm}[t,|\alpha|],
\end{equation}
where the integral operators $\textsf{K}^{\pm}$ are def\/ined as
\begin{equation}
\textsf{K}^{\pm}=\int_1^{h_I}dt\mathcal{K}^{\pm}(t,|\alpha|,h_I),
\end{equation}
then the solution can be written out straightforwardly via using of the Neumann series
\begin{equation}
f_{\pm}[h_I,|\alpha|]=g^{\pm}[h_I,|\alpha|]+\sum_{n=1}^\infty\epsilon^n{\textsf{K}^{\pm}}^ng^{\pm}[h_I,|\alpha|].
\end{equation}
It is easy to see now that in the limit $\epsilon\rightarrow0$ one obtains the result $f_{\pm}[h_I,|\alpha|]=g^{\pm}[h_I,|\alpha|]$ received earlier. The problem is, however, convergence of the series as well as def\/initeness of the integral operators $\textsf{K}^{\pm}$ because of in the presented situation the integral kernels $\textsf{K}^{\pm}(t,|\alpha|,h_I)$ are special functions. The presented method in general ref\/lects the typical problems arising from application of the improper boundary conditions.

\chapter{\emph{Ab Initio} Thermodynamics of Space Quanta {\AE}ther}

The quantum f\/ield theory formulated in terms of the static Fock rep\`{e}re\index{Fock rep\`{e}re} is the most natural approach to formulation of the state of thermodynamic equilibrium. The effectiveness of such a method is emphasized by the fact that despite that the system is manifestly nonequilibrium (For details see e.g. \cite{morozov}) application of the Fock space formulation leads to the properly defined equilibrium. In the particular situation related to the global one-dimensional quantum gravity, the equilibrium is strictly related to the ensemble of the quanta of space which in itself create the {\AE}ther\index{{\AE}ther}. Such a situation gives the most natural conditions for straightforward application of the f\/irst principles of statistical mechanics \cite{ker}, what will be resulting in \emph{ab initio} formulation of the thermodynamics of such an {\AE}ther of space quanta. In this section we shall present the constructive approach which uses the simplest possible approximation, but in itself is the most fundamental contribution to the thermodynamics. Namely, as the example of the thermodynamic strategy we shall apply so called \emph{one-particle approximation}\index{one-particle approximation}, based on the corresponding method of \emph{one-particle density matrix method}\index{one-particle density matrix}. The simplicity of the one-particle approximation\index{one-particle approximation} is the fact that in such a situation the density operator $\textsf{D}$ is equivalent to an occupation number operator. Then thermodynamic equilibrium is determined with respect to the static Fock rep\`{e}re\index{Fock rep\`{e}re} in the stable Bogoliubov vacuum, and therefore the one-particle density matrix\index{one-particle density matrix} in equilibrium $\mathbb{D}$ is established according to the von Neumann--Heisenberg picture\index{von Neumann--Heisenberg picture}.

\section{Entropy I: The Analytic Approach}\label{sec:4}

The non-equilibrium density operator $\textsf{D}$, which possesses a dynamical nature by its existence in a dynamical Fock rep\`{e}re\index{Fock rep\`{e}re}, is expressed in the static Fock rep\`{e}re\index{Fock rep\`{e}re} $\mathfrak{F}$ via the equilibrium density matrix $\mathbb{D}$ in the following way
\begin{equation}
\textsf{D}={\textsf{G}}^{\dagger}{\textsf{G}}=\mathfrak{F}^{\dagger}\mathbb{D}\mathfrak{F},
\end{equation}
where in the case of global one-dimensional quantum gravity one the density matrix has the form
\begin{equation}
  \mathbb{D}=\left[\begin{array}{cc}\dfrac{(\mu+1)^2}{{4\mu}}&\dfrac{1-\mu^2}{{4\mu}}\\
  \dfrac{1-\mu^2}{{4\mu}}&\dfrac{(\mu-1)^2}{{4\mu}}\end{array}\right]\label{den},
\end{equation}
here $\mu=\dfrac{\omega}{\omega_I}$ is the mass scale of the system of space quanta. Note that in the present situation
\begin{eqnarray}
\mathbb{D}^2&=&(\tr\mathbb{D})\mathbb{D},\\
\det\mathbb{D}&=&0.
\end{eqnarray}
Vanishing of the determinant means that in the one-particle approximation\index{one-particle approximation} the corresponding thermodynamics is \emph{irreversible}, or in other words that the thermodynamic processes in the {\AE}ther of space quanta are irreversible.\index{{\AE}ther}

Employing the density matrix (\ref{den}) one can establish the value of the occupation number
\begin{equation}
  N=\dfrac{\tr\left(\mathbb{D}^2\right)}{\tr\mathbb{D}}=\dfrac{\tr\left((\tr\mathbb{D})\mathbb{D}\right)}{\tr\mathbb{D}}=\dfrac{(\tr\mathbb{D})^2}{\tr\mathbb{D}}=\tr\mathbb{D}=\dfrac{\mu^2+1}{2\mu},
\end{equation}
and the entropy can be derived from its basic def\/inition
\begin{eqnarray}
  S=\dfrac{\tr(\mathbb{D}\ln\mathbb{D})}{\tr\mathbb{D}}.\label{entropia0}
\end{eqnarray}
The problem is to derive $\ln\mathbb{D}$. It can be performed effectively by application of the matrix Taylor series\index{Taylor series}
\begin{equation}
  \ln\mathbb{D}=\ln[(\mathbb{D}-\mathbb{I})+\mathbb{I}]=\sum_{n=1}^\infty\dfrac{(-1)^{n+1}}{n}(\mathbb{D}-\mathbb{I})^n,
\end{equation}
and the matrix Newton binomial series\index{Newton binomial series}
\begin{eqnarray}
(\mathbb{D}-\mathbb{I})^n&=&\sum_{k=1}^n\binom{n}{k}\mathbb{D}^k(-\mathbb{I})^{n-k}=\sum_{k=1}^{n}(-1)^{n-k}\binom{n}{k}\mathbb{D}^k=\\
&=&\sum_{k=1}^{n}(-1)^{n-k}\binom{n}{k}(\tr\mathbb{D})^{k-1}\mathbb{D},
\end{eqnarray}
where
\begin{equation}
\binom{n}{m}=\dfrac{n!}{m!(n-m)!},
\end{equation}
is the Newton binomial symbol\index{Newton binomial symbol}, which allows to write out the formula
\begin{equation}\label{cluser}
  S=\sum_{n=1}^{\infty}\sum_{k=1}^n\dfrac{(-1)^{k-1}}{n}\binom{n}{k-1}S_k,
\end{equation}
where $S_k$ are \emph{cluster entropies}\index{cluster entropies}
\begin{equation}
  S_k=\dfrac{\tr(\mathbb{D}^{k})}{\tr\mathbb{D}}=\dfrac{\tr\left((\tr\mathbb{D})^{k-1}\mathbb{D}\right)}{\tr\mathbb{D}}=\dfrac{(\tr\mathbb{D})^{k-1}\tr\mathbb{D}}{\tr\mathbb{D}}=N^{k-1}.
\end{equation}
We shall call the series expansion (\ref{cluser}) \emph{the cluster series}\index{cluster series}. Its summation is the method for obtaining entropy in the analytical way.

According to a certain mathematical tradition the series expansion (\ref{entropia0}) converges if and only if the spectral radius $\rho$ of the difference $\mathbb{D}-\mathbb{I}$, where $\mathbb{I}$ is $2\times2$ unit matrix, is
\begin{equation}
\rho(\mathbb{D}-\mathbb{I})<1.
\end{equation}
Straightforward calculation leads to the condition for the mass scale
\begin{equation}
  \mu\in[1;2+\sqrt{3}),
\end{equation}
and the result of the summation procedure is
\begin{equation}
  S=\dfrac{\zeta(1)}{2}\left(\dfrac{\mu^2-1}{\mu^2+1}\right)^2+\dfrac{\mu^4+6\mu^2+1}{(\mu^2+1)^2}\ln\dfrac{(\mu-1)^2}{2\mu},\label{entropia}
\end{equation}
where $\zeta(s)$ is the Riemann zeta function
\begin{equation}
\zeta(s)=\sum_{n=1}^{\infty}\dfrac{1}{n^s},
\end{equation}
and $\zeta(1)$ is manifestly divergent.

Note that by straightforward application of the Hagedorn hadronization formula $\omega\sim T_H$ \cite{hag}, where $m$ is the mass of the system, one can establish the hadronized temperature\index{temperature} as \begin{equation}
  \dfrac{T_H}{T_I}=\mu.
\end{equation}
By the relation $\Delta\omega\sim\Delta T_H$ one obtains the hadronized temperature\index{temperature} normalized to $T_I$ value
\begin{equation}\label{srednia}
\dfrac{\Delta T_H}{T_I}=\mu_{\max}-\mu_{\min}=1+\sqrt{3}\approx 2.732,
\end{equation}
so that one can establish the ratio
\begin{equation}
\dfrac{T_H}{\Delta T_H}\in\left[\dfrac{\sqrt{3}-1}{2},\dfrac{\sqrt{3}+1}{2}\right),
\end{equation}
what allows to determine the reciprocal
\begin{equation}
\dfrac{\Delta T_H}{T_H}\in\left(-(\sqrt{3}+1),\sqrt{3}+1\right].
\end{equation}
Def\/ining anisotropy as $\delta T_H=\Delta T_H-T_H$ one derives
\begin{equation}
  \dfrac{\delta T_H}{T_H}\in\left(-(\sqrt{3}+2),\sqrt{3}\right],
\end{equation}
so that half of the difference anisotropy is exactly
\begin{equation}\label{srednia2}
  \Delta\left(\dfrac{\delta T_H}{T_H}\right)=\left(\dfrac{\Delta T_H}{T_H}\right)_{\max}-\left(\dfrac{\Delta T_H}{T_H}\right)_{\min}=2(\sqrt{3}+1)=2\dfrac{\Delta T_H}{T_I}.
\end{equation}
In this manner one obtains
\begin{equation}\label{differn}
\Delta T_H=\dfrac{T_I}{2}\Delta\left(\dfrac{\delta T_H}{T_H}\right)\approx2.732T_I.
\end{equation}
The difference (\ref{differn}) can be identif\/ied with a background temperature\index{temperature}, i.e. $\Delta T_H\equiv T_B$. For the initial datum $T_I\sim1\mathrm{K}$ it is very close to the averaged cosmic microwave background radiation temperature\index{temperature}, $T_{CMB}\approx2.725\mathrm{K}$. Because of the quantity $|T_{CMB}-T_B|$ is small $7\cdot10^{-3}\mathrm{K}$, one can deduce that next approximations will be resulting in successive contributions to $T_{CMB}$. In other words
\begin{equation}
T_{CMB}=T_B+\ldots.
\end{equation}

In the dynamical Fock rep\`{e}re\index{Fock rep\`{e}re} Hamiltonian operator is
\begin{equation}
  \textsf{H}=\dfrac{m}{2}\left(\textsf{G}^\dagger\textsf{G}+\textsf{G}\textsf{G}^\dagger\right)=\mathfrak{F}^\dagger\mathbb{H}\mathfrak{F},
\end{equation}
where $\mathbb{H}$ is the Hamiltonian matrix in the static Fock rep\`{e}re\index{Fock rep\`{e}re}
\begin{equation}
  \mathbb{H}=\left[\begin{array}{cc} \dfrac{\omega_I}{4}\left(1+\mu^2\right)&\dfrac{\omega_I}{4}\left(1-\mu^2\right)\vspace*{5pt}\\
  \dfrac{\omega_I}{4}\left(1-\mu^2\right)&\dfrac{\omega_I}{4}\left(1+\mu^2\right)\end{array}\right],\label{shm}
\end{equation}
which for f\/ixed mass scale has discrete spectrum
\begin{equation}
\mathrm{Spec}~\mathbb{H}=\left\{\dfrac{\omega_I}{2}\mu^2,\dfrac{\omega_I}{2}\right\}.
\end{equation}
The internal energy calculated from the Hamiltonian matrix (\ref{shm}) is
\begin{equation}\label{int}
  U=\dfrac{\tr(\mathbb{D}\mathbb{H})}{\tr\mathbb{D}}=\dfrac{\omega_I}{4}(\mu^2+1).
\end{equation}
The Hamiltonian matrix $\mathbb{H}$, however, consists constant term $\mathbb{H}_{I}$
\begin{equation}
  \mathbb{H}_I=\left[\begin{array}{cc} \dfrac{\omega_I}{4}& \dfrac{\omega_I}{4}\vspace*{5pt}\\ \dfrac{\omega_I}{4}& \dfrac{\omega_I}{4}\end{array}\right]
\end{equation}
which can be eliminated by simple renormalization
\begin{equation}
  \mathbb{H}\rightarrow\mathbb{H}'=\mathbb{H}-\mathbb{H}_I=\left[\begin{array}{cc} \dfrac{\omega_I}{4}\mu^2&-\dfrac{\omega_I}{4}\mu^2\vspace*{5pt}\\-\dfrac{\omega_I}{4}\mu^2&\dfrac{\omega_I}{4}\mu^2\end{array}\right].
\end{equation}
The spectrum of the renormalized Hamiltonian matrix is
\begin{equation}
  \mathrm{Spec}~\mathbb{H}'=\left\{\dfrac{\omega_I}{2}\mu^2,0\right\},
\end{equation}
and straightforward computation of the renormalized internal energy yields the following result
\begin{equation}\label{int1}
  U'=\dfrac{\tr(\mathbb{D}\mathbb{H}')}{\tr\mathbb{D}}=\dfrac{\omega_I}{4}\mu^2\equiv U-U_I,
\end{equation}
where $U_I=\dfrac{\omega_I}{4}$ is the constant term, which possesses the property of the Eulerian homogeneity of degree $2$\index{Eulerian homogeneity of degree $2$}, i.e.
\begin{equation}
U'[\alpha\mu]=\alpha^2U'[\mu].
\end{equation}
In this manner the thermodynamics describing space quanta {\AE}ther \index{{\AE}ther}can be formulated in the standard way of the Eulerian systems\index{Eulerian systems}.

The three elementary and fundamental physical characteristics, i.e. the occupation number $N$, the internal energy $U$, and the entropy $S$, just were derived, and therefore one can conclude the thermodynamics by straightforward application of the f\/irst principles. Actually the entropy (\ref{entropia}) is manifestly divergent due to the presence of formal inf\/inity $\zeta(1)$. Straightforward calculation shows that the temperature\index{temperature} $T=\dfrac{\delta U}{\delta S}$ arising from the entropy (\ref{entropia}) contains the term with $\zeta(1)$. However, it is also visible that such a temperature\index{temperature} possesses f\/inite limit if and only if one re-scale initial data mass to formal inf\/inity, i.e. $\omega_I \rightarrow \omega_I\zeta(1)$. Because of the mass $m$ is related to length $l$ like $m\sim1/l$, performing of the limiting procedure $\omega_I\rightarrow\infty$ corresponds with introduction to the theory \emph{purely point object} $l_I\rightarrow0$.

Scaling of initial data is not good procedure, however, because of it has no well-def\/ined physical meaning. There is another possibility for reorganization of the troublesome divergence. Namely, it can be seen by straightforward calculation that the entropy renormalization $S\rightarrow \dfrac{S}{\zeta(1)}$ with performing the formal limit $\zeta(1)\rightarrow\infty$ leads to the equivalent result for the thermodynamics with no necessity of application of unclear scaling in initial data. Such an \emph{entropy renormalization}\index{entropy renormalization} corresponds to an initial quantum state of an embedded three-dimensional space being a purely point object, and yields perfect accordance with the second law of thermodynamics
\begin{equation}
  S\longrightarrow S'=\lim_{\zeta(1)\rightarrow\infty}\dfrac{S}{\zeta(1)}=\dfrac{1}{2}\left(\dfrac{\mu^2-1}{\mu^2+1}\right)^2\geqslant0.\label{sren}
\end{equation}

Calculating the temperature\index{temperature} $T'$ of space quanta one obtains the formula
\begin{equation}
  T'=\dfrac{\delta U'}{\delta S'}=\omega_I\dfrac{(\mu^2+1)^3}{8(\mu^2-1)},
\end{equation}
and one sees that initially, \emph{i.e.} for $\mu=1$, temperature\index{temperature} is inf\/inite. Such a situation describes the \emph{Hot Big Bang (HBB) phenomenon}\index{Hot Big Bang (HBB)}. It can be seen that after the HBB point the system is cooled right up until mass scale reaches the value
\begin{equation}
\mu_{PT}=\sqrt{2}\approx1.414,
\end{equation}
and then is warmed, what means that the value $\mu_{PT}$ is the phase transition point.

This phenomenon is better visible when one computes the energetic heat capacity\index{heat capacity} $C_{U}$
\begin{equation}\label{heat}
  C_U=T\dfrac{\delta S'}{\delta T}=\dfrac{\delta U'}{\delta T}=\dfrac{(\mu^2-1)^2}{(\mu^2-2)(\mu^2+1)^2},
\end{equation}
which possesses the singularity in the point $\mu_{PT}$. Application of the generalized law of equipartition
\begin{equation}
 \dfrac{\delta U'}{\delta T'}=\dfrac{f}{2},
\end{equation}
allows to establish the variability of the number of degrees of freedom\index{degrees of freedom}
\begin{equation}
  f=2C_{U}.
\end{equation}
The Helmholtz free energy\index{Helmholtz free energy} \mbox{$F=U'-T'S'$} that is
\begin{equation}
  F=-\dfrac{\omega_I}{16}(\mu^4-4\mu^2-1),
\end{equation}
is f\/inite for f\/inite value of initial data $\omega_I$, increases since the initial point $\mu=1$ until the phase transition $\mu_{PT}$, and then decreases. Therefore, the \emph{thermal equilibrium point} is the HBB\index{Hot Big Bang (HBB)} point related to the initial point $\mu_{eq}=1$. In the region of mass scales for which $1\leqslant\mu<\mu_{PT}$ the mechanical isolation is absent, but it is present after the phase transition, i.e. in the region $\mu>\mu_{PT}$.

Calculating the chemical potential\index{chemical potential}
\begin{equation}
  \varpi=\dfrac{\delta F}{\delta N}=-\omega_I\dfrac{\mu^3(\mu^2-2)}{2(\mu^2-1)},\label{chem}
\end{equation}
one can see straightforwardly that in the HBB\index{Hot Big Bang (HBB)} point $\mu_{eq}=1$ this potential diverges and in the phase transition point $\mu_{PT}$ it vanishes. Applying the chemical potential\index{chemical potential} (\ref{chem}) together with the occupation number $N$ and the Helmholtz free energy\index{Helmholtz free energy} $F$ yields the appropriate free energy def\/ined by the Landau grand potential\index{Landau grand potential} $\Omega$
\begin{equation}
  \Omega=F-\varpi N=\omega_I\dfrac{3\mu^6+\mu^4-11\mu^2-1}{16(\mu^2-1)},
\end{equation}
and therefore the corresponding Massieu--Planck free entropy\index{Massieu--Planck free entropy} $\Xi$ can be also derived straightforwardly
\begin{equation}
  \Xi=-\dfrac{\Omega}{T}=-\dfrac{3\mu^6+\mu^4-11\mu^2-1}{2(\mu^2+1)^3}.
\end{equation}
Consequently the grand partition function\index{grand partition function} $Z$ is established as
\begin{eqnarray}\label{gpf}
  Z=e^{\Xi}=\exp\left\{-\dfrac{3\mu^6+\mu^4-11\mu^2-1}{2(\mu^2+1)^3}\right\}.
\end{eqnarray}

The 2nd order Eulerian homogeneity of the system of space quanta yields the equation of state $\dfrac{PV}{T} =\ln Z=\Xi$ and determines the product of pressure $P$ and volume $V$ as
\begin{equation}\label{eos}
  PV=-\Omega,
\end{equation}
and together with the appropriate Gibbs--Duhem equation
\begin{equation}
V\delta P=S'\delta T+N\delta\varpi
\end{equation}
allows to establish the value of the pressure
\begin{equation}
  |P|=\exp\left\{-\int\left(\dfrac{S}{\Omega}\delta T+\dfrac{N}{\Omega}\delta\varpi\right)\right\}.
\end{equation}

Similarly, the f\/irst law of thermodynamics
\begin{equation}
-\delta\Omega=S'\delta T+P\delta V+N\delta\varpi,
\end{equation}
together with the equation of state (\ref{eos}) determine the volume
\begin{equation}
V=\dfrac{|\Omega|}{|P|},\label{volme}
\end{equation}
which is positive by def\/inition. Regarding the equation of state (\ref{eos}) for $\Omega=-|\Omega|<0$ the pressure is $P=|P|$, whereas for $\Omega=|\Omega|>0$ the pressure has the value $P=-|P|$. In this manner one receives
\begin{eqnarray}
P=P(\mu_0)\exp\left[\int_{\mu_0}^\mu\dfrac{8t\left(t^6-3t^2+4\right)}{3t^8-2t^6-12t^4+10t^2+1}dt\right],
\end{eqnarray}
where $\mu_0$ is some reference value of $\mu$. Regarding the relation (\ref{volme}), $V$ is a f\/ixed parameter and its value can be established as follows
\begin{equation}\label{vol}
V=\dfrac{\omega_I}{P(\mu_0)}\dfrac{3\mu^6+\mu^4-11\mu^2-1}{16(\mu^2-1)}\exp\left[-\int_{\mu_0}^\mu\dfrac{8t\left(t^6-3t^2+4\right)}{3t^8-2t^6-12t^4+10t^2+1}dt\right].
\end{equation}

One can also determine two another thermodynamical potentials: the Gibbs free energy\index{Gibbs free energy} $G=U-TS+PV$ and enthalpy\index{enthalpy} $H=U+PV$. The results are as follows
\begin{eqnarray}
G&=&-\dfrac{\omega_I\mu^2\left(\mu^4-\mu^2-2\right)}{4\left(\mu^2-1\right)},\\
H&=&-\dfrac{\omega_I\left(3\mu^6-3\mu^4-7\mu^2-1\right)}{16\left(\mu^2-1\right)}.
\end{eqnarray}

Equivalently, the thermodynamics of space quanta {\AE}ther \index{{\AE}ther}can be expressed by the size scale $\lambda=\dfrac{1}{\mu}$. There are the relations relating both the scales with an occupation number
\begin{eqnarray}
  \lambda&=&N\left(1\mp\sqrt{\strut{1-\dfrac{1}{N^2}}}\right),\\ \mu&=&N\left(1\pm\sqrt{\strut{1-\dfrac{1}{N^2}}}\right),
\end{eqnarray}
that in the limit of inf\/inite $N$ are equal
\begin{eqnarray}
  \lambda(N=\infty)&=&\left\{0,\infty\right\},\\
  \mu(N=\infty)&=&\left\{\infty,0\right\}.
\end{eqnarray}
Therefore, there are two possible asymptotic behaviors. The f\/irst situation is $\lambda=0$, $\mu=\infty$ which can be interpreted as a \emph{black hole}\index{black hole} as well as with HBB\index{Hot Big Bang (HBB)}. The second situation is $\lambda=\infty$ which def\/ined stable classical physical object - the classical space-time.

One can establish the number $n$ of space quanta generated from the stable Bogoliubov vacuum determined by the static Fock space\index{Fock space} related to initial data. We shall call such class of states \emph{vacuum space quanta}\index{vacuum space quanta}. By def\/inition this is the vacuum expectation value of the one-particle density operator\index{one-particle density operator}. In other words
\begin{equation}
 n=\dfrac{\lv\textsf{D}\rv}{\vv}=\dfrac{\lv{\textsf{G}}^{\dagger}{\textsf{G}}\rv}{\vv}.
\end{equation}
Straightforward application of the Bogoliubov transformation\index{Bogoliubov transformation}, the canonical commutation relations\index{canonical commutation relations} of the static Fock space\index{Fock space} leads to
\begin{eqnarray}
{\textsf{G}}^{\dagger}{\textsf{G}}&=&\left(u\textsf{G}_I^\dagger-v\textsf{G}_I\right)\left(-v^\ast\textsf{G}_I^\dagger+u^\ast\textsf{G}_I\right)=\nonumber\\
&=&|v|^2\textsf{G}_I\textsf{G}_I^\dagger+|u|^2\textsf{G}_I^\dagger\textsf{G}_I-v^\ast{u}\textsf{G}_I^\dagger\textsf{G}_I^\dagger-vu^\ast\textsf{G}_I\textsf{G}_I=\nonumber\\
&=&|v|^2+\left(|u|^2+|v|^2\right)\textsf{G}_I^\dagger\textsf{G}_I-v^\ast{u}\textsf{G}_I^\dagger\textsf{G}_I^\dagger-vu^\ast\textsf{G}_I\textsf{G}_I.
\end{eqnarray}
and using of the properties of the stable Bogoliubov vacuum\index{Bogoliubov vacuum} gives
\begin{eqnarray}
\lv{\textsf{G}}^{\dagger}{\textsf{G}}\rv&=&|v|^2\vv+\left(|u|^2+|v|^2\right)\lv\textsf{G}_I^\dagger\textsf{G}_I\rv-\\
&-&v^\ast{u}\lv\textsf{G}_I^\dagger\textsf{G}_I^\dagger\rv-vu^\ast\lv\textsf{G}_I\textsf{G}_I\rv=\\
&=&|v|^2\vv.\label{nof}
\end{eqnarray}
Therefore the number of vacuum space quanta is
\begin{equation}
  n=|v|^2=\dfrac{(\mu-1)^2}{4\mu}.
\end{equation}
In this manner the mass scale has two values
\begin{equation}\label{maas}
  \mu_{\pm}(n)=\left(\sqrt{n}\pm\sqrt{n+1}\right)^2,
\end{equation}
corresponding to two independent phases of the space quanta {\AE}ther\index{{\AE}ther}. For convenience we shall call the phase described by the sign $+$ \emph{the positive phase of the {\AE}ther}, and the phase described by the sign $-$ \emph{the negative phase of the {\AE}ther}\index{phase of the {\AE}ther!positive}\index{phase of the {\AE}ther!negative}. The phases have completely different asymptotic behaviour. Namely,
\begin{equation}
\lim_{n\rightarrow\infty}\mu_{\pm}(n)=\left\{\begin{array}{cc}\infty&\textrm{for~the~positive~phase}\\
0&~\textrm{for~the~negative~phase}\end{array}\right..
\end{equation}

Let us consider the asymptotic $n\rightarrow\infty$ thermodynamics of the {\AE}ther. The case of positive phase is
\begin{eqnarray}
T&\rightarrow&\infty,\\
C_U&\rightarrow&0,\\
f&\rightarrow&0,\\
F&\rightarrow&\infty,\\
\varpi&\rightarrow&\infty,\\
\Omega&\rightarrow&\infty,\\
\Xi&\rightarrow&-\dfrac{3}{2},\\
Z&\rightarrow&e^{-3/2},\\
G&\rightarrow&\infty,\\
H&\rightarrow&\infty,\\
P&\rightarrow&P_\infty,\\
V&\rightarrow&\infty,
\end{eqnarray}
where $P_\infty$ is the (constant) value of the pressure in the asymptotic value $\mu=\infty$ which can be assessed numerically. For instance in the trivial case $\mu_0=0$ one obtains $P_\infty\approx P(0)\exp(23.9527)$, whereas for $\mu_0=1$ one receives $P_\infty\approx P(1)\exp(510959)$.

Similarly one can analyse the negative phase
\begin{eqnarray}
T&\rightarrow-\dfrac{\omega_I}{8},\\
C_U&\rightarrow&-\dfrac{1}{2},\\
f&\rightarrow&-1,\\
F&\rightarrow&\dfrac{\omega_I}{16},\\
\varpi&\rightarrow&0,\\
\Omega&\rightarrow&\dfrac{\omega_I}{16},\\
\Xi&\rightarrow&\dfrac{1}{2},\\
Z&\rightarrow&e^{1/2},\\
G&\rightarrow&0,\\
H&\rightarrow&-\dfrac{\omega_I}{16},\\
P&\rightarrow&P_0,\\
V&\rightarrow&\dfrac{\omega_I}{16P_0},
\end{eqnarray}
where $P_0$ is the (constant) value of the pressure in $\mu=0$ which can be assessed numerically. For instance when $\mu_0=0$ one has $P_0=P(0)$. For $\mu_0=1$ one obtains $P_0\approx P(1)\exp(-151.536)$.

In this section we have presented the next implication of the global one-dimensional quantum gravity. It was shown that this algorithm yields constructive, consistent, and plausible phenomenology, that is thermodynamics of {\AE}ther\index{{\AE}ther}, in the discussed situation describing space quanta behavior. The theory of quantum gravity as well as the thermodynamics can be applied to any general relativistic space-times which metrics can be presented in the form of the $3+1$ splitting\index{$3+1$ splitting}. Such space-time satisfy the Mach principle, \emph{i.e.} are isotropic. Their importance for elementary particle physics, cosmology and high energy astrophysics is experimentally conf\/irmed; one can say that these are \emph{phenomenological space-times}\index{space-time!phenomenological}.

As the example of \emph{ab initio} formulation of thermodynamics we have employed the one-particle approximation\index{one-particle approximation} of the density matrix. Application of the renormalization method to the entropy and to the Hamiltonian matrix resulted in the second order Eulerian homogeneity property. The Landau grand potential\index{Landau grand potential} $\Omega$ and the Massieu--Planck free entropy\index{Massieu--Planck free entropy} $\Xi$ were employed to the consistent description. The grand partition function\index{grand partition function} $Z$ and thermodynamic volume $V$ were determined constructively. Another thermodynamical potentials were derived in frames of the \emph{entropic formalism}\index{entropic formalism}, which accords with the f\/irst and the second principles of thermodynamics. Physical information following from the thermodynamics of space quanta {\AE}ther \index{{\AE}ther}is the crucial point of the construction presented in this section. Actually the proposed approach differ from another ones (Cf. \emph{e.g.} \cite{ajl,amb,war,pad}) by \emph{ab initio} treatment of the quantum gravity phenomenology.

Studying of particular physical situations in frames of the proposed approach seems to be the most important prospective arising from the thermodynamics of space quanta {\AE}ther. From experimental point of view the presented considerations possess evident usefulness, because of bosonic systems are common in high energy physics.

\section{Entropy II: The Algebraic Approach}

Let us consider the space quanta {\AE}ther in the grand canonical ensemble. First of all let us express the static one-particle density matrix\index{one-particle density matrix!via the Bogoliubov coeff\/icients} via the Bogoliubov coeff\/icients\index{Bogoliubov coeff\/icients}
\begin{equation}\label{den}
\mathbb{D}=\left[\begin{array}{cc}|u|^2&-uv\\-u^{\ast}v^{\ast}&|v|^2\end{array}\right].
\end{equation}
The basic quantity is an entropy, which for an arbitrary quantum system is def\/ined by the standard Boltzmann--von Neumann formula\index{Boltzmann--von Neumann entropy}
\begin{equation}\label{entropia}
  S =\dfrac{\tr\left(\mathbb{D}\ln\mathbb{D}\right)}{\tr\mathbb{D}},
\end{equation}
and in the present situation can be immediately computed from the density matrix (\ref{den}).

The problem is to establish the logarithm of the one-particle density matrix\index{one-particle density matrix!logarithm} $\ln\mathbb{D}$. It can be performed by application of of the algebraic methods, particularly polynomial long division algorithm, the characteristic polynomial, and the Cayley--Hamilton theorem\index{Cayley--Hamilton theorem} (For some details of basic and advanced algebra see e.g. books in the Ref. \cite{algebra}). Let us present the method in detail for any analytical function of a matrix $\mathbb{D}$. Let us consider the characteristic polynomial $ch_{\mathbb{D}}(\lambda)$ of the matrix $\mathbb{D}$, where $\lambda$ is eigenvalue of $\mathbb{D}$. If $p(\lambda)$ is an analytical function, i.e. possesses power series expansion, then via using of the division transformation one can present $p(\lambda)$ as the dividend which divisor is the characteristic polynomial
\begin{equation}
  p(\lambda)=q(\lambda)ch_{\mathbb{D}}(\lambda)+r(\lambda).\label{plam}
\end{equation}
In other words the problem is to establish the remainder polynomial $r(\lambda)$ and the quotient polynomial $q(\lambda)$. Recall that according to the Cayley--Hamilton theorem\index{Cayley--Hamilton theorem} the characteristic polynomial of a matrix $\mathbb{D}$ evaluated on this matrix vanishes identically
\begin{equation}
  ch_{\mathbb{D}}(\mathbb{D})=0,
\end{equation}
so that consequently the evaluation $p(\mathbb{D})$ is exactly equal to the reminder polynomial evaluated on the matrix $\mathbb{D}$
\begin{equation}
  p(\mathbb{D})=r(\mathbb{D}).
\end{equation}
If a matrix $\mathbb{D}$ is a matrix of dimension $n\times n$ then its characteristic polynomial $ch_{\mathbb{D}}(\lambda)$ is a polynomial of degree $n$ which coeff\/icients are invariants of a matrix $\mathbb{D}$. Therefore the remainder polynomial $r(\lambda)$ must of the order $n-1$ at most. There is some problem when there are eigenvalues of $\mathbb{D}$ for which the function $p(\lambda)$ has singularity. Then, however, we shall not include such eigenvalues, and for determination of the quotient $q(\lambda)$ and the remainder polynomial $r(\lambda)$  we shall differentiate the relation (\ref{plam}) $n$ times and evaluate all $n+1$ relations on the non-singular eigenvalues of $\mathbb{D}$. Such a procedure generates the system of $n+1$ equations which allows to establish the coeff\/icients of the remainder polynomial $r(\lambda)$ as well as leads to the quotient $q(\lambda)$ as the result of solving an appropriate differential equation of degree $n$ at most. It must be emphasized that the quotient $q(\lambda)$ as a solution of ordinary differential equation is assured to be an analytical function.

Let us apply such a method to the situation given by the $2\times2$ matrix $\mathbb{D}$ having zero determinant $\det\mathbb{D}=0$, and the function $p(x)=\ln{x}$. The characteristic polynomial of the matrix $\mathbb{D}$ is
\begin{equation}
ch_{\mathbb{D}}(\lambda)=\det\left(\mathbb{D}-\lambda\mathbb{I}\right)=\lambda^2-(\tr\mathbb{D})\lambda,
\end{equation}
and its eigenvalues are $\lambda=0$ and $\lambda=\tr\mathbb{D}$. By the Cayley--Hamilton theorem\index{Cayley--Hamilton theorem} one has
\begin{equation}
\mathbb{D}^2-(\tr\mathbb{D})\mathbb{D}=0,
\end{equation}
i.e. $\mathbb{D}^2=(\tr\mathbb{D})\mathbb{D}$. Let us write out the relation (\ref{plam}) for this case
\begin{equation}
  \ln\lambda=q(\lambda)\left(\lambda^2-(\tr\mathbb{D})\lambda\right)+a_0+a_1\lambda,\label{plam1}
\end{equation}
where $a_0$ and $a_1$ are the coeff\/icients of the polynomial $r(\lambda)$. The problem is to establish $q(\lambda)$ and the coeff\/icients $a_0$ and $a_1$. First of all let us note that $\lambda=0$ is the singularity of $\ln\lambda$, what means that we shall not consider this eigenvalue. For determination of the three unknown quantities let us differentiate the relation (\ref{plam1}) two times. The results are as follows
\begin{eqnarray}
  \dfrac{1}{\lambda}&=&q'(\lambda)\left(\lambda^2-(\tr\mathbb{D})\lambda\right)+q(\lambda)\left(2\lambda-\tr\mathbb{D}\right)+a_1,\label{plam2}\\
  -\dfrac{1}{\lambda^2}&=&q''(\lambda)\left(\lambda^2-(\tr\mathbb{D})\lambda\right)+2q'(\lambda)\left(2\lambda-\tr\mathbb{D}\right)+2q(\lambda).\label{plam3}
\end{eqnarray}
Taking into account the fact that evaluation of the characteristic polynomial on an eigenvalue is zero. Application of this fundamental fact in the case of the eigenvalue $\lambda=\tr\mathbb{D}$ leads to signif\/icant simplif\/ication of the equations (\ref{plam1}), (\ref{plam2}) and (\ref{plam3})
\begin{eqnarray}
  \ln\lambda&=&a_0+a_1\lambda,\label{plam1a}\\
  \dfrac{1}{\lambda}&=&q(\lambda)\lambda+a_1,\label{plam2a}\\
  -\dfrac{1}{\lambda^2}&=&2q'(\lambda)\lambda+2q(\lambda),\label{plam3a}
\end{eqnarray}
which can be presented in the following form
\begin{eqnarray}
  &&a_0=\ln\lambda-a_1\lambda,\label{plam1b}\\
  &&a_1=\dfrac{1}{\lambda}-q(\lambda)\lambda,\label{plam2b}\\
  &&\lambda q'(\lambda)+q(\lambda)+\dfrac{1}{2\lambda^2}=0.\label{plam3b}
\end{eqnarray}
The equation (\ref{plam3b}) is the differential equation for the quotient $q(\lambda)$ and can be solved straightforwardly
\begin{equation}
  q(\lambda)=\dfrac{1}{2\lambda^2}+\dfrac{C}{\lambda},
\end{equation}
where $C$ is constant of integration. Because of we are still interested in the concrete eigenvalue $\lambda=\tr\mathbb{D}$ one receives
\begin{equation}
  q(\tr\mathbb{D})=\dfrac{1}{2(\tr\mathbb{D})^2}+\dfrac{C}{\tr\mathbb{D}},
\end{equation}
and by this reason one receives
\begin{eqnarray}
a_1&=&\dfrac{1}{\tr\mathbb{D}}-q(\tr\mathbb{D})\tr\mathbb{D}=\dfrac{1}{2\tr\mathbb{D}}-C,\\
a_0&=&\ln\tr\mathbb{D}-a_1\tr\mathbb{D}=\ln\tr\mathbb{D}-\dfrac{1}{2}+C\tr\mathbb{D}.
\end{eqnarray}
In this manner one can compute the function $\ln\mathbb{D}$ as follows
\begin{equation}
\ln\mathbb{D}=a_0\mathbb{I}+a_1\mathbb{D}=\left(\ln\tr\mathbb{D}-\dfrac{1}{2}+C\tr\mathbb{D}\right)\mathbb{I}+\left(\dfrac{1}{2\tr\mathbb{D}}-C\right)\mathbb{D},
\end{equation}
and consequently one obtains
\begin{eqnarray}
  \mathbb{D}\ln\mathbb{D}&=&\left(\ln\tr\mathbb{D}-\dfrac{1}{2}+C\tr\mathbb{D}\right)\mathbb{D}+\left(\dfrac{1}{2\tr\mathbb{D}}-C\right)\mathbb{D}^2=\\
  &=&\left(\ln\tr\mathbb{D}-\dfrac{1}{2}+C\tr\mathbb{D}\right)\mathbb{D}+\left(\dfrac{1}{2\tr\mathbb{D}}-C\right)(\tr\mathbb{D})\mathbb{D}=\\
&=&\left(\ln\tr\mathbb{D}-\dfrac{1}{2}+C\tr\mathbb{D}\right)\mathbb{D}+\left(\dfrac{1}{2}-C\tr\mathbb{D}\right)\mathbb{D}=\\
&=&\left(\ln\tr\mathbb{D}-\dfrac{1}{2}+C\tr\mathbb{D}+\dfrac{1}{2}-C\tr\mathbb{D}\right)\mathbb{D}=\\
&=&\left(\ln\tr\mathbb{D}\right)\mathbb{D}.
\end{eqnarray}
Now it is easy to establish the entropy
\begin{equation}
  S=\dfrac{\tr(\mathbb{D}\ln\mathbb{D})}{\tr\mathbb{D}}=\dfrac{\tr\left((\ln\tr\mathbb{D})\mathbb{D}\right)}{\tr\mathbb{D}}=\dfrac{(\ln\tr\mathbb{D})\tr\mathbb{D}}{\tr\mathbb{D}}=\ln\tr\mathbb{D}.
\end{equation}
In our situation the trace of the density matrix is
\begin{equation}
\tr\mathbb{D}=|u|^2+|v|^2=2|v|^2+1
\end{equation}
and by this reason one obtains
 \begin{equation}\label{entropy}
   S=\ln\left(2|v|^2+1\right)=-\ln\Sigma,
 \end{equation}
where $\Sigma$ is the quantum statistics of the system of space quanta
 \begin{equation}
   \Sigma=\dfrac{1}{2|v|^2+1}=\dfrac{1}{2n+1},
 \end{equation}
where $n$ is the number of vacuum space quanta (\ref{nof}).

Let us compute the thermodynamical potentials for the entropy received entropy (\ref{entropy}) and the same values of the internal energy $U$, and the occupation number $N$ established in the previous section, i.e.
\begin{eqnarray}
  S&=&\ln\left[\dfrac{\mu^2+1}{2\mu}\right],\\
  U&=&\dfrac{\omega_I}{4}\mu^2,\\
  N&=&\dfrac{\mu^2+1}{2\mu}.
\end{eqnarray}
The most important is of course the temperature\index{temperature} of the system
\begin{equation}
  T=\dfrac{\omega_I\mu^2}{2}\dfrac{\mu^2+1}{\mu^2-1},
\end{equation}
which for $\mu>1$ is manifestly negative. The heat capacity\index{heat capacity} has the form
\begin{equation}
C_U=\dfrac{1}{2}+\dfrac{1}{\mu^4-2\mu^2-1},
\end{equation}
and the number of degrees of freedom\index{degrees of freedom} is
\begin{equation}
f=1+\dfrac{2}{\mu^4-2\mu^2-1}.
\end{equation}
The Helmholtz free energy\index{Helmholtz free energy} can be also derived straightforwardly
\begin{equation}
F=\dfrac{\omega_I\mu^2}{4}\left(1-2\dfrac{\mu^2+1}{\mu^2-1}\ln\left[\dfrac{\mu^2+1}{2\mu}\right]\right),
\end{equation}
and the chemical potential\index{chemical potential} has the form
\begin{equation}
 \varpi=-\dfrac{2\omega_I\mu^3\left(\mu^4-2\mu^2-1\right)}{\left(\mu^2-1\right)^3}\ln\left[\dfrac{\mu^2+1}{2\mu}\right].
\end{equation}
The Landau grand potential\index{Landau grand potential} and the Massieu--Planck free entropy\index{Massieu--Planck free entropy} are respectively
\begin{eqnarray}
 \Omega&=&\dfrac{\omega_I\mu^2}{4}\left(1+2\dfrac{\left(\mu^2-3\right)\left(\mu^2+1\right)^2}{\left(\mu^2-1\right)^3}\ln\left[\dfrac{\mu^2+1}{2\mu}\right]\right),\\
 \Xi&=&-\dfrac{\mu^2-1}{2\left(\mu^2+1\right)}-\dfrac{\left(\mu^2-3\right)\left(\mu^2+1\right)}{\left(\mu^2-1\right)^2}\ln\left[\dfrac{\mu^2+1}{2\mu}\right].
\end{eqnarray}
The grand partition function\index{grand partition function} can be established as
\begin{equation}
 Z=\exp\left[-\dfrac{\mu^2-1}{2\left(\mu^2+1\right)}\right]\left(\dfrac{\mu^2+1}{2\mu}\right)^{-\dfrac{ \left(\mu^2-3\right)\left(\mu^2+1\right)}{\left(\mu^2-1\right)^2}}.
\end{equation}
The pressure is
\begin{equation}
 P=P(\mu_0)\exp\left[-4\int_{\mu_0}^\mu\dfrac{(t^2-1)^2\left(t^4-2t^2-1\right)+4\left(t^4+4t^2+1\right)\ln\left[\dfrac{t^2+1}{2t}\right]}{t\left(t^2-1\right)\left(\left(t^2-1\right)^3+2\left(t^2-3\right)\left(t^2+1\right)^2\ln\left[\dfrac{t^2+1}{2t}\right]\right)}dt\right],
\end{equation}
where $\mu_0$ is some reference value of $\mu$, and the thermodynamical volume has the form
\begin{eqnarray}
\!\!\!\!\!\!\!\!\!\!\!\!\!\!\!\!\!\!\!\!
 &&V=\dfrac{\omega_I\mu^2}{4P(\mu_0)}\left|1+2\dfrac{\left(\mu^2-3\right)\left(\mu^2+1\right)^2}{\left(\mu^2-1\right)^3}\ln\left[\dfrac{\mu^2+1}{2\mu}\right]\right|\times\\
\!\!\!\!\!\!\!\!\!\!\!\!\!\!\!\!\!\!\!\!
&&\times\exp\left[4\int_{\mu_0}^\mu\dfrac{(t^2-1)^2\left(t^4-2t^2-1\right)+4\left(t^4+4t^2+1\right)\ln\left[\dfrac{t^2+1}{2t}\right]}{t\left(t^2-1\right)\left(\left(t^2-1\right)^3+2\left(t^2-3\right)\left(t^2+1\right)^2\ln\left[\dfrac{t^2+1}{2t}\right]\right)}dt\right].\nonumber
\end{eqnarray}
The Gibbs free energy\index{Gibbs free energy} and enthalpy\index{enthalpy} are respectively
\begin{eqnarray}
 G&=&\dfrac{\omega_I\mu^2\left(\mu^2+1\right)\left(\mu^4-2\mu^2-1\right)}{\left(\mu^2-1\right)^3}\ln\left[\dfrac{\mu^2+1}{2\mu}\right],\\
 H&=&-\dfrac{\omega_I\mu^2\left(\mu^2-3\right)\left(\mu^2+1\right)^2}{2\left(\mu^2-1\right)^3}\ln\left[\dfrac{\mu^2+1}{2\mu}\right].
\end{eqnarray}

There is the question about asymptotic $n\rightarrow\infty$ thermodynamics of the {\AE}ther\index{{\AE}ther}. The positive phase is then
\begin{eqnarray}
T&\rightarrow\infty,\\
C_U&\rightarrow&\dfrac{1}{2},\\
f&\rightarrow&1,\\
F&\rightarrow&-\infty,\\
\varpi&\rightarrow&-\infty,
\end{eqnarray}
\begin{eqnarray}
\Omega&\rightarrow&\infty,\\
\Xi&\rightarrow&-\infty,\\
Z&\rightarrow&0,\\
G&\rightarrow&-\infty,\\
H&\rightarrow&-\infty,\\
P&\rightarrow&P_\infty,\\
V&\rightarrow&\infty,
\end{eqnarray}
where $P_\infty$ is the numerical value of the pressure in the asymptotic value $\mu=\infty$. For example in the trivial situation $\mu_0=0$ one obtains the value $P_\infty\approx{P}(0)\exp(473.822)$, whereas when $\mu_0=1$ one receives another result $P_\infty\approx P(1)\exp(0.208615)\approx1.23197P(1)$.

Similarly, the asymptotic thermodynamics of the negative phase of the space quanta {\AE}ther can be analyzed. The results are as follows
\begin{eqnarray}
T&\rightarrow0,\\
C_U&\rightarrow&-\dfrac{1}{2},\\
f&\rightarrow&-1,\\
F&\rightarrow&0,\\
\varpi&\rightarrow&0,\\
\Omega&\rightarrow&0,\\
\Xi&\rightarrow&-\infty,\\
Z&\rightarrow&-\infty,\\
G&\rightarrow&0,\\
H&\rightarrow&0,\\
P&\rightarrow&P_0,\\
V&\rightarrow&0,
\end{eqnarray}
where $P_0$ is the numerical value of the pressure for the value $\mu=0$. For example in the trivial situation $\mu_0=0$ one has $P_0=P(0)$, while when $\mu_0=1$ one obtains $P_0\approx P(1)\exp(-510956)$.

\chapter*{Epilogue}
\addcontentsline{toc}{part}{Epilogue}

\begin{flushright}
\textit{Science is a differential equation. Religion is a boundary condition.}\vspace*{10pt}\\
\textsl{Alan Turing}
\end{flushright}

This book presented the constructive model of physical Reality based on the realization of the fusion of two fundamental concepts of Antiquity: Aristotelian {\AE}ther and the Epicurean--Islamic Multiverse. The theory in itself creates the unif\/ied point of view which I proposed to call \emph{{\AE}thereal Multiverse}. There is evident opportunity and necessity to apply the proposed approach straightforwardly to the concrete problems which theoretical results could be compared with experimental and observational data. The only such a treatment guarantees physical consistency of \emph{{\AE}thereal Multiverse}.

I believe that \emph{{\AE}thereal Multiverse} governs Nature at comparatively small scales. In my opinion the Planck scale, i.e. the scale in which quantum physics meets classical physics, is the good candidate for such effects. However, this is the only my personal belief, and therefore there is justif\/ied necessity to verify its consequences empirically. Possibly, the scope of applicability is much more wide than I think presently, but this is also possible that the region of applicability is completely different then I have suggested. Another possible candidates for the new physics having a place at comparatively small scales include the Compton scale or introduced in this book the Compton--Planck scale.

Anyway, there is a certain knowledge which we have established in this book. Namely, we have proved that straightforward philosophical reasoning involving both the concepts of {\AE}ther and Multiverse can be productively performed, and applied to consistent construction of new physics. The philosophical spirit of the modal realism, which was the ideological fundament of our consideration, resulted in fruitful and fashionable development of two fundamental ideas of Antiquity, which at the f\/irst glance look like in an old-fashioned manner or manifestly fossilized than like the foundations of the new physics. The crucial point of the presented deductions was constructiveness of the applied theoretical approach based on the method of analogy. The central methodological background of our deductions was the principle of simplicity, which enabled to receive new physical description from well-established knowledge of abstract mathematics and mathematical physics.

From the philosophical point of view we have proven obvious non triviality. Namely, we showed that it is possible to joint manifestly the systems of Aristotelian and Epicurean--Islamic philosophy in the one unif\/ied and productive picture of physical Reality. This is in itself a paradoxical situation because of these philosophies have been understood as completely different ideological systems which are impossible to unify. This creates the strong belief that new physics can be effectively formulated via involving of the old and well-established philosophy and knowledge to the new applications. In other words, it is my deep conviction that the new physics is the only sophisticated structure hidden in a philosophical interpretation of a mathematical formalism of the old physics. The identif\/ication method, which we frequently applied in our studies, is the only straightforward purely logical consequence of a philosophical interpretation which play the fundamental role in all natural sciences.

In this book a number of results obtained in my earlier research work was def\/initely updated, improved or even rejected from the general physical scenario. However, I have a deep conviction that the mathematical truth was established in a great detail. This level of discussion allows to think that the presented scheme of philosophical reasoning is the most productive way for new constructive deductions and, moreover, in itself creates the new logical system of physics. At the f\/irst glance this logic can be diff\/icult to accept or even impossible to practical applications. However, it has been shown in this book that when the application is realized in a constructive way it results in the elegant picture of the physical Reality.

The physical scenario presented in this book is essentially novel, because of in a whole is a certain application of both the methods and the philosophical background of theory of relativity and quantum theory, which are fundamentally conf\/irmed as the fundamental theories of physics. Their possible deformations and modifications follow from the recent deductions of high energy physics and astroparticle physics. I will be satisf\/ied when \emph{{\AE}thereal Multiverse} will be turned out a productive approach to theoretical physics.

\backmatter

\newpage
\thispagestyle{empty}

\newpage
\thispagestyle{empty}
\addcontentsline{toc}{part}{Index}
\includepdf[pages=-]{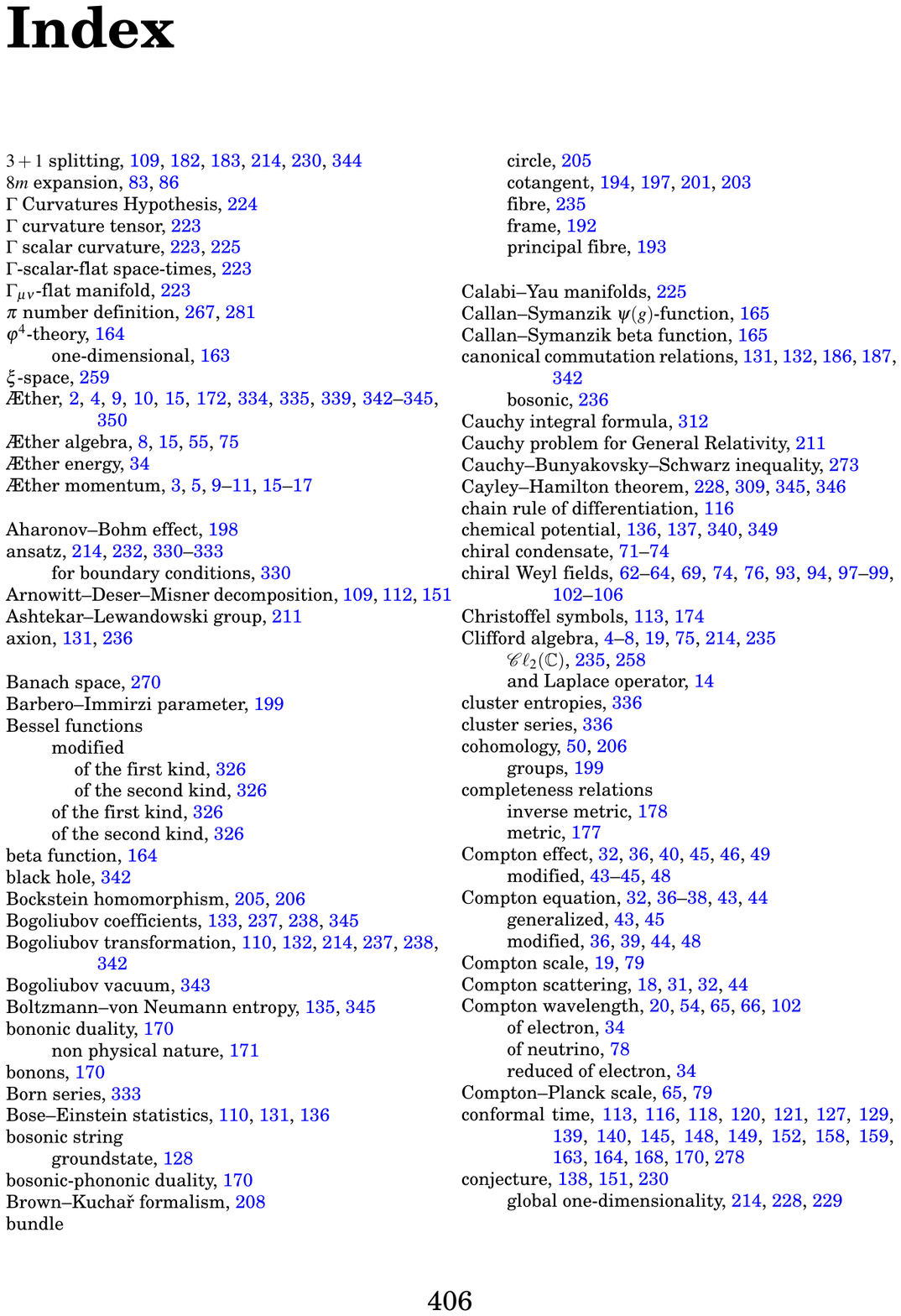}

\begin{thebibliography}{999}
\addcontentsline{toc}{part}{Bibliography}
\footnotesize{

\bibitem{rcarroll} R. Carroll, \emph{On the Emergence Theme of Physics} (World Scientific, 2010)
\bibitem{pais}
P.A. Schilpp, \emph{Albert Einstein: Philosopher-Scientist} (MJF Books, 1970);\\
J. Mehra, \emph{Einstein, Physics and Reality} (World Scientif\/ic, 1999);\\
A. Pais, \emph{'Subtle is the Lord\ldots'. The Science and the Life of Albert Einstein} (Oxford University Press, 2005)
\bibitem{aristotle} W.D. Ross and J.A. Smith (Eds.), \emph{The Works of Aristotle} Twelve volumes (Clarendon Press, 1908-1952)
\bibitem{boyle} T. Birch (Ed.), \emph{The Works of the Honourable Robert Boyle}, 6 vols. (London 1672)
\bibitem{opticks} I. Newton, \emph{Opticks: or a Treatise of the Ref\/lexions, Refractions, Inf\/lections, and Colours of Light} (1704)
\bibitem{lesage}
G.L. Le Sage, \emph{The Newtonian Lucretius}, in \emph{The Le Sage Theory fo Gravitation} (Translated by C.G. Abbot, with introductory note by S.P. Langley) (Annual Report of the Board of Regents of the Smithsonian Institution, 1898), pp. 139-160;\\
T. Thomson, \emph{Biographical Account of M. Le Sage}, Annals of Philosophy \textbf{11} (Baldwin, 1818), pp. 241-252;\\
M.R. Edwards, \emph{Pushing Gravity: New Perspectives on Le Sage's Theory of Gravitation} (C. Roy Keys, 2002)
\bibitem{kelvin}
W. Thomson, Phil. Mag. \textbf{45}, 321 (1873);\\
W. Thomson, \emph{Mathematical and Physical Papers} (Cambridge University Press, 1882-1911);\\
W. Thomson, \emph{Collected Papers in Physics and Engineering} (Cambridge University Press, 1912);\\
Lord Kelvin, \emph{Baltimore Lectures on Molecular Dynamics and the Wave Theory of Light} (C.J. Clay and Sons, 1904)
\bibitem{jcmaxwell}
J.C. Maxwell, \emph{On Physical lines of force}, Phil. Mag. \textbf{21} p. 161, 281, 338 (1861); Phil. Mag. \textbf{22}, p. 12, 85 (1861);\\
J.C. Maxwell, \emph{Ether}, Encyclop{\ae}dia Britannica (9th ed.) \emph{8} (1878), pp. 569-572;\\
J.C. Maxwell, \emph{A Treatise on Electricity and Magnetism} (Clarendeon Press, 1873)
\bibitem{otherae}
H. Fizeau, \emph{Sur les hypotheses relatives a l'ether lumineux}, Annales de Chemie et de Physique (3rd Series) \textbf{57}, 385 (1859);\\
T.R. Birks, \emph{On Matter and Ether or the Secret Laws of Physical Change} (MacMillan, 1862);\\
S.T. Preston, \emph{Physics of the Ether} (E \& F.N. Spon, 1875);\\
P.G. Tait, \emph{Lectures on Some Recent Advances in Physical Science with a Special Lecture on Force} (MacMillan, 1876);\\
W. Barlow, \emph{New Theories of Matter and of Force} (Sampson Low, 1885);\\
J.S. Russell, \emph{The Wave and Translation in the Oceans of Water, Air, and Ether} (Trubner, 1885);\\
H.W. Watson and S.H. Burbury, \emph{The Mathematical Theory of Electricity and Magnetism Vol 1 Electrostatics \& Vol 2 Magnetism and Electrodynamics} (Clarendon, 1885 \& 1889);\\
A.E. Dolbear, \emph{Matter, Ether, and Motion. The Factors and Relations of Physical Science} (Lee and Shepard, 1892);\\
J. Tyndall, \emph{New Fragments} (D. Appleton, 1896);\\
A.E. Dolbear, \emph{Modes of Motion. Mechanical Concepts of Physical Phenomena} (Lee and Shepard, 1897);\\
J. Larmor, Phil. Trans. Roy. Soc. \textbf{190}, 205 (1897);\\
D. Mendeleeff, \emph{Principles of Chemistry} (P.F. Collier and Son, 1897);\\
E. Wiechert, \emph{Grundlagen der Elektrodynamik} (Teubner, 1899);\\
J. Larmor, \emph{Aether and Matter. A Development of the Dynamical Relations of the Aether to Material Systems on the Basis of the Atomic Constitution of Matter Including a Discussion of the Inf\/luence of the Earth's Motion on Optical Phenomena} (Cambridge University Press, 1900);\\
N.E. Gilbert, \emph{Some Experiments Upon The Relations Between Ether, Matter, and Electricity}. Dissertation submitted to the Board of University Studies of the John Hopkins University for the Degree of Doctor of Philosophy (1901);\\
O. Reynolds, \emph{On an Inversion of Ideas as to the Structure of the Universe} (Cambridge University Press, 1902);\\
W.G. Hooper, \emph{Aether and Gravitation} (Chapman and Hall, 1903);\\
G. Adam, \emph{Electricity. The Chemistry of Ether} (Whitaker \& Ray, 1904);\\
J.E. Gore, \emph{Studies in Astronomy} (Chatto \& Windus, 1904);\\
D. Mendeleeff, \emph{An Attempt Towards a Chemical Conception of the Ether} (Longmans, 1904);\\
G. Mie, \emph{Molek\"{u}le, Atome, Welt\"{a}ther} (Teubner, 1904);\\
J.J. Thomson, \emph{Electricity and matter} (Charles Scribner's Sons, 1904);\\
A.M. Clerke, \emph{Modern Cosmogonies} (Adam and Charles Black, 1905);\\
W.A. Shenstone, \emph{The New Physics and Chemistry. A Series of Popular Essays on  Physical and Chemical Subjects} (Smith, Elder \& Co., 1906);\\
S. Arrhenius, \emph{Theories of Chemistry}. Being Delivered at the University of California, in Berkeley (Longmans, 1907);\\
O. Lodge, \emph{Modern Views of Electricity} (MacMillan, 1907);\\
    O. Lodge, \emph{The Ether of Space} (Harper \& Brothers, 1909);\\
    G.W. De Tunzelmann, \emph{A Treatise on Electrical Theory and the Problem of the Universe} (Griff\/in, 1910);\\
    J. Larmor, \emph{Aether}, in Encyclop{\ae}dia Britannica (11th ed.) (1911);\\
    F. Harris, \emph{Gravitation} (Longmans, 1912);\\
    W.W.R. Ball, \emph{Mathematics Recreations and Essays} (6th ed., MacMillan, 1914);\\
    M. Erwin, \emph{The Universe and the Atom. The Ether Constitution, Creation and Structure of Atoms, Gravitation, and Electricity, Kinetically Explained} (Constable, 1915);\\
    G.H. Darwin, \emph{Scientif\/ic Papers} (Cambridge University Press, 1907-1916);\\
    J.H. Jeans, \emph{The Mathematical Theory of Electricity and Magnetism} (Cambridge University Press, 1911);\\
    W.J. Spillman, \emph{A theory of gravitation and related phenomena} (The New Era, 1915);\\
    F.W. Very, \emph{The Luminiferous Ether. (I) Its Relation to the Electron and to a Universal Interstellar Medium (II) Its Relation to the Atom} (The Four Seas, 1919);\\
    B. Harrow, \emph{From Netwon to Einstein. Changing Conceptions of the Universe} (Van Nostrand, 1920);\\
    R.A. Sampson, \emph{On Gravitation and Relativity} being the Halley Lecture delivered on June 12, 1920 (Claredon, 1920);\\
    W. Bragg, \emph{Electrons \& Ether Waves}. Being the twenty-third Robert Boyle lecture on 11th May 1921 (Oxford University Press, 1921);\\
    P.A. Campbell, \emph{A Non-Euclidean Theory of Matter and Electricity} (Cambridge University Press, 1921);\\
    E. Cunningham, \emph{Relativity, The Electron Theory and Gravitation} (Longmans, 1921);\\
    L. Page, \emph{An Introduction to Electrodynamics From the Standpoint of the Electron Theory} (Ginn, 1922);\\
    H.S. Redgrove, \emph{Alchemy: Ancient and Modern} (William Rider \& Son, 1922);\\
    T.J.J. See, \emph{Electrodynamic Wave-Theory pf  Physical Forces Vol. 1 \& Vol. 2 New theory of the Aether} (Thos. P. Nichols \& Son, 1917 \& 1922)
\bibitem{whitt} E.T. Whittaker, \emph{A History of the Theories of Aether and Electricity} Vol. 1 The classical theories \& Vol. 2 The modern theories 1900-1926 (2nd ed., Thomas Nelson and Sons, 1951)
\bibitem{michmor}
    A.A. Michelson, Am. J. Sci. \textbf{22}, 120 (1881);\\
    A.A. Michelson and E.W. Morley, Am. J. Sci. \textbf{34}, 333 (1887)
\bibitem{michel} A.A. Michelson, \emph{Light waves and their uses} (The University of Chicago Press, 1903)
\bibitem{halorentz}
    H.A. Lorentz, Zittingsverlag akad. v. Wet. \textbf{1}, 74 (1892);\\
    H.A. Lorentz, \emph{The Theory of Electrons and its Applications to the Phenomena of Light and Radiant Heat}. A Course of Lectures Delivered in Columbia University, New York, in March and April 1906 (2nd ed., Teubner, 1916);\\
    H.A. Lorentz, \emph{Problems on Modern Physics}. A Course of Lectures Delivered in the California Institute of Technology (Ginn, 1927);\\
    H.A. Lorentz, \emph{Lectures on Theoretical Physics. Vol I-III}. Delivered at the University of Leiden (MacMillan, 1927-1931)
\bibitem{heavi}
O. Heaviside, The Electrician p. 23, 83, 147, 458 (1888-1889) \& Electrical papers, Vol 2. p. 490, 504 (1894); Phil. Mag. \textbf{27}(167), 324 (1889)
\bibitem{fitzg} G.F. FitzGerald, Letters to the Editor, Science \textbf{13}, 390 (1889)
\bibitem{othexp}
F.T. Trouton and H.R. Noble, Phil. Trans. Royal Soc. \textbf{A202}, 165 (1903);\\
F.T. Trouton and A.O. Rankine, Proc. Roy. Soc. \textbf{80} (420) (1908);\\
M.G. Sagnac, Comptes Rendus de L'Acad\'{e}mie des Sciences \textbf{157}, p. 708, 710, 1410 (1913);\\
M.G. Sagnac, Journale de Physique et le Radium \textbf{4}(5), 177 (1914);\\
A.A. Michelson, Astrophys. J. \textbf{61}, 137 (1925);\\
A.A. Michelson, H.G. Gale, and F. Pearson, Astrophys. J. \textbf{61}, 139 (1925);\\
R.J. Kennedy and E.M. Thorndike, Phys. Rev. \textbf{42}(2), 400 (1932);\\
G.W. Hammar, Phys. Rev. \textbf{48}(5), 462 (1935)
\bibitem{pauli} W. Pauli, \emph{Theory of Relativity} (Pergamon Press, 1958)
\bibitem{mvlaue} M. von Laue, M\"unchester Sitzungsberichte, 405 (1911)
\bibitem{miller}
E.W. Morley and D. Miller, Proc. Am. Acad. Arts \& Sci. \textbf{41}, 321 (1905); Science \textbf{21}(531), 339 (1905); Science \textbf{25}, 525 (1907);\\
D. Miller, Phys. Rev. \textbf{19}, 407 (1922); Proc. Nat. Acad. Sci. \textbf{11}, 306 (1925); Phys. Rev. \textbf{27}(6), 812 (1926); Science \textbf{63}, 433 (1926); Astrophys. J. \textbf{68}, 341 (1928); J. Roy. Ast. Soc. Canada \textbf{24}, 82 (1930); Science \textbf{77}(2007), 587 (1933); Phys. Rev. \textbf{43}, 1054 (1933); Rev. Mod. Phys. \textbf{5}(2), 203 (1933); Nature \textbf{133}, 162 (1934);\\
G. Joos and D. Miller, Phys. Rev. \textbf{45}, 114 (1934)
\bibitem{drift}
    L. Silberstein, Science Suppl. - Science News \textbf{62}(1596), 8 (1925);\\
    C.T. Chase, A repetition of the Trouton-Noble ether drift experiment. Master's thesis, California Instutute of Technology (1926);\\
    H. Mineur, J. Roy. Ast. Soc. Canada \textbf{21}, 206 (1927);\\
    A.A. Michelson, H.G. Gale, and F. Pearson, Nature \textbf{115}, 566 (1925);\\
    A.A. Michelson, F.G. Pease, and F. Pearson, Nature \textbf{123}, 88 (1929); J. Optical Soc. Am. \textbf{18}, 181 (1929); Astrophys. J. \textbf{82}, 26 (1935)\\
    \emph{Miller Challenges Einstein: Explains Ether Drift Research and Function of Interferometer}, The Case Alumns, 10 (December 1929);\\
    W.F.G. Swann, Phys. Rev. \textbf{35}, 336 (1930);\\
    W.L. Laurence, \emph{New Evidence held to Upset Einstein: Formula Based on Overlooked Optics Law Offered Here by Prof. Cartmel. Existence of Ether Seen. Difference in Speed of Light Observed From Earth Urged as Relativity Challenge.}, in New York Times, p. 1 (February 23, 1936);\\
    H. Fletcher, Nat. Acad. Sci. \textbf{23}, 60 (1943);\\
    W. Reich, \emph{Cosmic Superimposition} (Orgone Institute Press, 1951), republished as \emph{Ether, God, and Devil: Cosmic Superimposition} (Farrar, Strauss \& Giroux, 1973);\\
    G. Szekeres, Phys. Rev. \textbf{104}, 1791 (1956);\\
    L. Swenson, \emph{The Ethereal Aether: A History of the Michelson-Morley-Miller Aether-Drift Experiments} (U. Texas Press, 1972);\\
    J. DeMeo, \emph{Premilinary Analysis of Changes in Kansas Weather Coincidental to Experimental Operations with a Reich Cloudbuster}, Thesis, University of Kansas, Lawrence, geography-meteorology Dept., 1979; chapter in \emph{The Orgone Accumulator Handbook} (Natural Energy, 1989); Pulse of the Planet \textbf{1}(2), 3 (1989); in \emph{Proceedings of 72nd Annual Meeting of American Association for the Advancement of Science. Northern Arizona Univ., Flagstaff, Arizona, 2-6 June 1996} (SW \& Rocky Mountain Division, 1996), pp. 41-42; Pulse of the Planet \textbf{5}, 138 (2002); Pulse of the Planet \textbf{5}, 114 (2002); in C. Whitney (Ed.) \emph{Proceedings of the Natural Philosophy Alliance} \textbf{1}(1) (Spring 2004), pp. 15-20; James De Meo's Research Website: Orgone Biophysical Research Lab/Saharasia \href{http://www.orgonlab.org/index.htm}{http://www.orgonlab.org/index.htm};\\
    M. Allais, \emph{L'Anisotropie de l'espace: La n\'{e}cessaire r\'{e}vision de certains postulats des th\'{e}ories contemporaines} (Cl\'{e}ment Juglar, 1997); in \emph{21st Century Science and Technology}, (Spring, 1998), pp. 26-34; Pulse of the Planet \textbf{5}, 132 (2002); Comptes Rendus de L'Acad\'{e}mie des Sciences \textbf{327}(IIb), p. 1405, 1411 (1999); Comptes Rendus de L'Acad\'{e}mie des Sciences \textbf{1}(IV), 1205 (2000); Pulse of the Planet \textbf{5}, 132 (2002);\\
    H. Munera, Apeiron \textbf{5} (1-2), 37 (1998); Ann. de la Fond. Louis de Broglie \textbf{27}(3), 463 (2002);\\
    Yu.M. Galaev, Radiophysics and Electronics \textbf{5}(1), 119 (2000); Space-time and Substance \textbf{2}(5(10)), 211 (2001); Space-time and Substance \textbf{3}(5(15)), 207 (2002)
    R.P. Crease, \emph{Finding the f\/law in falsif\/iability}, Physics World (December 2002);\\
    C. L\"{a}mmerzahl, H. Dittus, A. Peters, and S. Schiller, Class. Quant. Grav. \textbf{18}, 2499 (2002);\\
    M. Consoli and E. Costanzo arXiv:astro-ph/0311576; Nuovo Cim. B \textbf{119}, 393 (2004) arXiv:gr-qc/0406065; arXiv:gr-qc/0511160; arXiv:gr-qc/0604009; arXiv:0710.5613[gr-qc]\\
    M. Consoli and L. Pappalardo, arXiv:0912.0103[gr-qc];\\
    C.M.L. de Aragao, M. Consoli, and A. Grillo, arXiv:gr-qc/0507048; arXiv:0509066;\\
    T.J. Roberts, arXiv:physics/0608238 [physics.gen-ph];\\
    R.T. Cahill, Apeiron \textbf{11}(1), 53 (2004); Prog. Phys., 60 (July 2006); Prog. Phys., 73 (October 2006); Prog. Phys., 63 (October 2007); arXiv:0804.0039 [physics.gen-ph];\\
    R.T. Cahill and F. Stokes Prog. Phys., 103 (April 2008);\\
    J. Leach, A.J. Wright, J.B. G\"otte, J.M. Girkin, L. Allen, S. Franke-Arnold, S.M. Barnett, and M.J. Padgett, Phys. Rev. Lett. \textbf{100}, 153902 (2008)
\bibitem{shankland} R.S. Shankland, S.W. McCuskey, F.C. Leone, and G. Kuerti, Rev. Mod. Phys. \textbf{27}, 167 (1955)
\bibitem{muller} R.A. Muller, Scientif\/ic American \textbf{238}, 64 (1978)
\bibitem{kostro} L. Kostro, \emph{Einstein and the Ether} (Apeiron, 2000)
\bibitem{korper1} A. Einstein, Ann. Phys. \textbf{17}, 891 (1905)
\bibitem{sidel}
    A. Einstein, Phys. Z. \textbf{10}, 817 (1909); Naturwissenschaften \textbf{6}(48), 697 (1918);\\
    A. Einstein, \emph{Ether and the Theory of Relativity}, in A. Einstein \emph{Sidelights on Relativity} (Methuen, 1922), pp. 3-24;\\
    A. Einstein, \emph{\"Uber den \"Ather}, Verhandl. der Schweizerischen Naturforsch. Gesellsch., pp. 85-93 (1924)
\bibitem{eininf} A. Einstein and L. Infeld, \emph{The Evolution of Physics. The Growth of Ideas from Early Concepts to Relativity and Quanta} (Simon and Schuster, 1938)
\bibitem{ecartan}
    E. Cartan, Annales scientif\/iques de l'Ecole. Normale Sup\`{e}rieure \textbf{40}, 325 (1923);\\
    E. Cartan, Annales scientif\/iques de l'Ecole. Normale Sup\`{e}rieure \textbf{41}, 1 (1924)
\bibitem{traut}
    A.M. Trautman, \emph{Lectures on General Relativity (Brandeis Suminer Institute in Theoretical Physics)} (Prentice Hall, 1964) ;\\
    A.M. Trautman, in B. Hoffmann (Ed.) \emph{Perspectives in Geometry and Relativity: Essays in Honor of V\'{a}clav Hlavat\'{y}} (Indiana University Press, 1966), pp. 413–425
\bibitem{mtw} C.W. Misner, K.S. Thorne, and J.A. Wheeler, \emph{Gravitation} (W.H. Freeman, 1973)
\bibitem{davies} P.C.W. Davies and J.R. Brown (Eds.), \emph{The Ghost in the Atom: A Discussion of the Mysteries of Quantum Physics} (Cambridge University Press, 1993)
\bibitem{bell} J.S. Bell, \emph{Speakable and Unspeakable in Quantum Mechanics: Collected Papers on Quantum Philosophy} (Cambridge University Press, 1987)
\bibitem{feynman} R.P. Feynman, R.B. Leighton, and M. Sands, \emph{The Feynman Lectures on Physics Vol. II Mainly electromagnetism and matter}, Sec. 12-7 (Addison-Wesley, 1965)
\bibitem{whefeyn} J.A. Wheeler and R.P. Feynman, Rev. Mod. Phys. \textbf{11}(3), 425 (1949)
\bibitem{dirae}
    P.A.M. Dirac, Nature \textbf{168}, 906 (1951); Proc. Roy. Soc. (London) \textbf{A 209}, 291 (1951); Sci. Month. \textbf{78}, 142 (1954)
\bibitem{indirin} L. Infeld, Nature \textbf{169}, 702 (Jan. 31, 1952);\\
    P.A.M. Dirac, Nature \textbf{169}, 702 (Feb. 16, 1952)
\bibitem{planck} M. Planck, Verhandl. der Deutsehen Physikal. Gesellsch. \textbf{2}, 237 (1900)
\bibitem{stern} A. Einstein and O. Stern, Ann. d. Phys. \textbf{14}, 489 (1913)
\bibitem{geoms} J.A. Wheeler, \emph{Geometrodynamics} (Academic Press, 1962)
\bibitem{aspden} H. Aspden, Phys. Lett. \textbf{85A}, 411 (1981)
\bibitem{puthof} H.E. Puthoff, Phys. Rev. \textbf{A40}, 4857 (1989)
\bibitem{sidbooks} B.G. Sidharth, \emph{When the Universe Took a U-turn} (World Scientif\/ic, 2010);\\
    B.G. Sidharth, \emph{The Thermodynamic Universe. Exploring the Limits of Physics} (World Scientif\/ic, 2008);\\
    B.G. Sidharth, \emph{The Universe of Fluctuations. The Architecure of Space-time and the Universe} (Springer, 2005)
\bibitem{aemod} M. Gomes, J.R. Nascimento, A.Yu. Petrov, and A.J. Da Silva, Phys. Rev. \textbf{D81}, 045018 (2010);\\
    W. Donnelly and T. Jacobson, Phys. Rev. \textbf{D82}, 064032 (2010);\\
    S.M. Carroll, T.R. Dulaney, M.I. Gresham, and H. Tam, Phys. Rev. \textbf{D79}, 065011 (2009); Phys. Rev. \textbf{D79}, 065012 (2009);\\
    S.M. Carroll and H. Tam, Phys. Rev. \textbf{D78}, 044047 (2008);\\
    B. Withers, Class. Quant. Grav. \textbf{26}, 225009 (2009;,\\
    X. Kuang and Y. Ling, JCAP 0910, 024 (2009);\\
    A. Chatrabhuti, P. Patcharamaneepakorn, and P. Wongjun, JHEP 0908, 019 (2009);\\
    C. Armendariz-Picon and A. Diez-Tejedor, JCAP 0912, 018 (2009);\\
    E.V. Linder and R.J. Scherrer, Phys. Rev. \textbf{D80}, 023008 (2009);\\
    D.-C. Dai, R. Matsuo, and G. Starkman, Phys. Rev. \textbf{D78}, 104004 (2008);\\
    C. Bonvin, R. Durrer, P.G. Ferreira, G. Starkman and T.G. Zlosnik, Phys. Rev. \textbf{D77}, 024037 (2008);\\
    Y. Xie and T.-Y. Huang, Phys. Rev. \textbf{D77}, 124049 (2008);\\
    T. Tamaki and U. Miyamoto, Phys. Rev. \textbf{D77}, 024026 (2008);\\
    T. Jacobson, PoSQG-Ph, 020 (2007);\\
    C. Eling, T. Jacobson, and M.C. Miller, Phys. Rev. \textbf{D76}, 042003 (2007);\\
    D. Garf\/inkle, C. Eling, and T. Jacobson, Phys. Rev. \textbf{D76}, 024003 (2007);\\
    R.A. Konoplya and A. Zhidenko, Phys. Lett. \textbf{B644}, 186 (2007); Phys. Lett. \textbf{B648}, 236 (2007);\\
    T.G. Zlosnik, P.G. Ferreira, and G.D. Starkman, Phys. Rev. \textbf{D75}, 044017 (2007);\\
    C. Eling, Phys. Rev. \textbf{D73}, 084026 (2006); Phys. Rev. \textbf{D76}, 084033 (2007);\\
    C. Eling and T. Jacobson, Phys. Rev. \textbf{D74}, 084027 (2006); Class. Quant. Grav. \textbf{23}, 5625-5642 (2006); Class. Quant. Grav. \textbf{23}, 5643 (2006);\\
    B.Z. Foster, Phys. Rev. \textbf{D73}, 024005 (2006); Phys. Rev. \textbf{D73}, 104012 (2006);\\
    B.Z. Foster and T. Jacobson, Phys. Rev. \textbf{D73}, 064015 (2006);\\
    M. Levin and X.-G. Wen, Phys. Rev. \textbf{D73}, 035122 (2006)
\bibitem{epicurus}
                Epicurus \& Epicurean Philosophy, \href{http://www.epicurus.net}{http://www.epicurus.net};\\
                R. Geer, \emph{Epicurus: Letters, Principal Doctrines, and Vatican Sayings} (Prentice Hall, 1997);\\
                E.M. O'Connor, \emph{The Essential Epicurus: Letters, Principal Doctrines, Vatican Sayings, and Fragments} (Prometheus Books, 1993);\\
                H. Usener, \emph{Epicurea} (Irvington, 1987);\\
                N.W. DeWitt, \emph{Epicurus and His Philosophy} (2nd ed., University of Minnesota Press, 1964);\\
                R.D. Hicks, \emph{Stoic and Epicurean} (Charles Scribner's Sons, 1925)
\bibitem{quran} Abdullah Yusuf Ali, \emph{The Qur'an: Text, Translation \& Commentary (English and Arabic Edition)} (Tahrike Tarsile Qur'an, 1987);\\
    Abdullah Yusuf Ali, \emph{The Meaning of the Holy Qur'an} (10th ed., Amana Publications, 2002)
\bibitem{setia} Adi Setia, \emph{Fakhr al-Din al-Razi on physics and the nature of the physical world: a preliminary survey}, Islam \& Science \textbf{2} (Winter, 2004) \href{http://f\/indarticles.com/p/articles/mi\_m0QYQ/is\_2\_2/ai\_n9532826/}{http://f\/indarticles.com/p/articles/mi\_m0QYQ/is\_2\_2/ai\_n9532826/}
\bibitem{james} W. James, \emph{The Varieties of Religious Experience: A Study in Human Nature}. Being the Clifford Lectures on Natural Religion delivered at Edinburgh in 1901-1902 (Longmans, Gree \& Co., 1909);\\
    W. James, \emph{A Pluralistic Universe}. Hibbert Lectures 1908 (University of Nebraska Press, 1909);\\
    \emph{William James: Writings 1902-1910} (Library of America, 1987)
\bibitem{wittgen} L. Wittgenstein, \emph{Tractatus Logico-Philosophicus} (Kegan Paul, Trench and Tr\"ubner, 1922)
\bibitem{frege} P. Geach and M. Black (Eds.) \emph{Translations from the Philosophical Writings of Gottlob Frege} (3rd ed., Blackwell, 1980)
\bibitem{suszko} R. Suszko, Notre Dame J. Formal Logic \textbf{9}, 7 (1968); Analele Universitatii Bucuresti, Acta Logica \textbf{11}, 105 (1968); Studia Logica \textbf{27}, 7 (1971); in R. Parikh (Ed.) \emph{Logic Colloquium: symposium on logic held at Boston, 1972-73} (Lect. Notes Math. \textbf{453}, Springer, 1975), pp. 169-239
\bibitem{omyla} M. Omy\l a, J. Symbolic Logic \textbf{55}, 422 (1990); Language and Ontology, 195 (1982); Studies in Logic, Grammar, and Rhetoric \textbf{10}(23), 21 (2007)
\bibitem{dlewis} D. Lewis, \emph{Counterfactuals} (Rev. printing, Blackwell, 1986);\\
    D. Lewis, \emph{On the Plurality of Worlds} (Blackwell, 1986);\\
    D. Lewis, \emph{Parts of Classes} (Blackwell, 1991)
\bibitem{kripke} S. Kripke, \emph{Naming and Necessity} (Harvard University Press, 1980)
\bibitem{tarski} A. Tarski, Phil. Phenomen. Research \textbf{4}, 341 (1944)
\bibitem{wright} C. Wright, \emph{Truth and Objectivity} (Harvard University Press, 1992)
\bibitem{lynch} M. Lynch, \emph{Truth as One and Many} (Oxford University Press, 2009);\\
M. Lynch, \emph{True to Life: Why Truth Matters} (MIT Press, 2004);\\
M. Lynch, \emph{Truth in Context: An Essay on Pluralism and Objectivity} (MIT Press, 1998)
\bibitem{goodman}
H.N. Goodman, \emph{Languages of Art: An Approach to a Theory of Symbols} (2nd ed., Hackett, 1976);\\
H.N. Goodman, \emph{The Structure of Appearance} (3rd ed., reidel, 1977);\\
H.N. Goodman, \emph{Ways of Worldmaking} (Hackett, 1978);\\
H.N. Goodman, \emph{Fact, Fiction, and Forecast} (4th ed., Harvard University Press, 1983);\\
H.N. Goodman, \emph{Of Mind and Other Matters} (Harvard University Press, 1984)
\bibitem{everett} H. Everett III, \emph{The Theory of the Universal Wavefunction}, PhD thesis, Princeton University (1957);\\
    H. Everett III, Rev. Mod. Phys. \textbf{29}, 454 (1957)
\bibitem{mwid}
    B.S. DeWitt, in C.M. DeWitt and J.A. Wheeler (Eds.), \emph{Battelle Rencontres 1967. Lectures in Mathematics and Physics} (W.A. Benjamin, 1968), pp. 318-332;\\
    B.S. DeWitt, Phys. Today \textbf{23}(9), 30 (1970);\\
    B.S. DeWitt, in B. D'Espagnat (Ed.) \emph{Proceddings of the International School of Physics ''Enrico Fermi'', Course IL: Foundations of Quantum Mechanics} (Academic Press, 1971), pp. 211-262
\bibitem{mwi}
    B.S. DeWitt and N. Graham (Eds.), \emph{The Many-Worlds Interpretation of Quantum Mechanics: A Fundamental Exposition by Hugh Everett, III, with Papers by J.A. Wheeler, B.S. DeWitt, L.N. Cooper and D. Van Vechten, and N. Graham} (Princeton University Press, 1973);\\
    J.A. Wheeler and W.H. Zurek (Eds.), \emph{Quantum Theory and Measurement} (Princeton University Press, 1983)
\bibitem{gleason} A.M. Gleason, J. Math. Mech. \textbf{6}, 885 (1957)
\bibitem{hartle} J.B. Hartle, Am. J. Phys. \textbf{36}(8), 704 (1968)
\bibitem{harthawk} J.B. Hartle and S.W. Hawking, Phys. Rev. D \textbf{28}(12), 2960 (1983)
\bibitem{alin}
    A.D. Linde, Phys. Lett. B \textbf{129}, 177 (1983); Mod. Phys. Lett. A \textbf{1}, 81 (1986); Phys. Lett. B \textbf{175}, 395 (1986);\\
    A.D. Linde, \emph{Inf\/lation and Quantum Cosmology} (Academic Press, 1990);\\
    A.D. Linde, \emph{Particle Physics and Inf\/lationary Cosmology} (Harwood Academic Publishers, 1990);\\
    A.D. Linde, Scientif\/ic American, 98 (March 1998)
\bibitem{bartip}
J.D. Barrow and F.J. Tipler, \emph{The Anthropic Cosmological Principle} (Oxford University Press, 1985)
\bibitem{tipp}
    F.J. Tipler, in R. Penrose and C.J. Isham (Eds.), \emph{Quantum Concepts in Space and Time} (Clarendon, 1986), pp. 204-214;\\
    F.J. Tipler, \emph{The Physics of Immortality. Modern Cosmology, God and the Resurrection of the Dead} (Anchor Books, 1994)
\bibitem{gribbin} J.R. Gribbin, \emph{In Search of Schr\"odinger's Cat: Quantum Physics and Reality} (Bantam, 1984);\\
    J.R. Gribbin, \emph{In Search of the Multiverse: Parallel Worlds, Hidden Dimensions, and the Ultimate Quest for the Frontiers of Reality} (John Wiley \& Sons, 2010)
\bibitem{lockwood} M. Lockwood, \emph{Mind, Brain, \& the Quantum} (Basil Blackwell, 1989)
\bibitem{gellhart} M. Gell-Mann and J.B. Hartle, in W.H. Zurek (Ed.), \emph{Complexity, Entropy and the Physics of Information} (Addison-Wesley, 1990), pp. 425-459
\bibitem{albert} D. Albert, \emph{Quantum Mechanics and Experience} (Harvard University Press, 1992)
\bibitem{penrose} R. Penrose, \emph{Shadows of the Mind} (Oxford University Press, 1994)
\bibitem{chalmers} D.J. Chalmers, \emph{The Conscious Mind} (Oxford University Press, 1996)
\bibitem{deutsch} D. Deutsch, \emph{The Fabric of Reality: The Science of Parallel Universes and Its Implications} (Penguin Press, 1997)
\bibitem{kaku}
    M. Kaku, \emph{Hyperspace: A Scientif\/ic Odyssey Through Parallel Universes, Time Warps, and the Tenth Dimension} (Oxford University Press, 1994);\\
    M. Kaku, \emph{Parallel Worlds: The Science of Alternative Universes and Our Future in the Cosmos} (Allen Lane, 2005)
\bibitem{plaga} R. Plaga, Found. Phys. \textbf{27}, 559 (1997)
\bibitem{barrett} J.A. Barrett, \emph{The Quantum Mechanics of Minds and Worlds} (Oxford University Press, 1999)
\bibitem{deutsch1} D. Deutsch, Int. J. Theor. Phys. \textbf{24}(1), 1 (1985); Proc. Roy. Soc. London A \textbf{440}, 97 (1985); Proc. Roy. Soc. London A \textbf{445}, 3129 (1999); arXiv:quant-ph/0104033
\bibitem{dpage} D. Page, AIP Conf. Proc. \textbf{493}, 225 (1999), arXiv:gr-qc/0001001
\bibitem{polley} L. Polley, arXiv:quant-ph/9906124; arXiv:quant-ph/0102113
\bibitem{zurek} W.H. Zurek, Phys. Rev. A \textbf{71}, 052105 (2005), arXiv:quant-ph/0405161
\bibitem{zurek1} W.H. Zurek, Nature Physics \textbf{5}, 181 (2009), arXiv:0903.5082 [quant-ph]; Phys. Rev. Lett. \textbf{90}, 120404 (2003), arXiv:quant-ph/0211037
\bibitem{wallace} D. Wallace, in S. Saunders, J. Barrett, A. Kent, and D. Wallace (Eds.), \emph{Many Worlds? Everett, Quantum Theory, and Reality} (Oxford University Press, 2010), pp. 227-262; arXiv:quant-ph/0312157; Stud. Hist. Philos. Mod. Phys. \textbf{34}, 415 (2003); arXiv:quant-ph/0211104
\bibitem{saunders} S. Saunders, Proc. Roy. Soc. London A \textbf{460}, 1771 (2004); arXiv:quant-ph/0412194
\bibitem{whebio} J.A. Wheeler and K.W. Ford, \emph{Geons, Black Holes, and Quantum Foam: A Life in Physics} (W.W. Norton, 1998)
\bibitem{smolin1}
    L. Smolin, \emph{The Life of the Cosmos} (Oxford University Press, 1999);\\
    L. Smolin, \emph{Three Roads to Quantum Gravity} (Basic Books, 2001)
\bibitem{greene}
    B. Greene, \emph{The Elegant Universe: Superstrings, Hidden Dimensions, and the Quest for the Ultimate Theory} (Vintage Books, 2000);\\
    B. Greene, \emph{The Fabric of the Cosmos: Space, Time, and the Texture of Reality} (Alfred A. Knopf, 2004)
\bibitem{gardner} M. Gardner, \emph{Are Universes Thicker Than Blackberries?: Discourses on G\"odel, Magic Hexagrams, Little Red Riding Hood, and Other Mathematical and Pseudoscientif\/ic Topics} (W.W. Norton, 2004)
\bibitem{bruce} C. Bruce, \emph{Schr\"odinger's Rabbits. The Many Worlds of Quantum} (Joseph Henry Press, 2004)
\bibitem{randall} L. Randall, \emph{Warped Passages: Unraveling the Mysteries of the Universe's Hidden Dimensions} (Harper Perennial, 2006)
\bibitem{carr} B. Carr (Ed.), \emph{Universe or Multiverse?} (Cambridge University Press 2007)
\bibitem{mtegm} M. Tegmark, Nature \textbf{448}, 23 (5 July 2007); in J.D. Barrow, P.C.W. Davies, and C.L. Harper (Eds.), \emph{Science and Ultimate Reality: Quantum Theory, Cosmology, and Complexity} (Cambridge University Press, 2004), pp. 459-491; Scientif\/ic American, 41 (May 2003); Fortsch. Phys. \textbf{46}, 855 (1998); Ann. Phys. \textbf{270}, 1 (1998)
\bibitem{byrne}
    P. Byrne, Scientif\/ic American, 98 (December 2007);\\
    P. Byrne, \emph{The Many Worlds of Hugh Everett III: Multiple Universes, Mutual Assured Destruction, and the Meltdown of a Nuclear Family} (Oxford University Press, 2010)
\bibitem{allori} V. Allori, S. Goldstein, R. Tumulka, and N. Zangh\`{i}, \emph{Many Worlds and Schr\"odinger First Quantum Theory}, arXiv:0903.2211[quant-ph], Br. J. Philos. Sci. \textbf{62}(1), 1 (2011), first published online June 30, 2010
\bibitem{osnaghi} S. Osnaghi, F. Freitas, and O. Freire Jr., Stud. Hist. Philos. Mod. Phys. \textbf{40}, 97 (2009)
\bibitem{jenper} A. Jenkins and G. Perez, Scientif\/ic American, 42 (January 2010)
\bibitem{fentro} J. Feng and M. Trodden, Scientif\/ic American, 38 (November 2010)
\bibitem{glin}
    L.A. Glinka, Prespacetime Journal \textbf{1}(9), 1395 (November 2010) viXra:1011.0007;\\
    L.A. Glinka, \emph{Natural Emergence Scheme: (Very) Early Universe as Static Multiverse of Superf\/luid Fermi-Bose Superstrings}, presentation within GEOSET, \href{http://geoset.fsu.edu/L1.swf}{http://geoset.fsu.edu/L1.swf}
\bibitem{kent} A. Kent, Int. J. Mod. Phys. A \textbf{5}, 1745 (1990); in S. Saunders, J. Barrett, A. Kent, and D. Wallace (Eds.), \emph{Many Worlds? Everett, Quantum Theory, and Reality} (Oxford University Press, 2010), pp. 307-354
\bibitem{landsman} N.P. Landsman, in F. Weinert, K. Hentschel, D. Greenberger and B. Falkenburg (Eds.) \emph{Compendium of Quantum Physics} (Springer, 2008), pp. 6-9
\bibitem{parfit} D. Parf\/it, \emph{Reasons and Persons} (Oxford University Press, 1986)
\bibitem{weinanth} S. Weinberg, Phys. Rev. Lett. \textbf{59}, 2607 (1987)
\bibitem{thooft} G. 't Hooft, arXiv:gr-qc/9310026
\bibitem{hawki} S.W. Hawking, \emph{Black Holes and Baby Universes and Other Essays} (Bantam, 1994)
\bibitem{rees}
    M.J. Rees, \emph{Before the Beginning: Our Universe and Others} (Addison-Wesley, 1997);\\
    M.J. Rees, \emph{Just six numbers: the deep forces that shape tyhe universe} (Basic Books, 2001)
\bibitem{beken} J.D. Bekenstein, Scientif\/ic American, 59 (August 2003)
\bibitem{susskind}
    L. Susskind, arXiv:hep-th/0302219;\\
    L. Susskind, \emph{The Cosmic Landscape: String Theory and the Illusion of Intelligent Design} (Little, Brown, 2005)
\bibitem{finalth} S. Weinberg, \emph{Dreams of a Final Theory: The Scientist's Search for the Ultimate Laws of Nature} (Vintage, 1994)
\bibitem{smolin} L. Smolin, \emph{The Trouble With Physics: The Rise of String Theory, the Fall of a Science, and What Comes Next} (Houghton Miff\/in Harcourt, 2006)
\bibitem{woit} P. Woit, \emph{Not Even Wrong: The Failure of String Theory \& the Continuing Challange to to Unify the Laws of Physics} (Basic Books, 2006)

\bibitem{laglinka1} L.A. Glinka, Apeiron \textbf{16}, 147 (2009)
\bibitem{paulidirac} W. Pauli, Ann. Inst. H. Poincar\'{e} \textbf{6}, 109 (1936)
\bibitem{standardth} J.D. Bjorken and S.D. Drell, \emph{Relativistic Quantum Mechanics} (McGraw-Hill, 1964);\\
    W. Greiner, \emph{Relativistic Quantum Mechanics. Wave Equations} (3rd ed., Springer, 2000);\\
    M. Peskin and D. Schroeder, \emph{An Introduction to Quantum Field Theory} (Westview Press, 1995);\\
    M. Kaku, \emph{Quantum Field Theory. A Modern Introduction} (Oxford University Press, 1993)
\bibitem{diraceqn} P.A.M. Dirac, Proc. Roy. Soc. A \textbf{117}, 610 (1928); Proc. Roy. Soc. A \textbf{126}, 360 (1929)

  \bibitem{snyder} H. S. Snyder, Phys. Rev. \textbf{71}, 38 (1947); \emph{Phys. Rev.} \textbf{72}, 68 (1947)
  \bibitem{nspac} M.V. Battisti and S. Meljanac, Phys. Rev. \textbf{D82}, 024028 (2010),\\
      M.V. Battisti and S. Meljanac, Phys. Rev. \textbf{D79}, 067505 (2009),\\
      S. Meljanac, D. Meljanac, A. Samsarov, and M. Stojic, Mod. Phys. Lett. \textbf{A25}, 579 (2010),\\
      S. Meljanac, D. Meljanac, A. Samsarov, and M. Stojic, arXiv:0909.1706[math-ph]
  \bibitem{lorvia} D.M. Mattingly, L. Maccione, M. Galaverni, S. Liberati, and G. Sigl, JCAP \textbf{1002}, 007 (2010),\\
      L. Maccione, A.M. Taylor, D.M. Mattingly and S. Liberati, JCAP \textbf{0904}, 022 (2009),\\
      L. Maccione, A.M. Taylor, D.M. Mattingly and S. Liberati, Int. J. Mod. Phys. \textbf{D18}, 1621 (2009),\\
      B.G. Sidharth, Found. Phys. \textbf{38}, 89 (2008)
  \bibitem{greiner1} W. Greiner, \emph{Relativistic Quantum Mechanics} (3rd ed., Springer, 2000)
  \bibitem{peskin} M.E. Peskin and D.V. Schroeder, \emph{An Introduction to Quantum Field Theory} (Addison-Wesley, 1995)
  \bibitem{kiefer} C. Kiefer, \emph{Quantum Gravity} (2nd ed., Oxford University Press, 2007)
  \bibitem{lee} T.D. Lee, \emph{Particle Physics and Introduction to Field Theory} (Harwood Academic Press, 1981); Phys. Lett. \textbf{122B}, 217 (1983)
  \bibitem{astpar}
L. Bergstr\"om and A. Goobar, \emph{Cosmology and Particle Astrophysics} (2nd. ed., Springer, 2006),\\
D.H. Perkins, \emph{Particle Astrophysics} (Oxford University Press, 2003),\\
H.V. Klapdor-Kleingrothaus and K. Zuber, \emph{Particle Astrophysics} (Rev. ed., IOP Publishing, 2000)
\bibitem{ahc} A.H. Compton, Phys. Rev. \textbf{21(5)}, 483 (1923)
\bibitem{tipler}
R. Ferraro, \emph{Einstein's Space-Time. An Introduction o Special and General Relativity} (Springer, 2007),\\
W. Rindler, \emph{Relativity. Special, General, and Cosmological} (2nd ed., Oxford University Press, 2006),\\
P.M. Schwarz and J.H. Schwarz, \emph{Special Relativity. From Einstein to Strings} (Cambridge University Press, 2004),\\
S. Gasiorowicz, \emph{Quantum Physics} (3rd ed., John Wiley \& Sons, 2003),\\
P.A. Tipler, \emph{Modern Physics} (10th ed., Worth Publishers, 1992)

  \bibitem{ng}
A. Connes, \emph{Noncommutative Geometry} (Academic Press, 1994);\\
G. Dito, Lett. Math. Phys. \textbf{48}, 307 (1999);\\
N. Seiberg and E. Witten, JHEP \textbf{09}, 032 (1999);\\
D.J. Gross, A. Hashimoto, and N. Itzhaki, Adv. Theor. Math. Phys. \textbf{4}, 893 (2000);\\
A.S. Cattaneo and G. Felder, Prog. Theor. Phys. Suppl. \textbf{144}, 38 (2001);\\
N. Bel Baraka, Int. J. Theor. Phys. \textbf{41}(4), 737 (2002);\\
G. Dito and D. Sternheimer, Lect. Math. Theor. Phys. \textbf{1}, 9 (2002);\\
M.R. Douglas and N.A. Nekrasov, Rev. Mod. Phys. \textbf{73}, 977 (2002);\\
G. Fiore, M. Maceda, and J. Madore, J. Math. Phys. \textbf{43}, 6307 (2002);\\
R.J. Szabo, Phys. Rep. \textbf{378}, 207 (2003);\\
M. Chaichian, K. Nishijima and A. Tureanu, Phys. Lett. B \textbf{568}, 146 (2003);\\
L. Alvarez-Gaume and M. A. Vazquez-Mozo, Nucl. Phys. B \textbf{668}, 293 (2003);\\
A. Berard and H. Mohrbach, Phys. Rev. D \textbf{69}, 127701 (2004);\\
A. Das and J. Frenkel, Phys. Rev. D \textbf{69}, 065017 (2004);\\
M. Chaichian, M. N. Mnatsakanova, K. Nishijima, A. Tureanu, and Yu. A. Vernov, arXiv:hep-th/0402212;\\
D.H.T. Franco and C.M.M. Polito, J. Math. Phys. \textbf{46}, 083503 (2005);\\
M. Chaichian, P.P. Kulish, K. Nshijima, and A. Tureanu, Phys. Lett. B \textbf{604}, 98 (2004);\\
C.D. Fosco and G. Torroba, Phys. Rev. D \textbf{71}, 065012 (2005);\\
C.-S. Chu, K. Furuta, and T. Inami, Int. J. Mod. Phys. A \textbf{21}, 67 (2006);\\
B. Schroer, Ann. Phys. \textbf{319}, 92 (2005) arXiv:hep-th/0405105; Ann. Phys. \textbf{321}, 435 (2006);\\
O.W. Greenberg, Phys. Rev. D \textbf{73}, 045014 (2006);\\
M.A. Soloviev, Theor. Math. Phys. \textbf{147}, 660 (2006); Theor. Math. Phys. \textbf{153}, 1351 (2007);\\
E. Harikumar and V.O. Rivelles, Class. Quant. Grav. \textbf{23}, 7551 (2006);\\
G. Fiore and J. Wess, Phys. Rev. D \textbf{75}, 105022 (2007);\\
M. Chaichian, M. N. Mnatsakanova, A. Tureanu, and Yu. S. Vernov, JHEP \textbf{0809}, 125 (2008);\\
M.V. Battisti and S. Meljanac, Phys. Rev. D \textbf{79}, 067505 (2009); Phys. Rev. D \textbf{82}, 024028 (2010);\\
M. Daszkiewicz, Mod. Phys. Lett. A \textbf{24}, 1325 (2009); Acta Phys. Polon. B \textbf{41}, 1881 (2010); Acta Phys. Polon. B \textbf{41}, 1889 (2010); Mod. Phys. Lett. A \textbf{25}, 1059 (2010);\\
A.H. Chamseddine and A. Connes, Forts. Phys. \textbf{58}, 553 (2010);\\
D. Kolodrubetz and M. Marcolli, Phys. Lett. B \textbf{693}, 166 (2010);\\
E. Brown and R.B. Mann, Phys. Lett. B \textbf{694}(4-5), 440 (2010)
\bibitem{kontsevich}  M. Kontsevich, in G. Dito and D. Sternheimer (Eds.), \emph{Proceedings of Euroconference Moshe Flato (CMF 2000), Dijon, France, 11-13 Sep 2000} (Lett. Math. Phys. \textbf{66}(3), 2003), pp. 157-216
\bibitem{pamd}  P.A.M. Dirac, \emph{The Principles of Quantum Mechanics} (Clarendon Press, 1958)
\bibitem{markov}  M.A. Markov, Prog. Theor. Phys. Suppl. \textbf{E65}, 85 (1965); Sov. Phys. JETP \textbf{24}, 584 (1967)
\bibitem{kadyshevsky}
  V.G. Kadyshevsky, Sov. Phys. JETP \textbf{14}, 1340(1962); Nucl. Phys. B \textbf{141}, 477 (1978); in W. Beiglb\"ock, A. B\"ohm, and E. Takasugi (Eds.) \emph{Group Theoretical Methods in Physics: Seventh International Colloquium and Integrative Conference on Group Theory and Mathematical Physics, Held in Austin, Texas, September 11–16, 1978} (Lect. Notes Phys. \textbf{94}, Springer, 1978), pp. 114-124; PEPAN \textbf{11}(1), 5 (1980);\\
  V.G. Kadyshevsky and M. D. Mateev, Phys. Lett. B \textbf{106}, 139 (1981); Nuovo Cim. A \textbf{87}(3), 324 (1985);\\
  M.V. Chizhov, A.D. Donkov, V.G. Kadyshevsky, and M.D. Mateev, Nuovo Cim. A \textbf{87}(3), 350 (1985); Nuovo Cim. A \textbf{87}(4), 373 (1985);\\
  V.G. Kadyshevsky, Phys. Part. Nucl. \textbf{29}(3), 227 (1998);\\
  V.G. Kadyshevsky, M.D. Mateev, V.N. Rodionov, and A.S. Sorin, Dokl. Phys. \textbf{51}(6), 287 (2006); CERN-TH/2007-150, arXiv:0708.4205 [hep-ph]
\bibitem{radionov} V.N. Rodionov, arXiv:0903.4420 [hep-ph]
\bibitem{book} A.E. Chubykalo, V.V. Dvoeglazov, D.J. Ernst, V.G. Kadyshevsky, and Y.S. Kim (Eds.), \emph{Lorentz Group, CPT and Neutrinos: Proceedings of the International Workshop, Zacatecas, Mexico, 23-26 June 1999} (World Scientif\/ic, 2000)
\bibitem{sidharth}     B.G. Sidharth, \emph{The Thermodynamic Universe} (World Scientif\/ic, 2008)
\bibitem{bgs2008foop1} B.G. Sidharth, Found. Phys. \textbf{38}, 89 (2008); Found. Phys. \textbf{38}, 695 (2008)
\bibitem{bgs2005ijmp1} B.G. Sidharth, Int. J. Mod. Phys. E \textbf{14}, 1 (2005); Int. J. Mod. Phys. E \textbf{14}, 927 (2005)
\bibitem{antimatter}
   M. Kaku and J.T. Thompson, \emph{Beyond Einstein: The Cosmic Quest for the Theory of the Universe} (Oxford University Press, 1997);\\
   G. Fraser, \emph{Antimatter: The Ultimate Mirror} (Cambridge University Press, 2000)
\bibitem{laglinka}
   L.A. Glinka, Apeiron \textbf{17}(4), 223 (2010); Apeiron \textbf{17}(4), 243 (2010)
\bibitem{maccione}
  L. Maccione, S. Liberati, and G. Sigl, DESY 10039, arXiv:1003.5468 [astro-ph.HE];\\
  D.M. Mattingly, L. Maccione, M. Galaverni, S. Liberati, and G. Sigl, JCAP \textbf{1002}, 007 (2010);\\
  S. Liberati and L. Maccione, Ann. Rev. Nucl. Part. Sci. \textbf{59}, 245 (2009);\\
  L. Maccione, A.M. Taylor, D.M. Mattingly, and S. Liberati, JCAP \textbf{0904}, 022 (2009);
  L. Maccione, S. Liberati, A. Celotti, J.G. Kirk, and P. Ubertini, Phys. Rev. D \textbf{78}, 103003 (2008)
\bibitem{sidpc} B.G. Sidharth, \emph{Private communication}, March-May 2009
\bibitem{bgs2005ijmp2} B.G. Sidharth, Int. J. Mod. Phys. E \textbf{14}, 927 (2005); arXiv:0811.4541 [physics.gen-ph]; arXiv:0902.3342 [physics.gen-ph]
\bibitem{planckscale}
  G. Amelino-Camelia and J Kowalski-Glikman (Eds.), \emph{Planck Scale Effects in Astrophysics and Cosmology} (Lect. Notes Phys. \textbf{669}, Springer, 2005);\\
  C. Callender and N. Huggett (Eds.), \emph{Physics Meets Philosophy at the Planck Scale: Contemporary Theories in Quantum Gravity} (Cambridge University Press, 2004);\\
  R.L. Amoroso, G. Hunter, N. Kafatos, and J.-P. Vigier (Eds.), \emph{Gravitation and Cosmology: From the Hubble Radius to the Planck Scale. Proceedings of a Symposium in Honour of the $80^{th}$ Birthday of Jean-Pierre Vigier} (Kluwer Academic Publishers, 2002)
\bibitem{gft}
  I.J.R. Aitchinson, \emph{An Informal Introduction to Gauge Field Theories} (Cambridge University Press, 1984);\\
  F. Halzen and A.D. Martin, \emph{Quarks and Leptons: An Introductory Course in Modern Particle Physics} (John Wiley \& Sons, 1984);\\
  P.D.B. Collins, A.D. Martin, and E.J. Squires, \emph{Particle Physics and Cosmology} (John Wiley \& Sons, 1989);\\
  G. Kane, \emph{Modern Elementary Particle Physics: The Fundamental Particles and Forces?} (Addison-Wesley, 1993);\\
  J.F. Donoghue, E. Golowich, and B.R. Holstein, \emph{Dynamics of The Standard Model} (Cambridge University Press, 1996);\\
  S. Weinberg, \emph{The Quantum Theory of Fields. Vol. II Modern Applications}, (Cambridge University Press, 1996)
  B.R. Martin and G. Shaw, \emph{Particle Physics} (2nd ed., John Wiley \& Sons, 1997);\\
  P.H. Frampton, \emph{Gauge Field Theories} (2nd ed., John Wiley \& Sons, 2000);\\
  S. Pokorski, \emph{Gauge Field Theories} (2nd ed., Cambridge Univeristy Press, 2000);\\
  I.J.R. Aitchinson and A.J.G. Hey, \emph{Gauge Theories in Particle Physics. Vols. I \& II} (3rd ed., IOP Publishing, 2003-2004);\\
  M. Guidry, \emph{Gauge Field Theories: An Introduction with Applications} (Wiley-VCH, 2004)
\bibitem{greiner} W. Greiner, S. Schramm, and E. Stein, \emph{Quantum Chromodynamics} (3rd ed., Springer, 2007)
\bibitem{uhep} G. 't Hooft, Phys. Lett. B \textbf{198}, 61 (1987)
\bibitem{uheap}
  C. Giunti and C.W. Kim, \emph{Fundamentals of Neutrino Physics and Astrophysics} (Oxford University Press, 2007);\\
  M. Lemoine and G. Sigl (Eds.), \emph{Physics and Astrophysics of Ultra-High-Energy Cosmic Rays} (Lect. Notes Phys. \textbf{576}, Springer, 2001);\\
  F.W. Stecker and S.T. Scully, New J. Phys. \textbf{11}, 085003 (2009)


\bibitem{ein1} A. Einstein, {Phys. Z.} \textbf{15}, 176 (1914); {Sitzungsber. Preuss. Akad. Wiss. Berlin} \textbf{2}, 1030 (1914); {Sitzungsber. Preuss. Akad. Wiss. Berlin} \textbf{44}, 778 (1915); {Sitzungsber. Preuss. Akad. Wiss. Berlin} \textbf{46}, 799 (1915); {Sitzungsber. Preuss. Akad. Wiss. Berlin} \textbf{47}, 831 (1915); {Sitzungsber. Preuss. Akad. Wiss. Berlin} \textbf{48}, 844 (1915); {Ann. d. Physik} \textbf{49}, 769 (1916); {Sitzungsber. Preuss. Akad. Wiss. Berlin} \textbf{2}, 1111 (1916); {Sitzungsber. Preuss. Akad. Wiss. Berlin} \textbf{1}, 142 (1917); {Ann. d. Physik} \textbf{55}, 241 (1918); {Sitzungsber. Preuss. Akad. Wiss. Berlin} \textbf{1}, 349 (1919); {Math. Ann.} \textbf{7}, 99 (1926)
\bibitem{rie} B. Riemann, {G\"ott. Abh. Ges. Wiss.} \textbf{13}, 1 (1867)
\bibitem{car} E. Cartan, \emph{Le\c{c}ons sur la G\'{e}om\'{e}trie des Espaces de Riemann} (Gauthier-Villars, 1928)
\bibitem{hil} D. Hilbert, {K\"onigl. Gesell. d. Wiss. G\"ottingen, Nachr., Math.-Phys. Kl. \textbf{27}}, 395 (1915); {K\"onigl. Gesell. d. Wiss. G\"ottingen, Nachr., Math.-Phys. Kl. \textbf{61}}, 53 (1917)
\bibitem{pal} A. Palatini, {Rend. Circ. Mat. Palermo \textbf{43}}, 203 (1919)
\bibitem{frw}
A.A. Friedmann, {Zeits. f. Physik} \textbf{10}, 377 (1922); {Zeits. f. Physik} \textbf{21}, 326 (1924);\\
G. Lema\^{\i}tre, {J. Math. Phys.} \textbf{4}, 188 (1925); {Phys. Rev.} \textbf{25}, 903 (1925); {Bull. Astron. Inst. Netherlands} \textbf{5}, N200, 273 (1930); {Monthly Not. Roy. Astron. Soc.} \textbf{91}, 490 (1931); Nature \textbf{127}, No. 3210, 706 (1931); {Ann. Soc. Sci. Brux.} \textbf{A47}, 49 (1927); {ibid.} \textbf{A53}, 51 (1933);\\
H.P. Robertson, {Phil. Mag.} \textbf{5}, 835 (1928); {Proc. Nat. Acad. Sci.} \textbf{15}, 822 (1929); {Science} \textbf{76}, 221 (1932); {Rev. Mod. Phys.} \textbf{5}, 62 (1933); {Astrophys. J.} \textbf{82}, 248 (1935); {Astrophys. J.} \textbf{83}, 187 (1936); {Astrophys. J.} \textbf{83}, 257 (1936);\\
A.G. Walker, {Proc. Lon. Math. Soc. 2} \textbf{42}, 90, (1937)
\bibitem{ein2} A. Einstein, \emph{The Meaning of Relativity} (6th ed., Routledge, 2003)
\bibitem{zel} A.L. Zelmanov, {Dokl. Acad. Nauk USSR} \textbf{227(1)}, 78 (1976)
\bibitem{orbifold} A. Adem, J. Leida, and Y. Ruan, \emph{Orbifolds and Stringy Topology} (Cambridge University Press, 2007)
\bibitem{dir} P.A.M. Dirac, \emph{Lectures on Quantum Mechanics} (Belfer Graduate School of Science, Yeshiva University Press, 1964); {Phys. Rev.} \textbf{114}, 924 (1959); Phys. Rev. Lett. \textbf{2}, 368 (1959); {Proc. Roy. Soc. (London) A} \textbf{246}, 326 (1958); {Proc. Roy. Soc. (London) A} \textbf{246}, 333 (1958); {Can. J. Math.} \textbf{2}, 129 (1950)
\bibitem{adm} R. Arnowitt, S. Deser and C.W. Misner, in L. Witten (Ed.) \emph{Gravitation: an introduction to current research} (John Wiley \& Sons, 1961), pp. 227-265
\bibitem{poisson}
S. Chandrasekhar, \emph{The Mathematical Theory of Black Holes} (Clarendon Press, 1983);\\
E. Poisson, \emph{A Relativist's Toolkit. The Mathematics of Black-Hole Mechanics}. (Cambridge University Press, 2004)
\bibitem{wein}
S. Weinberg, \emph{Cosmology} (Oxford University Press, 2008);\\
O. Gr{\o}n and S. Hervik, \emph{Einstein's General Relativity. With Modern Applications in Cosmology.} (Springer, 2007);\\
J. Pleba\'nski and A. Krasi\'nski, \emph{An Introduction to General Relativity and Cosmology} (Cambridge University Press, 2006);\\
P. Hoyng, \emph{Relativistic Astrophysics and Cosmology: A Primer} (Springer, 2006);\\
T.-P. Cheng, \emph{Relativity, Gravitation, and Cosmology: A Basic Introduction} (Oxford University Press, 2005);\\
S. Carroll, \emph{Space-time and Geometry. An Introduction to General Relativity} (Addison-Wesley, 2004);\\
J.B. Hartle, \emph{Gravity. An Introduction to Einstein's General Relativity} (Addison-Wesley, 2003);\\
A. Liddle, \emph{An Introduction to Modern Cosmology} (2nd, John Wiley \& Sons, 2003);\\
T. Padmanabhan, \emph{Theoretical Astrophysics Volume III: Galaxies and Cosmology} (Cambridge University Press, 2002);\\
B. Ryden, \emph{Introduction to Cosmology} (Addison-Wesley, 2002);\\
E.F. Taylor and J.A. Wheeler, \emph{Exploring Black Holes. Introduction to General Relativity} (Addison-Wesley, 2000);\\
R. d'Inverno, \emph{Introducing Einstein's Relativity} (Clarendon Press, 1998);\\
A. Krasi\'nski, \emph{Inhomogeneous Cosmological Models} (Cambridge University Press, 1997);\\
T. Padmanabhan, \emph{Cosmology and Astrophysics Through Problems} (Cambridge University Press, 1996);\\
M. Lachi\`{e}ze-Rey, \emph{Cosmology: A First Course} (Cambridge University Press, 1995);\\
L.D. Landau and E.M. Lifshitz, \emph{Course of Theoretical Physics, Vol 2. The Classical Theory of Fields} (4th English ed., Butterworth Heinemann, 1994);\\
P.J.E. Peebles, \emph{Principles of Physical Cosmology} (Princeton University Press, 1993);\\
E.W. Kolb and M.S. Turner, \emph{Early Universe} (Addison-Wesley, 1989);\\
M. Berry, \emph{Principles of Cosmology and Gravitation} (Cambridge University Press, 1976);\\
V. Fock, \emph{The Theory of Space, Time, and Gravitation} (2nd rev. ed., Pergamon Press, 1964)
\bibitem{str} D. L\"ust and S. Theisen, \emph{Lectures on String Theory} (Lect. Notes Phys. \textbf{346}, Springer, 1989)
\bibitem{whe} J.A. Wheeler, {Ann. Phys.} \textbf{2}, 604 (1957); \emph{Geometrodynamics} (Academic Press, 1962); in C. DeWitt and B. DeWitt (Eds.) \emph{Relativity, Groups, and Topology. Lectures Delivered at Les Houches During the 1963 Session of the Summer School of Theoretical Physics} (Gordon and Breach, 1964), pp. 317-501; in C.M. DeWitt and J.A. Wheeler (Eds.) \emph{Battelle Rencontres 1967 Lectures in Mathematics and Physics} (W.A. Benjamin, 1968), pp. 242-308; \emph{Einsteins Vision} (Springer, 1968); in R.P. Gilbert and R. Newton (Eds.), \emph{Analytic Methods in Mathematical Physics} (Gordon and Breach, 1970), pp. 335-378
\bibitem{dew} B.S. DeWitt, Phys. Rev. \textbf{160}, 1113 (1967)
\bibitem{bs} N.N. Bogoliubov and D.V. Shirkov, \emph{Introduction to the
	 Theory of Quantized Fields} (3rd English ed., John Wiley \& Sons, 1980)
\bibitem{ccr} J. von Neumann, Math. Ann. \textbf{104}, 570 (1931);\\
	 H. Araki and E.J. Woods, J. Math. Phys. \textbf{4}, 637 (1963)
\bibitem{bog} N.N. Bogoliubov, Sov. Phys. JETP \textbf{7}, 41 (1958)
\bibitem{ripka} J.-P. Blaizot and G. Ripka, \emph{Quantum theory of f\/inite systems} (Massachusetts Institute of Technology Press, 1986)
\bibitem{inflation} M. Lemoine, J. Martin, and P. Peter (Eds.), \emph{Inf\/lationary Cosmology} (Springer, 2008)
\bibitem{martin} J. Martin, in M. Lemoine, J. Martin, and P. Peter (Eds.), \emph{Inf\/lationary Cosmology} (Springer, 2008), pp. 193-242
\bibitem{mfb} V.F. Mukhanov, H.A. Feldman and R.H. Brandenberger, Phys. Rep. \textbf{215}, 203 (1992)
\bibitem{bardeen} J.A. Bardeen, Phys. Rev. D\textbf{22}, 1882 (1980)
\bibitem{martin1} J. Martin, Braz. J. Phys. \textbf{34}, 1307 (2004)
\bibitem{mukha} V. Mukhanov, \emph{Physical Foundations of Cosmology} (Cambridge University Press, 2005)
\bibitem{ising} G. Jug and B.N. Shalaev, J. Phys. A \textbf{32}, 7249 (1999)
\bibitem{suslov} I.M. Suslov, JETP Lett. \textbf{74}, 191 (2001)
\bibitem{solids} N.W. Ashcroft and N.D. Mermin, \emph{Solid State Physics} (Harcourt College Publishers, 1976);\\
    H. Haken, \emph{Quantum Field Theory of Solids: An Introduction} (Elsevier, 1983);\\
    G.D. Mahan, \emph{Many-Particle Physics} (2nd ed., Plenum Press, 1990);\\
    C. Kittel, \emph{Introduction to Sold State Physics} (8th ed., John  Wiley \& Sons, 2005)
\bibitem{qg01} J.R. Klauder (Ed.), \emph{Magic without magic: John Archibald Wheeler} (Freeman, 1972)
\bibitem{qg02} C.J. Isham, R. Penrose, and D.W. Sciama (Eds.), \emph{Quantum Gravity. An Oxford symposium} (Oxford University Press, 1975)
\bibitem{qg03} R. Balian and J. Zinn-Justin (Eds.), \emph{Methods in Field Theory. Les Houches, \'{E}cole D'\'{E}t\'{e} De Physique Th\'{e}orique. Session XXVIII} (North-Holland, 1976)
\bibitem{qg04} C.J. Isham, R. Penrose, and D.W. Sciama (Eds.), \emph{Quantum Gravity 2. A second Oxford symposium} (Oxford University Press, 1981)
\bibitem{qg05} S.M. Christensen (Ed.), \emph{Quantum Theory of Gravity. Essays in honor of the 60th birthday of Bryce S. DeWitt.} (Adam Hilger, 1984)
\bibitem{qg06} M.A. Markov and P.C. West (Eds.), \emph{Quantum Gravity. Proceedings of the Second Seminar, Oct 13-15, 1981, Moscow, USSR} (World Scientif\/ic, 1984)
\bibitem{qg07} M.A. Markov, V.A. Berezin, and V.P. Frolov (Eds.), \emph{Quantum Gravity. Proceedings of the Third Seminar, Oct 23-25, 1984, Moscow, USSR} (World Scientif\/ic, 1985)
\bibitem{qg08} R. Penrose and C.J. Isham (Eds.), \emph{Quantum Concepts in Space and Time} (Oxford University Press, 1986)
\bibitem{qg09} T. Padmanabhan and J.V. Narlikar, \emph{Gravity, Gauge Theories and Quantum Cosmology} (Kluwer Academic Publishers, 1986)
\bibitem{qg10} H.J. de Vega and N. S\'{a}nchez (Eds.), \emph{Field Theory, Quantum Gravity, and Strings. Proceedings of a Seminar Series Held at DAPHE, Observatoire de Meudon, and LPTHE, Universit\'{e} Pierre et Marie Curie, Paris, Between October 1984 and October 1985} (Lect. Notes Phys. \textbf{246}, Springer, 1986)
\bibitem{qg11} H.J. de Vega and N. S\'{a}nchez (Eds.), \emph{Field Theory, Quantum Gravity and Strings II Proceedings of a Seminar Series Held at DAPHE, Observatoire de Meudon, and LPTHE, Universit\'{e} Pierre et Marie Curie, Paris, Between October 1985 and October 1986} (Lect. Notes Phys. \textbf{280}, Springer, 1987)
\bibitem{qg12} M.A. Markov, V.A. Berezin, and V.P. Frolov (Eds.), \emph{Quantum Gravity. Proceedings of the Fourth Seminar, May 25-29, 1987, Moscow, USSR} (World Scientif\/ic, 1988)
\bibitem{qg13} J. Audretsch and V. de Sabbata (Eds.), \emph{Quantum Mechanics in Curved Space–Time} (Plenum Press, 1990)
\bibitem{qg14} A. Ashtekar and J. Stachel (Eds.), \emph{Conceptual Problems of Quantum Gravity} (Birkh\"auser, 1991)
\bibitem{qg15} S. Coleman, J.B. Hartle, T. Piran, and S. Weinberg (Eds.), \emph{Quantum Cosmology and Baby Universes} (World Scientif\/ic, 1991)
\bibitem{qg16} M.A. Markov, V.A. Berezin, and V.P. Frolov (Eds.), \emph{Quantum Gravity. Proceedings of the Fifth Seminar, May 28-June 1, 1990, Moscow, USSR} (World Scientif\/ic, 1991)
\bibitem{qg17} I.L. Buchbinder, S.D. Odintsov, and I.L. Shapiro, \emph{Effective Action in Quantum Gravity} (IOP Publishing, 1992)
\bibitem{qg18} D.J. Gross, T. Piran, and S. Weinberg (Eds.), \emph{Two Dimensional Quantum Gravity and Random Surfaces} (World Scientif\/ic, 1992)
\bibitem{qg19} M. Henneaux and C. Teitelboim, \emph{Quantization of Gauge Systems} (Princeton University Press, 1992)
\bibitem{qg20} M.C. Bento, O. Bertolami, J.M. Mour\~{a}o, and R.F. Picken (Eds.), \emph{Classical and Quantum Gravity} (World Scientif\/ic, 1993)
\bibitem{qg21} G.W. Gibbons and S.W. Hawking (Eds.), \emph{Euclidean Quantum Gravity} (World Scientif\/ic, 1993)
\bibitem{qg22} J.C. Baez (Ed.), \emph{Knots and Quantum Gravity} (Clarendon Press, 1994)
\bibitem{qg23} J. Ehlers and H. Friedrich (Eds.), \emph{Canonical Gravity: From Classical to Quantum} (Springer, 1994)
\bibitem{qg24} G. Esposito, \emph{Quantum Gravity, Quantum Cosmology and Lorentzian Geometries} (Springer, 1994)
\bibitem{qg25} E. Prugove\v{c}ki, \emph{Principles of Quantum General Relativity} (World Scientif\/ic, 1995)
\bibitem{qg26} P.D. D'Eath, \emph{Supersymmetric Quantum Cosmology} (Cambridge University Press, 1996)
\bibitem{qg27} R. Gambini and J. Pullin, \emph{Loops, Knots, Gauge Theories and Quantum Gravity} (Cambridge University Press, 1996)
\bibitem{qg28} P.G. Bergmann, V. De Sabbata, and H.J. Treder (Eds.), \emph{Quantum Gravity: International School of Cosmology and Gravitation XIV Course: 80th Birthday Dedication to Peter G. Bergmann, held 11-19 May, 1995 in Erice, Italy} (World Scientif\/ic, 1996)
\bibitem{qg29} G. Esposito, A.Yu. Kamenshchik, and G. Pollifrone, \emph{Euclidean Quantum Gravity on Manifolds with Boundary} (Springer, 1997)
\bibitem{qg30} G. 't Hooft, A. Jaffe, G. Mack, P.K. Mitter, R. Stora (Eds.) \emph{Quantum Fields and Quantum Space Time} (Plenum Press, 1997)
\bibitem{qg31} P. Fr\'{e}, V. Gorini, G. Magli, and U. Moschella, \emph{Classical and Quantum Black Holes} (IOP Publishing, 1999)
\bibitem{qg32} I.G. Avramidi, \emph{Heat Kernel and Quantum Gravity} (Springer, 2000)
\bibitem{qg33} J. Kowalski-Glikman (Ed.), \emph{Towards Quantum Gravity. Proceedings of the XXXV International Winter School on Theoretical Physics, Held in Polanica, Poland, 2-11 February 1999}. (Lect. Notes Phys. \textbf{541}, Springer, 2000)
\bibitem{qg34} B.N. Kursunoglu, S.L. Mintz, and A. Perlmutter (Eds.), \emph{Quantum Gravity, Generalized Theory of Gravitation and Superstring Theory-Based Unif\/ication} (Kluwer Academic Publishers, 2002)
\bibitem{qg35} S. Carlip, \emph{Quantum Gravity in 2+1 Dimensions} (Cambridge University Press, 2003)
\bibitem{qg36} G.W. Gibbons, E.P.S. Shellard, and S.J. Rankin (Eds.), \emph{The Future of Theoretical Physics and Cosmology} (Cambridge University Press, 2003)
\bibitem{qg37} D. Giulini, C. Kiefer, and C. L\"ammerzahl (Eds.), \emph{Quantum Gravity. From Theory To Experimental Search} (Lect. Notes Phys. 631, Springer, 2003)
\bibitem{qg38} C. Rovelli, \emph{Quantum Gravity} (Cambridge University Press, 2004)
\bibitem{qg39} A. Gomberoff and D. Marolf (Eds.), \emph{Lectures on Quantum Gravity} (Springer, 2005)
\bibitem{qg40} D. Rickles, S. French, and J. Saatsi (Eds.), \emph{The Structural Foundations of Quantum Gravity} (Clarendon Press, 2006)
\bibitem{qg41} B. Carr (Ed.), \emph{Universe of Multiverse?} (Cambridge University Press, 2007)
\bibitem{qg42} B. Fauser, J. Tolksdorf, and E. Zeidler (Eds.) \emph{Quantum Gravity. Mathematical Models and Experimental Bounds} (Birkh\"auser, 2007)
\bibitem{qg43} D. Gross, M. Henneaux, and A. Sevrin (Eds.), \emph{The Quantum Structure of Space and Time} (World Scientif\/ic, 2007)
\bibitem{qg44} V.F. Mukhanov and S. Winitzki, \emph{Introduction to Quantum Effects in Gravity} (Cambridge University Press, 2007)
\bibitem{qg45} T. Thiemann, \emph{Modern Canonical Quantum General Relativity} (Cambridge University Press, 2007)
\bibitem{qg45a} H.W. Hamber, \emph{Quantum Gravitation: The Feynman Path Integral Approach} (Springer, 2009)
\bibitem{qg46} D. Oriti, \emph{Approaches to Quantum Gravity. Toward a New Understanding of Space, Time, and Matter} (Cambridge University Press, 2009)
\bibitem{qg47} M. Bojowald, \emph{Canonical Gravity and Applications: Cosmology, Black Holes, and Quantum Gravity} (Cambridge University Press, 2010)
\bibitem{qg48} J. Murugan, A. Weltman, and G.F.R. Ellis, \emph{Foundations of Space and Time: Ref\/lections on Quantum Gravity} (Cambridge University Press, 2011)
\bibitem{gli}   L.A. Glinka, in B.G. Sidharth, F. Honsell, O. Mansutti, K. Sreenivasan, and A. De Angelis (Eds.), \emph{Frontiers of Fundamental and Computational Physics. 9$^{\mathrm{th}}$ International Symposium, Udine and Trieste, Italy 7-9 January 2008}, AIP Conf. Proc. \textbf{1018}, 94 (2008); Grav. Cosmol. \textbf{15}(4), 317 (2009); arXiv:0711.1380 [gr-qc]; arXiv:0712.1674 [gr-qc]; SIGMA 3, 087 (2007); in E. Ivanov and S. Fedoruk (Eds.), \emph{Supersymmetries and Quantum Symmetries (SQS'07): Proceedings of International Workshop, held in Dubna, Russia, July 30 - August 4, 2007} (Dubna JINR 2008), pp. 406-411

\bibitem{hawk00} S.W. Hawking, {Comm. Math. Phys.} \textbf{43}, 199 (1975); {Phys. Rev. D} \textbf{13}, 191 (1976); {Phys. Rev. D} \textbf{14}, 2460 (1976); {Comm. Math. Phys.} \textbf{55}, 133 (1977); {Phys. Rev. D} \textbf{18}, 1747 (1978); {Pont. Acad. Sci. Scri. Varia} \textbf{48}, 563 (1982); {Ph ys. Lett. B} \textbf{134}, 403 (1984); {Nucl. Phys. B} \textbf{239}, 257 (1984); {Phys. Rev. D} \textbf{32}, 259 (1985); {Phys. Rev. D} \textbf{32}, 2489 (1985) {Phys. Rev. D} \textbf{37}, 904 (1988); Phys. Rev. D \textbf{53}, 3099 (1996); {Phys. Scr. T} \textbf{117}, 49 (2005); Phys. Rev. D \textbf{72}, 084013 (2005); in S.W. Hawking and W. Israel (Eds.), \emph{General Relativity: An Einstein centenary survey} (Cambridge University Press 1979), pp. 746-785; in M. Levy and S. Deser (Eds.), \emph{Recent Developments in Gravitation. Cargese 1978} (Plenum Press 1979), pp. 145-175; in B.S. DeWitt and R. Stora (Eds.), \emph{Relativity, Groups, and Topology II} (Elsevier 1984), pp. 333-381; in H.J. de Vega and N. S\'{a}nchez (Eds.), \emph{Field Theory, Quantum Gravity, and Strings. Proceedings of a Seminar Series Held at DAPHE, Observatoire de Meudon, and LPTHE, Universit\'{e} Pierre et Marie Curie, Paris, Between October 1984 and October 1985} (Springer 1986), pp. 1-46; in S.W. Hawking and W. Israel (Eds.), \emph{Three hundred years of gravitation} (Cambridge University Press 1987), pp. 631-652; in J.J. Halliwell, J. Perez-Marcader, and W.H. Zurek (Eds.), \emph{Physical Origins of Time Asymmetry} (Cambridge University Press 1992), pp. 346-356;
    \emph{Hawking on the Big Bang and Black Holes} (World Scientif\/ic, 1993)
\bibitem{hawk01} G.W. Gibbons and S.W. Hawking, {Phys. Rev. D} \textbf{15}, 2752 (1977)
\bibitem{hawk02} J.B. Hartle and S.W. Hawking, {Phys. Rev. D} \textbf{28}, 2960 (1983)
\bibitem{hawk03} S.W. Hawking and J.C. Luttrell, {Phys. Lett. B} \textbf{143}, 83 (1984); {Nucl. Phys. B} \textbf{247}, 250 (1984)
\bibitem{hawk04} S.W. Hawking and Z.C. Wu, Phys. Lett. B \textbf{151}, 15 (1985)
\bibitem{hawk05} J.J. Halliwell and S.W. Hawking, {Phys. Rev. D} \textbf{31}, 1777 (1985)
\bibitem{hawk06} S.W. Hawking and D. Page,  {Nucl. Phys. B} \textbf{264}, 185 (1986); {Nucl. Phys. B} \textbf{298}, 789 (1988); Phys. Rev. D \textbf{42}, 2655 (1992)
\bibitem{hawk07} S.W. Hawking, R. Laf\/lamme, and G.W. Lyons, {Phys. Rev. D} \textbf{47}, 5342 (1993)
\bibitem{hawk08} R. Bousso and S.W. Hawking, Phys. Rev. D \textbf{52}, 5660 (1995)
\bibitem{hawk09} S.W. Hawking and S.F. Ross, Phys. Rev. D \textbf{52}, 5862 (1995)
\bibitem{hawk10} M.J. Cassidy and S.W. Hawking, Phys. Rev. D \textbf{57}, 2372 (1998)
\bibitem{hawk11} S.W. Hawking and C.J. Hunter, Phys. Rev. D \textbf{59}, 044025 (1999)
\bibitem{hawk12} S.W. Hawking, T. Hertog, and H.S. Reall, Phys. Rev. D \textbf{62}, 043501 (2000); Phys. Rev. D \textbf{63}, 083504 (2001)
\bibitem{hawk13} S.W. Hawking, T. Hertog, and N. Turok, Phys. Rev. D \textbf{62}, 063502 (2000)
\bibitem{hawk14} S.W. Hawking and T. Hertog, Phys. Rev. D \textbf{66}, 123509 (2002); {Phys. Rev. D} \textbf{73}, 123527 (2006)
\bibitem{hawk15} J.B. Hartle, S.W. Hawking, and T. Hertog, {Phys. Rev. Lett.} \textbf{100}, 201301 (2008); {Phys. Rev. D} \textbf{77}, 123537 (2008)
\bibitem{krie}  M. Kriele, \emph{Space-time. Foundations of General Relativity and Differential Geometry} (Lect. Notes Phys. Monogr. \emph{59}, Springer, 1999)
\bibitem{pet} P. Petersen, \emph{Riemannian Geometry} (2nd ed., Grad. Texts Math. \textbf{171}, Springer, 2006)
\bibitem{gau} K.F. Gauss, Gottingae: Typis Di eterichiansis, (1828)
\bibitem{cod} D. Codazzi, Ann. math. pura applicata 2, 101, (1868-1869)
\bibitem{han} A. Hanson, T. Regge, and C. Teitelboim, \emph{Constrained Hamiltonian Systems} (Contributi del Centro Linceo Interdisciplinare di Scienze Matematiche e loro Applicazioni, n. \textbf{22}, Accademia Nazionale dei Lincei, 1976)
\bibitem{qft}  B. DeWitt, \emph{The Global Approach to Quantum Field Theory, Vols. 1-2} (Int. Ser. Monogr. Phys. \textbf{114}, Clarendon Press, 2003)
\bibitem{nash} J.F. Nash, {Ann. Math.} \textbf{56}, 405 (1952); Ann. Math. \textbf{63}, 20 (1956)
\bibitem{kowa} A. Kowalczyk, {Bull. Acad. Polon. Sci. Ser. Sci. Math.} \textbf{28}, 385 (1981)
\bibitem{masa} S. Masahiro, \emph{Nash Manifolds} (Lect. Notes Math. \textbf{1269}, Springer 1987)
\bibitem{gunt} M. G\"unther, {Ann. Global Anal. Geom.} \textbf{7}, 69 (1989); {Math. Nachr.} \textbf{144}, 165 (1989)
\bibitem{york}  J.W. York, {Phys. Rev. Lett.} \textbf{28}, 1082 (1972)
\bibitem{sachs}
    S. Weinberg, \emph{Gravitation and Cosmology. Principles and Applications of the General Theory of Relativity} (John Wiley \& Sons, 1972);\\
    R.K. Sachs and H. Wu, \emph{General Relativity for Mathematicians} (Springer, 1977);\\
    B.F. Schutz, \emph{Geometrical Methods of Mathematical Physics} (Cambridge University Press, 1980);\\
    M. Carmeli, \emph{Classical Fields: General Relativity and Gauge Theory} (John Wiley \& Sons, 1982);\\
    B. O'Neill, \emph{Semi-Riemannian Geometry with Applications to Relativity} (Academic Press, 1983);\\
    R.M. Wald, \emph{General Relativity} (University of Chicago, 1984);\\
    N. Straumann, \emph{General Relativity and Relativistic Astrophysics} (Springer, 1984);\\
    F. De Felice and C.J.S. Clarke, \emph{Relativity on Curved Manifolds} (Cambridge University Press, 1990);\\
    L.P. Hughston and K.P. Tod, \emph{An Introduction to General Relativity} (Cambridge University Press, 1994);\\
    J. Stewart, \emph{Advanced General Relativity} (Cambridge University Press, 1996);\\
    H. Stephani, D. Kramer, M. A. H. MacCallum, C. Hoenselaers, and E. Herlt, \emph{Exact Solutions of Einstein’s Field Equations} (2nd ed., Cambridge University Press, 2003);\\
    G.S. Hall, \emph{Symmetries and Curvature Structure in General Relativity} (World Scientif\/ic, 2004);\\
    N.M.J. Woodhouse, \emph{General Relativity} (Springer, 2006);\\
    L. Ryder, \emph{Introduction to General Relativity} (Cambridge University Press, 2009)
\bibitem{gol} H. Goldstein, Ch. Poole, and J. Safko, \emph{Classical Mechanics} (3rd ed., Pearson, 2002)
\bibitem{peres} A. Peres, Nuovo Cim. \textbf{26}, 53 (1962)
\bibitem{fad}   L.D. Faddeev, {Usp. F\/iz. Nauk} \textbf{136}, 435 (1982)
\bibitem{dewi} B.S. DeWitt, in M. Carmeli, S.I. F\/ickler, and L. Witten (Eds.) \emph{Relativity. Proceedings of the Relativity Conference in the Midwest held at Cincinnati, Ohio, June 2-6, 1969} (Plenum Press, 1970), pp. 359-374
\bibitem{fish} A.E. Fischer, Ann. Global Anal. Geom. \textbf{14}, 263 (1996); Fields Inst. Comm. \textbf{7}, 107 (1996); in M. Gotay, J. Marsden, and V. Moncrief (Eds.), \emph{Mathematical Aspects of Classical Field Theory, proceedings of the AMS-IMS-SIAM Joint Summer Research Conference, Seattle, Washington, July 20-26, 1991}, (Cont. Math. \textbf{132}, 1992), pp. 331-366; {J. Math. Phys.} \textbf{27}, 718 (1986); {Gen. Rel. Grav.} \textbf{15}, 1191 (1983)
\bibitem{pek} O. Pekonen, J. Geom. Phys. \textbf{4} 493 (1987)
\bibitem{giul} D. Giulini, {Gen. Rel. Grav.} \textbf{41}, 785 (2009); in H. Falcke and F.W. Hehl, \emph{The Galactic Black Hole: Lectures on General Relativity and Astrophysics} (IOP Publishing, 2003), pp. 178-206; in H.-P. Breuer and F. Petruccione (Eds.), \emph{Relativistic quantum measurement and decoherence: Proceedings of Workshop On Relativistic Quantum Measurement And Decoherence, 9-10 Apr 1999, Naples, Italy} (Lect. Notes Phys. \textbf{559}, Springer, 2000), pp. 67-92; in F. Hehl, C. Kiefer, and R. Metzler (Eds.) \emph{Black Holes: Theory and Observation} (Lect. Notes Phys. \textbf{514}, Springer, 1998), pp. 224–243; Helv. Phys. Acta \textbf{69}, 333 (1996); Helv. Phys. Acta \textbf{68}, 438 (1995); Helv. Phys. Acta \textbf{68}, 87 (1995); Phys. Rev. D \textbf{5}, 5630 (1995); Comm. Math. Phys. \textbf{148}, 353 (1992)
\bibitem{fimon} A.E. Fischer and V.E. Moncrief, in S. Cotsakis and G.W. Gibbons (Eds.), \emph{Global Structure and Evolution in General Relativity: Proceedings of the First Samos Meeting on Cosmology, Geometry and Relativity, held at Karlovassi, Samos, Greece, September 5-7, 1994} (Lect. Notes Phys. \textbf{460}, Springer, 1996), pp. 111–173
\bibitem{steen} N. Steenrod, \emph{The Topology of Fibre Bundles} (Princeton University Press, 1951);\\
R. Sulanke and P. Wintgen, \emph{Differentialgeometrie und Faserb\"undel} (VEB Deutscher Verlag der Wissenschaften, 1972)
\bibitem{fisher} A.E. Fischer, in M. Carmeli, S.I. F\/ickler, and L. Witten (Eds.), \emph{Relativity. Proceedings of the Relativity Conference in the Midwest held at Cincinnati, Ohio, June 2-6, 1969} (Plenum Press, 1970), pp. 303-359
\bibitem{giulkief} D. Giulini and C. Kiefer, {Phys. Lett. A} \textbf{193}, 21 (1994)
\bibitem{kth} K.V. Kucha\v{r}, {J. Math. Phys.} \textbf{13}, 768 (1972);\\
C. Teitelboim, Lett. Nuovo Cim. \textbf{3}, 397 (1972);\\
S.A. Hojman, K. Kucha\v{r}, and C. Teitelboim, Nature (London), Phys. Sci. \textbf{97}, 245 (1973);\\
K.V. Kucha\v{r}, J. Math. Phys. \textbf{15}, 708 (1974)
\bibitem{ahaboh} Y. Aharonov and D. Bohm, Phys. Rev. \textbf{115}(3), 485 (1959)
\bibitem{barimi} F. Barbero, Phys. Rev. D \textbf{51}(10), 5507 (1995);\\
G. Immirzi, Class. Quant. Grav. \textbf{14}(10), L117 (1997)
\bibitem{witt} D. Witt, J. Math. Phys. \textbf{27}(2), 573 (1986)
\bibitem{frisor} J. Friedman and R. Sorkin, Phys. Rev. Lett. \textbf{44}, 1100 (1980)
\bibitem{isham} C.J. Isham, in J.J. Duff and C.J. Isham (Eds.), \emph{Quantum Structure of Space and Time. Proceedings of the Nuff\/ield Workshop, August 3–21 1981, Imperial College London} (Cambridge University Press, 1982), pp. 37–52
\bibitem{sorkin} R. Sorkin, in P.G. Bergmann and V. De Sabbata (Eds.), \emph{Topological
Properties and Global Structure of Space–Time} (NATO Advanced Study Institutes Series B\textbf{138}, Kluwer Academic Publishers, 1986), p. 249; in S. De Filippo, M. Marinaro, G. Marmo, and G. Vilasi (Eds.), \emph{Geometrical and Algebraic Aspects of Nonlinear Field Theory} (Elsevier, 1989), pp. 201–218
\bibitem{spin} F. Dowker and R. Sorkin, in R.C. Hilborn and G.M. Tino (Eds.), \emph{Spin-Statistics Connections and Commutation Relations: Experimental Tests and Theoretical Implications} (American Institute of Physics, 2000), pp. 205–218; Class. Quant. Grav. \textbf{15}, 1153 (1998);\\
    C. Aneziris, A.P. Balachandran, M. Bourdeau, S. Jo, T.R. Ramadas and R.D. Sorkin, Int. J. Mod. Phys. A \textbf{14}(20), 5459 (1989); Mod. Phys. Lett. A \textbf{4}(4), 331 (1989)
\bibitem{surgery}
    C.T.C. Wall, \emph{Surgery on Compact Manifolds} (2nd ed., edited and with a foreword by A.A. Ranicki, Mathematical Surveys and Monographs \textbf{69}, American Mathematical Society, 1999);\\
    J. Roe, \emph{Index Theory, Coarse Geometry, and Topology of Manifolds} (American Mathematical Society, 1996);\\
    A.A. Kosinski, \emph{Differential Manifolds} (Pure and Applied Mathematics \textbf{138}, Academic Press, 1993)
\bibitem{perel} G. Perelman, arXiv:math/0303109v1 [math.DG]
\bibitem{milnor} J.W. Milnor, Am. J. Math. \textbf{84}(1), 1 (1962)
\bibitem{fouxe} D.I. Fouxe-Rabinovitch, Rec. Math. \textbf{8}(50), 265 (1940); Rec. Math. \textbf{9}(51), 297 (1941)
\bibitem{mcmil} D. McCullough and A. Miller, Mem. Am. Math. Soc. \textbf{61}(344) (1986)
\bibitem{gilbert} N.D. Gilbert, Proc. Lond. Math. Soc. \textbf{54}, 115 (1987)
\bibitem{giul1} D. Giulini, in B. Fauser, J. Tolksdorf, and E. Zeidler (Eds.) \emph{Quantum Gravity. Mathematical Models and Experimental Bounds} (Birkh\"auser, 2007), pp. 161-201
\bibitem{lelong} J. Lelong-Ferrand, M\'{e}moires de la Classe Des Sciences de l’Acad\'{e}mie Royale Des Sciences, Des Lettres Et Des Beaux-Arts de Belgique \textbf{39}(5), 3 (1971)
\bibitem{geroch} P. Breitenlohner and D. Maison, Ann. Inst. Henri Poincar\'{e} A\textbf{46}(2), 215 (1987)
\bibitem{lens}
    P.J. Hilton and S. Wylie, \emph{Homology Theory: An Introduction to Algebraic Topology} (Cambridge University Press, 1965);\\
    S.-T. Hu, \emph{Homology Theory: A First Course in Algebraic Topology} (Holden-Day, 1966);\\
    J.W. Milnor, Bull. Amer. Math. Soc. \textbf{72}, 358 (1966);\\
    M.M. Cohen, \emph{A Course in Simple-Homotopy Theory} (Springer, 1973);\\
    H. Seifert and W. Threlfall, \emph{A Textbook of Topology} (Pure Appl. Math. \textbf{89}, Translated from the German edition of 1934, Academic Press, 1980);\\
    M. Audin, \emph{The Topology of Torus Action on Symplectic Manifolds} (Birkh\"auser, 1991);\\
    G.E. Bredon, \emph{Topology and Geometry} (Graduate Texts in Mathematics \textbf{139}, Springer, 1993);\\
    J. Stillwell, \emph{Classical Topology and Combinatorial Group Theory} (2nd ed., Springer, 1993);\\
    S.P. Novikov, \emph{Topology I: General Survey} (Springer, 1995);\\
    I.M. James, \emph{Handbook of Algebraic Topology} (North-Holland, 1995);\\
    W.P. Thurston, \emph{Three-Dimensional Geometry and Topology} (Princeton University Press, 1997);\\
    I.M. James, \emph{History of Topology} (North-Holland, 1999);\\
    A. Hatcher, \emph{Algebraic Topology} (Cambridge University Press, 2002);\\
    A.V. Bolsinov and A.T. Fomenko, \emph{Integrable Hamiltonian Systems: Geometry, Topology, Classif\/ication} (Chapman \& Hall/CRC, 2004)
\bibitem{tietze} H. Tietze, Monatsh. f\"ur Math. und Phys. \textbf{19}, 1 (1908)
\bibitem{alex} J.W. Alexander II, Trans. Amer. Math. Soc. \textbf{20}, 339 (1919)
\bibitem{seth} W. Threlfall and H. Seifert, Math. Ann. \textbf{104}, 1 (1930)
\bibitem{white} J.H.C. Whitehead, Ann. Math. \textbf{42} (5), 1197 (1941)
\bibitem{reid} K. Reidemeister, Abh. Math. Sem. Univ. Hamburg \textbf{11}, 102 (1935)
\bibitem{franz} W. Franz, J. Reine Angew. Math. \textbf{173}, 245 (1935)
\bibitem{piece} C.P. Rourke and B.J. Sanderson, \emph{Introduction to Piecewise-Linear Topology} (Springer, 1972)
\bibitem{brody} E.J. Brody, Ann. Math. \textbf{71}, 163 (1960)
\bibitem{przyas} J.H. Przytycki and A. Yasuhara, Geom. Ded. \textbf{98}(1) (2003)
\bibitem{sallon} P. Salvatore and R. Longoni, Riccardo, Topology \textbf{44}, 375 (2005)
\bibitem{rueff} M. Rueff, Compositio Math. \textbf{6}, 161 (1938)
\bibitem{seif} H. Seifert, Acta Math. \textbf{60}, 147 (1933);\\
P. Orlik, \emph{Seifert manifolds} (Lect. Notes Math. \textbf{291}, Springer 1972)
\bibitem{moise} E. Moise, Ann. Math. \textbf{56}, 96 (1952);\\
E.E. Moise, \emph{Geometric Topology in Dimensions 2 and 3} (Springer, 1977)
\bibitem{haup}
Yu.B. Rudyak, arXiv:math/0105047v1 [math.AT];\\
A. Ranicki (Ed.), \emph{The Hauptvermutung Book} (K-Monographs in
Mathematics \textbf{1}, Kluwer Academic Publishers, 1996);\\
D.P. Sullivan, Bull. Amer. Math. Soc. \textbf{73}, 59 (1967);\\
J.W. Milnor, Ann. Math. \textbf{74}(2), 575 (1961)
R.H. Bing, Ann. Math. \textbf{69}, 37 (1959)
\bibitem{kuch} K.V. Kucha\v{r}, in B. Hu, M. Ryan, and C. Vishveshvara (Eds.), \emph{Directions in General Relativity} (Cambridge University Press, 1993), pp. 201-221; in G. Kunstatter, D. Vincent, and J. Williams (Eds.), \emph{Proceedings of the 4th Canadian Conference on General Relativity and Relativistic Astrophysics} (World Scientif\/ic, 1992), pp. 211-314
\bibitem{tomo} S. Tomonaga, Prog. Theor. Phys. \textbf{1}, 27 (1946); Prog. Theor. Phys. \textbf{2}, 101 (1946)
\bibitem{schw} J. Schwinger, Phys. Rev. \textbf{74}, 1439 (1948)
\bibitem{brku} J.D. Brown and K.V. Kucha\v{r}, Phys. Rev. D \textbf{51}, 5600 (1995)
\bibitem{unru} W.G. Unruh, Phys. Rev. D \textbf{40}, 104852 (1989)
\bibitem{qgr0a} D.A. Craig and P. Singh, Phys. Rev. D \textbf{82}, 12352 (2010); AIP Conf. Proc. \textbf{1232}, 275 (2010)
\bibitem{qgr0}  A. Kleinschmidt and H. Nicolai, Int. J. Mod. Phys. D \textbf{19}(14), 2305 (2010)
\bibitem{qgr0b}  L.A. Glinka, Grav. Cosmol. \textbf{16}(1) 7, (2010); {Concepts Phys.} \textbf{6}, 19 (2009); {New Adv. Phys.} \textbf{2}, 1 (2008); arXiv:0804.3516 [gr-qc]
\bibitem{ham41d} C. Kiefer, {Gen. Rel. Grav.} \textbf{41}, 877 (2009); in J. Kowalski-Glikman (Ed.) \emph{Towards Quantum Gravity: Proceedings of the XXXV International Winter School on Theoretical Physics, Held in Polanica, Poland, 2-11 February 1999} ({Lect. Notes Phys.} \textbf{541}, Springer 2000), pp. 158-187; {in} J. Ehlers and H. Friedrich (Eds.) \emph{Canonical Gravity: From Classical to Quantum. Proceedings of the 117th WE Heraeus Seminar Held at Bad Honnef, Germany, 13-17 September 1993} (Springer, 1994), pp. 170–212; Ann. d. Phys. \textbf{15}, 129 (2006)
\bibitem{ham41e} H. Yoshino and M. Shibata, Phys. Rev. D \textbf{80}, 084025 (2009)
\bibitem{brewin} L. Brewin, Phys. Rev. D \textbf{80}, 084030 (2009); Gen. Rel. Grav. \textbf{39}, 521 (2007); Class. Quant. Grav. \textbf{15}, 2427 (1998)
\bibitem{ham41c} T. Damour and H. Nicolai, {Int. J. Mod. Phys. D} \textbf{17}, 525 (2008)
\bibitem{qgr1}  P. Gusin, {Phys. Rev. D} \textbf{77}, 066017 (2008)
\bibitem{qgr1a} J.F. Barbero G., AIP Conf. Proc. \textbf{1023}, 3 (2008)
\bibitem{qgr1d} S. Zonetti and G. Montani, Int. J. Mod. Phys. A \textbf{23}, 1240 (2008)
\bibitem{qgr1e} F. Cianfrani and G. Montani, Int. J. Mod. Phys. A \textbf{23}, 1149 (2008)
\bibitem{qgr1f} E. Castellanos, A. Camacho, and J.I. Rivas, AIP Conf. Proc. \textbf{977}, 202 ( 2008)
\bibitem{qgr1g} J.F. Barbero G. and E.J.S. Villasenor, Phys. Rev. D \textbf{77}, 121502 (2008)
\bibitem{qgr1h} S. Carlip, Class. Quant. Grav. \textbf{25}, 154010 (2008)
\bibitem{qgr1i} M.S. El Naschie, Chaos Solitons Fractals \textbf{36}, 808 (2008)
\bibitem{qgr1j} G. Montani and F. Cianfrani, Class. Quant. Grav. \textbf{25}, 065007 (2008)
\bibitem{qgr1k} R. Benini and G. Montani, Int. J. Mod. Phys. A \textbf{23}, 1244 (2008)
\bibitem{qgr1l} R. Garattini, J. Phys. A \textbf{41}, 164057 (2008); Nucl. Phys. Proc. Suppl. \textbf{57}, 316 (1997)
\bibitem{a-v}   I.Ya. Aref'eva and I. Volovich, Int. J. Geom. Meth. Mod. Phys. \textbf{05}, 641 (2008)
\bibitem{qgr3b} W. Nelson and M. Sakellariadou, Phys. Lett. B \textbf{661}, 37 (2008)
\bibitem{ham27} J.A. Isenberg, Int. J. Mod. Phys. D \textbf{17}, 265 (2008); in N.T. Bishop and S.D. Maharaj (Eds.), \emph{General Relativity and Gravitation: Proceedings of the 16th International Conference}, (World Scientif\/ic, 2002); {in} J.P. Harnard and S. Shnider (Eds.), \emph{Geometrical and Topological Methods in Gauge Theories: Proceedings of the Canadian Mathematical Society Summer Research Institute, McGill University, Montreal, September 3-8, 1979}. (Annals of Physics \textbf{129}, Springer, 1980), pp. 223-248
\bibitem{ham41b} D. Giulini and C. Kiefer, in I.-O. Stamatescu and E. Seiler (Eds.), \emph{Approaches to Fundamental Physics: An Assessment of Current Theoretical Ideas} ({Lect. Notes Phys.} \textbf{721}, Springer, 2007), pp. 131-150; {Class. Quant. Grav.} \textbf{12}, 403 (1995)
\bibitem{ham41c} C. L\"ammerzahl, in B. Fauser, J. Tolksdorf, and E. Zeidler (Eds.), \emph{Quantum Gravity: Mathematical Models and Experimental Bounds} (Birkha\"user, 2007), pp. 15–39
\bibitem{ham41a} A.Y. Kamenshchik, C. Kiefer, and B. Sandh\"ofer, {Phys. Rev. D} \textbf{76}, 064032 (2007)
\bibitem{qgr3}  Ch. Soo, {Class. Quant. Grav.} \textbf{24}, 1547 (2007)
\bibitem{qgr3a} M. Thibeault and C. Simeone, Int. J. Mod. Phys. D \textbf{16}, 1303 (2007)
\bibitem{qgr3b} R. Carroll, {Theor. Math. Phys.} \textbf{152}, 904 (2007)
\bibitem{qgr3d} L. Hardy, J. Phys. A \textbf{40}, 3081 (2007)
\bibitem{ham39b} D. Oriti, in B. Fauser, J. Tolksdorf, and E. Zeidler (Eds.) \emph{Quantum Gravity. Mathematical Models and Experimental Bounds} (Birkh\"auser, 2007), pp. 101-126; Braz. J. Phys. \textbf{35}(2B), 481 (2005)
\bibitem{qgr1d} I. Rodnianski, in M. Sanz-Sol\'{e}, J. Soria-de Diego, J.L. Verona-Malumbres, and J.M. Verdera-Melench\'{o}n (Eds.), \emph{Proceedings of the International Congress of Mathematicians: Madrid, Spain, August 22-30, 2006} Vol 3 (European Mathematical Society, 2007), pp. 421-442
\bibitem{qgr1c} Y. Ma, Front. Phys. China \textbf{1}, 125, (2006)
\bibitem{qgr4}  D. Rickles, {in} D. Rickles, S. French, and J. Saatsi (Eds.) \emph{The structural foundations of quantum gravity} (Clarendon Press, 2006), pp. 152-195; Stud. Hist. Phil. Mod. Phys. \textbf{36}, 691 (2005)
\bibitem{qgr4a} J. Mattingly, in J. Eisenstaedt and A. Kox (Eds.), \emph{The Universe of General Relativity, Vol. 11 of Einstein Studies} (Birkha\"user, 2006), pp. 327–338
\bibitem{qgr5}  A.B. Henriques, {Gen. Rel. Grav.} \textbf{38}, 1645 (2006)
\bibitem{qgr5b} D. Maxwell, J. Reine, Angew. Math. \textbf{590}, 1 (2006)
\bibitem{qgr5a} M.P. D\c abrowski, C. Kiefer, and B. Sandh\"ofer, Phys. Rev. D \textbf{74}, 044022 (2006)
\bibitem{ham27a} R. Bartnik and J. Isenberg, Class. Quant. Grav. \textbf{23}, 2559 (2006)
\bibitem{ham39a} C. Kiefer, J. M\"uller-Hill, and C. Vaz, Phys. Rev. D \textbf{73}, 044025 (2006)
\bibitem{ham39b} S. Klainerman and I. Rodnianski, Geom. Funct. Anal. \textbf{16}(1), 164 (2006); Invent. Math. \textbf{159}, 437 (2005)
\bibitem{ham39} N. Pinto-Neto, {Found. Phys.} \textbf{35}, 577 (2005)
\bibitem{ham40} M.J.W. Hall, {Gen. Rel. Grav.} \textbf{37}, 1505 (2005)
\bibitem{qgr5b} C. Kiefer, T. L\"uck, and P. Moniz, Phys. Rev. D \textbf{72}, 045006 (2005)
\bibitem{qgr6a} T.P. Shestakova and C. Simeone, Grav. Cosmol. \textbf{10}(3(39)), 161 (2004); Grav. Cosmol. \textbf{10}(4(40)), 257 (2004)
\bibitem{qgr5c} A.P. Gentle, N.D. George, W.A. Miller, and A. Kheyfets, Int. J. Mod. Phys. A \textbf{19}(10), 1609 (2004)
\bibitem{qgr5d} T-Y. Cao, Stud. Hist. Phil. Mod. Phys. \textbf{32}(2), 181 (2004)
\bibitem{qgr5e} Y. Choqu\'{e}t-Bruhat, Class. Quant. Grav. \textbf{21}, S127 (2004)
\bibitem{qgr6}  T. Kubota, T. Ueno, and N. Yokoi, {Phys. Lett. B} \textbf{579}, 200 (2004)
\bibitem{qgr7}  K. Meissner, {Class. Quant. Grav.} \textbf{21}, 5245 (2004)
\bibitem{ham35a} J. Butterf\/ield and C.J. Isham, in C. Callender and N. Huggett (Eds.), \emph{Physics Meets Philosophy at the Planck Scale: Contemporary Theories in Quantum Gravity}(Cambridge University Press, 2004), pp. 33-89; Found. Phys. \textbf{30}, 1707 (2000); in J. Butterf\/ield (Ed.), \emph{The Arguments of Time} (Oxford University Press, 1999), pp. 111-168
\bibitem{ham27d} L. Andersson and V. Moncrief, in P.T. Chru\'sciel and H. Friedrich (Eds.) \emph{The Einstein Equations and the Large Scale Behavior of Gravitational Fields: 50 Years of the Cauchy Problem in General Relativity} (Birkh\"auser, 2004), pp. 299–330; Ann. Inst. H. Poincar\'{e} \textbf{4}, 1 (2003)
\bibitem{ham35c} S. Weinstein, in C. Callender and N. Huggett (Eds.), \emph{Physics Meets Philosophy at the Planck Scale: Contemporary Theories in Quantum Gravity} (Cambridge University Press, 2004), pp. 90-100; in T. Piran (Ed.), \emph{Proceedings of 8th Marcel Grossmann Meeting on Recent Developments in Theoretical and Experimental General Relativity, Gravitation and Relativistic Field Theories (MG 8), Jerusalem, Israel, 22-27 Jun 1997.} (World Scientif\/ic, 1999), pp. 875-877
\bibitem{ham20} C.J. Isham, Adv. Theor. Math. Phys. \textbf{8}, 797 (2004); Adv. Theor. Math. Phys. \textbf{7}, 807 (2004); Adv. Theor. Math. Phys. \textbf{7}, 331 (2003); in G.W. Gibbons, E.P.S. Shellard, and S.J. Rankin (Eds.), \emph{The Future of Theoretical Physics and Cosmology: Celebrating Stephen Hawking's 60th Birthday} (Cambridge University Press, 2003), pp. 384-408; in M. Francaviglia, G. Longhi, L. Lusanna and E. Sorace (Eds.) \emph{Proceedings of the 14th International Conference on General Relativity and Gravitation} (World Scientif\/ic, 1997), pp. 167-209; Class. Quant. Grav. \textbf{13}, A5 (1996); {in} J. Ehlers and H. Friedrich (Eds.) \emph{Canonical Gravity: From Classical to Quantum. Proceedings of the 117th WE Heraeus Seminar Held at Bad Honnef, Germany, 13-17 September 1993} (Springer, 1994), pp. 1-21; in L.A. Ibort and M.A. Rodriguez (Eds.), \emph{Integrable Systems, Quantum Groups and Quantum Field Theories} (Kluwer Academic Publishers, 1993), pp. 157-288; in M.A. del Olmo, M. Santander, and J. Mateos-Guilarte (Eds.),\emph{Group Theoretical Methods in Physics: Proceedings of the XIX International Colloquium, held in Salamanca, Spain, 29 Jun - 5 Jul 1992} (1992), pp. 157-288; in H. Mitter and H. Gausterer (Eds.), \emph{Proceedings of 30th Internationale Universitatswochen fur Kernphysik: Recent Aspect of Quantum Fields} (Lect. Notes Phys. \textbf{396}, Springer, 1991), pp. 123-229; in A. Ashtekar and J.J. Stachel (Eds.), \emph{2nd Osgood Hill Conference: Festschrift Paul Dirac} (Birkh\"auser, 1991), pp. 351-400; in H.C. Lee (Ed.) \emph{Proceedings of 1989 Banff NATO ASI: Physics, Geometry and Topology} (Plenum Press, 1990), pp. 129-190; Nucl. Phys. B Proc. Suppl. \textbf{6}, 349, (1989); in M.A.H. MacCallum (Ed.), \emph{Proceedings of 11th International Conference on General Relativity and Gravitation, 6 - 12 Jul 1986, Stockholm} (Cambridge University Press, 1987), pp. 99-149; in A.T. Davies and D.G. Sutherland (Eds.), \emph{Proceedings of 28th Scottish Universities Summer School In Physics: Supersymmetry and Supergravity, 28 Jul-17 Aug 1985, Edinburgh, Scotland} (SUSSP Publications \textbf{28}, Edinburgh University Press, 1986), p. 0001; {in} S.M. Christensen (Ed.) \emph{Quantum Theory of Gravity: Essays in honor of the 60th birthday of Bryce S. De Witt} (Adam Hilger, 1984), pp. 299-314; in B.S. DeWitt and R. Stora (Eds.), \emph{Relativity, Groups, and Topology II} (Elsevier 1984), pp. 1059-1290; Proc. Roy. Soc. Lond. A \textbf{368}, 33 (1979); Proc. Roy. Soc. Lond. A \textbf{351}, 209 (1976); in C.J. Isham, R. Penrose, and D.W. Sciama (Eds.) \emph{Quantum Gravity: an Oxford Symposium} (Clarendon Press, 1975), pp. 1-77
\bibitem{qgr9c} S. Klainerman and F. Nicolo, \emph{The Evolution Problem in General Relativity} (Progr. Math. Phys. \textbf{25}, Birkh\"auser, 2003)
\bibitem{qgr9d} H. Lindblad and I. Rodnianski, C. R. Math. Acad. Sci. Paris \textbf{336}(11), 901 (2003)
\bibitem{ham27b} J. Isenberg, R. Mazzeo, and D. Pollack, Ann. Inst. H. Poincar\'{e} \textbf{4}, 369 (2003)
\bibitem{qgr9}  E. Anderson, J. Barbour, B. Foster, and N. 'O Murchadha, Class. Quant. Grav. \textbf{20}, 1571 (2003)
\bibitem{qgr9a} R.J. Gleiser, C.N. Kozameh, and F. Parisi, Class. Quant. Grav. \textbf{20}, 4375 (2003)
\bibitem{ham11a} D. Christodoulou, in V.G. Gurzadyan, R.T. Jantzen, and R. Ruff\/ini (Eds.), \emph{Proceedings of the Ninth Marcel Grossmann Meeting on General Relativity, University of Rome "La Sapienza", July 2-8, 2000} (World Scientif\/ic, 2003), pp. 44–54; Ann. Math. \textbf{149}, 183 (1999); Comm. Pure Appl. Math. \textbf{46}, 1131 (1993); Phys. Rev. Lett. \textbf{25}, 1596 (1970)
\bibitem{qgr9b} P. H\'{a}j\'{i}\v{c}ek, in D. Giulini, C. Kiefer, and C. L\"ammerzahl (Eds.), \emph{Quantum Gravity. From Theory To Experimental Search} (Lect. Notes Phys. 631, Springer, 2003), pp. 255–299
\bibitem{ham38} M.J.W. Hall, K. Kumar, and M. Reginatto, {J. Phys A: Math. Gen.} \textbf{36}, 9779 (2003)
\bibitem{ham38a} J.B. Hartle, in G.W. Gibbons, E.P.S. Shellard, and S.J. Rankin (Eds.), \emph{The Future of Theoretical Physics and Cosmology: Celebrating Stephen Hawking's 60th Birthday} (Cambridge University Press, 2003), pp. 38-50, 615-620
\bibitem{ham38a} G. Gibbons, in G.W. Gibbons, E.P.S. Shellard, and S.J. Rankin (Eds.), \emph{The Future of Theoretical Physics and Cosmology: Celebrating Stephen Hawking's 60th Birthday} (Cambridge University Press, 2003), pp. 351-372
\bibitem{ham38c} F. Dowker, in G.W. Gibbons, E.P.S. Shellard, and S.J. Rankin (Eds.), \emph{The Future of Theoretical Physics and Cosmology: Celebrating Stephen Hawking's 60th Birthday} (Cambridge University Press, 2003), pp. 436-452
\bibitem{ham38d} D. Page, in G.W. Gibbons, E.P.S. Shellard, and S.J. Rankin (Eds.), \emph{The Future of Theoretical Physics and Cosmology: Celebrating Stephen Hawking's 60th Birthday} (Cambridge University Press, 2003), pp. 621-648
\bibitem{ham38e} A. Vilenkin, in G.W. Gibbons, E.P.S. Shellard, and S.J. Rankin (Eds.), \emph{The Future of Theoretical Physics and Cosmology: Celebrating Stephen Hawking's 60th Birthday} (Cambridge University Press, 2003), pp. 649-666
\bibitem{ham38f} J. Halliwell, in G.W. Gibbons, E.P.S. Shellard, and S.J. Rankin (Eds.), \emph{The Future of Theoretical Physics and Cosmology: Celebrating Stephen Hawking's 60th Birthday} (Cambridge University Press, 2003), pp. 675-692
\bibitem{ham36} N. Pinto-Neto and E.S. Santini, {Gen. Rel. Grav.} \textbf{34}, 505 (2002); {Phys. Rev. D} \textbf{59}, 123517 (1999)
\bibitem{qgr9e} J.A. Belinchon, Int. J. Mod. Phys. D \textbf{11}, 527 (2002)
\bibitem{qgr9b} M. Castagnino, G. Catren, and R. Ferraro, Class. Quant. Grav. \textbf{19}, 4729 (2002)
\bibitem{qgr9f} A.E. Fischer and V.E. Moncrief, in P. Holmes, P.K. Newton, and A. Weinstein (Eds.), \emph{Geometry, Dynamics, and Mechanics: 60th Birthday Volume for J.E. Marsden} (Springer, 2002), pp. 1-60; in V. De Alfaro, J. Nelson, M. Cadoni, M. Cavaglia, and A.T. Filippov (Eds.) \emph{Constrained Dynamics and Quantum Gravity 1999: Abstracts of the Third Meeting on Constrained Dynamics and Quantum Gravity, QC99, Villasimius, Italy, 13-17 September 1999} (Nucl. Phys. B Proc. Suppl. \textbf{88}, North-Holland, 2000), pp. 83-102; in B. Fiedler, K. Groger, and J. Sprekels (Eds.) \emph{Equadiff 99: International Conference on Differential Equations, Berlin, August 17, 1999} Vol. 1 (World Scientif\/ic, 1999), pp. 279-282; Class. Quant. Grav. \textbf{16}, L79 (1999);  in S. Cotsakis and G.W. Gibbons (Eds.) \emph{Mathematical and Quantum Aspects of Relativity, held at Pythagoreon, Samos, greece, 31 August -4 Septmeber 1998} (Lect. Notes Phys. \textbf{537}, Springer, 1998), pp. 70-101; in J. Nelson (Ed.) \emph{Proceedings of the Second meeting on Constrained Dynamics and Quatum Gravity QG96, Santa Margherita Ligure, Italy, 17 Spetember 1996} (Nucl. Phys. B Proc. Suppl. \textbf{57}, North-Holland, 1997), pp. 142-161; Gen. Rel. Grav. \textbf{28}, 221 (1996); Gen. Rel. Grav. \textbf{28}, 207 (1996); in M. Flato, R. Kerner, and A. Lichnerowicz (Eds.) {Physics on Manifolds, Proceedings of the International Colloquium in honour of Yvonne Choquetbruhat, Paris, June 3-5, 1992} (Kluwer Academic Publishers, 1994), pp. 11-151;
\bibitem{qgr10} G.F.R. Ellis, {in} S. Bonometto, V. Gorini, and U. Moschella (Eds.) \emph{Modern Cosmology} (Institite of Physics, 2002), Chapter 3
\bibitem{qgr9h} M. Kenmoku, H. Kubotani, E. Takasugi, and Y. Yamazaki, Prog. Theor. Phys. \textbf{105}, 897 (2001); Phys. Rev. D \textbf{59}, 124004 (1999)
\bibitem{qgr9i} S. Biswas, A. Shaw, and D. Biswas, Int. J. Mod. Phys. D \textbf{10}, 585 (2001)
\bibitem{qgr9c} H. Yamazaki and T. Hara, Prog. Theor. Phys. \textbf{106}, 323 (2001)
\bibitem{ham26b} S. Carlip, {Rep. Prog. Phys.} \textbf{64}, 885 (2001)
\bibitem{ham26j} M. Anderson, Comm. Math. Phys. \textbf{222}, 533 (2001)
\bibitem{ham26c} S. Goldstein and S. Teufel, in C. Callender and N. Huggett (Eds.), \emph{Physics Meets Philosophy at the Planck Scale} (Cambridge University Press, 2001), pp. 275–289
\bibitem{ham26a} H. Friedrich and A. Rendall, in B.G. Schmidt (Ed.), \emph{Einstein's Field Equations and Their Physical Implications: Selected Essays in Honour of J\"urgen Ehlers} (Lect. Notes Phys. \textbf{540}, Springer, 2000), pp. 127-223
\bibitem{qgr15a} A. Kheyfets and W.A. Miller, Int. J. Mod. Phys. A \textbf{15}, 4125 (2000); Phys. Rev. D \textbf{51}, 493 (1995)
\bibitem{qgr9f} M.A. Ahmed, Nuovo Cim. B \textbf{115}, 1127 (2000)
\bibitem{qgr9d} M.L. Filchenkov, Russ. Phys. J. \textbf{43}, 921 (2000)
\bibitem{agr9x} K.S. Stelle, Nucl. Phys. B Proc. Suppl. \textbf{88}, 3 (2000)
\bibitem{qgr9g} G.Lifschytz and V. Periwal, JHEP \textbf{0004}, 026 (2000)
\bibitem{ham27c} Y. Choqu\'{e}t-Bruhat, J. Isenberg, and J.W. York, Jr., Phys. Rev. D \textbf{61}, 084034 (2000)
\bibitem{qgr11b} M. Kenmoku, Grav. Cosmol. \textbf{5}, 289 (1999)
\bibitem{qgr11c} B.S. DeWitt, in T. Piran (Ed.), \emph{Proceedings of 8th Marcel Grossmann Meeting on Recent Developments in Theoretical and Experimental General Relativity, Gravitation and Relativistic Field Theories (MG 8), Jerusalem, Israel, 22-27 Jun 1997.} (World Scientif\/ic, 1999), pp. 6-25; in R.E. Allen (Ed.) \emph{Relativity, Particle Physics And Cosmology: Proceedings of Richard Arnowitt Fest: A Symposium On Supersymmetry And Gravitation, 5-7 Apr 1998, College Station, Texas} (World Scientif\/ic, 1999), pp. 70-92; in S.W. Hawking and W. Israel (Eds.), \emph{General Relativity: An Einstein centenary survey} (Cambridge University Press 1979), pp. 680-745; {Phys. Rep.} \textbf{19}, 295 (1975); {Gen. Rel. Grav.} \textbf{1}, 181 (1970); \emph{Dynamical Theory of Groups and Fields} (Gordon and Breach, 1965); in L. Witten (Ed.) \emph{Gravitation: An Introduction to Current Research} (John Wiley \& Sons, 1962), pp. 266–381; Phys. Rev. \textbf{162}, 1195 (1967); Phys. Rev. \textbf{162}, 1239 (1967)
\bibitem{ham18} K. Kucha\v{r}, in J. Butterf\/ield (Ed.), \emph{The Arguments of Time} (Oxford University Press, 1999), pp. 169-195; in M. Francaviglia, G. Longhi, L. Lusanna and E. Sorace (Eds.) \emph{Proceedings of the 14th International Conference on General Relativity and Gravitation} (World Scientif\/ic, 1997), pp. 511-514; Phys. Rev. D \textbf{50}, 3961 (1994); in R. Gleiser, C. Kozameh and O. Moreschi (Eds.), \emph{General Relativity and Gravitation 1992: Proceedings of the Thirteenth International Conference on General Relativity and Gravitation, held Huerta Grande, Cordoba, 28 June-4 July, 1992} (IOP Publishing, 1993), pp. 119-150; Phys. Rev. D \textbf{43}, 3332 (1991); Phys. Rev. D \textbf{39}, 2263 (1989); Cont. Math. \textbf{71}, 285 (1988); Phys. Rev. D \textbf{34}, 3044 (1986); Phys. Rev. D \textbf{34}, 3031 (1986); J. Math. Phys. \textbf{22}, 2640 (1981); in W. Israel (Ed.) \emph{Relativity, Astrophysics and Cosmology: Proceedings of The Summer School held 14-26 August, 1972 at The Banff Centre, Banff, Alberta} (Kluwer Academic Publishers, 1973), pp. 238-288; {J. Math. Phys.} \textbf{13}, 768 (1972); Phys. Rev. D \textbf{4}, 955 (1971); J. Math. Phys. \textbf{11}, 3322 (1970)
\bibitem{qgr12} J.W. Norbury, {Eur. J. Phys.} \textbf{19}, 143 (1998)
\bibitem{qgr13} A.O. Barvinsky and C. Kiefer, {Nucl. Phys. B} \textbf{526}, 509 (1998)
\bibitem{qgr13a} V.A. Berezin, A. Boyarsky, and A.Yu. Neronov, Phys. Rev. D \textbf{57}, 1118 (1998)
\bibitem{qgr13i} A. Mostafazadeh, J. Math. Phys. \textbf{39}, 4499 (1998)
\bibitem{qgr13f} T. Brotz, Phys. Rev. D \textbf{57}, 2349 (1998)
\bibitem{qgr13h} H. Lu, J. Maharana, S. Mukherji, and C.N. Pope, Phys. Rev. D \textbf{57}, 2219 (1998)
\bibitem{qgr12a} J. Makela and P. Repo, Phys. Rev. D \textbf{57}, 4899 (1998)
\bibitem{qgr14a} R. Gleiser, O. Nicasio, R. Price, and J. Pullin, Phys. Rev. D \textbf{57}, 3401 (1998)
\bibitem{qgr14} T. Horiguchi, Nuovo Cim. B \textbf{113}, 429 (1998); {Nuovo Cim. B} \textbf{112}, 1227 (1997); {Nuovo Cim. B} \textbf{112}, 1107 (1997); {Nuovo Cim. B} \textbf{111}, 293 (1996); {Nuovo Cim. B} \textbf{111}, 85 (1996); {Nuovo Cim. B} \textbf{111}, 49 (1996); {Nuovo Cim. B} \textbf{111}, 165 (1996); {Nuovo Cim. B} \textbf{110}, 839 (1995); Mod. Phys. Lett. A \textbf{9}, 1429 (1994); {Mod. Phys. Lett. A} \textbf{8}, 777 (1993); {Phys. Rev. D} \textbf{48}, 5764 (1993)
\bibitem{qgr13b} A. B\l aut and J. Kowalski-Glikman, Phys. Lett. B \textbf{406}, 33 (1997)
\bibitem{qgr13j} J.W. Barrett and L. Crane, Class. Quant. Grav. \textbf{14}, 2113 (1997); J. Math. Phys. \textbf{39}, 3296 (1998)
\bibitem{ham35b} R. Sorkin, Int. J. Theor. Phys. \textbf{36}, 2759 (1997)
\bibitem{ham27d} J. Isenberg and J. Park, Class. Quant. Grav. \textbf{14}, A189 (1997)
\bibitem{qgr13g} M.D. Pollock, Int. J. Mod. Phys. D \textbf{6}, 91 (1997); Mod. Phys. Lett. A \textbf{12}, 2057 (1997); Int. J. Mod. Phys. D \textbf{4}, 305 (1995); Int. J. Mod. Phys. D \textbf{3}, 579 (1994); {Int. J. Mod. Phys. A} \textbf{7}, 4149 (1992)
\bibitem{qgr20a} R. Parentani, Phys. Rev. D \textbf{56}, 4618 (1997)
\bibitem{agr35c} M.A. Scheel, T.W. Baumgarte, G.B. Cook, S.L. Shapiro, and S.A. Teukolsky, Phys. Rev. d \textbf{56}, 6320 (1997)
\bibitem{qgr13f} A. Ambj{\o}rn, M. Carfora, and A. Marzouli, \emph{The Geometry of Dynamical
Triangulations} (Springer, 1997); in F. David, P. Ginsparg and J. Zinn-Justin (Eds.), \emph{Fluctuating Geometries in Statistical Mechanics and Field Theory: Proceedings of the Les Houches Summer School} (North-Holland, 1996), pp. 77–195
\bibitem{qgr14c} T.W. Baumgarte, G.B. Cook, M.A. Scheel, S.L. Shapiro, and S.A. Teukolsky, Phys. Rev. Lett. \textbf{79}, 1182 (1997); Phys. Rev. D \textbf{54}, 4849 (1996)
\bibitem{qgr20x} J. Louko and S. N. Winters-Hilt, Phys. Rev. D \textbf{54}, 2647 (1996)
\bibitem{qgr20b} G. Lifschytz, S.D. Mathur, and M. Ortiz, Phys. Rev. D \textbf{53}, 766 (1996)
\bibitem{qgr13u} R. Penrose, Gen. Rel. Grav. \textbf{28}, 581 (1996)
\bibitem{qgr13s} P. H\"ubner, Phys. Rev. D \textbf{53}, 701 (1996)
\bibitem{qgr13t} S.R. Brandt and E. Seidel, Phys. Rev. D \textbf{54}, 1403 (1996)
\bibitem{qgr13c} J. Kowalski-Glikman and K. Meissner, Phys. Lett. B \textbf{376}, 48 (1996)
\bibitem{qgr13d} J. Kowalski-Glikman, in B. Jancewicz, J.T. Sobczyk, and J. Lukierski (Eds.) \emph{From Field Theory to Quantum Groups. Birthday volume dedicated to Jerzy Lukierski} (World Scientif\/ic, 1996), pp. 229-242
\bibitem{qgr13e} D.L. Wiltshire, in B. Robson, N. Visvanathan and W.S. Woolcock (eds.), \emph{Cosmology: The Physics of the Universe: Proceedings of the 8th Physics Summer School, Australian National University, Canberra, Australia, 16 January-3 February, 1995} (World Scientif\/ic, 1996), pp. 473-531
\bibitem{qgr13j} B. Robson, in B. Robson, N. Visvanathan and W.S. Woolcock (eds.), \emph{Cosmology: The Physics of the Universe: Proceedings of the 8th Physics Summer School, Australian National University, Canberra, Australia, 16 January-3 February, 1995} (World Scientif\/ic, 1996), pp. 473–531
\bibitem{13f} M. Perry, in G.S. Hall and J.R. Pulham, \emph{General Relativity:Proceedings of the Forty Sixth Scottish Universities Summer School in Physics, Aberdeen, July 1995} (IOP Publishing, 1996), pp. 377-406
\bibitem{14b} N.T. Bishop, R. Gomez, P.R. Holvorcem, R.A. Matzner, P. Papadopoulos, and J. Winicour, Phys. Rev. Lett. \textbf{76}, 4303 (1996)
\bibitem{14a} T.W. Baumgarte, S.L. Shapiro, and S.A. Teukolsky, Astrophys. J. \textbf{458}, 680 (1996); Astrophys. J. \textbf{443}, 717 (1995)
\bibitem{qgr15b} M. Varadarajan, Phys. Rev. D \textbf{52}, 7080 (1995)
\bibitem{qgr15c} S.P. Kim, Phys. Rev. D \textbf{52}, 3382 (1995)
\bibitem{qgr15d} N.P. Landsman, Class. Quant. Grav. \textbf{12}, L119 (1995)
\bibitem{qgr15e} J. Feinberg and Y. Peleg, Phys. Rev. D \textbf{52}, 1988 (1995)
\bibitem{qgr15f} T. Horiguchi, K. Maeda, and M. Sakamoto, Phys. Lett. B \textbf{344}, 105 (1995)
\bibitem{qgr15g} F. Embacher, Grav. Cosmol. \textbf{1}, 46 (1995)
\bibitem{qgr23p} C. Rovelli and L. Smolin, Phys. Rev. D \textbf{52}, 5743 (1995); in B.R. Lyer (Ed.), \emph{Highlights in Gravitation and Cosmology} (Cambridge University Press, 1988)
\bibitem{qgr15i} A. Ashtekar, in B. Julia and J. Zinn-Justin (Eds.), \emph{Les Houches, Session
LVII, 1992: Gravitation and Quantizations} (Elsevier, 1995), pp. 181-283
\bibitem{ham16c} A.M. Abrahams, A. Andreson, Y. Choqu\'{e}t-Bruhat, and J.W. York, Jr., Phys. Rev. Lett. \textbf{75}, 3377 (1995)
\bibitem{qgr16} S. Carlip, {Class. Quant. Grav.} \textbf{11}, 31 (1994)
\bibitem{qgr18} P. Mansf\/ield, {Nucl. Phys. B} \textbf{418}, 113 (1994)
\bibitem{qgr20j} E. Adi and S. Solomon, Phys. Lett. B \textbf{336}, 152 (1994)
\bibitem{qgr20k} A. Ishikawa, Phys. Rev. D \textbf{50}, 2609 (1994)
\bibitem{qgr20l} J.E. Lidsey, Class. Quant. Grav. \textbf{11}, 1211 (1994)
\bibitem{qgr16a} A.M. Abrahams, G.B. Cook, S.L. Shapiro, and. S.A. Teukolsky, Phys. Rev. D \textbf{49}, 5153 (1994)
\bibitem{ham22a} J.B. Barbour, Class. Quant. Grav. \textbf{11}, 2853 (1994); Class. Quant. Grav. \textbf{11}, 2875 (1994); in R. Penrose and C.J. Isham (Eds.), \emph{Quantum Concepts in Space and Time} (Oxford University Press, 1986), pp. 236-246
\bibitem{ham16b} A.M. Abrahams and C.R. Evans, Phys. Rev. Lett. \textbf{70}, 2980 (1993); Phys. Rev. D \textbf{37}, 318 (1988)
\bibitem{qgr20c} G. Hayward and K. Wong, Phys. Rev. D \textbf{47}, 4778 (1993); {Phys. Rev. D} \textbf{46}, 620 (1992)
\bibitem{qgr16n} M.W. Choptuik, Phys. Rev. Lett. \textbf{70}, 9 (1993)
\bibitem{qgr20g} S. Capozziello and R. de Ritis, Int. J. Mod. Phys. D \textbf{2}, 373 (1993)
\bibitem{qgr20h} N. Pinto-Neto and A.F. Velasco, Gen. Rel. Grav. \textbf{25}, 991 (1993)
\bibitem{qgr20i} J. Maekelae, Phys. Rev. D \textbf{48}, 1679 (1993)
\bibitem{qgr20m} S. Abe, Phys. Rev. D \textbf{47}, 718 (1993)
\bibitem{qgr20n} H. Anada, Y. Mizumoto, and T. Kitazoe, Mod. Phys. Lett. A \textbf{8}, 45 (1993); Mod. Phys. Lett. A \textbf{8}, 1065 (1993)
\bibitem{qgr20o} D.S. Salopek, J.M. Stewart, and J. Parry, Phys. Rev. D \textbf{48}, 719 (1993)
\bibitem{qgr20p} H.-J. Matschull, Class. Quant. Grav. \textbf{10}, L149 (1993)
\bibitem{qgr20q} T. Hori, Prog. Theor. Phys. \textbf{90}, 743 (1993)
\bibitem{qgr20r} P.D. D'Eath, S.W. Hawking, and O. Obregon, Phys. Lett. B \textbf{300}, 44 (1993)
\bibitem{ham33} N. Pinto-Neto and A. F. Velasco, Gen. Rel. Grav. \textbf{25}(10), 991 (1993)
\bibitem{ham16d} S.L. Shapiro and S.A. Teukolsky, Astrophys. J. \textbf{45}, 2739 (1992); Astrophys. J. \textbf{307}, 575 (1986); Astrophys. J. \textbf{335}, 199 (1980)
\bibitem{qgr16c} E. Seidel and W.-M. Suen, Phys. Rev. Lett. \textbf{69}, 1845 (1992)
\bibitem{qgr20s} S. Chakraborty, Int. J. Theor. Phys. \textbf{31}, 289 (1992)
\bibitem{qgr20d} F.D. Mazzitelli, Phys. Rev. D \textbf{46}, 4758 (1992)
\bibitem{qgr20t} K. Shimizu and S. Wada, Int. J. Mod. Phys. A \textbf{7}, 1627 (1992)
\bibitem{ham31b} C.L. Stone and K.V. Kucha\v{r}, Class. Quant. Grav. \textbf{9}, 757 (1992)
\bibitem{ham31a} M. Ferraris, M. Francaviglia, and I. Sinicco, Nuovo Cim. B \textbf{107}, 11 (1992)
\bibitem{ham31c} G. Gorelik, in J. Eisenstaedt (Ed.),\emph{Studies in the History of General Relativity, Vol. 3 of Einstein Studies} (Birkha\"user, 1992), pp. 364–379
\bibitem{qgr20f} M.J. Duncan, in P. Langacker and M. Cvetic (Eds.), \emph{Proceedings: Testing The Standard Model - TASI-90: Theoretical Advanced Study Inst. In Elementary Particle Physics, 3-29 Jun 1990, Boulder, Colorado} (World Scientif\/ic, 1991), pp. 743-770
\bibitem{qgr20u} G. Giampieri, Phys. Lett. B \textbf{261}, 411 (1991)
\bibitem{qgr20w} J.P. Dias and M. Figueira Ann. Poincare Phys. Theor. \textbf{54}, 17 (1991)
\bibitem{qgr20z} M.J. Duncan and L.G. Jensen, {Nucl. Phys. B} \textbf{361}, 695 (1991)
\bibitem{ham29} Z. Bern, S.K. Blau, and E. Mottola, {Phys. Rev. D} \textbf{33}, 1212 (1991)
\bibitem{ham30} P.O. Mazur, {Phys. Lett. B} \textbf{262}, 405 (1991)
\bibitem{qgr23b} G.T. Horowitz, Class. Quant. Grav. \textbf{8}, 587 (1991); Phys. Rev. D \textbf{31}, 1169 (1985)
\bibitem{ham31} C. Kiefer and T.P. Singh, {Phys. Rev. D} \textbf{44}, 1067 (1991)
\bibitem{ham25b} K.V. Kucha\v{r} and C.G. Torre,  Phys. Rev. D \textbf{43}, 419 (1991); Phys. Rev. D \textbf{44}, 3116 (1991)
\bibitem{ham25a} P. H\'{a}j\'{i}\v{c}ek and K.V. Kucha\v{r} Phys. Rev. D \textbf{41}, 1091 (1990)
\bibitem{qgr21c} G.B. Cook and J.W. York, Phys. Rev. D \textbf{41}, 1077 (1990)
\bibitem{qgr20x} T. Padmanabhan and T.P. Singh Class. Quant. Grav. \textbf{7}, 411 (1990)
\bibitem{qgr20y} A.O. Barvinsky, Phys. Lett. B \textbf{241}, 201 (1990)
\bibitem{qgr25} W. Fischler, D. Morgan, and J. Polchinski, Phys. Rev. D \textbf{42}, 4042 (1990)
\bibitem{agr21b} G.B. Cook, \emph{Initial Data for the Two-Body Problem of General Relativity}, PhD Thesis, University of North Carolina (1990)
\bibitem{qgr21a} C.R. Evans, L.S. Finn, and D.W. Hobill (Eds.), \emph{Frontiers in Numerical Relativity} (Cambridge University Press, 1989)
\bibitem{ham14d} J.W. York, Jr., in C.R. Evans, L.S. Finn, and D.W. Hobill (Eds.), \emph{Frontiers in Numerical Relativity} (Cambridge University Press, 1989), pp. 89-109; in F.J. Tipler, \emph{Essays in General Relativity} (Academic Press, 1980), pp. 39-58; in L. Smarr (Ed.), \emph{Sources of Gravitational Radiation} (Cambridge University Press, 1979), pp. 83-126; Ann. Inst. H. Poincar\'{e} \textbf{21}, 319 (1974); J. Math. Phys. \textbf{14}, 456 (1973); Phys. Rev. Lett. \textbf{26}, 1656 (1971)
\bibitem{qgr21} A. Vilenkin, {Phys. Rev. D} \textbf{39}, 1116 (1989)
\bibitem{qgr22} S. Weinberg, {Rev. Mod. Phys.} \textbf{61}, 1 (1989)
\bibitem{qgr22a} V.D. Ivashchuk, V.N. Melnikov and A.I. Zhuk Nuovo Cim. B \textbf{104}, 575 (1989)
\bibitem{th3}   W. Fishler, I. Klebanov, J. Polchinski, and L. Susskind, {Nucl. Phys. B} \textbf{327}, 157 (1989)
\bibitem{ham28a} E. Alvarez, {Rev. Mod. Phys.} \textbf{61} (3), 561 (1989)
\bibitem{th4}   S.B. Giddings and A. Strominger, {Nucl. Phys. B} \textbf{321}, 481 (1989)
\bibitem{th5}   A. Hosoya and M. Morikawa, {Phys. Rev. D} \textbf{39}, 1123 (1989)
\bibitem{qgr23} M. McGuigan, {Phys. Rev. D} \textbf{39}, 2229 (1989); {Phys. Rev. D} \textbf{38}, 3031 (1988)
\bibitem{ham23w} P.J. Schinder, S.A. Bludman, and T. Piran, Phys. Rev. D \textbf{37}, 2722 (1988)
\bibitem{th7}   V.A. Rubakov, {Phys. Lett. B} \textbf{214}, 503 (1988)
\bibitem{qgr23x} L. Smolin, in J.A. Isenberg (Ed.), \emph{Mathematics and General Relativity: Proceedings of the AMS-IMS-SIAM Joint Summer Research Conference held June 22—28,1986 with support from the National Science Foundation} (Contemporary Mathematics \textbf{71}, AMS, 1988), pp. 55-98
\bibitem{qgr23y} T. Jacobson, in J.A. Isenberg (Ed.), \emph{Mathematics and General Relativity: Proceedings of the AMS-IMS-SIAM Joint Summer Research Conference held June 22—28,1986 with support from the National Science Foundation} (Contemporary Mathematics \textbf{71}, AMS, 1988), pp. 99-104
\bibitem{qgr23z} J.L. Friedman and I. Jack, Phys. Rev. D \textbf{37}, 3495 (1988)
\bibitem{th7a} N.T. Bishop, Gen. Rel. Grav. \textbf{20} 573 (1988); Gen. Rel. Grav. \textbf{16}, 589 (1984); Gen. Rel. Grav. \textbf{14}, 717 (1982)
\bibitem{qgr24} H.C. Tsamis and R.P. Woodard, Phys. Rev. D \textbf{36}, 3641 (1987)
\bibitem{qgr23a} M. Gleiser, R. Holman, and N.P. Neto, Nucl. Phys. B \textbf{294}, 1164 (1987), FERMILAB-Pub-87/73-A
\bibitem{qgr24a} J. Thornburg, Class. Quant. Grav. \textbf{4}, 1119 (1987)
\bibitem{qgr24c} C.R. Evans, in M.P. Ulmer (Ed.), \emph{Proceedings of the 13th Texas Symposium on Relativistic Astrophysics} (World Scientif\/ic, 1987), pp. 152-156; in J.M. Centrella (Ed.) \emph{Dynamical Spacetimes and Numerical Relativity} (Cambridge University Press, 1986), pp. 3-39; \emph{A Method for Numerical Relativity: Simulation of Axisymmetric Gravitational Collapse and Gravitational Radiation Generation}, PhD thesis (University of Texas at Austin, 1984)
\bibitem{qgr24b} K.S. Thorne, in S.W. Hawking and W. Israel (Eds.), \emph{Three hundred years of gravitation} (Cambridge University Press 1987), pp. 330-458
\bibitem{qgr22c} A. Anderson and B. DeWitt, Found. Phys. \textbf{16}, 91 (1986)
\bibitem{qgr24d} M.W. Choptuik and W.G. Unruh, Gen. Rel. Grav. \textbf{18}, 813 (1986)
\bibitem{qgr24e} H. Friedrich, Comm. Math. Phys. \textbf{107}, 587 (1986),; J. Geom. Phys. \textbf{3}, 101 (1986)
\bibitem{qgr23d} L.Z. Fang and M. Li, Phys. Lett. B \textbf{169}, 28 (1986)
\bibitem{ham22b} A.O. Barvinsky and V.N. Ponomarev, Sov. Phys. J. \textbf{29}(3), 187 (1986); Phys. Lett. B \textbf{167}, 289 (1986)
\bibitem{qgr23c} W.Z. Chao, Phys. Rev. D \textbf{31}, 3079 (1985)
\bibitem{min3}  R. Brandenburger, {Rev. Mod. Phys.} \textbf{57}, 1 (1985)
\bibitem{min6}  D.N. Page, {Phys. Rev. D} \textbf{32}, 2496 (1985)
\bibitem{min7a} P. Amsterdamski, {Phys. Rev. D} \textbf{31}, 3073 (1985)
\bibitem{ham22} C.J. Isham and K.V. Kucha\v{r}, Annals Phys. \textbf{164}, 288 (1985); Annals Phys. \textbf{164}, 316 (1985)
\bibitem{ham22a} T. Nakamura, Y. Kojima, and K. Oohara, Phys. Lett. A \textbf{107}, 452 (1985); Phys. Lett. A \textbf{106}, 235 (1984)
\bibitem{ham21} C.J. Isham and A.C. Kakas, Class. Quant. Grav. \textbf{1}, 621 (1984); Class. Quant. Grav. \textbf{1}, 633 (1984)
\bibitem{min2}  A.D. Linde, {Rep. Prog. Phys.} \textbf{47}, 925 (1984)
\bibitem{min2b} J. Bowen, J. Rauber, and J.W. York, Jr., Class. Quant. Grav. \textbf{1}, 591 (1984)
\bibitem{kulk2} A.D. Kulkarni, J. Math. Phys. \textbf{25}, 1028 (1984)
\bibitem{kulk1} A.D. Kulkarni, L.C. Shepley, and J.W. York, Jr., Phys. Lett. A \textbf{96}, 228 (1983)
\bibitem{min2a} N. Deruelle and T. Piran (Eds.), \emph{Gravitational Radiation} (North-Holland, 1983)
\bibitem{ham26p} R. Schoen and S.T. Yau, Comm. Math. Phys. \textbf{90}, 575 (1983); Comm. Math. Phys. \textbf{65}, 45 (1979)
\bibitem{ham28b} J.M. Bowen, Gen. Rel. Grav. \textbf{14}, 1183 (1982); Gen. Rel. Grav. \textbf{11}, 227 (1979)
\bibitem{hamm} J.W. York, Jr. and T. Piran, in R. Matzner and L. Shepley, \emph{Spacetime and Geometry: The Alfred Schild Lectures} (University of Texas Press, 1982), pp. 147-176
\bibitem{hamm1} M. Cantor and A.D. Kulkarni, Phys. Rev. D \textbf{25}, 2521 (1982)
\bibitem{hamm2} J. Isenberg and J.E. Marsden, Phys. Rep. \textbf{89}, 179 (1982)
\bibitem{min2a} D.N. Page and C.D. Geilker, Phys. Rev. Lett. \textbf{47}, 979 (1981)
\bibitem{min2b} A.E. Fischer, J.E. Marsden, and V.E. Moncrief, Ann. Inst. H. Poincar\'{e} \textbf{33}, 147 (1980); in F.J. Tipler, \emph{Essays in General Relativity} (Academic Press, 1980), pp. 79-96
\bibitem{ham28a} J.M. Bowen and J.W. York, Jr., Phys. Rev. D \textbf{21}, 2047 (1980)
\bibitem{ham28}  J.A. Isenberg and J.M. Nester, {in} A. Held (Ed.) \emph{General Relativity and Gravitation. One Hundred Years After the Birth of Albert Einstein.} (Plenum Press, 1980), pp. 23-97
\bibitem{ham4} P.G. Bergmann and A.B. Komar, {in} A. Held (Ed.) \emph{General Relativity and Gravitation. One Hundred Years After the Birth of Albert Einstein.} (Plenum Press, 1980), pp. 227-254; Int. J. Theor. Phys. \textbf{5}, 15 (1972); in \emph{Recent Developments in General Relativity. A collection of papers dedicated to Leopold Infeld} (Pergamon Press, Polish Scientif\/ic Publishers, 1962), pp. 31-46
\bibitem{ham4a} Y. Choqu\'{e}t-Bruhat and J.W. York, Jr., {in} A. Held (Ed.) \emph{General Relativity and Gravitation. One Hundred Years After the Birth of Albert Einstein.} (Plenum Press, 1980), pp. 99-172
\bibitem{ham4b} V.G. Lapchinsky and V.A. Rubakov, Acta Phys. Pol. B \textbf{10}, 1041 (1979)
\bibitem{ham25} D. Christodoulou, M. Francaviglia, and W.M. Tulczyjew, {Gen. Rel. Grav.} \textbf{10}, 567 (1979)
\bibitem{ham25a} A.E. Fischer and J.E. Marsden, in J. Ehlers (Ed.), \emph{Isolated Gravitating Systems in General Relativity: Proceedings of The International School Of Physics ''Enrico Fermi'' Course LXVII} (North-Holland, 1979), pp. 396-456; in S.W. Hawking and W. Israel (Eds.), \emph{General Relativity: An Einstein centenary survey} (Cambridge University Press 1979), pp. 138-211; Can. J. Math. \textbf{XXIX}(1), 193 (1977); Gen. Rel. Grav. \textbf{7}, 915 (1976); Sympos. Math. \textbf{XIV}, 193 (1974); Gen. Rel. Grav. \textbf{5}, 73 (1974); Bull. Amer. Math. Soc. \textbf{80}, 479 (1974); Bull. Amer. Math. Soc. \textbf{79}, 997 (1973); Gen. Rel. Grav. \textbf{4}, 309 (1973); in D. Farnsworth, J. Fink, J. Porter and A. Thompson (Eds.), \emph{Methods of Local and Global Differential Geometry in General Relativity: Proceedings of the Regional Conference on Relativity held at the University of Pittsburgh, Pittsburgh, Pennsylvania, July 13–17, 1970} (Lect. Notes Phys. \textbf{14}, Springer, 1972), pp. 176-188; J. Math. Phys. \textbf{13}, 546 (1972); Comm. Math. Phys. \textbf{28}, 1 (1972)
\bibitem{ham25x} D. Christodoulou and M. Francaviglia, in J. Ehlers (Ed.), \emph{Isolated Gravitating Systems in General Relativity: Proceedings of The International School Of Physics ''Enrico Fermi'' Course LXVII} (North-Holland, 1979), pp. 480-495
\bibitem{ham25b} Y. Choqu\'{e}t-Bruhat, A.E. Fischer and J.E. Marsden, in J. Ehlers (Ed.), \emph{Isolated Gravitating Systems in General Relativity: Proceedings of The International School Of Physics <<Enrico Fermi>> Course LXVII} (North-Holland, 1979), pp. 322-395
\bibitem{ham26a} L. Smarr (Ed.), \emph{Sources of Gravitational Radiation} (Cambridge University Press, 1979); in M.D. Papagiannis (Ed.), \emph{Eighth Texas Symposium on Relativistic Astrophysics} (The New York Academy of Sciences, 1977), pp. 569-604
\bibitem{ham26} M. Francaviglia, {Riv. Nuovo Cim.} \textbf{1}, 1303 (1978)
\bibitem{qgr27c} H.-S. Tsao, Phys. Lett. B \textbf{68}, 79 (1977)
\bibitem{ham26b} K. Eppley, Phys. Rev. D \textbf{16}, 1609 (1977)
\bibitem{ham23b} D. Karmer, Acta Phys. Pol. B \textbf{7}, 117 (1976)
\bibitem{ham23} S.A. Hojman, K. Kucha\v{r}, and C. Teitelboim, Nature (London), {Ann. Phys. NY} \textbf{96}, 88 (1976)
\bibitem{ham23a} N. \'{O} Murchada and J.W. York, Jr., Gen. Rel. and Grav. \textbf{7}, 257 (1976); Phys. Rev. D \textbf{10}, 437 (1974); Phys. Rev. D \textbf{10}, 2345 (1974); J. Math. Phys. \textbf{14}, 1551 (1973)
\bibitem{ham19} M.A.H. MacCallum, {in} C.J. Isham, R. Penrose, and D.W. Sciama (Eds.) \emph{Quantum Gravity: an Oxford Symposium} (Clarendon Press, 1975), pp. 174-208
\bibitem{qgr27} P.B. Gilkey, {J. Diff. Geom.} \textbf{10}, 601 (1975); {Proc. Symp. Pure. Math.} \textbf{27}, 265 (1975)
\bibitem{ham15a} A. \v{C}ade\v{z}, Ann. Phys. \textbf{83}, 449 (1974)
\bibitem{ham15} A. Ashtekar and R. Geroch, {Rep. Progr. Phys.} \textbf{37}, 1211 (1974)
\bibitem{ham16} T. Regge and C. Teitelboim, {Annals Phys. NY} \textbf{88}, 286, (1974); {Phys. Lett. B} \textbf{53}, 101 (1974)
\bibitem{qgr27a} J. Tarski, Ann. Inst. H. Poincar\'{e} \textbf{20}(1), 95 (1974); Ann. Inst. H. Poincar\'{e} \textbf{17}, 171 (1972); Ann. Inst. H. Poincar\'{e} \textbf{17}, 313 (1972); Ann. Inst. H. Poincar\'{e} \textbf{11}, 331 (1969); J. Math. Phys. \textbf{7}, 560 (1966)
\bibitem{qgr27b} G. 't Hooft, Nucl. Phys. B \textbf{62}, 444 (1973); in M. L\'{e}vy and S. Deser (Eds.), \emph{Recent Developments in Gravitation, Proceedings from Carg\`{e}se 1978, Vol. B44 of NATO Advanced Study Institutes Series} (Kluwer Academic Publishers, 1979), pp. 323–345
\bibitem{ham14} C. Teitelboim, Phys. Rev. D \textbf{25}, 3159 (1982); {in} A. Held (Ed.) \emph{General Relativity and Gravitation. One Hundred Years After the Birth of Albert Einstein.} (Plenum Press, 1980), pp. 195-226; Phys. Lett. B \textbf{56}, 376 (1975); Ann. Phys. \textbf{79}(2), 542 (1973); Ann. Phys. \textbf{80}, 542 (1973); Phys. Rev. D \textbf{8}, 3266 (1973); Phys. Rev. D \textbf{5}, 2941 (1972)
\bibitem{ham14a} C.J. Isham, Abdus Salam, and J.A. Strathdee, Phys. Lett. B \textbf{46}, 407 (1973)
\bibitem{ham14c} F. Lund, Phys. Rev. D \textbf{8}, 3247 (1973)
\bibitem{ham17a} F. Estabrook, H. Wahlquist, S. Christensen, B. DeWitt, L. Smarr, and E. Tsiang, Phys. Rev. D \textbf{7}, 2814 (1973)
\bibitem{ham17c} S.W. Hawking and G.F.R. Ellis, \emph{The Large Scale Structure of Space-Time} (Cambridge
University Press, 1973)
\bibitem{ham14b} B.K. Berger, D.M. Chitre, V.E. Moncrief, and Y. Nutku, Phys. Rev. D \textbf{5}, 2467 (1972)
\bibitem{ham17} R. Geroch, J. Math. Phys. \textbf{13}, 956 (1972);  J. Math. Phys. \textbf{8}, 782 (1967)
\bibitem{ham11a} M. Ryan, \emph{Hamiltonian Cosmology} (Springer, 1972)
\bibitem{ham12} V. Moncrief and C. Teitelboim, {Phys. Rev. D} \textbf{6}, 966 (1972)
\bibitem{ham12a} V. Moncrief, Phys. Rev. D \textbf{5}, 277 (1972)
\bibitem{ham13a} G.W. Gibbons, Comm. Math. Phys. \textbf{27}, 87 (1972)
\bibitem{ham13b} G.W. Gibbons and B.F. Schutz, Mon. Not. Roy. Astr. Soc. \textbf{159}, 41P (1972)
\bibitem{qgr28a} V.K. Patodi, {J. Diff. Geom.} \textbf{5}, 233 (1971)
\bibitem{ham11b} J.R. Klauder and E.W. Aslaksen, Phys. Rev. D \textbf{2}, 272 (1970)
\bibitem{ham11} D.R. Brill and R.H. Gowdy, Rep. Prog. Phys. \textbf{33}, 413 (1970)
\bibitem{ham11c} Y. Choqu\'{e}t-Bruhat, S\'{e}minaire Jean Leray \textbf{2}, 1 (1969-1970)
\bibitem{qgr27e} C.W. Misner, in M. Carmeli, S.I. F\/ickler, and L. Witten (Eds.) \emph{Relativity. Proceedings of the Relativity Conference in the Midwest held at Cincinnati, Ohio, June 2-6, 1969} (Plenum Press, 1970), pp. 55-79; Phys. Rev. Lett. \textbf{22}, 1071 (1969); Phys. Rev. \textbf{186}, 1319 (1969); Phys. Rev. \textbf{186}, 1328 (1969); Ann. Phys. \textbf{24}, 102 (1963); Phys. Rev. \textbf{118}, 1110 (1960); Rev. Mod. Phys. \textbf{29}, 497 (1957)
\bibitem{qgr27d} U.H. Gerlach, Phys. Rev. \textbf{177}, 1929 (1969)
\bibitem{qgr28a} A. Peres, Phys. Rev. \textbf{171}, 1335 (1968)
\bibitem{qgr28}  H.P. McKean and I.M. Singer, {J. Diff. Geom.} \textbf{1}, 43 (1967)
\bibitem{ham9}  A.B. Komar, {Phys. Rev.} \textbf{153}, 1385 (1967); {Phys. Rev.} \textbf{164}, 1595 (1967)
\bibitem{ham2}  P.G. Bergmann, in M. Carmeli, S.I. F\/ickler, and L. Witten (Eds.) \emph{Relativity. Proceedings of the Relativity Conference in the Midwest held at Cincinnati, Ohio, June 2-6, 1969} (Plenum Press, 1970), pp. 43-53; Phys. Rev. \textbf{144}, 1078 (1966); Rev. Mod. Phys. \textbf{33}, 510 (1961); Rev. Mod. Phys. \textbf{29}, 352 (1957); Nuovo Cim. \textbf{3}, 1177 (1956)
\bibitem{ham8a} M. May and R.H. White, Phys. Rev. \textbf{141}, 1232 (1966)
\bibitem{ham8}  H. Leutwyler, Nuovo Cim. \textbf{42}, 159 (1966); Phys. Rev. \textbf{134}, B1155 (1964)
\bibitem{ham8b} S. Hahn and R.W. Lindquist, Ann. Phys. \textbf{29}, 304 (1964)
\bibitem{ham8a} R.W. Lindquist, J. Math. Phys. \textbf{4}, 938 (1963)
\bibitem{ham8b} D.R. Brill and R.W. Lindquist Phys. Rev. \textbf{131}, 471 (1963)
\bibitem{ham7}  R.F. Beierlein, D.H. Sharp, and J.A. Wheeler, {Phys. Rev.} \textbf{126}, 1864 (1962)
\bibitem{ham7a} S. Mandelstam, Proce. Roy. Soc. London A \textbf{270}(1342), 346 (1962)
\bibitem{ham6}  R. Arnowitt, S. Deser, and C.W. Misner, Phys. Rev. \textbf{122}, 997 (1961); Phys. Rev. \textbf{121}, 1556 (1961); Ann. Phys. \textbf{11}, 116, (1960); J. Math. Phys. \textbf{1}, 434 (1960); Phys. Rev. Lett. \textbf{4}, 375 (1960); Nuovo Cim. \textbf{15}, 487 (1960); Phys. Rev. \textbf{120}, 313 (1960); Phys. Rev. \textbf{118}, 1100 (1960); Phys. Rev. \textbf{117}, 1595 (1960); Phys. Rev. \textbf{116}, 1322 (1959); {Phys. Rev.} \textbf{113}, 745 (1959)
\bibitem{ham5} J.L. Anderson, Phys. Rev. \textbf{114}, 1182 (1959)
\bibitem{ham3}  P.W. Higgs, Phys. Rev. Lett. \textbf{3}, 66 (1959); Phys. Rev. Lett. \textbf{1}, 373 (1958)
\bibitem{ham1a} A. Lichnerowicz, \emph{Th\'{e}ories relativistes de la gravitation} (Masson, 1955)
\bibitem{ham1}  F.A.E. Pirani and A. Schild, {Phys. Rev.} \textbf{79}, 986 (1950)
\bibitem{ash} A. Ashtekar, Phys. Rev. Lett. \textbf{57}, 2244 (1986); Phys. Rev. D \textbf{36}, 1587 (1987)
\bibitem{rovsmol} C. Rovelli and L. Smolin, Nucl. Phys. B \textbf{331}, 80 (1990)
\bibitem{found}
T. Jacobson and L. Smolin, Nucl. Phys. B \textbf{299}, 295 (1988);
C. Rovelli and L. Smolin, Phys. Rev. Lett. \textbf{61}, 1155 (1988); Nucl. Phys. B \textbf{133}, 90 (1990); Nucl. Phys. B \textbf{442}, 593 (1995); Nucl. Phys. B \textbf{456}, 753 (1995);
A. Ashtekar, \emph{New Perspectives in Canonical Gravity} (Bibliopolis, 1988);
A. Ashtekar and R.S. Tate, \emph{Non-perturbative Canonical Gravity} (World Scientif\/ic, 1991);
C. Rovelli, Class. Quant. Grav. \textbf{6}, 911 (1989); Nucl. Phys. B \textbf{331}, 80 (1990); Class. Quant. Grav. \textbf{8}, 297 (1991); in A.I. Janis and J.R. Porter (Eds.), \emph{Recent Advances in General Relativity: Essays in Honor of Ted Newman} (Birkh\"auser, 1992), pp. 85-102; in N. Dadhich and J. Narlikar (Eds.), \emph{Gravitation and Relativity: At the Turn of the Millenium. Proceedings of the GR-15 Conference, held at IUCAA, Pune, India, December 16–21, 1997} (Inter-University Centre for Astronomy and Astrophysics, 1998), pp. 281–331;
A. Ashtekar, C. Rovelli and L. Smolin, Phys. Rev. D \textbf{44}, 1740 (1991);
A. Ashtekar and J. Lewandowski, Class. Quant. Grav. \textbf{14}, A55 (1997); Class. Quant. Grav. \textbf{21}, R53 (2004)
\bibitem{bojo} M. Bojowald, Class. Quant. Grav. \textbf{17}, 1489 (2000); Class. Quant. Grav. \textbf{17}, 1509 (2000); Class. Quant. Grav. \textbf{18}, 1055 (2001); Class. Quant. Grav. \textbf{18}, 1071 (2001); Class. Quant. Grav. \textbf{18}, L109 (2001); Phys. Rev. D \textbf{64}, 084018 (2001); Phys. Rev. Lett. \textbf{87}, 121301 (2001); Phys. Rev. Lett. \textbf{86}, 5227 (2001); Phys. Rev. Lett. \textbf{89}, 261301 (2002); Class. Quant. Grav. \textbf{19}, 5113 (2002); Class. Quant. Grav. \textbf{19}, 2717 (2002); Class. Quant. Grav. \textbf{20}, 2595 (2003); J. Phys. Conf. Ser. \textbf{24}, 77 (2005); Class. Quant. Grav. \textbf{23}, 987 (2006); Nature Phys. \textbf{3}(8), 523 (2007); Gen. Rel. Grav. \textbf{40}, 2659 (2008); Phys. Rev. Lett. \textbf{100}, 221301 (2008); Living Rev. Rel. \textbf{11}, 4 (2008); Class. Quant. Grav. \textbf{26}, 075020 (2009); AIP Conf. Proc. \textbf{1196}, 62 (2009); AIP Conf. Proc. \textbf{1256}, 66 (2010)
\bibitem{loop}
M. Bojowald, D. Mulryne, W. Nelson, and R. Tavakol, Phys. Rev. D \textbf{82}, 124055 (2010);
M. Bojowald, J.D. Reyes, R. Tibrewala, Phys. Rev. D \textbf{80}, 084002 (2009);
M. Bojowald, G.M. Hossain , M. Kagan, S. Shankaranarayanan, Phys. Rev. D \textbf{78}, 063547 (2008); Phys. Rev. D \textbf{79}, 043505 (2009), Phys. Rev. D \textbf{82}, 109903 (2010);
M. Bojowald and J.D. Reyes, Class. Quant. Grav. \textbf{26}, 035018 (2009);
M. Bojowald and R. Das, Class. Quant. Grav. \textbf{25}, 195006 (2008);
M. Bojowald, T. Harada, and R. Tibrewala, Phys. Rev. D \textbf{78}, 064057 (2008);
M. Bojowald, R. Das, and R.J. Scherrer, Phys. Rev. D \textbf{77}, 084003 (2008);
M. Bojowald and G.M. Hossain, Phys. Rev. D \textbf{77}, 023508 (2008);
M. Bojowald and R. Das, Phys. Rev. D \textbf{75}, 123521 (2007);
M. Bojowald, H. Hernandez, and A. Skirzewski, Phys. Rev. D \textbf{76}, 063511 (2007);
M. Bojowald, D. Cartin, and G. Khanna, Phys. Rev. D \textbf{76}, 064018 (2007);
M. Bojowald, H. Hernandez, M. Kagan, and A. Skirzewski, Phys. Rev. D \textbf{75}, 064022 (2007); M. Bojowald and A. Rej, Class. Quant. Grav. \textbf{22}, 3399 (2005);
M. Bojowald and H.A. Morales-Tecotl, in N. Breton, J.L. Cervantes-Cota, and M. Salgado (Eds.) \emph{Proceedings of 5th Mexican School On Gravitation And Mathematical Physics: The Early Universe And Observational Cosmology (DGFM 2002), 24-29 Nov 2002, Playa del Carmen, Quintana Roo, Mexico} (Lect. Notes Phys. \textbf{646}, Springer, 2004), pp. 421-462;
Ashtekar, M.Campiglia, and A. Henderson, Phys. Rev. D \textbf{82}, 124043 (2010);
A. Ashtekar, M. Bojowald, and J. Lewandowski, Adv. Theor. Math. Phys. \textbf{7}, 233 (2003);
A. Ashtekar, W. Kaminski, and J. Lewandowski. Phys. Rev. D \textbf{79}, 064030 (2009);
A. Henderson, C. Rovelli, F. Vidotto, and E. Wilson-Ewing, Class. Quant. Grav. \textbf{28}, 025003 (2011); M. Domagala, K. Giesel, W. Kaminski, and J. Lewandowski, Phys. Rev. D \textbf{82}, 104038 (2010);
W. Kaminski, M. Kisielowski, and J. Lewandowski, Class. Quantum Grav. \textbf{27}, 095006 (2010);
W. Kaminski, J. Lewandowski, and T. Pawlowski, Class. Quant. Grav. \textbf{26}, 035012 (2009); Class. Quant. Grav. \textbf{26}, 245016 (2009);
\L. Szulc, Phys. Rev. D \textbf{78}, 064035 (2008);
I. Agullo, G.J.F. Barbero, E.F. Borja, J. Diaz-Polo, and E.J.S. Villasenor, Phys. Rev. D \textbf{82}, 084029 (2010); Phys. Rev. D \textbf{80}, 084006 (2009);
D.-W. Choiu and L-F. Li, Phys. Rev. D \textbf{80}, 043512 (2009);
D.-W. Chiou and M. Geiller, Phys. Rev. D \textbf{82}, 064012 (2010);
A. Corichi, T. Vukasinac, and J.A. Zapata, AIP Conf. Proc. \textbf{1256}, 172 (2010);
C. Beetle and J. Engle, Class. Quant. Grav. \textbf{27}, 235024 (2010);
C. Rovelli and S. Speziale, Phys. Rev. D \textbf{82}, 044018 (2010);
Y. Ma, C. Soo, and J. Yang, Phys. Rev. D \textbf{81}, 124026 (2010);
L. Fatibene, M. Ferraris, and  M. Francaviglia, Class. Quant. Grav. \textbf{27}, 185016 (2010);
S. Mercuri, PoS ISFTG, 016 (2009);
R. Gambini and J. Pullin, J. Phys. Conf. Ser. \textbf{189}, 012034 (2009); Adv. Sci. Lett. \textbf{2}, 251 (2009); Class. Quant. Grav. \textbf{26}, 035002 (2009);
H. Sahlmann, Hanno Sahlmann, Class. Quant. Grav. \textbf{27}, 225007 (2010);
J. Brunnemann and D. Rideout, Class. Quant. Grav. \textbf{27}, 205008 (2010);
M. Han, Class. Quant. Grav. \textbf{27}, 215009 (2010);
F. Cianfrani and G. Montani, Phys. Rev. D \textbf{80}, 084045 (2009);
J.F. Barbero G., J. Phys. Conf. Ser. \textbf{175}, 012005 (2009);
A. Perez, AIP Conf. Proc. \textbf{1132}, 386 (2009);
Ch.-A. Li, J.-J. Jiang, and J.-Q. Su, Sci. China G \textbf{52}, 1179 (2009);
A. DeBenedictis, Can. J. Phys. \textbf{87}, 255 (2009);
J.M. Garcia-Islas, Can. J. Phys. \textbf{88}, 223 (2010);
S.J. Gates, Jr., S.V. Ketov, and N. Yunes, Phys. Rev. D \textbf{80}, 065003 (2009);
M. Bojowald, J.D. Reyes, and R. Tibrewala, Phys. Rev. D \textbf{80}, 084002 (2009);
D. Mamone and C. Rovelli, Class. Quant. Grav. \textbf{26}, 245013 (2009);
J. Grain and A. Barrau, Phys. Rev. Lett. \textbf{102}, 081301 (2009);
J. Engle, Class. Quant. Grav. \textbf{27}, 035003 (2010);
F. Cianfrani and G. Montani, Phys. Rev. Lett. \textbf{102}, 091301 (2009);
M. Bojowald and J.D. Reyes, Class. Quant. Grav. \textbf{26}, 035018 (2009);
S. Speziale, Adv. Sci. Lett. \textbf{2}, 280 (2009);
C, Perini, C. Rovelli, S. Speziale, Phys. Lett. B \textbf{682}, 78 (2009);
E. Bianchi, Nucl. Phys. B \textbf{807}, 591 (2009);
K. Giesel and T. Thiemann, Class. Quant. Grav. \textbf{27}, 175009 (2010); Class. Quant. Grav.\textbf{23}, 5667 (2006); Class. Quant. Grav. \textbf{23}, 5693 (2006);
J. Brunnemann and T. Thiemann, Class. Quant. Grav. \textbf{23}, 1429 (2006); Class. Quant. Grav. \textbf{23}, 1395 (2006); Class. Quant. Grav. \textbf{23}, 1289 (2006);
B. Bahr and T. Thiemann, Class. Quant. Grav. \textbf{26}, 235022 (2009); Class. Quant. Grav. \textbf{26}, 045011 (2009); Class. Quant. Grav. \textbf{26}, 045012 (2009); Class. Quant. Grav. \textbf{24}, 2109 (2007);
B. Dittrich and T. Thiemann, J. Math. Phys. \textbf{50}, 012503 (2009); Class. Quant. Grav. \textbf{23}, 1143 (2006); Class. Quant. Grav. \textbf{23}, 1143 (2006); Class. Quant. Grav. \textbf{23}, 1089 (2006); Class. Quant. Grav. \textbf{23}, 1067 (2006); Class. Quant. Grav. \textbf{23}, 1025 (2006);
T. Thiemann, in D. Oriti (Ed.), \emph{Approaches to Quantum Gravity. Toward a New Understanding of Space, Time, and Matter} (Cambridge University Press, 2009), pp. 235-252; Int. J. Mod. Phys. A \textbf{23}, 8, 1113 (2008); in I.-O. Stamatescu and E. Seiler (Eds.), \emph{Approaches to Fundamental Physics: An Assessment of Current Theoretical Ideas} ({Lect. Notes Phys.} \textbf{721}, Springer, 2007), pp. 185-263; Class. Quant. Grav. \textbf{23}, 2211 (2006); in D. Giulini, C. Kiefer, and C. L\"ammerzahl (Eds.), \emph{Quantum Gravity. From Theory To Experimental Search} (Lect. Notes Phys. 631, Springer, 2003), pp. 41-135;
A. Ashtekar, Gen. Rel. Grav. \textbf{41}, 707 (2009);
C. Fleischhack, in B. Fauser, J. Tolksdorf, and E. Zeidler (Eds.) \emph{Quantum Gravity. Mathematical Models and Experimental Bounds} (Birkh\"auser, 2007), pp. 203-220;
A. Perez, in D. Oriti (Ed.), \emph{Approaches to Quantum Gravity. Toward a New Understanding of Space, Time, and Matter} (Cambridge University Press, 2009), pp. 272-289
\bibitem{nrel}
    R. d’Inverno, in G.S. Hall and J.R. Pulham (Eds.), \emph{General Relativity} (IOP Publishing, 1996), pp. 331-376;\\
    B. Br\"ugmann, Ann. Phys. \textbf{9}, 227 (2000);\\
    L. Lehner, Class. Quant. Grav. \textbf{18}, R25 (2001); arXiv:gr-qc/0106072;\\
    U. Sperhake, \emph{Non-linear numerical Schemes in General Relativity}, PhD thesis, University of Southampton, arXiv:gr-qc/0201086;\\
    C. Bona and C. Palenzuela-Luque, \emph{Elements of Numerical Relativity: From Einstein’s Equations to Black Hole Simulations} (Springer, 2005);\\
    J. Frauendiener, D.J.W. Giulini, and V.Perlick (Eds.), \emph{Analytical and Numerical Approaches to Mathematical Relativity: With a Foreword by Roger Penrose} (Springer, 2006);\\
    E. Gourgoulhon, arXiv:gr-qc/0703035.
\bibitem{darm} G. Darmois, \emph{Les \'{e}quations de la gravitation einsteinienne}, M\'{e}morial des Sciences Math\'{e}matiques \textbf{25} (Gauthier-Villars, 1927)
\bibitem{lich39} A. Lichnerowicz, \emph{Sur certains probl\`{e}mes globaux relatifs au syst\`{e}me des \'{e}quations d’Einstein} (Hermann, 1939); Actual. Sci. Ind. \textbf{833}
\bibitem{lich44} A. Lichnerowicz, J. Math. Pures et Appl. \textbf{23}, 37 (1944)
\bibitem{bruh} Y. Four\`{e}s-Bruhat (Y. Choqu\'{e}t-Bruhat), Acta Math. \textbf{88}, 141 (1952); J. Rational Mech. Anal. \textbf{5}, 951 (1956)
\bibitem{york72} J.W. York, Phys. Rev. Lett. \textbf{28}, 1082 (1972); J. Math. Phys. \textbf{13}, 125 (1972)
\bibitem{omurd} N. \'{O} Murchadha and J.W. York, Phys. Rev. D \textbf{10}, 428 (1974)
\bibitem{smar77} L. Smarr, Ann. N.Y. Acad. of Sci. \textbf{302}, 569 (1977)
\bibitem{smaryork} L. Smarr and J.W. York, Jr., Phys. Rev. D \textbf{17}, 2529 (1978); Phys. Rev. D \textbf{17}, 1945 (1978)
\bibitem{barpir} J.M. Bardeen and T. Piran, Phys. Rep. \textbf{96}, 206 (1983)
\bibitem{naka} T. Nakamura, Prog. Theor. Phys. \textbf{65}, 1876 (1981)
\bibitem{stark} R.F. Stark and T. Piran, Phys. Rev. Lett. \textbf{55}, 891 (1985)
\bibitem{nok} T. Nakamura, K. Oohara, and Y. Kojima, Prog. Theor. Phys. Suppl. \textbf{90}, 1 (1987)
\bibitem{bona} C. Bona and J. Mass\'{o}, Phys. Rev. D \textbf{40}, 1022 (1989); Phys. Rev. Lett. \textbf{68}, 1097 (1992); Phys. Rev. D \textbf{38}, 2419 (1988)
\bibitem{chby} Y. Choqu\'{e}t-Bruhat and J.W. York, Jr., C.R. Acad. Sc. Paris \textbf{321}, S\'{e}rie I, 1089, (1995)
\bibitem{kst} L.E. Kidder, M.A. Scheel, and S.A. Teukolsky, Phys. Rev. D \textbf{64}, 064017 (2001)
\bibitem{shinak} M. Shibata and T. Nakamura, Phys. Rev. D \textbf{52}, 5428 (1995)
\bibitem{baum} T.W. Baumgarte and S.L. Shapiro, Phys. Rev. D \textbf{59}, 024007 (1999)
\bibitem{york99} J.W. York, Phys. Rev. Lett. \textbf{82}, 1350 (1999)
\bibitem{shib} M. Shibata, Prog. Theor. Phys. \textbf{101}, 1199 (1999); Phys. Rev. D \textbf{60}, 104052 (1999)
\bibitem{hayw}  S.A. Hayward, Phys. Rev. D \textbf{61}, 101503 (2000)
\bibitem{bonaz} S. Bonazzola, E. Gourgoulhon, P. Grandcl\'{e}ment, and J. Novak, Phys. Rev. D \textbf{70}, 104007 (2004)
\bibitem{stephani1} H. Stephani, \emph{Relativity: An Introduction to Special and General Relativity} (3rd ed., Cambridge University Press, 2004)
\bibitem{besse} A.L. Besse, \emph{Einstein Manifolds} (Springer, 1987)
\bibitem{calyau}
E. Calabi, Proc. Internat. Congress. Math. Amsterdam \textbf{2}, 206 (1954); in \emph{Algebraic Geometry and Topology. A symposium in honor of S. Lefschetz} (Princeton University Press, 1957), pp. 78–89; Trans. Amer. Math. Soc. \textbf{67}, 401 (1958)\\
S.T. Yau, Comm. Pure and Appl. Math. \textbf{31}(3), 339 (1978); Proc. Nat. Acad. Sci. USA \textbf{74}, 1798 (1977); in \emph{Surveys in Differential Geometry. Vol. XIII. Geometry, analysis, and algebraic geometry: forty years of the Journal of Differential Geometry}, Surv. Differ. Geom. \textbf{13}, 277 (2009)

\bibitem{hypkah}
E. Calabi, Ann. Ecol. Norm. Sup. \textbf{12}, 269 (1979);\\
A. Beauville, J. Diff. Geom. \textbf{18}, 755 (1983)

\bibitem{cyhk}
M.H. Emam, Class. Quant. Grav. \textbf{27}, 163001 (2010);
C.M. Hull, U. Lindstrom, M. Rocek, R. von Unge, and M. Zabzine, JHEP \textbf{1008}, 060 (2010);
H. Lu and Z.-L. Wang, J. Geom. Phys. \textbf{60}, 1741 (2010);
T. Eguchi and K. Hikami, Lett. Math. Phys. \textbf{92}, 269 (2010);
H. Lu, Y. Pang and Z.-L. Wang, Class. Quant. Grav. \textbf{27}, 155018 (2010);
Y.-H. He, S.-J. Lee, and A. Lukas, JHEP \textbf{1005}, 071 (2010);
J. Fu, Z. Wang, and D. Wu Math. Res. Lett. \textbf{17}, 887 (2010);
A.D. Popov, Nucl. Phys. B \textbf{828}, 594 (2010);
R. Heluani and M. Zabzine, Adv. Math. \textbf{223}, 1815 (2010);
I. Biswas and B. McKay J. Geom. Phys. \textbf{60}, 661 (2010);
A.Q. Velez and A. Boer, Comm. Math. Phys. \textbf{297}, 597 (2010);
M. Kreuzer, Ukr. J. Phys. \textbf{55}, 613 (2010);
F. Loran and H. Soltanpanahi, Adv. Theor. Math. Phys. \textbf{13}, 637 (2009);
M. Eto, T. Fujimori, S.B. Gudnason, M.Nitta, and K. Ohashi, Nucl. Phys. B \textbf{815}, 495 (2009);
H. Jockers and M. Soroush, Comm. Math. Phys. \textbf{290}, 249 (2009); Nucl. Phys. B \textbf{821}, 535 (2009);
A.S. Haupt, A. Lukas, K.S. Stelle, JHEP \textbf{0905}, 069 (2009);
S. Rollenske, R.P. Thomas, Journal of Topology \textbf{2}, 405 (2009);
Y.-G. Oh and K. Zhu, Asian J. Math. \textbf{13}, 323 (2009);
B. Haghighat and A. Klemm, JHEP \textbf{0901}, 029 (2009);
V. Tosatti, J. Eur. Math. Soc. \textbf{11}, 755 (2009);
F.P. Correia, JHEP \textbf{0912}, 004 (2009);
M. Kreuzer, Fortsch. Phys. \textbf{57}, 625 (2009);
H. Ooguri and M. Yamazaki, Phys. Rev. Lett. \textbf{102}, 161601 (2009); Comm. Math. Phys. \textbf{292}, 179 (2009);
E. Palti, JHEP \textbf{0904}, 099 (2009);
S.P. de Alwis Phys. Lett. B \textbf{675}, 377 (2009);
R.S. Garavuso, M. Kreuzer, and A. Noll JHEP \textbf{0903}, 007 (2009);
M.C. Brambilla, Rev. Mat. Compl. \textbf{22}, 1 (2009);
F. Witt, Rend. Sem. Mat. Univ. Politec. Torino \textbf{66}, 1 (2008);
K. Ueda and M. Yamazaki, JHEP \textbf{0812}, 045 (2008);
V. Braun, T. Brelidze, M.R. Douglas, and B.A. Ovrut, JHEP\textbf{0807}, 120 (2008); JHEP \textbf{0805}, 080 (2008);
A. Ricco, Int. J. Mod. Phys. A \textbf{23}, 2187 (2008);
L. Covi, M. Gomez-Reino, C. Gross, J. Louis, G.A. Palma, and C.A. Scrucca JHEP \textbf{0806}, 057 (2008);
S. McReynolds, Mod. Phys. Lett. A \textbf{23}, 1841 (2008);
S. Mizoguchi, Phys. Lett. B \textbf{669}, 352 (2008);
N. Halmagyi and T. Okuda, JHEP \textbf{0803}, 028 (2008);
M. Cvetic and T. Weigand, Phys. Rev. Lett. \textbf{100}, 251601 (2008);
M. Cicoli, J.P. Conlon, and F. Quevedo, JHEP \textbf{0801}, 052 (2008);
A. Misra and P. Shukla, Nucl. Phys. B \textbf{799}, 165 (2008);
M.R. Douglas, R.L. Karp, S. Lukic, and R. Reinbacher, J. Math. Phys. \textbf{49}, 032302 (2008);
C. Doran, B. Greene, S. Judes, Comm. Math. Phys. \textbf{280}, 675 (2008);
L. Grant and K. Narayan, Class. Quant. Grav. \textbf{25}, 045010 (2008);
V. Tosatti, B. Weinkove, S.-T. Yau, Proc. London Math. Soc. \textbf{97}, 401 (2008);
A. Klemm and R. Pandharipande, Comm. Math. Phys. \textbf{281}, 621 (2008);
V. Bouchard and R. Donagi, Comm. Numb. Theor. Phys. \textbf{2}, 1 (2008);
A. Klemm and M. Marino, Comm. Math. Phys. \textbf{280}, 27 (2008);
H. Fang, Z. Lu, and K.-I. Yoshikawa, J. Diff. Geom. \textbf{80}(2), 175 (2008);
O. Iyama and I. Reiten, Am. J. Math. \textbf{130}(4), 1087 (2008);
S. Cynk and C. Meyer, Rocky Mountain J. Math. \textbf{38}(6), 1937 (2008); Int. J. Math. \textbf{18}(3), 331 (2007);
P. Grange and S. Schafer-Nameki JHEP \textbf{0710}, 052 (2007);
O.A.P. Mac Conamhna, Phys. Rev. D \textbf{76}, 106010 (2007);
R. Sriharsha, JHEP \textbf{0703}, 095 (2007);
D. Conti, J. Geom. Phys. \textbf{57}(12), 2483 (2007);
J. Gomis and T. Okuda, JHEP \textbf{0707}, 005 (2007); JHEP \textbf{0702}, 083 (2007);
A. Fayyazuddin, Class. Quant. Grav. \textbf{24}, 3151 (2007);
D.Tsimpis, JHEP \textbf{0703}, 099 (2007);
R. D'Auria, S. Ferrara, and M. Trigiante, Nucl. Phys. B \textbf{780}, 28 (2007);
W. Chen, M. Cvetic, H. Lu, C.N. Pope, and J.F. Vazquez-Poritz, Nucl. Phys. B \textbf{785}, 74 (2007);
B. Andreas and G. Curio, J. Geom. Phys. \textbf{57}, 2249 (2007);
R. Ahl Laamara, A. Belhaj, L.B. Drissi, and E.H. Saidi, Nucl. Phys. B \textbf{776}, 287 (2007);
P.S. Aspinwall, J. Math. Phys. \textbf{48}, 082304 (2007);
J.P. Conlon, D. Cremades, and F. Quevedo, JHEP \textbf{0701}, 022 (2007);
D.-E. Diaconescu, A. Garcia-Raboso, R.L. Karp, and K. Sinha Adv. Theor. Math. Phys. \textbf{11}, 471 (2007);
S. Bellucci, S. Krivonos, and A. Shcherbakov, Phys. Lett. B \textbf{645}, 299 (2007);
T. Oota and Y. Yasui, Phys. Lett. B \textbf{639}, 54 (2006);
R. Blumenhagen, S. Moster, and T. Weigand, Nucl. Phys. B \textbf{751}, 186 (2006);
S. Okada, Int. Math. Res. Not. \textbf{2006}, 58743 (2006);
S. Bellucci, S. Ferrara, A. Marrani, and A. Yeranyan, Riv. Nuovo Cim. \textbf{29}(5), 1 (2006);
P. Kaura and A. Misra, Fortsch. Phys. \textbf{54}, 1109 (2006);
D.-E. Diaconescu, A. Garcia-Raboso, and K. Sinha, JHEP \textbf{0606}, 058 (2006);
R. Ahl Laamara, A. Belhaj, L.B. Drissi, and E.H. Saidi, J. Phys. A \textbf{39}, 5965 (2006);
M. Aganagic, D. Jafferis, and N. Saulina, JHEP \textbf{0612}, 018 (2006);
U. Bruzzo and A. Ricco, Lett. Math. Phys. \textbf{76}, 57 (2006);
B. Forbes and M. Jinzenji, JHEP \textbf{0603}, 061 (2006);
T. Eguchi and Y. Tachikawa, JHEP \textbf{0601},100 (2006);
Z.Lu and X. Sun, Comm. Math. Phys. \textbf{261}, 297 (2006); J. Inst. Math. Jussieu \textbf{3}(2), 185 (2004);
T. Bridgeland, Comm. Math. Phys. \textbf{266}, 715 (2006);
A. Fayyazuddin, T.Z. Husain, and I. Pappa, Phys. Rev. D \textbf{73}, 126004 (2006);
N. Halmagyi, A. Sinkovics, and P. Sulkowski, JHEP \textbf{0601}, 040 (2006);
T. Pantev and E. Sharpe, Nucl. Phys. B \textbf{733}, 233 (2006);
L.N. Lipatov, A. Sabio Vera, V.N. Velizhanin, and G.G. Volkov, Int. J. Mod. Phys. A \textbf{21}, 2953 (2006);
M.B. Schulz, JHEP \textbf{0605}, 023 (2006);
M. Schuett, Canad. Math. Bull. \textbf{49}(2), 296 (2006); Coll. Math. \textbf{55}(2), 219 (2004);
J. Bryan and R. Pandharipande, Geom. Topol. Monogr. \textbf{8}, 97 (2006);
H. Jockers and J. Louis, Nucl. Phys. B \textbf{705}, 167 (2005);
T.W. Grimm, Fortsch. Phys. \textbf{53}, 1179 (2005);
H. Jockers, Fortsch. Phys. \textbf{53}, 1087 (2005);
P. Koerber, JHEP \textbf{0508}, 099 (2005);
D. Berenstein, C.P. Herzog, P. Ouyang, and S. Pinansky, JHEP \textbf{0509}, 084 (2005);
A. Bilal and S. Metzger, JHEP \textbf{0508}, 097 (2005);
V. Balasubramanian, P. Berglund, J.P. Conlon, and F. Quevedo, JHEP \textbf{0503}, 007 (2005);
T.W. Grimm and J. Louis, Nucl. Phys. B \textbf{718}, 153 (2005);
T. Kimura, Nucl. Phys. B \textbf{711}, 163 (2005);
M. Rocek and N. Wadhwa, Adv. Theor. Math. Phys. \textbf{9}, 315 (2005);
A. Belhaj, L.B. Drissi, J. Rasmussen, E.H. Saidi, and A. Sebbar, J. Phys. A \textbf{38}, 6405 (2005);
B. Andreas and D.H. Ruiperez, Adv. Theor. Math. Phys. \textbf{9}, 253 (2005);
A. Klemm, M. Kreuzer, E. Riegler, and E. Scheidegger, JHEP \textbf{0505}, 023 (2005);
T. Okuda, JHEP \textbf{0503}, 047 (2005);
M. Arai, M. Nitta, and N. Sakai, Prog. Theor. Phys. \textbf{113}, 657 (2005);
D. M\"ulsch and B. Geyer, Int. J. Geom. Meth. Mod. Phys. \textbf{2}, 409 (2005);
A. Misra and A.Nanda, Fortsch. Phys. \textbf{53}, 246 (2005);
D. Huybrechts, Int. J. Math. \textbf{16}, 13 (2005);
A. Misra, Int. J. Mod. Phys. A \textbf{20}, 2059 (2005); Fortsch. Phys. \textbf{52}, 831 (2004)
B. Banos and A. Swann, Class. Quant. Grav. \textbf{21}, 3127 (2004);
M. Hssaini, M. Kessabi, B. Marouf\/i, and M.B.Sedra, Afr. J. Math. Phys. \textbf{1}, 301 (2004);
J.P. Conlon and F. Quevedo, JHEP \textbf{0410}, 039 (2004);
U.H. Danielsson, M.E. Olsson, and M. Vonk, JHEP \textbf{0411}, 007 (2004);
V. Bouchard, B. Florea, and M. Marino, JHEP \textbf{0412}, 035 (2004);
T. Eguchi and Y. Sugawara, JHEP \textbf{0405}, 014 (2004);
T.W. Grimm and J. Louis, Nucl. Phys. B \textbf{699}, 387 (2004);
G. Volkov, Int. J. Mod. Phys. A \textbf{19}, 4835 (2004);
M. Lynker, R. Schimmrigk, and S. Stewart, Nucl. Phys. B \textbf{700}, 463 (2004);
V. Braun, B.A. Ovrut, T. Pantev, and R. Reinbacher, JHEP \textbf{0412}, 062 (2004);
M. Grana, T.W. Grimm, H. Jockers, and J. Louis, Nucl. Phys. B \textbf{690}, 21 (2004);
A. Giryavets, S. Kachru, P.K. Tripathy, and S.P. Trivedi, JHEP \textbf{0404}, 003 (2004);
H. Lu, C.N. Pope, and K.S. Stelle, JHEP \textbf{0407}, 072 (2004);
L. Jarv, T. Mohaupt, and F. Saueressig, JCAP \textbf{0402}, 012 (2004);
F. Ferrari, Adv. Theor. Math. Phys. \textbf{7}, 619 (2004);
R. Donagi, B.A. Ovrut, T. Pantev, and R. Reinbacher, JHEP \textbf{0401}, 022 (2004);
S. Govindarajan and J. Majumder, Pramana \textbf{62}, 711 (2004);
M. Grana, R. Minasian, M. Petrini, and A. Tomasiello, Compt. Ren. Phys. \textbf{5}, 979 (2004);  JHEP \textbf{0408}, 046 (2004);
A.A. Malykh, Y. Nutku, M.B. Sheftel, J. Phys. A \textbf{36}, 10023 (2003);
M.L. Barberis, Math. Phys. Anal. Geom. \textbf{6}, 1 (2003);
T. Johnsen and A.L. Knutsen, Comm. Algebra \textbf{31}(8), 3917 (2003);
L.R. Huiszoon and K. Schalm, JHEP \textbf{0311}, 019 (2003);
D. Kastor, JHEP \textbf{0307}, 040 (2003);
M.Dunajski and L.J. Mason, J. Math. Phys. \textbf{44}, 3430 (2003); Comm. Math. Phys. \textbf{213}, 641 (2000);
W.-D. Ruan, J. Sympl. Geom. \textbf{1}(3), 435 (2002);
M. Dunajski and P. Tod, Differ. Geom. Appl. \textbf{14}, 39 (2001);
M. Cvetic, G.W. Gibbons, H. Lu, and C.N. Pope, Nucl. Phys. B \textbf{617}, 151 (2001);
Z.Lu, J. Geom. Anal. \textbf{11}(1), 103 (2001); J. Geom. Anal. \textbf{11}(4), 635 (2001);
G. Grantcharov and Y.S. Poon, Comm. Math. Phys. \textbf{213}, 19 (2000);
G. Papadopoulos and J. Gutowski, Nucl. Phys. B \textbf{551}, 650 (1999);
J.P. Gauntlett, G.W. Gibbons, G. Papadopoulos, and P.K. Townsend, Nucl. Phys. B \textbf{500}, 133 (1997);
L. Rozansky and  E. Witten, Selecta Math. \textbf{3}, 401 (1997);
B. Greene, arXiv:hep-th/9702155
\bibitem{clifford}
W.K. Clifford, Amer. J. Math. \textbf{1}, 350 (1878);\\
R. Brauer and H. Weyl, Amer. J. Math. \textbf{57}, 425 (1935);\\
\'{E}. Cartan, \emph{Th\'{e}orie des spineurs} (Actualit\'{e}s Scientif\/iques et Industrielles, No. \textbf{643} et 701, Hermann, 1938);\\
C. Chevalley, \emph{The Algebraic Theory of Spinors} (Columbia University Press, 1954);\\
C.T.C. Wall, J. Reine Ang. Math. \textbf{213}, 187 (1963);\\
M.F. Atiyah, R. Bott, and A. Shapiro, Topology \textbf{3} (Suppl. 1), 3 (1964);\\
M.M. Postnikov, \emph{Lie Groups and Lie Algebras} (Mir, 1986)\\
P. Budinich and A. Trautman, \emph{The Spinorial Chessboard} (Trieste Notes in Physics, Springer, 1988)
J.E. Gilbert and M.A.M. Murray, \emph{Clifford algebras and Dirac operators in harmonic analysis} (Cambridge University Press, 1991);\\
A. Trautman and K. Trautman, J. Geom. Phys. \textbf{15}, 1 (1994);\\
I.R. Porteous, \emph{Clifford Algebras and The Classical Groups} (Cambridge Studies in Advanced Mathematics vol. \textbf{50}, Cambridge University Press, 1995);\\
A. Trautman, Contemporary Math. \textbf{203}, 3 (1997);\\
P. Lounesto, \emph{Clifford Algebras and Spinors} (2nd ed., London Mathematical Society, Lect. Notes Series \textbf{286}, Cambridge University Press, 2001);\\
V.V. Fern\'{a}ndez, A.M. Moya, and W.A. Rodrigues Jr., Adv. Appl. Clifford Alg. \textbf{11}, 1 (2001);\\
J.C. Baez, Bull. Amer. Math. Soc. (N. S.) \textbf{39}, 145 (2002);\\
A. Tarutman, in J.-P. Fran\c{c}oise, G.L. Naber and Tsou S.T. (Eds.), \emph{Encyclopedia of Mathematical Physics} (Elsevier, 2006), Vol. \textbf{1}, pp. 518–530;\\
P. Angl\`{e}s, \emph{Conformal Groups in Geometry and Spin Structures} (Birkh\"auser, 2008);\\
J. Gallier, arXiv: 0805.0311 [math.GM]
\bibitem{spec}
    Yu.A. Brychkov, \emph{Handbook of Special Functions: Derivatives, Integrals, Series, and Other Formulas} (CRC Press, 2008);\\
    S. Kanemitsu and H. Tsukada, \emph{Vistas of Special Functions} (World Scientif\/ic, 2007);\\
    G.E. Andrews, R. Askey, and R. Roy, \emph{Special Functions} (Cambridge University Press 1999);\\
    A.F. Nikiforov and V.B. Uvarov, \emph{Special Functions of Mathematical Physics: A Unif\/ied Introduction with Applications} (Birkh\"auser, 1988);\\
    Z.X. Wang and D.R. Guo, \emph{Special Functions} (World Scientif\/ic, 1989);\\
    A. Wawrzy\'nczyk, \emph{Group Representations and Special Functions} (Kluwer Academic Publishers/PWN - Polish Scientif\/ic Publishers, 1984);\\
    W.W. Bell, \emph{Special Functions for Scientists and Engineers} (D. Van Nostrand, 1968);\\
    W. Miller, Jr., \emph{Lie Theory and Special Functions} (Academic Press, 1968);\\
    N.N. Lebedev, \emph{Special Functions and Their Applications} (Prentice-Hall, 1965)
\bibitem{funan}
B.P. Rynne and M.A. Youngson, \emph{Linear Functional Analysis} (2nd ed., Springer, 2008);\\
V.N. Hansen, \emph{Functional Analysis: Entering Hilbert Space} (World Scientif\/ic, 2006);\\
K. Saxe, \emph{Beginning Functional Analysis} (Springer, 2002);\\
K. Atkinson and W. Han, \emph{Theoretical Numerical Analysis - A Functional Analysis Framework} (Springer, 2001);\\
E. Kreyszig, \emph{Introductory Functional Analysis with Applications} (John Wiley \& Sons, 1978);\\
R.F. Curtain and A.J. Pritchard, \emph{Functional Analysis in Modern Applied Mathematics} (Academic Press, 1977);\\
P. Roman, \emph{Some Modern Mathematics for Physicists and Other Outsiders. An Introduction to Algebra, Topology,  and Functional Analysis} Vols. 1 \& 2 (Pergamon Press, 1975);\\
S.K. Berberian, \emph{Lectures in Functional Analysis and Operator Theory} (Springer, 1974);\\
W. Rudin, \emph{Functional Analysis} (McGraw-Hill, 1973)
\bibitem{lebe}
H. Lebesgue, \emph{Le\c{c}ons sur l'int\'{e}graion et le recherche des fonctions primitives} (Gauthier--Villars, 1904; 2nd ed. 1928); Ann. de Toulouse (3) I, 117 (1909);\\
N. Dunford and J.T. Schwartz, \emph{Linear Operators} (Wiley, 1958)
\bibitem{ibb} I. Bia\l ynicki-Birula and Z. Bia\l ynicka-Birula, \emph{Quantum Electrodynamics} (Pergamon Press/PWN -- Polish Scientif\/ic Publishers, 1975)
\bibitem{compan}
L.V. Ahlfors, \emph{Complex Analysis. An Introduction To The Theory Of Analytic Functions Of One Complex Variable} (McGraw--Hill, 1966);\\
R. Nevanlinna and V. Paatero, \emph{Introduction to Complex Analysis} (Addison--Wesley, 1969);\\
W. Rudin, \emph{Real and Complex Analysis} (3rd ed., McGraw--Hill, 1987);\\
T.W. Gamelin, \emph{Complex Analysis} (Springer, 2001);\\
M. Ablowitz and A.S. Fokas, \emph{Complex Variables: Introduction and Applications } (2nd ed., Cambridge University Press, 2003);\\
E. Freitag and R. Busam, \emph{Complex Analysis} (Springer, 2005)
\bibitem{radial}
    A. Messiah, \emph{Quantum Mechanics} (North-Holland, 1962);\\
    L.I. Schiff, \emph{Quantum Mechanics} (3rd ed., Mc-Graw-Hill, 1968);\\
    E. Merzbacher, \emph{Quantum Mechanics} (2nd ed., John Wiley \& Sons, 1970);\\
    J.E.G. Farine, \emph{Quantum Theory of Scattering Processes} (Pergamon Press, 1973);\\
    W.O. Amrein, J.M. Jauch, and K.B. Sinha, \emph{Scattering Theory in Quantum Mechanics} (W.A. Benjamin, 1977);\\
    R.L. Liboff, \emph{Introductory Quantum Mechanics} (Addison-Wesley, 1980);\\
    A. Galindo and P. Pascual, \emph{Quantum Mechanics} (Springer, 1990);\\
    L.D. Landau and E.M. Lifshitz, \emph{Quantum Mechancis: Non-Relativistic Theory} (3rd ed., Pergamon Press, 1991);\\
    I. Bia\l ynicki-Birula, M. Cieplak, and J. Kami\'nski, \emph{Theory of Quanta} (Oxford University Press, 1992);\\
    J.J. Sakurai, \emph{Modern Quantum Mechanics} (Addison-Wesley, 1994);\\
    R. Shankar, \emph{Quantum Mechanics} (2nd ed., Plenum Press, 1994);\\
    D.J. Griff\/its, \emph{Introduction to Quantum Mechanics} (Prentice Hill, 1995);\\
    H.A. Bethe and R. Jackiw, \emph{Intermediate Quantum Mechanics} (3rd ed., Westview, 1997);\\
    W. Greiner, \emph{Quantum Mechanics: An Introduction} (4th ed., Springer, 2001);\\
    J. Schwinger, \emph{Quantum Mechanics: Symbolism of Atomic Measurements} ed. by B.-G. Englert (Springer, 2001);\\
    J.-L. Basdevant and J. Dalibard, \emph{Quantum Mechanics} (Springer, 2002);\\
    K. Gottfried and T.-M. Yan, \emph{Quantum Mechanics: Fundamentals} (2nd ed., Springer, 2003);\\
    A.C. Phillips, \emph{Introduction to Quantum Mechanics} (John Wiley \& Sons, 2003);\\
    C. Cohen-Tannoudji, B. Diu, and F. Laloe, \emph{Quantum Mechanics} (Wiley, 2006);\\
    F. Schwabl, \emph{Quantum Mechanics} (4th ed., Springer, 2007);\\
    G. Auletta, M. Fortunato, and G. Parisi, \emph{Quantum Mechanics} (Cambridge University Press, 2009);\\
    N. Zettili, \emph{Quantum Mechanics: Concepts and Applications} (John Wiley \& Sons, 2009)
\bibitem{inteq}
M. B\^{o}cher, \emph{An Introduction to the Study of Integral Equations} (Cambridge University Press, 1909);\\
D. Hilbert, \emph{Grundz\"uge einer allgemeinen Theorie der linearen Integralgleischungen} (Teubner, 1912)\\
T. Lalesco, \emph{Introduction a la th\'{e}orie des \'{e}quations int\'{e}grales. Avec une pr\'{e}face de \'{E}. Picard} (A. Hermann et Fils, 1912);\\
V. Volterra, \emph{Le\c{c}ons sur les \'{e}quations int\'{e}grales et les \'{e}quations int\'{e}gro-diff\'{e}rentielles} (Gauthier-Villars, 1913); \emph{Theory of Functionals and of Integral and Integro-Differential Equations} (Blackie \& Son, 1930);\\
A. Kneser, \emph{Die Integralgleichungen und ihre Anwendungen in der mathematischen Physik} (Vieweg, 1922)\\
H. Hochstadt, \emph{Integral Equations} (John Wiley \& Sons, 1973);\\
M.L. Krasnov, \emph{Integral Equations: Introduction to the theory} (in Russian) (Nauka, 1975);\\
A.B. Mingarelli, \emph{Volterra-Stieltjes Integral Equations and Generalized Ordinary Differential Expressions} (Lect. Notes Math. \textbf{989}, Springer-Verlag, 1983);\\
D. Porter and D.S.G. Stirling, \emph{Integral equations: A practical treatment, from spectral theory to applications} (Cambridge University Press, 1990);\\
A.D. Polyanin and A.V. Manzhirov, \emph{Handbook of Integral Equations} (2nd ed., Chapman \& Hall/CRC, 2008)
\bibitem{morozov} D.N. Zubarev, V. Morozov, and G. R\"opke, \emph{Statistical Mechanics of Nonequilibrium Processes} (John Wiley \& Sons, 1996)
\bibitem{ker} K. Huang, \emph{Statistical Mechanics} (2nd ed., John Wiley \& Sons, 1987)
\bibitem{hag} R. Hagedorn, Suppl. Nuovo Cim. \textbf{3}, 147 (1965); Suppl. Nuovo Cim. \textbf{6}, 169 (1968)
\bibitem{ajl} J. Ambj{\o}rn, J. Jurkiewicz, and R. Loll, in D. Oriti (Ed.), \emph{Approaches to Quantum Gravity. Toward a New Understanding of Space, Time, and Matter} (Cambridge University Press, 2009), pp. 341-359; Int. J. Mod. Phys. D \textbf{17}, 2515 (2009); Contemp. Phys. \textbf{47}, 103 (2006); Phys. Rev. Lett. \textbf{95}, 171301 (2005); Phys. Lett. B \textbf{607}, 205 (2005); Phys. Rev. D \textbf{72}, 064014 (2005); Phys. Lett. B \textbf{581}, 255 (2004); Phys. Rev. Lett. \textbf{93}, 131 (2004); Nucl. Phys. Proc. Suppl. \textbf{106}, 980 (2002); Nucl. Phys. B \textbf{610}, 347 (2001); Phys. Rev. D \textbf{64}, 044011 (2001); Nucl. Phys. Proc. Suppl. \textbf{94}, 689 (2001); Phys. Rev. Lett. \textbf{85}, 924 (2000)
\bibitem{amb} J. Ambjorn, A. Goerlich, J. Jurkiewicz, and R. Loll, Phys. Rev. D \textbf{78}, 063544 (2008); Phys. Rev. Lett. \textbf{100}, 091304 (2008)
\bibitem{war} B.F.L. Ward, Int. J. Mod. Phys. D \textbf{17}, 627 (2008)
\bibitem{pad} T. Padmanabhan, Phys. Rev. D \textbf{28}, 745 (1983)
\bibitem{algebra}
G. Birkhoff and S. Mac Lane, \emph{A Survey of Modern Algebra} (4th ed., Macmillan, 1977);\\
L. Childs, \emph{A Concrete Introduction to Higher Algebra} (Springer, 1979);\\
L.C. Grove, \emph{Algebra} (Academic Press, 1983);\\
A. Kurosh, \emph{Higher Algebra} (5th print., Mir Publishers, 1988);\\
S.H. Friedberg, A.J. Insel, and L.E. Spence, \emph{Linear Algebra} (2nd ed., Prentice Hall, 1989);\\
M.L. Curtis, \emph{Abstract Linear Algebra} (Springer, 1990);\\
J.A. Gallian, \emph{Contemporary Abstract Algebra} (2nd ed., D.C. Heath and Company, 1990);\\
B. Fine and G. Rosenberger, \emph{The Fundamental Theorem of Algebra} (Springer, 1997);\\
J.B. Fraleigh, \emph{A First Course in Abstract Algebra} (7th ed., Addison-Wesley, 2003);\\
T.W. Hungerford, \emph{Algebra} (Graduate Texts in Mathematics \textbf{73}, Springer, 2003);\\
D.S. Dummit and R.M. Foote, \emph{Abstract Algebra} (3rd ed., John Wiley \& Sons, 2004);\\
W.J. Gilbert and W.K. Nicholson, \emph{Modern Algebra with Applications} (2nd ed., John Wiley \& Sons, 2004);\\
S. Barnard, \emph{Higher Algebra} (Read Books, 2008)
}
\end{thebibliography}
\end{document}